\documentclass[11pt, a4paper]{article}
\usepackage{url}
\usepackage{geometry}
\usepackage{a4wide}
\usepackage{graphicx}
\usepackage{natbib}
\usepackage{enumitem}
\usepackage{hyperref}
\usepackage{setspace}
\usepackage{booktabs}
\usepackage{amsmath, amsthm, amssymb, amsfonts}
\usepackage{bm, bbm}
\usepackage[ruled, vlined]{algorithm2e}
\usepackage{xcolor}
\usepackage{pifont}
\usepackage{multirow, makecell}
\usepackage{xr}

\theoremstyle{plain}
\newtheorem{assumption}{Assumption}
\newtheorem{example}{Example}
\newtheorem{remark}{Remark}
\newtheorem{theorem}{Theorem}
\newtheorem{lemma}{Lemma}
\newtheorem{proposition}{Proposition}
      
\allowdisplaybreaks

\DeclareMathAlphabet\mathbfcal{OMS}{cmsy}{b}{n}

\newcommand{\mbf}{\mathbf}
\newcommand{\mc}{\mathcal}
\newcommand{\bmx}{\begin{bmatrix}}
\newcommand{\emx}{\end{bmatrix}}

\newcommand{\eps}{\epsilon}

\renewcommand{\l}{\left}
\renewcommand{\r}{\right}

\def\wh{\widehat}
\def\wt{\widetilde}

\newcommand{\E}[0]{\mathsf{E}}
\newcommand{\Var}[0]{\mathsf{Var}}
\newcommand{\Cov}[0]{\mathsf{Cov}}

\newcommand{\p}{\mathsf{P}}
\newcommand{\R}{\mathbb{R}}
\newcommand{\Z}{\mathbb{Z}}
\newcommand{\N}{\mathbb{N}}

\newcommand{\iid}{\text{\upshape iid}}
\newcommand{\nn}{\nonumber}
\newcommand{\trunc}[0]{\dagger}
\newcommand{\mat}{\mathsf{mat}}
\newcommand{\vecop}{\mathsf{vec}}

\newcommand{\kk}{{(k)}}
\newcommand{\ii}{{[\iota]}}

\newcommand{\cgr}{\color{black}}

\DeclareFontFamily{U}{mathx}{}
\DeclareFontShape{U}{mathx}{m}{n}{<-> mathx10}{}
\DeclareSymbolFont{mathx}{U}{mathx}{m}{n}
\DeclareMathAccent{\wc}{0}{mathx}{"71}

\graphicspath{{./figs/}}

\title{Tail-robust factor modelling of vector and tensor time series in high dimensions}
\author{Matteo Barigozzi$^1$ and Haeran Cho$^2$ and Hyeyoung Maeng$^3$}

\begin{document}

\maketitle

\footnotetext[1]{Department of Economics, Universit\`a di Bologna.
%, Piazza Scaravilli 2, 40126, Bologna, Italy. 
Email: \url{matteo.barigozzi@unibo.it}. Supported by MIUR (PRIN2022, Grant 2022H2STF2\_001).}

\footnotetext[2]{School of Mathematics, University of Bristol.
Email: \url{haeran.cho@bristol.ac.uk}. Partially supported by Engineering and Physical Sciences Research Council (EP/Z531327/1).}

\footnotetext[3]{Department of Mathematical Sciences, Durham University.
Email: \url{hyeyoung.maeng@durham.ac.uk}.}

%%%%%%%%%%%%%%%%%%%%%%%%%%%%%%%%%%%%%%%%%%%%

\begin{abstract}
We study the problem of factor modelling vector- and tensor-valued time series in the presence of heavy tails in the data, which produce extreme observations with non-negligible probability. We propose to combine a two-step procedure for tensor decomposition with data truncation, which is easy to implement and does not require an iterative search for a numerical solution. Departing away from the light-tail assumptions often adopted in the time series factor modelling literature, we derive the consistency and asymptotic normality of the proposed estimators while assuming the existence of the $(2 + 2\epsilon)$-th moment only for some $\epsilon \in (0, 1)$. Our rates explicitly depend on $\eps$ characterising the effect of heavy tails, and on the chosen level of truncation. We also propose a consistent criterion for determining the number of factors. Simulation studies and applications to two macroeconomic datasets demonstrate the good performance of the proposed estimators.
\end{abstract}

\noindent%
{\it Keywords:} tensor factor model, heavy tail, high dimensionality, robustness

\section{Introduction}

Factor modelling is a popular approach to dimension reduction in high-dimensional time series analysis. 
It has been successfully applied to various tasks involving large panels of time series, from forecasting macroeconomic variables \citep{stock2002forecasting}, to building low-dimensional indicators of the whole economic activity \citep{stock2002macroeconomic}.
With rapid technological developments, tensor (multi-dimensional array) time series datasets are routinely collected, which are large in volume as well as being of high dimensions. 
Consequently, the demand is greater than ever for learning a parsimonious yet flexible representation that preserves the array structure of the original data, driving a renewed in interest in factor modelling of tensor-valued time series.

Characterisation of a low-rank object for tensor data is not trivial, as extensions of the matrix singular value decomposition are not unique for tensors of higher order.
Based on the two popular tensor decompositions \citep{kolda2009tensor}, two approaches exist to tensor factor analysis, Tucker factor models and its special case, CanDecomp/PARAFAC (CP, \citeauthor{han2022tensor}, \citeyear{han2022tensor}; \citeauthor{chang2023modelling}, \citeyear{chang2023modelling}).
We focus on the former which, having gained vast popularity in the literature on high-dimensional statistics, has been adopted for tensor factor modelling under the assumptions of factors accounting for pervasive serial \citep{wang2019factor, chen2022factor,han2020tensor} or cross-sectional \citep{yu2022projected, barigozzi2022statistical, chen2023statistical, he2023matrix, chen2024rank,zhang2022tucker} dependence. 
Various estimation methods of Tucker tensor factor models exist which typically exploit a large gap in the eigenvalues of second moment matrices that is present between the eigenvalues attributed to the factors and the remainder. 
As such, they naturally involve the principal component (PC) analysis and in turn, their theoretical properties have been studied assuming the existence of the fourth moment in the data.

Datasets exhibiting tail behaviour that does not warrant such fourth moment conditions, are frequently observed in economics and finance \citep{cont2001empirical, ibragimov2015heavy}, neuroscience \citep{eklund2016cluster} and genomics \citep{purdom2005error}, to name a few. 
In fact, heavy-tailedness is one of the stylised features of high-dimensional data which may arise from the increased chance of extreme events, or the complexity of the data generating mechanism \citep{fan2021shrinkage}.
It is well-documented that sample estimators for the second moments are highly sensitive to anomalous observations and consequently, the PC-based methods suffer from the heavy-tailedness \citep{kristensen2014factor}.
% and it is well-documented that many datasets exhibit heavy tails \citep{ke2019user}. Heavy-tailed distribution is a viable model for data contaminated by outliers that are typically encountered in applications. Due to heavy-tailedness, the probability that some observations are sampled far away from the `true' parameter of the population is non-negligible. We refer to these outlying data points as stochastic outliers. A procedure that is robust against such outliers, evidenced by its better finite-sample performance than a non-robust method, is called a tail-robust procedure.

Despite the growing interest in the factor modelling of tensor time series, few papers address the problem of tail-robust estimation, i.e.\ estimation of factor structures in the absence of a fourth moment, by means of the Huber loss \citep{he2023robust, he2023huber, wang2023adaptively, barigozzi2023robust}.
% Existing efforts in this direction include \cite{he2023robust}, \cite{barigozzi2023robust} and \cite{wang2023adaptively} where, noting the connection between PCA and least squares estimation, the problem is cast as that of $M$-estimation based on Huber loss.
% They propose numerical solutions to the $M$-estimation problems due to the multilinear nature of the factor decomposition, and investigate the theoretical properties of the proposed estimators under the restrictive assumption imposing serial independence.
In this paper, we study the problem of tail-robust estimation of a Tucker-type factor model for tensor-valued time series, where the factor tensor accounts for pervasive cross-sectional dependence.
A key ingredient of the proposed method is a data truncation step combined with a projection-based iterative procedure for Tucker decomposition of the tensor data which, while computationally and conceptually simple, leads to the estimators that are robust to the presence of heavy tails. 
Throughout, we address the more challenging case of tensor-valued time series but the proposed methodology is directly applicable to vector-valued time series corresponding to a trivial, single-mode tensor. 

In Table~\ref{tab:comparison}, we provide a summary of the theoretical and computational aspects of our proposal in comparison with the existing ones. 
Theoretically, our contributions are two-fold and, to the best of our knowledge, are new to the literature. 
First, we characterise the tail behaviour of the data under a weak moment assumption on the existence of the $(2 + 2\epsilon)$-th moment for some $\epsilon \in (0, 1)$, through which the effect of heavy tails on the choice of the truncation parameter and the rates of estimation is made explicit; in particular, as $\epsilon \to 1$, our rates nearly match the best rate known in light-tailed situations (namely $\min(\sqrt{np_{-k}}, p)$, see the caption of Table~\ref{tab:comparison}).
This is further complemented by the asymptotic normality of the proposed estimator, also first in its kind in the tail-robust tensor factor modelling literature. 
Second, provided that the cross-sections are suitably ordered, our theoretical analysis permits both serial and spatial dependence in the idiosyncratic component within a framework of strongly mixing random field, moving away from the prevalent approach in robust factor modelling that assumes serial and spatial independence.
Computationally, our estimators require at most two iterations without any numerical optimisation and thus are straightforward to compute. 
% They propose numerical solutions to the $M$-estimation problems due to the multilinear nature of the factor decomposition, and investigate the theoretical properties of the proposed estimators under the restrictive assumption imposing serial independence.

\begin{table}[t!]
\caption{Comparison of Tucker factor modelling methods for matrix- and tensor-valued time series, where factors drive pervasive cross-sectional dependence. 
For each, we state the rate of estimation attainable for mode-$k$ loading space and the conditions on the permitted order of the tensor-valued time series ($K$), whether temporal and/or spatial dependence is allowed on the idiosyncratic component (see Assumptions~\ref{assum:indep} and~\ref{assum:rf}) and the assumption on the tail behaviour; specifically, the `tail' column reports the considered range of the largest $\nu$ such that the data have a finite $\nu$-th (centered) moment.
Additionally we give the number of required iterations, where `$\infty$' means that the specific number of iterations has not been given.
Here, $n$ denotes the sample size, $p_k$ the dimension of the $k$-th mode, $p = \prod_{k = 1}^K p_k$ and $p_{-k} = p/p_k$, and $M_n$ denotes an upper bound on the factor elements, see Assumption~\ref{assum:heavy}~\ref{cond:factor:bound}.
}
\vskip .1cm
\label{tab:comparison}
\resizebox{\textwidth}{!}{
\footnotesize
\begin{tabular}{l c c c c c c}
\toprule
Method & $K$ & Tail & Temporal & Spatial & Rate of estimation & Iteration \#  \\
\cmidrule(lr){1-1} \cmidrule(lr){2-2} \cmidrule(lr){3-3} \cmidrule(lr){4-4} \cmidrule(lr){5-5} \cmidrule(lr){6-6} \cmidrule(lr){7-7} 
\multirow{2}{*}{This paper} & \multirow{2}{*}{$\ge 1$} & \multirowcell{2}{$2 + 2\eps$ \\ $\eps\in (0, 1)$} & \ding{52} & \ding{56} & $\tfrac{M_n^{1 - \eps}}{\sqrt{np_{-k}}} \vee \tfrac{1}{p}$ & 2 % & \ding{52} & \ding{56}
\\
 & & & \ding{52} & \ding{52} & $\tfrac{M_n^{1 - \eps}}{\sqrt{np_{-k}}} \vee \l( \tfrac{\log^K(np)}{np_{-k}} \r)^{\frac{\eps}{1 + \eps}} \vee \tfrac{1}{p}$ & 1 % & \ding{52} & \ding{56}
\\
\cmidrule(lr){1-1} \cmidrule(lr){2-2} \cmidrule(lr){3-3} \cmidrule(lr){4-4} \cmidrule(lr){5-5} \cmidrule(lr){6-6} \cmidrule(lr){7-7} 
\cite{wang2023adaptively} & $2$ & $(2, \infty)$ & \ding{56} & \ding{56} & $\frac{1}{\sqrt{n p_{-k}}} \vee \frac{1}{\sqrt{n p_{k}}} \vee \frac{1}{\sqrt p}$ & $\infty$  % & \ding{52}  & \ding{52}
\\
\cite{he2023robust} & 2 & $(2, \infty)$ & \ding{56} & \ding{56} & $\frac{1}{\sqrt{n p_{-k}}} \vee \frac{1}{\sqrt{n p_{k}}} \vee \frac{1}{\sqrt p}$ & $\infty$ % \ding{52} 
\\
\cite{barigozzi2023robust} & $\ge 2$ & $(2, 4)$ & \ding{56} & \ding{56} & $\min_{k' \in [K]} \frac{1}{\sqrt{p_{-k'}}}$ & $\infty$ % & \ding{52}  & \ding{52}
\\
\cmidrule(lr){1-1} \cmidrule(lr){2-2} \cmidrule(lr){3-3} \cmidrule(lr){4-4} \cmidrule(lr){5-5} \cmidrule(lr){6-6} \cmidrule(lr){7-7} 
\cite{chen2023statistical} & 2 & $[8, \infty)$ & \ding{52} & \ding{52} & $\frac{1}{\sqrt{np_{-k}}} \vee \frac{1}{\sqrt{p_k}}$ & 0 % & \ding{56}  & \ding{56}
\\
\cite{yu2022projected} & 2 & $[8, \infty)$ & \ding{52} & \ding{52} & $\frac{1}{\sqrt{n p_{-k}}}  \vee \frac{1}{np_k} \vee \frac{1}{p}$ & 1
 % & \ding{52} & \ding{52}
\\
\cite{barigozzi2022statistical} & $\ge 1$ & $[4, \infty)$ & \ding{52} & \ding{52} & $\frac{1}{\sqrt{n p_{-k}}}  \vee \frac{1}{np_k} \vee \frac{1}{p}$ & $1$
 % & \ding{52} & \ding{52}
\\
\cite{zhang2022tucker} & $\ge 1$ & $[8, \infty)$ & \ding{52} & \ding{52} & $\frac{1}{\sqrt{n p_{-k}}} \vee \frac{1}{p}$ & $\infty$  % & \ding{52} & \ding{52}
\\
\cite{chen2024rank} & $\ge 2$ & $[4, \infty)$ & \ding{52} & \ding{52} & $\frac{1}{\sqrt{np_{-k}}} \vee \frac{1}{p} \vee \frac{1}{n}$ & $\infty$ % & \ding{52} & \ding{52}
\\
\bottomrule 
\end{tabular}}
\end{table}

% The arguments given in \cite{chen2024rank} are somewhat limited; they focus on the situation where $n \asymp p_k$. The minimax optimality discussed in their Section~3.6 is also in this setting, and it is limited to applying the existing lower bound for the eigenvector of a square matrix from the pre-averaging step. I believe that often you cannot improve the bound obtained from Davis-Kahan theorem, and the rate $\frac{1}{\sqrt{n p_{-k}}} \vee \frac{1}{p}$ derived with the known projection direction would be the `optimal' rate. In their response to referee's reports, they argue that if signs of loadings are homogeneous, their assumptions are met.

We briefly mention alternative robust approaches to time series factor modelling where the aims are related yet distinct from ours.
There are procedures designed to attain a high breakdown point under Huber's contamination model \citep{huber1964}, see, e.g.\ \cite{baragona2007outliers}, \cite{alonso2020robust} and \cite{trucios2021robustness} in the context of time series factor modelling, and \cite{maronna2008robust}, \cite{she2016robust} and \cite{pena2016generalized} on robust PCA.
Methods for quantile factor modelling \citep{chen2021quantile, he2022quantile} enable the estimation of quantile-dependent factors; in particular, at the quantile level $0.5$, these methods can be considered as a form of robust factor analysis.
% Chen et al. (2021) proposed the Quantile Factor Model (QFM) for extracting quantile-dependent factors, and the corresponding estimation procedure at quantile level $\tau = 0.5$ can be considered as a form of robust factor analysis, denoted by QFA (Quantile Factor Analysis) in this paper. He et al. (2020) provided a theoretical analysis of the iterative estimators, whereas Chen et al. (2021) focused on the theoretical minimizers. Neither of these methods requires moment conditions on the idiosyncratic errors. Compared to QFM, the mean factor model is more suitable for practical financial problems such as portfolio selection, as the Mean-Variance (MV) framework introduced by Markowitz (1952) forms the foundation of modern portfolio theory.
There are also papers on elliptical factor models \citep{han2018eca, fan2018large, he2022large, qiu2023robust} where scatter matrices such as Kendall's tau or Spearman correlation matrices, are employed for the estimation of factor structures. 
% han2018eca second moment, fan2018large fourth moment, he2022large no moment
% \cite{bai2022robust} also replaces sample (auto)covariance with its robust counterpart based on robust scale estimator, operate under the factor model of \cite{lam2012} where dominant serial correlations are captured by factors, consistency of autocovariance matrices rather than establishing this under some moment assumptions with an assumption that $pn^{-1/2} \to 0$. 
All the aforementioned papers consider vector time series factor modelling with the exception of \cite{he2025new}. 
% On the other hand, we set out to achieve tail-robustness in time series factor modelling while thoroughly accounting for the multi-array structure of the data, and fully specify the effect of the tail behaviour on the performance of the proposed estimators.

%\paragraph{Organisation of the paper.}
% TO BE COMMENTED FOR BIOMETRIKA
In what follows, Section~\ref{sec:model} introduces the tensor factor model. 
Section~\ref{sec:method} describes the suite of methods for estimating the mode-wise loadings, factors and the numbers of factors, and Section~\ref{sec:theor} establishes the theoretical consistency and asymptotic normality of the proposed estimators. Through numerical experiments on simulated (Section~\ref{sec:sim}) and a macroeconomic dataset (Section~\ref{sec:euro}), we demonstrate the competitive performance of the proposed methods. Proofs of all theoretical results and complete numerical results including an additional real data application, are given in a Supplementary Appendix. An implementation of the proposed methods is available at \url{https://github.com/haeran-cho/robustTFM}.
\vspace{10pt}

\textbf{Notations.} 
We write $[n] = \{1, \ldots, n\}$ for any positive integer $n$.
For a random variable $X$ and $\nu \ge 1$, we write $\Vert X \Vert_\nu = \{\E(\vert X \vert^\nu) \}^{1/\nu}$.
%For a vector $\mbf a = (a_1, \ldots, a_p)^\top$,
%we denote its $\ell_1$- and $\ell_2$-norm by 
%$\vert \mbf a \vert_1 = \sum_{i = 1}^p \vert a_i \vert$
%and $\vert \mbf a \vert_2 = \sqrt{\sum_{i = 1}^p a_i^2}$.
For a matrix $\mbf A = [a_{ii'}, \, i \in [m], \, i' \in [n]] \in\mathbb R^{m\times n}$, 
we denote by $\mbf A^\top$ its transpose, and by $\mbf A_{i \cdot}$ and $\mbf A_{\cdot i}$ the $i$-th row and column vectors.
We write $\vert \mbf A \vert_2 = \sqrt{\sum_{i \in [m]} \sum_{i' \in [n]} \vert a_{ii'} \vert^2}$ and denote its spectral norm by $\Vert \mbf A \Vert$. By $\mbf I$, we denote an identity matrix. 
For $\mc X = [X_{i_1 \ldots i_K}, \, i_k\in[p_k],\, k\in[K] ]\in \R^{p_1 \times \cdots \times p_K}$,
%$\mc X = [X_{i_1 \ldots i_K}] \in \R^{p_1 \times \cdots \times p_K}$ \footnote{\matt{in this case we do not write the range of $i_k$ as we did for the matrix, so either we add it here and then in the following we always use such notation (while now we do only sometime) or we only indicate $\in\mathbb R^{p_k etc}$ and we remove the more detailed notation from the matrix notation above and in the following.}} be an order-$K$ tensor with $p = \prod_{k = 1}^K p_k$ and $p_{-k} = p / p_k$.
its mode-$k$ unfolding matrix denoted by $\mat_k(\mc X)$, % \in \R^{p_k \times p_{-k}}$ 
is the $p_k \times p_{-k}$-matrix arranging all the $p_{-k}$ mode-$k$ fibers of $\mc X$ in its columns.
We write the mode-$k$ product of $\mc X$ with an $m \times p_k$-matrix $\mbf A = [a_{ij}]$ by $\mc X \times_k \mbf A \in \R^{p_1 \times \ldots \times p_{k - 1} \times m \times p_{k + 1} \times \ldots \times p_K}$, which has as its $(i_1, \ldots, i_{k - 1}, j, i_{k + 1}, \ldots, i_K)$-th element given by $\sum_{i_k = 1}^{p_k} X_{i_1 \ldots i_k \ldots i_K} a_{j i_k}$.
We denote the vectorisation of $\mc X$ as $\vecop(\mc X) \in \R^p$, which stacks the columns of $\mat_1(\mc X)$, and write $\vert \mc X \vert_2 = \vert \vecop(\mc X) \vert_2$.
With $\otimes$ we denote the Kronecker product. 
% Let $\mbf A_{i \cdot}$ and $\mbf A_{\cdot k}$ denote the $i$-th row and the $k$-th column of a matrix $\mbf A$.
For two real numbers, set $a \vee b = \max(a, b)$ and $a \wedge b = \min(a, b)$.
Given two sequences $\{a_n\}$ and $\{b_n\}$, we write $a_n = O(b_n)$ if, for some finite positive constant $C$
there exists $N \in \N_0 = \N \cup \{0\}$ such that
$|a_n| |b_n|^{-1} \le C$ for all $n \ge N$, and $a_n \asymp b_n$ if $a_n = O(b_n)$ and $b_n = O(a_n)$.
We write $a_n = o(b_n)$ if, for every $\varepsilon > 0$, there exists $N \in \N_0$ such that $|a_n| |b_n|^{-1} \le \varepsilon$ for all $n \ge N$.
By $O_P$ and $o_P$, we denote the probabilistic extensions of $O$ and~$o$, respectively.

\section{Tensor time series factor model}
\label{sec:model}

Under a Tucker decomposition-based factor model, a time series $\{\mc X_t\}_{t \in [n]}$ of the $K$-dimensional arrays (tensors) with $\mc X_t = [X_{i_1 \ldots i_K, t}, \, i_k \in [p_k], \, k \in [K]] \in \R^{p_1 \times \cdots \times p_K}$, satisfies
\begin{align}
\mc X_t &= \bm\chi_t + \bm\xi_t, \text{ \ where \ } \bm\chi_t = \mc F_t \times_1 \bm\Lambda_1 \times_2 \ldots \times_K \bm\Lambda_K,
\label{eq:model} 
\end{align}
with $\mc F_t = [f_{j_1 \ldots j_K, t}, \, j_k \in [r_k], \, k \in [r_k]] \in \R^{r_1 \times \cdots \times r_K}$ and $\bm\Lambda_k = [\lambda_{k, i_k j_k}, \, i_k \in [p_k], \, j_k \in [r_k]] \in \R^{p_k \times r_k}$ for some integers $r_k \ll p_k$. 
Then, the $(i_1, \ldots, i_K)$-th element of $\bm\chi_t$ is % with any $i_k \in [p_k], \, k \in [K]$, is
\begin{align*}
\chi_{i_1 \ldots i_K, t}=
\sum_{j_1 \in [r_1]} \cdots \sum_{j_K \in [r_K]} \lambda_{1, i_1 j_1} \cdots \lambda_{K, i_K j_K} f_{j_1 \ldots j_K, t}.
\end{align*}
The core tensor~$\mc F_t$ serves the role of the latent common factor with its dimensions $r_k, \, k \in [K]$, fixed and independent of $p_1, \ldots, p_K$ and $n$, and is loaded by mode-wise loading matrices $\bm\Lambda_k$ to drive pervasive cross-sectional dependence in the tensor data as specified later.
The idiosyncratic component $\bm\xi_t = [\xi_{i_1 \ldots i_K, t}]$ % \in \R^{p_1 \times \ldots \times p_K}$ 
is a tensor-valued time series with $\E(\xi_{i_1 \ldots i_K, t}) = 0$, and is permitted to be serially and cross-sectionally dependent as specified later.
%\citep{kolda2009tensor}, 
%where $\otimes$ denotes the Kronecker product {\matt move this definition above to the notation section}.
Under~\eqref{eq:model}, the mode-$k$ unfolding of $\bm\chi_t$ admits the following decomposition with $\bm\Delta_k := \otimes_{l = K, \, l \ne k}^1 \bm\Lambda_l$:
\begin{align}
\label{eq:model:unfold}
\mat_k(\bm\chi_t) = \bm\Lambda_k \mat_k(\mc F_t) \l( \bm\Lambda_K \otimes \cdots \otimes \bm\Lambda_{k + 1} \otimes \bm\Lambda_{k - 1} \otimes \cdots \otimes \bm\Lambda_1 \r)^\top  = \bm\Lambda_k \mat_k(\mc F_t) \bm\Delta_k^\top.
\end{align}

\begin{example}
\label{ex:matrix:model}
When $K = 1$, the model~\eqref{eq:model} reduces to a vector time series factor model such that $\bm\chi_t = \bm\Lambda_1 \mc F_t$ with $\mc F_t \in \R^{r_1}$, and its $i$-th element admits the decomposition $\chi_{it} = \sum_{j \in [r_1]} \lambda_{1, i j} f_{j, t}$ for any $i \in [p_1]$.
Typical examples of its application include the modelling of macroeconomic indicators of a given country (\citeauthor{mccracken2016fred}, \citeyear{mccracken2016fred}, see Appendix~\ref{sec:fredmd}). 
When $K = 2$, model~\eqref{eq:model} is a matrix time series factor model where $\bm\chi_t = \bm\Lambda_1 \mc F_t \bm\Lambda_2^\top$ with $\mc F_t\in \R^{r_1 \times r_2}$, so that $\mat_1(\bm\chi_t) = \bm\chi_t = \bm\Lambda_1 \mc F_t \bm\Lambda_2^\top$ and $\mat_2(\bm\chi_t) = \bm\chi_t^\top = \bm\Lambda_2 \mc F_t^\top \bm\Lambda_1^\top$; its $(i_1, i_2)$-th element is written as $\chi_{i_1 i_2, t} = \sum_{j_1 \in [r_1]} \sum_{j_2 \in [r_2]} \lambda_{1, i_1j_1} \lambda_{2, i_2 j_2} f_{j_1 j_2, t}$ for any $i_k \in [p_k], \, k \in [2]$.
In Section~\ref{sec:euro}, we analyse a matrix time series consisting of macroeconomic indicators (collected in mode~1, i.e.\ rows) observed in Euro Area countries (collected in mode~2, i.e.\ columns) \citep{barigozzi2024large}. 
\end{example}

\section{Tail-robust estimation}
\label{sec:method}

Various methods exist for the estimation of the low-rank approximation of tensor data under a Tucker decomposition; we refer to \cite{kolda2009tensor} and \cite{luo2023low} for an overview.
In the time series factor modelling literature, a popular approach for the estimation under~\eqref{eq:model} is to obtain a pre-estimator via higher-order singular value decomposition (HOSVD, \citealp{de2000multilinear}), followed by one or more iterations of mode-wise SVD of the data projected onto the pre-estimated loading space (a.k.a.\ higher-order orthogonal iteration, HOOI, \citealp{de2000best}), see e.g.\ \cite{yu2022projected} and \cite{he2023matrix} for the case of matrix factor modelling, which is extended to higher-order tensors by \cite{barigozzi2022statistical} and \cite{zhang2022tucker}.

To address the heavy-tail behaviour of the data, we propose to combine the projection-based iterative procedure with a data truncation step which, despite its simplicity, has not been explored in the context of tail-robust estimation of factor models of any order $K \ge 1$. 
This allows us to accurately estimate the model even in presence of extreme events, such as deep economic recessions, or the recent pandemic and wars and the subsequent related rise of inflation.
We first present the proposed methods for estimating the loadings (Section~\ref{sec:pc}) and the core tensor factor (Section~\ref{sec:core}) supposing that the factor numbers $r_k, \, k \in [K]$, are known, with the special case of matrix factor modelling discussed in Example~\ref{ex:matrix:est}.
Section~\ref{sec:r:est} presents our approach to estimating $r_k$.

\subsection{Factor loading estimation}
\label{sec:pc}

Let us write $X_{\mbf i, t} = X_{i_1 \ldots i_K, t}$ with $\mbf i = (i_1, \ldots, i_K)^\top$.
Given a truncation parameter $\tau > 0$, we denote the element-wise truncated data 
\begin{align}
\label{eq:trunc}
X^{\trunc}_{\mbf i, t}(\tau) := \mathsf{sign}(X_{\mbf i, t}) \cdot (\vert X_{\mbf i, t}\vert \wedge \tau), \quad \text{i.e.} \quad
X^{\trunc}_{\mbf i, t}(\tau) = \l\{\begin{array}{ll}
X_{\mbf i, t} & \text{if } \vert X_{\mbf i, t} \vert \le \tau, \\
\mathsf{sign}(X_{\mbf i, t}) \cdot \tau & \text{if } \vert X_{\mbf i, t} \vert > \tau,
\end{array} \r.
\end{align}
and $\mc X^{\trunc}_t(\tau) = [X^{\trunc}_{\mbf i, t}(\tau), \, \mbf i \in \prod_{k = 1}^K [p_k] ]$.
Then, the sample counterpart of the mode-$k$ second moment matrix of $\mc X_t$, defined as $\bm\Gamma^{\kk} := (np_{-k})^{-1} \sum_{t \in [n]} \E[ \mat_k(\mc X_t) \mat_k(\mc X_t)^\top ]$, is given by
\begin{align}
\label{eq:gamma:hat}
\wh{\bm\Gamma}^\kk(\tau) := \frac{1}{n p_{-k}} \sum_{t = 1}^n \mat_k(\mc X^{\trunc}_t(\tau)) \mat_k(\mc X^{\trunc}_t(\tau))^\top.
\end{align}
We denote % the eigendecomposition of $\wh{\bm\Gamma}^\kk(\tau)$ by $\wh{\bm\Gamma}^\kk(\tau) = \sum_{j = 1}^{p_k \wedge n} \wh\mu_j^\kk(\tau) \wh{\mbf e}_j^\kk(\tau) (\wh{\mbf e}_j^\kk(\tau))^\top$, where 
by $(\wh\mu_j^\kk(\tau), \wh{\mbf e}_j^\kk(\tau)), \, j \in [\min(p_k, n)]$, the pairs of eigenvalues and (normalised) eigenvectors of $\wh{\bm\Gamma}^\kk(\tau)$, where $\wh\mu^\kk_j(\tau)$ are ordered in the decreasing order.
Under~\eqref{eq:model}, for the identifiability between the latent $\bm\chi_t$ and $\bm\xi_t$ (asymptotically), and that between $\bm\Lambda_k$'s and $\mc F_t$, it is commonly assumed that $p_k^{-1} \bm\Lambda_k^\top \bm\Lambda_k = \mbf I_{r_k}$ (see Assumption~\ref{assum:loading} below), and that the entries of $\bm\xi_t$ are weakly correlated within and across the mode (see Assumptions~\ref{assum:indep}--\ref{assum:rf}).
These lead to a gap between the $r_k$ largest eigenvalues of $\bm\Gamma^\kk$ that diverge linearly in $p_k$, and the remaining ones.
Such observations motivate the choice $\wh{\bm\Lambda}_k(\tau) := \sqrt{p_k} \, \wh{\mbf E}_k(\tau) = \sqrt{p_k} [\wh{\mbf e}_j^\kk(\tau), \, 1 \le j \le r_k]$, as an (initial) estimator of $\bm\Lambda_k$, which amounts to HOSVD performed on the truncated data matrix.
For vector time series ($K = 1$), we regard $\wh{\bm\Lambda}_1(\tau)$ as the final~estimator.

For $K \ge 2$, the representation in~\eqref{eq:model:unfold} for the mode-$k$ unfolding of $\bm\chi_t$, suggests that further refinement can be achieved by projecting $\mat_k(\mc X^{\trunc}_t(\tau))$ onto the column space of $\bm\Delta_k$ followed by SVD. 
This leads to the following iterative estimator for some $\iota \ge 1$, 
\begin{align}
\label{eq:lambda:check}
\wc{\bm\Lambda}^\ii_k(\tau) := \sqrt{p_k} \, \wc{\mbf E}^\ii_k(\tau) \text{ \ with \ } \wc{\mbf E}^\ii_k(\tau) = [\wc{\mbf e}^{\kk, \ii}_j(\tau), \, 1 \le j \le r_k],
\end{align}
where $(\wc{\mu}^{\kk, \ii}_j(\tau), \wc{\mbf e}^{\kk, \ii}_j(\tau)), \, j \in [\min(p_k, n)]$, denote the pairs of eigenvalues and eigenvectors of
\begin{align}
\wc{\bm\Gamma}^{\kk, \ii}(\tau) &:= \frac{1}{np_{-k}} \sum_{t \in [n]} \mat_k(\mc X^{\trunc}_t(\tau)) \wc{\mbf D}^{[\iota - 1]}_k(\tau) (\wc{\mbf D}^{[\iota - 1]}_k(\tau))^\top \mat_k(\mc X^{\trunc}_t(\tau))^\top, % \quad \text{with}
\label{eq:gamma:wc}
%\\
%\wh{\mbf D}_k(\tau) &:= \wh{\mbf E}_K(\tau) \otimes \cdots \otimes \wh{\mbf E}_{k + 1}(\tau) \otimes \wh{\mbf E}_{k - 1}(\tau) \otimes \cdots \otimes \wh{\mbf E}_1(\tau).
%\nn
\end{align}
with $\wc{\mbf D}^\ii_k(\tau) := \wc{\mbf E}^\ii_K(\tau) \otimes \cdots \otimes \wc{\mbf E}^\ii_{k + 1}(\tau) \otimes \wc{\mbf E}^\ii_{k - 1}(\tau) \otimes \cdots \otimes \wc{\mbf E}^\ii_1(\tau)$; for convenience, we sometimes write the initial estimator as $\wh{\mbf E}_k(\tau) = \wc{\mbf E}^{[0]}_k(\tau)$. 

Algorithm~\ref{algo:lambda} summarises the steps for the estimation of factor loadings.
Our theoretical investigation shows that at most $\iota = 2$ iterations are sufficient for the resultant estimator to achieve asymptotic normality with a rate comparable to those attainable under stronger moment conditions.
If $p_1 = \ldots = p_k = p_0 \asymp n$ and $r_1 = \ldots = r_K = r$, the overall cost to obtain the twice-iterated estimator for given $\tau$ is $O(np_0^{K + 1} + 2 K nr p_0^K)$ \citep{luo2021sharp}.
Section~\ref{sec:tuning} describes a cross validation (CV) procedure for the selection of $\tau$.

\begin{algorithm}[h!t!b!]
\label{algo:lambda}
\DontPrintSemicolon
\KwIn{ Data $\{\mc X_t\}_{t \in [n]}$, factor numbers $r_k, \, k \in [K]$, truncation parameter $\tau$ \;}
\BlankLine

Perform data truncation and obtain $\mc X^\trunc_t(\tau) = [ X^\trunc_{\mbf i, t}(\tau)]$ \;

\For{$k \in [K]$}{
	\BlankLine 
	
	Compute $\wh{\bm\Gamma}^\kk(\tau)$ using $\mat_k(\mc X^{\trunc}_t(\tau))$ as in~\eqref{eq:gamma:hat} \;

	Obtain $\wh{\mbf E}_k(\tau) = \wc{\mbf E}^{[0]}_k(\tau) = [\wh{\mbf e}^\kk_j(\tau), \, j \in [r_k]]$, from the SVD of $\wh{\bm\Gamma}^\kk(\tau)$ \;
}

Set $\iota \leftarrow 1$ \;

\While{$\iota \le 2$}{
\For{$k \in [K]$}{
	\BlankLine 
	
	Compute $\wc{\bm\Gamma}^{\kk, \ii}(\tau)$ using $\mat_k(\mc X^{\trunc}_t(\tau)) \wc{\mbf D}^{[\iota - 1]}_k(\tau)$ as in~\eqref{eq:gamma:wc} \;

	Obtain $\wc{\bm\Lambda}^\ii_k(\tau) = \sqrt{p_k} [\wc{\mbf e}^{\kk, \ii}_j(\tau), \, j \in [r_k]]$ from the SVD of~$\wc{\bm\Gamma}^{\kk, \ii}(\tau)$ \;
    }
    Set $\iota \leftarrow \iota + 1$
}

\KwOut{$\wc{\bm\Lambda}^\ii_k(\tau), \, k \in [K], \, \iota \in \{0, 1, 2\}$}
\caption{Projection-based iterative estimation of $\bm\Lambda_k, \, k \in [K]$.}
\end{algorithm}

\begin{remark}
\label{rem:methods} 
\upshape{
We propose to truncate the elements of $\mc X_t$ in order to lessen the influence of extreme observations on the estimators of the second moment matrices.
In the literature on high-dimensional vector time series analysis, \cite{wang2022rate} study the estimation of (auto)covariance matrices and show the near-minimax optimality for the truncation-based estimator (see also Remark~\ref{rem:one} below). 
\cite{ke2019user} and \cite{fan2021shrinkage} consider spectrum-wise truncation estimators that truncate $\ell_2$- or $\ell_4$-norms of the random vectors; their theoretical consistency is established under a fourth moment condition.
Alternatively to the proposed truncation, which is applied symmetrically around zero to each $X_{\mbf i, t}$, we may adopt winsorisation or element-wise Huber regression; we refer to \cite{zhang2021robust} where the consistency of the latter estimator is established under a condition comparable to Assumption~\ref{assum:heavy} below.}
Unlike in the $M$-estimation framework, the truncation step in~\eqref{eq:trunc} allows for explicitly decomposing the estimation errors which % ensures that $\wh{\bm\Gamma}^{\kk}(\tau)$ is non-negative definite, which may be useful for downstream inferential tasks.
facilitates the theoretical analysis of our proposed estimators.
% This $M$-estimation approach has the advantage over element-wise truncation since the former truncates symmetrically around the true expectation, whereas the latter do so around zero. note that the difference becomes insignificant when the sample size $n$ is large; quoted from \cite{ke2019user}. Due to smaller bias, $M$-estimators are expected to outperform the simple truncation estimators. However, since the optimal choice of the robustification parameter is often much larger than the population moments in magnitude, either element- or spectrum-wise, the difference between truncation estimators and $M$-estimators becomes insignificant when the sample size $n$ is large. Therefore, we advocate using the simple truncated estimator primarily due to its simplicity and computational efficiency
\end{remark}

% Do one more iteration: We obtain the third-stage estimator
% \begin{align}
% \label{eq:lambda:tilde}
% \wt{\bm\Lambda}_k(\tau) := \sqrt{p_k} \, \wt{\mbf E}_k(\tau) \text{ \ with \ } \wt{\mbf E}_k(\tau) = [\wt{\mbf e}^\kk_j(\tau), \, 1 \le j \le r_k],
% \end{align}
% where $(\wc{\mu}^{\kk, [2]}_j(\tau), \wt{\mbf e}^\kk_j(\tau))$ denote the pairs of eigenvalues and eigenvectors of
% \begin{align}
% \wc{\bm\Gamma}^{\kk, [2]}(\tau) &:= \frac{1}{np_{-k}} \sum_{t = 1}^n \mat_k(\mc X^{\trunc}_t(\tau)) \wc{\mbf D}^{[1]}_k(\tau) \wc{\mbf D}^{[1]}_k(\tau)^\top \mat_k(\mc X^{\trunc}_t(\tau))^\top, 
% \end{align}
% with $\wc{\mbf D}^{[1]}_k(\tau) := \wc{\mbf E}_K(\tau) \otimes \cdots \otimes \wc{\mbf E}_{k + 1}(\tau) \otimes \wc{\mbf E}_{k - 1}(\tau) \otimes \cdots \otimes \wc{\mbf E}_1(\tau)$.

\subsection{Core tensor factor estimation}
\label{sec:core}

For the estimation of the core tensor $\mc F_t$, we adopt an analogue of the PC estimator in vector time series factor modelling that takes the form of a cross-sectional weighted average of $X_{\mbf i, t}, \, \mbf i \in \prod_{k = 1}^K [p_k]$, with the weights determined by the estimators of $\bm\Lambda_k$.
Empirically, such an estimator may be sensitive to some $X_{\mbf i, t}$ taking extremely large values. %  which, due to heavy tails and high dimensionality, has a non-negligible probability of occurrence.
Therefore, we consider an estimator of $\mc F_t$ by taking the weighted average of the truncated observations with a truncation parameter $\kappa > 0$, i.e.\
\begin{align}
\label{eq:f:hat}
\wh{\mc F}_t(\tau, \kappa) := \frac{1}{p} \mc X^{\trunc}_t(\kappa) \times_1 \wc{\bm\Lambda}_1(\tau)^\top \times_2 \ldots \times_K \wc{\bm\Lambda}_K(\tau)^\top = \frac{1}{p} \mc X^{\trunc}_t(\kappa) \times_{k = 1}^K \wc{\bm\Lambda}_k(\tau)^\top,
\end{align}
where $\wc{\bm\Lambda}_k(\tau) = \wc{\bm\Lambda}^\ii_k(\tau)$ for some $\iota \in \{1, 2\}$.
Theoretically, the estimator without any additional truncation, namely $\wh{\mc F}_t(\tau) \equiv \wh{\mc F}_t(\tau, \infty)$, achieves consistency both in terms of a point-wise error or its $\ell_2$-aggregation over time, see Theorem~\ref{thm:factor} below. 
However numerically, there is strong evidence supporting the truncation of observations prior to cross-sectional aggregation, which we explore on simulated datasets in Section~\ref{sec:sim}. 

\begin{example}
\label{ex:matrix:est}
In the special case of matrix factor model ($K = 2$), the steps of estimation given in Sections~\ref{sec:pc}--\ref{sec:core} can be described without using tensor-related operators.
For the loading matrix estimation, we first perform the element-wise truncation with $\tau > 0$ as the truncation parameter, and obtain $\mc X^{\trunc}_t(\tau) = [X^{\trunc}_{ii', t}, \, i \in [p_1],\, i' \in [p_2]]$ with $X^{\trunc}_{ii', t} = \mathsf{sign}(X_{ii', t}) \cdot (\vert X_{ii', t} \vert \wedge \tau)$.
Then, we estimate the two second moment matrices of $\mc X_t$, defined as $\bm\Gamma^{(1)} = (np_2)^{-1} \sum_{t \in [n]} \E(\mc X_t \mc X_t^\top)$ and $\bm\Gamma^{(2)} = (np_1)^{-1} \sum_{t \in [n]} \E(\mc X_t^\top \mc X_t)$, by $\wh{\bm\Gamma}^{(1)}(\tau) = (np_2)^{-1} \sum_{t \in [n]} \mc X^\trunc_t(\tau) \mc X^\trunc_t(\tau)^\top$ and $\wh{\bm\Gamma}^{(2)}(\tau) = (np_1)^{-1} \sum_{t \in [n]} \mc X^\trunc_t(\tau)^\top \mc X^\trunc_t(\tau)$, respectively. From this, the initial estimator $\wh{\bm\Lambda}_1(\tau)$ of the row loadings $\bm\Lambda_1$ is obtained from the $r_1$ leading eigenvectors of $\wh{\bm\Gamma}^{(1)}(\tau)$ multiplied by $\sqrt{p_1}$, and the column loadings are estimated by $\wh{\bm\Lambda}_2(\tau)$ analogously.
Then at the $\iota$-th iteration for $\iota \in \{1, 2\}$, we project $\mc X^\trunc_t(\tau)$ onto the column space of $\wc{\bm\Lambda}^{[\iota - 1]}_2(\tau)$ (with $\wc{\bm\Lambda}^{[0]}_2(\tau) = \wh{\bm\Lambda}_2(\tau)$), compute 
$$
\wc{\bm\Gamma}^{(1), \ii}(\tau) = \frac{1}{np_2} \sum_{t \in [n]} \mc X^\trunc_t(\tau) \l( \frac{\wc{\bm\Lambda}^{[\iota - 1]}_2(\tau)}{\sqrt{p_2}} \r) \l( \frac{\wc{\bm\Lambda}^{[\iota - 1]}_2(\tau)}{\sqrt {p_2}} \r)^\top \mc X^\trunc_t(\tau)^\top
$$ 
and derive $\wc{\bm\Lambda}_1^\ii(\tau)$ as its $r_1$ leading eigenvectors multiplied by $\sqrt{p_1}$, and $\wc{\bm\Lambda}_2^\ii(\tau)$ is obtained analogously.
For estimating $\mc F_t \in \R^{r_1 \times r_2}$, we project $\mc X_t$ after truncation onto the estimated row and column loadings as $\wh{\mc F}_t(\tau, \kappa) = (p_1p_2)^{-1} \wc{\bm\Lambda}_1^\ii(\tau)^\top \mc X^\trunc_t(\kappa) \wc{\bm\Lambda}_2^\ii(\tau)$, for some $\kappa > 0$.
\end{example}

\subsection{Factor number estimation}
\label{sec:r:est}

%Factor number estimation has been extensively studied in the vector \citep{bai2002, alessi2010improved, onatski10, ahn2013, fan2022estimating} as well as in the higher-order tensor \citep{han2022rank, chen2024rank} settings.
Commonly adopted estimators of the factor number exploit the presence of a large gap in the eigenvalues of $\bm\Gamma^{\kk}$ attributed to the presence of factors, see, among others, \citet{bai2002, alessi2010improved, onatski10, ahn2013, fan2022estimating} in the vector setting and \citet{han2022rank, chen2024rank} in the tensor setting.
Here, we propose to infer $r_k, \, k \in [K]$, by screening the ratio of eigenvalues as
\begin{align*}
\wh{r}_k(\tau) := {\arg\max}_{ 1 \le j \le \bar{r}_k } \l( \wc\mu^\kk_{j + 1}(\tau) + \rho \r)^{-1} \wc\mu^\kk_j(\tau),
% \wh r_k(\tau) = \min\l\{ 1 \le j \le \min(n, p_k): \, (\wh\mu^\kk_{j + 1}(\tau))^{-1} \wh\mu^\kk_j(\tau) > \rho \r\}
\end{align*}
with some fixed $1 \le \bar{r}_k \le p_k - 1$, where for simplicity, we write $\wc{\mu}^\kk_j(\tau)$ to denote the $j$-th largest eigenvalue of the first-iteration estimator $\wc{\bm\Gamma}^{\kk, [1]}(\tau)$.
We add a small constant $\rho$ in the denominator to ensure that the ratio is well-defined. % with the eigenvalues obtained from non-negative definite matrices.
The calculation of $\wc\mu^\kk_j(\tau)$ itself requires projecting the truncated data onto the low-dimensional space spanned by the pre-estimators of $\bm\Lambda_k$.
Therefore, we adopt the approach of \cite{barigozzi2022statistical} that iteratively updates the factor number estimators and the second moment matrices of the projected data; see Algorithm~\ref{alg:r} for its full description.
Proposition~\ref{prop:fn} below shows that Algorithm~\ref{alg:r} converges after one iteration provided that $\bar{r}_k$ is chosen sufficiently large.

\begin{algorithm}[h!t!b!]
\label{alg:r}
\DontPrintSemicolon
\KwIn{ Data $\{\mc X_t\}_{t \in [n]}$, $\rho > 0$, maximum allowed factor numbers $\bar{r}_k, \, k \in [K]$, truncation parameter $\tau$, maximum number of iterations $N$ \;}
\BlankLine

\For{$k \in [K]$}{
	Obtain $(\wh\mu_k^\kk(\tau), \wh{\mbf e}^\kk_j(\tau)), \, j \ge 1$, from $\wh{\bm\Gamma}^\kk(\tau)$ computed as in~\eqref{eq:gamma:hat} \;

	Initialise $\wh r^{(0)}_k(\tau) = \bar{r}_k$ \;
}
\BlankLine

Initialise $m = 1$ \;
\BlankLine

\While{$m \le N$}{
	\For{$k \in [K]$}{
		Obtain the eigenvalues $\wc\mu^{\kk, (m)}_j(\tau), \, j \ge 1$, of $\wc{\bm\Gamma}^{\kk, (m)}(\tau) \equiv \wc{\bm\Gamma}^{\kk, (m), [1]}(\tau)$ computed as in~\eqref{eq:gamma:wc} with $\wc{\mbf E}_k^{(m), [0]}(\tau) = [\wh{\mbf e}^\kk_j(\tau), \, 1 \le j \le \wh r_k^{(m - 1)}(\tau)]$ \;
		\BlankLine 
		
		Find the ratio-based estimator
		\begin{align}
		\label{eq:r:est}
		\wh r^{(m)}_k(\tau) = {\arg\max}_{1 \le j \le \bar{r}_k} 
		\wc\mu^{\kk, (m)}_j(\tau) \l( \wc\mu^{\kk, (m)}_{j + 1}(\tau) + \rho \r)^{-1}
		\end{align}
	}
	\If{$\wh r^{(m - 1)}_k(\tau) = \wh r^{(m)}_k(\tau)$ for all $k \in [K]$}{ 
	\BlankLine 
		Set $\wh r^{(M)}_k(\tau) = \wh r^{(m)}_k(\tau), \, k \in [K]$, and $m \leftarrow M + 1$ \;
	} \lElse{
		Set $m \leftarrow m + 1$ 
	}}
\KwOut{$\wh r^{(M)}_k(\tau), \, k \in [K]$}
\caption{Iterative estimation of $r_k, \, k \in [K]$.}
\end{algorithm}

%Our proposed estimator in~\eqref{eq:r:est} is distinguished from the existing ratio-based estimators \citep{ahn2013, barigozzi2022statistical, barigozzi2023robust}.
%Specifically, $\wh{r}_k(\tau)$ is identified as the smallest index at which the eigenvalue ratio exceeds a threshold $\rho$, which needs to be selected a priori, while the existing estimators involve maximisation of the ratio over a fixed set of indices with the maximum permitted number of factors as a tuning parameter. 

% Considering the setting with $K = 1$, \cite{ahn2013} impose a specific structure on $\bm\xi_t$ that enables controlling the $\bar{r}$ largest eigenvalues of the sample covariance matrix by means of random matrix theory, provided that the dimensions satisfy $\min(n, p_1)/\max(n, p_1) \in (0, 1]$. Such an approach is not feasible in our problem due to the use of the non-linear, element-wise truncation of the data, and the factor number selection rule in~\eqref{eq:r:est} requires the selection of a tuning parameter $\pi_{n, p_k}$. On the other hand, our estimator avoids selecting an arbitrary upper bound $\bar{r}$, and we do not restrict the dimension provided that $p_k = o(\log(n))$, as discussed later below Theorem~\ref{thm:first}.

\subsection{Tuning parameter selection}
\label{sec:tuning}

We propose to select the truncation parameters $\tau$ and $\kappa$ via CV.
% We first de-mean and standardise the data using the median and the median absolute deviation (MAD) of each variable time series; we denote the thus-transformed data by $X_{\mathbf i, t}$.
First, a sequence of possible truncation parameter values are generated as $\bm{\mathfrak{t}} = \{ \mathfrak{t}_m, \, m \in [M] \}$, where $\mathfrak{t}_1 = \max_{\,\mathbf i, t} \vert X_{\mathbf i, t} \vert$, $\mathfrak{t}_M = \text{median\,}_{\mathbf i, t} \vert X_{\mathbf i, t} \vert$, and $\mathfrak{t}_m, \, 2 \le m \le M - 1$, are chosen such that $\{\log(\mathfrak{t}_m), \, m \in [M]\}$ is a sequence of equi-distanced elements.
We then partition the data into $L$ parts with the index sets $\mc I_\ell = \{\lceil n/L \rceil (\ell - 1) + 1, \ldots, \min(\lceil n/L \rceil \ell, n)\}, \, \ell \in [L]$, and compute the CV measure as
\begin{align}
\label{eq:tau:cv}
\text{CV}(\mathfrak{t}_l) := \sum_{k  = 1}^K \sum_{\ell = 1}^L \l[ 1 - \frac{1}{r_k} \text{tr}\l( \wc{\mbf E}^\ii_{-\ell, k}(\mathfrak{t}_l) (\wc{\mbf E}^\ii_{-\ell, k}(\mathfrak{t}_l))^\top  \wc{\mbf E}^\ii_{\ell, k}(\mathfrak{t}_l) (\wc{\mbf E}^\ii_{\ell, k}(\mathfrak{t}_l))^\top \r) \r],
\end{align}
where $\wc{\mbf E}^\ii_{-\ell, k}(\mathfrak{t}_l)$ (resp.\ $\wc{\mbf E}^\ii_{\ell, k}(\mathfrak{t}_l)$) denotes the estimator of $\bm\Lambda_k/\sqrt{p_k}$ in~\eqref{eq:lambda:check} obtained with $\mc X_t, \, t \in [n] \setminus \mc I_\ell$ (resp.\ $t \in \mc I_\ell$). % we recommend $\iota = 2$ in agreement with our theoretical findings. 
Then, we find $\tau_{\text{\tiny CV}} = \arg\min_{\,\mathfrak{t}_l \in \mathfrak{t}} \text{CV}(\mathfrak{t}_l)$ for the subsequent estimation steps.
The factor estimator in~\eqref{eq:f:hat} involves the sum of $p$ cross-sections of $\mc X_t$.
% unlike the mode-wise second moment estimators that involve the sum of $np_{-k}$ `observations' for each $k$. 
Therefore, the construction of a CV procedure for the selection of $\kappa$ requires partitioning the cross-sections while preserving the tensor structure, which may result in substantial data loss.
Empirically we find that setting $\kappa = \tau_{\text{\tiny CV}}$ leads to good numerical performance, and adopt this choice in all our numerical experiments which are performed with $\iota = 2$, $L = 3$ and $M = 50$.

For factor number estimation, we set $\bar{r}_k = \min(\lfloor p_k / 2 \rfloor, 20)$ and $\rho = 1 / \wc{\mu}^{\kk, [1]}_1(\tau)$.
The estimation of the factor numbers faces the additional difficulty that there is an interplay between the choice of truncation parameter $\tau$ and the choice of $r_k, \, k \in [K]$. 
In practice, we integrate the CV step to the iterative procedure outlined in Algorithm~\ref{alg:r}; specifically, we set $\tau = \mathfrak{t}_1$, the largest value from the grid, and run Algorithm~\ref{alg:r} to estimate $r_k$'s.
These estimators are fed into the CV selection and vice versa, until the estimators of $r_k$'s stabilise.
In Section~\ref{sec:euro} and Appendix~\ref{sec:fredmd}, when analysing the real datasets, we explore an approach that identifies a `region of stability' of the factor number estimators over varying truncation parameters.

\section{Theory}
\label{sec:theor}

\subsection{Assumptions}
\label{sec:assum}

We introduce the assumptions which ensure (asymptotic) identifiability of the factor model, as well as characterising the tail behaviour of $\{\mc X_t\}_{t \in [n]}$ and the dependence therein.
\begin{assumption}
\label{assum:loading}
\it{ 
For all $k \in [K]$, $\bm\Lambda_k = [\lambda_{k, ij}, \, i \in [p_k], \, j \in [r_k]]$ satisfy: (i) $p_k^{-1} \bm\Lambda_k^\top \bm\Lambda_k = \mbf I_{r_k}$ for all $p_k \ge r_k$, and (ii) $\max_{i \in [p_k]} \max_{j \in [r_k]} \vert \lambda_{k, ij} \vert \le \bar{\lambda} < \infty$.
}
\end{assumption}

\begin{assumption}
\label{assum:factor}
\it{ $\{\mc F_t\}_{t \in [n]}$ is a sequence of deterministic $K$-dimensional arrays of dimensions $r_1 \times \cdots \times r_K$ that are independent of $p_1, \ldots, p_K$ and $n$. 
For each $k \in [K]$ there exists a positive definite matrix $\bm\Gamma_f^\kk \in \R^{r_k \times r_k}$ so that $\Vert n^{-1} \sum_{t \in [n]} \mat_k(\mc F_t) \mat_k(\mc F_t)^\top - \bm\Gamma_f^\kk \Vert = o(1)$ as $n \to \infty$, and the eigenvalues of $\bm\Gamma_f^\kk$ are distinct from one another.}
\end{assumption}
Under the Tucker factor model in~\eqref{eq:model}, $\bm\Lambda_k$ and $\mc F_t$ are not identifiable in that for any invertible $\mbf R \in \R^{r_k \times r_k}$, we have $\mat_k(\bm\chi_t) = \bm\Lambda_k \mbf R \mbf R^{-1} \mat_k(\mc F_t) \bm\Delta_k$ (see~\eqref{eq:model:unfold}).
% In Assumptions~\ref{assum:loading} and \ref{assum:factor}, we impose orthonormality of the columns of $\bm\Lambda_k$ for all~$p_k$, and orthogonality of the factors for all~$n$. 
Assumptions~\ref{assum:loading}--\ref{assum:factor} are the tensor analogues of the conditions found in \citet{AR56} and \citet{bai2013principal}, and similar conditions are frequently found in the literature, see e.g.\ \citet{stock2002forecasting}, \citet{yu2022projected} and \citet{barigozzi2022statistical} in vector, matrix and tensor settings.
% Although slightly stronger than the related conditions found in e.g.\ \citet{stock2002forecasting}, \citet{yu2022projected} and \citet{barigozzi2022statistical} in vector, matrix and tensor settings, which only require the asymptotic orthogonality, Assumptions~\ref{assum:loading}~(i) is a reasonable condition often found in the classical factor analysis literature \citep{AR56}.
%{\cgr Assumptions~\ref{assum:loading}~(i) and~\ref{assum:factor} are a tensor analogue of the conditions found in \citet{bai2012statistical} and \citet{bai2013principal}.
%Compared to those conditions requiring asymptotic orthogonality of $\bm\Lambda_k$ (see
%\citet{stock2002forecasting}, \citet{yu2022projected} and \citet{barigozzi2022statistical} for related conditions in vector, matrix and tensor settings, respectively), Assumption~\ref{assum:loading} is slightly stronger; we refer to \cite{barigozzi2022estimation} for the discussion on identifiability conditions in the factor model literature.}
These assumptions ensure that our loading estimators approximate $\bm\Lambda_k$ up to an (asymptotically) orthogonal matrix, see Theorems~\ref{thm:first}--\ref{thm:third} below.
Following the classical approach in factor analysis \citep{amemiya1987asymptotic,anderson2003introduction}, we treat factors as being deterministic, see also \cite{bai2012statistical} and \citet{onatski2012asymptotics} in the context of vector-valued time series and \cite{he2023robust} and \citet{barigozzi2023robust} in matrix or tensor settings. 
This is a technical assumption adopted for handling the dependence in the data after a non-linear transform is performed to tackle the heavy-tailedness; see Remark~\ref{rem:stochastic} below for its possible relaxation.
% adopting the notion of the conditional mixingness \citep{prakasa2009conditional}.}
%\footnote{\cite{he2023huber} also treat factors as fixed in their proof and further, bounded, and $\xi_{it}$ independent over $i$ and $t$, same approach taken in the tensor setting in \cite{barigozzi2023robust}}
% The assumption of deterministic factors requires a care since $\{\mc X_t\}_{t \in \Z}$ is no longer second-order stationary even if $\{\bm\xi_t\}_{t \in \Z}$ is.

\begin{assumption}
\label{assum:heavy}
{\it There exist constants $\eps \in (0, 1)$ and $\omega, C > 0$ which satisfy:
\begin{enumerate}[wide, itemsep = 0pt, label = (\roman*)] 
\item \label{cond:heavy:idio} $\max_{\,\mbf i \in \prod_{k = 1}^K [p_k]} \max_{t \in \Z} \Vert \xi_{\mbf i, t} \Vert_{2 + 2\epsilon} \le \omega$, where $\Vert \xi_{\mbf i, t} \Vert_{2 + 2\epsilon} = \{ \E( \vert \xi_{\mbf i, t} \vert^{2 + 2\epsilon}) \}^{\frac{1}{2 + 2\epsilon}}$.

\item \label{cond:heavy:factor} For all $n \ge 1$, $n^{-1} \sum_{t = 1}^n \vert \mc F_t \vert_2^\nu \le \omega^\nu$ for $\nu \in \{1 + \epsilon, 2, 2 + 2\epsilon\}$.

\item \label{cond:factor:bound} For all $n \ge 1$, there exists $M_n > 0$ which satisfies $\max( \max_{t \in [n]} \vert \mc F_t \vert_2, \omega) \le M_n$ and may diverge with $n \to \infty$.
\end{enumerate}
}
\end{assumption}

Assumption~\ref{assum:heavy}~\ref{cond:heavy:idio} relaxes the widely found requirements of the existence of the fourth % \citep{barigozzi2022statistical, chen2024rank} 
or higher % \citep{chen2023statistical, yu2022projected, zhang2022tucker} 
moments of $\xi_{\mbf i, t}$ (see the references in Table~\ref{tab:comparison}), or even (sub-)Gaussianity \citep{chen2022factor, han2022tensor}.
Assumption~\ref{assum:heavy}~\ref{cond:heavy:factor} places an analogous condition on the factors while accommodating for their deterministic nature. 
We do not rule out the situation where $\vert \mc F_t \vert_2$ takes an extreme value at some time point, see Assumption~\ref{assum:heavy}~\ref{cond:factor:bound}, whereas Huber loss-based estimation requires the much stronger condition of bounded $\max_t \vert \mc F_t \vert_2$ \citep{he2023huber, wang2023adaptively, barigozzi2023robust}. 
% while making a comparable assumption on the tail behaviour of $\xi_{\mbf i, t}$.
The rates of estimation we derive make explicit their dependence on $\eps$, $M_n$ and $\vert \mc F_t \vert_2$, see Theorems~\ref{thm:first}--\ref{thm:factor} below.

Before presenting the condition on the cross-sectional and serial dependence in $\{\mc X_t\}_{t \in [n]}$, we introduce two definitions.
First, for two $\sigma$-algebras $\mc A$ and $\mc B$, we denote by $\alpha(\mc A, \mc B) = \sup_{A \in \mc A, B \in \mc B} \vert \p(A \cap B) - \p(A) \p(B) \vert$ the $\alpha$-mixing coefficient. 
Second, we define the semi-distance between two sets $\mc E_i \in \N^K \times \Z, \, i = 1, 2$, by 
\begin{align*}
\rho(\mc E_1, \mc E_2) = \sup\l\{ \max\l( \max_{1 \le k \le K} \vert i_k - j_k \vert, \, \vert t - u \vert \r) : \, (\mbf i, t) \in \mc E_1, \, (\mbf j, u) \in \mc E_2 \r\}.
\end{align*}
Then, we make \underline{either} of the following two assumptions on serial and cross-sectional dependence in $\bm\xi = \{\bm\xi_{\mbf i, t}, \, (\mbf i, t) \in \N^K \times \Z\}$.

\begin{assumption}
\label{assum:indep}
{\it 
\begin{enumerate}[wide, itemsep = 0pt, label = (\roman*)] 
\item \label{cond:indep:mixing} For any $\mbf i, \mbf i' \in \prod_{k = 1}^K [p_k]$, the two time series $\{\xi_{\mbf i, t}\}_{t \in \Z}$ and $\{\xi_{\mbf i', t}\}_{t \in \Z}$ are independently distributed when $\mbf i \ne \mbf i'$. Also letting $\alpha_{\mbf i}(m) = \sup_{t \in \Z} \alpha(\sigma\{\xi_{\mbf i, u}, \, u \le t\}, \sigma\{\xi_{\mbf i, v}, \, v \ge t  + m\})$ for each $\mbf i \in \prod_{k = 1}^K [p_k]$, we have $\alpha_{\mbf i}(m) \le \exp(- c_0 m)$ for all $\mbf i$, for some constant $c_0 > 0$.

\item \label{cond:indep:factor:mixing} For all $n \ge 1$ and $k \in [K]$, there exists some constant $c_\eps > 0$ such that
\begin{align*}
\frac{1}{n} \sum_{t, u \in [n]} \vert \mc F_t \vert_2^{1 + \eps} \vert \mc F_u \vert_2^{1 + \eps} \exp\l( - \frac{c_0 \eps \vert t - u \vert}{1 + \eps} \r) &\le \omega^{2 + 2\eps} \, c_\eps,
\\
\frac{1}{n} \sum_{t, u \in [n]} \vert \mc F_t \vert_2^{1 + \eps} \vert \mc F_u \vert_2^{1 + \eps} \exp\l( - \frac{c_0 \vert t - u \vert}{3 \log(np_{-k})} \r) &\le \omega^{2 + 2\eps} \, c_\eps \log(np_{-k}).
\end{align*}
\end{enumerate}
}
\end{assumption}

\begin{assumption}
\label{assum:rf}
{\it 
\begin{enumerate}[wide, itemsep = 0pt, label = (\roman*)] 
\item \label{cond:rf:mixing} Let $\bm\xi$ be a measurable random field  with its strong mixing coefficient defined as $\alpha(m, \ell_1, \ell_2) = \sup_{\mc E_1, \mc E_2} \{ \alpha(\mc S(\mc E_1), \mc S(\mc E_2)); \, \vert \mc E_i \vert \le \ell_i, \, i = 1, 2, \, \rho(\mc E_1, \mc E_2) \ge m \}$, where for each set $\mc E_i \subset \N^{K} \times \Z$, $\vert \mc E_i \vert$ denotes its cardinality and $\mc S(\mc E_i)$ is the $\sigma$-algebra generated by $\{ \bm\xi_{\mbf i, t}, \, (\mbf i, t) \in \mc E_i\}$.
Then, there exist some constant $c_0 > 0$ such that
$\alpha(m) := \alpha(m, \infty, \infty) \le \exp(- c_0 m)$ for all $m \ge 0$.

\item \label{cond:rf:factor:mixing} For all $n \ge 1$,  there exists some constant $c_\eps > 0$ such that
\begin{align*}
\frac{1}{n} \sum_{t, u \in [n]} \vert \mc F_t \vert_2^{1 + \eps} \vert \mc F_u \vert_2^{1 + \eps} \exp\l( - \frac{c_0 \eps \vert t - u \vert}{K(1 + \eps)} \r) &\le \omega^{2 + 2\eps} \, c_\eps K,
\\
\frac{1}{n} \sum_{t, u \in [n]} \vert \mc F_t \vert_2^{1 + \eps} \vert \mc F_u \vert_2^{1 + \eps} \exp\l( - \frac{c_0 \vert t - u \vert}{3K \log(np)} \r) &\le \omega^{2 + 2\eps} \, c_\eps K \log(np).
\end{align*}
\end{enumerate}
}
\end{assumption}

As seen in Table~\ref{tab:comparison}, in the existing literature, the analysis of tail-robust factor modelling methods is restricted to the case of temporal and spatial independence, % \citep{he2023huber, wang2023adaptively, barigozzi2023robust}
and weak cross-sectional assumptions are permitted only under stronger assumptions on the tail behaviour. % \citep{chen2023statistical, han2022tensor, barigozzi2022statistical}
We make a first attempt at tail-robust factor modelling of serially and cross-sectionally dependent tensor data under Assumption~\ref{assum:rf}~\ref{cond:rf:mixing} which, adopting the notion of strongly mixing random field \citep{doukhan1995mixing}, pre-supposes that across the multiple modes, the cross-sections of tensors are arranged in a meaningful order.
A natural setting for such an assumption is when the data are collected on a spatial grid, e.g.\ as in neuroimaging applications \citep{tzourio-mazoyer2002automated}. % obtained from the Alzheimer’s Disease Neuroimaging Initiative (ADNI) (\citeauthor{tzourio-mazoyer2002automated}, \citeyear{tzourio-mazoyer2002automated}, \url{https://adni.loni.usc.edu})}
Assumption~\ref{assum:indep}~\ref{cond:indep:mixing}, a special instance of Assumption~\ref{assum:rf}~\ref{cond:rf:mixing}, permits serial dependence in $\{\bm\xi_{\mbf i, t}\}_{t \in \Z}$ while imposing independence across~$\mbf i$.

\begin{remark}
\label{rem:stochastic}    
We may allow for stochastic $\mc F_t$ and relax Assumption~\ref{assum:factor} by replacing the above assumptions with those imposed on the moments of $\mc F_t$ and the conditional distribution of $\bm\xi_t$ given $\mc F := \{ \mc F_t \}_{t \in [n]}$; e.g.\ Assumption~\ref{assum:heavy} becomes (i)~$\max_{\mbf i \in \prod_{k = 1}^K [p_k]} \max_{t \in \Z} [ \E( \vert \xi_{\mbf i, t} \vert^{2 + 2\epsilon} \vert \mc F) ]^{\frac{1}{2 + 2\eps}} \le \omega$, and (ii)~$\max_{\mbf j \in \prod_{k = 1}^K [r_k]} \max_{t \in [n]} [\E( \vert \mc F_{\mbf j, t} \vert_2^{2 + 2\eps})]^{\frac{1}{2 + 2\eps}} \le \omega$.
The second conditions in Assumptions~\ref{assum:indep}--\ref{assum:rf} 
% place a strong mixing-like condition on $\{\mc F_t\}_{t \in [n]}$ while accounting for their deterministic nature; they 
are akin to the condition bounding $n^{-1} \E(\vert \sum_{t \in [n]} \vert \mc F_t \vert_2^{1 + \eps} \vert^2)$ for stochastic~$\mc F_t$, and thus can be replaced by a strong mixing condition placed on $\{\mc F_t\}_{t \in \Z}$. 
% Similarly, in place of the second conditions in Assumptions~\ref{assum:indep}-\ref{assum:rf}, we may directly impose mixingness on the stochastic $\{\mc F_t\}_{t \in \Z}$. 
\end{remark}

Under the above assumptions, the latent common and idiosyncratic components of the Tucker factor model are asymptotically identifiable as $\min(p_1, \ldots, p_K) \to \infty$, since the leading $r_k$ eigenvalues of $\bm\Gamma^\kk$ (mode-$k$ second moment matrix of $\mc X_t$) are distinct and diverge linearly in~$p_k$ while the remaining ones are bounded; we refer to Appendix~\ref{sec:identify} for the precise statement of asymptotic identifiability of $\bm\chi_t$ and $\bm\xi_t$. 

\subsection{Asymptotic properties}
\label{sec:asymp}

For investigating the theoretical properties of the proposed estimators, let us define 
\begin{align}
\tau^\kk_{n, p} &:= \l\{ \begin{array}{ll}
\omega \l( \frac{np_{-k}}{\log(np_{-k})} \r)^{\frac{1}{2 + 2\eps}} & \text{under Assumption~\ref{assum:indep}}, \\
\omega \l( \frac{np_{-k}}{\log^K(n p)} \r)^{\frac{1}{2 + 2\eps}} & \text{under Assumption~\ref{assum:rf}}, \\
\end{array}\r. 
\label{eq:tau}
\end{align} 
for each $k \in [K]$, where $\epsilon \in (0, 1)$ and $\omega > 0$ are defined in Assumption~\ref{assum:heavy}, and 
\begin{align}
\psi^\kk_{n, p} &= \l\{\begin{array}{ll}
\omega^2 \l(\frac{\log(np_{-k})}{np_{-k}} \r)^{\frac{\eps}{1 + \eps}} = (\tau^\kk_{n, p})^2 \cdot \frac{\log(np_{-k})}{np_{-k}} & \text{under Assumption~\ref{assum:indep},}
\\
\omega^2 \l(\frac{\log^K(np)}{np_{-k}} \r)^{\frac{\eps}{1 + \eps}} = (\tau^\kk_{n, p})^2 \cdot \frac{\log^K(np)}{np_{-k}} & \text{under Assumption~\ref{assum:rf}.}
\end{array}\r.
\label{eq:psi}
\end{align}
Also, let us write $\bar{\psi}^\kk_{n, p} = \sum_{k' \in [K] \setminus \{k\}} \psi^{(k')}_{n, p}$ and $\bar{\psi}_{n, p} = \sum_{k \in [K]} \psi^\kk_{n, p}$.
The following three theorems study the initial and the iteratively projected estimators of $\bm\Lambda_k$. For vector time series, Theorem~\ref{thm:first} gives the final estimation rate, while Theorems~\ref{thm:second}--\ref{thm:third} investigate the consistency and asymptotic distribution of the iterative estimators for tensor time series with $K \ge 2$.
% Theorem~\ref{thm:third} establishes the asymptotic normality of the second-iteration estimator under Assumption~\ref{assum:indep}.

\begin{theorem}[Initial loading estimator]
\label{thm:first}
{\it Suppose that Assumptions~\ref{assum:loading}, \ref{assum:factor} and \ref{assum:heavy} hold.
We set $\tau \asymp \tau^\kk_{n, p}$ for each $k \in [K]$ as in~\eqref{eq:tau}, and assume that $M_n$ in Assumption~\ref{assum:heavy}~\ref{cond:factor:bound} satisfies
\begin{align}
\label{eq:tau:M}
M_n \max_{k \in [K]} \, (\tau^{(k)}_{n, p})^{-1} \log^{\frac{1}{2\eps}}(np_{-k}) = o(1).
\end{align}
Then, there exists a matrix $\wh{\mbf H}_k \in \R^{r_k \times r_k}$ satisfying $\wh{\mbf H}_k^\top \wh{\mbf H}_k = \mbf I_{r_k} + o_P(1)$ (in the element-wise sense) such that as $\min(n, p_1, \ldots, p_K) \to \infty$,
\begin{align*}
\frac{1}{\sqrt{p_k}} \l\Vert \wh{\bm\Lambda}_k(\tau) - \bm\Lambda_k \wh{\mbf H}_k \r\Vert 
= \l\{ \begin{array}{ll}
O_P \l( \frac{M_n^{1 - \eps}}{\sqrt{np_{-k}}} \vee \frac{1}{p_k} \vee \frac{\psi^\kk_{n, p}}{\sqrt{p_k}} \r)
& \text{under Assumption~\ref{assum:indep},}
\\
O_P\l( \psi^\kk_{n, p} \vee \frac{1}{p_k} \r) & \text{under Assumption~\ref{assum:rf}.}
\end{array}\r.
\end{align*}
}
\end{theorem}

\begin{theorem}[First iteration loading estimator]
\label{thm:second}
{\it Suppose that Assumptions~\ref{assum:loading}, \ref{assum:factor} and \ref{assum:heavy} hold, we set $\tau \asymp \tau^\kk_{n, p}$ as in~\eqref{eq:tau} for each $k \in [K]$, and~\eqref{eq:tau:M} is met.
Then, there exists some $\wc{\mbf H}^{[1]}_k \in \R^{r_k \times r_k}$ satisfying $(\wc{\mbf H}^{[1]}_k)^\top \wc{\mbf H}^{[1]}_k = \mbf I_{r_k} + o_P(1)$ such that as $\min(n, p_1, \ldots, p_K) \to \infty$,
\begin{align*}
& \frac{1}{\sqrt{p_k}} \l\Vert \wc{\bm\Lambda}^{[1]}_k(\tau) - \bm\Lambda_k \wc{\mbf H}^{[1]}_k \r\Vert 
\\
= & \, \l\{ \begin{array}{ll}
O_P\l[ 
\frac{M_n^{1 - \eps}}{\sqrt{np_{-k}}} \vee \frac{1}{p} 
\vee \bar{\psi}^\kk_{n, p} \l( \frac{\psi^\kk_{n, p}}{\sqrt{p_k}} + \frac{M_n^{1 - \eps}}{\sqrt n}
\r) 
\vee \frac{\bar{\psi}_{n, p}}{\sqrt p} \r] 
& \text{under Assumption~\ref{assum:indep},}
\\
O_P\l( \frac{M_n^{1 - \eps}}{\sqrt{np_{-k}}} \vee \frac{1}{p} \vee \psi^\kk_{n, p} \vee  \frac{M_n^{1 - \eps} \bar{\psi}^\kk_{n, p}}{\sqrt n} \r) & \text{under Assumption~\ref{assum:rf}.}
\end{array}\r.
\end{align*}
% \begin{align*}
% & \frac{1}{\sqrt{p_k}} \l\Vert \wc{\bm\Lambda}^{[1]}_k(\tau) - \bm\Lambda_k \wc{\mbf H}^{[1]}_k \r\Vert 
% = 
% O_P\l[ 
% \frac{M_n^{1 - \eps}}{\sqrt{np_{-k}}} \vee \frac{1}{p} 
% \vee \bar{\psi}^\kk_{n, p} \l( \frac{\psi^\kk_{n, p}}{\sqrt{p_k}} + \frac{M_n^{1 - \eps}}{\sqrt n}
% \r) 
% \vee \frac{\bar{\psi}_{n, p}}{\sqrt p} \r] 
% \end{align*}
% under Assumption~\ref{assum:indep}, while under Assumption~\ref{assum:rf},
% \begin{align*}
% & \frac{1}{\sqrt{p_k}} \l\Vert \wc{\bm\Lambda}^{[1]}_k(\tau) - \bm\Lambda_k \wc{\mbf H}^{[1]}_k \r\Vert 
% = O_P\l( \frac{M_n^{1 - \eps}}{\sqrt{np_{-k}}} \vee \frac{1}{p} \vee \psi^\kk_{n, p} \vee  \frac{M_n^{1 - \eps} \bar{\psi}^\kk_{n, p}}{\sqrt n} \r).
% \end{align*}
}
\end{theorem}

\begin{theorem}[Second iteration loading estimator]
\label{thm:third}
{\it Suppose that Assumptions~\ref{assum:loading}, \ref{assum:factor}, \ref{assum:heavy} and~\ref{assum:indep} hold, $\tau \asymp \tau^\kk_{n, p}$ as in~\eqref{eq:tau} for each $k \in [K]$, and~\eqref{eq:tau:M} is met.
Additionally, assume that 
\begin{align}
\label{eq:np}
\frac{\psi^\kk_{n, p}}{\sqrt{p}}
\vee
\bar{\psi}^\kk_{n, p} \l( \frac{\psi^\kk_{n, p}}{\sqrt{p_k}} + \frac{M_n^{1 - \eps}}{\sqrt n} + \frac{1}{\sqrt p} \r) = O\l( \frac{M_n^{1 - \eps}}{\sqrt{np_{-k}}} \vee \frac{1}{p} \r)
\end{align}
for all $k \in [K]$, as $\min(n, p_1, \ldots, p_K) \to \infty$. 
\begin{enumerate}[wide, itemsep = 0pt, label = (\roman*)] 
\item \label{thm:third:one} There exists some $\wc{\mbf H}^{[2]}_k \in \R^{r_k \times r_k}$ satisfying $(\wc{\mbf H}^{[2]}_k)^\top \wc{\mbf H}^{[2]}_k = \mbf I_{r_k} + o_P(1)$ such that 
\begin{align*}
\frac{1}{\sqrt{p_k}} \l\Vert \wc{\bm\Lambda}^{[2]}_k(\tau) - \bm\Lambda_k \wc{\mbf H}^{[2]}_k \r\Vert = O_P\l( \frac{M_n^{1 - \eps}}{\sqrt{np_{-k}}} \vee \frac{1}{p} \r).
% = O_P\l[ \frac{M_n^{1 - \eps}}{\sqrt{np_{-k}}} \vee \frac{1}{p} 
% \vee \sum_{k' \in [K] \setminus \{k\}} \frac{M_n^{1 - \eps}}{\sqrt{np_{-k'}}} \l( \frac{M_n^{1 - \eps}}{\sqrt n} + \frac{\psi^\kk_{n, p}}{\sqrt{p_k}} \r)
% \vee \frac{\psi^\kk_{n, p}}{\sqrt{p}}
% \r].
\end{align*}

\item \label{thm:third:two} Further assume that $M_n = M \in (0, \infty)$ for all $n \ge 1$, and $\sqrt{np_{-k}} = o(p)$ for given $k \in [K]$.
Then for any $i \in [p_k]$, we have as $\min(n, p_1, \ldots, p_K) \to \infty$,
\begin{align*}
& \sqrt{np_{-k}} \l(\wc{\bm\Lambda}^{[2]}_{k, i\cdot}(\tau) - \bm\Lambda_{k, i\cdot} \wc{\mbf H}^{[2]}_k \r)^\top \to \mc N_{r_k}\l( \mbf 0, \bm\Phi^\kk_i(\tau) \r), \text{ \ where}
\\
& \bm\Phi^\kk_i(\tau) :=(\bm\Gamma^\kk_f)^{-1} \l( \frac{1}{np_{-k}} \sum_{t, u \in [n]}  \mat_k(\mc F_t) \bm\Delta_k^\top \bm\Psi^\kk_{i, tu}(\tau) \bm\Delta_k \mat_k(\mc F_u)^\top \r) (\bm\Gamma^\kk_f)^{-1}, 
\end{align*}
and $\bm\Psi^\kk_{i, tu}(\tau) := \text{\upshape diag}( \Cov(X^\trunc_{k, i\ell, t}(\tau), X^\trunc_{k, i\ell, u}(\tau)), \, \ell \in [p_{-k}] )$ with $X^\trunc_{k, i\ell, t}(\tau)$ denoting the element of $\mat_k(\mc X^\trunc_t(\tau))$, such that $\Vert \bm\Phi^\kk_i(\tau) \Vert = O( M^{2 - 2\eps} \omega^{2 + 2\eps})$.
\end{enumerate}
}
\end{theorem}

Most notably, the above results make explicit the effect of heavy tails on the rates of estimation through $\epsilon$ (see the definition of $\psi^\kk_{n, p}$ in~\eqref{eq:psi}) and $M_n$ (defined in Assumption~\ref{assum:heavy}~\ref{cond:factor:bound}), which is distinguished from the existing work on tail-robust factor modelling.
Under the mild condition in~\eqref{eq:tau:M}, which permits $M_n$ to grow polynomially in $n$, we have $M_n^{1 - \eps}/\sqrt{np_{-k}} = o(1)$ as $\min(n, p_1, \ldots, p_K) \to \infty$.
As an illustration, let $r_k = 1$ for all $k \in [K]$ and suppose that $\{\mc F_t\}_{t \in [n]}$ is a sequence of i.i.d.\ regularly varying random variables with the index $\alpha > 0$. Then we can relate $M_n$ to the data generating process as $M_n \asymp n^{1 / \alpha}$ \citep{mikosch2010limit}. 
With $\alpha = 2 + 2\eps$, the condition in~\eqref{eq:tau:M} is readily met, e.g.\ if $K \ge 2$ and $p_k \asymp n^\gamma$ for all $k$ with any $\gamma > 0$.
Comparing Theorems~\ref{thm:first} and~\ref{thm:second} shows that the first-iteration estimator reduces the rate of the estimation error attributed to the latency of the factor-driven common component thanks to the projection step, from $p_k^{-1}$ to $p^{-1}$.
Theorem~\ref{thm:second} reveals that as $\epsilon \to 1$, the rates attained by $\wc{\bm\Lambda}^{[1]}_k(\tau)$ are comparable to those available in light-tailed settings.
Also, their dependence on the cross-sectional dimensions is sharper ($p^{-1}$) than those attainable by Huber loss-based methods ($p^{-1/2}$ or $p_k^{-1/2}$), even under the more general Assumption~\ref{assum:rf} permitting temporal and spatial dependence, see Table~\ref{tab:comparison}.

In Theorem~\ref{thm:third} analysing $\wc{\bm\Lambda}^{[2]}_k(\tau)$, we impose~\eqref{eq:np}, a mild condition
% and is satisfied by commonly encountered economic applications (see Section~\ref{sec:euro}); 
which is readily met if e.g.\ $\log(np) = o(\min(n, p_k))$ and $\max_{k' \in [K] \setminus \{k\}} p_{k'} = o(np_k)$ when $\eps = 1$.
This is for the ease of presentation, since any gain in the rate of estimation from the extra iterations is limited to the reduction of non-leading terms.
In fact, under Assumption~\ref{assum:indep} and~\eqref{eq:np}, the rates reported in Theorems~\ref{thm:second} and~\ref{thm:third} for $\wc{\bm\Lambda}^\ii_k(\tau), \, \ii \in \{1, 2\}$, are comparable.
This is in line with \cite{luo2021sharp} who, investigating the problem of Tucker decomposition reconstruction, remark that `{\it tensor reconstruction error rate of HOOI with only one iteration is optimal}' (see their Remark~6). % in the sense that further iteration improves only the constant applied to the rate.
At the same time, the second iteration allows us to establish the asymptotic normality of $\wc{\bm\Lambda}^{[2]}_{k, i\cdot}(\tau)$, which is a first result in its kind in the context of tail-robust tensor factor modelling. 
The necessity for an extra iteration stems from the fact that for $\wc{\bm\Lambda}^{[1]}_{k, i\cdot}(\tau)$, the projection involved in~\eqref{eq:gamma:wc} is from the initial estimator which attains a sub-optimal rate of estimation compared to that involved in $\wc{\bm\Lambda}^{[2]}_{k, i\cdot}(\tau)$. % (evident from comparing Theorems~\ref{thm:first} and~\ref{thm:second}).
We refer to Appendix~\ref{app:sim:add} for the numerical investigation into the convergence behaviour of $\wc{\bm\Lambda}^{\ii}_k(\tau)$.

\begin{remark}
\label{rem:one}
\upshape{
\begin{enumerate}[wide, itemsep = 0pt, label = (\roman*)] 
\item \label{rem:one:two} 
In the case of vector time series ($K = 1$), we have $p_{-1} = 1$ and the first-stage (and final) estimator $\wh{\bm\Lambda}_1(\tau)$ amounts to the popularly adopted PC-based estimator combined with the data truncation. Consequently, as $\epsilon \to 1$, the rates in Theorem~\ref{thm:first} match (up to a logarithmic factor in the case of spatial dependence) the rates derived in light-tailed settings, see \citet[Theorem 2]{bai2003}.
Also, an intermediate result (Proposition~\ref{prop:first:init}) shows that $\wh{\bm\Gamma}^{\kk}(\tau)$ is near-minimax optimal for vector time series, see \citet[Proposition~7]{wang2022rate}.

\item In Theorem~\ref{thm:third}, we make the more stringent condition in~\ref{thm:third:two} that requires a fixed upper bound on $\vert \mc F_t \vert_2$.
If we suppose that $\vecop(\mc F_t)$ is on an $\ell_2$-ball of radius $M_n$, we can relax the condition and still establish the asymptotic normality of $\wc{\bm\Lambda}^{[2]}_{k, i \cdot}(\tau)$ with the rate of convergence suitably adjusted, namely to $\sqrt{np_{-k}} M_n^{-1 + \eps}$. 
\end{enumerate}
}
\end{remark}

\begin{theorem}[Tensor factor estimator]
\label{thm:factor}
{\it Suppose that Assumptions~\ref{assum:loading}, \ref{assum:factor} and \ref{assum:heavy} hold, $\tau \asymp \tau^\kk_{n, p}$ is set as in~\eqref{eq:tau} in producing $\wc{\bm\Lambda}^\ii_k(\tau)$ for each $k \in [K]$, and~\eqref{eq:tau:M} is met.
With $\iota =~2$ under Assumption~\ref{assum:indep} and $\iota = 1$ under Assumption~\ref{assum:rf}, we have as $\min(n, p_1, \ldots, p_K) \to \infty$.
\begin{enumerate}[wide, label = (\roman*)]
\item \label{thm:factor:one} 
For each $t \in [n]$, 
\begin{align}
& \l\vert \wh{\mc F}_t(\tau) -  \mc F_t \times_{k = 1}^K (\wc{\mbf H}^\ii_k)^{-1} \r\vert_2
\nn \\
&= \l\{\begin{array}{ll}
O_P\l[ (\vert \mc F_t \vert_2 + \omega) \l( \sum_{k \in [K]} \frac{M_n^{1 - \eps}}{\sqrt{np_{-k}}} \vee \frac{1}{\sqrt p} \r) \r] & \text{under Assumption~\ref{assum:indep} and~\eqref{eq:np},}
\\
O_P\l[ (\vert \mc F_t \vert_2 + \omega) \l( \sum_{k \in [K]} \frac{M_n^{1 - \eps}}{\sqrt{np_{-k'}}} \vee \bar{\psi}_{n, p} \vee \frac{1}{\sqrt p} \r) \r] & \text{under Assumption~\ref{assum:rf},}
\end{array}\r.
\label{eq:thm:factor:one}
\\
& \frac{1}{n} \sum_{t \in [n]} \l\vert \wh{\mc F}_t(\tau) - \mc F_t \times_{k = 1}^K (\wc{\mbf H}^\ii_k)^{-1} \r\vert_2^2
\nn \\
&= 
\l\{ \begin{array}{ll}
O_P\l[ \omega^2 \l( \sum_{k \in [K]} \l( \frac{M_n^{1 - \eps}}{\sqrt{np_{-k}}} \r)^2 \vee \frac{1}{p} \r) \r] & \text{under Assumption~\ref{assum:indep} and~\eqref{eq:np},}
 \\
O_P\l[ \omega^2 \l( \sum_{k \in [K]} \l( \frac{M_n^{1 - \eps}}{\sqrt{np_{-k}}} \r)^2 \vee \bar{\psi}_{n, p}^2 \vee \frac{1}{p} \r) \r] & \text{under Assumption~\ref{assum:rf}.}
\end{array}\r.
\label{eq:thm:factor:two} 
\end{align}

\item \label{thm:factor:two}
Let Assumption~\ref{assum:indep} hold, and assume that $\max_{k \in [K]} p_k = o(n)$ and $M_n = M \in (0, \infty)$ for all $n \ge 1$. % and that there exists some constant $c_1 \in (0, \infty)$ satisfying $\min_{\mbf i \in [p_1] \times \ldots \times [p_K]} \Var(X^\trunc_{\mbf i, t}(\kappa)) \ge c_1 > 0$.
Then for given $t \in [n]$, as $\min(n, p_1, \ldots, p_K) \to \infty$,
\begin{align*}
& \sqrt{p} \l( \vecop(\wh{\mc F}_t(\tau)) - (\wc{\mbf H}^{[2]})^{-1} \vecop(\mc F_t) \r) \to \mc N_r(\mbf 0, \bm\Upsilon_t), 
\end{align*}
with $\wc{\mbf H}^{[2]} = \wc{\mbf H}^{[2]}_K \otimes \cdots \otimes \wc{\mbf H}^{[2]}_1$.
Here, $\bm\Upsilon_t = \mbf E_\chi^\top \Cov(\vecop(\bm\xi_t)) \mbf E_\chi$ where $\mbf E_\chi := \otimes_{k = K}^1 \mbf E_{\chi, k}$ with $\mbf E_{\chi, k}$ containing the $r_k$ largest leading eigenvectors of $\bm\Gamma^\kk_\chi := \bm\Lambda_k (n^{-1} \sum_{t \in [n]} \mat_k(\mc F_t) \mat_k(\mc F_t)^\top ) \bm\Lambda_k^\top$, and satisfies $\Vert \bm\Upsilon_t \Vert \le \omega^2$. 
% is defined in~\eqref{eq:upsilon} fulfilling $\Vert \bm\Upsilon_t \Vert \le \omega^2$.
\end{enumerate}
}
\end{theorem}

Theorem~\ref{thm:factor} shows that when the interest lies in estimating the factor tensor at a given $t$, the rate of estimation scales with $\vert \mc F_t \vert_2$ which may grow with $\min(n, p_1, \ldots, p_K)$, unlike when investigating the averaged $\ell_2$-error over all $t \in [n]$.
When $\vert \mc F_t \vert_2 = O(1)$, the results in~\eqref{eq:thm:factor:one} are comparable to those found in light-tail settings, see \citet[Theorem~3.6]{barigozzi2022statistical} who report the rate $O_P(p^{-1/2})$ when $n$ is sufficiently large.
% \footnote{Theorem~3 of \cite{chen2023statistical} deriving the rate $O_P(\min(p_1, p_2)^{-1})$ when $K = 2$}
In heavy-tailed situations, \citet[Theorem~3.3]{barigozzi2023robust} derive the rate of $O_P(\max_{k \in [K]} p_{-k}^{-1})$ compared to which~\eqref{eq:thm:factor:two} reports a far more competitive rate.
A careful inspection of the proof of Theorem~\ref{thm:factor} reveals that the rate does not change when we additionally truncate $\mc X_t$ with a truncation parameter $\kappa = \kappa_p \to \infty$ as $\min(p_1, \ldots, p_K) \to \infty$.
In Section~\ref{sec:sim}, we show that numerically, additional truncation contributes to reducing the error in estimating the common component, and thus we advocate such an approach in practice. 

Finally, the following Proposition~\ref{prop:fn} shows that, provided that the initial estimator $\wh r^{(0)}_k = \bar{r}_k$ is chosen appropriately large, Algorithm~\ref{alg:r} iteratively gives a consistent estimator of $r_k$.

\begin{proposition}[Factor number estimator]
\label{prop:fn}
{\it Suppose that Assumptions~\ref{assum:loading}, \ref{assum:factor}, \ref{assum:heavy} and~\ref{assum:indep} or~\ref{assum:rf} hold, and $\tau \asymp \tau^\kk_{n, p}$ is set as in~\eqref{eq:tau} and~\eqref{eq:tau:M} is met.
Also, let $\rho = \rho_{n, p} \to 0$ satisfy, as $\min(n, p_1, \ldots, p_K) \to \infty$,
\begin{align*}
\sum_{k' \in [K]} \l( \psi^{(k')}_{n, p} \vee \frac{1}{p_{k'}} \r) = O(\rho_{n, p}).
\end{align*}
Then, provided that $r_k \le r^{(m - 1)}_k \le \bar{r}_k$, 
the estimator in~\eqref{eq:r:est} satisfies $\p(\wh r_k^{(m)}(\tau) = r_k) \to 1$. 
}
\end{proposition}

\section{Simulation studies}
\label{sec:sim}

In this section, we focus on the tensor-valued time series scenarios and defer the vector case to Appendix~\ref{sec:sim:vec}.
% \footnote{An implementation of the proposed methods is available at \url{https://github.com/haeran-cho/robustTFM}.}
Following \cite{barigozzi2023robust}, we generate tensor time series with $K = 3$ and $(r_1, r_2, r_3) = (3, 3, 3)$, according to the following three scenarios while varying $n \in \{100, 200, 500\}$:
\ref{f:one} $(p_1, p_2, p_3) = (10, 10, 10)$, \ref{f:two} $(p_1, p_2, p_3) = (100, 10, 10)$, and \ref{f:three} $(p_1, p_2, p_3) = (20, 30, 40)$.
%\begin{enumerate}[wide, itemsep = 0pt, label = (T\arabic*)]
%\item \label{f:one} $p_1 = p_2 = p_3 = 10$;
%\item \label{f:two} $p_1 = 100$ and $p_2 = p_3 = 10$;
%\item \label{f:three} $p_1 = 20$, $p_2 = 30$ and $p_3 = 40$;
%\end{enumerate}
For the generation of $\mc F_t$ and $\bm\xi_t$, we consider Gaussian and $t_3$-distributions and introduce both temporal and spatial dependence. 
We additionally consider the situations where either the factors or the idiosyncratic components are contaminated by outliers at random, setting $\varrho \in \{0, 0.1, 0.5, 1\} \times 10^{-2}$ as the proportion of outliers out of the $nr$ entries of $\{\mc F_t\}_{t \in [n]}$ or the $np$ entries of $\{\bm\xi_t\}_{t \in [n]}$.
This exercise is motivated by \cite{raymaekers2024challenges} who, while noting the differences between the models for (cellwise) outliers and heavy tails and the objectives thereof, also remark that `{\it estimators for heavy-tailed data can still perform reasonably well under cellwise contamination}'.

In comparison with our proposed estimator ($\wc{\bm\Lambda}^{[2]}_k(\tau)$, referred to as `Trunc'), we include the iterative projection procedure of \cite{barigozzi2022statistical} (iPE) and the pre-averaging-based estimator of \cite{chen2024rank} (PreAve), both of which are designed for light-tailed situations and included as a benchmark only, to gauge the advantage of adopting Trunc in the presence of heavy tails and outliers.
We also consider the Huber loss-based estimator of \cite{barigozzi2023robust} (RTFA) which has been shown to perform competitively against a variety of existing methods for tensor factor analysis \citep{chen2022factor, zhang2022tucker}, both under light- and heavy-tailed settings.
When investigating the performance in common component estimation, we look into the role of additional truncation in tensor factor estimation by comparing Trunc (truncation applied to factor estimation with $\kappa = \tau$) and `noTrunc' ($\kappa = \infty$), see the discussions below Theorem~\ref{thm:factor}. 
When investigating the loading and common component estimation performance, we treat the factor numbers as known, and the performance of the factor number estimator is separately examined in Appendix~\ref{app:sim:tensor:r}; we refer to Appendix~\ref{app:sim:tensor} for complete descriptions and results.

\begin{figure}[h!t!b!]
\centering
\begin{tabular}{c}
\includegraphics[width = .9\textwidth]{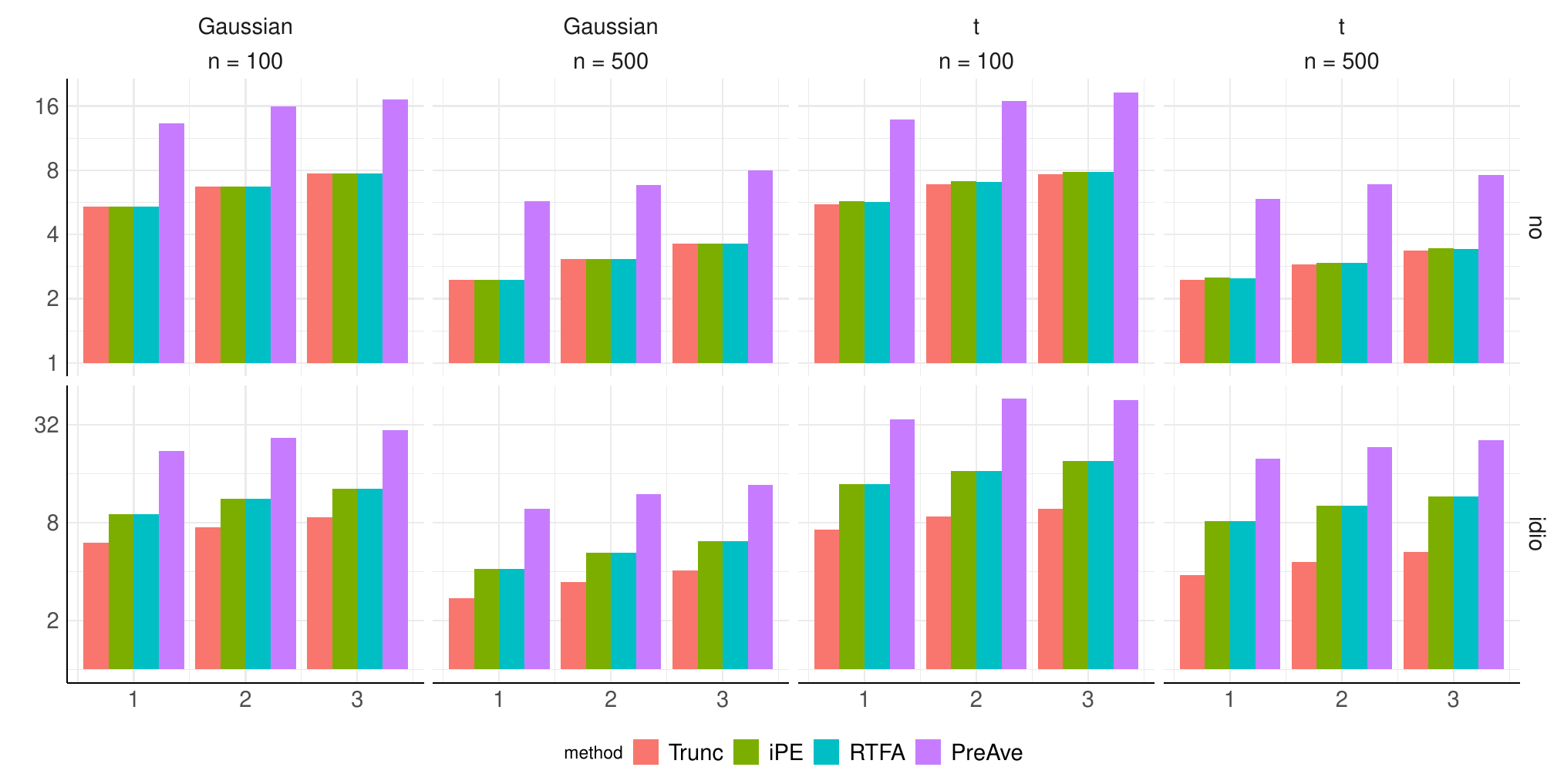} \\
\includegraphics[width = .9\textwidth]{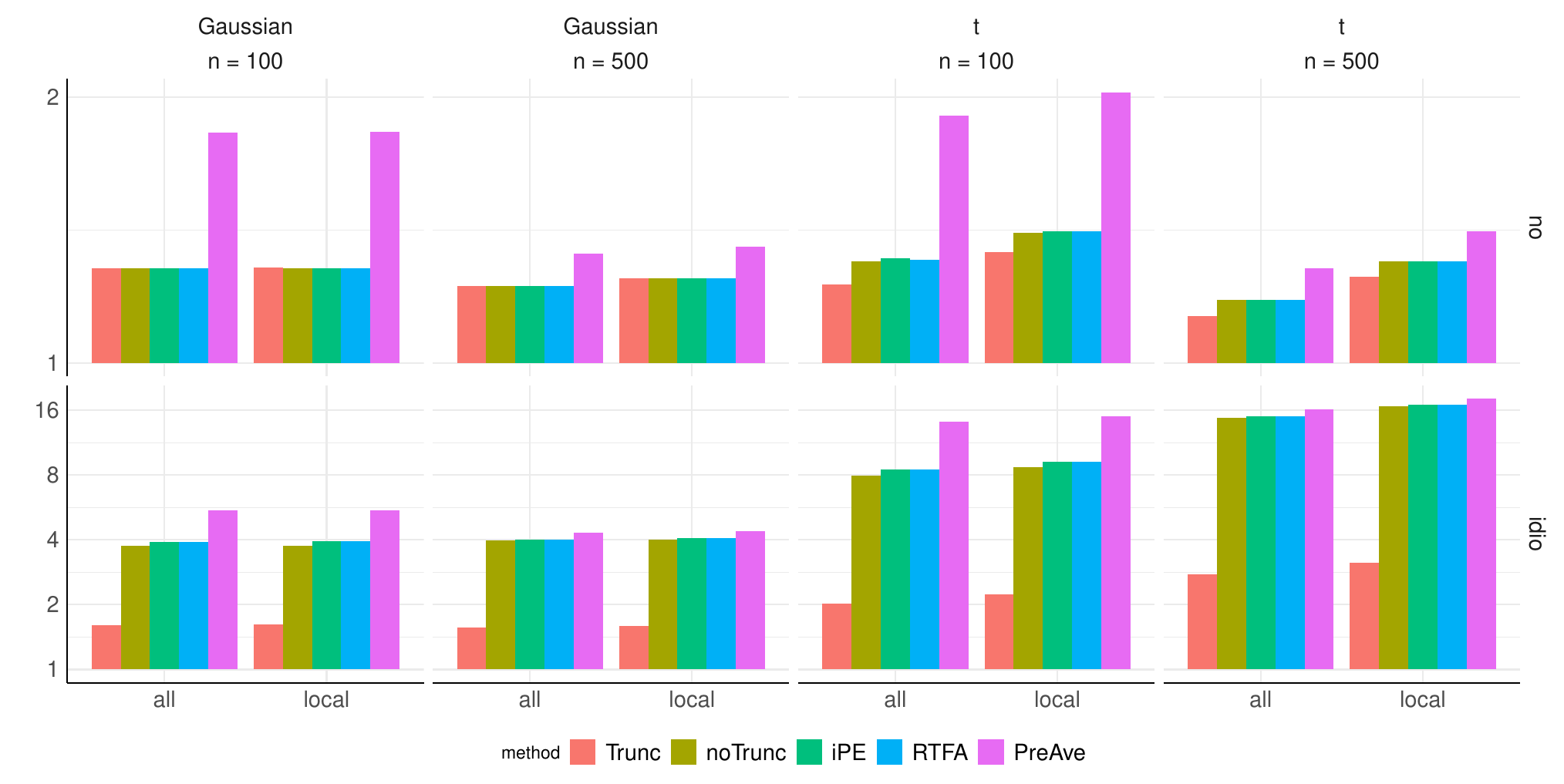} 
\end{tabular}
\caption{\ref{f:three} with $(p_1, p_2, p_3) = (20, 30, 40)$.
Top: Loading estimation errors measured as in~\eqref{eq:err:loading:tensor} for each mode ($x$-axis) for Trunc, iPE, RTFA and PreAve averaged over $100$ realisations per setting, over varying $n \in \{100, 500\}$ and distributions for $\mc F_t$ and $\bm\xi_t$ (Gaussian and $t_3$).
Bottom: Common component estimation errors measured as in~\eqref{eq:err:chi:tensor} with $\mc T = [n]$ (`all') and $\mc T = \{n - 10 + 1, \ldots, n\}$ (`local') ($x$-axis) where we additionally include `noTrunc' (see the text).
Within each plot, we consider the cases of no outlier (`no'), and when the outliers are in the $0.5\%$ of entries of $\bm\xi_t$ (`idio').
The $y$-axis is in the log-scale and all errors have been scaled for better presentation.}
\label{fig:tensor:f:three:est}
\end{figure}

\begin{figure}[h!t!b!]
\centering
\includegraphics[width = 1\textwidth]{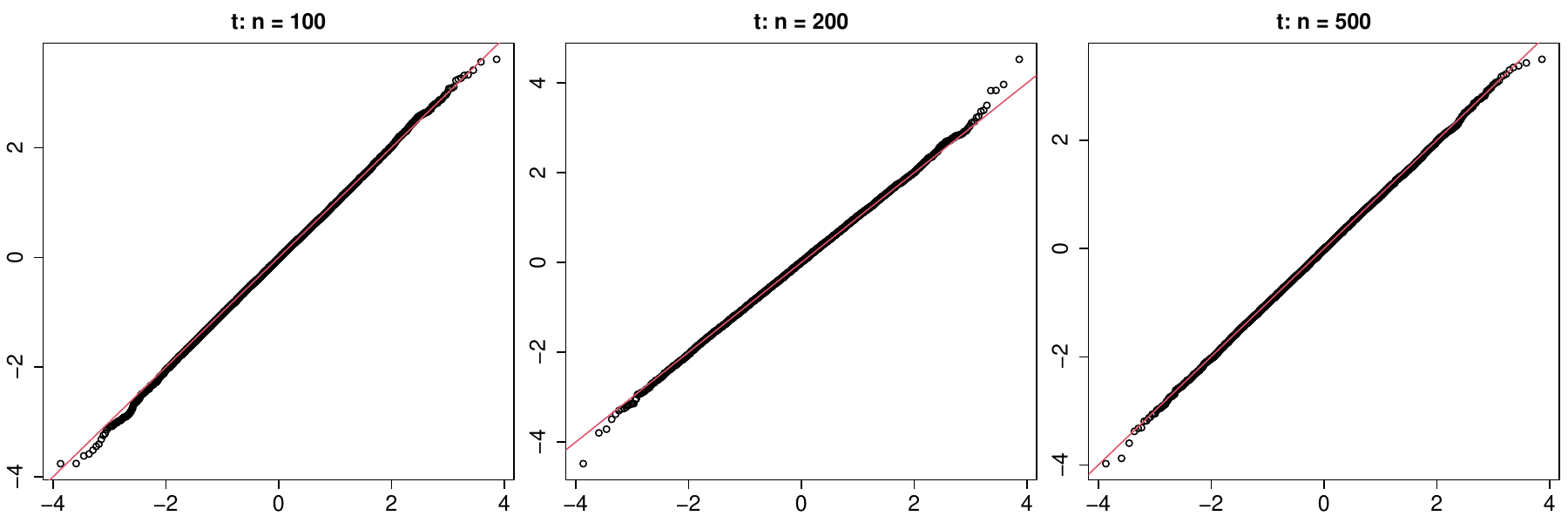}
\caption{\ref{f:three}: Plots of the sample quantiles of the scaled and centered entries of $\wc{\bm\Lambda}^{[2]}_k(\tau)$ ($y$-axis) against the quantiles from the standard normal distribution ($x$-axis) over varying $n \in \{100, 200, 500\}$ (left to right) when the data are generated from the $t_3$ distribution. In each plot, the $y = x$ line is given in red. 
See Appendix~\ref{app:sim:infer} for full details.}
\label{fig:tensor:f:three:infer}
\end{figure}

As a representative example, Figure~\ref{fig:tensor:f:three:est} reports some estimation results obtained under~\ref{f:three}.
% For each setting, we report the evaluation metrics in Equations~\eqref{eq:err:loading:tensor}--\eqref{eq:err:chi:tensor} (see Appendix~\ref{app:sim:tensor:setting}) averaged over $100$ realisations respectively measuring the loading and the common component estimation errors.
In loading estimation, all methods show marginally improved performance with growing $n$, and the mode-wise error is larger for the mode $k$ corresponding to smaller $p_k$ (hence larger $np_{-k}$), confirming the theoretical findings.
Comparing the performance under~\ref{f:one} ($p = 10^3$), \ref{f:two} ($p = 10^4$) and \ref{f:three} ($p = 2.4 \times 10^4$), estimation errors decrease considerably with increasing $p$.
Under Gaussianity with no outliers, iPE and RTFA perform the best but with the introduction of heavy tails and idiosyncratic outliers, Trunc performs as well as, or superior to iPE, RTFA and PreAve according to all metrics. 

Noteworthy differences are observed when the outliers are present in $\bm\xi_t$: Trunc is not affected by the growing proportion of outliers thanks to the data-driven truncation parameter selection via CV (Section~\ref{sec:tuning}), which is effective in loading space estimation as well as factor estimation (and together, the estimation of the common component).
RTFA, while attaining tail-robustness in loading space estimation via Huber loss minimisation, is sensitive to the presence of anomalous observations when estimating the common component.
This is attributed to the fact that RTFA (as well as iPE and PreAve) simply takes a weighted average of the raw data for the estimation of $\mc F_t$ unlike Trunc, and the benefit of data truncation in factor estimation is also apparent when comparing Trunc and noTrunc, see the bottom panel of Figure~\ref{fig:tensor:f:three:est}.
The local estimation error is greater than the global one, which conforms to Theorem~\ref{thm:factor}~\ref{thm:factor:one} showing that the estimation error at given $t$ scales with $\vert \mc F_t \vert_2$.
Overall, the performance of Trunc in most settings does not deviate far from its performance in the Gaussian setting without any outlier.
We defer the case when the outliers are present in $\mc F_t$ to Appendix~\ref{app:sim:tensor}; while it is not possible to recover the uncontaminated factors as all cross-sections of $\mc X_t$ are contaminated by the outliers, reasonable recovery of loadings and $\bm\chi_t$ (post-contamination) is achieved by most methods. 

Finally, Figure~\ref{fig:tensor:f:three:infer} verifies the asymptotic normality of $\wc{\bm\Lambda}^{[2]}_k(\tau)$ derived in Theorem~\ref{thm:third}, see Appendix~\ref{app:sim:infer} for the complete results.

\section{Euro Area macroeconomic data}
\label{sec:euro}

We analyse the EA-MD \citep{barigozzi2024large}, a collection of $37$ macroeconomic indicators that have been collected at a monthly frequency for $8$ Euro Area (EA) countries (Austria, Belgium, Germany, Greece, Spain, France, Italy and the Netherlands). The data form a matrix-valued ($K = 2$) time series of dimensions $(p_1, p_2) = (8, 37)$; the data span the period from 2002-02 to 2023-09 ($n = 257$). % (AT, BE, DE, EL, ES, FR, IT, NL)
All time series are individually transformed to stationarity based on standard unit-root tests as suggested by \citet{barigozzi2024large}, and are further centered and standardised using the median and the mean absolute deviation, respectively. 
In this dataset, Covid-19 pandemic is likely to play a major role since such a large outlier is likely to drive most of the co-movement due to the short sample size~$n$, and thus possibly bias the factor analysis, which motivates the use of our robust approach.
% Given the relative smallness of the data, we choose not to perform a forecasting exercise; rather, we adopt the proposed methods to investigate the presence of common factors in the EA-MD. 

%\begin{figure}[h!t!b!]
%\centering
%\includegraphics[width = 1\textwidth]{euro_ht.pdf}
%\caption{ EA-MD: The $37$ macroeconomic indicators observed for eight countries between 2002-02 and 2023-09, after transformation, centering and standardisation.}
%\label{fig:er:x}
%\end{figure}
% Unlike in Section~\ref{sec:fredmd}, where the dimensions of the panel data are of comparable orders, 

% For the mode-$1$ (resp.\ mode-$2$) unfolded data, the dimension $p_1 = 8$ (resp.\ $p_2 = 37$) is considerably smaller than the corresponding `sample size' $np_2 = 9509$ (resp.\ $np_1 = 2056$), a situation that does not favour the use of information criteria for factor number estimation as pointed out by \cite{onatski2024}.
We adopt the ratio-based estimator discussed in Section~\ref{sec:r:est} with $\bar{r}_k = \min(\lfloor p_k/2 \rfloor, 20)$ which, for the most choice of the truncation parameter, returns $(\wh r_1, \wh r_2) = (1, 3)$ as where the `regions of stability' are achieved over varying $\tau$ (see Figure~\ref{fig:er:r} in Appendix~\ref{app:euro}); this agrees with the output from the method of \cite{barigozzi2023robust} while that of \cite{chen2024rank} returns $\wh r_1 = 1$ and $\wh r_2 \in \{2, 3, 4\}$ due to the random projections adopted therein.

Setting $(\wh r_1, \wh r_2) = (1, 3)$, we investigate the efficacy of truncation by comparing the proposed truncation-based estimator (`Trunc') against the non-tail-robust counterpart (`noTrunc' with $\tau = \kappa = \infty$); we note the similarity between noTrunc and the approach taken in \cite{yu2022projected} but the former involves two iterations rather than one. 
The CV procedure in Section~\ref{sec:tuning} returns $\tau_{\text{\tiny CV}} \approx 5.306$ (Figure~\ref{fig:er:cv}). 
Assumptions~\ref{assum:loading} and~\ref{assum:factor} allow for the identification of factors up to an orthogonal transformation, but they lack economic meaning, i.e.\ our estimation method can be used for exploratory factor analysis but not for confirmatory one in general. 
Therefore, we apply the Varimax rotation \citep{mardia1979} to $\wc{\bm\Lambda}^{[2]}_2(\tau) \in \R^{37 \times 3}$, which results in a rotation matrix close to the identity matrix. 
With $\wh r_1 = 1$, the column vector $\wc{\bm\Lambda}^{[2]}_1(\tau)$ does not suffer from the identification issue.  
% \citet{bai2013principal} discuss alternative possibilities but they are applicable when $K = 1$ only.

\begin{figure}[h!t!p!]
\centering
\begin{tabular}{cc}
\includegraphics[width = 0.45\linewidth]{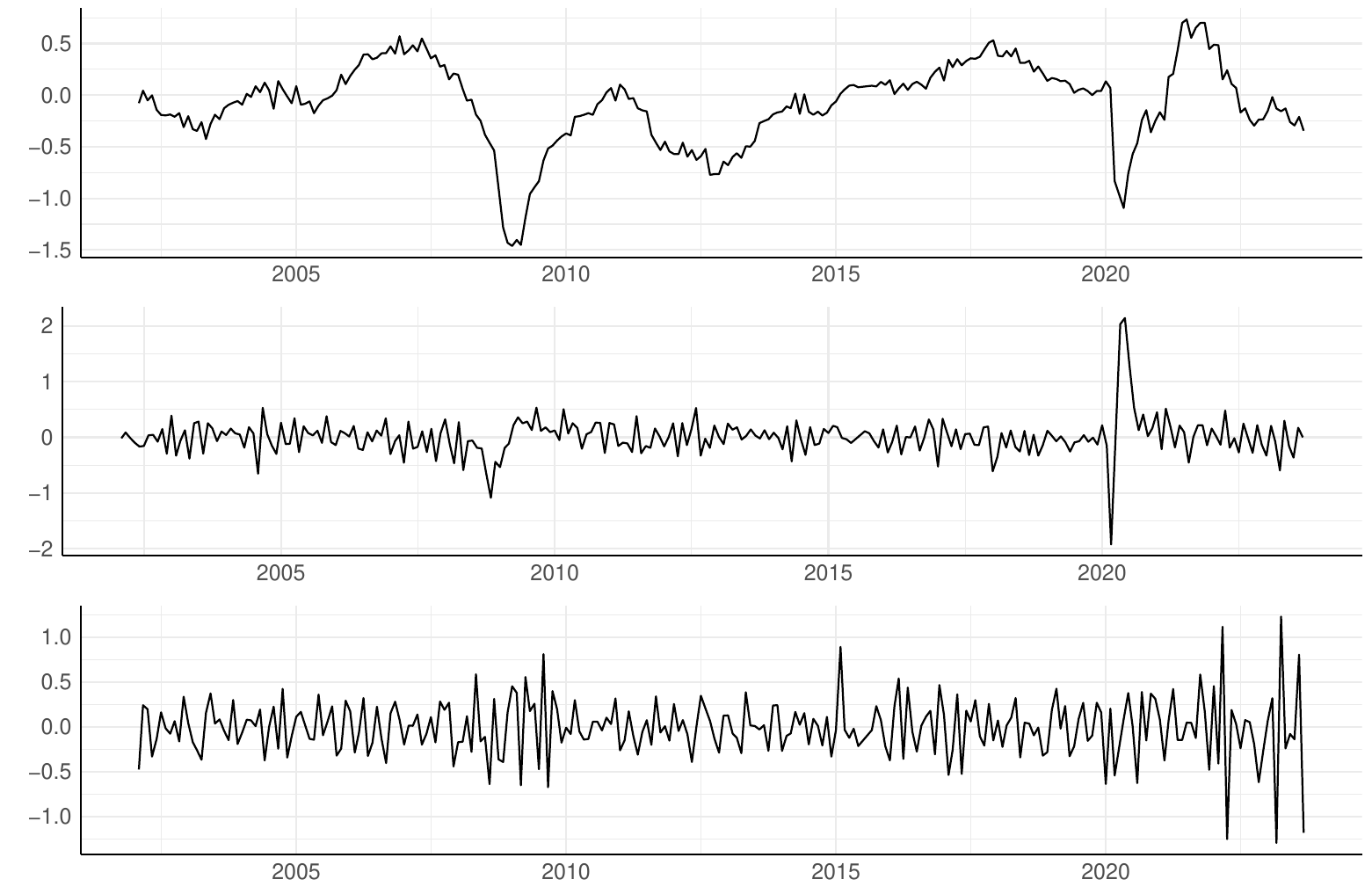} &
\includegraphics[width = 0.45\linewidth]{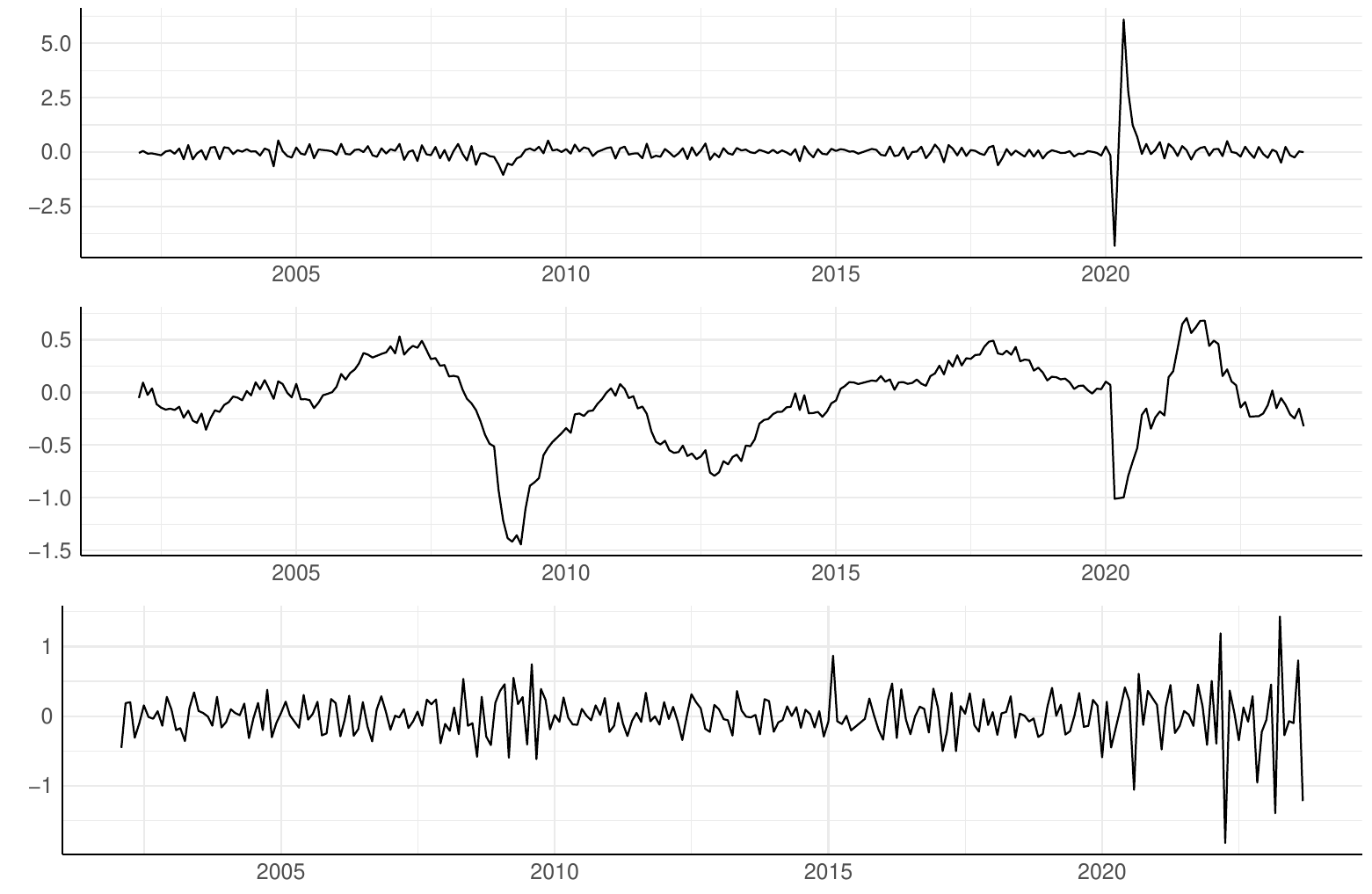}
\end{tabular}
\caption{EA-MD: Factor time series $\wh f_{1j, t}(\tau, \kappa)$ for $j = 1, 2, 3$ (top to bottom) with (left) and without (right) the truncation.}
\label{fig:ea:factor:trunc}
\end{figure}

The noTrunc approach over-represents the potential outliers around Covid-19, which results in the leading factor estimates (top right of Figure~\ref{fig:ea:factor:trunc}) with sharp peaks around early 2020.
While the effect of these outliers is also observable from the second factor returned by Trunc, its extent is limited and the leading factor returned by Trunc clearly represents the economic cycle characterised by lower frequency oscillations typical of the business cycle, with downturns during recessions. 
We may regard the results from Trunc as more realistic, indeed the main drivers of macroeconomic datasets are often found to be those factors associated with the real economic activity, such as industrial production, which also drive the business cycle \citep{barigozzi2024large}.

\begin{figure}[h!t!p!]
\centering
\begin{tabular}{cc}
\includegraphics[width = 0.7\linewidth]{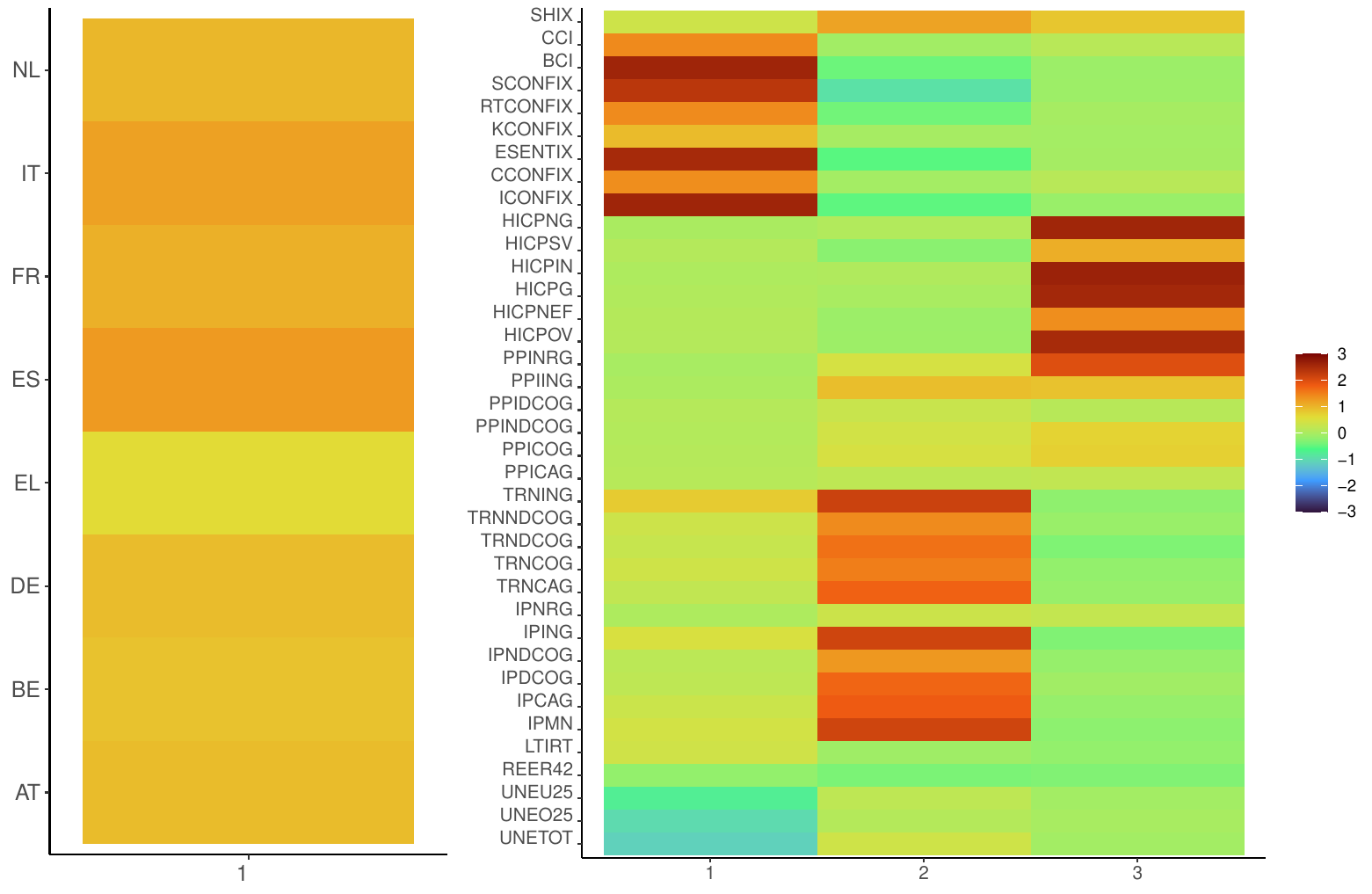} \\
\includegraphics[width = 0.7\linewidth]{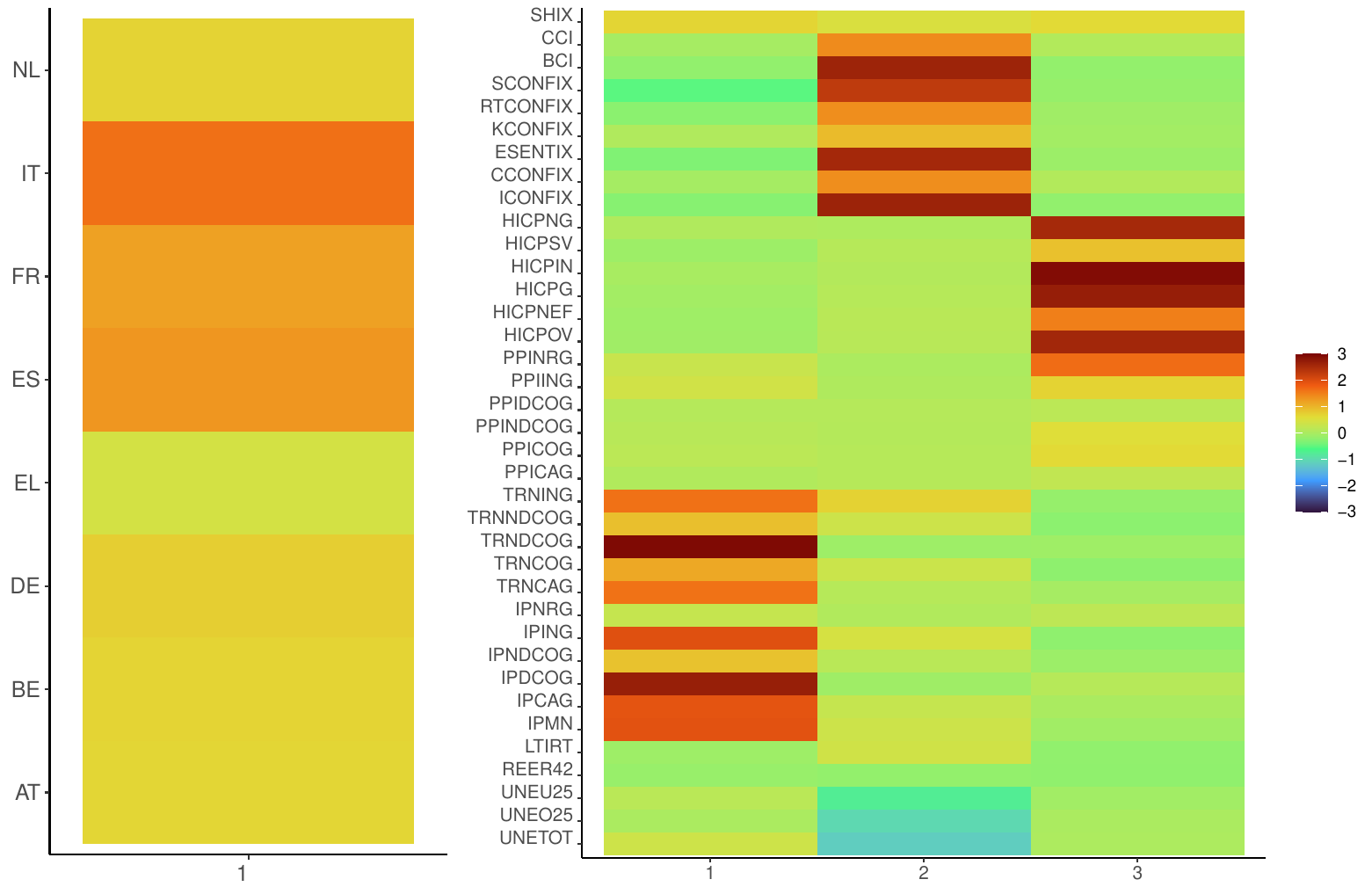} 
\end{tabular}
\caption{EA-MD: Estimated loading matrices $\wc{\bm\Lambda}^{[2]}_k(\tau)$ for $k = 1$ (left, $(p_1, \wh r_1) = (8, 1)$) and $k = 2$ (right, $(p_2, \wh r_2) = (37, 3)$) with and without the truncation (top to bottom).}
\label{fig:ea:loading:trunc}
\end{figure}

Overall, the loading matrices estimated with or without truncation exhibit similar patters, see Figure~\ref{fig:ea:loading:trunc}; namely, the elements of $\wc{\bm\Lambda}^{[2]}_1(\tau)$ are of the same sign across the $8$ countries, and the columns of $\wc{\bm\Lambda}^{[2]}_2(\tau)$ exhibit clusterings based on the grouping of the macroeconomic indicators. 
At the same time, recalling that each column of $p_2^{-1/2} \wc{\bm\Lambda}^{[2]}_2(\tau)$ has the unit $\ell_2$-norm, truncation of extreme observations enables us to better recover the loadings of smaller magnitude, particularly among the indicators associated with industrial production (IDs starting with IP), see also Table~\ref{tab:ea:loading:trunc} which shows that the loading estimates from noTrunc are more extreme. 

These results have an implication on the forecasting performance, as confirmed in the following forecasting exercise which shows that the forecasts from Trunc, being driven by the business cycle factor, are marginally superior to those from noTrunc.
Denoting by $N = 236$ the number of observations prior to 2022-01, we sequentially produce a one-step ahead forecast for the two indicators, the growth rate of the industrial production manufacturing index (IPMN) and the difference of the core consumer prices inflation (HICPNEF), at $t \in \{N + 1, \ldots, n\}$ (recall that $n = 257$), each time using all the preceding observations as the training data, see Appendix~\ref{app:euro} for complete details. 
Table~\ref{tab:ea:forecast} reports the summary of the forecasting errors for each indicator, where Trunc marginally outperforms noTrunc for both indicators in all metrics. 
As expected from Figure~\ref{fig:ea:loading:trunc}, where the loading estimates differ the most for France and Italy, these countries contribute the most to the difference between the two estimators' forecasting performance. % over the period (2023 April to June).

\begin{table}[h!t!]
\caption{EA-MD: We report the mean, median and the standard deviation of the one-step ahead forecasting errors across the 8 countries over time for IPMN and HICPNEF, with and without truncation. 
% $\text{Err}_{\mbf i, t}(\star), \, t \in \{N + 1, \ldots, n\}, \, \mbf i \in [p_1] \times \{\text{IPMN, HICPNEF}\}$, with and without truncation for each indicator.
Additionally, we report the percentage of the instances when 
% $\text{Err}_{\mbf i, t}(\text{Trunc}) < \text{Err}_{\mbf i, t}(\text{noTrunc})$ for each indicator 
the forecasting error of Trunc is smaller than that of noTrunc for each indicator (`Perc'). 
See~\eqref{eq:ea:fe} for the definition of the forecasting error.}
\label{tab:ea:forecast}
\centering
\begin{tabular}{r cccc cccc}
\toprule
& \multicolumn{4}{c}{IPMN (in $10^{-2}$)} & \multicolumn{4}{c}{HICPNEF (in $10^{-3}$)} \\ 
& Mean & Median & SE & Perc & Mean & Median & SE & Perc \\
\cmidrule(lr){1-1} \cmidrule(lr){2-5} \cmidrule(lr){6-9}
Trunc & 1.524 & 1.242 & 1.222 & 56.548 & 
4.247 & 2.706 & 4.248 & 54.762 
\\
noTrunc & 1.632 & 1.323 & 1.422 & -- &
4.273 & 2.765 & 4.184 & --
\\
\bottomrule
\end{tabular}
\end{table}

% \section{Conclusions}
% \label{sec:conc}

% We propose a new, tail-robust procedure for the estimation of factor structures for vector- and tensor-valued time series.
% The proposed method based on data truncation is easy to implement and does not need to seek for a numerical solution for an optimisation problem.
% Under a weak assumption imposing the existence of $(2 + 2\eps)$-th moment, we establish the consistency and asymptotic normality of the proposed estimators where the effect of heavy tails is explicitly characterised in the level of truncation and the rates of estimation through $\eps \in (0, 1)$.
% In particular, thus-formulated results reveal that as $\eps \to 1$, the performance of the truncation-based estimator matches that of their non-robust counterpart in light-tailed settings.
% These aspects set our proposal apart from the existing tail-robust factor modelling methods based on Huber regression. 
% Simulation studies show that it works well in the presence of both heavy tails as well as outliers, and the applications to macroeconomic datasets further demonstrate its promising performance.

\bibliographystyle{apalike}
\bibliography{fbib_full}

\clearpage
\appendix

\numberwithin{equation}{section}
\numberwithin{figure}{section}
\numberwithin{table}{section}
\numberwithin{theorem}{section}
\numberwithin{assumption}{section}
\numberwithin{lemma}{section}
\numberwithin{proposition}{section}

\section{Asymptotic identifiability}
\label{sec:identify}

From~\eqref{eq:model:unfold}, we have the mode-$k$ second moment matrix $\bm\Gamma^\kk$ decomposed as
\begin{align}
\bm\Gamma^{\kk} &= \frac{1}{p_{-k}} \bm\Lambda_k \l( \frac{1}{n} \sum_{t = 1}^n \mat_k(\mc F_t) \bm\Delta_k^\top \bm\Delta_k \mat_k(\mc F_t)^\top \r) \bm\Lambda_k^\top + 
\frac{1}{n p_{-k}} \sum_{t = 1}^n \E\l[ \mat_k(\bm\xi_t) \mat_k(\bm\xi_t)^\top \r]
\nn \\
&= \bm\Lambda_k \l( \frac{1}{n} \sum_{t = 1}^n \mat_k(\mc F_t) \mat_k(\mc F_t)^\top \r) \bm\Lambda_k^\top + 
\frac{1}{n p_{-k}} \sum_{t = 1}^n \E\l[ \mat_k(\bm\xi_t) \mat_k(\bm\xi_t)^\top \r]
\nn \\
&=: \bm\Gamma_\chi^\kk + \bm\Gamma_\xi^\kk.
% &= \bm\Lambda_k \bm\Gamma_f^\kk \bm\Lambda_k^\top +  \frac{1}{n p_{-k}} \sum_{t = 1}^n \E\l[ \mat_k(\bm\xi_t) \mat_k(\bm\xi_t)^\top \r]
%&= \frac{1}{p_{-k}} \bm\Lambda_k \l( \frac{1}{n} \sum_{t = 1}^n \mat_k(\mc F_t) \mat_k(\mc F_t)^\top \r) \bm\Lambda_k + \frac{1}{n p_{-k}} \sum_{t = 1}^n \E\l[ \mat_k(\bm\xi_t) \mat_k(\bm\xi_t)^\top \r]
% =: \bm\Gamma_\chi^\kk + \bm\Gamma_\xi^\kk.
\label{eq:gamma:decomp}
\end{align}

Let us define $\wt{\bm\Gamma}_\chi^\kk := \bm\Lambda_k \bm\Gamma_f^\kk \bm\Lambda_k^\top$.
Under Assumption~\ref{assum:loading}~(i), we have for all $k \in [K]$,
\begin{align*}
\frac{1}{p_{-k}} \bm\Delta_k^\top \bm\Delta_k = \mbf I_{r_{-k}}
\end{align*}
for all $p_k$'s.
This, together with Assumption~\ref{assum:factor}, leads to 
\begin{align}
\frac{1}{p_k} \l\Vert \bm\Gamma^\kk_{\chi} - \wt{\bm\Gamma}^\kk_\chi \r\Vert \le & \,
\frac{1}{p_k} \l\Vert \bm\Lambda_k \l( \frac{1}{n} \sum_{t = 1}^n \mat_k(\mc F_t) \mat_k(\mc F_t)^\top - \bm\Gamma^\kk_f \r) \bm\Lambda_k^\top \r\Vert = o(1).
% \nn \\
% \le & \,  C\l( \sum_{l \in [K] \setminus \{k\}} \frac{1}{\sqrt{p_l}} + \frac{1}{\sqrt{n}} \r).
\label{eq:approx}
\end{align}
Assumptions~\ref{assum:loading}~(i) and~\ref{assum:factor} indicate that the $r_k$ non-zero eigenvalues of $\wt{\bm\Gamma}_\chi^\kk$ are distinct and diverging linearly in $p_k$ as $p_k \to \infty$, which is inherited by the $r_k$ leading eigenvalues of $\bm\Gamma^\kk_\chi$ from~\eqref{eq:approx} and Weyl's inequality (see Lemma~\ref{lem:common:cov} below).
From this, it follows that the $r_k$ common factors are pervasive across the~$p_k$ cross-sections of $\mat_k(\bm\chi_t)$ for all $k \in [K]$.
On the other hand, $\bm\Gamma_\xi^\kk$ has bounded eigenvalues for all $p_k$, either under Assumptions~\ref{assum:indep} or~\ref{assum:rf}, see Lemma~\ref{lem:idio:cov}.
Then, thanks to Weyl's inequality, these observations lead to a diverging gap between the leading $r_k$ eigenvalues of $\bm\Gamma^\kk$ and the remainder, which ensures that the latent components $\bm\chi_t$ and $\bm\xi_t$ are separable in all $K$ modes asymptotically as $p_k \to \infty$.

For any non-negative definite matrix $\mbf A$, let $\mu_j(\mbf A)$ denote its $j$-th largest eigenvalue.
Also, let us write $\mu^\kk_{f, j} = \mu_j(\bm\Gamma^\kk_f)$.
\begin{lemma}
\label{lem:common:cov} Suppose that Assumptions~\ref{assum:loading}--\ref{assum:factor} hold.
Then for each $k \in [K]$, the $r_k$ non-zero eigenvalues of $\bm\Gamma^{\kk}_\chi$ and $\wt{\bm\Gamma}^{\kk}_\chi$, denoted by $\mu^{\kk}_{\chi, j}$ and $\wt{\mu}^\kk_{\chi, j}, \, j \in [r_k]$, respectively, satisfy
\begin{align*}
\l\vert \frac{1}{p_k} \wt{\mu}^\kk_{\chi, j} - \mu^\kk_{f, j} \r\vert &= o(1), % O\l( \frac{1}{\sqrt{p_k}} \r),
\\
\l\vert \frac{1}{p_k} \mu^\kk_{\chi, j} - \mu^\kk_{f, j} \r\vert &= o(1). % O\l( \frac{1}{\sqrt{n}} + \sum_{l \in [K]} \frac{1}{\sqrt{p_l}} \r).
\end{align*}
\end{lemma}

\begin{proof}
By Assumption~\ref{assum:loading}~(i),
\begin{align*}
 \frac{1}{p_k} \bm\Gamma^\kk_f \bm\Lambda_k^\top \bm\Lambda_k = \bm\Gamma^\kk_f
\end{align*}
and thus for all $j \in [r_k]$,
\begin{align*}
\frac{1}{p_k} \mu_j\l( \bm\Gamma^\kk_f \bm\Lambda_k^\top \bm\Lambda_k \r) = \mu^\kk_{f, j}. % \le \frac{C \mu^\kk_{f, 1}}{\sqrt{p_k}}.
\end{align*}
Since $r_k$ non-zero eigenvalues of $\wt{\bm\Gamma}^\kk_\chi$ are identical to those of $\bm\Gamma^\kk_f \bm\Lambda_k^\top \bm\Lambda_k$, the first statement holds.
The second statement follows from evoking~\eqref{eq:approx} with the first one.
\end{proof}

\begin{lemma}
\label{lem:idio:cov} Suppose that Assumption~\ref{assum:heavy}~\ref{cond:heavy:idio} and either of Assumptions~\ref{assum:indep}~\ref{cond:indep:mixing} or~\ref{assum:rf}~\ref{cond:rf:mixing} hold.
Then, there exists a constant $C_\epsilon > 0$ that may depend on $\epsilon$, such that
$\Vert \bm\Gamma^{\kk}_\xi \Vert \le C_\epsilon \omega^2$ for all $k \in [K]$.
\end{lemma}

\begin{proof}
Firstly, under Assumptions~\ref{assum:indep}~\ref{cond:indep:mixing}, $\bm\Gamma_\xi^\kk$ is diagonal such that
\begin{align*}
\l\Vert \bm\Gamma_\xi^{\kk} \r\Vert 
% &\le \max_{i \in [p_k]} \frac{1}{n p_{-k}} \sum_{t = 1}^n \sum_{\mbf i, \mbf i' \in \prod_{k' \in [K]} [p_{k'}]: \, i_k = i'_k = i} \Cov(\xi_{\mbf i, t}, \xi_{\mbf i', t}) \nn \\
&\le \max_{i \in [p_k]} \frac{1}{n p_{-k}} \sum_{t = 1}^n \sum_{\mbf i \in \prod_{k' \in [K]} [p_{k'}]: \, i_k = i} \Vert \xi_{\mbf i, t} \Vert_2^2 \le \omega^2.
\end{align*}
Next, let us write the elements of $\mat_k(\bm\xi_t)$ by $\xi_{k, i\ell, t}$ for $i \in [p_k]$ and $\ell \in [p_{-k}]$, and denote by $\Vert \cdot \Vert_1$ the matrix norm induced by the vector $\ell_1$-norm.
Under Assumption~\ref{assum:rf}~\ref{cond:rf:mixing}, we have for all $i \in [p_k]$, 
\begin{align*}
\Vert \bm\Gamma^{\kk}_\xi \Vert_1 \le & \, \max_{i \in [p_k]} \frac{1}{np_{-k}} \sum_{j \in [p_k]} \sum_{t \in [n]} \sum_{\ell \in [p_{-k}]} \l\vert \E\l( \xi_{k, i\ell, t} \xi_{k, j\ell, t} \r) \r\vert 
\\
\le & \, \max_{i \in [p_k]}  \frac{8}{np_{-k}} \sum_{j \in [p_k]} \sum_{t \in [n]} \sum_{\ell \in [p_{-k}]} \Vert \xi_{k, i\ell, t} \Vert_{2 + 2\eps} \Vert \xi_{k, j\ell, t} \Vert_{2 + 2\eps} \exp\l( - \frac{c_0 \eps \vert i - j \vert }{1 + \eps} \r) 
\\
\le & \, \max_{i \in [p_k]}  8 \omega^2 \sum_{j \in [p_k]} \exp\l( - \frac{c_0 \eps \vert i - j \vert }{1 + \eps} \r) \le C_\epsilon \omega^2,
\end{align*}
where the first inequality follows from Lemma~\ref{lem:hall} and the second from Assumption~\ref{assum:heavy}~\ref{cond:heavy:idio}.
This establishes that $\Vert \bm\Gamma^{\kk}_\xi \Vert \le \Vert \bm\Gamma^{\kk}_\xi \Vert_1 \le C_\epsilon \omega^2$.
\end{proof}

\section{Proofs}

Throughout, for a matrix $\mbf A = [a_{ii'}] \in \R^{m \times n}$, we write its Frobenius norm by $\Vert \mbf A \Vert_F = \vert \mbf A \vert_2$ and write $\vert \mbf A \vert_\infty = \max_{i \in [m]} \max_{i' \in [n]} \vert a_{ii'} \vert$.
By $\mbf O$ and $\mbf 0$, we denote a matrix or a vector of zeros whose dimensions depend on the context. 
We write $a_n \lesssim b_n$ and $a_n = O(b_n)$ interchangeably.
Also, with some abuse of notation, we write $\mbf A = \mbf B + o(1)$ for two matrices $\mbf A$ and $\mbf B$ of compatible, fixed dimensions, if the equality holds element-wise.

We write the pairs of eigenvalues and eigenvectors of $\bm\Gamma^\kk$ by $(\mu^\kk_j, \mbf e^\kk_j)$, with $\mu^\kk_1 \ge \ldots \ge \mu^\kk_{p_k}$, and similarly define $(\mu^\kk_{\chi, j}, \mbf e^\kk_{\chi, j})$ for $\bm\Gamma^\kk_\chi$ in~\eqref{eq:gamma:decomp}.
We also write $\mbf E_k = [\mbf e^\kk_1, \ldots, \mbf e^{\kk}_{r_k}]$ and $\mbf E_{\chi, k} = [\mbf e^\kk_{\chi, 1}, \ldots, \mbf e^{\kk}_{\chi, r_k}]$. [!!!]
Assumption~\ref{assum:factor} indicates that there exist pairs of fixed, positive constants $(\alpha^\kk_j, \beta^\kk_j), \, j \in [r_k]$, such that
$\beta^\kk_1 \ge \mu^\kk_{f, 1} \ge \alpha^\kk_1 > \beta^\kk_2 \ge \ldots \ge \alpha^\kk_{r_k - 1} > \beta^\kk_{r_k} \ge \mu^\kk_{f, r_k} \ge \alpha^\kk_{r_k}$, where $\mu^\kk_{f, j}$ denotes the $j$-th largest eigenvalue of $\bm\Gamma^\kk_f$.
We frequently write 
\begin{align*}
\mbf X_{k, t} &:= \mat_k(\mc X_t) = [X_{k, ii', t}, \, i \in [p_k], \, i' \in [p_{-k}]],
\text{ \ and \ }
\\
\mbf X^{\trunc}_{k, t}(\tau) &:= \mat_k(\mc X^{\trunc}_t(\tau)) = [X^{\trunc}_{k, ii', t}(\tau), \, i \in [p_k], \, i' \in [p_{-k}]],
\end{align*}
and $\mbf Z_{k, t}(\tau) = \mbf X^{\trunc}_{k, t}(\tau) - \E(\mbf X^{\trunc}_{k, t}(\tau))$.
% Let $\bm\varphi_i$ denote a vector of zeros except for its $i$-th element set at one; its dimension is determined by the context.
By $C_\eps > 0$, we denote a constant that depends only on $\eps$ which may differ from one instance to another.
% We suppress the dependence of the estimator on $\tau$ or $\kappa$ where there is no confusion.
We write $r_{-k} = r/r_k$ with $r = \prod_{k = 1}^K r_k$.
% Also, we sometimes write $\sum_{i_1 \in \mc A_1} \cdots \sum_{i_K \in \mc A_K} a_{i_1, \ldots, i_K} = \otimes_{k = 1}^K \sum_{i_k \in \mc A_k} a_{i_1, \ldots, i_K}$.

\subsection{Preliminary lemmas}

\begin{lemma}
\label{lem:one}
Suppose that Assumptions~\ref{assum:loading}~(ii) and~\ref{assum:heavy} hold.
Then for any $\mbf i, \mbf i' \in \prod_{k = 1}^K [p_k]$ and $t \in [n]$:
\begin{enumerate}[label = (\roman*)]
\item \label{lem:one:one} $\Vert X^{\trunc}_{\mbf i, t}(\tau) \Vert_\nu \le \Vert X_{\mbf i, t} \Vert_\nu \lesssim \vert \mc F_t \vert_2 + \omega$ for any $\nu \in [1, 2 + 2\eps]$.
\item \label{lem:one:four} $\E( \vert X^{\trunc}_{\mbf i, t}(\tau) X^{\trunc}_{\mbf i', t}(\tau) \vert^\nu ) \lesssim \tau^{2(\nu - 1 - \eps)} (\vert \mc F_t \vert_2^{2 + 2\eps} + \omega^{2 + 2\eps})$ for any $\nu \in [2, \infty)$.
\item \label{lem:one:two} $\vert \E(X^\trunc_{\mbf i, t}(\tau)) - \E(X_{\mbf i, t}) \vert \lesssim \tau^{- 1 - 2\eps} (\vert \mc F_t \vert_2^{2 + 2\eps} + \omega^{2 + 2\eps})$.
\item \label{lem:one:three} $\vert \E(X^\trunc_{\mbf i, t}(\tau) X^\trunc_{\mbf i', t}(\tau)) - \E(X_{\mbf i, t} X_{\mbf i', t}) \vert \lesssim \tau^{-2\eps} (\vert \mc F_t \vert_2^{2 + 2\eps} + \omega^{2 + 2\eps})$.
\end{enumerate}
\end{lemma}
\begin{proof}
For~\ref{lem:one:one}, the first inequality follows by construction. 
Under Assumption~\ref{assum:loading}~(ii),
\begin{align}
\label{eq:chiit:bound}
\vert \chi_{i_1 \ldots i_K, t} \vert = \l\vert \sum_{j_1 \in [r_1]} \ldots \sum_{j_K \in [r_K]} f_{j_1 \ldots j_K, t} \cdot \prod_{k = 1}^K \lambda_{i_k j_k} \r\vert
\le \prod_{k = 1}^K r_k \bar{\lambda}^K \vert \mc F_t \vert_2
\end{align}
by Cauchy-Schwarz inequality, for any $(i_1, \ldots, i_K)^\top \in \prod_{k = 1}^K [p_k]$.
Combined with Assumption~\ref{assum:heavy}~\ref{cond:heavy:idio}, we have
$\Vert X_{\mbf i, t} \Vert_\nu \le \vert \chi_{\mbf i, t} \vert + \Vert \xi_{\mbf i, t} \Vert_\nu \lesssim \vert \mc F_t \vert_2 + \omega$ by Minkowski inequality.

For~\ref{lem:one:four}, note that 
\begin{align*}
\E\l( \l\vert X^{\trunc}_{\mbf i, t}(\tau) X^{\trunc}_{\mbf i', t}(\tau) \r\vert^\nu \r) &\le \E\l[ \min\l( \l\vert X^{\trunc}_{\mbf i, t}(\tau) X^{\trunc}_{\mbf i', t}(\tau) \r\vert^\nu, \tau^{2\nu} \r) \r]
\\
&= \tau^{2\nu} \E\l[ \min\l( \l\vert \frac{X^{\trunc}_{\mbf i, t}(\tau) X^{\trunc}_{\mbf i', t}(\tau)}{\tau^2} \r\vert^\nu, 1 \r) \r]
\\
&\le \tau^{2\nu} \E\l( \l\vert \frac{X^{\trunc}_{\mbf i, t}(\tau) X^{\trunc}_{\mbf i', t}(\tau)}{\tau^2} \r\vert^{1 + \eps} \r)
\le \tau^{2(\nu - 1 - \eps)} \sqrt{\Vert X_{\mbf i, t} \Vert_{2 + 2\eps}^{2 + 2\eps} \Vert X_{\mbf i', t} \Vert_{2 + 2\eps}^{2 + 2\eps}}
\\
&\le C_\eps \tau^{2(\nu - 1 - \eps)} (\vert \mc F_t \vert_2^{2 + 2\eps} + \omega^{2 + 2\eps}),
\end{align*}
where the last inequality follows from~\ref{lem:one:one} and $C_r$ inequality.

For~\ref{lem:one:two}, we have
\begin{align*}
\l\vert \E(X^\trunc_{\mbf i, t}(\tau)) - \E(X_{\mbf i, t}) \r\vert
&\le \E\l[ (X_{\mbf i, t} - \mathsf{sign}(X_{\mbf i, t})) \cdot \mathbb{I}_{\{ \vert X_{\mbf i, t} \vert > \tau\}} \r] \le \E\l[ \vert X_{\mbf i, t} \vert \cdot \mathbb{I}_{\{ \vert X_{\mbf i, t} \vert > \tau\}} \r] 
\\
&\le \Vert X_{\mbf i, t} \Vert_{2 + 2\eps} \cdot \p(\vert X_{\mbf i, t} \vert > \tau)^{\frac{1 + 2\eps}{2 + 2\eps}}
\le \tau^{-1 - 2\eps} \Vert X_{\mbf i, t} \Vert_{2 + 2\eps}^{2 + 2\eps}
\\
&\le C_\eps \tau^{-1 - 2\eps} (\vert \mc F_t \vert_2^{2 + 2\eps} + \omega^{2 + 2\eps}),
\end{align*}
which follows from H\"{o}lder's, Markov and $C_r$ inequalities combined with~\ref{lem:one:one}.

For~\ref{lem:one:three}, note that for all $\mbf i, \mbf i' \in \prod_{k = 1}^K [p_k]$ and $t \in [n]$, we have
\begin{align*}
& \l\vert X^{\trunc}_{\mbf i, t} X^{\trunc}_{\mbf i', t} - X_{\mbf i, t} X_{\mbf i', t} \r\vert 
\le 
\l\vert \l(X_{\mbf i, t}X_{\mbf i', t} - \mathsf{sign}(X_{\mbf i, t}X_{\mbf i', t}) \tau^2\r) \mathbb{I}_{\{\vert X_{\mbf i, t} \vert > \tau\}} \mathbb{I}_{\{\vert X_{\mbf i', t} \vert > \tau\}} \r\vert
\\
& +
\l\vert X_{\mbf i, t} \mathbb{I}_{\{\vert X_{\mbf i, t} \vert \le \tau\}} \l( X_{\mbf i', t} - \mathsf{sign}(X_{\mbf i', t}) \tau \r) \mathbb{I}_{\{\vert X_{\mbf i', t} \vert > \tau\}} \r\vert
+
\l\vert X_{\mbf i', t} \mathbb{I}_{\{\vert X_{\mbf i', t} \vert \le \tau\}} \l( X_{\mbf i, t} - \mathsf{sign}(X_{\mbf i, t}) \tau \r) \mathbb{I}_{\{\vert X_{\mbf i, t} \vert > \tau\}} \r\vert
\\
\le & \, \vert X_{\mbf i, t}X_{\mbf i', t} \vert \l( \mathbb{I}_{\{\vert X_{\mbf i, t} \vert > \tau\}} \mathbb{I}_{\{\vert X_{\mbf i', t} \vert > \tau\}}  + \mathbb{I}_{\{\vert X_{\mbf i, t} \vert > \tau\}} \mathbb{I}_{\{\vert X_{\mbf i', t} \vert \le \tau\}} + \mathbb{I}_{\{\vert X_{\mbf i, t} \vert \le \tau\}} \mathbb{I}_{\{\vert X_{\mbf i', t} \vert > \tau\}} \r)
\\
\le & \, \vert X_{\mbf i, t}X_{\mbf i', t} \vert \l( \mathbb{I}_{\{\vert X_{\mbf i, t} \vert > \tau\}} + \mathbb{I}_{\{\vert X_{\mbf i', t} \vert > \tau\}} \r).
\end{align*}
Then by~\ref{lem:one:one}, we have
\begin{align*}
& \, \l\vert \E(X^\trunc_{\mbf i, t}(\tau) X^\trunc_{\mbf i', t}(\tau)) - \E(X_{\mbf i, t} X_{\mbf i', t}) \r\vert
\le \E\l[ \l\vert X_{\mbf i, t}X_{\mbf i', t} \r\vert \l( \mathbb{I}_{\{\vert X_{\mbf i, t} \vert > \tau\}} + \mathbb{I}_{\{\vert X_{\mbf i', t} \vert > \tau\}} \r) \r]
\\
&\le \Vert X_{\mbf i, t} \Vert_{2 + 2\eps} \Vert X_{\mbf i', t} \Vert_{2 + 2\eps} 
\l[ \p( \vert X_{\mbf i, t} \vert > \tau ) + \p( \vert X_{\mbf i', t} \vert > \tau ) \r]^{\frac{\eps}{1 + \eps}}
\\
&\lesssim \Vert X_{\mbf i, t} \Vert_{2 + 2\eps} \Vert X_{\mbf i', t} \Vert_{2 + 2\eps} \cdot \frac{\Vert X_{\mbf i, t} \Vert_{2 + 2\eps}^{2\eps} + \Vert X_{\mbf i', t} \Vert_{2 + 2\eps}^{2\eps}}{\tau^{2\eps}}
\\
&\le C_\eps \tau^{-2\eps} (\vert \mc F_t \vert_2^{2 + 2\eps} + \omega^{2 + 2\eps}).
\end{align*}
\end{proof}

\begin{lemma}[Lemma~1 of \cite{wang2022rate}]
\label{lem:mixing}
For any $\alpha$-mixing time series $\{Y_t\}_{t \in \Z}$ and measurable function $f(\cdot)$, the sequence of the transformed process $\{ f(Y_t) \}_{t \in \Z}$ is also $\alpha$-mixing with its mixing coefficients bounded by those of the original sequence.
\end{lemma}

\begin{lemma}[Corollary A.2 of \cite{hall1980}]
\label{lem:hall}
Suppose that $X$ and $Y$ are random variables which are $\mc G$ and $\mc H$-measurable, respectively, for $\sigma$-algebra $\mc G$ and $\mc H$, and that $\Vert X \Vert_{\nu_1}, \Vert Y \Vert_{\nu_2} < \infty$ for some $\nu_1, \nu_2 \in (1, \infty)$ with $\nu_1^{-1} + \nu_2^{-1} < 1$.
Then,
\begin{align*}
\vert \Cov(X, Y) \vert \le 8 \Vert X \Vert_{\nu_1} \Vert Y \Vert_{\nu_2} [\alpha(\mc G, \mc H)]^{1 - \nu_1^{-1} - \nu_2^{-1}},
\end{align*}
where $\alpha(\mc G, \mc H) = \sup_{A \in \mc G, B \in \mc H} \vert \p(A \cap B) - \p(A) \p(B) \vert$.
\end{lemma}

\begin{lemma}
\label{lem:dk}
Let $\mbf S, \wt{\mbf S}, \wh{\mbf S} \in \R^{p \times p}$ % \mbf S = \bm\Gamma^\kk_\chi, \wt{\mbf S} = \E( XX'), \wt{\mbf S} = XX'
denote symmetric, non-negative definite matrices fulfilling (a subset of) the following conditions.
\begin{enumerate}[label = (C\arabic*)]
\item \label{lem:dk:cond:one} $\mbf S$ has $r$ non-zero eigenvalues $\mu_j, \, j \in [r]$ with $r \le p$, satisfying $\vert p^{-1} \mu_j - \gamma_j \vert = o(1)$, where
\begin{align*}
\beta_1 \ge \frac{\gamma_1}{p} \ge \alpha_1 > \beta_2 \ge \ldots \ge \alpha_{r - 1} > \beta_r \ge \frac{\gamma_r}{p} \ge \alpha_r > 0,
\end{align*}
with some pairs of positive constants $(\alpha_j, \beta_j), \, j \in [r]$.

\item \label{lem:dk:cond:two} $p^{-1} \Vert \wh{\mbf S} - \wt{\mbf S} \Vert_F = O_P(\zeta_{n, p})$ where $\zeta_{n, p} \to 0$ as $n, p \to \infty$.

% \item \label{lem:dk:cond:three} $\vert \wh{\mbf S} - \wt{\mbf S} \vert_\infty = O_P(\bar{\zeta}_{n, p})$ where $\bar{\zeta}_{n, p} \to 0$ as $n, p \to \infty$.

\item \label{lem:dk:cond:four} $\Vert \wt{\mbf S} - \mbf S \Vert \le C_1 < \infty$.

\item \label{lem:dk:cond:five} $\max(\vert \mbf S \vert_\infty, \vert \wt{\mbf S} \vert_\infty) \le C_2 < \infty$.
\end{enumerate}
Denoting the pairs of eigenvalues and eigenvectors of $\wh{\mbf S}$ by $(\wh\mu_j, \wh{\mbf e}_j), \, j \ge 1$, let us write $\wh{\mbf M} = \text{diag}(\wh\mu_j, \, j \in [r])$ and $\wh{\mbf E} = [\wh{e}_{ij}, \, i \in [p], \, j \in [r]]$, and analogously define $(\mu_j, \mbf e_j)$, $\mbf M$ and $\mbf E$ (resp.\ $(\wt\mu_j, \wt{\mbf e}_j)$, $\wt{\mbf M}$ and $\wt{\mbf E}$) for $\mbf S$ (resp.\ $\wt{\mbf S}$).

\begin{enumerate}[wide, label = (\roman*)]
\item \label{lem:dk:one} 
Suppose that~\ref{lem:dk:cond:one}, \ref{lem:dk:cond:two} and~\ref{lem:dk:cond:four} hold.
Then, there exist diagonal matrices $\mbf J, \wt{\mbf J} \in \R^{p \times p}$ with $\pm 1$ on their diagonal such that 
\begin{align*}
\Vert \wh{\mbf E} - \wt{\mbf E} \wt{\mbf J} \Vert_F = O_P(\zeta_{n, p}) \text{ \ and \ } \Vert \wh{\mbf E} - \mbf E \mbf J \Vert_F = O_P(\zeta_{n, p} \vee p^{-1}).
\end{align*}

\item \label{lem:dk:two} Suppose that~\ref{lem:dk:cond:one}, \ref{lem:dk:cond:two} and~\ref{lem:dk:cond:four} hold.
Then, 
\begin{align*}
p^{-1} \Vert \wh{\mbf M} - \wt{\mbf M} \Vert = O_P(\zeta_{n, p}), \quad p^{-1} \Vert \wh{\mbf M} - \mbf M \Vert = O_P(\zeta_{n, p} \vee p^{-1}), 
\\
p \Vert \wh{\mbf M}^{-1} - \wt{\mbf M}^{-1} \Vert = O_P(\zeta_{n, p}) \text{ \ and \ } 
p \Vert \wh{\mbf M}^{-1} - \mbf M^{-1} \Vert = O_P(\zeta_{n, p} \vee p^{-1}).
\end{align*}

\item \label{lem:dk:three} Suppose that~\ref{lem:dk:cond:one}, \ref{lem:dk:cond:four} and~\ref{lem:dk:cond:five} hold. Then, there exists a constant $C_3 > 0$ such that
\begin{align*}
\max_{j \in [r]} \max_{i \in [p]} \vert \wt e_{ij} \vert \le C_3 p^{-1/2} \text{ \ and \ } \max_{j \in [r]} \max_{i \in [p]} \vert e_{ij} \vert \le C_3 p^{-1/2}.    
\end{align*}

%\item \label{lem:dk:four} Suppose that~\ref{lem:dk:cond:one}--\ref{lem:dk:cond:five} hold. Then, $\max_{j \in [r]} \max_{i \in [p]} \vert \wh e_{ij} \vert = O_P(p^{-1/2})$.
\end{enumerate}
\end{lemma}

\begin{proof}
For~\ref{lem:dk:one}, note that by Chebyshev's inequality and~\ref{lem:dk:cond:two}, we have
\begin{align*}
& \frac{1}{p} \Vert \wh{\mbf S} - \wt{\mbf S} \Vert
\le 
\frac{1}{p} \Vert \wh{\mbf S} - \wt{\mbf S} \Vert_F
= O_P\l( \zeta_{n, p} \r).
\end{align*}
Also by Weyl's inequality and~\ref{lem:dk:cond:four}, we have $\vert \wt\mu_j - \mu_j \vert \le C_1$ for all $j \in [p]$.
Then by Theorem~2 of \cite{yu2015} and~\ref{lem:dk:cond:one}, there exists such $\wt{\mbf J}$ satisfying
\begin{align*}
\Vert \wh{\mbf E} - \wt{\mbf E} \wt{\mbf J} \Vert_F &\le 
\frac{2 \sqrt{2r} \Vert \wh{\mbf S} - \wt{\mbf S} \Vert}
{\min(\wt\mu_1 - \wt\mu_2, \wt\mu_r - \wt\mu_{r + 1})}
\\
&= O_P\l( \frac{p \zeta_{n, p}}{\min( (\alpha_1 - \beta_2)p - 2C_1, \alpha_r p - 2C_1)} \r) = O_P(\zeta_{n, p}).
\end{align*}
The second result follows from the analogous arguments.

For~\ref{lem:dk:two}, by Weyl's inequality and~\ref{lem:dk:cond:two}, for all $j \in [p]$,
\begin{align*}
\frac{1}{p} \vert \wh\mu_j - \wt\mu_j \vert
\le \frac{1}{p} \Vert \wh{\mbf S} - \wt{\mbf S} \Vert
= O_P\l(\zeta_{n, p}\r),
\end{align*}
from which the first statement follows.
Also from~\ref{lem:dk:cond:one} and~\ref{lem:dk:cond:four}, $p^{-1} \mu_r \ge \alpha_r + o(1)$ and thus $p^{-1} \wt{\mu}_r \ge \alpha_r + o(1) - C_1 p^{-1}$ and $p^{-1} \wh{\mu}_r \ge \alpha_r + o(1) - C_1 p^{-1} + O_P(\zeta_{n, p})$, which imply that the matrices $p^{-1} \wt{\mbf M}$ and $p^{-1}\wh{\mbf M}$ are asymptotically invertible with
\begin{align*}
\l\Vert \l( p^{-1} \wt{\mbf M} \r)^{-1} \r\Vert \le \frac{1}{\alpha_r(1 + o(1))}
\text{ \ and \ }
\l\Vert \l( p^{-1} \wh{\mbf M} \r)^{-1} \r\Vert \le \frac{1}{\alpha_r(1 + o_P(1)}.
\end{align*}
Further, we have
\begin{align*}
& \l\Vert \l( p^{-1} \wh{\mbf M} \r)^{-1} - \l( p^{-1} \wt{\mbf M} \r)^{-1} \r\Vert
\le \l\Vert \l( p^{-1} \wh{\mbf M} \r)^{-1} - \l( p^{-1} \wt{\mbf M} \r)^{-1} \r\Vert_F
= \sqrt{p^2 \sum_{j \in [r]} \l( \frac{1}{\wh\mu_j} - \frac{1}{\wt\mu_j} \r)^2}
\\
\le \, & \sum_{j \in [r]} \frac{p^{-1}\vert \wt\mu_j - \mu_j\vert}
{p^{-1} \wt\mu_j \cdot p^{-1}\mu_j}
= O_P\l( \frac{r p^2}{\mu_r^2} \zeta_{n, p} \r).
\end{align*}
The second and the fourth claims follow similarly.

To see that~\ref{lem:dk:three} holds, note that
\begin{align*}
\wt s_{ii} = \sum_{j \in [r]} \wt\mu_j \vert \wt e_{ij} \vert^2 \le C_2,
\end{align*}
which implies that $\max_{j \in [r]} \max_{i \in [p]} \vert \wt e_{ij} \vert \le C_2/\sqrt{p(\alpha_r + o(1)) - C_1} \le C_3/\sqrt{p}$ for some $C_3 > 0$.
Similarly, the second claim follows. 

% As for~\ref{lem:dk:four}, note that $\max_{i \in [p]} \wh s_{ii} = \sum_{j \in [p]} \wh\mu_j \vert \wh e_{ij} \vert^2 \le C_2 + O_P(\bar{\zeta}_{n, p})$ and that for all $j \in [r]$, by~\ref{lem:dk:two},
% \begin{align*}
% \frac{1}{\sqrt{\wh\mu_j}} \le \frac{1}{\sqrt{p}} \cdot \frac{1}{\sqrt{\alpha_j + O_P(\zeta_{n, p} \vee p^{-1})}}
% = O_P\l(\frac{1}{\sqrt p}\r).
% \end{align*}
% Hence, it follows that $\max_{j \in [r]} \max_{i \in [p]} \vert \wh e_{ij} \vert = O(\wh\mu_j^{-1/2}) = O_P(p^{-1/2})$.
\end{proof}

\subsection{Proof of Theorem~\ref{thm:first}}

We first derive an initial rate of estimation for $\wh{\bm\Gamma}^\kk(\tau)$, from which that of $\wh{\bm\Lambda}_k(\tau)$ follows; {\cgr under Assumption~\ref{assum:rf}, the second claim in~\eqref{eq:prop:one:two} gives the rate in Theorem~\ref{thm:first} together with Lemma~\ref{lem:e:lambda}.}
\begin{proposition}
\label{prop:first:init}
{\it Suppose that Assumptions~\ref{assum:loading}, \ref{assum:factor} and \ref{assum:heavy} hold, as well as either of Assumption~\ref{assum:indep} or Assumption~\ref{assum:rf}. 
For each $k \in [K]$, recalling the definition of $\tau^{\kk}_{n, p}$ in~\eqref{eq:tau}, we set $\tau \asymp \tau^\kk_{n, p}$.
Then, there exist diagonal matrices $\mbf J_k, \wh{\mbf J}_k \in \R^{r_k \times r_k}$ with $\pm 1$ on their diagonal entries such that as $\min(n, p_1, \ldots, p_K) \to \infty$,
\begin{align}
& \frac{1}{p_k} \l\Vert \wh{\bm\Gamma}^\kk(\tau) - \bm\Gamma^\kk_\chi \r\Vert = O_P\l( \psi^{\kk}_{n, p} \vee \frac{1}{p_k} \r),
\label{eq:prop:one:one}
\\
& \l\Vert \wh{\mbf E}_k(\tau) - \mbf E_k \mbf J_k \r\Vert = O_P\l( \psi^\kk_{n, p} \r) \text{ \ and \ }
\l\Vert \wh{\mbf E}_k(\tau) - \mbf E_{\chi, k} \wh{\mbf J}_k \r\Vert = O_P\l( \psi^\kk_{n, p}  \vee \frac{1}{p_k} \r),
\label{eq:prop:one:two}
\end{align}
with $\psi^{\kk}_{n, p}$ defined in~\eqref{eq:psi}.}
\end{proposition}

From here on, we make Assumption~\ref{assum:indep}.
For each $k \in [K]$, denote by $\wh{\mbf M}_k(\tau) \in \R^{r_k \times r_k}$ the diagonal matrix containing the eigenvalues $\wh{\mu}^\kk_j(\tau), \, j \in [r_k]$, of $\wh{\bm\Gamma}^\kk(\tau)$ on its diagonal. 
From now on, we suppress the dependence on $\tau$ where there is no confusion.
By Weyl's inequality,~\eqref{eq:prop:one:one}, Lemma~\ref{lem:common:cov} and Assumption~\ref{assum:factor}, we have
\begin{align}
\label{eq:wh:M:bound:one}
\Lambda_{\max}\l( p_k^{-1} \wh{\mbf M}_k \r) \le 
\beta^\kk_1 + o_P(1) 
% O_P\l( \psi^\kk_{n, p} \vee \frac{1}{p_k} \r) 
\text{ \, and \, }
\Lambda_{\min}\l( p_k^{-1} \wh{\mbf M}_k \r) \ge \alpha^\kk_{r_k} + o_P(1),
% O_P\l( \psi^\kk_{n, p} \vee \frac{1}{p_k} \r),
\end{align}
which ensures the asymptotic invertibility of $p_k^{-1} \wh{\mbf M}_k$. 
Also Lemma~\ref{lem:first:eigval} shows that
\begin{align}
\label{eq:wh:M:bound:two}
\l\Vert \l( p_k^{-1} \wh{\mbf M}_k \r)^{-1} \r\Vert \le 
\l\Vert \l( p_k^{-1} \mbf M_{\chi, k} \r)^{-1} \r\Vert
+ O_P\l( \psi^\kk_{n, p} \vee \frac{1}{p_k} \r) = \frac{1}{\alpha^\kk_{r_k}} + o_P(1) = O_P(1),
\end{align}
where $ \mbf M_{\chi, k} = \text{diag}(\mu^\kk_{\chi, 1}, \ldots, \mu^\kk_{\chi, r_k})$.
Next, we decompose $\wh{\bm\Gamma}^\kk$ as
\begin{align}
\frac{1}{p_k} \wh{\bm\Gamma}^\kk % &= \frac{1}{np} \sum_{t \in [n]} \mbf X^{\trunc}_{k, t} \l( \mbf X^{\trunc}_{k, t} \r)^\top \nn \\
&= \frac{1}{np} \sum_{t \in [n]} \mbf Z_{k, t} \mbf Z_{k, t}^\top 
\nn \\
\quad & + \frac{1}{np} \sum_{t \in [n]} \mbf Z_{k, t} \E\l( \mbf X^{\trunc}_{k, t} \r)^\top + \frac{1}{np} \sum_{t \in [n]} \E\l( \mbf X^{\trunc}_{k, t} \r) \mbf Z_{k, t}^\top
\nn \\
\quad & + \frac{1}{np} \sum_{t \in [n]} \E\l( \mbf X^{\trunc}_{k, t} \r) \E\l( \mbf X^{\trunc}_{k, t} - \mbf X_{k, t} \r)^\top + \frac{1}{np} \sum_{t \in [n]} \E\l( \mbf X^{\trunc}_{k, t} - \mbf X_{k, t} \r) \E\l( \mbf X^{\trunc}_{k, t} \r)^\top
\nn \\
\quad & + \frac{1}{np} \sum_{t \in [n]} \E(\mbf X_{k, t}) \E(\mbf X_{k, t})^\top
=: T_1 + T_{2, 1} + T_{2, 2} + T_{3, 1} + T_{3, 2} + \frac{1}{p_k} \bm\Gamma^\kk_\chi,
\label{eq:wh:gamma:decomp}
\end{align}
where the last equality follows from Assumption~\ref{assum:loading}~(i).
Then, noting that $\wh{\bm\Gamma}^\kk \wh{\mbf E}_k = \wh{\mbf E}_k \wh{\mbf M}_k$, making use of the decomposition in~\eqref{eq:wh:gamma:decomp}, we have
\begin{align*}
& \wh{\mbf E}_k - {\cgr \frac{1}{\sqrt{p_k}} \bm\Lambda_k} \wh{\mbf H}_k = \l( T_1 + T_{2, 1} + T_{2, 2} + T_{3, 1} + T_{3, 2} \r) \wh{\mbf E}_k \l( \frac{1}{p_k} \wh{\mbf M}_k \r)^{-1}, 
\\
& \text{with \ } \wh{\mbf H}_k = {\cgr \frac{1}{n \sqrt{p_k}} \sum_{t \in [n]} \mat_k(\mc F_t) \mat_k(\mc F_t)^\top \bm\Lambda_k^\top \wh{\mbf E}_k \l( \frac{1}{p_k} \wh{\mbf M}_k \r)^{-1},}
\end{align*}
by Assumptions~\ref{assum:loading}~(i).
This, together with Lemmas~\ref{lem:thm:first:t:one}--\ref{lem:thm:first:t:three} and~\eqref{eq:wh:M:bound:two}, shows that
\begin{align}
\frac{1}{p_k} \l\Vert \wh{\bm\Gamma}^\kk - \bm\Gamma^\kk_\chi \r\Vert
&= O_P \l( \frac{M_n^{1 - \eps}}{\sqrt{np_{-k}}} \vee \frac{1}{p_k} \vee \frac{\psi^\kk_{n, p}}{\sqrt{p_k}} \r),
\label{eq:first:cov:rate}
\\
\l\Vert \wh{\mbf E}_k - {\cgr \frac{1}{\sqrt{p_k}} \bm\Lambda_k} \wh{\mbf H}_k \r\Vert
&= O_P \l( \frac{M_n^{1 - \eps}}{\sqrt{np_{-k}}} \vee \frac{1}{p_k} \vee \frac{\psi^\kk_{n, p}}{\sqrt{p_k}} \r).
\label{eq:first:loading:rate}
\end{align}
We conclude the proof by noting that
\begin{align*}
\Vert \wh{\mbf H}_k \Vert &\le \frac{1}{n} \sum_{t \in [n]} \vert \mc F_t \vert_2^2 \cdot \frac{1}{\sqrt{p_k}} \Vert \bm\Lambda_k \Vert \Vert \wh{\mbf E}_k \Vert \l\Vert \l( \frac{1}{p_k} \wh{\mbf M}_k \r)^{-1} \r\Vert \le (\alpha^\kk_{r_k})^{-1} \omega^2 (1 + o_P(1)) 
\end{align*}
by Assumptions~\ref{assum:heavy}~\ref{cond:heavy:factor} and~\eqref{eq:wh:M:bound:two}, and hence
\begin{align}
\label{eq:hh}
\mbf I_{r_k} = \wh{\mbf E}_k^\top \wh{\mbf E}_k = {\cgr \frac{1}{\sqrt{p_k}} \wh{\mbf E}_k^\top \bm\Lambda_k} \wh{\mbf H}_k + o_P(1) = {\cgr \frac{1}{p_k} \wh{\mbf H}_k^\top \bm\Lambda_k^\top \bm\Lambda_k \wh{\mbf H}_k + o_P(1)} = \wh{\mbf H}_k^\top \wh{\mbf H}_k + o_P(1),
\end{align}
by Assumptions~\ref{assum:loading}~(i).

\subsubsection{Proof of Proposition~\ref{prop:first:init}}

\begin{proof}[Proof of~\eqref{eq:prop:one:one}]
We suppress the dependence on $\tau$ where there is no confusion.
Denote by $\bm\Gamma^\kk = [\gamma^\kk_{ij}]_{i, j \in [p_k]}$ and $\wh{\bm\Gamma}^\kk = [\wh\gamma^\kk_{ij}]_{i, j \in [p_k]}$.
Then, 
\begin{align*}
\l\vert \wh\gamma^\kk_{ij} - \gamma^\kk_{ij} \r\vert &\le 
\l\vert \E(\wh\gamma^\kk_{ij}) - \gamma^\kk_{ij} \r\vert + \l\vert \wh\gamma^\kk_{ij} - \E(\wh\gamma^\kk_{ij}) \r\vert =: T_{1, ij} + T_{2, ij}.
\end{align*}
By Lemma~\ref{lem:one}~\ref{lem:one:three} and Assumption~\ref{assum:heavy}~\ref{cond:heavy:factor}, we have
\begin{align}
T_{1, ij} &\le \frac{1}{np_{-k}} \sum_{t \in [n]} \sum_{i' \in [p_{-k}]} \l\vert \E(X^\trunc_{k, ii', t} X^\trunc_{k, ji', t}) - \E(X_{k, ii', t} X_{k, ji', t}) \r\vert
\nn \\
&\lesssim \frac{1}{n \tau^{2\eps}} \sum_{t \in [n]} ( \vert \mc F_t \vert_2^{2 + 2\eps} + \omega^{2 + 2\eps} ) \le \frac{2\omega^{2 + 2\eps}}{\tau^{2\eps}}.
\label{eq:t1}
\end{align}
% Then, with $\tau = \tau^{\kk}_{n, p}$ given in~\eqref{eq:tau} and recalling the definition of $\psi^{\kk}_{n, p}$ in~\eqref{eq:psi}, we have % from~\eqref{eq:t1},
% \begin{align}
% \label{eq:t1}
% \max_{i, j \in [p]} T_{1, ij} \lesssim \psi^\kk_{n, p}.
% \end{align}
Next, we proceed to bound $T_{2, ij}$ separately under Assumptions~\ref{assum:indep} and~\ref{assum:rf}.

\paragraph{Under Assumption~\ref{assum:indep}.}
By Lemma~\ref{lem:mixing}, the sequence $\{Y_{ij, \ell, t}, \, t \in [n], \, \ell \in [p_{-k}]\}$ formed by concatenating $\{Y_{ij, \ell, t}\}_{t \in [n]}$ with $Y_{ij, \ell, t} = X^{\trunc}_{k, i\ell, t} X^{\trunc}_{k, j\ell, t} - \E(X^{\trunc}_{k, i\ell, t} X^{\trunc}_{k, j\ell, t})$, is also $\alpha$-mixing with the mixing coefficient as in Assumption~\ref{assum:indep}~\ref{cond:indep:mixing}. 
%, in addition to being bounded as $\vert Y_{ij, \ell, t} \vert \le 2\tau^2$.
Then for some $\nu > 2$, by Lemma~\ref{lem:hall}, we have
\begin{align}
& \vert \Cov(Y_{ij, \ell, t}, Y_{ij, \ell, u}) \vert \le 8 \Vert Y_{ij, \ell, t} \Vert_\nu \Vert Y_{ij, \ell, u} \Vert_\nu (\alpha(\vert t- u \vert))^{\frac{\nu - 2}{\nu}}, 
\text{ \ where} \label{eq:vn:one}
\\
& \Vert Y_{ij, \ell, t} \Vert_\nu^\nu = \E\l( \l\vert X^{\trunc}_{i\ell, t} X^{\trunc}_{j\ell, t} - \E(X^{\trunc}_{i\ell, t} X^{\trunc}_{j\ell, t}) \r\vert^\nu \r) \le 2^\nu \E\l( \l\vert X^{\trunc}_{i\ell, t} X^{\trunc}_{i'j\ell, t} \r\vert^\nu \r) \label{eq:vn:two}
\end{align}
by $C_r$ inequality.
Noting that 
\begin{align*}
T_{2, ij} = \frac{1}{np_{-k}} \sum_{\ell \in [p_{-k}]} \sum_{t \in [n]} Y_{ij, \ell, t},
\end{align*}
we combine~\eqref{eq:vn:one}--\eqref{eq:vn:two} with Lemma~\ref{lem:one}~\ref{lem:one:four} and Assumption~\ref{assum:indep} and obtain
\begin{align*}
\Vert T_{2, ij} \Vert_2^2 &\le \frac{1}{(np_{-k})^2} \sum_{t, u \in [n]} \sum_{\ell \in [p_{-k}]} \vert \Cov(Y_{ij, \ell, t}, Y_{ij, \ell, u}) \vert
\\
&\lesssim \frac{\tau^{\frac{4(\nu - 1 - \eps)}{\nu}}}{n^2p_{-k}} \sum_{t, u \in [n]} (\vert \mc F_t \vert_2 + \omega)^{\frac{2 + 2\eps}{\nu}} (\vert \mc F_u \vert_2 + \omega)^{\frac{2 + 2\eps}{\nu}} \exp\l( - \frac{c_0 (\nu - 2) \vert t - u \vert}{\nu} \r)
\\
&\lesssim \frac{\tau^{2 - 2\eps}}{n^2p_{-k}} \sum_{t, u \in [n]} (\vert \mc F_t \vert_2 + \omega)^{1 + \eps} (\vert \mc F_u \vert_2 + \omega)^{1 + \eps} \exp\l( - \frac{c_0 \vert t - u \vert}{3\log(np_{-k})} \r)
\\
&\lesssim \frac{\omega^{2 + 2\eps} c_\eps \tau^{2 - 2\eps} \log(np_{-k})}{np_{-k}},
\end{align*}
by setting $\nu = 2 + \log^{-1}(np_{-k})$. 
In third inequality above, we use that 
\begin{align*}
\tau^{\frac{4(\nu - 1 - \eps)}{\nu}} \le \tau^{2(1 - \eps + \log^{-1}(np_{-k}))} \lesssim \tau^{2 - 2\eps}.
\end{align*}
Noting that the upper bounds on $\vert T_{1, ij} \vert$ and $\Vert T_{2, ij} \Vert_2$ do not depend on $i$, $j$ or $k$, it follows that
\begin{align*}
\frac{1}{p_k} \l\Vert \wh{\bm\Gamma}^\kk - \bm\Gamma^\kk \r\Vert_F &\le \sqrt{\frac{1}{p_k^2} \sum_{i, j \in [p_k]} T_{1, ij}^2} + \frac{1}{p_k} \l\Vert \wh{\bm\Gamma}^\kk - \E(\wh{\bm\Gamma}^\kk) \r\Vert_F 
\\
&= O_P\l( \frac{\omega^{2 + 2\eps}}{\tau^{2\eps}} + \omega^{1 + \eps} \tau^{1 - \eps} \sqrt{\frac{\log(np_{-k})}{np_{-k}}} \r) 
= O_P\l( \omega^2 \l( \frac{\log(np_{-k})}{np_{-k}} \r)^{\frac{\eps}{1 + \eps}} \r)
\end{align*}
by Markov's inequality, with $\tau \asymp \tau^\kk_{n, p}$ in~\eqref{eq:tau}. 

\paragraph{Under Assumption~\ref{assum:rf}.}
WLOG, for notational convenience, we fix $k = 1$ and denote by $\mbf i_{2:K} = (i_2, \ldots, i_K)$.
Let us define $Y_{ij, \mbf i_{2:K}, t} = X^{\trunc}_{1, (i, \mbf i_{2:K}), t} X^{\trunc}_{1, (j, \mbf i_{2:K}), t} - \E(X^{\trunc}_{1, (i, \mbf i_{2:K}), t} X^{\trunc}_{1, (j, \mbf i_{2:K}), t})$ for $i, j \in [p_1]$.
By Lemma~\ref{lem:mixing}, we have $\mbf Y = \{Y_{ij, \mbf i_{2:K}, t}, \, i_l \in [p_l], \, 2 \le l \le K, \,  t \in [n]\}$, a strongly mixing random field with the mixing coefficient as in Assumption~\ref{assum:rf}~\ref{cond:rf:mixing}.
Then by Lemma~\ref{lem:hall} and Assumption~\ref{assum:rf}~\ref{cond:rf:mixing}, we have for some $\nu > 2$,
\begin{align*}
& \l\vert \Cov(Y_{ij, \mbf i_{2:K}, t}, Y_{ij, \mbf i'_{2:K}, u}) \r\vert 
\\
\le & \, 8 \Vert Y_{ij, \mbf i_{2:K}, t} \Vert_\nu \Vert Y_{ij, \mbf i'_{2:K}, u} \Vert_\nu \exp\l( - \frac{c_0(\nu - 2)(\vert t - u \vert + \sum_{l = 2}^K \vert i_l - i'_l \vert)}{K\nu} \r),
\end{align*}
from the observation that $\max(\max_{2 \le l \le K} \vert i_l - i'_l \vert, \vert t - u \vert) \ge K^{ - 1} (\sum_{l = 2}^K \vert i_l - i'_l \vert + \vert t - u \vert)$.
Further, 
\begin{align*}
\Vert Y_{ij, \mbf i_{2:K}, t} \Vert_\nu^\nu \lesssim \tau^{2(\nu - 1 - \eps)} (\vert \mc F_t \vert_2^{2 + 2\eps} + \omega^{2 + 2\eps})
\end{align*}
for all $i, j, \mbf i_{2:K}$ and $t$, see~\eqref{eq:vn:two} and Lemma~\ref{lem:one}~\ref{lem:one:four}.
Noticing that
\begin{align*}
T_{2, ij} = \frac{1}{np_{-1}} \sum_{t \in [n]} \otimes_{l = 2}^K \sum_{i_l \in [p_l]} Y_{ij, \mbf i_{2:K}, t},
\end{align*}
we have with $\nu = 2 + \log^{-1}(np)$,
\begin{align*}
& \Vert T_{2, ij} \Vert_2^2 
\le \frac{1}{(np_{-1})^2} \sum_{t, u \in [n]} \otimes_{l = 2}^K \sum_{i_l, i'_l \in [p_l]} \Cov(Y_{ij, \mbf i_{2:K}, t}, Y_{ij, \mbf i'_{2:K}, u})
\\
\lesssim & \, \frac{1}{(np_{-1})^2} \sum_{t, u \in [n]} \otimes_{l = 2}^K \sum_{i_l, i'_l \in [p_l]} \Vert Y_{ij, \mbf i_{2:K}, t} \Vert_\nu \Vert Y_{ij, \mbf i'_{2:K}, u} \Vert_\nu 
\\
& \qquad \qquad \times \exp\l( - \frac{c_0(\nu - 2)(\vert t - u \vert + \sum_{l = 2}^K \vert i_l - i'_l \vert)}{K\nu} \r)
\\
\lesssim & \, \frac{\tau^{\frac{4(\nu - 1 - \eps)}{\nu}}}{np_{-1}} \cdot \frac{1}{n} \sum_{t, u \in [n]} ( \vert \mc F_t \vert_2 + \omega)^{\frac{2 + 2\eps}{\nu}} ( \vert \mc F_u \vert_2 + \omega)^{\frac{2 + 2\eps}{\nu}} \exp\l( - \frac{c_0 (\nu - 2) \vert t - u \vert}{K \nu} \r) 
\\ 
& \qquad \qquad \qquad \times \prod_{l = 2}^K \frac{1}{p_l} \sum_{i_l, i'_l \in [p_l]} \exp\l( -\frac{c_0 (\nu - 2) \vert i_l - i'_l \vert}{K \nu} \r)
\\
\lesssim & \, \frac{\tau^{2 - 2\eps}}{np_{-1}} \cdot \frac{1}{n} \sum_{t, u \in [n]} ( \vert \mc F_t \vert_2 + \omega)^{1 + \eps} ( \vert \mc F_u \vert_2 + \omega)^{1 + \eps} \exp\l( - \frac{c_0 \vert t - u \vert}{3 K \log(np)} \r) 
\\ 
& \qquad \qquad \qquad \times \prod_{l = 2}^K \frac{1}{p_l} \sum_{i_l, i'_l \in [p_l]} \exp\l( -\frac{c_0 \vert i_l - i'_l \vert}{3K\log(np)} \r)
\lesssim \omega^{2 + 2\eps} \tau^{2 - 2\eps} c_\eps \frac{(K\log(np))^K}{np_{-1}},
\end{align*}
where the final inequality follows from Assumption~\ref{assum:rf}~\ref{cond:rf:factor:mixing}. 
The above upper bound does not depend on $i, j$ or $k$ and thus it follows that
\begin{align*}
\frac{1}{p_k} \l\Vert \wh{\bm\Gamma}^\kk - \bm\Gamma^\kk \r\Vert_F &\le \sqrt{\frac{1}{p_k^2} \sum_{i, j \in [p_k]} T_{1, ij}^2} + \frac{1}{p_k} \l\Vert \wh{\bm\Gamma}^\kk - \E(\wh{\bm\Gamma}^\kk) \r\Vert_F 
\\
&= O_P\l( \frac{\omega^{2 + 2\eps}}{\tau^{2\eps}} + \omega^{1 + \eps} \tau^{1 - \eps} \sqrt{\frac{\log^K(np_{-k})}{np_{-k}}} \r) 
= O_P\l( \omega^2 \l( \frac{\log^K(np)}{np_{-k}} \r)^{\frac{\eps}{1 + \eps}} \r)
\end{align*}
by Markov's inequality, with $\tau \asymp \tau^\kk_{n, p}$ in~\eqref{eq:tau}. 
\medskip

Finally, combining the bound on $p_k^{-1} \Vert \wh{\bm\Gamma}^\kk - \bm\Gamma^\kk \Vert_F$ with Lemma~\ref{lem:idio:cov}, the proof of the first claim is complete.
\end{proof}

\begin{proof}[Proof of~\eqref{eq:prop:one:two}]
% The following lemma establishes the consistency of the first step estimator of $\bm\Lambda_k$, deriving a preliminary rate in the case of cross-sectional independence assumed in Assumption~\ref{assum:indep}.
% \begin{lemma}
% \label{lem:first}
% {\it
% Let the conditions in Proposition~\ref{prop:first:init} hold.
% Then for each $k \in [K]$, there exist diagonal matrices $\mbf J_k, \wh{\mbf J}_k \in \R^{r_k \times r_k}$ with $\pm 1$ on their diagonal entries such that 
% \begin{align*}
% & \l\Vert \wh{\mbf E}_k(\tau) - \mbf E_k \mbf J_k \r\Vert = O_P\l( \psi^\kk_{n, p} \r), \text{ \ and \ }
% \l\Vert \wh{\mbf E}_k(\tau) - \mbf E_{\chi, k} \wh{\mbf J}_k \r\Vert = O_P\l( \psi^\kk_{n, p}  \vee \frac{1}{p_k} \r)
% %\\
% %& \frac{1}{\sqrt{p_k}} \l\Vert \wh{\bm\Lambda}_k(\tau) - \bm\Lambda_k \wh{\mbf J}_k \r\Vert = O_P\l( \psi^\kk_{n, p}  \vee \frac{1}{p_k} \r).
% \end{align*}
% with $\tau = \tau^\kk_{n, p}$.
% }
% \end{lemma}
Let us set $\mbf S = \bm\Gamma^\kk_\chi$, $\wt{\mbf S} = \bm\Gamma^\kk$ and $\wh{\mbf S} = \wh{\bm\Gamma}^\kk$.
Then thanks to {\cgr Lemma~\ref{lem:common:cov}, we have $\mu^\kk_{\chi, j}, \, j \in [r_k]$, fulfil the condition~\ref{lem:dk:cond:one} in Lemma~\ref{lem:dk}.}
Also, $\Vert \bm\Gamma^\kk - \bm\Gamma^\kk_\chi \Vert = \Vert \bm\Gamma^\kk_\xi \Vert \lesssim \omega^2$ from Lemma~\ref{lem:idio:cov}. 
These establish that $\bm\Gamma^\kk_\chi$ and $\bm\Gamma^\kk$ meet~\ref{lem:dk:cond:four} in place of $\mbf S$ and $\wt{\mbf S}$.
Combining this with~\eqref{eq:prop:one:one} (playing the role of \ref{lem:dk:cond:two}), the claim follows from Lemma~\ref{lem:dk}~\ref{lem:dk:one}. % and the last from the identification that $\bm\Lambda_k = \sqrt{p_k} \E_{\chi, k}$.
\end{proof}

\subsubsection{Supporting lemmas}

\begin{lemma}
\label{lem:first:eigval}
{\it Let Assumptions~\ref{assum:loading}, \ref{assum:factor}, \ref{assum:heavy}, and~\ref{assum:indep} or~\ref{assum:rf} hold. 
For each $k \in [K]$, we have
\begin{align*}
\l\Vert \l( p_k^{-1} \wh{\mbf M}_k \r)^{-1} - \l( p_k^{-1} \mbf M_{\chi, k} \r)^{-1} \r\Vert = O_P\l( \psi^\kk_{n, p} \vee \frac{1}{p_k} \r),
\end{align*}
where $\mbf M_{\chi,  k} = \text{diag}(\mu^\kk_{\chi, 1}, \ldots, \mu^\kk_{\chi, r_k})$.
}
\end{lemma}

\begin{proof}
As noted in the proof of~\eqref{eq:prop:one:two}, $\bm\Gamma^\kk$ and $\bm\Gamma^\kk_\chi$ fulfil the conditions~\ref{lem:dk:cond:one} and~\ref{lem:dk:cond:four} in Lemma~\ref{lem:dk} in place of $\wt{\mbf S}$ and $\mbf S$, respectively. The conclusions follow from~\eqref{eq:prop:one:one} and Lemma~\ref{lem:dk}~\ref{lem:dk:two}.
\end{proof}

\begin{lemma}
\label{lem:thm:first:t:one}
{\it
Let Assumptions~\ref{assum:loading}, \ref{assum:factor}, \ref{assum:heavy} and~\ref{assum:indep} hold. 
For each $k \in [K]$, we have
\begin{align*}
\frac{1}{np} \l\Vert \sum_{t \in [n]} \mbf Z_{k, t} \mbf Z_{k, t}^\top \r\Vert = O_P \l(  \frac{M_n^{1 - \eps}}{\sqrt{np_{-k}}} \vee \frac{1}{p_k} \vee \frac{\psi^\kk_{n, p}}{\sqrt{p_k}} \r).
\end{align*}
}
\end{lemma}

\begin{proof}
Note that
\begin{align*}
\frac{1}{n^2p^2} \l\Vert \sum_{t \in [n]} \mbf Z_{k, t} \mbf Z_{k, t}^\top \r\Vert^2
\le & \, \frac{2}{n^2p^2} \l\Vert \sum_{t \in [n]} \l( \mbf Z_{k, t} \mbf Z_{k, t}^\top - \E\l( \mbf Z_{k, t} \mbf Z_{k, t}^\top \r) \r) \r\Vert^2
\\
& + \frac{2}{n^2p^2} \l\Vert \sum_{t \in [n]} \E\l( \mbf Z_{k, t} \mbf Z_{k, t}^\top \r) \r\Vert^2 =: U_1 + U_2.
\end{align*}
Then with $\mbf Z_{k, t} = [Z_{k, i\ell, t}, \, i \in [p_k], \, \ell \in [p_{-k}]]$, we have
\begin{align*}
\E(U_1) &\le \frac{2}{n^2p^2} \sum_{i, j \in [p_k]} \E\l[ \l( \sum_{t \in [n]} \sum_{\ell \in [p_{-k}]} \l( Z_{k, i\ell, t} Z_{k, j\ell, t} - \E(Z_{k, i\ell, t} Z_{k, j\ell, t}) \r) \r)^2 \r]
\\
&\le \frac{2}{n^2p^2} \sum_{i, j \in [p_k]} \sum_{\ell, m \in [p_{-k}]} \sum_{t, u \in [n]} \l\vert \Cov(Z_{k, i\ell, t} Z_{k, j\ell, t}, Z_{k, im, u} Z_{k, jm, u}) \r\vert
\\
&= \frac{2}{n^2p^2} \sum_{i, j \in [p_k]} \sum_{\ell \in [p_{-k}]} \sum_{t, u \in [n]} \l\vert \Cov(Z_{k, i\ell, t} Z_{k, j\ell, t}, Z_{k, i\ell, u} Z_{k, j\ell, u}) \r\vert
\\
&\le \frac{2}{n^2p^2} \sum_{i, j \in [p_k]} \sum_{\ell \in [p_{-k}]} \sum_{t, u \in [n]} \l\Vert Z_{k, i\ell, t} Z_{k, j\ell, t} \r\Vert_\nu \l\Vert Z_{k, i\ell, u} Z_{k, j\ell, u} \r\Vert_\nu \exp\l( - \frac{c_0(\nu - 2) \vert t - u \vert}{\nu} \r)
\\
&\lesssim \frac{(\tau^\kk_{n, p})^{\frac{4(\nu - 1 - \eps)}{\nu}}}{n^2p^2} \sum_{i \in [p_k]} \sum_{\ell \in [p_{-k}]} \sum_{t, u \in [n]} (\vert \mc F_t \vert_2 + \omega)^{\frac{2 + 2\eps}{\nu}} (\vert \mc F_u \vert_2 + \omega)^{\frac{2 + 2\eps}{\nu}} \exp\l( - \frac{c_0 \vert t - u \vert}{3\log(np_{-k})} \r)
\\
& \quad + \frac{1}{n^2p^2} \sum_{\substack{i, j \in [p_k] \\ i \ne j}} \sum_{\ell \in [p_{-k}]} \sum_{t, u \in [n]} (\vert \mc F_t \vert_2 + \omega)^2 (\vert \mc F_u \vert_2 + \omega)^2 \exp\l( - \frac{c_0 \eps \vert t - u \vert}{1 + \eps} \r)
\\
&\lesssim \frac{(\tau^\kk_{n, p})^{2 - 2\eps}}{np} \log(np_{-k}) + \frac{M_n^{2 - 2\eps}}{np_{-k}} \lesssim \l( \frac{\psi^\kk_{n, p}}{\sqrt{p_k}}  + \frac{M_n^{1 - \eps}}{\sqrt{np_{-k}}} \r)^2
\end{align*}
% when i = j and ell = m, cov(Z_{il, t} Z_{il, t}, Z_{im, u}Z_{im, u}) = E(Z_{il, t} Z_{il, t}) E(Z_{im, u}Z_{im, u}) - E(Z_{il, t} Z_{il, t}) E(Z_{im, u}Z_{im, u}) = 0
with $\nu \in \{ 2 + \log^{-1}(np_{-k}), 2 + 2\eps \}$ for the case of $i = j$ and $i \ne j$, respectively, the first equality is due to the cross-sectional independence (Assumption~\ref{assum:indep}~\ref{cond:indep:mixing}), the third inequality holds due to Lemmas~\ref{lem:mixing} and~\ref{lem:hall} and Assumption~\ref{assum:indep}~\ref{cond:indep:mixing}, the fourth due to Lemma~\ref{lem:one}~\ref{lem:one:one} and~\ref{lem:one:four}, and the penultimate one follows from Assumptions~\ref{assum:heavy}~\ref{cond:factor:bound} and~\ref{assum:indep}~\ref{cond:indep:factor:mixing}.
As for $U_2$, notice that $\sum_{t \in [n]} \E(\mbf Z_{k, t} \mbf Z_{k, t}^\top)$ is a diagonal matrix such that
\begin{align*}
U_2 &\lesssim \max_{i \in [p_k]} \l( \frac{1}{np} \sum_{t \in [n]} \sum_{\ell \in [p_{-k}]} \E(Z_{k, i\ell, t}^2) \r)^2
\lesssim \max_{i \in [p_k]} \l( \frac{1}{np} \sum_{t \in [n]} \sum_{\ell \in [p_{-k}]} (\vert \mc F_t \vert_2^2 + \omega^2) \r)^2 \lesssim \l( \frac{\omega^2}{p_k} \r)^2
\end{align*}
by Lemma~\ref{lem:one}~\ref{lem:one:one} and Assumption~\ref{assum:heavy}~\ref{cond:heavy:factor}.
Collecting the bounds on $U_1$ and $U_2$ and by Markov's inequality, the conclusion follows.
\end{proof}

\begin{lemma}
\label{lem:thm:first:t:two}
{\it
Let Assumptions~\ref{assum:loading}, \ref{assum:factor}, \ref{assum:heavy} and~\ref{assum:indep} hold. 
For each $k \in [K]$, we have
\begin{align*}
\frac{1}{np} \l\Vert \sum_{t \in [n]} \mbf Z_{k, t} \E( \mbf X^\trunc_{k, t} )^\top \r\Vert &= O_P \l( \frac{M_n^{1 - \eps}}{\sqrt{np_{-k}}} \r),
\\
\frac{1}{np} \l\Vert \sum_{t \in [n]} \E( \mbf X^\trunc_{k, t} ) \mbf Z_{k, t}^\top \r\Vert &= O_P \l( \frac{M_n^{1 - \eps}}{\sqrt{np_{-k}}} \r).
\end{align*}
}
\end{lemma}

\begin{proof}
With $\E(\mbf X^\trunc_{k, t}) = [\E(X^\trunc_{k, i\ell, t}), \, i \in [p_k], \, \ell \in [p_{-k}]]$, we have
\begin{align*}
& \E\l( \l\Vert \frac{1}{np} \sum_{t \in [n]} \mbf Z_{k, t} \E( \mbf X^\trunc_{k, t} )^\top \r\Vert^2 \r)
\\
\le & \, \frac{1}{n^2p^2} \sum_{i, j \in [p_k]} \sum_{\ell, m \in [p_{-k}]} \sum_{t, u \in [n]} \E(X^\trunc_{k, j\ell, t}) \E(X^\trunc_{k, j m, u}) \Cov(Z_{k, i\ell, t}, Z_{k, i m, u} )
\\
= & \, \frac{1}{n^2p^2} \sum_{i, j \in [p_k]} \sum_{\ell \in [p_{-k}]} \sum_{t, u \in [n]} \E(X^\trunc_{k, j\ell, t}) \E(X^\trunc_{k, j \ell, u}) \Cov(Z_{k, i\ell, t}, Z_{k, i \ell, u} )
\\
\le & \, \frac{1}{n^2p^2} \sum_{i, j \in [p_k]} \sum_{\ell \in [p_{-k}]} \sum_{t, u \in [n]} \l\vert \E(X^\trunc_{k, j\ell, t}) \E(X^\trunc_{k, j \ell, u}) \r\vert \l\Vert Z_{k, i\ell, t} \r\Vert_\nu \l\Vert Z_{k, i \ell, u} \r\Vert_\nu \exp\l( - \frac{c_0(\nu - 2) \vert t - u \vert}{\nu} \r)
\\
\lesssim & \, \frac{1}{n^2p^2} \sum_{i, j \in [p_k]} \sum_{\ell \in [p_{-k}]} \sum_{t, u \in [n]} (\vert \mc F_t \vert_2 + \omega)^2 (\vert \mc F_u \vert_2 + \omega)^2 \exp\l( - \frac{c_0 \eps \vert t - u \vert}{1 + \eps} \r)
\lesssim \frac{M_n^{2 - 2\eps}}{np_{-k}}
\end{align*}
with $\nu = 2 + 2\eps$, where the first equality is due to the cross-sectional independence (Assumption~\ref{assum:indep}~\ref{cond:indep:mixing}), the second inequality holds due to Lemmas~\ref{lem:mixing} and~\ref{lem:hall} and Assumption~\ref{assum:indep}~\ref{cond:indep:mixing}, the third due to Lemma~\ref{lem:one}~\ref{lem:one:one}, and the last one follows from Assumptions~\ref{assum:heavy}~\ref{cond:factor:bound} and~\ref{assum:indep}~\ref{cond:indep:factor:mixing}.
The second result follows by the analogous arguments.
\end{proof}

\begin{lemma}
\label{lem:thm:first:t:three}
{\it
Let Assumptions~\ref{assum:loading}, \ref{assum:factor}, \ref{assum:heavy}, and~\ref{assum:indep} or~\ref{assum:rf} hold. 
For each $k \in [K]$, we have
\begin{align*}
\frac{1}{np} \l\Vert \sum_{t \in [n]} \E( \mbf X^\trunc_{k, t} ) \E( \mbf X^\trunc_{k, t} - \mbf X_t )^\top \r\Vert &= O_P \l(  \frac{M_n^\eps \sqrt{\log(np_{-k})}}{(\tau^\kk_{n, p})^\eps} \cdot  \frac{M_n^{1 - \eps}}{\sqrt{np_{-k}}} \r),
\\
\frac{1}{np} \l\Vert \sum_{t \in [n]} \E( \mbf X^\trunc_{k, t} - \mbf X_t ) \E(\mbf X_{k, t})^\top \r\Vert &= O_P \l(  \frac{M_n^\eps \sqrt{\log(np_{-k})}}{(\tau^\kk_{n, p})^\eps} \cdot  \frac{M_n^{1 - \eps}}{\sqrt{np_{-k}}} \r).
\end{align*}
}
\end{lemma}

\begin{proof}
With $\E(\mbf X^\trunc_{k, t} - \mbf X_{k, t}) = [\E(X^\trunc_{k, i\ell, t} - X_{k, i\ell, t}), \, i \in [p_k], \, \ell \in [p_{-k}]]$, we have
\begin{align*}
& \l\Vert \frac{1}{np} \sum_{t \in [n]} \E( \mbf X^\trunc_{k, t} ) \E( \mbf X^\trunc_{k, t} - \mbf X_t )^\top \r\Vert_F^2
\\
% = & \, \sum_{i, j \in [p_k]} \frac{1}{n^2p^2} \l( \sum_{t \in [n]} \sum_{\ell \in [p_{-k}]} \E(X^{\trunc}_{k, i\ell, t}) \E( X^\trunc_{k, j \ell, t} - X_{k, j \ell, t}) \r)^2
= & \, \sum_{i, j \in [p_k]} \frac{1}{n^2p^2} \sum_{\ell, m \in [p_{-k}]} \sum_{t, u \in [n]} 
 \E(X^{\trunc}_{k, i\ell, t}) \E(X^{\trunc}_{k, i m, u})  \E( X^\trunc_{k, j \ell, t} - X_{k, j \ell, t}) \E( X^\trunc_{k, j m, u} - X_{k, j m, u})
\\
\lesssim & \sum_{i, j \in [p_k]} \frac{1}{n^2p^2} \sum_{\ell, m \in [p_{-k}]} \sum_{t, u \in [n]} \frac{(\vert \mc F_t \vert_2 + \omega)^{3 + 2\eps} (\vert \mc F_u \vert_2 + \omega)^{3 + 2\eps}}{(\tau^\kk_{n, p})^{2 + 4\eps}} 
\\
\lesssim & \, \frac{\omega^{4 + 4\eps} M_n^2}{(\tau^\kk_{n, p})^{2 + 4\eps}} 
\lesssim \l(  \frac{M_n^\eps \sqrt{\log(np_{-k})}}{(\tau^\kk_{n, p})^\eps} \cdot  \frac{M_n^{1 - \eps}}{\sqrt{np_{-k}}} \r)^2,
\end{align*}
where the first inequality is due to Lemma~\ref{lem:one}~\ref{lem:one:one} and~\ref{lem:one:two} and the second from Assumption~\ref{assum:heavy}~\ref{cond:heavy:factor} and~\ref{cond:factor:bound}.
The second result follows by the analogous arguments.
\end{proof}

{\cgr
\begin{lemma}
\label{lem:e:lambda}
{\it Let Assumptions~\ref{assum:loading} and~\ref{assum:factor} hold. 
For each $k \in [K]$, there exists a matrix $\mbf H_k \in \R^{r_k \times r_k}$ satisfying $\mbf H_k^\top \mbf H_k = \mbf I_{r_k}$, such that
\begin{align*}
\frac{1}{\sqrt{p_k}} \bm\Lambda_k \mbf H_k = \mbf E_{\chi, k}.
\end{align*}
}
\end{lemma}
\begin{proof}
By definition,
\begin{align*}
\mbf E_{\chi, k} \mbf M_{\chi, k} = \bm\Gamma^\kk_\chi \mbf E_{\chi, k} &= \bm\Lambda_k \underbrace{ \l( \frac{1}{n} \sum_{t \in [n]} \mat_k(\mc F_t) \mat_k(\mc F_t)^\top \r) }_{=: \bar{\bm\Gamma}^\kk_f} \bm\Lambda_k^\top \mbf E_{\chi, k},
\end{align*}
where $\Vert \bar{\bm\Gamma}^\kk_f \Vert = O(1)$ under Assumptions~\ref{assum:loading} and~\ref{assum:factor}.
Let us set
\begin{align*}
\mbf H_k = \sqrt{p_k} \bar{\bm\Gamma}^\kk_f \bm\Lambda_k^\top \mbf E_{\chi, k} (\mbf M_{\chi, k})^{-1}.
\end{align*}
Then, we have
\begin{align*}
\mbf H_k^\top \mbf H_k &= p_k (\mbf M_{\chi, k})^{-1} \mbf E_{\chi, k}^\top \bm\Lambda_k  \l( \bar{\bm\Gamma}^\kk_f \r)^2 \bm\Lambda_k^\top \mbf E_{\chi, k} (\mbf M_{\chi, k})^{-1}
\\
&= (\mbf M_{\chi, k})^{-1} \mbf E_{\chi, k}^\top \bm\Lambda_k  \bar{\bm\Gamma}^\kk_f \bm\Lambda_k^\top \bm\Lambda_k \bar{\bm\Gamma}^\kk_f \bm\Lambda_k^\top \mbf E_{\chi, k} (\mbf M_{\chi, k})^{-1} 
= \mbf I_{r_k}.
\end{align*}
\end{proof}
}

\subsection{Proof of Theorem~\ref{thm:second}}

Throughout, we suppress the dependence on $\tau$ where there is no confusion.
For each $k \in [K]$, we decompose $\wc{\bm\Gamma}^{\kk, [1]}$ as
\begin{align}
\frac{1}{p_k} \wc{\bm\Gamma}^{\kk, [1]} =& \, \frac{1}{np} \sum_{t \in [n]} \mbf X^{\trunc}_{k, t} \wh{\mbf D}_k \wh{\mbf D}_k^\top (\mbf X^{\trunc}_{k, t})^\top 
\nn
\\
= & \, \frac{1}{np} \sum_{t \in [n]} \mbf Z_{k, t} \wh{\mbf D}_k \wh{\mbf D}_k^\top \mbf Z_{k, t}^\top
\nn \\
& + \frac{1}{np} \sum_{t \in [n]} \mbf Z_{k, t} \wh{\mbf D}_k \wh{\mbf D}_k^\top \E\l( \mbf X^{\trunc}_{k, t} \r)^\top
+ \frac{1}{np} \sum_{t \in [n]} \E\l( \mbf X^{\trunc}_{k, t} \r) \wh{\mbf D}_k \wh{\mbf D}_k^\top \mbf Z_{k, t}^\top
\nn \\
& + \frac{1}{np} \sum_{t \in [n]} \E\l( \mbf X^{\trunc}_{k, t} - \mbf X_{k, t} \r) \wh{\mbf D}_k \wh{\mbf D}_k^\top \E\l( \mbf X^{\trunc}_{k, t} \r)^\top 
\nn \\
& + \frac{1}{np} \sum_{t \in [n]} \E(\mbf X_{k, t}) \wh{\mbf D}_k \wh{\mbf D}_k^\top \E\l( \mbf X^{\trunc}_{k, t} - \mbf X_{k, t} \r)^\top
\nn \\
& + \frac{1}{np} \sum_{t \in [n]} \E(\mbf X_{k, t}) \l( \wh{\mbf D}_k \wh{\mbf D}_k^\top - {\cgr \frac{1}{p_{-k}} \bm\Delta_k \bm\Delta_k^\top} \r) \E(\mbf X_{k, t})^\top
\nn \\
& + {\cgr \frac{1}{np_kp_{-k}^2} \sum_{t \in [n]} \E(\mbf X_{k, t}) \bm\Delta_k \bm\Delta_k^\top \E(\mbf X_{k, t})^\top }
\nn \\
=: & \, T_1 + T_{2, 1} + T_{2, 2} + T_{3, 1} + T_{3, 2} + T_4 + {\cgr \frac{1}{p_k} \bm\Gamma^\kk_\chi}.
\label{eq:wc:gamma:decomp}
\end{align}
where $\bm\Gamma^\kk_\chi$ is defined in~\eqref{eq:gamma:decomp}.
Based on this, we derive the following rate of estimation for $\wc{\bm\Gamma}^{\kk, [1]}(\tau)$:
\begin{proposition}
\label{prop:gamma:wc}
{\it 
Let Assumptions~\ref{assum:loading}, \ref{assum:factor} and~\ref{assum:heavy} hold. 
Then for each $k \in [K]$, 
\begin{align*}
& \frac{1}{p_k} \l\Vert \wc{\bm\Gamma}^{\kk, [1]}(\tau) - \bm\Gamma^\kk_\chi \r\Vert 
\\
= & \, \l\{\begin{array}{l}
O_P\l[ 
\frac{M_n^{1 - \eps}}{\sqrt{np_{-k}}} 
\vee
\bar{\psi}^\kk_{n, p} \l( \frac{\psi^\kk_{n, p}}{\sqrt{p_k}}
+ \frac{M_n^{1 - \eps}}{\sqrt n}
\r)
\vee \frac{\bar{\psi}_{n, p}}{\sqrt{p}}
\vee \sum_{k' \in [K] \setminus \{k\}} \l( \frac{M_n^{1 - \eps}}{\sqrt{np_{-k'}}} \vee \frac{1}{p_{k'}} \vee \frac{\psi^{(k')}_{n, p}}{\sqrt p_{k'}} \r)
\r] 
\\
\text{\qquad under Assumption~\ref{assum:indep},}  \\
O_P\l( \bar{\psi}_{n, p} \vee \sum_{k' \in [K] \setminus \{k\}} \frac{1}{p_{k'}} \r) 
\\
\text{\qquad under Assumption~\ref{assum:rf},}
\end{array}
\r.
\end{align*}
with $\tau = \tau^\kk$ chosen as in~\eqref{eq:tau}. 
} 
\end{proposition}
\begin{proof}
Noting the decomposition in~\eqref{eq:wc:gamma:decomp}, we obtain the desired rates by collecting the bounds on $T_1$--$T_4$ derived in Lemmas~\ref{lem:T:one}--\ref{lem:T:four}. 
\end{proof}

For each $k \in [K]$, denote by $\wc{\mbf M}^{[1]}_k(\tau) \in \R^{r_k \times r_k}$ the diagonal matrix containing the eigenvalues $\wc{\mu}^{\kk, [1]}_j(\tau), \, j \in [r_k]$, of $\wc{\bm\Gamma}^{\kk, [1]}(\tau)$ on its diagonal. 
By Proposition~\ref{prop:gamma:wc}, Lemma~\ref{lem:second:rough} and the arguments analogous to those adopted in proving~\eqref{eq:wh:M:bound:one} and~\eqref{eq:wh:M:bound:two}, we have
$p_k^{-1} \wc{\mbf M}^{[1]}_k$ asymptotically invertible and 
\begin{align}
\label{eq:wc:M:bound}
\l\Vert \l( p_k^{-1} \wc{\mbf M}^{[1]}_k \r)^{-1} \r\Vert \le \frac{1}{\alpha^\kk_{r_k}} + o_P(1) = O_P(1).
\end{align}
Let us set 
{\cgr
\begin{align*} 
\wc{\mbf H}^{[1]}_k = \frac{1}{n\sqrt{p_k}p_{-k}} \sum_{t \in [n]} \mat_k(\mc F_t) \bm\Delta_k^\top \wh{\mbf D}_k \wh{\mbf D}_k^\top \bm\Delta_k \mat_k(\mc F_t)^\top \bm\Lambda_k^\top \wc{\mbf E}^{[1]}_k \l( \frac{1}{p_k} \wc{\mbf M}^{[1]}_k \r)^{-1},
\end{align*}}
% with $\mbf D_{\chi, k}$ defined in Lemma~\ref{lem:D} below, such that by Assumption~\ref{assum:loading}~(i),
such that
\begin{align*}
{\cgr \frac{1}{\sqrt{p_k}} \bm\Lambda_k} \wc{\mbf H}^{[1]}_k = \frac{1}{n p} \sum_{t \in [n]} \E(\mbf X_{k, t}) \wh{\mbf D}_k \wh{\mbf D}_k^\top \E(\mbf X_{k, t})^\top \wc{\mbf E}^{[1]}_k \l( \frac{1}{p_k} \wc{\mbf M}^{[1]}_k \r)^{-1}.
\end{align*}
Then noting that $\wc{\bm\Gamma}^{\kk, [1]} \wc{\mbf E}^{[1]}_k = \wc{\mbf E}^{[1]}_k \wc{\mbf M}^{[1]}_k$, we have
\begin{align*}
\wc{\mbf E}^{[1]}_k - {\cgr \frac{1}{\sqrt{p_k}} \bm\Lambda_k} \wc{\mbf H}^{[1]}_k = \l( T_1 + T_{2, 1} + T_{2, 2} + T_{3, 1} + T_{3, 2} \r) \wc{\mbf E}^{[1]}_k \l( \frac{1}{p_k} \wc{\mbf M}^{[1]}_k \r)^{-1}.
\end{align*}
By Lemmas~\ref{lem:T:one}--\ref{lem:T:three} and~\eqref{eq:wc:M:bound},
\begin{align*}
& \l\Vert \wc{\mbf E}^{[1]}_k - {\cgr \frac{1}{\sqrt{p_k}} \bm\Lambda_k} \wc{\mbf H}^{[1]}_k \r\Vert 
= \l\{ \begin{array}{l}
O_P\l[ 
\frac{M_n^{1 - \eps}}{\sqrt{np_{-k}}} \vee \frac{1}{p} 
\vee
\bar{\psi}^\kk_{n, p} \l( \frac{\psi^\kk_{n, p}}{\sqrt{p_k}}
+ \frac{M_n^{1 - \eps}}{\sqrt n}
\r)
\vee \frac{\bar{\psi}_{n, p}}{\sqrt p}
\r] 
\\
\text{\quad under Assumption~\ref{assum:indep},}
\\
O_P\l( \frac{M_n^{1 - \eps}}{\sqrt{np_{-k}}} \vee \frac{1}{p} \vee \psi^\kk_{n, p} \vee  \frac{M_n^{1 - \eps} \bar{\psi}^\kk_{n, p}}{\sqrt n} \r)
\\
\text{\quad under Assumption~\ref{assum:rf}.}
\end{array}\r. 
\end{align*}
We conclude the proof by noting that
{\cgr
\begin{align*}
\Vert \wc{\mbf H}^{[1]}_k \Vert \le \frac{1}{n\sqrt{p_k}p_{-k}} \sum_{t \in [n]} \vert \mc F_t \vert_2^2 \Vert \bm\Delta_k \Vert^2 \Vert \bm\Lambda_k \Vert \l\Vert \l( \frac{1}{p_k} \wc{\mbf M}^{[1]}_k \r)^{-1} \r\Vert \le (\alpha^\kk_{r_k})^{-1} \omega^2 (1 + o_P(1))
\end{align*}}
by Assumptions~\ref{assum:loading} and~\ref{assum:heavy}~\ref{cond:heavy:factor} and~\eqref{eq:wc:M:bound}, and hence
{\cgr 
\begin{align}
% \label{eq:HH:wh}
\mbf I_{r_k} = (\wc{\mbf E}^{[1]}_k)^\top \wc{\mbf E}^{[1]}_k = \frac{1}{p_k} (\wc{\mbf H}^{[1]}_k)^\top \bm\Lambda_k^\top \bm\Lambda_k \wc{\mbf H}^{[1]}_k + o_P(1) = (\wc{\mbf H}^{[1]}_k)^\top \wc{\mbf H}^{[1]}_k + o_P(1). \nn
\end{align}}

\subsubsection{Supporting lemmas}

Throughout, we suppress the dependence on $\tau$ where there is no confusion.
%% do we need all these? (i) (ii) (v)
\begin{lemma}
\label{lem:D}
{\it 
Let Assumptions~\ref{assum:loading}, \ref{assum:factor}, \ref{assum:heavy}, and~\ref{assum:indep} or~\ref{assum:rf} hold, and define
\begin{align*}
\mbf D_k &= \mbf E_K \otimes \cdots \otimes \mbf E_{k + 1} \otimes \mbf E_{k - 1} \otimes \cdots \otimes \mbf E_1. % \text{ \ and} \\ 
% \mbf D_{\chi, k} &= \mbf E_{\chi, K} \otimes \cdots \otimes \mbf E_{\chi, k + 1} \otimes \mbf E_{\chi, k - 1} \otimes \cdots \otimes \mbf E_{\chi, 1}.
\end{align*}
Recall $\mbf J_k$ and $\wh{\mbf J}_k$ from Proposition~\ref{prop:first:init} and $\wh{\mbf H}_k$ from Theorem~\ref{thm:first}.
Then for all $k \in [K]$, with $\tau = \tau^\kk$ chosen as in~\eqref{eq:tau}, we have:
\begin{enumerate}[wide, label = (\roman*)]
\item \label{lem:D:one}
Letting $\mbf J_{-k} = \mbf J_K \otimes \cdots \otimes \mbf J_{k + 1} \otimes \mbf J_{k - 1} \otimes \cdots \otimes \mbf J_1$, 
\begin{align*}
& \l\Vert \wh{\mbf D}_k - \mbf D_k \mbf J_{-k} \r\Vert = O_P \l( \bar{\psi}^\kk_{n, p} \r).
\end{align*}

\item \label{lem:D:two} Letting $\wh{\mbf J}_{-k} = \wh{\mbf J}_{k + 1} \otimes \wh{\mbf J}_{k - 1} \otimes \cdots \otimes \wh{\mbf J}_1$,
\begin{align*}
& \l\Vert \wh{\mbf D}_k - {\cgr \frac{1}{\sqrt{p_{-k}}} \bm\Delta_k} \wh{\mbf J}_{-k} \r\Vert = O_P \l( \sum_{k' \in [K] \setminus \{k\}} \l( \psi^{(k')}_{n, p} \vee \frac{1}{p_{k'}}\r) \r).
\end{align*}
\item \label{lem:D:two:new} 
Suppose that Assumption~\ref{assum:indep} holds.
Letting $\wh{\mbf H}_{-k} = \wh{\mbf H}_K \otimes \cdots \otimes \wh{\mbf H}_{k + 1} \otimes \wh{\mbf H}_{k - 1} \otimes \cdots \otimes \wh{\mbf H}_1$, 
\begin{align*}
& \l\Vert \wh{\mbf D}_k - {\cgr \frac{1}{\sqrt{p_{-k}}} \bm\Delta_k } \wh{\mbf H}_{-k} \r\Vert = O_P \l( \sum_{k' \in [K] \setminus \{k\}} \l( \frac{\psi^{(k')}_{n, p}}{\sqrt{p_{k'}}} \vee 
\frac{M_n^{1 - \eps}}{\sqrt{np_{-k'}}} \vee 
\frac{1}{p_{k'}} \r) \r).
\end{align*}
\item \label{lem:D:three} There exists some constant $C > 0$ such that $\vert \mbf D_k \vert_\infty \le Cp_{-k}^{-1/2}$ and {\cgr $\vert \bm\Delta_k \vert_\infty \le C$.}
% \item \label{lem:D:four} $\vert \wh{\mbf D}_k \vert_\infty = O_P(p_{-k}^{- 1/2})$.
\end{enumerate}}
\end{lemma}

\begin{proof}
WLOG, we consider the case where $k = 1$.
Note that,
\begin{align*}
\wh{\mbf D}_1 - \mbf D_1 \mbf J_{-1}
=& \, (\wh{\mbf E}_K - \mbf E_K \mbf J_K) \otimes_{k' = K - 1}^2 \wh{\mbf E}_{k'} 
+ (\mbf E_K \mbf J_K) \otimes (\wh{\mbf E}_{K - 1} - \mbf E_{K - 1} \mbf J_{K - 1}) \otimes_{k' = K - 2}^2 \wh{\mbf E}_{k'} 
\\
& + \ldots + \otimes_{k' = K}^3 \mbf E_k \mbf J_k \otimes (\wh{\mbf E}_2 - \mbf E_2 \mbf J_2)
\end{align*}
such that
\begin{align*}
\l\Vert \wh{\mbf D}_1 - \mbf D_1 \mbf J_{-1} \r\Vert 
& \le \l\Vert \wh{\mbf E}_K - \mbf E_K \mbf J_K \r\Vert \prod_{k' = K - 1}^2 \l\Vert \wh{\mbf E}_{k'} \r\Vert
\\
& + \l\Vert \mbf E_K \mbf J_K \r\Vert \; \l\Vert \wh{\mbf E}_{K - 1} - \mbf E_{K - 1} \mbf J_{K - 1} \r\Vert \prod_{k' = K - 2}^2 \l\Vert \wh{\mbf E}_{k'} \r\Vert
\\
& + \ldots + \prod_{k' = K}^3 \l\Vert \mbf E_k \mbf J_k \r\Vert \; \l\Vert \wh{\mbf E}_2 - \mbf E_2 \mbf J_2 \r\Vert
= O_P\l( \sum_{k' = 2}^K \psi^{(k')}_{n, p} \r)
\end{align*}
by~\eqref{eq:prop:one:two} in Proposition~\ref{prop:first:init}, which proves~\ref{lem:D:one}. The proofs of~\ref{lem:D:two} and~\ref{lem:D:two:new} take the analogous steps, the former utilising~\eqref{eq:prop:one:two} and the latter~\eqref{eq:first:loading:rate}, and thus is omitted.
For~\ref{lem:D:three}, observe that by Lemma~\ref{lem:one}~\ref{lem:one:one}, Assumption~\ref{assum:heavy} and Cauchy-Schwarz inequality, $\vert \bm\Gamma^\kk \vert_\infty \lesssim n^{-1} \sum_{t \in [n]} (\vert \mc F_t \vert_2^2 + \omega^2) \le 2 \omega^2$. % and we can similarly show that $\vert \bm\Gamma^\kk_\chi \vert_\infty \lesssim \omega^2$.
Combining this with the arguments used in the proof of~\eqref{eq:prop:one:two}, by Lemma~\ref{lem:dk}~\ref{lem:dk:three}, we have
\begin{align}
% \label{eq:E:inf}
\vert \mbf E_k \vert_\infty \le \frac{C'}{\sqrt{p_k}} % \text{ \ and \ } \vert \mbf E_{\chi, k} \vert_\infty \le \frac{C'}{\sqrt{p_k}}
\nn
\end{align}
for all $k \in [K]$ and some constant $C' > 0$, which proves the first claim with $C = (C')^{K - 1}$.
{\cgr The second claim follows with $C = \bar{\lambda}^{K - 1}$ due to Assumption~\ref{assum:loading}~(ii).}
% Similarly, by Lemma~\ref{lem:dk}~\ref{lem:dk:four}, we have $\vert \wh{\mbf E}_k \vert_\infty = O_P(p_k^{-1/2})$ which leads to~\ref{lem:D:four}.
\end{proof}

\begin{lemma}
\label{lem:DD}
{\it Let Assumptions~\ref{assum:loading}, \ref{assum:factor}, \ref{assum:heavy}, and~\ref{assum:indep} or~\ref{assum:rf} hold. 
Then for all $k \in [K]$,
\begin{align}
\l\Vert \wh{\mbf D}_k \wh{\mbf D}_k^\top - \mbf D_k \mbf D_k^\top \r\Vert &=  O_P\l( \bar{\psi}^\kk_{n, p} \r).
\label{eq:DD:bound}
\end{align}
Also, under Assumption~\ref{assum:indep},
\begin{align}
\l\Vert \wh{\mbf D}_k \wh{\mbf D}_k^\top -  {\cgr \frac{1}{p_{-k}} \bm\Delta_k\bm\Delta_k^\top } \r\Vert &= O_P \l( \sum_{k' \in [K] \setminus \{k\}} \l( \frac{M_n^{1 - \eps}}{\sqrt{np_{-k'}}} \vee \frac{1}{p_{k'}} \vee \frac{\psi^{(k')}_{n, p}}{\sqrt p_{k'}} \r) \r),
\label{eq:DD:indep}
\end{align}
while under Assumption~\ref{assum:rf},
\begin{align}
\l\Vert \wh{\mbf D}_k \wh{\mbf D}_k^\top -  {\cgr \frac{1}{p_{-k}} \bm\Delta_k\bm\Delta_k^\top } \r\Vert = O_P\l( \sum_{k' \in [K] \setminus \{k\}} \l( \psi^{(k')}_{n, p} \vee \frac{1}{p_{k'}} \r) \r).
\label{eq:DD:rf}
\end{align}
}
\end{lemma}

\begin{proof}
By Lemma~\ref{lem:D}~\ref{lem:D:one}, we have
\begin{align*}
\l\Vert \wh{\mbf D}_k \wh{\mbf D}_k^\top - \mbf D_k \mbf D_k^\top \r\Vert &\le \l\Vert \wh{\mbf D}_k \l(\wh{\mbf D}_k - \mbf D_k \mbf J_{-k} \r)^\top \r\Vert + \l\Vert \l(\wh{\mbf D}_k \ - \mbf D_k \mbf J_{-k} \r) \mbf J_{-k} \mbf D_k^\top \r\Vert 
= O_P\l( \bar{\psi}^\kk_{n, p} \r),
\end{align*}
which proves~\eqref{eq:DD:bound}.
Similarly under Assumption~\ref{assum:rf},
\begin{align*}
\l\Vert \wh{\mbf D}_k \wh{\mbf D}_k^\top -  {\cgr \frac{1}{p_{-k}} \bm\Delta_k\bm\Delta_k^\top } \r\Vert 
\le &\, \l\Vert \wh{\mbf D}_k \l( \wh{\mbf D}_k - {\cgr \frac{1}{\sqrt{p_{-k}}} \bm\Delta_k} \wh{\mbf J}_{-k} \r)^\top \r\Vert
\\
& \quad + \l\Vert \wh{\mbf D}_k - {\cgr \frac{1}{\sqrt{p_{-k}}} \bm\Delta_k} \wh{\mbf J}_{-k} \r\Vert \l\Vert \wh{\mbf J}_{-k} \r\Vert \l\Vert {\cgr \frac{1}{\sqrt{p_{-k}}} \bm\Delta_k} \r\Vert
\\
= &\, O_P\l( \sum_{k' \in [K] \setminus \{k\}} \l( \psi^{(k')}_{n, p} \vee \frac{1}{p_{k'}} \r) \r)
\end{align*}
which follows from Lemma~\ref{lem:D}~\ref{lem:D:two}, Proposition~\ref{prop:first:init} and Assumption~\ref{assum:loading}~(i), thus proving~\eqref{eq:DD:rf}. 

Recall that $\wh{\mbf H}_k \in \R^{r_k \times r_k}$ is asymptotically invertible with $\Vert \wh{\mbf H}_k \Vert = O_P(1)$. 
Further, under Assumption~\ref{assum:indep}, by Theorem~\ref{thm:first}, we have 
\begin{align*}
\mbf I_{r_k} &= \wh{\mbf E}_k^\top \wh{\mbf E}_k % = {\cgr \frac{1}{\sqrt{p_k}} \wh{\mbf E}_k^\top \bm\Lambda_k} \wh{\mbf H}_k + O_P\l( \frac{M_n^{1 - \eps}}{\sqrt{np_{-k}}} \vee \frac{1}{p_k} \vee \frac{\psi^\kk_{n, p}}{\sqrt p_k} \r) \\
= \wh{\mbf H}_k^\top \wh{\mbf H}_k + O_P\l( \frac{M_n^{1 - \eps}}{\sqrt{np_{-k}}} \vee \frac{1}{p_k} \vee \frac{\psi^\kk_{n, p}}{\sqrt p_k} \r),
\end{align*}
taking steps analogous to those used in~\eqref{eq:hh}.
From the above, it follows that $\Vert \wh{\mbf H}_k^{-1} \Vert = O_P(1)$ and by the same token, we have $\Vert \wh{\mbf H}_{-k}^{-1} \Vert = O_P(1)$.
Then, we have
{\cgr
\begin{align*}
& \l\Vert \wh{\mbf D}_k \wh{\mbf H}_{-k}^\top - \frac{1}{\sqrt{p_{-k}}} \bm\Delta_k \r\Vert
= \l\Vert \l( \wh{\mbf D}_k \wh{\mbf H}_{-k}^\top \wh{\mbf H}_{-k} - \frac{1}{\sqrt{p_{-k}}} \bm\Delta_k \wh{\mbf H}_{-k} \r) \wh{\mbf H}_{-k}^{-1} \r\Vert
\\
\le &\, \l\Vert \wh{\mbf D}_k\l\{ \mbf I_{r_{-k}} + O_P\l[ \sum_{k' \in [K] \setminus \{k\} } \l( \frac{M_n^{1 - \eps}}{\sqrt{np_{-k'}}} \vee \frac{1}{p_{k'}} \vee \frac{\psi^{(k')}_{n, p}}{\sqrt p_{k'}} \r) \r] \r\} - \frac{1}{\sqrt{p_{-k}}} \bm\Delta_k \wh{\mbf H}_{-k} \r\Vert \; \l\Vert \wh{\mbf H}_{-k}^{-1} \r\Vert 
\\
= &\, O_P\l( \sum_{k' \in [K] \setminus \{k\} } \l( \frac{M_n^{1 - \eps}}{\sqrt{np_{-k'}}} \vee \frac{1}{p_{k'}} \vee \frac{\psi^{(k')}_{n, p}}{\sqrt p_{k'}} \r) \r)
\end{align*}}
by Lemmas~\ref{lem:D}~\ref{lem:D:two:new}.
Then,
\begin{align*}
& \l\Vert \wh{\mbf D}_k \wh{\mbf D}_k^\top - {\cgr \frac{1}{p_{-k}} \bm\Delta_k \bm\Delta_k^\top } \r\Vert 
\\
\le &\, \l\Vert \wh{\mbf D}_k \l( \wh{\mbf D}_k - \frac{1}{\sqrt{p_{-k}}} \bm\Delta_k \wh{\mbf H}_k \r)^\top \r\Vert 
+ \l\Vert \l( \wh{\mbf D}_k \wh{\mbf H}_k^\top - \frac{1}{\sqrt{p_{-k}}} \bm\Delta_k \r) \frac{1}{\sqrt{p_{-k}}} \bm\Delta_k^\top \r\Vert 
\\
= &\, O_P \l( \sum_{k' \in [K] \setminus \{k\}} \l( \frac{M_n^{1 - \eps}}{\sqrt{np_{-k'}}} \vee \frac{1}{p_{k'}} \vee \frac{\psi^{(k')}_{n, p}}{\sqrt p_{k'}} \r) \r),
\end{align*}
which proves~\eqref{eq:DD:indep}.
\end{proof}

Lemmas~\ref{lem:T:one}--\ref{lem:T:four} analyse the terms involved in~\eqref{eq:wc:gamma:decomp}.

\begin{lemma}
\label{lem:T:one}
{\it Let Assumptions~\ref{assum:loading}, \ref{assum:factor} and~\ref{assum:heavy} hold. 
For each $k \in [K]$, we have the followings:
\begin{enumerate}[wide, itemsep = 0pt, label = (\roman*)] 
\item \label{lem:T:one:one} Under Assumption~\ref{assum:indep}, 
% \footnote{
% \begin{align*}
% \frac{1}{\sqrt{np_{-k}}} \l(
% \frac{M^{1 - \eps}}{\sqrt{p_{-k}}} + \l( \frac{n^{1 - \eps} p_{\max}^{2\eps} \log^{2\eps}(np)}{p_k^{1 + 3\eps} p_{-k}^{2\eps}} \r)^{\frac{1}{2 + 2\eps}}
% + 
% \l( \frac{p_{\max}^{2\eps} \log^{4 + 4\eps}(np)}{n^{3\eps - 1} p_{-k}^{3 \eps - 1} p_k^{3\eps + 1}} \r)^{\frac{1}{2 + 2\eps}}
% +
% M_n^{1 - \eps}\l( \frac{p_{\max}^{2\eps} p_{-k}^{1 - \eps} \log^{2\eps}(np) }{ n^{2\eps} p_k^{2\eps}} 
% \r)^{\frac{1}{2 + 2\eps}} \r)
% \vee
% \frac{1}{p}
% \end{align*}}
\begin{align*}
\frac{1}{np} \l\Vert \sum_{t \in [n]} \mbf Z_{k, t} \wh{\mbf D}_k \wh{\mbf D}_k^\top \mbf Z_{k, t}^\top \r\Vert = O_P\l[ \frac{\psi^\kk_{n, p} \bar{\psi}^\kk_{n, p}}{\sqrt{p_k}}
\vee \frac{\bar{\psi}_{n, p}}{\sqrt p}
\vee \frac{M_n^{1 - \eps}}{\sqrt n} \l( \frac{1}{p_{-k}} \vee \bar{\psi}^\kk_{n, p} \r)
\vee \frac{1}{p}
\r].
\end{align*}

\item \label{lem:T:one:three} Under Assumption~\ref{assum:rf},
\begin{align*}
\frac{1}{np} \l\Vert \sum_{t \in [n]} \mbf Z_{k, t} \wh{\mbf D}_k \wh{\mbf D}_k^\top \mbf Z_{k, t}^\top \r\Vert = O_P\l( \psi^\kk_{n, p} \vee \frac{1}{p} \r).
\end{align*}
\end{enumerate}
}
\end{lemma}

\begin{proof}[Proof of Lemma~\ref{lem:T:one}~\ref{lem:T:one:one}] 
Let us write
\begin{align}
\frac{1}{np} \l\Vert \sum_{t \in [n]} \mbf Z_{k, t} \wh{\mbf D}_k \wh{\mbf D}_k^\top \mbf Z_{k, t}^\top \r\Vert 
&\le \frac{1}{np} \l\Vert \sum_{t \in [n]} \mbf Z_{k, t} \l( \wh{\mbf D}_k \wh{\mbf D}_k^\top - \mbf D_k \mbf D_k^\top \r) \mbf Z_{k, t}^\top \r\Vert
\nn \\
& \qquad
+ \frac{1}{np} \l\Vert \sum_{t \in [n]} \mbf Z_{k, t} \mbf D_k \mbf D_k^\top \mbf Z_{k, t}^\top \r\Vert =: U_1 + U_2. 
\label{eq:t:one:u}
\end{align}
By Lemma~\ref{lem:abc},
\begin{align}
\label{eq:t:one:u:one}
U_1^2 \le \underbrace{\frac{1}{n^2p^2} \sum_{i, j \in [p_k]} \sum_{\ell, m \in [p_{-k}]} \l( \sum_{t \in [n]} Z_{k, i\ell, t} Z_{k, jm, t} \r)^2}_{V_1} \l\Vert \wh{\mbf D}_k \wh{\mbf D}_k^\top - \mbf D_k \mbf D_k^\top \r\Vert^2,
\end{align}
where
\begin{align*}
\E(V_1) \le & \, \frac{2}{n^2p^2} \sum_{i, j \in [p_k]} \sum_{\ell, m \in [p_{-k}]} \sum_{t, u \in [n]} \Cov(Z_{k, i\ell, t} Z_{k, jm, t}, Z_{k, i\ell, u} Z_{k, jm, u})
\\
& + \frac{2}{n^2p^2} \sum_{i, j \in [p_k]} \sum_{\ell, m \in [p_{-k}]} \l( \sum_{t \in [n]} \E(Z_{k, i\ell, t} Z_{k, jm, t}) \r)^2
=: V_{1, 1} + V_{1, 2}.
\end{align*}
From Lemmas~\ref{lem:one}~\ref{lem:one:one} and~\ref{lem:one:four}, \ref{lem:mixing} and~\ref{lem:hall} and Assumptions~\ref{assum:heavy}~\ref{cond:factor:bound} and~\ref{assum:indep},
\begin{align*}
V_{1, 1} &\lesssim \frac{2}{n^2p^2} \sum_{i, j \in [p_k]} \sum_{\ell, m \in [p_{-k}]} \sum_{t, u \in [n]} \Vert Z_{k, i\ell, t} Z_{k, jm, t} \Vert_\nu \Vert Z_{k, i\ell, u} Z_{k, jm, u} \Vert_\nu \exp\l( - \frac{c_0(\nu - 2)\vert t - u \vert}{\nu} \r)
\\
&\le \frac{2}{n^2p^2} \sum_{i \in [p_k]} \sum_{\ell \in [p_{-k}]} \sum_{t, u \in [n]} \Vert Z_{k, i\ell, t}^2 \Vert_\nu \Vert Z_{k, i\ell, u}^2 \Vert_\nu \exp\l( - \frac{c_0(\nu - 2)\vert t - u \vert}{\nu} \r)
\\
&+ \frac{2}{n^2p^2} \sum_{\substack{i, j \in [p_k] \\  i \ne j}} \sum_{\substack{\ell, m \in [p_{-k}] \\ \ell \ne m}} \sum_{t, u \in [n]} \Vert Z_{k, i\ell, t} \Vert_\nu \Vert Z_{k, jm, t} \Vert_\nu
\\
& \qquad \qquad \qquad \qquad \qquad \qquad \times \Vert Z_{k, i\ell, u} \Vert_\nu \Vert Z_{k, jm, u} \Vert_\nu \exp\l( - \frac{c_0(\nu - 2)\vert t - u \vert}{\nu} \r)
\\
&\lesssim \frac{(\tau^\kk_{n, p})^{\frac{4(\nu - 1 - \eps)}{\nu}}}{n^2p} \sum_{t, u \in [n]} (\vert \mc F_t \vert_2 + \omega)^{\frac{2 + 2\eps}{\nu}} (\vert \mc F_u \vert_2 + \omega)^{\frac{2 + 2\eps}{\nu}} \exp\l( - \frac{c_0\vert t - u \vert}{3\log(np_{-k})} \r) 
\\
&+ \frac{1}{n^2} \sum_{t, u \in [n]} (\vert \mc F_t \vert_2 + \omega)^2 (\vert \mc F_u \vert_2 + \omega)^2 \exp\l( - \frac{c_0 \eps \vert t - u \vert}{1 + \eps} \r)
\\
&\lesssim \frac{c_\eps (\tau^\kk_{n, p})^{2 - 2\eps}}{p_k} \cdot \frac{\log(np_{-k})}{np_{-k}} + \frac{M_n^{2 - 2\eps}c_\eps}{n}
\lesssim \l( \frac{\psi^\kk_{n, p}}{\sqrt{p_k}} + \frac{M_n^{1 - \eps}}{\sqrt{n}} \r)^2
\end{align*}
with $\nu \in \{2 + \log^{-1}(np_{-k}), 2 + 2\eps\}$ for the case of $i = j$ and $i \ne j$, respectively. 
Also by Lemma~\ref{lem:one}~\ref{lem:one:one} and Assumptions~\ref{assum:heavy}~\ref{cond:heavy:factor} and~\ref{assum:indep}~\ref{cond:indep:mixing}, we have
\begin{align*}
V_{1, 2} = \frac{2}{n^2p^2} \sum_{i \in [p_k]} \sum_{\ell \in [p_{-k}]} \l( \sum_{t \in [n]} \E(Z_{k, i\ell, t}^2) \r)^2
\lesssim \frac{1}{p} \l( \frac{1}{n} \sum_{t \in [n]} (\vert \mc F_t \vert_2^2 + \omega^2) \r)^2
\lesssim \frac{\omega^4}{p}.
\end{align*}
Combining the bounds on $V_{1, 1}$ and $V_{1, 2}$ with~\eqref{eq:DD:bound} and~\eqref{eq:t:one:u:one}, by Markov's inequality,
\begin{align*}
U_1 = O_P\l[ \l( \frac{\psi^\kk_{n, p}}{\sqrt{p_k}} \vee \frac{M_n^{1 - \eps}}{\sqrt n} \vee \frac{\omega^2}{\sqrt p} \r) \bar{\psi}^\kk_{n, p} \r].
\end{align*}
As for $U_2$, writing $\mbf D_k = [d^\kk_{\ell q}, \, \ell \in [p_{-k}], \, q \in [r_{-k}]]$,
\begin{align*}
\E(U_2^2) \le & \, \frac{2}{n^2p^2} \sum_{i, j \in [p_k]} \E\l[ \l( \sum_{t \in [n]} \sum_{\ell, m \in [p_{-k}]} \sum_{q \in [r_{-k}]} d^\kk_{\ell q} d^{\kk}_{m q} \l( Z_{k, i\ell, t} Z_{k, j m, t} - \E(Z_{k, i\ell, t} Z_{k, j m, t}) \r) \r)^2 \r] 
\\ & + \frac{2}{n^2p^2} \l\Vert \sum_{t \in [n]} \E\l( \mbf Z_{k, t} \mbf D_k \mbf D_k^\top \mbf Z_{k, t}^\top \r) \r\Vert^2
=: V_{2, 1} + V_{2, 2}.
\end{align*}
Then, by Assumptions~\ref{assum:heavy}~\ref{cond:factor:bound} and~\ref{assum:indep}, Lemmas~\ref{lem:one}~\ref{lem:one:one} and~\ref{lem:one:four}, \ref{lem:mixing}, \ref{lem:hall} and~\ref{lem:D}~\ref{lem:D:three},
\begin{align*}
V_{2, 1} \lesssim &\, \frac{1}{n^2p^2} \sum_{i, j \in [p_k]} \sum_{\ell, \ell', m, m' \in [p_{-k}]} \sum_{q, q' \in [r_{-k}]} \sum_{t, u \in [n]} d^\kk_{\ell q} d^{\kk}_{m q} d^\kk_{\ell' q'} d^{\kk}_{m' q'} \Cov( Z_{k, i\ell, t} Z_{k, j m, t}, Z_{k, i\ell', u} Z_{k, j m', u})
\\
\le &\, \frac{1}{n^2p^2} \sum_{i \in [p_k]} \sum_{\substack{\ell, \ell', m, m' \in [p_{-k}] \\ (\ell, m) = (\ell', m') \text{ \ or} \\ (\ell, m) = (m', \ell') \text{ \ or} \\ (\ell, \ell') = (m, m')}} \sum_{q, q' \in [r_{-k}]} \sum_{t, u \in [n]} d^\kk_{\ell q} d^{\kk}_{m q} d^\kk_{\ell' q'} d^{\kk}_{m' q'} \Vert Z_{k, i\ell, t} Z_{k, i m, t} \Vert_\nu \Vert Z_{k, i\ell', u} Z_{k, i m', u} \Vert_\nu 
\\
& \quad \times \exp\l( - \frac{c_0(\nu - 2)\vert t - u \vert}{\nu} \r)
+ \frac{1}{n^2p^2} \sum_{\substack{i, j \in [p_k] \\  i \ne j}} \sum_{\ell, m \in [p_{-k}]} \sum_{q, q' \in [r_{-k}]} \sum_{t, u \in [n]} d^\kk_{\ell q} d^{\kk}_{m q} d^\kk_{\ell q'} d^{\kk}_{m q'} \\
& \quad \times \Vert Z_{k, i\ell, t} \Vert_\nu \Vert Z_{k, j m, t} \Vert_\nu \Vert Z_{k, i\ell, u} \Vert_\nu \Vert Z_{k, j m, u} \Vert_\nu \exp\l( - \frac{c_0(\nu - 2)\vert t - u \vert}{\nu} \r)
\\
\lesssim & \, \frac{(\tau^\kk_{n, p})^{2 - 2\eps}}{n^2 p_k p_{-k}^2} \sum_{t, u \in [n]} (\vert \mc F_t \vert_2 + \omega)^{1 + \eps} (\vert \mc F_u \vert_2 + \omega)^{1 + \eps} \exp\l( - \frac{c_0\vert t - u \vert}{3\log(np_{-k})} \r)
\\
& + \frac{M_n^{2 - 2\eps}}{n^2p_{-k}^2} \sum_{t, u \in [n]} (\vert \mc F_t \vert_2 + \omega)^{1 + \eps} (\vert \mc F_u \vert_2 + \omega)^{1 + \eps} \exp\l( - \frac{c_0 \eps \vert t - u \vert}{1 + \eps} \r)
\\
\lesssim & \, \l( \frac{\psi^\kk_{n, p}}{\sqrt{p}} + \frac{M_n^{1 - \eps}}{\sqrt n p_{-k}} \r)^2
\end{align*}
with $\nu \in \{ 2 + \log^{-1}(np_{-k}), 2 + 2\eps \}$.
Also, since $\E(\sum_{\ell, m \in [p_{-k}]} \sum_{q \in [r_{-k}]} d^\kk_{\ell q} d^\kk_{m q} Z_{k, i\ell, t} Z_{k, jm, t}) = 0$ for $i \ne j$ due to Assumption~\ref{assum:indep}~\ref{cond:indep:mixing}, we have
\begin{align*}
V_{2, 2} &\lesssim \max_{i \in [p_k]} \frac{1}{p^2} \l( \frac{1}{n} \sum_{t \in [n]} \sum_{\ell \in [p_{-k}]} \sum_{q \in [r_{-k}]} d^\kk_{\ell q} d^\kk_{\ell q} \E( Z_{k, i\ell, t}^2 ) \r)^2
\\
&\lesssim \max_{i \in [p_k]} \frac{1}{p^2} \l( \frac{1}{n} \sum_{t \in [n]} (\vert \mc F_t \vert_2^2 + \omega^2) \r)^2 \lesssim \frac{\omega^4}{p^2},
\end{align*}
due to Assumption~\ref{assum:heavy}~\ref{cond:heavy:factor} and Lemma~\ref{lem:D}~\ref{lem:D:three}.
Collecting the bounds on $V_{2, 1}$ and $V_{2, 2}$, we obtain by Markov's inequality, 
\begin{align*}
U_2 = O_P\l( \frac{\psi^\kk_{n, p}}{\sqrt p} \vee \frac{M_n^{1 - \eps}}{\sqrt{n} p_{-k}}\vee \frac{\omega^2}{p} \r),
\end{align*}
and thus
\begin{align*}
U_1 + U_2 = O_P\l[ \frac{\psi^\kk_{n, p} \bar{\psi}^\kk_{n, p}}{\sqrt{p_k}}
\vee \frac{\bar{\psi}_{n, p}}{\sqrt p}
\vee \frac{M_n^{1 - \eps}}{\sqrt n} \l( \frac{1}{p_{-k}} \vee \bar{\psi}^\kk_{n, p} \r)
\vee \frac{1}{p}
\r]
\end{align*}
which completes the proof. 
\end{proof}

\begin{proof}[Proof of Lemma~\ref{lem:T:one}~\ref{lem:T:one:three}]
We continue with the decomposition in~\eqref{eq:t:one:u}.
WLOG, we fix $k = 1$.
Then from Lemma~\ref{lem:one}~\ref{lem:one:four}, \ref{lem:mixing} and~\ref{lem:hall} and Assumption~\ref{assum:rf}, 
\begin{align*}
V_{1, 1} &\lesssim \frac{1}{n^2p^2} \sum_{i, j \in [p_k]} \otimes_{i = 2}^K \sum_{i_l, i_l' \in [p_l]} \sum_{t, u \in [n]} \Vert Z_{1, (i, \mbf i_{2:K}), t} Z_{1, (j, \mbf i'_{2:K}), t} \Vert_\nu \Vert Z_{1, (i, \mbf i_{2:K}), u} Z_{1, (j, \mbf i'_{2:K}), u} \Vert_\nu 
\\
& \qquad \qquad \qquad \times \exp\l( - \frac{c_0(\nu - 2)(\vert t - u \vert + \sum_{l = 2}^K \vert i_l - i'_l \vert)}{K\nu} \r)
\\
&\lesssim \frac{(\tau^{(1)}_{n, p})^{\frac{4(\nu - 1 - \eps)}{\nu}}}{np_{-k}} \cdot \frac{1}{n} \sum_{t, u \in [n]} (\vert \mc F_t \vert_2 + \omega)^{\frac{2 + 2\eps}{\nu}} (\vert \mc F_u \vert_2 + \omega)^{\frac{2 + 2\eps}{\nu}} \exp\l( - \frac{c_0\vert t - u \vert}{3K\log(np)} \r)
\\ 
& \qquad \qquad \times \prod_{i = 2}^K \frac{1}{p_l} \sum_{i_l, i_l' \in [p_l]} \exp\l( - \frac{c_0\vert i_l - i'_l \vert}{3K\log(np)} \r)
\\
&\lesssim \frac{c_\eps K^K (\tau^{(1)}_{n, p})^{2 - 2\eps}}{np_{-k}} \log^K(np)
\lesssim ( \psi^{(1)}_{n, p} )^2
\end{align*}
with $\nu = 2 + \log^{-1}(np_{-k})$.
Similarly, from Lemma~\ref{lem:one}~\ref{lem:one:one} and Assumption~\ref{assum:heavy}~\ref{cond:heavy:factor},
\begin{align*}
V_{1, 2} &= \frac{2}{n^2p^2} \otimes_{l \in [K]} \sum_{i_l, i'_l \in [p_l]} \l( \sum_{t \in [n]} \E(Z_{1, \mbf i, t} Z_{1, \mbf i', t}) \r)^2
\\
&\lesssim \frac{1}{n^2p^2} \otimes_{l \in [K]} 
\sum_{i_l, i'_l \in [p_l]} \l[ \sum_{t \in [n]} \Vert Z_{1, \mbf i, t} \Vert_\nu \Vert Z_{1, \mbf i'_{2:K}, t} \Vert \exp\l( - \frac{c_0 (\nu - 2) \sum_{l \in [K]} \vert i_l - i'_l \vert}{K\nu} \r) \r]^2
\\
&\lesssim \frac{1}{p} \l[ \frac{1}{n} \sum_{t \in [n]} (\vert \mc F_t \vert_2 + \omega)^2 \r]^2 \prod_{l \in [K]} \frac{1}{p_l} \sum_{i_l, i'_l \in [p_l]} \exp\l( - \frac{2 c_0 \eps \vert i_l - i'_l \vert}{K(1 + \eps)} \r) 
\lesssim \frac{C_\eps \omega^4}{p}.
\end{align*}
with $\nu = 2 + 2\eps$.
These arguments apply to all $k \in [K]$ and thus combining the bounds on $V_{1, 1}$ and $V_{1, 2}$ with~\eqref{eq:DD:bound} and~\eqref{eq:t:one:u:one}, by Markov's inequality,
\begin{align*}
U_1 = O_P\l[ \l( \psi^\kk_{n, p} \vee \frac{1}{p} \r) \bar{\psi}^\kk_{n, p} \r].
\end{align*}
% Note that
% \begin{align}
% \frac{1}{n} \sum_{t \in [n]} \E\l( \l\Vert \mbf Z_{k, t} \r\Vert_F^2 \r) &= \frac{1}{n} \sum_{t \in [n]} \sum_{i \in [p_k]} \sum_{j \in [p_{-k}]} \Vert Z_{k, ij, t} \Vert_2^2 \lesssim \frac{1}{n} \sum_{t \in [n]} \sum_{i \in [p_k]} \sum_{j \in [p_{-k}]} \Vert X_{k, ij, t} \Vert_2^2 
% \nn \\
% &\le \frac{2p}{n} \sum_{t \in [n]} (\vert \mc F_t \vert_2^2 + \omega^2) \lesssim p\omega^2,
% \label{eq:z:bound}
% \end{align}
% from Lemma~\ref{lem:one}~\ref{lem:one:one} and Assumption~\ref{assum:heavy}. Combining~\eqref{eq:z:bound} and~\eqref{eq:DD:bound},
% \begin{align*}
% U_1 \le \frac{1}{np} \sum_{t \in [n]} \l\Vert \mbf Z_{k, t} \r\Vert_F^2 \l\Vert \wh{\mbf D}_k \wh{\mbf D}_k^\top - \mbf D_k \mbf D_k^\top \r\Vert 
% = O_P\l( \omega^2 \bar{\psi}_{n, p}^\kk \r).
% \end{align*}
As for $U_2$, note that
\begin{align*}
\E(U_2^2) \le & \, 2 \; \E\l( \l\Vert \frac{1}{np} \sum_{t \in [n]} \l( \mbf Z_{1, t} \mbf D_1 \mbf D_1^\top \mbf Z_{1, t}^\top - \E\l( \mbf Z_{1, t} \mbf D_1 \mbf D_1^\top \mbf Z_{1, t}^\top \r) \r) \r\Vert^2 \r)
\\
& + 2 \l\Vert \frac{1}{np} \sum_{t \in [n]} \E\l( \mbf Z_{1, t} \mbf D_1 \mbf D_1^\top \mbf Z_{1, t}^\top \r) \r\Vert^2 =: 2 \E(\Vert \mbf V_1 \Vert^2) + 2 \E(\Vert \mbf V_2 \Vert^2),
\end{align*}
with $\mbf V_\ell = [V_{\ell, ij}, \, i, j \in [p_k]]], \, \ell = 1, 2$. 
Setting $\nu = 2 + \log^{-1}(np)$, we have
\begin{align*}
\Vert V_{1, ij} \Vert_2^2 &\le \frac{C^4r_{-1}^2}{n^2p^2p_{-1}^2} \sum_{t, u \in [n]} \otimes_{l = 2}^K \sum_{a_l, b_l, c_l, d_l \in [p_l]} \l\vert \Cov\l( Z_{1, (i, \mbf a_{2:K}), t} Z_{1, (j, \mbf b_{2:K}), t}, Z_{1, (i, \mbf c_{2:K}), u} Z_{1, (j, \mbf d_{2:K}) u} \r) \r\vert
\\
&\le \frac{8C^4r_{-1}^2}{n^2p^2p_{-1}^2} \sum_{t, u \in [n]} \otimes_{l = 2}^K \sum_{a_l, b_l, c_l, d_l \in [p_l]} \l\Vert Z_{1, (i, \mbf a_{2:K}), t} Z_{1, (j, \mbf b_{2:K}), t} \r\Vert_\nu \l\Vert Z_{1, (i, \mbf c_{2:K}), u} Z_{1, (j, \mbf d_{2:K}) u} \r\Vert_\nu \times
\\
& \exp\l\{ - \frac{c_0 (\nu - 2) [\vert t - u \vert +  \min( \sum_{l = 2}^K \vert a_l - c_l \vert, \sum_{l = 2}^K \vert b_l - c_l \vert, \sum_{l = 2}^K \vert a_l - d_l \vert, \sum_{l = 2}^K \vert b_l - d_l \vert)]}{K\nu} \r\}
\\
&\lesssim \frac{(\tau^{(1)}_{n, p})^{\frac{4 (\nu - 1 - \eps)}{\nu}}}{n^2p^2p_{-1}^2} \sum_{t, u \in [n]} \l( \vert \mc F_t \vert_2^{\frac{2 + 2\eps}{\nu}} + \omega^{\frac{2 + 2\eps}{\nu}} \r) \l( \vert \mc F_u \vert_2^{\frac{2 + 2\eps}{\nu}} + \omega^{\frac{2 + 2\eps}{\nu}} \r) \exp\l( - \frac{c_0 \vert t - u \vert (\nu - 2)}{K\nu} \r) \times 
\\
& \qquad \prod_{l = 2}^K \sum_{a_l, b_l, c_l, d_l \in [p_l]} \exp\l( - \frac{c_0 (\nu - 2) \min( \vert a_l - c_l \vert, \vert b_l - c_l \vert, \vert a_l - d_l \vert, \vert b_l - d_l \vert)}{K\nu} \r)
\\
&\lesssim \frac{(\tau^{(1)}_{n, p})^{2 - 2\eps}}{n^2p^2p_{-1}^2} \sum_{t, u \in [n]} (\vert \mc F_u \vert_2^{1 + \eps} + \omega^{1 + \eps}) (\vert \mc F_u \vert_2^{1 + \eps} + \omega^{1 + \eps}) \exp\l( - \frac{c_0 \vert t - u \vert)}{3 K \log(np)} \r) \times
\\
& \qquad \prod_{l = 2}^K \sum_{a_l, b_l, c_l, d_l \in [p_l]} \exp\l( - \frac{c_0 \min( \vert a_l - c_l \vert, \vert b_l - c_l \vert, \vert a_l - d_l \vert, \vert b_l - d_l \vert)}{3K\log(np)} \r)
\\
&\lesssim \frac{\omega^{2 + 2\eps} (\tau^{(1)}_{n, p})^{2 - 2\eps} p_{-1} \log^K(np)}{np^2} \lesssim \frac{(\psi^{(1)}_{n, p})^2}{p_1^2}
\end{align*}
where the first inequality follows from Lemma~\ref{lem:D}~\ref{lem:D:three},
the second from Assumption~\ref{assum:rf}~\ref{cond:rf:mixing} and Lemmas~\ref{lem:mixing} and~\ref{lem:hall},
the third from Lemma~\ref{lem:one}~\ref{lem:one:four} and the penultimate one from Assumption~\ref{assum:rf}~\ref{cond:rf:factor:mixing}.
Analogous arguments apply to all $k \in [K]$ such that
\begin{align}
\label{lem:t:one:v:one}
\E(\Vert \mbf V_1 \Vert^2) \le  \sum_{i, j \in [p_k]} \Vert V_{1, ij} \Vert_2^2 \lesssim ( \psi^\kk_{n, p} )^2.
\end{align}
Next, continuing to fix $k = 1$, for all $i, j \in [p_1]$, it holds that
\begin{align*}
V_{2, ij} 
&\le \frac{1}{np} \sum_{t \in [n]} \otimes_{l = 2}^K \sum_{i_l, i'_l \in [p_l]} \sum_{q \in [r_{-1}]} d^{(1)}_{\ell q} d^{(1)}_{m q} \l\vert \Cov\l( Z_{1, (i, \mbf i_{2:K}), t}, Z_{1, (j, \mbf i'_{2:K}), t} \r) \r\vert
\\
&\le \frac{C^2r_{-1}}{npp_{-1}} \sum_{t \in [n]} \otimes_{l = 2}^K \sum_{i_l, i'_l \in [p_l]} \Vert Z_{1, (j, \mbf i_{2:K}), t} \Vert_\nu \Vert Z_{1, (i, \mbf i'_{2:K}), t} \Vert_\nu 
\\
& \qquad \qquad \times \exp\l( - \frac{c_0(\nu - 2)(\vert i - j \vert + \sum_{l = 2}^K \vert i_l - i'_l \vert)}{K \nu} \r)
\\
&\le \frac{C^2r_{-1}}{npp_{-1}} \exp\l( - \frac{c_0 \eps \vert i - j \vert}{K (1 + \eps) \log(np)} \r) \sum_{t \in [n]} (\vert \mc F_t \vert_2^2 + \omega^2) \prod_{l = 2}^K \sum_{i_l, i'_l \in [p_l]} \exp\l( - \frac{c_0 \eps \vert i_l - i'_l \vert}{K (1 + \eps)} \r)
\\
&\lesssim \frac{C_\eps \omega^2}{p} \exp\l( - \frac{c_0 \eps \vert i - j \vert}{K(1 + \eps) } \r)
\end{align*}
with $\nu = 2 + 2\eps$, where the second inequality follows from Lemma~\ref{lem:D}~\ref{lem:D:three} and Lemma~\ref{lem:hall}, the third from Lemma~\ref{lem:one}~\ref{lem:one:one} and the last from Assumption~\ref{assum:heavy}~\ref{cond:heavy:factor}.
Similar arguments hold for all $k \in [K]$, and thus 
\begin{align}
\label{lem:t:one:v:two}
\Vert \mbf V_2 \Vert \le \Vert \mbf V_2 \Vert_1 \le \sum_{j \in [p_k]} \vert V_{2, ij} \vert \lesssim \frac{C_\eps \omega^2}{p}.
\end{align}
Putting together the bounds on $\Vert \mbf V_\ell \Vert, \, \ell = 1, 2$, we have
\begin{align*}
U_2 = O_P \l( \psi^\kk_{n, p} \vee \frac{1}{p} \r).
\end{align*}
Combining the bound on $U_2$ with that on $U_1$, the proof is complete.
\end{proof}

\begin{lemma}
\label{lem:T:two}
{\it Let Assumptions~\ref{assum:loading}, \ref{assum:factor} and~\ref{assum:heavy} hold.  
For each $k \in [K]$, we have the followings:
\begin{enumerate}[wide, itemsep = 0pt, label = (\roman*)]
\item \label{lem:T:two:one} Under Assumption~\ref{assum:indep},
\begin{align*}
\frac{1}{np} \l\Vert \sum_{t \in [n]} \mbf Z_{k, t} \wh{\mbf D}_k \wh{\mbf D}_k^\top \E(\mbf X^{\trunc}_{k, t})^\top \r\Vert = O_P\l( M_n^{1 - \eps} \l( \frac{\bar{\psi}^\kk_{n, p}}{\sqrt n} \vee \frac{1}{\sqrt{np_{-k}}} \r) \r),
\\
\frac{1}{np} \l\Vert \sum_{t \in [n]} \E(\mbf X^{\trunc}_{k, t}) \wh{\mbf D}_k \wh{\mbf D}_k^\top \mbf Z_{k, t}^\top \r\Vert = O_P\l( M_n^{1 - \eps} \l( \frac{\bar{\psi}^\kk_{n, p}}{\sqrt n} \vee \frac{1}{\sqrt{np_{-k}}} \r) \r).
\end{align*}

\item \label{lem:T:two:three} Under Assumption~\ref{assum:rf},
\begin{align*}
\frac{1}{np} \l\Vert \sum_{t \in [n]} \mbf Z_{k, t} \wh{\mbf D}_k \wh{\mbf D}_k^\top \E(\mbf X^{\trunc}_{k, t})^\top \r\Vert = O_P\l( M_n^{1 - \eps} \l( \frac{\bar{\psi}^\kk_{n, p}}{\sqrt n} \vee \frac{1}{\sqrt{np_{-k}}} \r) \r),
\\
\frac{1}{np} \l\Vert \sum_{t \in [n]} \E(\mbf X^{\trunc}_{k, t}) \wh{\mbf D}_k \wh{\mbf D}_k^\top \mbf Z_{k, t}^\top \r\Vert = O_P\l( M_n^{1 - \eps} \l( \frac{\bar{\psi}^\kk_{n, p}}{\sqrt n} \vee \frac{1}{\sqrt{np_{-k}}} \r) \r).
\end{align*}
\end{enumerate}
}
\end{lemma}

\begin{proof}[Proof of Lemma~\ref{lem:T:two}~\ref{lem:T:two:one}]
Let us write
\begin{align}
\frac{1}{np} \l\Vert \sum_{t \in [n]} \mbf Z_{k, t} \wh{\mbf D}_k \wh{\mbf D}_k^\top \E(\mbf X^{\trunc}_{k, t})^\top \r\Vert \le& \, 
\frac{1}{np} \l\Vert \sum_{t \in [n]} \mbf Z_{k, t} \l( \wh{\mbf D}_k \wh{\mbf D}_k^\top - \mbf D_k \mbf D_k^\top \r) \E(\mbf X^{\trunc}_{k, t})^\top \r\Vert 
\nn \\
& + \frac{1}{np} \l\Vert \sum_{t \in [n]} \mbf Z_{k, t} \mbf D_k \mbf D_k^\top \E(\mbf X^{\trunc}_{k, t})^\top \r\Vert =: U_1 + U_2.
\label{eq:lem:t:two:u}
\end{align}
By Lemma~\ref{lem:abc}, we have
\begin{align*}
U_1^2 \le \underbrace{\frac{1}{n^2p^2} \sum_{i, j \in [p_k]} \sum_{\ell, m \in [p_{-k}]} \l( \sum_{t \in [n]} Z_{k, i\ell, t} \E(X^\trunc_{k, jm, t}) \r)^2}_{V_1} \l\Vert \wh{\mbf D}_k \wh{\mbf D}_k^\top - \mbf D_k \mbf D_k^\top \r\Vert^2
\end{align*}
where, by Lemmas~\ref{lem:one}~\ref{lem:one:one}, \ref{lem:mixing}, and~\ref{lem:hall} and Assumptions~\ref{assum:heavy}~\ref{cond:factor:bound} and~\ref{assum:indep},
\begin{align*}
\E(V_1) &= \frac{1}{n^2p^2} \sum_{i, j \in [p_k]} \sum_{\ell, m \in [p_{-k}]} \sum_{t, u \in [n]} \E(X^\trunc_{k, jm, t}) \E(X^\trunc_{k, jm, u}) \Cov(Z_{k, i\ell, t}, Z_{k, i\ell, u})
\\
&\lesssim \frac{1}{n^2p^2} \sum_{i, j \in [p_k]} \sum_{\ell, m \in [p_{-k}]} \sum_{t, u \in [n]} \l\vert \E(X^\trunc_{k, jm, t}) \E(X^\trunc_{k, jm, u}) \r\vert
\\
& \qquad \times \Vert Z_{k, i\ell, t} \Vert_\nu \Vert Z_{k, i\ell, u} \Vert_\nu \exp\l( - \frac{c_0 (\nu - 2) \vert t - u \vert}{\nu} \r)
\\
& \lesssim \frac{1}{n^2} \sum_{t, u \in [n]} (\vert \mc F_t \vert_2 + \omega)^2 (\vert \mc F_u \vert_2 + \omega)^2 \exp\l( - \frac{c_0 \eps \vert t - u \vert}{1 + \eps} \r)
\le \frac{c_\eps M_n^{2 - 2\eps} \omega^{2 + 2\eps}}{n}
\end{align*}
with $\nu = 2 + 2\eps$, which leads to
\begin{align*}
U_1 = O_P\l( \frac{M_n^{1 - \eps}}{\sqrt n} \bar{\psi}^\kk_{n, p} \r)   
\end{align*}
by Markov's inequality and~\eqref{eq:DD:bound}.
Similarly, additionally evoking Lemma~\ref{lem:D}~\ref{lem:D:three},
\begin{align*}
\E(U_2^2) &\le \frac{1}{n^2p^2} \sum_{i, j \in [p_k]} \E\l[ \l( \sum_{\ell, m \in [p_{-k}]} \sum_{q \in [r_{-k}]} \sum_{t \in [n]} d^\kk_{\ell q} d^\kk_{m q} \E(X^\trunc_{k, jm, t}) Z_{k, i\ell, t} \r)^2 \r]
\\
&= \frac{1}{n^2p^2} \sum_{i, j \in [p_k]} \sum_{\ell, m, m' \in [p_{-k}]} \sum_{q, q' \in [r_{-k}]} \sum_{t, u \in [n]} d^\kk_{\ell q} d^\kk_{m q} d^\kk_{\ell q'} d^\kk_{m' q'} \\
& \qquad \times \E(X^\trunc_{k, jm, t}) \E(X^\trunc_{k, jm', u}) \Cov(Z_{k, i\ell, t}, Z_{k, i\ell, u})
\\
&\lesssim \frac{1}{n^2 p_{-k}} \sum_{t, u \in [n]} (\vert \mc F_t \vert_2 + \omega)^2 (\vert \mc F_u \vert_2 + \omega)^2 \exp\l( - \frac{c_0 \eps \vert t - u \vert}{1 + \eps} \r)
\lesssim \frac{c_\eps M_n^{2 - 2\eps}}{n p_{-k}},
\end{align*}
leading to
\begin{align*}
U_2 = O_P\l( \frac{M_n^{1 - \eps}}{\sqrt{np_{-k}}} \r),
\end{align*}
which completes the proof of the first claim. 
The second claim follows analogously.
\end{proof}

\begin{proof}[Proof of Lemma~\ref{lem:T:two}~\ref{lem:T:two:three}]
We continue with the decomposition in~\eqref{eq:lem:t:two:u} and fix $k = 1$.
By Lemmas~\ref{lem:one}~\ref{lem:one:one}, \ref{lem:mixing}, and~\ref{lem:hall} and Assumptions~\ref{assum:heavy}~\ref{cond:factor:bound} and~\ref{assum:rf}, 
\begin{align*}
\E(V_1) &= \frac{1}{n^2p^2} \sum_{i, j \in [p_1]} \otimes_{l = 2}^K \sum_{i_l, i'_l \in [p_l]} \sum_{t, u \in [n]} \E(X^\trunc_{1, (j, \mbf i'_{2:K}), t}) \E(X^\trunc_{1, (j, \mbf i'_{2:K}), u}) \Cov(Z_{1, (i, \mbf i_{2:K}), t}, Z_{1, (i, \mbf i_{2:K}), u})
\\
&\lesssim \frac{1}{n^2p^2} \sum_{i, j \in [p_1]} \otimes_{l = 2}^K \sum_{i_l, i'_l \in [p_l]} \sum_{t, u \in [n]} \l\vert \E(X^\trunc_{1, (j, \mbf i'_{2:K}), t}) \E(X^\trunc_{1, (j, \mbf i'_{2:K}), u}) \r\vert
\\
& \qquad \times \Vert Z_{1, (i, \mbf i_{2:K}), t} \Vert_\nu \Vert Z_{1, (i, \mbf i_{2:K}), u} \Vert_\nu \exp\l( - \frac{c_0 (\nu - 2) \vert t - u \vert}{K\nu} \r)
\\
& \lesssim \frac{1}{n^2} \sum_{t, u \in [n]} (\vert \mc F_t \vert_2 + \omega)^2 (\vert \mc F_u \vert_2 + \omega)^2 \exp\l( - \frac{c_0 \eps \vert t - u \vert}{K(1 + \eps)} \r)
\le \frac{c_\eps K M_n^{2 - 2\eps} \omega^{2 + 2\eps}}{n}
\end{align*}
with $\nu = 2 + 2\eps$, which leads to
\begin{align*}
U_1 = O_P\l( \frac{M_n^{1 - \eps}}{\sqrt n} \bar{\psi}^\kk_{n, p} \r)   
\end{align*}
% Note that by Lemma~\ref{lem:one}~\ref{lem:one:one} and Assumption~\ref{assum:heavy},
% \begin{align}
% \label{eq:chit:bound}
% \frac{1}{n} \sum_{t \in [n]} \Vert \E(\mbf X^{\trunc}_{k, t}) \Vert_F^2
% \le \frac{p}{n} \sum_{t \in [n]} \l( \vert \mc F_t \vert_2^2 + \omega^2 \r) \lesssim p \omega^2.
% \end{align}
% This, together with \eqref{eq:z:bound} and~\eqref{eq:DD:bound}, gives
% \begin{align*}
% U_1 &\le \frac{1}{np} \sum_{t \in [n]} \l\Vert \mbf Z_{k, t} \r\Vert_F \l\Vert \E(\mbf X^{\trunc}_{k, t}) \r\Vert_F \l\Vert \wh{\mbf D}_k \wh{\mbf D}_k^\top - \mbf D_k \mbf D_k^\top \r\Vert 
% \\
% &\le \frac{1}{p} \sqrt{\frac{1}{n} \sum_{t \in [n]} \l\Vert \mbf Z_{k, t} \r\Vert_F^2 \cdot \frac{1}{n} \sum_{t \in [n]} \l\Vert \E(\mbf X^{\trunc}_{k, t}) \r\Vert_F^2} \l\Vert \wh{\mbf D}_k \wh{\mbf D}_k^\top - \mbf D_k \mbf D_k^\top \r\Vert
% = O_P\l( \omega^2 \bar{\psi}^\kk_{n, p} \r).
% \end{align*}
As for $U_2$, WLOG, we fix $k = 1$ for notational convenience.
Then,
\begin{align*}
\E(U_2^2) &\lesssim \frac{1}{n^2p^2 p_{-1}^2} \sum_{i, j \in [p_1]} \otimes_{l = 2}^K \sum_{a_l, b_l, c_l, d_l \in [p_l]} \sum_{t, u \in [n]}  \E(X^\trunc_{1, (j, \mbf c_{2:K}), t}) \E(X^\trunc_{1, (j, \mbf d_{2:K}), t})
\\
& \qquad \qquad \qquad \qquad \times \Cov(Z_{1, (i, \mbf a_{2:K}), t}, Z_{1, (i, \mbf b_{2:K}), u})
\\
&\lesssim \frac{1}{n^2p^2p_{-1}^2}  \sum_{i, j \in [p_1]} \otimes_{l = 2}^K \sum_{a_l, b_l, c_l, d_l \in [p_l]} \sum_{t, u \in [n]} \E(X^\trunc_{1, (j, \mbf c_{2:K}), t}) \E(X^\trunc_{1, (j, \mbf d_{2:K}), t})
\\
& \qquad \times \Vert Z_{1, (i, \mbf a_{2:K}), t} \Vert_\nu \Vert Z_{1, (i, \mbf b_{2:K}), u} \Vert_\nu \exp\l( - \frac{c_0 (\nu - 2) ( \vert t - u \vert + \sum_{l = 2}^K \vert i_l - i'_l \vert)}{K (\nu - 2)}\r)
\\
&\lesssim \frac{1}{n p_{-1}} \cdot \frac{1}{n} \sum_{t, u \in [n]} (\vert \mc F_t \vert_2 + \omega)^2 (\vert \mc F_u \vert_2 + \omega)^2 \exp\l( - \frac{c_0 \eps \vert t - u \vert}{K (1 + \eps)} \r) 
\\
& \qquad \times \prod_{l = 2}^K \frac{1}{p_l} \sum_{i_l, i_l' \in [p_l]} \exp\l( - \frac{c_0 \eps \vert i_l - i'_l \vert}{K (1 + \eps)} \r) 
\lesssim \frac{c_\eps K^K \omega^{2 + 2\eps} M_n^{2 - 2\eps}}{n p_{-k}},
\end{align*}
with $\nu = 2 + 2\eps$, by Assumptions~\ref{assum:heavy}~\ref{cond:factor:bound} and~\ref{assum:rf} and Lemmas~\ref{lem:one}~\ref{lem:one:one}, \ref{lem:hall} and~\ref{lem:D}~\ref{lem:D:three}.
Hence by Markov's inequality,
\begin{align*}
U_2 = O_P\l( \frac{M_n^{1 - \eps}}{\sqrt{np_{-k}}} \r).
\end{align*}
Combining the bound on $U_2$ with that on $U_1$, the proof of the first claim is complete.
The second claim is proved analogously.
\end{proof}

\begin{lemma}
\label{lem:T:three}
{\it Let Assumptions~\ref{assum:loading}, \ref{assum:factor} and~\ref{assum:heavy} hold. 
For each $k \in [K]$, we have the followings under either Assumptions~\ref{assum:indep} or~\ref{assum:rf}:
\begin{align*}
\frac{1}{np} \l\Vert \sum_{t \in [n]} \E\l( \mbf X^{\trunc}_{k, t} - \mbf X_{k, t} \r) \wh{\mbf D}_k \wh{\mbf D}_k^\top \E\l( \mbf X^{\trunc}_{k, t}\r)^\top \r\Vert &= 
O_P\l( \frac{M_n}{\tau^\kk_{n, p}} \psi^\kk_{n, p} \r),
% = O_P\l( \frac{M_n^\eps \sqrt{\log(np_{-k})}}{(\tau^\kk_{n, p})^\eps} \cdot \frac{M_n^{1 - \eps}}{\sqrt{np_{-k}}} \r),
\\
\frac{1}{np} \l\Vert \sum_{t \in [n]} \E\l( \mbf X_{k, t} \r) \wh{\mbf D}_k \wh{\mbf D}_k^\top \E\l( \mbf X^{\trunc}_{k, t} - \mbf X_{k, t} \r) ^\top \r\Vert &= 
O_P\l( \frac{M_n}{\tau^\kk_{n, p}} \psi^\kk_{n, p} \r).
% O_P\l( \frac{M_n^\eps \sqrt{\log(np_{-k})}}{(\tau^\kk_{n, p})^\eps} \cdot \frac{M_n^{1 - \eps}}{\sqrt{np_{-k}}} \r).
\end{align*}
}
\end{lemma}

\begin{proof}
Let us write
\begin{align}
& \frac{1}{np} \l\Vert \sum_{t \in [n]} \E(\mbf X^{\trunc}_{k, t} - \mbf X_{k, t}) \wh{\mbf D}_k \wh{\mbf D}_k^\top \E(\mbf X^{\trunc}_{k, t})^\top \r\Vert 
\nn \\
\le& \, 
\frac{1}{np} \l\Vert \sum_{t \in [n]} \E(\mbf X^{\trunc}_{k, t} - \mbf X_{k, t}) \l( \wh{\mbf D}_k \wh{\mbf D}_k^\top - \mbf D_k \mbf D_k^\top \r) \E(\mbf X^{\trunc}_{k, t})^\top \r\Vert 
\nn \\
& + \frac{1}{np} \l\Vert \sum_{t \in [n]} \E(\mbf X^{\trunc}_{k, t} - \mbf X_{k, t}) \mbf D_k \mbf D_k^\top \E(\mbf X^{\trunc}_{k, t})^\top \r\Vert =: U_1 + U_2.
\nn % \label{eq:lem:t:three:u}
\end{align}
By Lemma~\ref{lem:abc}, we have
\begin{align*}
U_1^2 \le \underbrace{\frac{1}{n^2p^2} \sum_{i, j \in [p_k]} \sum_{\ell, m \in [p_{-k}]} \l( \sum_{t \in [n]} \E(X^\trunc_{k, i\ell, t} - X_{k, i\ell, t}) \E(X^\trunc_{k, jm, t}) \r)^2}_{V_1} \l\Vert \wh{\mbf D}_k \wh{\mbf D}_k^\top - \mbf D_k \mbf D_k^\top \r\Vert^2
\end{align*}
where, by Lemmas~\ref{lem:one}~\ref{lem:one:one} and~\ref{lem:one:two} and Assumption~\ref{assum:heavy}~\ref{cond:heavy:factor} and~\ref{cond:factor:bound},
\begin{align*}
V_1 &\lesssim % \frac{1}{p^2} \sum_{i, j \in [p_k]} \sum_{\ell, m \in [p_{-k}]} 
\l( \frac{1}{n (\tau^\kk_{n, p})^{1 + 2\eps}} \sum_{t \in [n]} (\vert \mc F_t \vert_2 + \omega)^{3 + 2\eps} \r)^2
\lesssim \l( \frac{M_n \omega^{2 + 2\eps}}{ (\tau^\kk_{n, p})^{1 + 2\eps} } \r)^2 = \l( \frac{M_n}{\tau^\kk_{n, p}} \psi^\kk_{n, p} \r)^2
\end{align*}
which, together with~\eqref{eq:DD:bound}, leads to
\begin{align*}
U_1 = O_P\l( \frac{M_n}{\tau^\kk_{n, p}} \psi^\kk_{n, p} \bar{\psi}^\kk_{n, p} \r).   
\end{align*}
Similarly, additionally evoking Lemma~\ref{lem:D}~\ref{lem:D:three},
\begin{align*}
U_2^2 &\le \frac{1}{n^2p^2} \sum_{i, j \in [p_k]} \l( \sum_{\ell, m \in [p_{-k}]} \sum_{q \in [r_{-k}]} \sum_{t \in [n]} d^\kk_{\ell q} d^\kk_{m q} \E(X^\trunc_{k, i\ell, t} - X_{k, i\ell, t})\E(X^\trunc_{k, jm, t}) \r)^2
\\
&\lesssim \l( \frac{1}{n(\tau^\kk_{n, p})^{1 + 2\eps}} \sum_{t \in [n]} (\vert \mc F_t \vert_2 + \omega)^{3 + 2\eps} \r)^2
\lesssim \l( \frac{M_n}{\tau^\kk_{n, p}} \psi^\kk_{n, p} \r)^2,
\end{align*}
leading to
\begin{align*}
U_2 = O\l( \frac{M_n}{\tau^\kk_{n, p}} \psi^\kk_{n, p} \r).
% = O\l( \frac{M_n^\eps \sqrt{\log(np_{-k})}}{(\tau^\kk_{n, p})^\eps} \cdot \frac{M_n^{1 - \eps}}{\sqrt{np_{-k}}} \r).
\end{align*}
which completes the proof of the first claim. 
The second claim is proved analogously. 
\end{proof}

\begin{lemma}
\label{lem:T:four}
{\it Let Assumptions~\ref{assum:loading}, \ref{assum:factor} and~\ref{assum:heavy} hold. 
For each $k \in [K]$, we have the followings:
\begin{enumerate}[wide, itemsep = 0pt, label = (\roman*)]
\item \label{lem:T:four:one} Under Assumption~\ref{assum:indep},
\begin{align*}
& \l\Vert \frac{1}{np} \sum_{t \in [n]} \E(\mbf X_{k, t}) \l( \wh{\mbf D}_k \wh{\mbf D}_k^\top - {\cgr \frac{1}{p_{-k}} \bm\Delta_k\bm\Delta_k^\top } \r) \E(\mbf X_{k, t})^\top \r\Vert 
\\
= &\, O_P \l( \sum_{k' \in [K] \setminus \{k\}} \l( \frac{M_n^{1 - \eps}}{\sqrt{np_{-k'}}} \vee \frac{1}{p_{k'}} \vee \frac{\psi^{(k')}_{n, p}}{\sqrt p_{k'}} \r) \r).
\end{align*}

\item \label{lem:T:four:two} Under Assumption~\ref{assum:rf},
\begin{align*}
\l\Vert \frac{1}{np} \sum_{t \in [n]} \E(\mbf X_{k, t}) \l( \wh{\mbf D}_k \wh{\mbf D}_k^\top - {\cgr \frac{1}{p_{-k}} \bm\Delta_k\bm\Delta_k^\top } \r) \E(\mbf X_{k, t})^\top \r\Vert = O_P\l( \sum_{k' \in [K] \setminus \{k\}} \l( \psi^{(k')}_{n, p} \vee \frac{1}{p_{k'}} \r) \r).
\end{align*}
\end{enumerate}
}
\end{lemma}

\begin{proof}[Proof of Lemma~\ref{lem:T:four}~\ref{lem:T:four:one}]
By Lemma~\ref{lem:abc}, we have
\begin{align*}
& \frac{1}{n^2p^2} \l\Vert \sum_{t \in [n]} \E(\mbf X_{k, t}) \l( \wh{\mbf D}_k \wh{\mbf D}_k^\top -  {\cgr \frac{1}{p_{-k}} \bm\Delta_k\bm\Delta_k^\top } \r) \E(\mbf X_{k, t})^\top \r\Vert^2_F
\\
\le &\, \underbrace{\frac{1}{n^2p^2} \sum_{i, j \in [p_k]} \sum_{\ell, m \in [p_{-k}]} \l( \sum_{t \in [n]} \E(X^\trunc_{k, i\ell, t}) \E(X^\trunc_{k, jm, t}) \r)^2}_{U_1} \l\Vert \wh{\mbf D}_k \wh{\mbf D}_k^\top -  {\cgr \frac{1}{p_{-k}} \bm\Delta_k\bm\Delta_k^\top } \r\Vert_F^2.
\end{align*}
By Lemma~\ref{lem:one}~\ref{lem:one:one} and Assumption~\ref{assum:heavy}~\ref{cond:heavy:factor}, we have
\begin{align*}
U_1 &\lesssim \l( \frac{1}{n} \sum_{t \in [n]} (\vert \mc F_t \vert_2 + \omega)^2 \r)^2 \lesssim \omega^4
\end{align*}
which, in combination with~\eqref{eq:DD:indep}, leads to the claim.
\end{proof}

\begin{proof}[Proof of Lemma~\ref{lem:T:four}~\ref{lem:T:four:two}]
The proof takes analogous steps as in that of Lemma~\ref{lem:T:four}~\ref{lem:T:four:one} except that we evoke~\eqref{eq:DD:rf} in place of~\eqref{eq:DD:indep}.
\end{proof}

{\cgr
\begin{lemma}
\label{lem:second:rough}
{\it
Let Assumptions~\ref{assum:loading}, \ref{assum:factor}, \ref{assum:heavy}, and~\ref{assum:indep} or~\ref{assum:rf} hold.
For $\wc{\mbf M}^{[1]}_k(\tau) \in \R^{r_k \times r_k}$ denotes the diagonal matrix containing the eigenvalues $\wc{\mu}^\kk_j(\tau), \, j \in [r_k]$, of $\wc{\bm\Gamma}^{\kk, [1]}(\tau)$ on its diagonal, we have
\begin{align*}
\l\Vert \l( p_k^{-1} \wc{\mbf M}^{[1]}_k(\tau) \r)^{-1} - \l( p_k^{-1} \mbf M_{\chi, k} \r)^{-1} \r\Vert = O_P\l( \frac{1}{p_k} \l\Vert \wc{\bm\Gamma}^{\kk, [1]} - \bm\Gamma^\kk_\chi \r\Vert \r).
\end{align*}
}
\end{lemma}}

\begin{proof} 
Having $\bm\Gamma^\kk_\chi$ fulfil the condition~\ref{lem:dk:cond:one} in Lemma~\ref{lem:dk} in place of $\mbf S$ and Proposition~\ref{prop:gamma:wc} take the role of~\ref{lem:dk:cond:two}, the conclusions follow from Lemma~\ref{lem:dk}~\ref{lem:dk:two}.
\end{proof}

\begin{lemma}
\label{lem:abc}
{\it
For some sequence of matrices $\mbf A_t = [a_{ij, t}], \mbf C_t, \, t \in [n]$, and some matrix $\mbf B$ of compatible dimensions,
\begin{align*}
\l\Vert \sum_{t \in [n]} \mbf A_t \mbf B \mbf C_t \r\Vert_F^2 \le \sum_{i, j} \l\Vert \sum_{t \in [n]} a_{ij, t} \mbf C_t \r\Vert_F^2 \Vert \mbf B \Vert_F^2.
\end{align*}
}
\end{lemma}

\begin{proof}
Writing $\mbf B = [b_{j \ell}]$ and $\mbf C_t = [c_{\ell m, t}]$, by Cauchy-Schwarz inequality,
\begin{align*}
\l\Vert \sum_{t \in [n]} \mbf A \mbf B \mbf C \r\Vert_F^2 
&= \sum_i \sum_m \l( \sum_j \sum_\ell \sum_{t \in [n]} a_{ij, t} b_{j \ell} c_{\ell m, t} \r)^2
\\
&\le \sum_i \sum_j \sum_\ell \sum_m  \l( \sum_{t \in [n]} a_{ij, t} c_{\ell m, t} \r)^2 \cdot \Vert \mbf B \Vert_F^2
= \sum_{i, j} \l\Vert \sum_{t \in [n]} a_{ij, t} \mbf C_t \r\Vert_F^2 \Vert \mbf B \Vert_F^2.
\end{align*}
\end{proof}

\subsection{Proof of Theorem~\ref{thm:third}}

Throughout, we suppress the dependence on $\tau$ where there is no confusion.

\subsubsection{Proof of Theorem~\ref{thm:third}~\ref{thm:third:one}}

For each $k \in [K]$, analogously as in~\eqref{eq:wc:gamma:decomp}, we may write
\begin{align}
\frac{1}{p_k} \wc{\bm\Gamma}^{\kk, [2]}
=& \, \frac{1}{np} \sum_{t \in [n]}
\mbf Z_{k, t} \wc{\mbf D}^{[1]}_k (\wc{\mbf D}^{[1]}_k)^\top \mbf Z_{k, t}^\top 
\nn \\
& + \frac{1}{np} \sum_{t \in [n]} \mbf Z_{k, t} \wc{\mbf D}^{[1]}_k (\wc{\mbf D}^{[1]}_k)^\top \E\l( \mbf X^{\trunc}_{k, t} \r)^\top + 
\frac{1}{np} \sum_{t \in [n]} \E\l( \mbf X^{\trunc}_{k, t} \r) \wc{\mbf D}^{[1]}_k (\wc{\mbf D}^{[1]}_k)^\top \mbf Z_{k, t}^\top 
\nn \\
& + \frac{1}{np} \sum_{t \in [n]} \E\l( \mbf X^{\trunc}_{k, t} - \mbf X_{k, t} \r) \wc{\mbf D}^{[1]}_k (\wc{\mbf D}^{[1]}_k)^\top \E\l( \mbf X^{\trunc}_{k, t} \r)^\top 
\nn \\
& + \frac{1}{np} \sum_{t \in [n]} \E(\mbf X_{k, t}) \wc{\mbf D}^{[1]}_k (\wc{\mbf D}^{[1]}_k)^\top \E\l( \mbf X^{\trunc}_{k, t} - \mbf X_{k, t} \r)^\top
\nn \\
& + \frac{1}{np} \sum_{t \in [n]} \E(\mbf X_{k, t}) \l( \wc{\mbf D}^{[1]}_k (\wc{\mbf D}^{[1]}_k)^\top - {\cgr \frac{1}{p_{-k}} \bm\Delta_k \bm\Delta_k^\top} \r) \E(\mbf X_{k, t})^\top + \frac{1}{p_k} \bm\Gamma^\kk_\chi
\nn \\
=: & \, T_1 + T_{2, 1} + T_{2, 2} + T_{3, 1} + T_{3, 2} + T_4 + \frac{1}{p_k} \bm\Gamma^\kk_\chi,
\label{eq:wt:gamma:decomp}
\end{align}
where $\bm\Gamma^\kk_\chi$ is defined in~\eqref{eq:gamma:decomp}.
Then by~\eqref{eq:np}, \eqref{eq:lem:T:one:wt:one}, \eqref{eq:lem:T:two:wt:one}, \eqref{eq:lem:T:three:wt:one} and Lemma~\ref{lem:T:four:wt}, we have
\begin{align}
\frac{1}{p_k} \l\Vert \wc{\bm\Gamma}^{\kk, [2]}- {\cgr \bm\Gamma^\kk_\chi } \r\Vert
= O_P\l( \sum_{k' \in [K]} \frac{M_n^{1 - \eps}}{\sqrt{np_{-k'}}} \vee \frac{1}{p} \r).
% \vee \frac{\psi^\kk_{n, p}}{\sqrt p} \r). 
\label{eq:wt:gamma}
\end{align}

Next, denote by $\wc{\mbf M}^{[2]}_k(\tau) \in \R^{r_k \times r_k}$ the diagonal matrix containing the eigenvalues $\wc{\mu}^{\kk, [2]}_j(\tau), \, j \in [r_k]$, of $\wc{\bm\Gamma}^{\kk, [2]}(\tau)$ on its diagonal. 
By~\eqref{eq:wt:gamma} and Lemma~\ref{lem:third:rough} and the arguments analogous to those adopted in proving~\eqref{eq:wh:M:bound:one} and~\eqref{eq:wh:M:bound:two}, we have
$p_k^{-1} \wc{\mbf M}^{[2]}_k$ asymptotically invertible and  
\begin{align}
\label{eq:wt:M:bound}
\l\Vert \l( p_k^{-1} \wc{\mbf M}^{[2]}_k \r)^{-1} \r\Vert \le \frac{1}{\alpha^\kk_{r_k}} + o_P(1) = O_P(1).
\end{align}
Let us set 
{\cgr
\begin{align}
\wc{\mbf H}^{[2]}_k = \frac{1}{n\sqrt{p_k}p_{-k}} \sum_{t \in [n]} \mat_k(\mc F_t) \bm\Delta_k^\top \wc{\mbf D}^{[1]}_k (\wc{\mbf D}^{[1]}_k)^\top \bm\Delta_k \mat_k(\mc F_t)^\top \bm\Lambda_k^\top \wc{\mbf E}^{[2]}_k \l( \frac{1}{p_k} \wc{\mbf M}^{[2]}_k \r)^{-1},
\nn %\label{eq:H:tilde}
\end{align}}
such that by Assumption~\ref{assum:loading}~(i),
\begin{align*}
{\cgr \frac{1}{\sqrt{p_k}} \bm\Lambda_k} \wc{\mbf H}^{[2]}_k = \frac{1}{n p} \sum_{t \in [n]} \E(\mbf X_{k, t}) \wc{\mbf D}^{[1]}_k (\wc{\mbf D}^{[1]}_k)^\top \E(\mbf X_{k, t})^\top \wc{\mbf E}^{[2]}_k \l( \frac{1}{p_k} \wc{\mbf M}^{[2]}_k \r)^{-1}.
\end{align*}
Then from~\eqref{eq:wt:gamma:decomp}, we may write
\begin{align*}
\wc{\mbf E}^{[2]}_k - {\cgr \frac{1}{\sqrt{p_k}} \bm\Lambda_k} \wc{\mbf H}^{[2]}_k = \l( T_1 + T_{2, 1} + T_{2, 2} + T_{3, 1} + T_{3, 2} \r) \wc{\mbf E}^{[2]}_k \l( \frac{1}{p_k} \wc{\mbf M}^{[2]}_k \r)^{-1},
\end{align*}
from which we derive that
\begin{align*}
\l\Vert \wc{\mbf E}^{[2]}_k - {\cgr \frac{1}{\sqrt{p_k}} \bm\Lambda_k} \wc{\mbf H}^{[2]}_k \r\Vert 
&= O_P \l[ \frac{M_n^{1 - \eps}}{\sqrt{np_{-k}}} \vee \frac{1}{p} \vee \frac{\psi^\kk_{n, p}}{\sqrt p} \vee \sum_{k' \in [K] \setminus \{k\}} \frac{M_n^{1 - \eps}}{\sqrt{np_{-k'}}}\l( \frac{M_n^{1 - \eps}}{\sqrt n} + \frac{\psi^\kk}{\sqrt{p_k}} \r) \r]
\\
&= O_P\l( \sum_{k' \in [K]} \frac{M_n^{1 - \eps}}{\sqrt{np_{-k'}}} \vee \frac{1}{p} \r)
\end{align*}
by~\eqref{eq:wt:M:bound}, \eqref{eq:lem:T:one:wt:one}, \eqref{eq:lem:T:two:wt:one}, \eqref{eq:lem:T:three:wt:one} and~\eqref{eq:np}.
% since
% \begin{align*}
% \frac{M_n}{\tau^\kk_{n, p}} \psi^\kk_{n, p}
% = \l( \frac{M_n}{\tau^\kk_{n, p}} \r)^{1 - \eps} \frac{1}{\sqrt{np_{-k}}}
% \end{align*}
Finally, the conclusion follows from that
\begin{align}
\label{eq:wc:h:two}
\Vert \wc{\mbf H}^{[2]}_k \Vert &\le \frac{1}{n\sqrt{p_k}p_{-k}} \sum_{t \in [n]} \vert \mc F_t \vert_2^2 \Vert \bm\Delta_k \Vert^2 \Vert \bm\Lambda_k \Vert \l\Vert \l( \frac{1}{p_k} \wc{\mbf M}^{[2]}_k \r)^{-1} \r\Vert \le (\alpha^\kk_{r_k})^{-1} \omega^2 (1 + o_P(1)) 
\end{align}
by Assumptions~\ref{assum:loading} and~\ref{assum:heavy}~\ref{cond:heavy:factor} and~\eqref{eq:wt:M:bound}, and hence
\begin{align*}
\mbf I_{r_k} = (\wc{\mbf E}^{[2]}_k)^\top \wc{\mbf E}^{[2]}_k = \frac{1}{p_k} (\wc{\mbf H}^{[2]}_k)^\top \bm\Lambda_k^\top \bm\Lambda_k \wc{\mbf H}^{[2]}_k + o_P(1) = (\wc{\mbf H}^{[2]}_k)^\top \wc{\mbf H}^{[2]}_k + o_P(1).
\end{align*}

\subsubsection{Proof of Theorem~\ref{thm:third}~\ref{thm:third:two}}
\label{app:pf:thm:third:two}

For any $k \in [K]$ and $i \in [p_k]$, we have 
\begin{align}
& \sqrt{np_{-k}} \l( \wc{\bm\Lambda}^{[2]}_{k, i\cdot} - \bm\Lambda_{k, i\cdot} \wc{\mbf H}^{[2]}_k \r)
\nn \\
=& \, \frac{1}{\sqrt{np}} \sum_{t \in [n]} \l[
\mbf Z_{k, i\cdot t} \wc{\mbf D}^{[1]}_k (\wc{\mbf D}^{[1]}_k)^\top \mbf Z_{k, t}^\top 
+ \mbf Z_{k, i\cdot, t} \wc{\mbf D}^{[1]}_k (\wc{\mbf D}^{[1]}_k)^\top \E\l( \mbf X^{\trunc}_{k, t} \r)^\top 
\r.
\nn \\
& 
+ \E\l( \mbf X^{\trunc}_{k, i\cdot, t} \r) \wc{\mbf D}^{[1]}_k (\wc{\mbf D}^{[1]}_k)^\top \mbf Z_{k, t}^\top 
+ \E\l( \mbf X^{\trunc}_{k, i\cdot, t} - \mbf X_{k, i\cdot, t} \r) \wc{\mbf D}^{[1]}_k (\wc{\mbf D}^{[1]}_k)^\top \E\l( \mbf X^{\trunc}_{k, t} \r)^\top 
\nn \\
& \l. 
+ \E(\mbf X_{k, i\cdot, t}) \wc{\mbf D}^{[1]}_k (\wc{\mbf D}^{[1]}_k)^\top \E\l( \mbf X^{\trunc}_{k, t} - \mbf X_{k, t} \r)^\top \r]
\l( \wc{\mbf E}^{[2]}_k \l( \frac{1}{p_k} \wc{\mbf M}^{[2]}_k \r)^{-1}
- 
{\cgr \mbf E_{\chi, k} \wc{\mbf J}^{[2]}_k \l( \frac{1}{p_k} \mbf M_{\chi, k} \r)^{-1}} \r)
\nn \\
+ & \frac{1}{\sqrt{np}} \sum_{t \in [n]} \l[
\mbf Z_{k, i\cdot t} \wc{\mbf D}^{[1]}_k (\wc{\mbf D}^{[1]}_k)^\top \mbf Z_{k, t}^\top 
+ \E\l( \mbf X^{\trunc}_{k, i\cdot, t} \r) \wc{\mbf D}^{[1]}_k (\wc{\mbf D}^{[1]}_k)^\top \mbf Z_{k, t}^\top \r. 
\nn \\
& + \E\l( \mbf X^{\trunc}_{k, i\cdot, t} - \mbf X_{k, i\cdot, t} \r) \wc{\mbf D}^{[1]}_k (\wc{\mbf D}^{[1]}_k)^\top \E\l( \mbf X^{\trunc}_{k, t} \r)^\top 
\nn \\ 
& \l.
+ \E(\mbf X_{k, i\cdot, t}) \wc{\mbf D}^{[1]}_k (\wc{\mbf D}^{[1]}_k)^\top \E\l( \mbf X^{\trunc}_{k, t} - \mbf X_{k, t} \r)^\top \r] {\cgr \mbf E_{\chi, k} \wc{\mbf J}^{[2]}_k \l( \frac{1}{p_k} \mbf M_{\chi, k} \r)^{-1}}
\nn \\
+ & 
\frac{1}{\sqrt{np}} \sum_{t \in [n]} \mbf Z_{k, i\cdot, t} \l( \wc{\mbf D}^{[1]}_k (\wc{\mbf D}^{[1]}_k)^\top - {\cgr \frac{1}{p_{-k}} \bm\Delta_k \bm\Delta_k^\top } \r) \E\l( \mbf X^{\trunc}_{k, t} \r)^\top {\cgr \mbf E_{\chi, k} \wc{\mbf J}^{[2]}_k \l( \frac{1}{p_k} \mbf M_{\chi, k} \r)^{-1}}
\nn \\
+ & \frac{1}{\sqrt{np}p_{-k}} \sum_{t \in [n]} \mbf Z_{k, i\cdot, t} {\cgr \bm\Delta_k \bm\Delta_k^\top } \E\l( \mbf X^{\trunc}_{k, t} - \mbf X_{k, t} \r)^\top {\cgr \mbf E_{\chi, k} \wc{\mbf J}^{[2]}_k \l( \frac{1}{p_k} \mbf M_{\chi, k} \r)^{-1}}
\nn \\
+ & \frac{1}{\sqrt{np} p_{-k}} \sum_{t \in [n]} \mbf Z_{k, i\cdot, t} {\cgr \bm\Delta_k \bm\Delta_k^\top } \E\l( \mbf X_{k, t} \r)^\top {\cgr \mbf E_{\chi, k} \wc{\mbf J}^{[2]}_k \l( \frac{1}{p_k} \mbf M_{\chi, k} \r)^{-1}}
\nn \\
=:& \, U_1 + U_2 + U_3 + U_4 + {\cgr \frac{1}{\sqrt{np_{-k}}} \sum_{t \in [n]} \mbf Z_{k, i\cdot, t} \bm\Delta_k \mat_k(\mc F_t)^\top (\bm\Lambda^\circ_k)^\top \wc{\mbf J}^{[2]}_k \l( \frac{1}{p_k} \mbf M_{\chi, k} \r)^{-1},}
\label{eq:wt:lambda:decomp}
\end{align}
{\cgr where $\mbf E_{\chi, k}$ and $\wc{\mbf J}^{[2]}_k$ are defined in Lemma~\ref{lem:third:rough:combined}, and $\bm\Lambda^\circ_k = p_k^{-1/2} \mbf E_{\chi, k}^\top \bm\Lambda_k \in \R^{r_k \times r_k}$.}
% Thanks to Assumption~\ref{assum:loading} and Lemmas~\ref{lem:D}~\ref{lem:D:three} and~\ref{lem:third:rough:combined}~(i), 
% \begin{align}
% \vert \bm\Lambda^\circ_k \vert_\infty = O(1) \text{ \ and \ } \vert \bm\Delta^\bullet_k \vert_\infty = O(1).
% \label{eq:dl:inf}
% \end{align}}
%% to continue
Under the conditions made in~\eqref{eq:np}, by~\eqref{eq:lem:T:one:wt:two}, \eqref{eq:lem:T:two:wt:two}, \eqref{eq:lem:T:three:wt:two} 
and Lemma~\ref{lem:third:rough:combined}~(iii), we have
\begin{align*}
\frac{U_1}{\sqrt{np_{-k}}} = O_P\l[ \l( \frac{M_n^{1 - \eps}}{\sqrt{np_{-k}}} \vee \frac{1}{p}\r) \l( \sum_{k' \in [K]} \frac{M_n^{1 - \eps}}{\sqrt{np_{-k'}}} \vee \frac{1}{p} \r)\r]. % \vee \frac{\psi^\kk_{n, p}}{\sqrt p} \r)\r].
% O_P\l[ 
% \l( \frac{M_n^{1 - \eps}}{\sqrt{np_{-k}}} \vee \frac{1}{p} 
% \vee \frac{\psi^\kk_{n, p}}{\sqrt p} 
% \vee \sum_{k' \in [K] \setminus \{k\}} \frac{M_n^{1 - \eps}}{\sqrt{np_{-k'}}} \l( \frac{M_n^{1 - \eps}}{\sqrt n} \vee \frac{1}{\sqrt p} \vee \frac{\psi^\kk_{n, p}}{\sqrt{p_k}} \r)
% \r)
% \l( \sum_{k' \in [K]} \frac{M_n^{1 - \eps}}{\sqrt{np_{-k'}}} \vee \frac{1}{p} \vee \frac{\psi^\kk_{n, p}}{\sqrt p} \r) \r].
\end{align*}
Similarly,~\eqref{eq:lem:T:one:wt:three}, \eqref{eq:lem:T:two:wt:three}, and~\eqref{eq:lem:T:three:wt:three} give
\begin{align*}
\frac{U_2}{\sqrt{np_{-k}}} = 
O_P\l[ \frac{M_n^{1 - \eps}}{\sqrt{np_{-k}}} \l( \frac{1}{\sqrt{p_k}} \vee \frac{M_n^\eps \sqrt{\log(np_{-k})}}{(\tau^\kk_{n, p})^\eps} \r) 
\vee \frac{1}{p} \r],
\end{align*} 
\eqref{eq:lem:T:two:wt:four} leads to 
\begin{align*}
\frac{U_3}{\sqrt{np_{-k}}}=  O_P\l[ \frac{M_n^{1 - \eps}}{\sqrt n} \l( \sum_{k' \in [K] \setminus \{k\}} \frac{M_n^{1 - \eps}}{\sqrt{np_{-k'}}} \vee \frac{1}{p} \r) \r]
\end{align*}
% the RHS is O(M^{1-eps}/sqrt(np_{-k}} thanks to (12) and since sqrt(np_{-k}) = o(p)
and finally, by~\eqref{eq:lem:T:two:wt:five},
\begin{align*}
\frac{U_4}{\sqrt{np_{-k}}} &= O_P\l[ \frac{M_n^{1 - \eps}}{\sqrt{np_{-k}}} \l( \frac{M_n}{\tau^\kk_{n, p}} \r)^{1 + 2\eps} \r].
\end{align*}
Then under the additional conditions made in Theorem~\ref{thm:third}~\ref{thm:third:two}, namely that $\sqrt{np_{-k}} = o(p)$ and $M_n = M$, we have
\begin{align}
\max( U_1, U_2, U_3, U_4 ) = o_P(1) \text{ \ as \ } \min(n, p_1, \ldots, p_K) \to \infty.
\label{eq:thm:third:two:op}
\end{align}
Next, let us define $\mbf Y_t \in \R^{r_k}$ as
\begin{align*}
\mbf Y_t &= \l( \frac{1}{p_k} \mbf M_{\chi, k} \r)^{-1}
\wc{\mbf J}^{[2]}_k \bm\Lambda^\circ_k \mat_k(\mc F_t) \l( \frac{1}{\sqrt{p_{-k}}} \bm\Delta_k \r)^\top \mbf Z_{k, i\cdot, t}^\top.
\end{align*}
In what follows, we derive the asymptotic distribution of $n^{-1/2} \sum_{t \in [n]} \mbf Y_t$ which, in view of the decomposition in~\eqref{eq:wt:lambda:decomp}, determines the asymptotic distribution of $\wc{\bm\Lambda}^{[2]}_{k, i \cdot}$ after appropriate centering and scaling.

Recalling that $\bm\Psi^\kk_{i, tu} = \text{diag}(\Cov(Z_{k, i\ell, t}, Z_{k, i\ell, u}), \, \ell \in [p_{-k}])$, we have
\begin{align*}
\l\Vert \bm\Psi^\kk_{i, tu} \r\Vert \le \max_{\ell \in [p_{-k}]} \vert \Cov(Z_{k, i\ell, t}, Z_{k, i\ell, u}) \vert 
\lesssim (\vert \mc F_t \vert_2 + \omega) (\vert \mc F_u \vert_2 + \omega) \exp\l( - \frac{c_0\eps\vert t - u \vert}{1 + \eps} \r),
\end{align*}
by Lemmas~\ref{lem:one}~\ref{lem:one:one}, \ref{lem:mixing} and~\ref{lem:hall}.
% such that
% \begin{align*}
% \l\Vert \frac{1}{n} \sum_{t, u} \bm\Psi^\kk_{i, tu}(\tau) \r\Vert \le \omega^{2 + 2\eps} c_\eps.
% \end{align*}
Then, 
\begin{align*}
& \bm\Phi^\kk_i := \Cov\l( \frac{1}{\sqrt{n}} \sum_{t \in [n]} \mbf Y_t \r) 
\\
=& \, \frac{1}{n p} \sum_{t, u \in [n]} \l( \frac{1}{p_k} \mbf M_{\chi, k} \r)^{-1}
\wc{\mbf J}^{[2]}_k \mbf E_{\chi, k}^\top \bm\Lambda_k \mat_k(\mc F_t) \bm\Delta_k^\top \bm\Psi^\kk_{i, tu} \bm\Delta_k \mat_k(\mc F_u)^\top \bm\Lambda_k^\top \mbf E_{\chi, k} \wc{\mbf J}^{[2]}_k \l( \frac{1}{p_k} \mbf M_{\chi, k} \r)^{-1}
\\
=& \, \frac{1}{p_k} (\bm\Gamma^\kk_f)^{-1} \mbf E_{\chi, k} \bm\Lambda_k \l( \frac{1}{np_{-k}} \sum_{t, u \in [n]} \mat_k(\mc F_t) \bm\Delta_k^\top \bm\Psi^\kk_{i, tu} \bm\Delta_k \mat_k(\mc F_u)^\top \r) \bm\Lambda_k^\top \mbf E_{\chi, k} (\bm\Gamma^\kk_f)^{-1} + o(1)
\\
=& \, (\bm\Gamma^\kk_f)^{-1} \l( \frac{1}{np_{-k}} \sum_{t, u \in [n]} \mat_k(\mc F_t) \bm\Delta_k^\top \bm\Psi^\kk_{i, tu} \bm\Delta_k \mat_k(\mc F_u)^\top \r) (\bm\Gamma^\kk_f)^{-1} + o(1)
\end{align*}
by Assumptions~\ref{assum:loading}~(i), \ref{assum:factor} and~\ref{assum:indep}~\ref{cond:indep:factor:mixing}, and Lemmas~\ref{lem:common:cov} and~\ref{lem:third:rough:combined}~(iv).
Also, it can be shown that
\begin{align*}
\l\Vert \bm\Phi^\kk_i \r\Vert &\le \l\Vert \bm\Gamma^\kk_f \r\Vert^{-2} \cdot \frac{1}{n} \sum_{t, u \in [n]} \vert \mc F_t \vert_2 \vert \mc F_u \vert_2 (\vert \mc F_t \vert_2 + \omega) (\vert \mc F_u \vert_2 + \omega) \exp\l( - \frac{c_0\eps\vert t - u \vert}{1 + \eps} \r) + o(1)
\\
&\lesssim M^{2 - 2\eps} \omega^{2 + 2\eps} c_\eps,
\end{align*}
from Assumptions~\ref{assum:factor} and~\ref{assum:heavy}~\ref{cond:factor:bound}.
% For any $j \in [r_k]$, we have (omitting $\tau$) by Lemmas~\ref{lem:one}~\ref{lem:one:one}, \ref{lem:mixing}, \ref{lem:hall}, \ref{lem:D}~\ref{lem:D:three} and Assumptions~\ref{assum:heavy}~\ref{cond:factor:bound} and~\ref{assum:indep}~\ref{cond:indep:factor:mixing},
% \begin{align*}
% & \frac{1}{n} \sum_{t, u \in [n]} \sum_{q, q' \in [r_{-k}]} \sum_{\ell, \ell' \in [p_{-k}]} \mat_k(\mc F_t)_{j q} \mat_k(\mc F_t)_{j q'} \delta_{k, \ell q} \delta_{k, \ell' q'} \Cov(Z_{k, i\ell, t}, Z_{k, i\ell', u})
% \\
% =& \, \frac{1}{n} \sum_{t, u \in [n]} \sum_{q, q' \in [r_{-k}]} \sum_{\ell \in [p_{-k}]} \mat_k(\mc F_t)_{j q} \mat_k(\mc F_t)_{j q'} \delta_{k, \ell q} \delta_{k, \ell q'} \Cov(Z_{k, i\ell, t}, Z_{k, i\ell, u})
% \\
% \lesssim & \, \frac{1}{n} \sum_{t, u \in [n]} \vert \mc F_t \vert_2^2 \vert \mc F_u \vert_2^2 \exp\l( -\frac{c_0 \eps \vert t - u \vert}{1 + \eps} \r) 
% \le M^{2 - 2\eps} \omega^{2 + 2\eps} c_\eps
% \end{align*}
% which, together with Assumption~\ref{assum:factor}, ensures that the $j$-th diagonal elements of $\bm\Phi^\kk_i(\tau)$ are bounded for all $j \in [r_k]$.
We now verify the conditions in Equations (3), (7) and (8) of \cite{merlevede2020functional} %, omitting $\tau$ where there is no confusion.
for $\mbf a^\top \mbf Y_t$ with any $\mbf a \in \mathbb{B}_2(1)$, where $\mathbb{B}_2(1) := \{ \mbf b \in \R^{r_k} : \, \vert \mbf b \vert_2 = 1 \}$.

\paragraph{Equation (3).} 
% Let us write $\bm\Lambda^\circ_k = [\lambda^\circ_{k, \ell q}, \, \ell \in [p_k], \, q \in [r_k]]$ and $\bm\Delta^\bullet_k = [\delta^\bullet_{k, \ell q}, \, \ell \in [p_{-k}], \, q \in [r_{-k}]]$.
By Assumptions~\ref{assum:loading}, \ref{assum:heavy}~\ref{cond:heavy:factor} and~\ref{cond:factor:bound} and~\ref{assum:indep}~\ref{cond:indep:mixing} and Lemmas~\ref{lem:common:cov}, \ref{lem:one}~\ref{lem:one:one} and~\ref{lem:third:rough:combined}~(iv), for any $j \in [r_k]$ and $\mbf a \in \mathbb{B}_2(1)$,
\begin{align}
\frac{1}{n} \sum_{t \in [n]} \Vert \mbf a^\top \mbf Y_t \Vert_2^2 
& \le \frac{1}{n p} \l\Vert \l( \frac{1}{p_k} \mbf M_{\chi, k} \r)^{-1} \r\Vert^2 \l\Vert \mbf E_{\chi, k}^\top \bm\Lambda_k \r\Vert^2 \cdot \sum_{t \in [n]}
\vert \mc F_t \vert_2^2 \l\Vert \bm\Delta_k^\top \mbf Z_{k, i \cdot, t}^\top \r\Vert_2^2
\nn \\
&\lesssim \frac{1}{n p_{-k}} \sum_{t \in [n]} \vert \mc F_t \vert_2^2 \sum_{\ell \in [p_{-k}]} \sum_{q \in [r_{-k}]} \delta_{k, \ell q}^2 \Vert Z_{k, i\ell, t} \Vert_2^2
\nn \\
&\lesssim \frac{1}{n} \sum_{t \in [n]} \vert \mc F_t \vert^2 (\vert \mc F_t \vert_2 + \omega)^2
\lesssim \omega^{2 + 2\eps} M^{2 - 2\eps}.
\label{eq:mp:three:one}
\end{align}
% for all $n \ge 1$.
By using the arguments analogous to those leading to~\eqref{eq:mp:three:one}, we derive that
\begin{align}
\Vert \mbf a^\top \mbf Y_t \Vert_4^4 \le & \, \l\Vert \l( \frac{1}{p_k} \mbf M_{\chi, k} \r)^{-1} \r\Vert^4 \l\Vert \frac{1}{\sqrt{p_k}} \mbf E_{\chi, k}^\top \bm\Lambda_k \r\Vert^4 \vert \mc F_t \vert_2^4 \E\l\{ \frac{1}{p_{-k}^2} \l[ \sum_{q \in [r_{-k}]} \l( \sum_{\ell \in [p_{-k}]} \delta_{\chi, \ell q} Z_{k, i\ell, t} \r)^2 \r]^2 \r\}
\nn \\
\lesssim & \, \frac{\vert \mc F_t \vert_2^4}{p_{-k}^2} \sum_{q, q' \in [r_{-k}]} \sum_{\substack{\ell, \ell', \ell'', \ell''' \in [p_{-k}]
\\
\{\ell, \ell'\} = \{\ell'', \ell'''\} \text{ \ or }
\\
\{\ell, \ell''\} = \{\ell', \ell'''\} \text{ \ or }
\\
\{\ell, \ell'''\} = \{\ell', \ell''\} 
}} \delta_{k, \ell q} \delta_{k, \ell' q} \delta_{k, \ell'' q'} \delta_{k, \ell''' q'} \E( Z_{k, i\ell, t} Z_{k, i\ell', t} Z_{k, i\ell'', t} Z_{k, i\ell''', t})
\nn \\
\lesssim & \, \vert \mc F_t \vert_2^4 (\vert \mc F_t \vert_2 + \omega)^4.
\label{eq:y:fourth}
\end{align}
Also for any $\varepsilon > 0$,
\begin{align*}
\p\l( \vert \mbf a^\top \mbf Y_t \vert > \sqrt{n} \varepsilon \r) \le \frac{\Vert \mbf a^\top \mbf Y_t \Vert_2^2}{n \varepsilon^2} 
\lesssim \frac{ \vert \mc F_t \vert_2^2 ( \vert \mc F_t \vert_2 + \omega)^2}{n \varepsilon^2}.
\end{align*}
Combined, we derive that 
\begin{align*}
\frac{1}{n} \sum_{t \in [n]} \E\l(\vert \mbf a^\top \mbf Y_t \vert^2 \cdot \mathbb{I}_{\{\vert \mbf a^\top \mbf Y_t \vert > \sqrt{n} \varepsilon\}} \r)
&\le \frac{1}{n} \sum_{t \in [n]} \sqrt{ \Vert \mbf a^\top \mbf Y_t \Vert_4^4 \cdot \p( \vert \mbf a^\top \mbf Y_t \vert > \sqrt{n} \varepsilon ) }
\\
& \lesssim \frac{1}{n} \sum_{t \in [n]} \frac{ \vert \mc F_t \vert_2^3 ( \vert \mc F_t \vert_2 + \omega)^3 }{\sqrt{n} \varepsilon} \lesssim \frac{M^{4 - 2\eps} \omega^{2 + 2\eps}}{\sqrt n \varepsilon} \to 0
\end{align*}
as $n \to \infty$, where the first inequality follows from  H\"{o}lder's inequality.
% Using similar arguments, we have for any $\varepsilon > 0$ and $j \in [r_k]$,
% \begin{align*}
% & \frac{1}{n} \sum_{t \in [n]} \E\l( Y_{jt}^2 \cdot \mathbb{I}_{\{\vert Y_{jt} \vert > \sqrt{n} \varepsilon\}} \r)
% \\
% = & \, \frac{1}{n(\mu^\kk_{f, j})^2} \sum_{t \in [n]} \sum_{q, q' \in [r_{-k}]} \mat_k(\mc F_t)_{j q} \mat_k(\mc F_t)_{j q'} \sum_{\ell, \ell' \in [p_{-k}]} \delta_{k, \ell q} \delta_{k, \ell q'}  \E\l(Z_{k, i\ell, t} Z_{k, i\ell', t} \cdot \mathbb{I}_{\{\vert Y_{jt} \vert > \sqrt{n} \varepsilon\}} \r)
% \\\
% \lesssim & \, \frac{1}{np_{-k}} \sum_{t \in [n]} \vert \mc F_t \vert_2^2 \sum_{\ell, \ell' \in [p_{-k}]}
% \E\l( \l\vert Z_{k, i\ell, t} Z_{k, i\ell', t} \r\vert^{1 + \eps} \r)^\frac{1}{1 + \eps} \p\l( \vert Y_{jt} \vert > \sqrt{n} \varepsilon \r)^{\frac{\eps}{1 + \eps}}
% \end{align*}
% where the first inequality follows from H\"{o}lder's inequality.
% Then, by Lemma~\ref{lem:one}
% \begin{align*}
% \sum_{\ell, \ell' \in [p_{-k}]} \l\Vert Z_{k, i\ell, t} Z_{k, i\ell', t} \r\Vert_{1 + \eps}
% = \sum_{\ell \in [p_{-k}]} \l\Vert Z_{k, i\ell, t} \r\Vert_{2 + 2\eps}^{\frac{1}{1 + \eps}}
% + \sum_{\substack{\ell, \ell' \in [p_{-k}] \\  \ell \ne \ell'}} \l\Vert Z_{k, i\ell, t} \r\Vert_{1 + \eps} \l\Vert Z_{k, i\ell', t} \r\Vert_{1 + \eps}
% \\
% p_{-k}
% +
% p_{-k}^2 (\vert \mc F_t \vert_2 + \omega)^2
% \end{align*}

\paragraph{Equations (7) and (8).}
Using the arguments adopted in~\eqref{eq:y:fourth}, for any $j \in [r_k]$ with $\delta = 2$,
\begin{align*}
\frac{1}{n} \sum_{t \in [n]} \Vert \mbf a^\top \mbf Y_t \Vert_{2 + \delta}^2 
&\lesssim \frac{1}{n} \sum_{t \in [n]} \vert \mc F_t \vert_2^2 (\vert \mc F_t \vert_2 + \omega)^2
\lesssim \omega^{2 + 2\eps} M^{2 - 2\eps}
\end{align*}
for all $n \ge 1$.
The second claim in Equation~(7) and that in (8) are met by Assumption~\ref{assum:indep}~\ref{cond:indep:mixing} and Lemma~\ref{lem:mixing}, due to the discussions in Section~2.1.1 of \cite{merlevede2020functional}.

Equipped with the above, we can show that by Corollary~2.2 of \cite{merlevede2020functional}, we have for any $\mbf a \in \mathbb{B}_2(1)$,
\begin{align*}
\frac{1}{\sqrt{n}} \sum_{t \in [n]} \mbf a^\top \mbf Y_t \to \mc N_{r_k}\l( \mbf 0, \mbf a^\top \bm\Phi_i^\kk \mbf a \r)
\end{align*}
as $\min(n, p_1, \ldots, p_K) \to \infty$.
This, combined with Cram\'{e}r-Wold theorem (cf.\ Theorem~29.4 of \citeauthor{billingsley1995}, \citeyear{billingsley1995}), completes the proof in combination with~\eqref{eq:thm:third:two:op}.

\subsubsection{Supporting lemmas}

% \begin{lemma}
% \label{lem:H:wt}
% {\it 
% Let the conditions in Theorem~\ref{thm:third} hold, and recall $\wc{\mbf H}^{[1]}_k$ defined therein and define $\wc{\mbf H}_{-k} = \wc{\mbf H}_K \otimes \cdots \otimes \wc{\mbf H}_{k + 1} \otimes \wc{\mbf H}_{k - 1} \otimes \cdots \otimes \wc{\mbf H}_1$. 
% Then for all $k \in [K]$,
% \begin{align*}
% & \l\Vert \mbf I_{r_k} - (\wc{\mbf H}^{[1]}_k)^\top \wc{\mbf H}^{[1]}_k \r\Vert = O_P\l( \frac{M_n^{1 - \eps}}{\sqrt{np_{-k}}} \vee \frac{1}{p} \r),
% \\
% & \l\Vert \mbf I_{r_{-k}} - \wc{\mbf H}_{-k}^\top \wc{\mbf H}_{-k} \r\Vert = O_P \l( \sum_{k' \in [K] \setminus \{k\}} \frac{M_n^{1 - \eps}}{\sqrt{np_{-k'}}} \vee \frac{1}{p} \r).
% \end{align*}
% }
% \end{lemma}

% \begin{proof}
% The proof proceeds analogously as in the proof of Lemma~\ref{lem:H} with Theorem~\ref{thm:second} in place of Theorem~\ref{thm:first}.
% \end{proof}

\begin{lemma}
\label{lem:DD:wc}
{\it Let the conditions in Theorem~\ref{thm:third} hold. Then for all $k \in [K]$,
\begin{align*}
\l\Vert \wc{\mbf D}^{[1]}_k (\wc{\mbf D}^{[1]}_k)^\top - {\cgr \frac{1}{p_{-k}} \bm\Delta_k \bm\Delta_k^\top} \r\Vert &= O_P \l( \sum_{k' \in [K] \setminus \{k\}} \frac{M_n^{1 - \eps}}{\sqrt{np_{-k'}}} \vee \frac{1}{p} \r).
\end{align*}
}
\end{lemma}

\begin{proof}
The proof proceeds analogously as in the proof of Lemmas~\ref{lem:D}~\ref{lem:D:two:new} and~\ref{lem:DD} with Theorem~\ref{thm:second} (and~\eqref{eq:np}) in place of Theorem~\ref{thm:first}.
\end{proof}

In proving Lemmas~\ref{lem:T:one:wt}--\ref{lem:T:three:wt}, we omit that the results are derived under the conditions made in Theorem~\ref{thm:third} and also the dependence on $\tau$.

%% LEMMA 1
\begin{lemma}
\label{lem:T:one:wt}
{\it For each $k \in [K]$, we have 
\begin{align}
& \frac{1}{np} \l\Vert \sum_{t \in [n]} \mbf Z_{k, t} \wc{\mbf D}^{[1]}_k (\wc{\mbf D}^{[1]}_k)^\top \mbf Z_{k, t}^\top \r\Vert 
\nn \\
& \qquad = O_P\l[ \l( \frac{M_n^{1 - \eps}}{\sqrt n} \vee \frac{1}{\sqrt p} \vee \frac{\psi^\kk_{n, p}}{\sqrt{p_k}} \r) \sum_{k' \in [K] \setminus \{k\}} \frac{M_n^{1 - \eps}}{\sqrt{np_{-k'}}} 
\vee
\frac{\psi^\kk_{n, p}}{\sqrt p} \vee \frac{M_n^{1 - \eps}}{\sqrt{n} p_{-k}}\vee \frac{1}{p}
\r].
\label{eq:lem:T:one:wt:one}
\end{align}
Also, for any $i \in [p_k]$, 
\begin{align}
& \frac{\sqrt{p_k}}{np} \l\Vert \sum_{t \in [n]} \mbf Z_{k, i\cdot, t} \wc{\mbf D}^{[1]}_k (\wc{\mbf D}^{[1]}_k)^\top \mbf Z_{k, t}^\top \r\Vert 
\nn \\
& \qquad = O_P\l[ \l( \frac{M_n^{1 - \eps}}{\sqrt n} \vee \frac{1}{\sqrt p} \vee \frac{\psi^\kk_{n, p}}{\sqrt{p_k}} \r) \sum_{k' \in [K] \setminus \{k\}} \frac{M_n^{1 - \eps}}{\sqrt{np_{-k'}}} 
\vee
\frac{\psi^\kk_{n, p}}{\sqrt p} \vee \frac{M_n^{1 - \eps}}{\sqrt{n} p_{-k}}\vee \frac{1}{p}
\r],
\label{eq:lem:T:one:wt:two}
\\
& {\cgr \frac{\sqrt{p_k}}{np} \l\Vert \sum_{t \in [n]} \mbf Z_{k, i\cdot, t} \wc{\mbf D}^{[1]}_k (\wc{\mbf D}^{[1]}_k)^\top \mbf Z_{k, t}^\top \mbf E_{\chi, k} \wc{\mbf J}^{[2]}_k \l(\frac{1}{p_k} \mbf M_{\chi, k} \r)^{-1} \r\Vert }
\nn \\
& = O_P\l[ \frac{1}{\sqrt{p_k}} \l( \frac{M_n^{1 - \eps}}{\sqrt n} \vee \frac{1}{\sqrt p}  \vee \frac{\psi^\kk_{n, p}}{\sqrt{p_k}} \r) \sum_{k' \in [K] \setminus \{k\}} \frac{M_n^{1 - \eps}}{\sqrt{np_{-k'}}} 
+
\frac{1}{\sqrt{p}} \l( \frac{\psi^\kk_{n, p}}{\sqrt{p_k}} \vee \frac{M_n^{1 - \eps}}{\sqrt{np_{-k}}} \r) \vee \frac{1}{p}
\r].
\label{eq:lem:T:one:wt:three}
\end{align}
}
\end{lemma}

\begin{proof}
To prove~\eqref{eq:lem:T:one:wt:one}, let us write
\begin{align*}
\frac{1}{np} \l\Vert \sum_{t \in [n]} \mbf Z_{k, t} \wc{\mbf D}^{[1]}_k (\wc{\mbf D}^{[1]}_k)^\top \mbf Z_{k, t}^\top \r\Vert 
&\le \frac{1}{np} \l\Vert \sum_{t \in [n]} \mbf Z_{k, t} \l( \wc{\mbf D}^{[1]}_k (\wc{\mbf D}^{[1]}_k)^\top - {\cgr \frac{1}{p_{-k}} \bm\Delta_k \bm\Delta_k^\top} \r) \mbf Z_{k, t}^\top \r\Vert
\\
& \qquad
+ \frac{1}{np} \l\Vert \frac{1}{p_{-k}} \sum_{t \in [n]} \mbf Z_{k, t} \bm\Delta_k \bm\Delta_k^\top \mbf Z_{k, t}^\top \r\Vert =: U_1 + U_2. 
\end{align*}
By Lemma~\ref{lem:abc},
\begin{align*}
U_1^2 \le \underbrace{\frac{1}{n^2p^2} \sum_{i, j \in [p_k]} \sum_{\ell, m \in [p_{-k}]} \l( \sum_{t \in [n]} Z_{k, i\ell, t} Z_{k, jm, t} \r)^2}_{V_1} \l\Vert \wc{\mbf D}^{[1]}_k (\wc{\mbf D}^{[1]}_k)^\top - {\cgr \frac{1}{p_{-k}} \bm\Delta_k \bm\Delta_k^\top} \r\Vert^2,
\end{align*}
where $V_1$ is bounded analogously as the corresponding term in the proof of Lemma~\ref{lem:T:one}~\ref{lem:T:one:one}. 
Hence, with Lemma~\ref{lem:DD:wc}, by Markov's inequality,
\begin{align*}
U_1 = O_P\l[ \l( \frac{\psi^\kk_{n, p}}{\sqrt{p_k}} \vee \frac{M_n^{1 - \eps}}{\sqrt n} \vee \frac{\omega^2}{\sqrt p} \r)
\l( \sum_{k' \in [K] \setminus \{k\}} \frac{M_n^{1 - \eps}}{\sqrt{np_{-k'}}} \vee \frac{1}{p} \r)
\r].
\end{align*}
As for $U_2$, we adopt the analogous arguments as those employed in bounding $U_2$ in the proof of Lemma~\ref{lem:T:one}~\ref{lem:T:one:one} in combination with Lemma~\ref{lem:D}~\ref{lem:D:three}, which yields
\begin{align*}
U_2 = O_P\l( \frac{\psi^\kk_{n, p}}{\sqrt p} \vee \frac{M_n^{1 - \eps}}{\sqrt{n} p_{-k}}\vee \frac{\omega^2}{p} \r),
\end{align*}
and thus
\begin{align*}
U_1 + U_2 = O_P\l[ \l( \frac{M_n^{1 - \eps}}{\sqrt n} \vee \frac{1}{\sqrt p} \vee \frac{\psi^\kk_{n, p}}{\sqrt{p_k}} \r) \sum_{k' \in [K] \setminus \{k\}} \frac{M_n^{1 - \eps}}{\sqrt{np_{-k'}}} 
\vee
\frac{\psi^\kk_{n, p}}{\sqrt p} \vee \frac{M_n^{1 - \eps}}{\sqrt{n} p_{-k}}\vee \frac{1}{p}
\r],
\end{align*}
which completes the proof of the first claim.
The claim in~\eqref{eq:lem:T:one:wt:two} follows following the analogous steps.

For~\eqref{eq:lem:T:one:wt:three}, we proceed similarly with some modifications.
Let us write
\begin{align*}
& {\cgr \frac{\sqrt{p_k}}{np} \l\Vert \sum_{t \in [n]} \mbf Z_{k, i\cdot, t} \wc{\mbf D}^{[1]}_k (\wc{\mbf D}^{[1]}_k)^\top \mbf Z_{k, t}^\top \mbf E_{\chi, k} \r\Vert }
\\
\le & \,\frac{\sqrt{p_k}}{np} \l\Vert \sum_{t \in [n]} \mbf Z_{k, i\cdot, t} \l( \wc{\mbf D}^{[1]}_k (\wc{\mbf D}^{[1]}_k)^\top - {\cgr \frac{1}{p_{-k}} \bm\Delta_k \bm\Delta_k^\top} \r) \mbf Z_{k, t}^\top \bm\Lambda_k \r\Vert
\\
& \qquad
+ \frac{\sqrt{p_k}}{n p p_{-k}} \l\Vert \sum_{t \in [n]} \mbf Z_{k, i\cdot, t} {\cgr \bm\Delta_k \bm\Delta_k^\top} \mbf Z_{k, t}^\top \mbf E_{\chi, k}\r\Vert =: U_3 + U_4. 
\end{align*}
By Lemma~\ref{lem:abc}, with {\cgr $\mbf E_{\chi, k} = [e^\kk_{\chi, ij}, i \in [p_k], \, j \in [r_k]]$,}
\begin{align*}
U_3^2 \le \underbrace{\frac{1}{n^2 p_k p_{-k}^2} \sum_{q \in [r_k]} \sum_{\ell, m \in [p_{-k}]} \l( \sum_{j \in [p_k]} \sum_{t \in [n]} Z_{k, i\ell, t} Z_{k, jm, t} e^\kk_{\chi, jq} \r)^2}_{V_3} \l\Vert \wc{\mbf D}^{[1]}_k (\wc{\mbf D}^{[1]}_k)^\top - {\cgr \frac{1}{p_{-k}} \bm\Delta_k \bm\Delta_k^\top} \r\Vert^2,
\end{align*}
where
\begin{align*}
\E(V_3) \le & \, \frac{2}{n^2 p_k p_{-k}^2} \sum_{q \in [r_k]} \sum_{\ell, m \in [p_{-k}]} \sum_{j, j' \in [p_k]} \sum_{t, u \in [n]} e^\kk_{\chi, jq} e^\kk_{\chi, j'q} \Cov(Z_{k, i\ell, t} Z_{k, jm, t}, Z_{k, i\ell, u} Z_{k, j'm, u})
\\
& + \frac{2}{n^2 p_k p_{-k}^2} \sum_{q \in [r_k]} \sum_{\ell, m \in [p_{-k}]} \l( \sum_{j \in [p_k]} \sum_{t \in [n]} e^\kk_{\chi, jq} \E(Z_{k, i\ell, t} Z_{k, jm, t}) \r)^2
=: V_{3, 1} + V_{3, 2}.
\end{align*}
From Lemmas~\ref{lem:one}~\ref{lem:one:one} and~\ref{lem:one:four}, \ref{lem:mixing}, \ref{lem:hall} and~\ref{lem:third:rough:combined}~(i) and Assumptions~\ref{assum:heavy}~\ref{cond:factor:bound} and~\ref{assum:indep},
\begin{align*}
V_{3, 1} &\lesssim \frac{1}{n^2p^2} \sum_{\ell \in [p_{-k}]} \sum_{t, u \in [n]} \Vert Z_{k, i\ell, t}^2 \Vert_\nu \Vert Z_{k, i\ell, u}^2 \Vert_\nu \exp\l( - \frac{c_0(\nu - 2)\vert t - u \vert}{\nu} \r)
\\
& \quad + \frac{1}{n^2p^2} \sum_{j \in [p_k]} \sum_{\ell, m \in [p_{-k}]} \sum_{t, u \in [n]} \Vert Z_{k, i\ell, t} \Vert_\nu \Vert Z_{k, jm, t} \Vert_\nu \Vert Z_{k, i\ell, u} \Vert_\nu \Vert Z_{k, jm, u} \Vert_\nu \exp\l( - \frac{c_0(\nu - 2)\vert t - u \vert}{\nu} \r)
\\
&\lesssim \frac{(\tau^\kk_{n, p})^{2 - 2\eps}}{n^2 p_k^2 p_{-k}} \sum_{t, u \in [n]} (\vert \mc F_t \vert_2 + \omega)^{1 + \eps} (\vert \mc F_u \vert_2 + \omega)^{1 + \eps} \exp\l( - \frac{c_0 \vert t - u \vert}{3 \log(np_{-k})} \r)
\\
& \quad + \frac{1}{n^2 p_k} \sum_{t, u \in [n]} (\vert \mc F_t \vert_2 + \omega)^2 (\vert \mc F_u \vert_2 + \omega)^2 \exp\l( - \frac{c_0 \eps \vert t - u \vert}{1 + \eps} \r)
\\
&\lesssim \frac{c_\eps (\tau^\kk_{n, p})^{2 - 2\eps}}{p_k^2} \cdot \frac{\log(np_{-k})}{np_{-k}} + \frac{M_n^{2 - 2\eps}c_\eps}{np_k}
\lesssim \l( \frac{\psi^\kk_{n, p}}{p_k} + \frac{M_n^{1 - \eps}}{\sqrt{np_k}} \r)^2
\end{align*}
with $\nu \in \{2 + \log^{-1}(np_{-k}), 2 + 2\eps\}$ for the case of $j = i$ and $j \ne i$, respectively. Also by Lemma~\ref{lem:one}~\ref{lem:one:one} and Assumption~\ref{assum:indep}, we have
\begin{align*}
V_{3, 2} \lesssim \frac{1}{n^2p^2} \sum_{\ell \in [p_{-k}]} \l( \sum_{t \in [n]} \E(Z_{k, i\ell, t}^2) \r)^2
\lesssim \frac{1}{p_k p} \l( \frac{1}{n} \sum_{t \in [n]} (\vert \mc F_t \vert_2^2 + \omega^2) \r)^2
\lesssim \frac{\omega^4}{p_k p}.
\end{align*}
Combining the bounds on $V_{3, 1}$ and $V_{3, 2}$ with Lemma~\ref{lem:DD:wc}, we have by Markov's inequality,
\begin{align*}
U_3 = O_P\l[ \frac{1}{\sqrt{p_k}} \l( \frac{\psi^\kk_{n, p}}{\sqrt{p_k}} \vee \frac{M_n^{1 - \eps}}{\sqrt n} \vee \frac{\omega^2}{\sqrt p} \r) \l( \sum_{k' \in [K] \setminus \{k\}} \frac{M_n^{1 - \eps}}{\sqrt{np_{-k'}}} \vee \frac{1}{p} \r) \r].
\end{align*}
As for $U_4$, writing {\cgr $\bm\Delta_k = [\delta_{k, \ell q}, \, \ell \in [p_{-k}], \, q \in [r_{-k}]]$}, we can upper bound $\E(U_4^2)$ by
\begin{align*}
& \, \frac{2}{n^2 p_k p_{-k}^4} \sum_{q \in [r_k]} \E\l[ \l( \sum_{t \in [n]} \sum_{j \in [p_k]} \sum_{\ell, m \in [p_{-k}]} \sum_{q' \in [r_{-k}]} e^\kk_{\chi, j q} \delta_{k, \ell q'} \delta_{k, m q'} \l( Z_{k, i\ell, t} Z_{k, j m, t} - \E(Z_{k, i\ell, t} Z_{k, j m, t}) \r) \r)^2 \r] 
\\ 
& + \frac{2}{n^2 p_k p_{-k}^4} \l\Vert \sum_{t \in [n]} \E\l( \mbf Z_{k, t} \bm\Delta_k \bm\Delta_k^\top \mbf Z_{k, t}^\top \mbf E_{\chi, k} \r) \r\Vert^2
=: V_{4, 1} + V_{4, 2}.
\end{align*}
Then, by Assumptions~\ref{assum:heavy}~\ref{cond:factor:bound} and~\ref{assum:indep}, Lemmas~\ref{lem:one}~\ref{lem:one:one} and~\ref{lem:one:four}, \ref{lem:mixing}, \ref{lem:hall}, \ref{lem:D}~\ref{lem:D:three} and~\ref{lem:third:rough:combined}~(i), % j \ne j' excluded since if so the value is zero
\begin{align*}
V_{4, 1} \lesssim &\, \frac{1}{n^2 p_k^2 p_{-k}^4} \sum_{j \in [p_k]} \sum_{\ell, \ell', m, m' \in [p_{-k}]} \sum_{t, u \in [n]} \Cov( Z_{k, i\ell, t} Z_{k, j m, t}, Z_{k, i\ell', u} Z_{k, j m', u})
\\
\le &\, \frac{1}{n^2 p_k^2 p_{-k}^4} \sum_{\substack{\ell, \ell', m, m' \in [p_{-k}] \\ (\ell, m) = (\ell', m') \text{ \ or} \\ (\ell, m) = (m', \ell') \text{ \ or} \\ (\ell, \ell') = (m, m')}} \sum_{t, u \in [n]} \Vert Z_{k, i\ell, t} Z_{k, i m, t} \Vert_\nu \Vert Z_{k, i\ell', u} Z_{k, i m', u} \Vert_\nu \exp\l( - \frac{c_0(\nu - 2)\vert t - u \vert}{\nu} \r)
\\
& + \frac{1}{n^2 p_k^2 p_{-k}^4} \sum_{\substack{j \in [p_k] \\  j \ne i}} \sum_{\ell, m \in [p_{-k}]} \sum_{t, u \in [n]} \Vert Z_{k, i\ell, t} \Vert_\nu \Vert Z_{k, j m, t} \Vert_\nu \Vert Z_{k, i\ell, u} \Vert_\nu \Vert Z_{k, j m, u} \Vert_\nu \exp\l( - \frac{c_0(\nu - 2)\vert t - u \vert}{\nu} \r)
\\
\lesssim & \, \frac{(\tau^\kk_{n, p})^{2 - 2\eps}}{n^2 p_k^2 p_{-k}^2} \sum_{t, u \in [n]} (\vert \mc F_t \vert_2 + \omega)^{1 + \eps} (\vert \mc F_u \vert_2 + \omega)^{1 + \eps} \exp\l( - \frac{c_0\vert t - u \vert}{3\log(np_{-k})} \r)
\\
& + \frac{M_n^{2 - 2\eps}}{n^2 p_k p_{-k}^2} \sum_{t, u \in [n]} (\vert \mc F_t \vert_2 + \omega)^{1 + \eps} (\vert \mc F_u \vert_2 + \omega)^{1 + \eps} \exp\l( - \frac{c_0 \eps \vert t - u \vert}{1 + \eps} \r)
\\
\lesssim & \, \l( \frac{\psi^\kk_{n, p}}{\sqrt{p_k p}} + \frac{M_n^{1 - \eps}}{\sqrt{np_{-k}p}} \r)^2
\end{align*}
with $\nu \in \{ 2 + \log^{-1}(np_{-k}), 2 + 2\eps \}$ for when $i = j$ and $i \ne j$, respectively.
Besides, we have
\begin{align*}
V_{4, 2} &\lesssim \frac{1}{p^2} \l[ \frac{1}{np_{-k}}
\sum_{t \in [n]} \sum_{\ell \in [p_{-k}]} \E( Z_{k, i\ell, t}^2) \r]^2
\lesssim \frac{1}{p^2} \l( \frac{1}{n} \sum_{t \in [n]} (\vert \mc F_t \vert_2^2 + \omega^2) \r)^2 \lesssim \frac{\omega^4}{p^2},
\end{align*}
due to Assumptions~\ref{assum:loading}, \ref{assum:heavy}~\ref{cond:heavy:factor},~\ref{assum:indep}~\ref{cond:indep:mixing} and Lemma~\ref{lem:D}~\ref{lem:D:three}.
Collecting the bounds on $V_{4, 1}$ and $V_{4, 2}$, we obtain by Markov's inequality, 
\begin{align*}
U_4 = O_P\l[ \frac{1}{\sqrt{p}} \l( \frac{\psi^\kk_{n, p}}{\sqrt{p_k}} \vee \frac{M_n^{1 - \eps}}{\sqrt{np_{-k}}} \r) \vee \frac{\omega^2}{p} \r].
\end{align*}
Combining the bounds on $U_3$ and $U_4$ and evoking Lemma~\ref{lem:common:cov} and~\eqref{eq:wc:h:two} completes the proof. 
\end{proof}

%% LEMMA 2
\begin{lemma}
\label{lem:T:two:wt}
{\it For each $k \in [K]$, we have
\begin{subequations}
\begin{align}
\frac{1}{np} \l\Vert \sum_{t \in [n]} \mbf Z_{k, t} \wc{\mbf D}^{[1]}_k (\wc{\mbf D}^{[1]}_k)^\top \E(\mbf X^{\trunc}_{k, t})^\top \r\Vert = O_P\l( \frac{M_n^{1 - \eps}}{\sqrt n} \l( \sum_{k' \in [K] \setminus \{k\}} \frac{M_n^{1 - \eps}}{\sqrt{np_{-k'}}} \vee \frac{1}{\sqrt{p_{-k}}} \r) \r),
\\
\frac{1}{np} \l\Vert \sum_{t \in [n]} \E(\mbf X^{\trunc}_{k, t}) \wc{\mbf D}^{[1]}_k (\wc{\mbf D}^{[1]}_k)^\top \mbf Z_{k, t}^\top \r\Vert = O_P\l( \frac{M_n^{1 - \eps}}{\sqrt n} \l( \sum_{k' \in [K] \setminus \{k\}} \frac{M_n^{1 - \eps}}{\sqrt{np_{-k'}}} \vee \frac{1}{\sqrt{p_{-k}}} \r) \r).
\end{align}
\label{eq:lem:T:two:wt:one}
\end{subequations}
Also, for any $i \in [p_k]$, 
\begin{subequations}
\begin{align}
& \frac{\sqrt{p_k}}{np} \l\Vert \sum_{t \in [n]} \mbf Z_{k, i \cdot, t} \wc{\mbf D}^{[1]}_k (\wc{\mbf D}^{[1]}_k)^\top \E(\mbf X^{\trunc}_{k, t})^\top \r\Vert = O_P\l( \frac{M_n^{1 - \eps}}{\sqrt n} \l( \sum_{k' \in [K] \setminus \{k\}} \frac{M_n^{1 - \eps}}{\sqrt{np_{-k'}}} \vee \frac{1}{\sqrt{p_{-k}}} \r) \r),
\\
& \frac{\sqrt{p_k}}{np} \l\Vert \sum_{t \in [n]} \E(\mbf X^{\trunc}_{k, i\cdot, t}) \wc{\mbf D}^{[1]}_k (\wc{\mbf D}^{[1]}_k)^\top \mbf Z_{k, t}^\top \r\Vert = O_P\l( \frac{M_n^{1 - \eps}}{\sqrt n} \l( \sum_{k' \in [K] \setminus \{k\}} \frac{M_n^{1 - \eps}}{\sqrt{np_{-k'}}} \vee \frac{1}{\sqrt{p_{-k}}} \r) \r),
\end{align}
\label{eq:lem:T:two:wt:two}
\end{subequations}
and
\begin{align}
& \frac{\sqrt{p_k}}{np} \l\Vert \sum_{t \in [n]} \E(\mbf X^{\trunc}_{k, i\cdot, t}) \wc{\mbf D}^{[1]}_k (\wc{\mbf D}^{[1]}_k)^\top \mbf Z_{k, t}^\top {\cgr \mbf E_{\chi, k} \wc{\mbf J}^{[2]}_k } \l( \frac{1}{p_k} \mbf M_{\chi, k} \r)^{-1} \r\Vert 
\nn \\
& \qquad = O_P\l( \frac{M_n^{1 - \eps}}{\sqrt{np_k}} \l( \sum_{k' \in [K] \setminus \{k\}} \frac{M_n^{1 - \eps}}{\sqrt{np_{-k'}}} \vee \frac{1}{\sqrt{p_{-k}}} \r) \r),
\label{eq:lem:T:two:wt:three}
\\
& \frac{\sqrt{p_k}}{np} \l\Vert \sum_{t \in [n]} \mbf Z_{k, i \cdot, t} \l( \wc{\mbf D}^{[1]}_k (\wc{\mbf D}^{[1]}_k)^\top - {\cgr \frac{1}{p_{-k}} \bm\Delta_k \bm\Delta_k^\top } \r) \E(\mbf X^{\trunc}_{k, t})^\top {\cgr \mbf E_{\chi, k} \wc{\mbf J}^{[2]}_k } \l( \frac{1}{p_k} \mbf M_{\chi, k} \r)^{-1} \r\Vert  
\nn \\
& \qquad = O_P\l( \frac{M_n^{1 - \eps}}{\sqrt n} \l( \sum_{k' \in [K] \setminus \{k\}} \frac{M_n^{1 - \eps}}{\sqrt{np_{-k'}}} \vee \frac{1}{p} \r) \r),
\label{eq:lem:T:two:wt:four} \\
& \frac{\sqrt{p_k}}{np} \l\Vert \sum_{t \in [n]} \mbf Z_{k, i \cdot, t} {\cgr \bm\Delta_k \bm\Delta_k^\top } \E(\mbf X^{\trunc}_{k, t} - \mbf X_{k, t})^\top {\cgr \mbf E_{\chi, k} \wc{\mbf J}^{[2]}_k } \l( \frac{1}{p_k} \mbf M_{\chi, k} \r)^{-1} \r\Vert 
\nn \\
& \qquad = O_P\l( \frac{M_n^{1 - \eps}}{\sqrt{np_{-k}}} \l( \frac{M_n}{\tau^\kk_{n, p}} \r)^{1 + 2\eps} \r).
\label{eq:lem:T:two:wt:five}
\end{align}
}
\end{lemma}

\begin{proof}
For the proof of the first claim in~\eqref{eq:lem:T:two:wt:one}, let us write
\begin{align*}
\frac{1}{np} \l\Vert \sum_{t \in [n]} \mbf Z_{k, t} \wc{\mbf D}^{[1]}_k (\wc{\mbf D}^{[1]}_k)^\top \E(\mbf X^{\trunc}_{k, t})^\top \r\Vert \le& \, 
\frac{1}{np} \l\Vert \sum_{t \in [n]} \mbf Z_{k, t} \l( \wc{\mbf D}^{[1]}_k (\wc{\mbf D}^{[1]}_k)^\top - {\cgr \frac{1}{p_{-k}} \bm\Delta_k \bm\Delta_k^\top } \r) \E(\mbf X^{\trunc}_{k, t})^\top \r\Vert 
\\
& + \frac{1}{np} \l\Vert \frac{1}{p_{-k}} \sum_{t \in [n]} \mbf Z_{k, t} {\cgr \bm\Delta_k \bm\Delta_k^\top } \E(\mbf X^{\trunc}_{k, t})^\top \r\Vert =: U_1 + U_2.
\end{align*}
By Lemma~\ref{lem:abc}, we have
\begin{align*}
U_1^2 \le \underbrace{\frac{1}{n^2p^2} \sum_{i, j \in [p_k]} \sum_{\ell, m \in [p_{-k}]} \l( \sum_{t \in [n]} Z_{k, i\ell, t} \E(X^\trunc_{k, jm, t}) \r)^2}_{V_1} \l\Vert \wc{\mbf D}^{[1]}_k (\wc{\mbf D}^{[1]}_k)^\top - \frac{1}{p_{-k}} \r\Vert^2.
\end{align*}
As shown in the proof of Lemma~\ref{lem:T:two}~\ref{lem:T:two:one},
\begin{align*}
\E(V_1) \le \frac{c_\eps M_n^{2 - 2\eps} \omega^{2 + 2\eps}}{n}
\end{align*}
which, together with Lemma~\ref{lem:DD:wc}, leads to
\begin{align*}
U_1 = O_P\l( \frac{M_n^{1 - \eps}}{\sqrt n} \l( \sum_{k' \in [K] \setminus \{k\}} \frac{M_n^{1 - \eps}}{\sqrt{np_{-k'}}} \vee \frac{1}{p} \r) \r).
\end{align*}
Also, we follow the steps in bounding $U_2$ appearing in the proof of Lemma~\ref{lem:T:two}~\ref{lem:T:two:one}, to show
\begin{align*}
U_2 = O_P\l( \frac{M_n^{1 - \eps}}{\sqrt{np_{-k}}} \r),
\end{align*}
which completes the proof.
The second claim therein as well as those in \eqref{eq:lem:T:two:wt:two} follow analogously.

For the claim in~\eqref{eq:lem:T:two:wt:three}, let us write
\begin{align*}
& \frac{\sqrt{p_k}}{np} \l\Vert \sum_{t \in [n]} \E(\mbf X^{\trunc}_{k, i\cdot, t}) \wc{\mbf D}^{[1]}_k (\wc{\mbf D}^{[1]}_k)^\top \mbf Z_{k, t}^\top {\cgr \mbf E_{\chi, k}} \r\Vert 
\\
\le& \, 
\frac{\sqrt{p_k}}{np} \l\Vert \sum_{t \in [n]} \E(\mbf X^{\trunc}_{k, i\cdot, t}) \l( \wc{\mbf D}^{[1]}_k (\wc{\mbf D}^{[1]}_k)^\top - {\cgr \frac{1}{p_{-k}} \bm\Delta_k \bm\Delta_k} \r) \mbf Z_{k, t}^\top {\cgr \mbf E_{\chi, k}} \r\Vert 
\\
& + \frac{\sqrt{p_k}}{npp_{-k}} \l\Vert \sum_{t \in [n]} \E(\mbf X^{\trunc}_{k, i\cdot, t}) {\cgr \bm\Delta_k \bm\Delta_k} \mbf Z_{k, t}^\top {\cgr \mbf E_{\chi, k}} \r\Vert =: U_3 + U_4.
\end{align*}
By Lemma~\ref{lem:abc}, we have
\begin{align*}
U_3^2 \le \underbrace{\frac{1}{n^2p_kp_{-k}^2} \sum_{q \in [r_k]} \sum_{\ell, m \in [p_{-k}]} \l( \sum_{j \in [p_k]} \sum_{t \in [n]} e^\kk_{\chi, jq} \E(X^\trunc_{k, i\ell, t}) Z_{k, jm, t} \r)^2}_{V_3} \l\Vert \wc{\mbf D}^{[1]}_k (\wc{\mbf D}^{[1]}_k)^\top - {\cgr \frac{1}{p_{-k}} \bm\Delta_k \bm\Delta_k} \r\Vert^2
\end{align*}
where, by Lemmas~\ref{lem:one}~\ref{lem:one:one}, \ref{lem:mixing}, \ref{lem:hall} and~\ref{lem:third:rough:combined}~(i) and Assumptions~\ref{assum:heavy}~\ref{cond:factor:bound} and~\ref{assum:indep}~\ref{cond:indep:mixing},
\begin{align*}
\E(V_3) &= \frac{1}{n^2p_kp_{-k}^2} \sum_{q \in [r_k]} \sum_{\ell, m \in [p_{-k}]} \sum_{j \in [p_k]} \sum_{t, u \in [n]} (e^\kk_{\chi, jq})^2 \E(X^\trunc_{k, i\ell, t}) \E(X^\trunc_{k, i\ell, u}) \Cov(Z_{k, jm, t}, Z_{k, jm, u})
\\
&\lesssim \frac{1}{n^2p_k} \sum_{t, u \in [n]} (\vert \mc F_t \vert_2 + \omega)^2 (\vert \mc F_u \vert_2 + \omega)^2 \exp\l( - \frac{c_0 \eps \vert t - u \vert}{1 + \eps} \r)
\le \frac{c_\eps M_n^{2 - 2\eps} \omega^{2 + 2\eps}}{np_k}.
\end{align*}
Then from Lemma~\ref{lem:DD:wc}, 
\begin{align*}
U_3 = O_P\l( \frac{M_n^{1 - \eps}}{\sqrt{np_k}} \l( \sum_{k' \in [K] \setminus \{k\}} \frac{M_n^{1 - \eps}}{\sqrt{np_{-k'}}} \vee \frac{1}{p} \r) \r). 
\end{align*}
Similarly, additionally evoking Lemma~\ref{lem:D}~\ref{lem:D:three},
\begin{align*}
\E(U_4^2) &\le \frac{1}{n^2p_kp_{-k}^4} \sum_{q \in [r_k]} \E\l[ \l( \sum_{j \in [p_k]} \sum_{\ell, m \in [p_{-k}]} \sum_{q' \in [r_{-k}]} \sum_{t \in [n]} e^\kk_{\chi, j q} \delta_{k, \ell q'} \delta_{k, m q'} \E(X^\trunc_{k, i\ell, t}) Z_{k, jm, t} \r)^2 \r]
\\
&= \frac{1}{n^2p_kp_{-k}^4} \sum_{q \in [r_k]} \sum_{j \in [p_k]} \sum_{\ell, \ell', m \in [p_{-k}]} \sum_{q', q'' \in [r_{-k}]} \sum_{t, u \in [n]} 
(e^\kk_{\chi, j q})^2 \delta_{k, \ell q'} \delta_{k, m q'} \delta_{k, \ell' q''} \delta_{k, m q''} \\
& \qquad \times \E(X^\trunc_{k, i\ell, t}) \E(X^\trunc_{k, i\ell', u}) \Cov(Z_{k, jm, t}, Z_{k, jm, u})
\\
&\lesssim \frac{1}{n^2 p} \sum_{t, u \in [n]} (\vert \mc F_t \vert_2 + \omega)^2 (\vert \mc F_u \vert_2 + \omega)^2 \exp\l( - \frac{c_0 \eps \vert t - u \vert}{1 + \eps} \r)
\lesssim \frac{c_\eps M_n^{2 - 2\eps} \omega^{2 + 2\eps}}{n p},
\end{align*}
which leads to
\begin{align*}
U_4 = O_P\l( \frac{M_n^{1 - \eps}}{\sqrt{np}} \r).
\end{align*}
Collecting the bounds on $U_3$ and $U_4$ and evoking Lemma~\ref{lem:common:cov} and~\eqref{eq:wc:h:two} completes the proof.

For the claim in~\eqref{eq:lem:T:two:wt:four}, let us write
\begin{align*}
U_5 := \frac{\sqrt{p_k}}{np} \l\Vert \sum_{t \in [n]} \mbf Z_{k, i\cdot, t} \l( \wc{\mbf D}^{[1]}_k (\wc{\mbf D}^{[1]}_k)^\top - {\cgr \frac{1}{p_{-k}} \bm\Delta_k \bm\Delta_k^\top } \r) \E(\mbf X^{\trunc}_{k, t})^\top {\cgr \mbf E_{\chi, k}} \r\Vert.
\end{align*}
By Lemma~\ref{lem:abc}, we have
\begin{align*}
U_5^2 \le \underbrace{\frac{1}{n^2p_kp_{-k}^2} \sum_{q \in [r_k]} \sum_{\ell, m \in [p_{-k}]} \l( \sum_{j \in [p_k]} \sum_{t \in [n]} e^\kk_{\chi, jq} Z_{k, i\ell, t} \E(X^\trunc_{k, jm, t}) \r)^2}_{V_5} \l\Vert \wc{\mbf D}^{[1]}_k (\wc{\mbf D}^{[1]}_k)^\top - {\cgr \frac{1}{p_{-k}} \bm\Delta_k \bm\Delta_k^\top } \r\Vert^2
\end{align*}
where, by Lemmas~\ref{lem:one}~\ref{lem:one:one}, \ref{lem:mixing}, \ref{lem:hall} and~\ref{lem:third:rough:combined}~(i) and Assumptions~\ref{assum:heavy}~\ref{cond:factor:bound} and~\ref{assum:indep}~\ref{cond:indep:mixing},
\begin{align*}
\E(V_5) &= \frac{1}{n^2p_kp_{-k}^2} \sum_{q \in [r_k]} \sum_{\ell, m \in [p_{-k}]} \sum_{j, j' \in [p_k]} \sum_{t, u \in [n]} e^\kk_{\chi, jq} e^\kk_{\chi, j'q} \E(X^\trunc_{k, jm, t}) \E(X^\trunc_{k, j'm, u}) \Cov(Z_{k, i\ell, t}, Z_{k, i\ell, u})
\\
&\lesssim \frac{1}{n^2} \sum_{t, u \in [n]} (\vert \mc F_t \vert_2 + \omega)^2 (\vert \mc F_u \vert_2 + \omega)^2 \exp\l( - \frac{c_0 \eps \vert t - u \vert}{1 + \eps} \r)
\\
&\le \frac{c_\eps M_n^{2 - 2\eps} \omega^{2 + 2\eps}}{n} \lesssim \l( \frac{M_n^{1 - \eps}}{\sqrt n} \r)^2,
\end{align*}
% with $\nu = 2 + 2\eps$, 
which leads to
\begin{align*}
U_5 = O_P\l[ \frac{M_n^{1 - \eps}}{\sqrt n} \l( \sum_{k' \in [K] \setminus \{k\}} \frac{M_n^{1 - \eps}}{\sqrt{np_{-k'}}} \vee \frac{1}{p} \r) \r]
\end{align*}
by Markov's inequality and Lemma~\ref{lem:DD:wc}. 
Combined with Lemma~\ref{lem:common:cov}, the proof is complete.

Finally, for~\eqref{eq:lem:T:two:wt:five}, we write
\begin{align*}
U_6 := \frac{\sqrt{p_k}}{npp_{-k}} \l\Vert \sum_{t \in [n]} \mbf Z_{k, i\cdot, t} {\cgr \bm\Delta_k \bm\Delta_k^\top } \E(\mbf X^{\trunc}_{k, t} - \mbf X_{k, t})^\top {\cgr \mbf E_{\chi, k}} \r\Vert.
\end{align*}
By Lemmas~\ref{lem:one}~\ref{lem:one:two} and~\ref{lem:D}~\ref{lem:D:three} and those arguments adopted in bounding $U_5$,
\begin{align*}
\E(U_6^2) &\le \frac{1}{n^2p_kp_{-k}^4} \sum_{q \in [r_k]} \E\l[ \l( \sum_{j \in [p_k]} \sum_{\ell, m \in [p_{-k}]} \sum_{q' \in [r_{-k}]} \sum_{t \in [n]} e^\kk_{\chi, j q} \delta_{k, \ell q'} \delta_{k, m q'} \E(X^\trunc_{k, jm, t} - X_{k, jm, t}) Z_{k, i\ell, t} \r)^2 \r]
\\
&= \frac{1}{n^2p_kp_{-k}^4} \sum_{q \in [r_k]} \sum_{j, j' \in [p_k]} \sum_{\ell, m, m' \in [p_{-k}]} \sum_{q', q'' \in [r_{-k}]} \sum_{t, u \in [n]} 
e^\kk_{\chi, j q} e^\kk_{\chi, j' q} \delta_{k, \ell q'} \delta_{k, m q'} \delta_{k, \ell q''} \delta_{k, m' q''} \\
& \qquad \times \E(X^\trunc_{k, jm, t} - X_{k, jm, t}) \E(X^\trunc_{k, j'm', u} - X_{k, j'm', u}) \Cov(Z_{k, i\ell, t}, Z_{k, i\ell, u}) 
\\
&\lesssim \frac{1}{n^2 p_{-k} (\tau^\kk_{n, p})^{2 + 4\eps}} \sum_{t, u \in [n]} (\vert \mc F_t \vert_2 + \omega)^{3 + 2\eps} (\vert \mc F_u \vert_2 + \omega)^{3 + 2\eps} \exp\l( - \frac{c_0 \eps \vert t - u \vert}{1 + \eps} \r)
\\
&\lesssim \frac{c_\eps \omega^{2 + 2\eps} M_n^{4 + 2\eps}}{n p_{-k} (\tau^\kk_{n, p})^{2 + 4\eps}} 
\lesssim \l( \frac{M_n^{1 - \eps}}{\sqrt{np_{-k}}} \l( \frac{M_n}{\tau^\kk_{n, p}} \r)^{1 + 2\eps} \r)^2,
\end{align*}
which leads to
\begin{align*}
U_6 = O_P\l( \frac{M_n^{1 - \eps}}{\sqrt{np_{-k}}} \l( \frac{M_n}{\tau^\kk_{n, p}} \r)^{1 + 2\eps} \r).
\end{align*}
Evoking Lemma~\ref{lem:common:cov} and~\eqref{eq:wc:h:two} completes the proof of~\eqref{eq:lem:T:two:wt:five}.
\end{proof}

\begin{lemma}
\label{lem:T:three:wt}
{\it For each $k \in [K]$, we have 
\begin{subequations}
\begin{align}
\frac{1}{np} \l\Vert \sum_{t \in [n]} \E\l( \mbf X^{\trunc}_{k, t} - \mbf X_{k, t} \r) \wc{\mbf D}^{[1]}_k (\wc{\mbf D}^{[1]}_k)^\top \E\l( \mbf X^{\trunc}_{k, t}\r)^\top \r\Vert &= % O_P\l( \frac{M_n}{\tau^\kk_{n, p}} \psi^\kk_{n, p} \r) =
O_P\l( \frac{M_n^\eps \sqrt{\log(np_{-k})}}{(\tau^\kk_{n, p})^\eps} \cdot \frac{M_n^{1 - \eps}}{\sqrt{np_{-k}}} \r),
\\
\frac{1}{np} \l\Vert \sum_{t \in [n]} \E\l( \mbf X_{k, t} \r) \wc{\mbf D}^{[1]}_k (\wc{\mbf D}^{[1]}_k)^\top \E\l( \mbf X^{\trunc}_{k, t} - \mbf X_{k, t} \r) ^\top \r\Vert &= O_P\l( \frac{M_n^\eps \sqrt{\log(np_{-k})}}{(\tau^\kk_{n, p})^\eps} \cdot \frac{M_n^{1 - \eps}}{\sqrt{np_{-k}}} \r).
\end{align}
\label{eq:lem:T:three:wt:one}    
\end{subequations}
Also, for any $i \in [p_k]$, 
\begin{subequations}
\begin{align}
& \frac{\sqrt{p_k}}{np} \l\Vert \sum_{t \in [n]} \E\l( \mbf X^{\trunc}_{k, i \cdot, t} - \mbf X_{k, i \cdot, t} \r) \wc{\mbf D}^{[1]}_k (\wc{\mbf D}^{[1]}_k)^\top \E\l( \mbf X^{\trunc}_{k, t}\r)^\top \r\Vert = O_P\l( \frac{M_n^\eps \sqrt{\log(np_{-k})}}{(\tau^\kk_{n, p})^\eps} \cdot \frac{M_n^{1 - \eps}}{\sqrt{np_{-k}}} \r),
\\
& \frac{\sqrt{p_k}}{np} \l\Vert \sum_{t \in [n]} \E\l( \mbf X_{k, i\cdot, t} \r) \wc{\mbf D}^{[1]}_k (\wc{\mbf D}^{[1]}_k)^\top \E\l( \mbf X^{\trunc}_{k, t} - \mbf X_{k, t} \r)^\top \r\Vert = O_P\l( \frac{M_n^\eps \sqrt{\log(np_{-k})}}{(\tau^\kk_{n, p})^\eps} \cdot \frac{M_n^{1 - \eps}}{\sqrt{np_{-k}}} \r),
\end{align}
\label{eq:lem:T:three:wt:two} 
\end{subequations}
and
\begin{subequations}
\begin{align}
& \frac{\sqrt{p_k}}{np} \l\Vert \sum_{t \in [n]} \E\l( \mbf X^{\trunc}_{k, i \cdot, t} - \mbf X_{k, i \cdot, t} \r) \wc{\mbf D}^{[1]}_k (\wc{\mbf D}^{[1]}_k)^\top \E\l( \mbf X^{\trunc}_{k, t}\r)^\top {\cgr \mbf E_{\chi, k} \wc{\mbf J}^{[2]}_k } \l(\frac{1}{p_k} \mbf M_{\chi, k} \r)^{-1} \r\Vert
\nn
\\
& \qquad =O_P\l( \frac{M_n^\eps\sqrt{\log(np_{-k})}}{(\tau^\kk_{n, p})^\eps} \cdot \frac{M_n^{1 - \eps}}{\sqrt{np_{-k}}} \r),
\\
& \frac{\sqrt{p_k}}{np} \l\Vert \sum_{t \in [n]} \E\l( \mbf X_{k, i\cdot, t} \r) \wc{\mbf D}^{[1]}_k (\wc{\mbf D}^{[1]}_k)^\top \E\l( \mbf X^{\trunc}_{k, t} - \mbf X_{k, t} \r)^\top {\cgr \mbf E_{\chi, k} \wc{\mbf J}^{[2]}_k } \l(\frac{1}{p_k} \mbf M_{\chi, k} \r)^{-1} \r\Vert
\nn
\\
& \qquad =O_P\l( \frac{M_n^\eps\sqrt{\log(np_{-k})}}{(\tau^\kk_{n, p})^\eps} \cdot \frac{M_n^{1 - \eps}}{\sqrt{np_{-k}}} \r).
\end{align}
\label{eq:lem:T:three:wt:three}   
\end{subequations}
}
\end{lemma}

\begin{proof}
For the proof of~\eqref{eq:lem:T:three:wt:one}, let us write
\begin{align*}
& \frac{1}{np} \l\Vert \sum_{t \in [n]} \E(\mbf X^{\trunc}_{k, t} - \mbf X_{k, t}) \wc{\mbf D}^{[1]}_k (\wc{\mbf D}^{[1]}_k)^\top \E(\mbf X^{\trunc}_{k, t})^\top \r\Vert 
\\
\le& \, 
\frac{1}{np} \l\Vert \sum_{t \in [n]} \E(\mbf X^{\trunc}_{k, t} - \mbf X_{k, t}) \l( \wc{\mbf D}^{[1]}_k (\wc{\mbf D}^{[1]}_k)^\top - {\cgr \frac{1}{p_{-k}} \bm\Delta_k \bm\Delta_k^\top } \r) \E(\mbf X^{\trunc}_{k, t})^\top \r\Vert 
\\
& + \frac{1}{np_kp_{-k}^2} \l\Vert \sum_{t \in [n]} \E(\mbf X^{\trunc}_{k, t} - \mbf X_{k, t}) {\cgr \bm\Delta_k \bm\Delta_k^\top } \E(\mbf X^{\trunc}_{k, t})^\top \r\Vert =: U_1 + U_2.
\end{align*}
By Lemma~\ref{lem:abc}, we have
\begin{align*}
U_1^2 \le \frac{1}{n^2p^2} \sum_{i, j \in [p_k]} \sum_{\ell, m \in [p_{-k}]} \l( \sum_{t \in [n]} \E(X^\trunc_{k, i\ell, t} - X_{k, i\ell, t}) \E(X^\trunc_{k, jm, t}) \r)^2 \l\Vert \wc{\mbf D}^{[1]}_k (\wc{\mbf D}^{[1]}_k)^\top - {\cgr \frac{1}{p_{-k}} \bm\Delta_k \bm\Delta_k^\top } \r\Vert^2,
\end{align*}
where the proof of Lemma~\ref{lem:T:three} and Lemma~\ref{lem:DD:wc} show that
\begin{align*} 
U_1 = O_P\l( \frac{M_n}{\tau^\kk_{n, p}} \psi^\kk_{n, p} \l( \sum_{k' \in [K] \setminus \{k\}} \frac{M_n^{1 - \eps}}{\sqrt{np_{-k'}}} \vee \frac{1}{p} \r) \r).   
\end{align*}
Also, proceeding similarly as in the arguments in the proof of Lemma~\ref{lem:T:three} for bounding $U_2$ therein, we obtain
\begin{align*}
U_2 = O\l( \frac{M_n}{\tau^\kk_{n, p}} \psi^\kk_{n, p} \r) = O_P\l( \frac{M_n^\eps\log(np_{-k})}{(\tau^\kk_{n, p})^\eps} \cdot \frac{1}{\sqrt{np_{-k}}} \r),
\end{align*}
which completes the proof of the first claim. 
The second claim in~\eqref{eq:lem:T:three:wt:one} as well as those in~\eqref{eq:lem:T:three:wt:two} are proved following the analogous steps. 

For the the proof of~\eqref{eq:lem:T:three:wt:three}, let us write
\begin{align}
& \frac{\sqrt{p_k}}{np} \l\Vert \sum_{t \in [n]} \E(\mbf X^{\trunc}_{k, i\cdot, t} - \mbf X_{k, i\cdot, t}) \wc{\mbf D}^{[1]}_k (\wc{\mbf D}^{[1]}_k)^\top \E(\mbf X^{\trunc}_{k, t})^\top {\cgr \mbf E_{\chi, k}} \r\Vert 
\nn \\
\le& \, 
\frac{\sqrt{p_k}}{np} \l\Vert \sum_{t \in [n]} \E(\mbf X^{\trunc}_{k, i\cdot, t} - \mbf X_{k, i\cdot, t}) \l( \wc{\mbf D}^{[1]}_k (\wc{\mbf D}^{[1]}_k)^\top - {\cgr \frac{1}{p_{-k}} \bm\Delta_k \bm\Delta_k^\top } \r) \E(\mbf X^{\trunc}_{k, t})^\top {\cgr \mbf E_{\chi, k}} \r\Vert 
\nn \\
& + \frac{\sqrt{p_k}}{npp_{-k}} \l\Vert \sum_{t \in [n]} \E(\mbf X^{\trunc}_{k, i\cdot, t} - \mbf X_{k, i\cdot, t}) {\cgr \bm\Delta_k \bm\Delta_k^\top } \E(\mbf X^{\trunc}_{k, t})^\top {\cgr \mbf E_{\chi, k}} \r\Vert =: U_3 + U_4.
\nn
\end{align}
By Lemma~\ref{lem:abc}, we have
\begin{multline*}
U_3^2 \le \underbrace{\frac{1}{n^2p_kp_{-k}^2} \sum_{q \in [r_k]} \sum_{\ell, m \in [p_{-k}]} \l( \sum_{j \in [p_k]} \sum_{t \in [n]} e^\kk_{\chi, jq} \E(X^\trunc_{k, i\ell, t} - X_{k, i\ell, t}) \E(X^\trunc_{k, jm, t}) \r)^2}_{V_3} 
\\
\times \l\Vert \wc{\mbf D}^{[1]}_k (\wc{\mbf D}^{[1]}_k)^\top - {\cgr \frac{1}{p_{-k}} \bm\Delta_k \bm\Delta_k^\top } \r\Vert^2
\end{multline*}
where, by Lemmas~\ref{lem:one}~\ref{lem:one:one}, \ref{lem:one:two} and~\ref{lem:third:rough:combined}~(i) and Assumption~\ref{assum:heavy}~\ref{cond:heavy:factor} and~\ref{cond:factor:bound},
\begin{align*}
V_3 &= \frac{1}{n^2p_kp_{-k}^2} \sum_{q \in [r_k]} \sum_{\ell, m \in [p_{-k}]} \sum_{j, j' \in [p_k]} \sum_{t, u \in [n]} e^\kk_{\chi, jq} e^\kk_{\chi, j'q} 
\\
& \qquad \times \E(X^\trunc_{k, i\ell, t} - X_{k, i\ell, t}) \E(X^\trunc_{k, i\ell, u} - X_{k, i\ell, u}) \E(X^\trunc_{k, jm, t}) \E(X^\trunc_{k, j'm, u})
\\
&\lesssim 
\l( \frac{1}{n (\tau^\kk_{n, p})^{1 + 2\eps}} \sum_{t \in [n]} (\vert \mc F_t \vert_2 + \omega)^{3 + 2\eps} \r)^2
\lesssim \l( \frac{M_n \omega^{2 + 2\eps}}{ (\tau^\kk_{n, p})^{1 + 2\eps} } \r)^2 = \l( \frac{M_n}{\tau^\kk_{n, p}} \psi^\kk_{n, p} \r)^2
\end{align*}
which, together with Lemma~\ref{lem:DD:wc}, leads to
\begin{align*}
U_3 = O_P\l( \frac{M_n}{\tau^\kk_{n, p}} \psi^\kk_{n, p} \l( \sum_{k' \in [K] \setminus \{k\}} \frac{M_n^{1 - \eps}}{\sqrt{np_{-k'}}} \vee \frac{1}{p} \r) \r).   
\end{align*}
Similarly, additionally evoking Lemma~\ref{lem:D}~\ref{lem:D:three},
\begin{align*}
U_4^2 &\le \frac{1}{n^2p_kp_{-k}^4} \sum_{q \in [r_k]} \l( \sum_{j \in [p_k]} \sum_{\ell, m \in [p_{-k}]} \sum_{q' \in [r_{-k}]} \sum_{t \in [n]} e^\kk_{\chi, jq} \delta_{k, \ell q'} \delta_{k, m q'} \E(X^\trunc_{k, i\ell, t} - X_{k, i\ell, t})\E(X^\trunc_{k, jm, t}) \r)^2
\\
&\lesssim \l( \frac{1}{n(\tau^\kk_{n, p})^{1 + 2\eps}} \sum_{t \in [n]} (\vert \mc F_t \vert_2 + \omega)^{3 + 2\eps} \r)^2
\lesssim \l( \frac{M_n}{\tau^\kk_{n, p}} \psi^\kk_{n, p} \r)^2 \lesssim \l( \frac{M_n^\eps\sqrt{\log(np_{-k}}}{(\tau^\kk_{n, p})^\eps} \cdot \frac{M_n^{1 - \eps}}{\sqrt{np_{-k}}} \r)^2,
\end{align*}
leading to
\begin{align*}
U_4 = O\l( \frac{M_n}{\tau^\kk_{n, p}} \psi^\kk_{n, p} \r).
\end{align*}
Collecting the bounds on $U_3$ and $U_4$ and evoking Lemma~\ref{lem:common:cov} and~\eqref{eq:wc:h:two} completes the proof. 
The remaining claim is proved following the analogous steps. 
\end{proof}

\begin{lemma}
\label{lem:T:four:wt}
{\it For each $k \in [K]$, we have 
\begin{align*}
& \l\Vert \frac{1}{np} \sum_{t \in [n]} \E(\mbf X_{k, t}) \l( \wc{\mbf D}^{[1]}_k (\wc{\mbf D}^{[1]}_k)^\top - {\cgr \frac{1}{p_{-k}} \bm\Delta_k \bm\Delta_k^\top } \r) \E(\mbf X_{k, t})^\top \r\Vert 
= O_P \l( \sum_{k' \in [K] \setminus \{k\}} \frac{M_n^{1 - \eps}}{\sqrt{np_{-k'}}} \vee \frac{1}{p} \r).
\end{align*}
}
\end{lemma}

\begin{proof}
By Lemma~\ref{lem:abc}, we have
\begin{align*}
& \frac{1}{n^2p^2} \l\Vert \sum_{t \in [n]} \E(\mbf X_{k, t}) \l( \wc{\mbf D}^{[1]}_k (\wc{\mbf D}^{[1]}_k)^\top - {\cgr \frac{1}{p_{-k}} \bm\Delta_k \bm\Delta_k^\top } \r) \E(\mbf X_{k, t})^\top \r\Vert^2_F
\\
\le &\, \underbrace{\frac{1}{n^2p^2} \sum_{i, j \in [p_k]} \sum_{\ell, m \in [p_{-k}]} \l( \sum_{t \in [n]} \E(X^\trunc_{k, i\ell, t}) \E(X^\trunc_{k, jm, t}) \r)^2}_{U_1} \l\Vert \wc{\mbf D}^{[1]}_k (\wc{\mbf D}^{[1]}_k)^\top - {\cgr \frac{1}{p_{-k}} \bm\Delta_k \bm\Delta_k^\top } \r\Vert_F^2.
\end{align*}
By Lemma~\ref{lem:one}~\ref{lem:one:one} and Assumption~\ref{assum:heavy}~\ref{cond:heavy:factor}, we have
\begin{align*}
U_1 &\lesssim \l( \frac{1}{n} \sum_{t \in [n]} (\vert \mc F_t \vert_2 + \omega)^2 \r)^2 \lesssim \omega^4
\end{align*}
which, in combination with Lemma~\ref{lem:DD:wc}, leads to the claim.
\end{proof}

{\cgr
\begin{lemma}
\label{lem:third:rough}
{\it
Let the conditions in Theorem~\ref{thm:third} hold.
% Then for each $k \in [K]$, there exists a diagonal matrix $\wc{\mbf J}^{[2]}_k \in \R^{r_k \times r_k}$ with $\pm 1$ on its diagonal entries such that with $\tau = \tau^\kk$ given in~\eqref{eq:tau}, we have
% \begin{align*}
% \l\Vert \wc{\mbf E}^{[2]}_k(\tau) - \mbf E_{\chi, k} \wc{\mbf J}^{[2]}_k \r\Vert 
% = O_P\l( \sum_{k' \in [K]} \frac{M_n^{1 - \eps}}{\sqrt{np_{-k'}}} \vee \frac{1}{p} \vee \frac{\psi^\kk_{n, p}}{\sqrt p} \r).
% \end{align*}
For $\wc{\mbf M}^{[2]}_k(\tau) \in \R^{r_k \times r_k}$ denoting the diagonal matrix containing the eigenvalues $\wc{\mu}^{\kk, [2]}_j(\tau), \, j \in [r_k]$, of $\wc{\bm\Gamma}^{\kk, [2]}(\tau)$ on its diagonal, we have
\begin{align*}
\l\Vert \l( p_k^{-1} \wc{\mbf M}^{[2]}_k(\tau) \r)^{-1} - \l( p_k^{-1} \mbf M_{\chi, k} \r)^{-1} \r\Vert 
= O_P\l( \sum_{k' \in [K]} \frac{M_n^{1 - \eps}}{\sqrt{np_{-k'}}} \vee \frac{1}{p} \r). % \vee \frac{\psi^\kk_{n, p}}{\sqrt p} \r).
\end{align*} 
}
\end{lemma}}

\begin{proof}
The proof proceeds analogously as in Lemma~\ref{lem:second:rough} with~\eqref{eq:wt:gamma} in place of Proposition~\ref{prop:gamma:wc}, and thus is omitted.  
% As noted in the proof of Proposition~\ref{prop:first:init}, $\bm\Gamma^\kk$ and $\bm\Gamma^\kk_\chi$ fulfil the conditions~\ref{lem:dk:cond:one} and~\ref{lem:dk:cond:four} in Lemma~\ref{lem:dk} in place of $\wt{\mbf S}$ and $\mbf S$, respectively. 
% Then~\eqref{eq:wt:gamma} taking the role of~\ref{lem:dk:cond:two}--\ref{lem:dk:cond:four}, the conclusions follow from Lemma~\ref{lem:dk}~\ref{lem:dk:one} and~\ref{lem:dk:two}.
\end{proof}

{\cgr 
\begin{lemma}
\label{lem:third:rough:combined}
{\it
Let the conditions in Theorem~\ref{thm:third} hold.
\begin{enumerate}[label = (\roman*)]
\item $\vert \mbf E_{\chi, k} \vert_\infty \le C p_k^{-1/2}$ with some constant $C \in (0, \infty)$.
\item There exists a diagonal matrix $\wc{\mbf J}^{[2]}_k \in \R^{r_k \times r_k}$ with $\pm 1$ on its diagonal, such that
\begin{align*} 
\l\Vert \wc{\mbf E}^{[2]}_k(\tau) - \mbf E_{\chi, k} \wc{\mbf J}^{[2]}_k \r\Vert 
= O_P\l( \sum_{k' \in [K]} \frac{M_n^{1 - \eps}}{\sqrt{np_{-k'}}} \vee \frac{1}{p} \r). % \vee \frac{\psi^\kk_{n, p}}{\sqrt p} \r).
\end{align*}
\item With $\wc{\mbf J}^{[2]}_k \in \R^{r_k \times r_k}$ from (ii), it holds that 
\begin{align*} 
\l\Vert \wc{\mbf E}^{[2]}_k(\tau) \l( p_k^{-1} \wc{\mbf M}^{[2]}_k(\tau) \r)^{-1} - \mbf E_{\chi, k} \wc{\mbf J}^{[2]}_k \l( p_k^{-1} \mbf M_{\chi, k} \r)^{-1} \r\Vert 
= O_P\l( \sum_{k' \in [K]} \frac{M_n^{1 - \eps}}{\sqrt{np_{-k'}}} \vee \frac{1}{p} \r). % \vee \frac{\psi^\kk_{n, p}}{\sqrt p} \r).
\end{align*}
\item There exists a diagonal matrix $\bar{\mbf J}_k \in \R^{r_k \times r_k}$ with $\pm 1$ on its diagonal, such that we have  $p_k^{-1/2} \mbf E_{\chi, k}^\top \bm\Lambda_k = \bar{\mbf J}_k + o(1)$.
\end{enumerate}
}
\end{lemma}}

\begin{proof}
Firstly, note that 
\begin{align*}
\vert \bm\Gamma^\kk_\chi \vert_\infty \le r_k \bar{\lambda}^2 \omega^2 + o(1)
\end{align*}
under Assumptions~\ref{assum:loading}, \ref{assum:factor} and~\ref{assum:heavy}~\ref{cond:heavy:factor}.
Then, due to~\eqref{eq:wt:gamma}, the conditions~\ref{lem:dk:cond:one}--\ref{lem:dk:cond:four} are met with $\mbf S = \wt{\mbf S} = \bm\Gamma^\kk_\chi$ and $\wh{\mbf S} = \wc{\bm\Gamma}^{(k), [2]}$, which leads to~(i) and~(ii) from Lemma~\ref{lem:dk}~\ref{lem:dk:one} and~\ref{lem:dk:three}. 
As for~(iii), since
\begin{align*}
& \l\Vert \wc{\mbf E}^{[2]}_k(\tau) \l( p_k^{-1} \wc{\mbf M}^{[2]}_k(\tau) \r)^{-1} - \mbf E_{\chi, k} \wc{\mbf J}^{[2]}_k \l( p_k^{-1} \mbf M_{\chi, k} \r)^{-1} \r\Vert
\\
\le &\, \l\Vert \wc{\mbf E}^{[2]}_k(\tau) - \mbf E_{\chi, k} \wc{\mbf J}^{[2]}_k \r\Vert \; \l\Vert \l( p_k^{-1} \wc{\mbf M}^{[2]}_k(\tau) \r)^{-1} \r\Vert + \Vert \mbf E_{\chi, k} \Vert \; \l\Vert \l( p_k^{-1} \wc{\mbf M}^{[2]}_k(\tau) \r)^{-1} - \l( p_k^{-1} \mbf M_{\chi, k} \r)^{-1} \r\Vert,
\end{align*}
the claim follows from Lemma~\ref{lem:third:rough} and~\eqref{eq:wt:M:bound}.
Finally, from repeated application of Assumptions~\ref{assum:loading} and~\ref{assum:factor}, 
\begin{align*}
& \frac{1}{\sqrt p_k} \bm\Lambda_k^\top \mbf E_{\chi, k} \l( \frac{1}{p_k} \mbf M_{\chi, k} \r) = 
\frac{1}{p_k^{3/2}} \bm\Lambda_k^\top \bm\Gamma^\kk_\chi \mbf E_{\chi, k}
\\
=& \, \frac{1}{p_k} \bm\Lambda_k^\top \bm\Lambda_k
\cdot \frac{1}{n} \sum_{t \in [n]} \mat_k(\mc F_t) \mat_k(\mc F_t)^\top \l( \frac{1}{\sqrt p_k} \bm\Lambda_k^\top \mbf E_{\chi, k} \r)
\\
=& \, \bm\Gamma^\kk_f \l( \frac{1}{\sqrt p_k} \bm\Lambda_k^\top \mbf E_{\chi, k} \r) + o(1).
\end{align*} 
Evoking Lemma~\ref{lem:common:cov} , we have $p_k^{-1/2} \bm\Lambda_k^\top \mbf E_{\chi, k} = \bar{\mbf J}_k + o(1)$.
% suppose it is a \bar{\mbf J}_k with some constant $a$, since by construction, $p^{-1/2} \bm\Lambda_k = \mbf E_{\chi, k} \mbf C$, it means $\mbf C = a \bar{\mbf J}_k$, then we evoke Assumption 1
\end{proof}

\subsection{Proof of Theorem~\ref{thm:factor}}

For notational convenience, we omit the dependence on $\tau$ and $\iota$, where $\iota = 2$ under Assumption~\ref{assum:indep} and $\iota = 1$ under Assumption~\ref{assum:rf} throughout the proof.

\subsubsection{Proof of \eqref{eq:thm:factor:one}}

% \begin{align*}
% \wh{\mc F}_t = \frac{1}{p} \wc{\bm\Lambda}_1^\top \mat_1(\mc X^\trunc_t) \l( \wc{\bm\Lambda}_K \otimes \ldots \otimes \wc{\bm\Lambda}_2 \r),
% \\
% \vecop(\mc X^\trunc_t) = \bm\Lambda \vecop\l( \mc F_t \r) + \vecop(\bm\xi_t)
% \end{align*}
Let define $\bm\Lambda := \otimes_{k = K}^1 \bm\Lambda_k$, $\wc{\bm\Lambda} := \otimes_{k = K}^1 \wc{\bm\Lambda}_k$ and $\wc{\mbf H} := \otimes_{k = K}^1 \wc{\mbf H}_k$.
Then, it follows that $\wc{\mbf H}^\top \wc{\mbf H} = \mbf I_r + o_P(1)$ such that $\wc{\mbf H}$ is asymptotically invertible. 
Combining this with Assumption~\ref{assum:loading}, Theorems~\ref{thm:second} (under Assumption~\ref{assum:rf}) and~\ref{thm:third} (under Assumption~\ref{assum:indep}), we have
\begin{align}
\frac{1}{\sqrt{p}} \l\Vert \wc{\bm\Lambda} - \bm\Lambda \wc{\mbf H} \r\Vert = 
\begin{cases}
O_P \l( \sum_{k \in [K]} \frac{M_n^{1 - \eps}}{\sqrt{np_{-k}}} \vee \frac{1}{p} \r) & \text{under Assumption~\ref{assum:indep}},
\\
O_P \l( \sum_{k \in [K]} \frac{M_n^{1 - \eps}}{\sqrt{np_{-k}}} \vee \frac{1}{p} \vee \bar{\psi}_{n, p} \r) & \text{under Assumption~\ref{assum:rf}},
\end{cases}
\label{eq:wt:lambda:bound:one}
\end{align}
by the arguments adopted in Lemma~\ref{lem:D}~\ref{lem:D:one}.
Further, the same error bounded is inherited to $\Vert \wc{\bm\Lambda} \wc{\mbf H}^{-1} - \bm\Lambda \Vert$, as
\begin{align}
\frac{1}{\sqrt{p}} \l\Vert \wc{\bm\Lambda} \wc{\mbf H}^{-1} - \bm\Lambda \r\Vert = 
\begin{cases}
O_P \l( \sum_{k \in [K]} \frac{M_n^{1 - \eps}}{\sqrt{np_{-k}}} \vee \frac{1}{p} \r) & \text{under Assumption~\ref{assum:indep}},
\\
O_P \l( \sum_{k \in [K]} \frac{M_n^{1 - \eps}}{\sqrt{np_{-k}}} \vee \frac{1}{p} \vee \bar{\psi}_{n, p} \r) & \text{under Assumption~\ref{assum:rf}},
\end{cases}
\label{eq:wt:lambda:bound:two}
\end{align}
Next, noting that $\vecop(\wh{\mc F}_t) = p^{-1} \wc{\bm\Lambda}^\top \vecop(\mc X^\trunc_t)$, %p^{-1} (\wc{\bm\Lambda}_K^\top \otimes \ldots \otimes \wc{\bm\Lambda}_1^\top) \vecop(\mc X^\trunc_t)$,
we write
\begin{align}
\label{eq:f:decomp:one}
\vecop(\wh{\mc F}_t) - \wc{\mbf H}^{-1} \vecop(\mc F_t) = \frac{1}{p} \wc{\bm\Lambda}^\top \l(\bm\Lambda - \wc{\bm\Lambda} \wh{\mbf H}^{-1} \r) \vecop\l( \mc F_t \r) + \frac{1}{p} \wc{\bm\Lambda}^\top \vecop(\bm\xi_t) =: U_1 + U_2.
\end{align}
Then by~\eqref{eq:wt:lambda:bound:two},
\begin{align*}
\vert U_1 \vert_2 &\le \frac{1}{p} \Vert \wc{\bm\Lambda} \Vert \l\Vert \bm\Lambda - \wc{\bm\Lambda} \wh{\mbf H}^{-1} \r\Vert \vert \mc F_t \vert_2 
\\
&= \begin{cases}
O_P \l( \vert \mc F_t \vert_2 \cdot \sum_{k \in [K]} \frac{M_n^{1 - \eps}}{\sqrt{np_{-k}}} \vee \frac{1}{p} \r) & \text{under Assumption~\ref{assum:indep}},
\\
O_P \l( \vert \mc F_t \vert_2 \cdot \sum_{k \in [K]} \frac{M_n^{1 - \eps}}{\sqrt{np_{-k}}} \vee \frac{1}{p} \vee \bar{\psi}_{n, p} \r) & \text{under Assumption~\ref{assum:rf}}.
\end{cases}
\end{align*}
Also, writing
\begin{align}
% \label{eq:f:decomp:two}
U_2 = \frac{1}{p} \l( \wc{\bm\Lambda} - \bm\Lambda \wc{\mbf H} \r)^\top \vecop(\bm\xi_t) + \frac{1}{p} \wc{\mbf H}^\top \bm\Lambda^\top \vecop(\bm\xi_t) =: V_{2, 1} + V_{2, 2},
\nn
\end{align}
we have from~\eqref{eq:wt:lambda:bound:one}, 
\begin{align*}
\vert V_{2, 1} \vert_2 &\le \frac{1}{p} \l\Vert \wc{\bm\Lambda} - \bm\Lambda \wc{\mbf H} \r\Vert \vert \vecop(\bm\xi_t) \vert_2 
\\
&= \begin{cases}
O_P \l( \omega \sum_{k \in [K]} \frac{M_n^{1 - \eps}}{\sqrt{np_{-k}}} \vee \frac{1}{p} \r) & \text{under Assumption~\ref{assum:indep}},
\\
O_P \l( \omega \sum_{k \in [K]} \frac{M_n^{1 - \eps}}{\sqrt{np_{-k}}} \vee \frac{1}{p} \vee \bar{\psi}_{n, p} \r) & \text{under Assumption~\ref{assum:rf}},
\end{cases}
\end{align*}
from the observation that under Assumption~\ref{assum:heavy}~\ref{cond:heavy:idio},
\begin{align}
\label{eq:xi:l2}
\E\l( \vert \vecop(\bm\xi_t) \vert_2^2 \r) \le p \omega^2.
\end{align}
With some notational abuse, denote by $\bm\Lambda = [\lambda_{\mbf i \mbf j}, \, \mbf i \in \prod_{k = 1}^K [p_k], \, \mbf j \in \prod_{k = 1}^K [r_k]]$.
Under Assumptions~\ref{assum:indep}, we have
\begin{align*}
\frac{1}{p^2} \E\l( \l\vert \bm\Lambda^\top \vecop(\bm\xi_t) \r\vert_2^2 \r) &= 
\frac{1}{p^2} \sum_{\mbf j \in \prod_{k = 1}^K [r_k]} \E\l( \l\vert \sum_{\mbf i \in \prod_{k = 1}^K [p_k]} \lambda_{\mbf i \mbf j} \xi_{\mbf i, t} \r\vert^2 \r)
\\
&= \frac{1}{p^2} \sum_{\mbf j \in \prod_{k = 1}^K [r_k]} \sum_{\mbf i \in \prod_{k = 1}^K [p_k]} \lambda_{\mbf i \mbf j}^2 \E(\xi_{\mbf i, t}^2)
\lesssim \frac{r \omega^2}{p}
\end{align*}
by Assumption~\ref{assum:loading}~(ii), which leads to $\vert V_{2, 2} \vert_2 = O_P(\omega p^{-1/2})$.
Under Assumptions~\ref{assum:rf},
\begin{align*}
\frac{1}{p^2} \E\l( \l\vert \bm\Lambda^\top \vecop(\bm\xi_t) \r\vert_2^2 \r) &= \frac{1}{p^2} \sum_{\mbf j \in \prod_{k = 1}^K [r_k]} \sum_{\mbf i, \mbf i' \in \prod_{k = 1}^K [p_k]} \lambda_{\mbf i \mbf j} \lambda_{\mbf i' \mbf j} \Cov(\xi_{\mbf i, t}, \xi_{\mbf i', t})
\\
&\lesssim \frac{r}{p^2} \sum_{\mbf i, \mbf i' \in \prod_{k = 1}^K [p_k]} \Vert \xi_{\mbf i, t} \Vert_\nu \Vert \xi_{\mbf i', t} \Vert_\nu \exp\l( - \frac{c_0(\nu - 2) \vert \mbf i - \mbf i' \vert_2}{K\nu} \r)
\\
&\lesssim \frac{r \omega^2}{p} \prod_{k = 1}^K \frac{1}{p_k} \sum_{i_k, i'_k \in [p_K]} \exp\l( -\frac{c_0 \eps \vert i_k - i'_k \vert}{K(1 + \eps)} \r) \lesssim \frac{r\omega^2}{p}
\end{align*}
with $\nu = 2 + 2\eps$, due to Lemma~\ref{lem:hall}, and thus we have $\vert V_{2, 2} \vert_2 = O_P(\omega p^{-1/2})$.

\subsubsection{Proof of \eqref{eq:thm:factor:two}}

From the proof of~\eqref{eq:thm:factor:one},
\begin{align*}
& \frac{1}{n} \sum_{t \in [n]} \l\vert \vecop(\wh{\mc F}_t) - \wc{\mbf H}^{-1} \vecop(\mc F_t) \r\vert_2^2 
\\
\le & \, \frac{2}{n} \sum_{t \in [n]} \l\vert \frac{1}{p} \wc{\bm\Lambda}^\top \l(\bm\Lambda - \wc{\bm\Lambda} \wh{\mbf H}^{-1} \r) \vecop\l( \mc F_t \r) \r\vert_2^2 + \frac{2}{n} \sum_{t \in [n]} \l\vert \frac{1}{p} \wc{\bm\Lambda}^\top \vecop(\bm\xi_t) \r\vert_2^2 =: U_1 + U_2.
\end{align*}
Then by~\eqref{eq:wt:lambda:bound:two} and Assumption~\ref{assum:heavy}~\ref{cond:heavy:factor},
\begin{align*}
\vert U_1 \vert &\lesssim \frac{1}{p^2} \Vert \wc{\bm\Lambda} \Vert^2 \l\Vert \bm\Lambda - \wc{\bm\Lambda} \wh{\mbf H}^{-1} \r\Vert^2  \cdot \frac{1}{n} \sum_{t \in [n]} \vert \mc F_t \vert_2^2 
\\
&= \begin{cases}
O_P \l[ \omega^2 \sum_{k \in [K]} \l( \frac{M_n^{1 - \eps}}{\sqrt{np_{-k}}} \r)^2 \vee \frac{1}{p^2} \r] & \text{under Assumption~\ref{assum:indep}},
\\
O_P \l[ \omega^2 \cdot \sum_{k \in [K]} \l( \frac{M_n^{1 - \eps}}{\sqrt{np_{-k}}} \r)^2 \vee \frac{1}{p^2} \vee \bar{\psi}_{n, p}^2 \r] & \text{under Assumption~\ref{assum:rf}}.
\end{cases}
\end{align*}
As for $U_2$, by the arguments analogous to those adopted in the proof of~\eqref{eq:thm:factor:one},
\begin{align*}
U_2 &\lesssim \frac{1}{n} \sum_{t \in [n]} \vert \vecop(\bm\xi_t) \vert_2^2 \cdot \l\Vert \frac{1}{p} \l( \wc{\bm\Lambda} - \bm\Lambda \wc{\mbf H} \r)^\top \r\Vert + \frac{1}{n} \sum_{t \in [n]} \l\Vert \frac{1}{p} \bm\Lambda^\top \vecop(\bm\xi_t) \r\Vert^2 
\\
&= \begin{cases}
O_P \l[ \omega^2 \sum_{k \in [K]} \l(\frac{M_n^{1 - \eps}}{\sqrt{np_{-k}}}\r)^2 \vee \frac{1}{p} \r] & \text{under Assumption~\ref{assum:indep}},
\\
O_P \l[ \omega^2 \sum_{k \in [K]} \l( \frac{M_n^{1 - \eps}}{\sqrt{np_{-k}}} \r)^2 \vee \frac{1}{p} \vee \bar{\psi}_{n, p}^2 \r) & \text{under Assumption~\ref{assum:rf}},
\end{cases}
\end{align*}
which completes the proof.

\subsubsection{Proof of Theorem~\ref{thm:factor}~\ref{thm:factor:two}}

Due the condition that $M_n = M$ and $\max_{k \in [K]} p_k = o(n)$, we have $\sqrt{p} = o(\min_{k \in [K]} \sqrt{np_{-k}})$.
From Lemma~\ref{lem:third:rough:combined}~(ii), analogously as in Lemma~\ref{lem:D}~\ref{lem:D:one}, we have
\begin{align}
\label{eq:wc:E:dk}
\l\Vert \wc{\mbf E} - \mbf E_\chi \wc{\mbf J} \r\Vert = O_P\l( \sum_{k \in [K]} \frac{1}{\sqrt{np_{-k}}} \vee \frac{1}{p} \r),
\end{align}
where $\mbf E_\chi := \mbf E_{\chi, K} \otimes \ldots \otimes \mbf E_{\chi, 1}$, $\wc{\mbf E} := \wc{\mbf E}_K \otimes \ldots \otimes \wc{\mbf E}_1$, and $\wc{\mbf J} := \wc{\mbf J}_K \otimes \ldots \otimes \wc{\mbf J}_1 \in \R^{r \times r}$, denotes a diagonal matrix with $\pm 1$ on its diagonal.
Also, from Lemma~\ref{lem:third:rough:combined}~(i), it follows that 
\begin{align}
\label{eq:bar:E:inf}
\vert \mbf E_\chi \vert_\infty \lesssim p^{-1/2}.
\end{align}
As in~\eqref{eq:f:decomp:one}, we write
\begin{align*}
\sqrt{p} \l( \vecop(\wh{\mc F}_t) - \wc{\mbf H}^{-1} \vecop(\mc F_t) \r) =& \, \frac{1}{\sqrt{p}} \wc{\bm\Lambda}^\top \l( \bm\Lambda -\wc{\bm\Lambda} \wc{\mbf H}^{-1} \r) \vecop(\mc F_t) + \l( \wc{\mbf E} - \mbf E_\chi \wc{\mbf J} \r)^\top \vecop(\bm\xi_t)
\\
& + \wc{\mbf J} \mbf E_\chi^\top \vecop(\bm\xi_t) =: U_1 + U_2 + \wc{\mbf J} \mbf E_\chi^\top \vecop(\bm\xi_t).
\end{align*}
By the arguments analogous to those adopted in the proof of~\eqref{eq:thm:factor:one}, we have $U_1 = o_P(1)$. Also, from~\eqref{eq:xi:l2} and~\eqref{eq:wc:E:dk},
\begin{align*}
U_2 = \l( \wc{\mbf E} - \mbf E_\chi \wc{\mbf J} \r)^\top \vecop(\bm\xi_t) 
= O_P\l[ \omega \l( \sum_{k \in [K]} \frac{\sqrt{p}}{\sqrt{np_{-k}}} \vee \frac{1}{\sqrt p} \r) \r] = o_P(1).
\end{align*}
The leading term, $\wc{\mbf J} \mbf E_\chi^\top \vecop(\bm\xi_t)$, has the covariance matrix
\begin{align}
\label{eq:upsilon}
\bm\Upsilon_t := \Cov\l( \wc{\mbf J} \mbf E_\chi^\top \vecop(\bm\xi_t) \r) = \mbf E_\chi^\top \Cov\l( \vecop(\bm\xi_t) \r) \mbf E_\chi,
\end{align}
such that under Assumption~\ref{assum:heavy}~\ref{cond:heavy:idio},
\begin{align*}
\Vert \bm\Upsilon_t \Vert \le \omega^2.
\end{align*}
With some abuse of notation, let us write $\mbf E_\chi = [e_{\chi, \mbf i \mbf j}, \, \mbf i \in \prod_{k = 1}^K [p_k], \, \mbf j \in \prod_{k = 1}^K [r_k]]$.
Then,
\begin{align*}
\mbf E_\chi^\top \vecop(\bm\xi_t)
= \sum_{\mbf i \in \prod_{k = 1}^K [p_k]} \l( e_{\chi, \mbf i \mbf j} \xi_{\mbf i, t}, \, \mbf j \in \prod_{k = 1}^K [r_k] \r)^\top
=: \sum_{\mbf i \in \prod_{k = 1}^K [p_k]} \mbf Y_{\mbf i, t}.
\end{align*}
For any $\mbf a = (a_{\mbf j}, \, j \in \prod_{k = 1}^K [r_k]])^\top$ with $\vert \mbf a \vert_2 = 1$, due to Assumption~\ref{assum:indep},
\begin{align*}
\sum_{\mbf i \in \prod_{k = 1}^K [p_k]} \Var(\mbf a^\top \mbf Y_{\mbf i, t}) = 
\Var\l( \mbf a^\top \mbf E_\chi^\top \vecop(\bm\xi_t) \r) = \Var\l( \mbf a^\top \mbf E_\chi^\top \vecop(\bm\xi_t) \r) \ge c_\xi \mbf a^\top \mbf E_\chi^\top \mbf E_\chi \mbf a = c_\xi,
\end{align*}
where $c_\xi \in (0, \infty)$ is a constant satisfying $\min_{\mbf i \in \prod_{k = 1}^K [p_k]} \Var(\xi_{\mbf i, t}) \ge c_\xi$.
Also thanks to~\eqref{eq:bar:E:inf},
\begin{align*}
\sum_{\mbf i \in \prod_{k = 1}^K [p_k]} \Vert \mbf a^\top \mbf Y_{\mbf i, t} \Vert_{2 + 2\eps}^{2 + 2\eps} 
% = \sum_{\mbf i \in \prod_{k = 1}^K [p_k]} \E\l[ \l\vert \l( \sum_{\mbf j \in \prod_{k = 1}^K [r_k]} a_{\mbf j} e_{\chi, \mbf i\mbf j} \r) \xi_{\mbf i, t} \r\vert^{2 + 2\eps} \r]
= \sum_{\mbf i \in \prod_{k = 1}^K [p_k]} \l\vert \l( \sum_{\mbf j \in \prod_{k = 1}^K [r_k]} a_{\mbf j} e_{\chi, \mbf i\mbf j} \r) \r\vert^{2 + 2\eps} \Vert \xi_{\mbf i, t} \Vert_{2 + 2\eps}^{2 + 2\eps}
\lesssim \omega^{2 + 2\eps} p^{-\eps} % since p * (1/sqrt(p))^(2 + 2\eps)
\end{align*}
under Assumption~\ref{assum:heavy}~\ref{cond:heavy:idio}.
Altogether, we have
\begin{align*}
\frac{ \sum_{\mbf i \in \prod_{k = 1}^K [p_k]} \Vert \mbf a^\top \mbf Y_{\mbf i, t} \Vert_{2 + 2\eps}^{2 + 2\eps} }{ \l[ \sum_{\mbf i \in \prod_{k = 1}^K [p_k]} \Var(\mbf a^\top \mbf Y_{\mbf i, t}) \r]^{1 + \eps} } \lesssim \frac{\omega^{2 + 2\eps} p^{-\eps}}{c_\xi^{1 + \eps}} \to 0
\end{align*}
as $\min(p_1, \ldots, p_K) \to \infty$.
Hence, for any $\mbf a \in \R^r$ with $\vert \mbf a \vert_2 = 1$, we have
\begin{align*}
\sum_{\mbf i \in \prod_{k = 1}^K [p_k]} \mbf a^\top \mbf Y_{\mbf i, t} \to \mc N_r(\mbf 0, \mbf a^\top \bm\Upsilon_t \mbf a)  
\end{align*}
by the Lyapunov condition (cf.\ Theorems~27.3 and~29.4 of \citeauthor{billingsley1995}, \citeyear{billingsley1995}) which, in combination with Cram\'{e}r-Wold theorem, leads to
\begin{align*}
\mbf E_\chi^\top \vecop(\bm\xi_t) \to_d \mc N_r(\mbf 0, \bm\Upsilon_t)
\end{align*}
as $\min(p_1, \ldots, p_K) \to \infty$.
This, combined with the bounds on $U_1$ and $U_2$ and that $\wc{\mbf J}$ is a diagonal matrix of signs, completes the proof. 

\subsection{Proof of Proposition~\ref{prop:fn}}

Below, we omit the dependence on the truncation parameter $\tau$ where there is no confusion.
The proof takes analogous steps as the proof of Theorem~3.9 of \cite{barigozzi2022statistical}.
Throughout, we denote the $j$-th largest eigenvalue of a square matrix $\mbf A$ by $\lambda_j(\mbf A)$.

Note that for $\wh r^{(m - 1)}_k > r_k$, it holds that for
\begin{align*}
\wh{\mbf D}^{(m)}_k &= \wh{\mbf E}^{(m)}_K \otimes \cdots \otimes \wh{\mbf E}^{(m)}_{k + 1} \otimes \wh{\mbf E}^{(m)}_{k - 1} \otimes \cdots \otimes \wh{\mbf E}^{(m)}_1,
\end{align*}
we have $\wh{\mbf D}^{(m)}_k ( \wh{\mbf D}^{(m)}_k )^\top - \wh{\mbf D}_k ( \wh{\mbf D}_k )^\top$ non-negative definite since the difference is composed of the outer products of non-negative definite matrices, which in turn follows from that $\wh{\mbf E}_k^{(m)} = [\wh{\mbf E}_k, \wt{\mbf E}_k^{(m)}]$ with $\wh{\mbf E}_k^\top \wt{\mbf E}_k^{(m)} = \mbf O$.
% since $([A, a] \otimes [B, b]) ([A, a] \otimes [B, b])^\top = ([A, a][A, a]') \otimes ([B, b][B, b]') = (AA' + aa') \otimes (BB' + bb') = (A \otimes B) (A \otimes B)' + (a \otimes b) (a \otimes b)'$
Then, noting that
\begin{align*}
\wc{\bm\Gamma}^{\kk, (m)} = \wc{\bm\Gamma}^\kk +
\frac{1}{np_{-k}} \sum_{t = 1}^n \mbf X^\trunc_{k, t} \l( \wh{\mbf D}^{(m)}_k ( \wh{\mbf D}^{(m)}_k )^\top - \wh{\mbf D}_k ( \wh{\mbf D}_k )^\top \r) (\mbf X^\trunc_{k, t})^\top,
\end{align*}
we have for all $j \in [r_k]$,
\begin{align}
\frac{1}{p_k} \lambda_j\l( \wc{\bm\Gamma}^{\kk, (m)} \r) \ge \frac{1}{p_k} \lambda_j\l( \wc{\bm\Gamma}^\kk \r) \ge \alpha^\kk_j + O_P(\rho_{n, p}),
\label{eq:prop:fn:one}
\end{align}
where the first inequality follows from Weyl's inequality, and the second from Assumption~\ref{assum:factor}, Proposition~\ref{prop:gamma:wc} and the condition on $\rho_{n, p}$.
Also, since
\begin{align*}
\wh{\bm\Gamma}^\kk - \wc{\bm\Gamma}^{\kk, (m)} 
= \frac{1}{np_{-k}} \sum_{t = 1}^n \mbf X^\trunc_{k, t} \l(\mbf I -  \wh{\mbf D}^{(m)}_k ( \wh{\mbf D}^{(m)}_k )^\top \r) (\mbf X^\trunc_{k, t})^\top,
\end{align*}
is also non-negative definite, it follows that
\begin{align}
\frac{1}{p_k} \lambda_j\l( \wc{\bm\Gamma}^{\kk, (m)} \r) \le \frac{1}{p_k} \lambda_j\l( \wh{\bm\Gamma}^\kk \r) \le 
\l\{ \begin{array}{ll}
\beta^\kk_j + O_P(\rho_{n, p}) & \text{for \ } 1 \le j \le r_k, \\
O_P(\rho_{n, p}) & \text{for \ } j \ge r_k + 1, 
\end{array}\r.
\label{eq:prop:fn:two}
\end{align}
by Weyl's inequality, \eqref{eq:prop:one:one} and Assumption~\ref{assum:factor}.
From~\eqref{eq:prop:fn:one} and~\eqref{eq:prop:fn:two}, we have
\begin{align*}
\frac{\wc{\mu}^{\kk, (m)}_j}{\wc{\mu}^{\kk, (m)}_{j + 1} + \rho_{n, p}} \le \l\{ \begin{array}{ll}
\frac{\beta^\kk_j + O_P(\rho_{n, p})}{\alpha^\kk_j + O_P(\rho_{n, p})} = O_P(1) & \text{for \ } 1 \le j \le r_k - 1, \\
\frac{O_P(\rho_{n, p})}{\rho_{n, p}} = O_P(1) & \text{for \ } j \ge r_k + 1,
\end{array}\r.
\end{align*}
and
\begin{align*}
\frac{\wc{\mu}^{\kk, (m)}_{r_k}}{\wc{\mu}^{\kk, (m)}_{r_k + 1} + \rho_{n, p}} \ge 
\frac{\alpha^\kk_{r_k} + O_P(\rho_{n, p})}{\rho_{n, p}} \asymp \rho_{n, p}^{-1} \to \infty
\end{align*}
as $\min(n, p_1, \ldots, p_k) \to \infty$, which completes the proof. 

\clearpage

\section{Complete simulation results}
\label{app:sim}

\subsection{Tensor time series}
\label{app:sim:tensor}

\subsubsection{Set-up}
\label{app:sim:tensor:setting}

\paragraph{Data generation.}
The following data generating process is taken from \cite{barigozzi2023robust} with modifications to augment the difficulty of the estimation problem.
We draw entries of $\bm\Lambda_k$ independently from $\text{Unif}[-1, 1]$.
With $\phi = \psi = 0.3$, we introduce serial dependence to $\mc F^\circ_t$ and $\bm\xi^\circ_t$ as
\begin{align}
\vecop(\mc F^\circ_t) &= \phi \cdot \vecop(\mc F^\circ_{t - 1}) + \sqrt{1 - \phi^2} \cdot \mbf e_t,
\label{eq:ar:f}
\\
\vecop(\bm\xi^\circ_t) &= \psi \cdot \vecop(\bm\xi^\circ_{t - 1}) + \sqrt{1 - \psi^2}  \cdot \vecop(\mc V_t).
\nn
\end{align}
% (cf.\ \cite{barigozzi2023robust} consider $\phi = \psi = 0.1$.
Here, we generate $\mc V_t \in \R^{p_1 \times p_2 \times p_3}$ such that $\vecop(\mc V_t) = \otimes_{k = K}^1 \bm\Sigma_k^{1/2} \mbf v_t$,
% $\vecop(\mc V_t) = \bm\Sigma_3^{1/2} \otimes \bm\Sigma_2^{1/2} \otimes \bm\Sigma_1^{1/2} \mbf v_t$, 
where each $\bm\Sigma_k$ has $p_k^{-1}$ in the off-diagonals and ones on the diagonal.
Also, $\mbf e_t \in \R^r$ and $\mbf v_t \in \R^p$ have i.i.d.\ zero-mean random elements drawn from either the standard normal distribution and the scaled~$t_3$ distribution such that $\Var(e_{jt}) = \Var(v_{it}) = 1$.
Note that the cross-sectional correlations induced by the above $\bm\Sigma_k$ do not conform to the strong mixing condition made in Assumption~\ref{assum:rf}.
Nonetheless, the results show that the proposed tail-robust methods are able to handle mild cross-sectional correlations, see also Appendix~\ref{app:sim:add} for the results obtained under the scenarios where $\bm\Sigma_k$ are Toeptliz matrices.

We consider
\begin{enumerate}[wide, itemsep = 0pt, label = (T\arabic*)]
\item \label{f:one} $p_1 = p_2 = p_3 = 10$,
\item \label{f:two} $p_1 = 100$ and $p_2 = p_3 = 10$,
\item \label{f:three} $p_1 = 20$, $p_2 = 30$ and $p_3 = 40$,
\end{enumerate}
while varying $n \in \{100, 200, 500\}$.
Once $\mc X^\circ_t = \bm\chi^\circ_t + \bm\xi^\circ_t$ is generated, we additionally consider the situations where the observed data $\mc X_t$ are contaminated by outliers:
\begin{enumerate}[wide, itemsep = 0pt, label = (O\arabic*)]
\item \label{o:one} Outliers are introduced to the idiosyncratic component.
Specifically, we randomly select $\mc O \subset \prod_{k = 1}^K [p_k] \times [n]$ with its cardinality $\vert \mc O \vert = [\varrho n p]$.
Then for $(\mbf i, t) \in \mc O$, we set $X_{\mbf i, t} = s_{\mbf i, t} \cdot U_{\mbf i, t}$ with $s_{\mbf i, t} \sim_{\iid} \text{Unif}\{-1, 1\}$ and $U_{\mbf i, t} \sim_{\iid} \text{Unif}[Q + 12, Q + 15]$ with $Q$ set to be the $\max(1 - 100/(np), 0.999)$-quantile of $\vert X^\circ_{\mbf i, t} \vert$; otherwise $X_{\mbf i, t} = X^\circ_{\mbf i, t}$ if $(\mbf i, t) \notin \mc O$.

\item \label{o:two} Outliers are introduced to the factors.
Specifically, we randomly select $\mc O \subset \prod_{k = 1}^K [r_k] \times [n]$ with its cardinality $\vert \mc O \vert = [\varrho n r]$.
Then for $(\mbf j, t) \in \mc O$, we set $f_{\mbf j, t} = s_{\mbf j, t} \cdot U_{\mbf j, t}$ with $s_{\mbf j, t} \sim_{\iid} \text{Unif}\{-1, 1\}$ and $U_{\mbf j, t} \sim_{\iid} \text{Unif}[Q + 12, Q + 15]$ with $Q$ set to be the $\max(1 - 100/(nr), 0.999)$-quantile of $\vert f^\circ_{\mbf j, t} \vert$, while $f_{\mbf j, t} = f^\circ_{\mbf j, t}$ otherwise.
\end{enumerate}

Either under~\ref{o:one} or~\ref{o:two}, when $\varrho = 0$, there are no outliers and we have $X_{\mbf i, t} = X^\circ_{\mbf i, t}$ for all $\mbf i$ and $t$.

\paragraph{Performance assessment.}
To assess the performance of any estimator $\wh{\bm\Lambda}_k \in \R^{p_k \times r_k}$ in loading space estimation, we compute
\begin{align}
\label{eq:err:loading:tensor}
\text{Err}_{\Lambda_k} = \sqrt{1 - \text{tr}\l( \Pi_{\wh{\bm\Lambda}_k} \Pi_{\bm\Lambda_k} \r) / r_k}, \text{ \ where \ } \Pi_{\mbf A} = \mbf A (\mbf A^\top \mbf A)^{-1} \mbf A^\top.
\end{align}
To assess the quality in common component estimation, for any estimator $\wh{\bm\chi}_t$, we evaluate 
\begin{align}
\label{eq:err:chi:tensor}
\text{Err}_\chi(\mc T) = \frac{\sum_{t \in \mc T} \vert \wh{\bm\chi}_t - \bm\chi_t \vert_2^2}{\sum_{t \in \mc T} \vert \bm\chi_t \vert_2^2} 
\end{align}
with $\mc T = [n]$ (`all') and $\mc T = \{n - 10 + 1, \ldots, n\}$ (`local'). 

\clearpage

\newgeometry{top = 1cm, bottom = 3cm}

\subsubsection{Estimation of loadings and common component}

\paragraph{No outlier.}

\begin{figure}[h!t!p!]
\centering
\includegraphics[width = 1\textwidth]{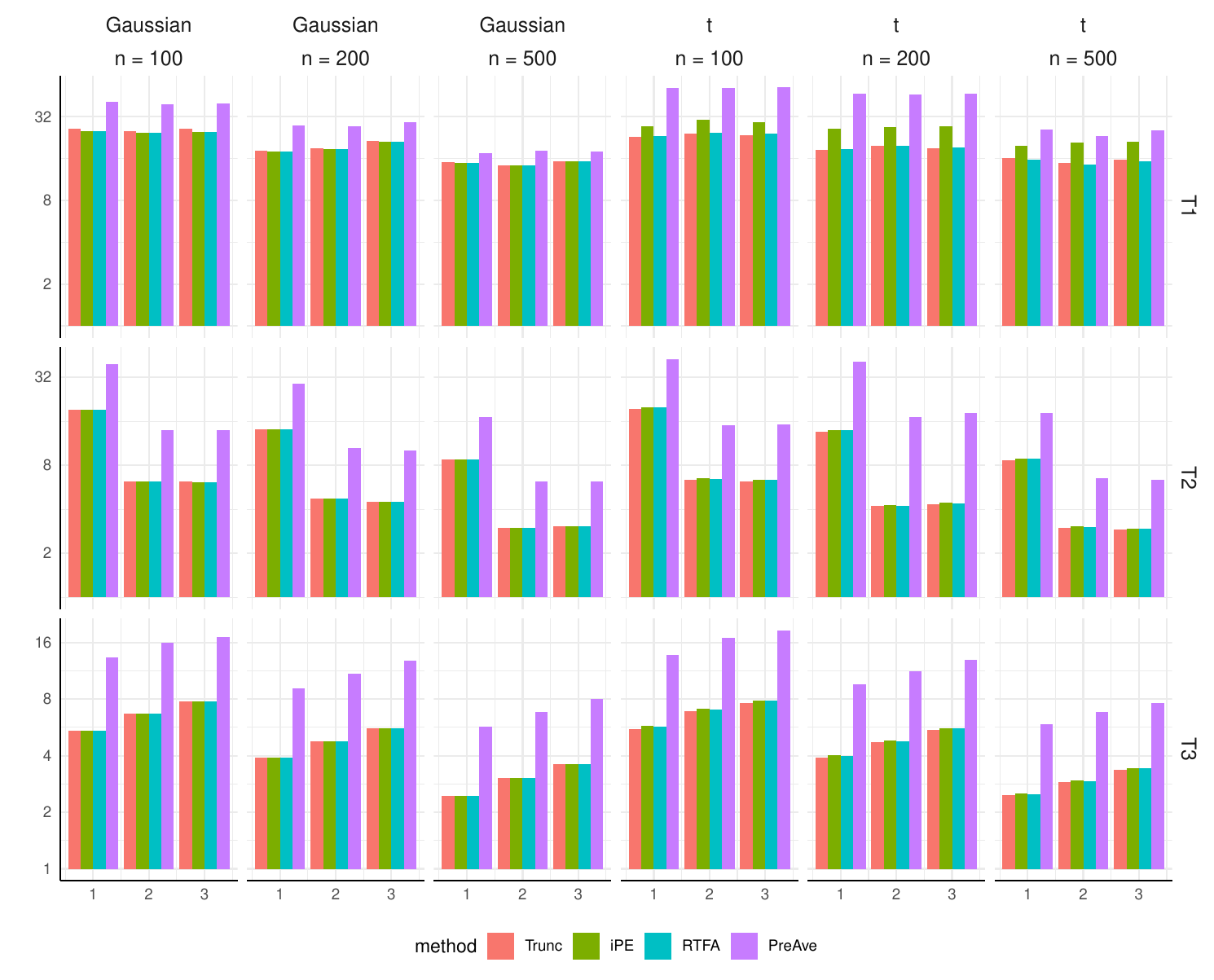}
\caption{Loading estimation errors measured as in~\eqref{eq:err:loading:tensor} for each mode ($x$-axis) for Trunc, iPE, RTFA and PreAve over varying $n \in \{100, 200, 500\}$ and the distributions for $\mc F_t$ and $\bm\xi_t$ (Gaussian and $t_3$) in the absence of any outlier, averaged over $100$ realisations per setting, for \ref{f:one}--\ref{f:three} (top to bottom).
In each plot, the $y$-axis is in the log-scale and all errors have been scaled for the ease of presentation.}
\label{fig:tensor:le:no}
\end{figure}

\begin{figure}[h!t!p!]
\centering
\includegraphics[width = 1\textwidth]{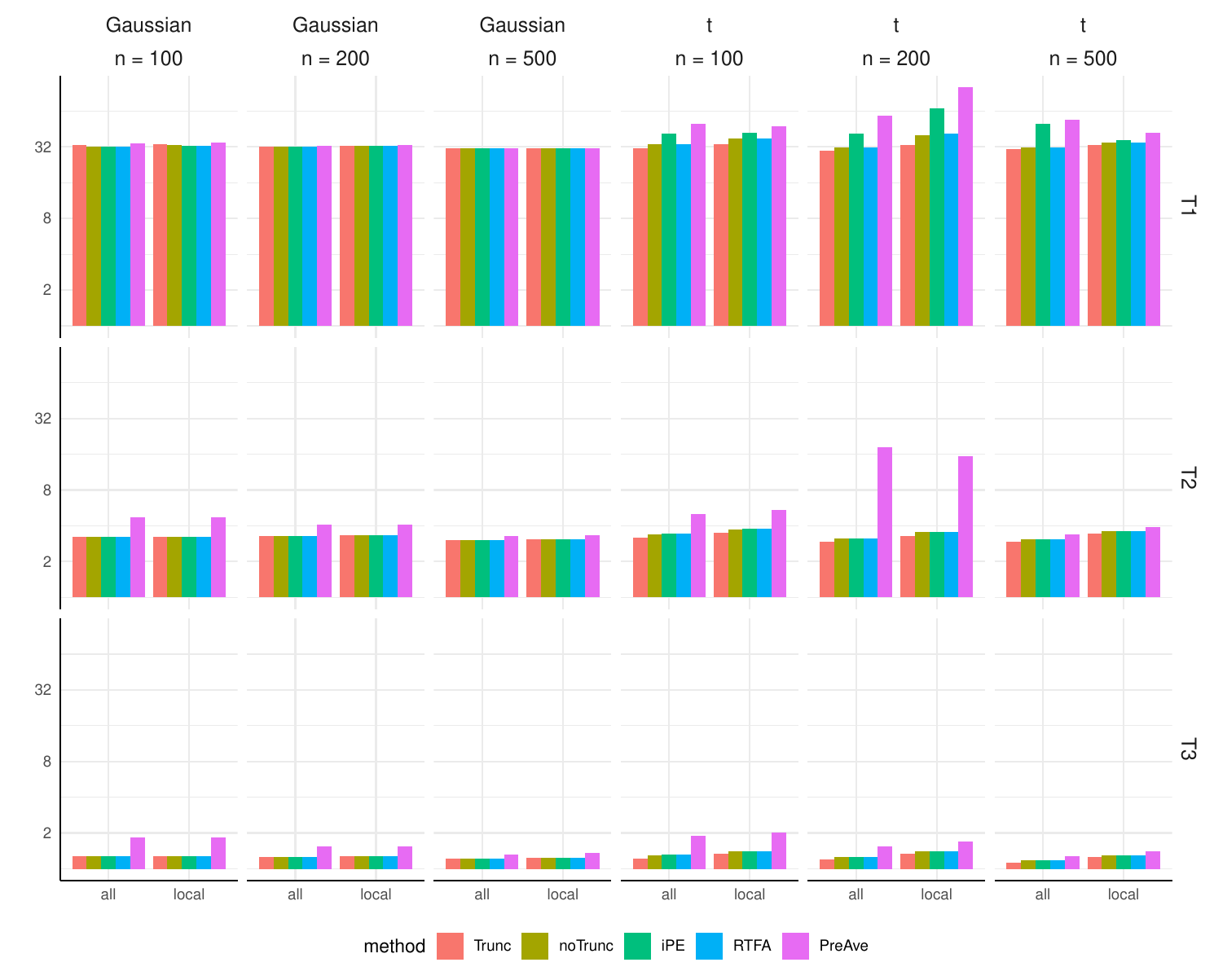}
\caption{Common component estimation errors measured as in~\eqref{eq:err:chi:tensor} with $\mc T = [n]$ (`all') and $\mc T = \{n - 10 + 1, \ldots, n \}$ (`local') for Trunc, noTrunc, iPE, RTFA and PreAve over varying $n \in \{100, 200, 500\}$ and the distributions for $\mc F_t$ and $\bm\xi_t$ (Gaussian and $t_3$) in the absence of any outlier, averaged over $100$ realisations per setting, for \ref{f:one}--\ref{f:three} (top to bottom). In each plot, the $y$-axis is in the log-scale and all errors have been scaled for the ease of presentation.}
\label{fig:tensor:ce:no}
\end{figure}

See Figures~\ref{fig:tensor:le:no}--\ref{fig:tensor:ce:no} and Tables~\ref{tab:tensor:le:no}--\ref{tab:tensor:ce:no} for the results from the loading and the common component estimation obtained under \ref{f:one}--\ref{f:three}, in the absence of any outlier.

\begin{table}[h!t!b!p!]
\caption{Loading estimation errors of Trunc, noTrunc, iPE, RTFA and PreAve measured as in~\eqref{eq:err:loading:tensor} for each mode scaled by $100$, over varying $n \in \{100, 200, 500\}$ and the distributions for $\mc F_t$ and $\bm\xi_t$ (Gaussian and $t_3$) in the absence of any outlier.
We report the mean and the standard deviation over $100$ realisations for each setting.}
\label{tab:tensor:le:no}
\centering
% \resizebox{\textwidth}{!}
{\scriptsize
\begin{tabular}{rrrr cc cc cc cc}
\toprule
&	&	&	&	\multicolumn{2}{c}{Trunc} &		\multicolumn{2}{c}{iPE} &		\multicolumn{2}{c}{RTFA} &		\multicolumn{2}{c}{PreAve} 		\\	
Model &	$n$ &	Dist &	Mode &	Mean &	SD &	Mean &	SD &	Mean &	SD &	Mean &	SD	\\	\cmidrule(lr){1-4} \cmidrule(lr){5-6} \cmidrule(lr){7-8} \cmidrule(lr){9-10} \cmidrule(lr){11-12}
\ref{f:one} &	$100$ &	Gaussian &	1 &	2.628 &	1.723 &	2.536 &	1.209 &	2.537 &	1.211 &	4.122 &	1.678	\\	
&	&	&	2 &	2.528 &	1.305 &	2.469 &	1.216 &	2.469 &	1.214 &	3.953 &	1.168	\\	
&	&	&	3 &	2.616 &	1.908 &	2.488 &	1.15 &	2.487 &	1.145 &	3.976 &	1.387	\\	\cmidrule(lr){4-4} \cmidrule(lr){5-6} \cmidrule(lr){7-8} \cmidrule(lr){9-10} \cmidrule(lr){11-12}
&	&	$t_3$ &	1 &	2.288 &	1.018 &	2.733 &	3.586 &	2.319 &	1.036 &	5.139 &	5.921	\\	
&	&	&	2 &	2.425 &	1.045 &	3.065 &	5.603 &	2.438 &	1.042 &	5.189 &	6.777	\\	
&	&	&	3 &	2.368 &	0.811 &	2.924 &	4.276 &	2.421 &	0.864 &	5.199 &	5.565	\\	\cmidrule(lr){3-4} \cmidrule(lr){5-6} \cmidrule(lr){7-8} \cmidrule(lr){9-10} \cmidrule(lr){11-12}
&	$200$ &	Gaussian &	1 &	1.821 &	0.852 &	1.81 &	0.839 &	1.81 &	0.841 &	2.79 &	0.913	\\	
&	&	&	2 &	1.892 &	0.789 &	1.879 &	0.778 &	1.879 &	0.779 &	2.741 &	0.839	\\	
&	&	&	3 &	2.141 &	1.653 &	2.124 &	1.623 &	2.124 &	1.626 &	2.94 &	1.347	\\	\cmidrule(lr){4-4} \cmidrule(lr){5-6} \cmidrule(lr){7-8} \cmidrule(lr){9-10} \cmidrule(lr){11-12}
&	&	$t_3$ &	1 &	1.86 &	0.819 &	2.637 &	4.933 &	1.874 &	0.844 &	4.672 &	10.607	\\	
&	&	&	2 &	1.969 &	0.947 &	2.692 &	4.92 &	1.973 &	1.013 &	4.604 &	9.930	\\	
&	&	&	3 &	1.909 &	0.75 &	2.743 &	5.998 &	1.916 &	0.753 &	4.67 &	8.578	\\	\cmidrule(lr){3-4} \cmidrule(lr){5-6} \cmidrule(lr){7-8} \cmidrule(lr){9-10} \cmidrule(lr){11-12}
&	$500$ &	Gaussian &	1 &	1.504 &	0.769 &	1.497 &	0.76 &	1.497 &	0.76 &	1.751 &	0.600	\\	
&	&	&	2 &	1.439 &	0.658 &	1.433 &	0.65 &	1.433 &	0.65 &	1.825 &	0.697	\\	
&	&	&	3 &	1.53 &	0.928 &	1.523 &	0.919 &	1.523 &	0.92 &	1.789 &	0.598	\\	\cmidrule(lr){4-4} \cmidrule(lr){5-6} \cmidrule(lr){7-8} \cmidrule(lr){9-10} \cmidrule(lr){11-12}
&	&	$t_3$ &	1 &	1.619 &	0.937 &	1.984 &	3.444 &	1.576 &	0.899 &	2.594 &	7.028	\\	
&	&	&	2 &	1.496 &	0.901 &	2.091 &	5.507 &	1.451 &	0.855 &	2.315 &	5.518	\\	
&	&	&	3 &	1.574 &	1.146 &	2.124 &	5.016 &	1.535 &	1.124 &	2.548 &	6.563	\\	\cmidrule(lr){1-4} \cmidrule(lr){5-6} \cmidrule(lr){7-8} \cmidrule(lr){9-10} \cmidrule(lr){11-12}
\ref{f:two} &	$100$ &	Gaussian &	1 &	1.91 &	0.296 &	1.91 &	0.297 &	1.91 &	0.297 &	3.948 &	0.882	\\	
&	&	&	2 &	0.618 &	0.184 &	0.616 &	0.182 &	0.616 &	0.182 &	1.398 &	0.472	\\	
&	&	&	3 &	0.616 &	0.172 &	0.613 &	0.17 &	0.613 &	0.17 &	1.398 &	0.475	\\	\cmidrule(lr){4-4} \cmidrule(lr){5-6} \cmidrule(lr){7-8} \cmidrule(lr){9-10} \cmidrule(lr){11-12}
&	&	$t_3$ &	1 &	1.932 &	0.302 &	2 &	0.338 &	1.991 &	0.334 &	4.263 &	0.990	\\	
&	&	&	2 &	0.634 &	0.192 &	0.649 &	0.207 &	0.647 &	0.205 &	1.501 &	0.533	\\	
&	&	&	3 &	0.623 &	0.198 &	0.636 &	0.198 &	0.633 &	0.196 &	1.514 &	0.524	\\	\cmidrule(lr){3-4} \cmidrule(lr){5-6} \cmidrule(lr){7-8} \cmidrule(lr){9-10} \cmidrule(lr){11-12}
&	$200$ &	Gaussian &	1 &	1.404 &	0.183 &	1.404 &	0.183 &	1.404 &	0.183 &	2.883 &	0.617	\\	
&	&	&	2 &	0.474 &	0.144 &	0.471 &	0.143 &	0.472 &	0.143 &	1.048 &	0.458	\\	
&	&	&	3 &	0.452 &	0.121 &	0.45 &	0.119 &	0.45 &	0.119 &	1.006 &	0.285	\\	\cmidrule(lr){4-4} \cmidrule(lr){5-6} \cmidrule(lr){7-8} \cmidrule(lr){9-10} \cmidrule(lr){11-12}
&	&	$t_3$ &	1 &	1.358 &	0.179 &	1.398 &	0.19 &	1.39 &	0.188 &	4.061 &	7.863	\\	
&	&	&	2 &	0.421 &	0.106 &	0.427 &	0.106 &	0.424 &	0.106 &	1.705 &	6.855	\\	
&	&	&	3 &	0.434 &	0.117 &	0.442 &	0.123 &	0.439 &	0.121 &	1.816 &	7.468	\\	\cmidrule(lr){3-4} \cmidrule(lr){5-6} \cmidrule(lr){7-8} \cmidrule(lr){9-10} \cmidrule(lr){11-12}
&	$500$ &	Gaussian &	1 &	0.873 &	0.123 &	0.872 &	0.123 &	0.872 &	0.123 &	1.711 &	0.341	\\	
&	&	&	2 &	0.3 &	0.105 &	0.299 &	0.105 &	0.299 &	0.105 &	0.62 &	0.189	\\	
&	&	&	3 &	0.306 &	0.104 &	0.305 &	0.103 &	0.305 &	0.103 &	0.62 &	0.169	\\	\cmidrule(lr){4-4} \cmidrule(lr){5-6} \cmidrule(lr){7-8} \cmidrule(lr){9-10} \cmidrule(lr){11-12}
&	&	$t_3$ &	1 &	0.87 &	0.125 &	0.889 &	0.128 &	0.885 &	0.129 &	1.815 &	0.388	\\	
&	&	&	2 &	0.298 &	0.111 &	0.304 &	0.114 &	0.303 &	0.113 &	0.655 &	0.207	\\	
&	&	&	3 &	0.293 &	0.093 &	0.296 &	0.093 &	0.294 &	0.094 &	0.632 &	0.245	\\	\cmidrule(lr){1-4} \cmidrule(lr){5-6} \cmidrule(lr){7-8} \cmidrule(lr){9-10} \cmidrule(lr){11-12}
\ref{f:three} &	$100$ &	Gaussian &	1 &	0.541 &	0.09 &	0.541 &	0.091 &	0.541 &	0.091 &	1.329 &	0.276	\\	
&	&	&	2 &	0.671 &	0.093 &	0.671 &	0.092 &	0.671 &	0.092 &	1.591 &	0.276	\\	
&	&	&	3 &	0.775 &	0.093 &	0.774 &	0.093 &	0.774 &	0.093 &	1.719 &	0.260	\\	\cmidrule(lr){4-4} \cmidrule(lr){5-6} \cmidrule(lr){7-8} \cmidrule(lr){9-10} \cmidrule(lr){11-12}
&	&	$t_3$ &	1 &	0.554 &	0.087 &	0.574 &	0.096 &	0.569 &	0.092 &	1.381 &	0.288	\\	
&	&	&	2 &	0.69 &	0.108 &	0.71 &	0.11 &	0.704 &	0.102 &	1.69 &	0.374	\\	
&	&	&	3 &	0.764 &	0.095 &	0.787 &	0.102 &	0.783 &	0.099 &	1.856 &	0.329	\\	\cmidrule(lr){3-4} \cmidrule(lr){5-6} \cmidrule(lr){7-8} \cmidrule(lr){9-10} \cmidrule(lr){11-12}
&	$200$ &	Gaussian &	1 &	0.39 &	0.066 &	0.39 &	0.066 &	0.39 &	0.066 &	0.913 &	0.155	\\	
&	&	&	2 &	0.478 &	0.063 &	0.478 &	0.063 &	0.478 &	0.063 &	1.092 &	0.163	\\	
&	&	&	3 &	0.563 &	0.069 &	0.563 &	0.069 &	0.563 &	0.069 &	1.278 &	0.237	\\	\cmidrule(lr){4-4} \cmidrule(lr){5-6} \cmidrule(lr){7-8} \cmidrule(lr){9-10} \cmidrule(lr){11-12}
&	&	$t_3$ &	1 &	0.392 &	0.065 &	0.401 &	0.068 &	0.4 &	0.067 &	0.963 &	0.221	\\	
&	&	&	2 &	0.47 &	0.07 &	0.48 &	0.074 &	0.479 &	0.074 &	1.124 &	0.180	\\	
&	&	&	3 &	0.547 &	0.073 &	0.56 &	0.077 &	0.558 &	0.076 &	1.292 &	0.225	\\	\cmidrule(lr){3-4} \cmidrule(lr){5-6} \cmidrule(lr){7-8} \cmidrule(lr){9-10} \cmidrule(lr){11-12}
&	$500$ &	Gaussian &	1 &	0.245 &	0.032 &	0.245 &	0.032 &	0.245 &	0.032 &	0.573 &	0.092	\\	
&	&	&	2 &	0.306 &	0.039 &	0.306 &	0.039 &	0.306 &	0.039 &	0.682 &	0.094	\\	
&	&	&	3 &	0.362 &	0.043 &	0.361 &	0.043 &	0.361 &	0.043 &	0.801 &	0.104	\\	\cmidrule(lr){4-4} \cmidrule(lr){5-6} \cmidrule(lr){7-8} \cmidrule(lr){9-10} \cmidrule(lr){11-12}
&	&	$t_3$ &	1 &	0.246 &	0.036 &	0.251 &	0.036 &	0.25 &	0.036 &	0.586 &	0.110	\\	
&	&	&	2 &	0.29 &	0.032 &	0.295 &	0.033 &	0.294 &	0.033 &	0.686 &	0.097	\\	
&	&	&	3 &	0.338 &	0.039 &	0.344 &	0.04 &	0.343 &	0.04 &	0.763 &	0.110	\\	\bottomrule
\end{tabular}}
\end{table}

\begin{table}[h!t!b!p!]
\caption{Common component estimation errors of Trunc, noTrunc, iPE, RTFA and PreAve measured as in~\eqref{eq:err:loading:tensor} with $\mc T = [n]$ (`all') and $\mc T = \{n - 10 + 1, \ldots, n\}$ (`local') scaled by $1000$, over varying $n \in \{100, 200, 500\}$ and the distributions for $\mc F_t$ and $\bm\xi_t$ (Gaussian and $t_3$).
We report the mean and the standard deviation over $100$ realisations for each setting.}
\label{tab:tensor:ce:no}
\centering
\resizebox{\textwidth}{!}
{\scriptsize
\begin{tabular}{rrrr cc cc cc cc cc}
\toprule
&	&	&	&	\multicolumn{2}{c}{Trunc} &		\multicolumn{2}{c}{noTrunc} &		\multicolumn{2}{c}{iPE} &		\multicolumn{2}{c}{RTFA} &		\multicolumn{2}{c}{PreAve} 		\\	
Model &	$n$ &	Dist &	Range &	Mean &	SD &	Mean &	SD &	Mean &	SD &	Mean &	SD &	Mean &	SD	\\	\cmidrule(lr){1-4} \cmidrule(lr){5-6} \cmidrule(lr){7-8} \cmidrule(lr){9-10} \cmidrule(lr){11-12} \cmidrule(lr){13-14}
\ref{f:one} &	$100$ &	Gaussian &	All &	33.059 &	15.354 &	32.308 &	10.698 &	32.055 &	10.013 &	32.054 &	10.011 &	34.086 &	10.340	\\	
&	&	&	Local &	33.672 &	15.405 &	33.029 &	11.869 &	32.774 &	11.258 &	32.773 &	11.256 &	34.76 &	11.408	\\	\cmidrule(lr){4-4} \cmidrule(lr){5-6} \cmidrule(lr){7-8} \cmidrule(lr){9-10} \cmidrule(lr){11-12} \cmidrule(lr){13-14}
&	&	$t_3$ &	All &	31.183 &	13.386 &	33.805 &	15.601 &	41.629 &	78.458 &	33.892 &	15.701 &	50.089 &	120.027	\\	
&	&	&	Local &	33.997 &	18.401 &	37.496 &	25.511 &	41.886 &	47.967 &	37.619 &	25.818 &	47.986 &	66.670	\\	\cmidrule(lr){3-4} \cmidrule(lr){5-6} \cmidrule(lr){7-8} \cmidrule(lr){9-10} \cmidrule(lr){11-12} \cmidrule(lr){13-14}
&	$200$ &	Gaussian &	All &	32.088 &	12.048 &	32.097 &	12.054 &	32.076 &	12.021 &	32.076 &	12.022 &	32.781 &	11.908	\\	
&	&	&	Local &	32.546 &	13.063 &	32.553 &	13.08 &	32.532 &	13.049 &	32.533 &	13.051 &	33.24 &	12.926	\\	\cmidrule(lr){4-4} \cmidrule(lr){5-6} \cmidrule(lr){7-8} \cmidrule(lr){9-10} \cmidrule(lr){11-12} \cmidrule(lr){13-14}
&	&	$t_3$ &	All &	29.567 &	8.576 &	31.541 &	10.577 &	41.377 &	80.622 &	31.617 &	10.966 &	58.644 &	161.159	\\	
&	&	&	Local &	33.361 &	14.938 &	40.306 &	58.31 &	67.13 &	291.068 &	41.535 &	70.138 &	101.211 &	522.470	\\	\cmidrule(lr){3-4} \cmidrule(lr){5-6} \cmidrule(lr){7-8} \cmidrule(lr){9-10} \cmidrule(lr){11-12} \cmidrule(lr){13-14}
&	$500$ &	Gaussian &	All &	30.972 &	9.427 &	30.979 &	9.43 &	30.971 &	9.42 &	30.971 &	9.421 &	30.965 &	9.211	\\	
&	&	&	Local &	31.368 &	9.513 &	31.374 &	9.514 &	31.367 &	9.509 &	31.367 &	9.509 &	31.376 &	9.441	\\	\cmidrule(lr){4-4} \cmidrule(lr){5-6} \cmidrule(lr){7-8} \cmidrule(lr){9-10} \cmidrule(lr){11-12} \cmidrule(lr){13-14}
&	&	$t_3$ &	All &	30.688 &	11.532 &	31.927 &	13.292 &	50.174 &	188.124 &	31.918 &	13.429 &	54.479 &	233.345	\\	
&	&	&	Local &	33.396 &	15.469 &	34.785 &	16.551 &	36.763 &	24.055 &	34.744 &	16.527 &	42.108 &	74.449	\\	\cmidrule(lr){1-4} \cmidrule(lr){5-6} \cmidrule(lr){7-8} \cmidrule(lr){9-10} \cmidrule(lr){11-12} \cmidrule(lr){13-14}
\ref{f:two} &	$100$ &	Gaussian &	All &	3.239 &	0.968 &	3.239 &	0.968 &	3.238 &	0.967 &	3.238 &	0.967 &	4.703 &	1.638	\\	
&	&	&	Local &	3.239 &	1.062 &	3.24 &	1.063 &	3.239 &	1.062 &	3.239 &	1.062 &	4.703 &	1.693	\\	\cmidrule(lr){4-4} \cmidrule(lr){5-6} \cmidrule(lr){7-8} \cmidrule(lr){9-10} \cmidrule(lr){11-12} \cmidrule(lr){13-14}
&	&	$t_3$ &	All &	3.189 &	0.958 &	3.385 &	1.061 &	3.416 &	1.081 &	3.412 &	1.079 &	5.048 &	1.759	\\	
&	&	&	Local &	3.488 &	1.246 &	3.73 &	1.463 &	3.769 &	1.511 &	3.76 &	1.488 &	5.46 &	2.130	\\	\cmidrule(lr){3-4} \cmidrule(lr){5-6} \cmidrule(lr){7-8} \cmidrule(lr){9-10} \cmidrule(lr){11-12} \cmidrule(lr){13-14}
&	$200$ &	Gaussian &	All &	3.295 &	0.86 &	3.295 &	0.86 &	3.295 &	0.859 &	3.295 &	0.859 &	4.07 &	1.154	\\	
&	&	&	Local &	3.323 &	0.996 &	3.323 &	0.996 &	3.323 &	0.996 &	3.323 &	0.996 &	4.102 &	1.274	\\	\cmidrule(lr){4-4} \cmidrule(lr){5-6} \cmidrule(lr){7-8} \cmidrule(lr){9-10} \cmidrule(lr){11-12} \cmidrule(lr){13-14}
&	&	$t_3$ &	All &	2.959 &	0.733 &	3.133 &	0.811 &	3.146 &	0.817 &	3.143 &	0.816 &	18.284 &	118.031	\\	
&	&	&	Local &	3.273 &	1.189 &	3.54 &	1.459 &	3.556 &	1.476 &	3.552 &	1.47 &	15.347 &	86.214	\\	\cmidrule(lr){3-4} \cmidrule(lr){5-6} \cmidrule(lr){7-8} \cmidrule(lr){9-10} \cmidrule(lr){11-12} \cmidrule(lr){13-14}
&	$500$ &	Gaussian &	All &	3.037 &	0.867 &	3.037 &	0.868 &	3.037 &	0.867 &	3.037 &	0.867 &	3.3 &	0.954	\\	
&	&	&	Local &	3.077 &	0.976 &	3.077 &	0.976 &	3.077 &	0.976 &	3.077 &	0.976 &	3.34 &	1.062	\\	\cmidrule(lr){4-4} \cmidrule(lr){5-6} \cmidrule(lr){7-8} \cmidrule(lr){9-10} \cmidrule(lr){11-12} \cmidrule(lr){13-14}
&	&	$t_3$ &	All &	2.932 &	0.863 &	3.068 &	0.917 &	3.073 &	0.919 &	3.072 &	0.919 &	3.371 &	1.025	\\	
&	&	&	Local &	3.421 &	1.267 &	3.599 &	1.485 &	3.604 &	1.489 &	3.602 &	1.486 &	3.91 &	1.577	\\	\cmidrule(lr){1-4} \cmidrule(lr){5-6} \cmidrule(lr){7-8} \cmidrule(lr){9-10} \cmidrule(lr){11-12} \cmidrule(lr){13-14}
\ref{f:three} &	$100$ &	Gaussian &	All &	1.28 &	0.252 &	1.28 &	0.251 &	1.28 &	0.251 &	1.28 &	0.251 &	1.824 &	0.385	\\	
&	&	&	Local &	1.281 &	0.29 &	1.281 &	0.291 &	1.281 &	0.291 &	1.281 &	0.291 &	1.825 &	0.416	\\	\cmidrule(lr){4-4} \cmidrule(lr){5-6} \cmidrule(lr){7-8} \cmidrule(lr){9-10} \cmidrule(lr){11-12} \cmidrule(lr){13-14}
&	&	$t_3$ &	All &	1.227 &	0.267 &	1.304 &	0.305 &	1.313 &	0.311 &	1.31 &	0.305 &	1.905 &	0.485	\\	
&	&	&	Local &	1.334 &	0.411 &	1.403 &	0.432 &	1.41 &	0.433 &	1.408 &	0.433 &	2.024 &	0.587	\\	\cmidrule(lr){3-4} \cmidrule(lr){5-6} \cmidrule(lr){7-8} \cmidrule(lr){9-10} \cmidrule(lr){11-12} \cmidrule(lr){13-14}
&	$200$ &	Gaussian &	All &	1.265 &	0.269 &	1.265 &	0.269 &	1.265 &	0.269 &	1.265 &	0.269 &	1.542 &	0.342	\\	
&	&	&	Local &	1.272 &	0.333 &	1.272 &	0.333 &	1.272 &	0.333 &	1.272 &	0.333 &	1.552 &	0.399	\\	\cmidrule(lr){4-4} \cmidrule(lr){5-6} \cmidrule(lr){7-8} \cmidrule(lr){9-10} \cmidrule(lr){11-12} \cmidrule(lr){13-14}
&	&	$t_3$ &	All &	1.194 &	0.245 &	1.257 &	0.269 &	1.26 &	0.271 &	1.259 &	0.271 &	1.538 &	0.330	\\	
&	&	&	Local &	1.339 &	0.433 &	1.406 &	0.45 &	1.409 &	0.45 &	1.409 &	0.45 &	1.698 &	0.510	\\	\cmidrule(lr){3-4} \cmidrule(lr){5-6} \cmidrule(lr){7-8} \cmidrule(lr){9-10} \cmidrule(lr){11-12} \cmidrule(lr){13-14}
&	$500$ &	Gaussian &	All &	1.222 &	0.215 &	1.222 &	0.215 &	1.222 &	0.215 &	1.222 &	0.215 &	1.328 &	0.231	\\	
&	&	&	Local &	1.246 &	0.28 &	1.246 &	0.28 &	1.246 &	0.28 &	1.246 &	0.28 &	1.353 &	0.295	\\	\cmidrule(lr){4-4} \cmidrule(lr){5-6} \cmidrule(lr){7-8} \cmidrule(lr){9-10} \cmidrule(lr){11-12} \cmidrule(lr){13-14}
&	&	$t_3$ &	All &	1.13 &	0.205 &	1.178 &	0.218 &	1.179 &	0.218 &	1.178 &	0.218 &	1.279 &	0.237	\\	
&	&	&	Local &	1.252 &	0.459 &	1.303 &	0.49 &	1.304 &	0.49 &	1.304 &	0.49 &	1.408 &	0.504	\\	\bottomrule
\end{tabular}}
\end{table}

\clearpage 

\paragraph{Outliers in the idiosyncratic component.}

\begin{figure}[h!t!p!]
\centering
\includegraphics[width = 1\textwidth]{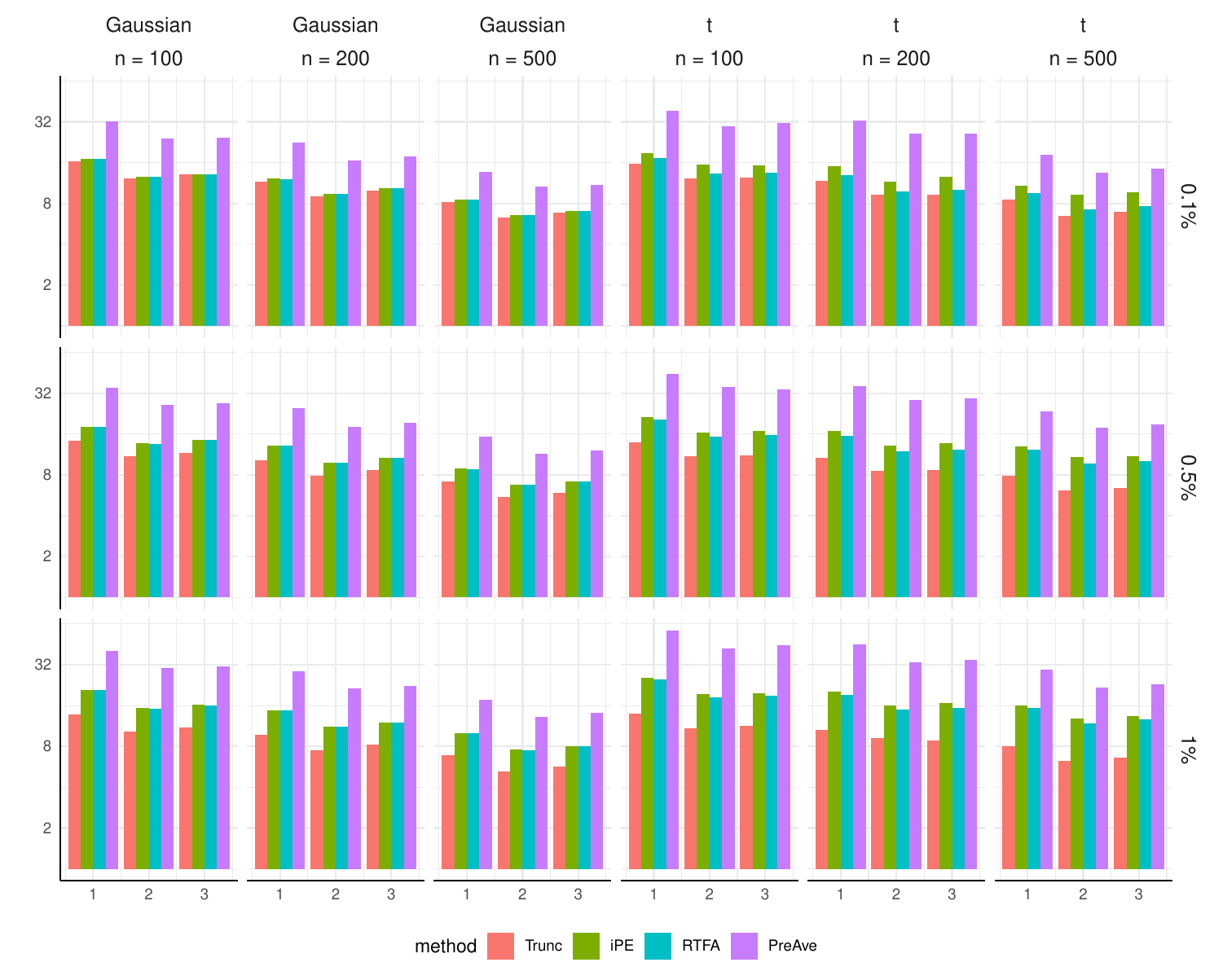}
\caption{\ref{f:one} Loading estimation errors measured as in~\eqref{eq:err:loading:tensor} for each mode ($x$-axis) for Trunc, iPE, RTFA and PreAve over varying $n \in \{100, 200, 500\}$, distributions for $\mc F_t$ and $\bm\xi_t$ (Gaussian and $t_3$) and the percentages of outliers in the idiosyncratic component under~\ref{o:one} ($\varrho \in \{0.1, 0.5, 1\}$, top to bottom), averaged over $100$ realisations per setting. In each plot, the $y$-axis is in the log-scale and all errors have been scaled for the ease of presentation.}
\label{fig:tensor:le:idio:f:one}
\end{figure}

\begin{figure}[h!t!p!]
\centering
\includegraphics[width = 1\textwidth]{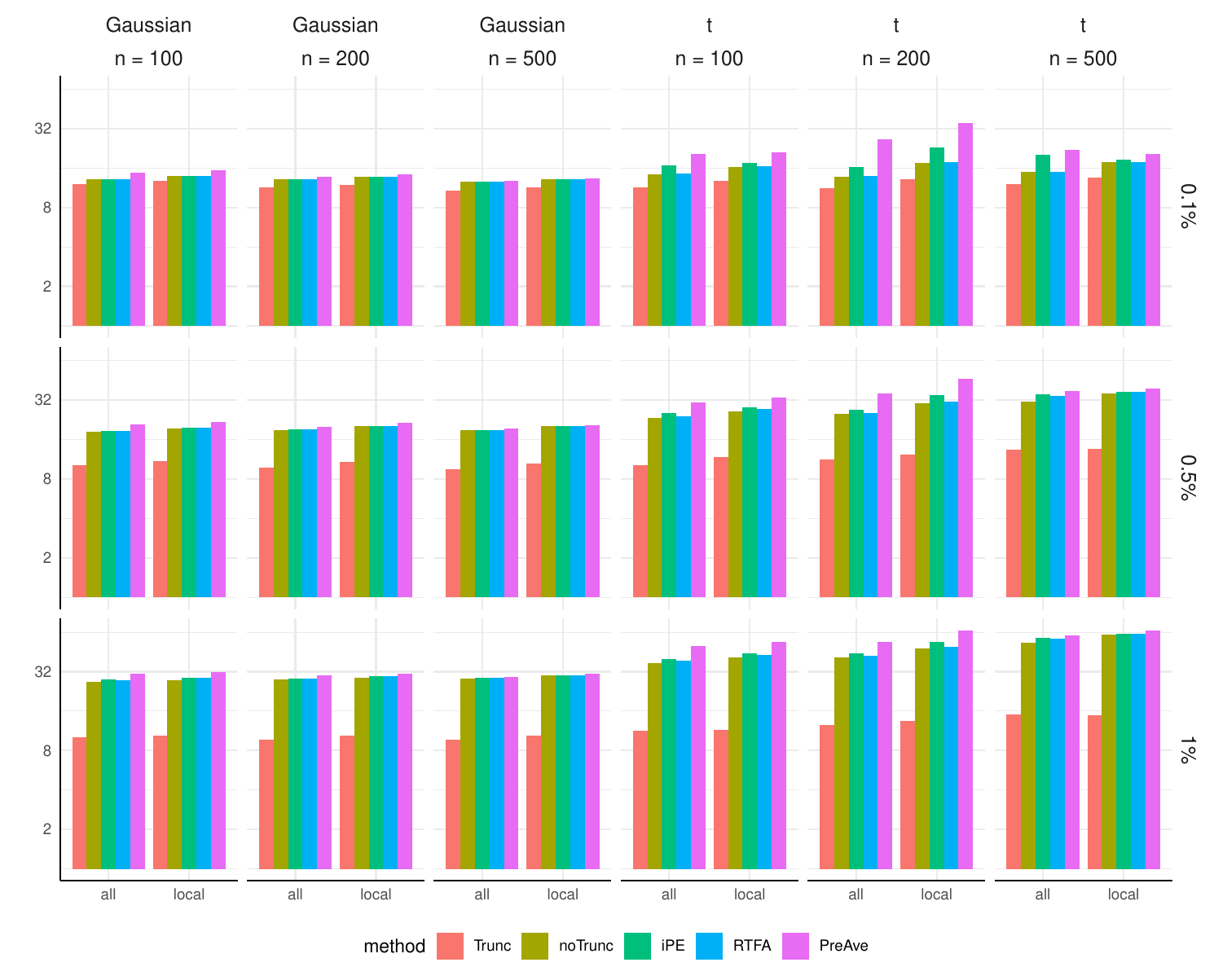}
\caption{\ref{f:one} Common component estimation errors measured as in~\eqref{eq:err:chi:tensor} with $\mc T = [n]$ (`all') and $\mc T = \{n - 10 + 1, \ldots, n \}$ (`local') for Trunc, noTrunc, iPE, RTFA and PreAve over varying $n \in \{100, 200, 500\}$, distributions for $\mc F_t$ and $\bm\xi_t$ (Gaussian and $t_3$) and the percentages of outliers in the idiosyncratic component under~\ref{o:one} ($\varrho \in \{0.1, 0.5, 1\}$, top to bottom), averaged over $100$ realisations per setting. In each plot, the $y$-axis is in the log-scale and all errors have been scaled for the ease of presentation.}
\label{fig:tensor:ce:idio:f:one}
\end{figure}

\begin{figure}[h!t!p!]
\centering
\includegraphics[width = 1\textwidth]{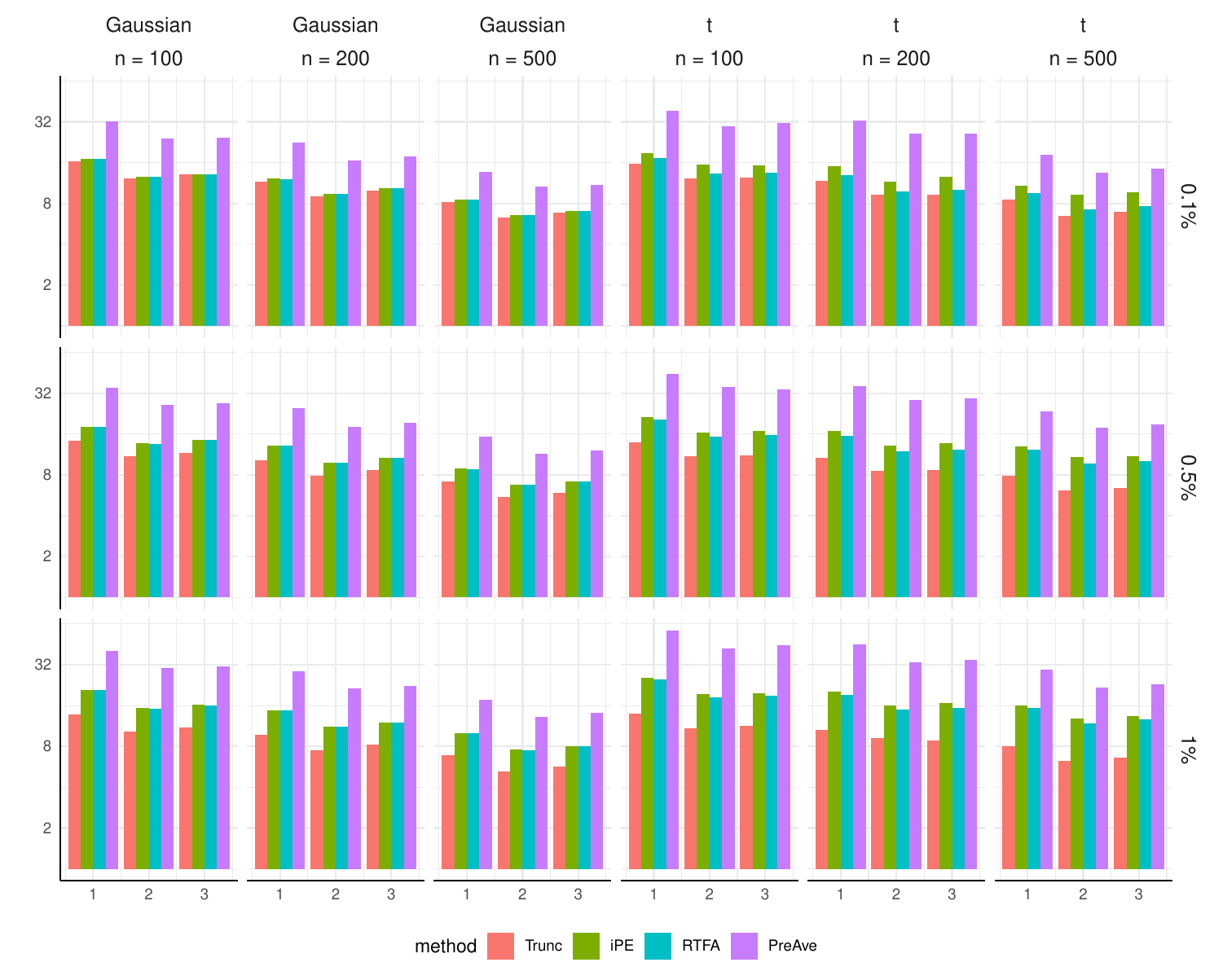}
\caption{\ref{f:two} Loading estimation errors measured as in~\eqref{eq:err:loading:tensor} for each mode ($x$-axis) for Trunc, iPE, RTFA and PreAve over varying $n \in \{100, 200, 500\}$, distributions for $\mc F_t$ and $\bm\xi_t$ (Gaussian and $t_3$) and the percentages of outliers in the idiosyncratic component under~\ref{o:one} ($\varrho \in \{0.1, 0.5, 1\}$, top to bottom), averaged over $100$ realisations per setting. In each plot, the $y$-axis is in the log-scale and all errors have been scaled for the ease of presentation.}
\label{fig:tensor:le:idio:f:two}
\end{figure}

\begin{figure}[h!t!p!]
\centering
\includegraphics[width = 1\textwidth]{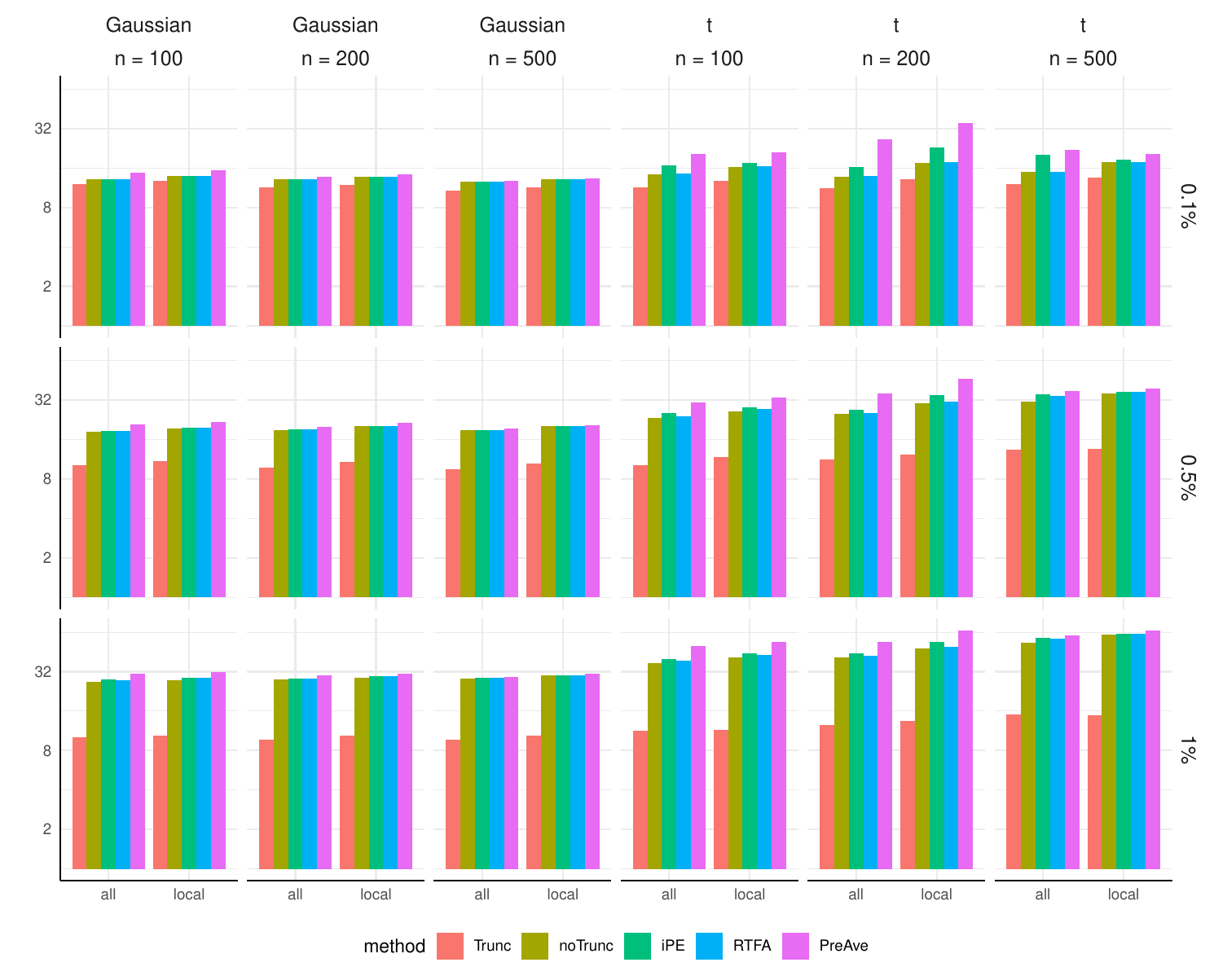}
\caption{\ref{f:two} Common component estimation errors measured as in~\eqref{eq:err:chi:tensor} with $\mc T = [n]$ (`all') and $\mc T = \{n - 10 + 1, \ldots, n \}$ (`local') for Trunc, noTrunc, iPE, RTFA and PreAve over varying $n \in \{100, 200, 500\}$, distributions for $\mc F_t$ and $\bm\xi_t$ (Gaussian and $t_3$) and the percentages of outliers in the idiosyncratic component under~\ref{o:one} ($\varrho \in \{0.1, 0.5, 1\}$, top to bottom), averaged over $100$ realisations per setting. In each plot, the $y$-axis is in the log-scale and all errors have been scaled for the ease of presentation.}
\label{fig:tensor:ce:idio:f:two}
\end{figure}

\begin{figure}[h!t!p!]
\centering
\includegraphics[width = 1\textwidth]{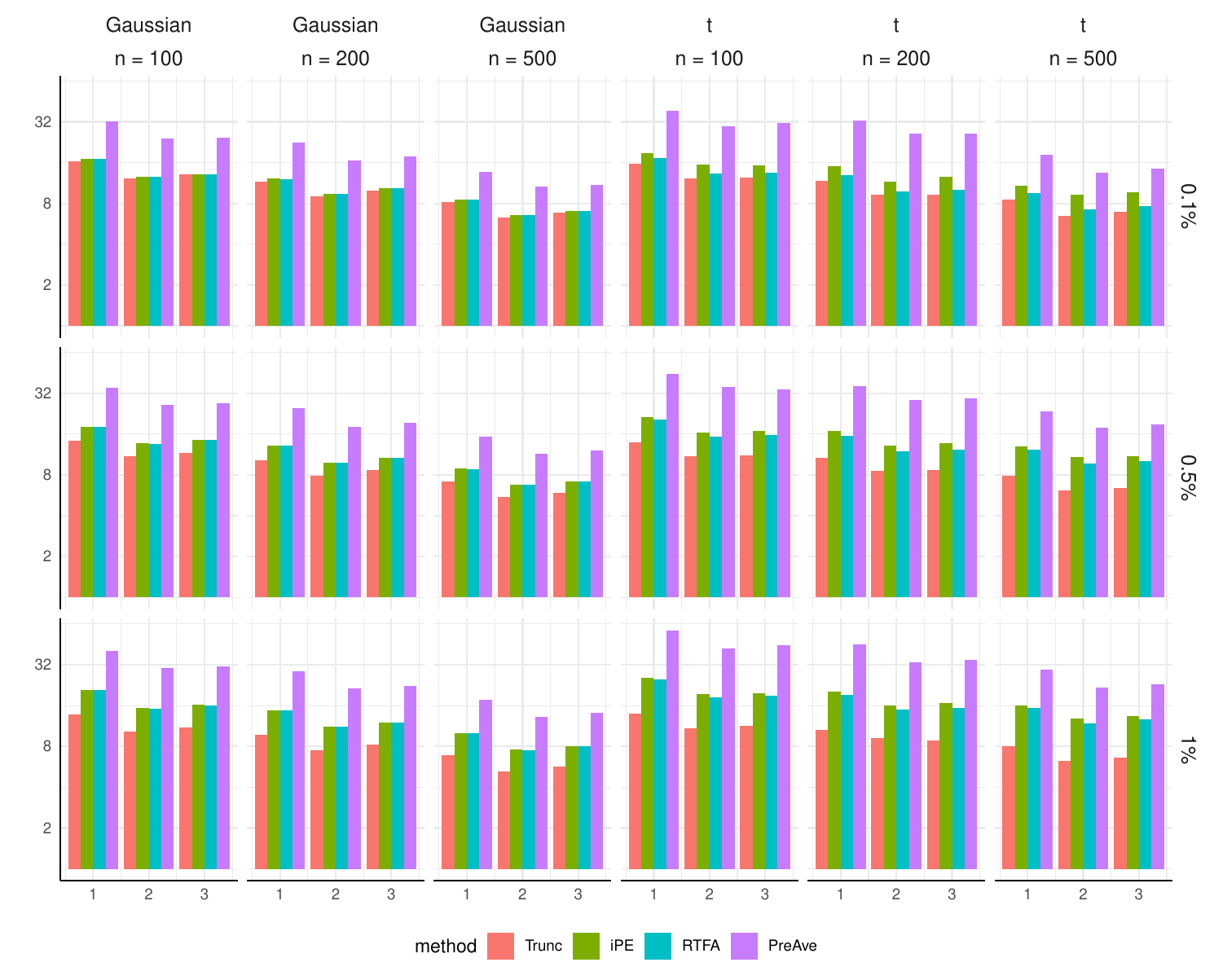}
\caption{\ref{f:three} Loading estimation errors measured as in~\eqref{eq:err:loading:tensor} for each mode ($x$-axis) for Trunc, iPE, RTFA and PreAve over varying $n \in \{100, 200, 500\}$, distributions for $\mc F_t$ and $\bm\xi_t$ (Gaussian and $t_3$) and the percentages of outliers in the idiosyncratic component under~\ref{o:one} ($\varrho \in \{0.1, 0.5, 1\}$, top to bottom), averaged over $100$ realisations per setting. In each plot, the $y$-axis is in the log-scale and all errors have been scaled for the ease of presentation.}
\label{fig:tensor:le:idio:f:three}
\end{figure}

\begin{figure}[h!t!p!]
\centering
\includegraphics[width = 1\textwidth]{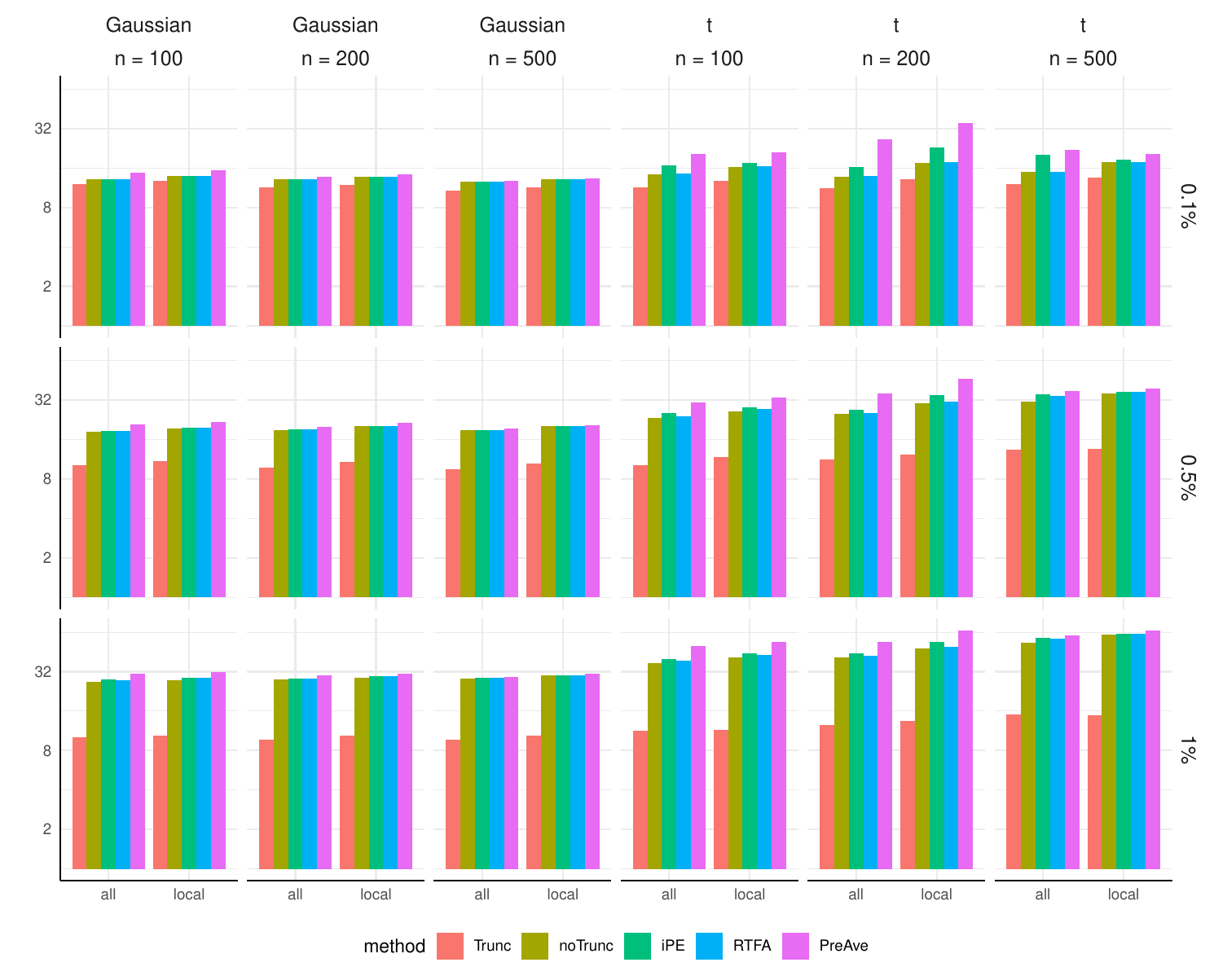}
\caption{\ref{f:three} Common component estimation errors measured as in~\eqref{eq:err:chi:tensor} with $\mc T = [n]$ (`all') and $\mc T = \{n - 10 + 1, \ldots, n \}$ (`local') for Trunc, noTrunc, iPE, RTFA and PreAve over varying $n \in \{100, 200, 500\}$, distributions for $\mc F_t$ and $\bm\xi_t$ (Gaussian and $t_3$) and the percentages of outliers in the idiosyncratic component under~\ref{o:one} ($\varrho \in \{0.1, 0.5, 1\}$, top to bottom), averaged over $100$ realisations per setting. In each plot, the $y$-axis is in the log-scale and all errors have been scaled for the ease of presentation.}
\label{fig:tensor:ce:idio:f:three}
\end{figure}

See Figures~\ref{fig:tensor:le:idio:f:one}--\ref{fig:tensor:ce:idio:f:three} and Tables~\ref{tab:tensor:le:idio:one}--\ref{tab:tensor:ce:idio:three} for the results from loading and common component estimation obtained under \ref{f:one}--\ref{f:three} with outliers in the idiosyncratic component under~\ref{o:one}.

\begin{table}[h!t!b!p!]
\caption{\ref{f:one} Loading estimation errors of Trunc, iPE, RTFA and PreAve measured as in~\eqref{eq:err:loading:tensor} for each mode scaled by $100$, over varying $n \in \{100, 200, 500\}$, the distributions for $\mc F_t$ and $\bm\xi_t$ (Gaussian and $t_3$) and the percentages of outliers in the idiosyncratic component under~\ref{o:one} ($\varrho \in \{0.1, 0.5, 1\}$).
We report the mean and the standard deviation over $100$ realisations for each setting.}
\label{tab:tensor:le:idio:one}
\centering
% \resizebox{\textwidth}{!}
{\scriptsize
\begin{tabular}{rrrr cc cc cc cc}
\toprule
&	&	&	&	\multicolumn{2}{c}{Trunc} &		\multicolumn{2}{c}{iPE} &		\multicolumn{2}{c}{RTFA} &		\multicolumn{2}{c}{PreAve} 		\\	
$n$ &	Dist &	\% &	Mode &	Mean &	SD &	Mean &	SD &	Mean &	SD &	Mean &	SD	\\	\cmidrule(lr){1-4} \cmidrule(lr){5-6} \cmidrule(lr){7-8} \cmidrule(lr){9-10} \cmidrule(lr){11-12}
$100$ &	Gaussian &	$0.1$ &	1 &	2.627 &	1.391 &	2.827 &	1.382 &	2.803 &	1.358 &	4.656 &	1.872	\\	
&	&	&	2 &	2.515 &	1.298 &	2.694 &	1.29 &	2.669 &	1.27 &	4.417 &	1.274	\\	
&	&	&	3 &	2.619 &	1.651 &	2.799 &	1.576 &	2.77 &	1.532 &	4.426 &	1.421	\\	\cmidrule(lr){3-4} \cmidrule(lr){5-6} \cmidrule(lr){7-8} \cmidrule(lr){9-10} \cmidrule(lr){11-12}
&	&	$0.5$ &	1 &	2.786 &	1.834 &	3.718 &	1.426 &	3.663 &	1.41 &	6.316 &	2.185	\\	
&	&	&	2 &	2.717 &	1.391 &	3.709 &	1.434 &	3.654 &	1.413 &	6.173 &	1.471	\\	
&	&	&	3 &	3.057 &	4.338 &	4.125 &	4.559 &	4.025 &	4.117 &	6.316 &	2.449	\\	\cmidrule(lr){3-4} \cmidrule(lr){5-6} \cmidrule(lr){7-8} \cmidrule(lr){9-10} \cmidrule(lr){11-12}
&	&	$1$ &	1 &	2.955 &	1.611 &	4.985 &	2.104 &	4.922 &	2.046 &	8.471 &	3.368	\\	
&	&	&	2 &	2.796 &	1.354 &	4.641 &	1.729 &	4.577 &	1.703 &	8.111 &	2.426	\\	
&	&	&	3 &	3.146 &	3.55 &	5.012 &	3.651 &	4.931 &	3.557 &	8.23 &	3.057	\\	\cmidrule(lr){2-4} \cmidrule(lr){5-6} \cmidrule(lr){7-8} \cmidrule(lr){9-10} \cmidrule(lr){11-12}
&	$t_3$ &	$0.1$ &	1 &	2.445 &	1.13 &	3.18 &	3.723 &	2.724 &	1.137 &	6.257 &	6.588	\\	
&	&	&	2 &	2.532 &	1.046 &	3.475 &	5.56 &	2.84 &	1.071 &	5.865 &	6.362	\\	
&	&	&	3 &	2.492 &	0.811 &	3.304 &	4.258 &	2.791 &	0.857 &	6.574 &	7.600	\\	\cmidrule(lr){3-4} \cmidrule(lr){5-6} \cmidrule(lr){7-8} \cmidrule(lr){9-10} \cmidrule(lr){11-12}
&	&	$0.5$ &	1 &	2.704 &	1.117 &	4.467 &	3.796 &	4 &	1.403 &	8.454 &	7.001	\\	
&	&	&	2 &	2.853 &	1.29 &	4.781 &	5.555 &	4.178 &	1.584 &	8.909 &	8.310	\\	
&	&	&	3 &	2.822 &	1.014 &	4.787 &	4.504 &	4.251 &	1.726 &	8.275 &	6.292	\\	\cmidrule(lr){3-4} \cmidrule(lr){5-6} \cmidrule(lr){7-8} \cmidrule(lr){9-10} \cmidrule(lr){11-12}
&	&	$1$ &	1 &	3.07 &	1.5 &	5.939 &	4.143 &	5.491 &	2.282 &	11.769 &	9.398	\\	
&	&	&	2 &	3.064 &	1.276 &	6.105 &	5.561 &	5.479 &	1.935 &	11.781 &	7.872	\\	
&	&	&	3 &	3.174 &	1.221 &	5.925 &	4.357 &	5.45 &	1.821 &	12.107 &	8.759	\\	\cmidrule(lr){1-4} \cmidrule(lr){5-6} \cmidrule(lr){7-8} \cmidrule(lr){9-10} \cmidrule(lr){11-12}
$200$ &	Gaussian &	$0.1$ &	1 &	1.867 &	0.879 &	2.033 &	0.924 &	2.014 &	0.925 &	3.157 &	1.030	\\	
&	&	&	2 &	1.917 &	0.803 &	2.052 &	0.828 &	2.037 &	0.829 &	3.087 &	0.895	\\	
&	&	&	3 &	2.176 &	1.716 &	2.304 &	1.61 &	2.293 &	1.623 &	3.303 &	1.474	\\	\cmidrule(lr){3-4} \cmidrule(lr){5-6} \cmidrule(lr){7-8} \cmidrule(lr){9-10} \cmidrule(lr){11-12}
&	&	$0.5$ &	1 &	2.001 &	1.087 &	2.763 &	1.386 &	2.72 &	1.378 &	4.52 &	1.341	\\	
&	&	&	2 &	2.02 &	0.817 &	2.699 &	1.027 &	2.67 &	1.013 &	4.28 &	1.242	\\	
&	&	&	3 &	2.276 &	1.822 &	2.943 &	2.068 &	2.906 &	2.01 &	4.456 &	1.793	\\	\cmidrule(lr){3-4} \cmidrule(lr){5-6} \cmidrule(lr){7-8} \cmidrule(lr){9-10} \cmidrule(lr){11-12}
&	&	$1$ &	1 &	2.028 &	0.894 &	3.314 &	1.379 &	3.278 &	1.357 &	5.835 &	1.885	\\	
&	&	&	2 &	2.114 &	0.818 &	3.454 &	1.186 &	3.417 &	1.174 &	5.874 &	1.824	\\	
&	&	&	3 &	2.356 &	1.905 &	3.67 &	2.54 &	3.638 &	2.494 &	5.941 &	2.680	\\	\cmidrule(lr){2-4} \cmidrule(lr){5-6} \cmidrule(lr){7-8} \cmidrule(lr){9-10} \cmidrule(lr){11-12}
&	$t_3$ &	$0.1$ &	1 &	2.059 &	1.179 &	2.975 &	4.911 &	2.224 &	1.011 &	5.241 &	10.557	\\	
&	&	&	2 &	2.181 &	1.781 &	2.955 &	4.812 &	2.24 &	0.969 &	5.217 &	10.078	\\	
&	&	&	3 &	2.128 &	1.11 &	3.083 &	5.96 &	2.243 &	0.861 &	5.318 &	8.649	\\	\cmidrule(lr){3-4} \cmidrule(lr){5-6} \cmidrule(lr){7-8} \cmidrule(lr){9-10} \cmidrule(lr){11-12}
&	&	$0.5$ &	1 &	2.261 &	1.009 &	3.811 &	4.923 &	3.097 &	1.229 &	7.055 &	11.282	\\	
&	&	&	2 &	2.475 &	1.76 &	4.13 &	5.13 &	3.459 &	2.118 &	7.449 &	11.591	\\	
&	&	&	3 &	2.425 &	1.194 &	4.143 &	5.903 &	3.345 &	1.417 &	7.345 &	9.450	\\	\cmidrule(lr){3-4} \cmidrule(lr){5-6} \cmidrule(lr){7-8} \cmidrule(lr){9-10} \cmidrule(lr){11-12}
&	&	$1$ &	1 &	2.428 &	0.984 &	4.839 &	4.977 &	4.15 &	1.606 &	8.912 &	10.954	\\	
&	&	&	2 &	2.951 &	3.819 &	5.378 &	6.324 &	4.685 &	4.441 &	9.095 &	11.277	\\	
&	&	&	3 &	2.634 &	1.539 &	5.365 &	6.085 &	4.565 &	1.867 &	9.091 &	9.797	\\	\cmidrule(lr){1-4} \cmidrule(lr){5-6} \cmidrule(lr){7-8} \cmidrule(lr){9-10} \cmidrule(lr){11-12}
$500$ &	Gaussian &	$0.1$ &	1 &	1.517 &	0.767 &	1.588 &	0.756 &	1.576 &	0.757 &	2.044 &	0.697	\\	
&	&	&	2 &	1.455 &	0.667 &	1.554 &	0.693 &	1.542 &	0.693 &	2.068 &	0.710	\\	
&	&	&	3 &	1.544 &	0.942 &	1.62 &	0.957 &	1.61 &	0.959 &	2.055 &	0.659	\\	\cmidrule(lr){3-4} \cmidrule(lr){5-6} \cmidrule(lr){7-8} \cmidrule(lr){9-10} \cmidrule(lr){11-12}
&	&	$0.5$ &	1 &	1.562 &	0.762 &	1.981 &	0.82 &	1.955 &	0.811 &	2.842 &	0.827	\\	
&	&	&	2 &	1.504 &	0.658 &	1.951 &	0.776 &	1.93 &	0.762 &	2.765 &	0.810	\\	
&	&	&	3 &	1.598 &	0.92 &	2.027 &	0.987 &	2.005 &	0.97 &	2.861 &	0.891	\\	\cmidrule(lr){3-4} \cmidrule(lr){5-6} \cmidrule(lr){7-8} \cmidrule(lr){9-10} \cmidrule(lr){11-12}
&	&	$1$ &	1 &	1.66 &	0.875 &	2.48 &	1.033 &	2.46 &	1.023 &	3.655 &	1.212	\\	
&	&	&	2 &	1.579 &	0.684 &	2.42 &	0.928 &	2.396 &	0.927 &	3.551 &	1.206	\\	
&	&	&	3 &	1.687 &	0.91 &	2.514 &	1.088 &	2.482 &	1.082 &	3.736 &	0.997	\\	\cmidrule(lr){2-4} \cmidrule(lr){5-6} \cmidrule(lr){7-8} \cmidrule(lr){9-10} \cmidrule(lr){11-12}
&	$t_3$ &	$0.1$ &	1 &	1.711 &	0.982 &	2.19 &	3.427 &	1.785 &	0.936 &	3.043 &	7.266	\\	
&	&	&	2 &	1.585 &	0.904 &	2.375 &	5.595 &	1.749 &	1.493 &	2.782 &	5.557	\\	
&	&	&	3 &	1.679 &	1.216 &	2.462 &	5.175 &	1.843 &	1.571 &	3.045 &	7.105	\\	\cmidrule(lr){3-4} \cmidrule(lr){5-6} \cmidrule(lr){7-8} \cmidrule(lr){9-10} \cmidrule(lr){11-12}
&	&	$0.5$ &	1 &	1.875 &	0.934 &	2.928 &	3.469 &	2.827 &	3.296 &	4.46 &	7.377	\\	
&	&	&	2 &	1.823 &	1.043 &	3.178 &	5.649 &	3.057 &	5.625 &	4.391 &	7.050	\\	
&	&	&	3 &	1.855 &	1.401 &	3.284 &	5.537 &	3.167 &	5.334 &	4.795 &	8.137	\\	\cmidrule(lr){3-4} \cmidrule(lr){5-6} \cmidrule(lr){7-8} \cmidrule(lr){9-10} \cmidrule(lr){11-12}
&	&	$1$ &	1 &	2.06 &	1.082 &	3.811 &	3.677 &	3.73 &	3.579 &	5.948 &	7.491	\\	
&	&	&	2 &	1.958 &	1.034 &	3.937 &	5.828 &	3.846 &	5.846 &	5.741 &	7.186	\\	
&	&	&	3 &	2.063 &	1.351 &	4.336 &	5.745 &	4.233 &	5.614 &	6.323 &	8.622	\\	\bottomrule
\end{tabular}
}
\end{table}

\begin{table}[h!t!b!p!]
\caption{\ref{f:one} Common component estimation errors of Trunc, noTrunc, iPE, RTFA and PreAve measured as in~\eqref{eq:err:chi:tensor} with $\mc T = [n]$ (`all') and $\mc T = \{n - 10 + 1, \ldots, n\}$ (`local') scaled by $1000$, over varying $n \in \{100, 200, 500\}$, the distributions for $\mc F_t$ and $\bm\xi_t$ (Gaussian and $t_3$) and the percentages of outliers in the idiosyncratic component under~\ref{o:one} ($\varrho \in \{0.1, 0.5, 1\}$).
We report the mean and the standard deviation over $100$ realisations for each setting.}
\label{tab:tensor:ce:idio:one}
\centering
\resizebox{\textwidth}{!}
{\scriptsize
\begin{tabular}{rrrr cc cc cc cc cc}
\toprule
&	&	&	&	\multicolumn{2}{c}{Trunc} &		\multicolumn{2}{c}{noTrunc} &		\multicolumn{2}{c}{iPE} &		\multicolumn{2}{c}{RTFA} &		\multicolumn{2}{c}{PreAve} 		\\	
$n$ &	Dist &	\% &	Range &	Mean &	SD &	Mean &	SD &	Mean &	SD &	Mean &	SD &	Mean &	SD	\\	\cmidrule(lr){1-4} \cmidrule(lr){5-6} \cmidrule(lr){7-8} \cmidrule(lr){9-10} \cmidrule(lr){11-12} \cmidrule(lr){13-14}
$100$ &	Gaussian &	$0.1$ &	All &	33.455 &	11.008 &	41.982 &	12.705 &	42.268 &	12.783 &	42.231 &	12.762 &	44.875 &	13.268	\\	
&	&	&	Local &	34.236 &	12.184 &	43.343 &	14.76 &	43.629 &	14.814 &	43.59 &	14.804 &	46.221 &	15.120	\\	\cmidrule(lr){3-4} \cmidrule(lr){5-6} \cmidrule(lr){7-8} \cmidrule(lr){9-10} \cmidrule(lr){11-12} \cmidrule(lr){13-14}
&	&	$0.5$ &	All &	39.472 &	28.619 &	81.07 &	22.955 &	82.76 &	23.121 &	82.629 &	23.004 &	87.389 &	23.500	\\	
&	&	&	Local &	40.134 &	28.032 &	83.248 &	26.692 &	84.784 &	26.419 &	84.643 &	26.272 &	89.314 &	26.639	\\	\cmidrule(lr){3-4} \cmidrule(lr){5-6} \cmidrule(lr){7-8} \cmidrule(lr){9-10} \cmidrule(lr){11-12} \cmidrule(lr){13-14}
&	&	$1$ &	All &	43.663 &	22.63 &	129.657 &	34.885 &	133.993 &	36.913 &	133.83 &	36.82 &	142.188 &	38.378	\\	
&	&	&	Local &	44.045 &	23.667 &	133.009 &	40.867 &	137.319 &	42.679 &	137.169 &	42.612 &	145.811 &	44.147	\\	\cmidrule(lr){2-4} \cmidrule(lr){5-6} \cmidrule(lr){7-8} \cmidrule(lr){9-10} \cmidrule(lr){11-12} \cmidrule(lr){13-14}
&	$t_3$ &	$0.1$ &	All &	34.229 &	14.747 &	47.738 &	19.901 &	56.128 &	80.231 &	48.256 &	20.241 &	68.019 &	123.378	\\	
&	&	&	Local &	36.736 &	19.468 &	52.354 &	29.875 &	57.33 &	50.403 &	52.955 &	30.455 &	68.558 &	88.547	\\	\cmidrule(lr){3-4} \cmidrule(lr){5-6} \cmidrule(lr){7-8} \cmidrule(lr){9-10} \cmidrule(lr){11-12} \cmidrule(lr){13-14}
&	&	$0.5$ &	All &	42.013 &	20.211 &	103.713 &	38.781 &	114.855 &	87.771 &	106.823 &	40.888 &	131.397 &	133.746	\\	
&	&	&	Local &	44.458 &	22.906 &	113.305 &	49.078 &	121.201 &	62.55 &	116.917 &	51.737 &	136.548 &	90.990	\\	\cmidrule(lr){3-4} \cmidrule(lr){5-6} \cmidrule(lr){7-8} \cmidrule(lr){9-10} \cmidrule(lr){11-12} \cmidrule(lr){13-14}
&	&	$1$ &	All &	51.305 &	28.153 &	171.686 &	60.751 &	186.443 &	99.973 &	178.438 &	64.033 &	216.495 &	159.991	\\	
&	&	&	Local &	50.076 &	35.75 &	189.954 &	90.185 &	201.948 &	99.06 &	197.386 &	93.25 &	231.41 &	130.627	\\	\cmidrule(lr){1-4} \cmidrule(lr){5-6} \cmidrule(lr){7-8} \cmidrule(lr){9-10} \cmidrule(lr){11-12} \cmidrule(lr){13-14}
$200$ &	Gaussian &	$0.1$ &	All &	33.446 &	12.403 &	42.314 &	14.657 &	42.462 &	14.629 &	42.45 &	14.632 &	43.417 &	14.457	\\	
&	&	&	Local &	34.026 &	13.991 &	43.589 &	16.943 &	43.732 &	16.928 &	43.722 &	16.937 &	44.685 &	16.797	\\	\cmidrule(lr){3-4} \cmidrule(lr){5-6} \cmidrule(lr){7-8} \cmidrule(lr){9-10} \cmidrule(lr){11-12} \cmidrule(lr){13-14}
&	&	$0.5$ &	All &	37.051 &	12.683 &	83.436 &	25.433 &	84.441 &	25.899 &	84.382 &	25.846 &	86.53 &	25.590	\\	
&	&	&	Local &	38.104 &	15.155 &	86.765 &	33.922 &	87.772 &	34.597 &	87.72 &	34.54 &	89.781 &	34.020	\\	\cmidrule(lr){3-4} \cmidrule(lr){5-6} \cmidrule(lr){7-8} \cmidrule(lr){9-10} \cmidrule(lr){11-12} \cmidrule(lr){13-14}
&	&	$1$ &	All &	41.301 &	14.369 &	134.554 &	39.238 &	136.884 &	40.526 &	136.812 &	40.472 &	141 &	41.095	\\	
&	&	&	Local &	42.153 &	16.987 &	137.415 &	50.171 &	139.829 &	51.761 &	139.755 &	51.687 &	144.087 &	51.772	\\	\cmidrule(lr){2-4} \cmidrule(lr){5-6} \cmidrule(lr){7-8} \cmidrule(lr){9-10} \cmidrule(lr){11-12} \cmidrule(lr){13-14}
&	$t_3$ &	$0.1$ &	All &	36.882 &	29.685 &	47.316 &	13.574 &	56.977 &	81.289 &	47.332 &	13.821 &	75.979 &	167.662	\\	
&	&	&	Local &	39.758 &	24.284 &	58.658 &	62.066 &	84.756 &	283.443 &	60.097 &	75.538 &	122.24 &	530.052	\\	\cmidrule(lr){3-4} \cmidrule(lr){5-6} \cmidrule(lr){7-8} \cmidrule(lr){9-10} \cmidrule(lr){11-12} \cmidrule(lr){13-14}
&	&	$0.5$ &	All &	48.936 &	33.266 &	109.872 &	27.741 &	120.934 &	88.181 &	111.324 &	29.002 &	145.495 &	191.421	\\	
&	&	&	Local &	47.946 &	28.949 &	129.931 &	85.113 &	156.03 &	279.577 &	134.074 &	106.16 &	199.005 &	535.579	\\	\cmidrule(lr){3-4} \cmidrule(lr){5-6} \cmidrule(lr){7-8} \cmidrule(lr){9-10} \cmidrule(lr){11-12} \cmidrule(lr){13-14}
&	&	$1$ &	All &	58.339 &	33.636 &	187.416 &	45.39 &	201.597 &	98.858 &	191.49 &	47.834 &	226.3 &	185.902	\\	
&	&	&	Local &	59.732 &	61.851 &	215.859 &	115.161 &	252.076 &	353.252 &	224.423 &	145.083 &	292.548 &	571.622	\\	\cmidrule(lr){1-4} \cmidrule(lr){5-6} \cmidrule(lr){7-8} \cmidrule(lr){9-10} \cmidrule(lr){11-12} \cmidrule(lr){13-14}
$500$ &	Gaussian &	$0.1$ &	All &	32.224 &	9.539 &	41.677 &	11.489 &	41.748 &	11.518 &	41.74 &	11.518 &	41.879 &	11.299	\\	
&	&	&	Local &	32.563 &	9.652 &	41.692 &	12.974 &	41.763 &	13.007 &	41.755 &	13.006 &	41.873 &	12.870	\\	\cmidrule(lr){3-4} \cmidrule(lr){5-6} \cmidrule(lr){7-8} \cmidrule(lr){9-10} \cmidrule(lr){11-12} \cmidrule(lr){13-14}
&	&	$0.5$ &	All &	36.617 &	10.658 &	84.782 &	20.29 &	85.231 &	20.485 &	85.208 &	20.469 &	85.817 &	20.326	\\	
&	&	&	Local &	37.521 &	12.05 &	88.152 &	26.985 &	88.615 &	27.183 &	88.593 &	27.177 &	89.182 &	27.014	\\	\cmidrule(lr){3-4} \cmidrule(lr){5-6} \cmidrule(lr){7-8} \cmidrule(lr){9-10} \cmidrule(lr){11-12} \cmidrule(lr){13-14}
&	&	$1$ &	All &	42.041 &	13.695 &	138.582 &	31.401 &	139.616 &	31.787 &	139.582 &	31.779 &	140.812 &	31.727	\\	
&	&	&	Local &	43.236 &	13.997 &	144.556 &	37.509 &	145.626 &	38 &	145.595 &	37.996 &	146.83 &	38.043	\\	\cmidrule(lr){2-4} \cmidrule(lr){5-6} \cmidrule(lr){7-8} \cmidrule(lr){9-10} \cmidrule(lr){11-12} \cmidrule(lr){13-14}
&	$t_3$ &	$0.1$ &	All &	42.516 &	55.806 &	52.435 &	17.619 &	70.818 &	189.486 &	52.702 &	18.566 &	75.683 &	238.988	\\	
&	&	&	Local &	41.268 &	20.601 &	58.996 &	27.027 &	61.259 &	31.348 &	59.187 &	27.09 &	67.386 &	82.774	\\	\cmidrule(lr){3-4} \cmidrule(lr){5-6} \cmidrule(lr){7-8} \cmidrule(lr){9-10} \cmidrule(lr){11-12} \cmidrule(lr){13-14}
&	&	$0.5$ &	All &	59.966 &	69.622 &	134.113 &	40.508 &	153.303 &	197.561 &	152.764 &	194.281 &	159.805 &	254.390	\\	
&	&	&	Local &	53.404 &	37.459 &	151.241 &	73.12 &	155.068 &	74.906 &	154.86 &	74.706 &	162.092 &	105.748	\\	\cmidrule(lr){3-4} \cmidrule(lr){5-6} \cmidrule(lr){7-8} \cmidrule(lr){9-10} \cmidrule(lr){11-12} \cmidrule(lr){13-14}
&	&	$1$ &	All &	69.85 &	71.463 &	235.957 &	69.586 &	257.054 &	211.767 &	256.694 &	210.392 &	264.881 &	270.685	\\	
&	&	&	Local &	65.513 &	49.217 &	264.334 &	129.118 &	270.066 &	131.648 &	269.859 &	131.429 &	280.553 &	158.214	\\	\bottomrule
\end{tabular}
}
\end{table}

\begin{table}[h!t!b!p!]
\caption{\ref{f:two} Loading estimation errors of Trunc, iPE, RTFA and PreAve measured as in~\eqref{eq:err:loading:tensor} for each mode scaled by $100$, over varying $n \in \{100, 200, 500\}$, the distributions for $\mc F_t$ and $\bm\xi_t$ (Gaussian and $t_3$) and the percentages of outliers in the idiosyncratic component under~\ref{o:one} ($\varrho \in \{0.1, 0.5, 1\}$).
We report the mean and the standard deviation over $100$ realisations for each setting.}
\label{tab:tensor:le:idio:two}
\centering
% \resizebox{\textwidth}{!}
{\scriptsize
\begin{tabular}{rrrr cc cc cc cc}
\toprule
&	&	&	&	\multicolumn{2}{c}{Trunc} &		\multicolumn{2}{c}{iPE} &		\multicolumn{2}{c}{RTFA} &		\multicolumn{2}{c}{PreAve} 		\\	
$n$ &	Dist &	\% &	Mode &	Mean &	SD &	Mean &	SD &	Mean &	SD &	Mean &	SD	\\	\cmidrule(lr){1-4} \cmidrule(lr){5-6} \cmidrule(lr){7-8} \cmidrule(lr){9-10} \cmidrule(lr){11-12}
$100$ &	Gaussian &	$0.1$ &	1 &	1.949 &	0.302 &	2.197 &	0.335 &	2.194 &	0.335 &	4.528 &	0.977	\\	
&	&	&	2 &	0.637 &	0.19 &	0.7 &	0.219 &	0.7 &	0.219 &	1.635 &	0.650	\\	
&	&	&	3 &	0.631 &	0.165 &	0.69 &	0.156 &	0.69 &	0.156 &	1.579 &	0.494	\\	\cmidrule(lr){3-4} \cmidrule(lr){5-6} \cmidrule(lr){7-8} \cmidrule(lr){9-10} \cmidrule(lr){11-12}
&	&	$0.5$ &	1 &	2.073 &	0.305 &	3.122 &	0.452 &	3.117 &	0.451 &	6.471 &	1.558	\\	
&	&	&	2 &	0.689 &	0.191 &	0.975 &	0.252 &	0.974 &	0.252 &	2.296 &	0.763	\\	
&	&	&	3 &	0.681 &	0.176 &	0.955 &	0.243 &	0.953 &	0.243 &	2.296 &	0.755	\\	\cmidrule(lr){3-4} \cmidrule(lr){5-6} \cmidrule(lr){7-8} \cmidrule(lr){9-10} \cmidrule(lr){11-12}
&	&	$1$ &	1 &	2.197 &	0.295 &	3.972 &	0.541 &	3.968 &	0.54 &	8.227 &	1.691	\\	
&	&	&	2 &	0.764 &	0.216 &	1.307 &	0.404 &	1.305 &	0.404 &	3.025 &	1.103	\\	
&	&	&	3 &	0.737 &	0.183 &	1.241 &	0.302 &	1.239 &	0.302 &	2.964 &	0.897	\\	\cmidrule(lr){2-4} \cmidrule(lr){5-6} \cmidrule(lr){7-8} \cmidrule(lr){9-10} \cmidrule(lr){11-12}
&	$t_3$ &	$0.1$ &	1 &	2.037 &	0.284 &	2.595 &	0.396 &	2.583 &	0.394 &	5.632 &	1.295	\\	
&	&	&	2 &	0.691 &	0.2 &	0.838 &	0.252 &	0.836 &	0.252 &	1.905 &	0.582	\\	
&	&	&	3 &	0.681 &	0.218 &	0.832 &	0.257 &	0.829 &	0.257 &	2.023 &	0.671	\\	\cmidrule(lr){3-4} \cmidrule(lr){5-6} \cmidrule(lr){7-8} \cmidrule(lr){9-10} \cmidrule(lr){11-12}
&	&	$0.5$ &	1 &	2.31 &	0.333 &	4.212 &	0.528 &	4.201 &	0.527 &	9.186 &	2.544	\\	
&	&	&	2 &	0.825 &	0.218 &	1.354 &	0.349 &	1.35 &	0.347 &	3.228 &	1.243	\\	
&	&	&	3 &	0.842 &	0.27 &	1.325 &	0.416 &	1.322 &	0.418 &	3.153 &	0.933	\\	\cmidrule(lr){3-4} \cmidrule(lr){5-6} \cmidrule(lr){7-8} \cmidrule(lr){9-10} \cmidrule(lr){11-12}
&	&	$1$ &	1 &	2.568 &	0.433 &	5.682 &	0.761 &	5.672 &	0.758 &	13.284 &	4.824	\\	
&	&	&	2 &	0.988 &	0.343 &	1.819 &	0.513 &	1.816 &	0.511 &	4.324 &	1.715	\\	
&	&	&	3 &	1.013 &	0.357 &	1.835 &	0.492 &	1.833 &	0.491 &	4.428 &	1.492	\\	\cmidrule(lr){1-4} \cmidrule(lr){5-6} \cmidrule(lr){7-8} \cmidrule(lr){9-10} \cmidrule(lr){11-12}
$200$ &	Gaussian &	$0.1$ &	1 &	1.438 &	0.188 &	1.634 &	0.207 &	1.631 &	0.206 &	3.321 &	0.696	\\	
&	&	&	2 &	0.487 &	0.141 &	0.536 &	0.146 &	0.535 &	0.145 &	1.206 &	0.471	\\	
&	&	&	3 &	0.463 &	0.125 &	0.517 &	0.14 &	0.515 &	0.14 &	1.169 &	0.313	\\	\cmidrule(lr){3-4} \cmidrule(lr){5-6} \cmidrule(lr){7-8} \cmidrule(lr){9-10} \cmidrule(lr){11-12}
&	&	$0.5$ &	1 &	1.541 &	0.188 &	2.347 &	0.267 &	2.343 &	0.267 &	4.738 &	0.825	\\	
&	&	&	2 &	0.535 &	0.153 &	0.759 &	0.242 &	0.759 &	0.241 &	1.663 &	0.622	\\	
&	&	&	3 &	0.501 &	0.128 &	0.716 &	0.171 &	0.714 &	0.17 &	1.658 &	0.480	\\	\cmidrule(lr){3-4} \cmidrule(lr){5-6} \cmidrule(lr){7-8} \cmidrule(lr){9-10} \cmidrule(lr){11-12}
&	&	$1$ &	1 &	1.628 &	0.191 &	2.966 &	0.33 &	2.963 &	0.329 &	6.047 &	1.004	\\	
&	&	&	2 &	0.568 &	0.152 &	0.936 &	0.221 &	0.936 &	0.221 &	2.181 &	0.700	\\	
&	&	&	3 &	0.548 &	0.141 &	0.945 &	0.237 &	0.943 &	0.237 &	2.194 &	0.571	\\	\cmidrule(lr){2-4} \cmidrule(lr){5-6} \cmidrule(lr){7-8} \cmidrule(lr){9-10} \cmidrule(lr){11-12}
&	$t_3$ &	$0.1$ &	1 &	1.445 &	0.188 &	1.912 &	0.245 &	1.9 &	0.242 &	5.136 &	8.030	\\	
&	&	&	2 &	0.47 &	0.108 &	0.592 &	0.153 &	0.588 &	0.151 &	2.197 &	7.773	\\	
&	&	&	3 &	0.478 &	0.125 &	0.598 &	0.141 &	0.592 &	0.139 &	2.173 &	7.713	\\	\cmidrule(lr){3-4} \cmidrule(lr){5-6} \cmidrule(lr){7-8} \cmidrule(lr){9-10} \cmidrule(lr){11-12}
&	&	$0.5$ &	1 &	1.673 &	0.299 &	3.277 &	0.42 &	3.267 &	0.42 &	7.996 &	8.033	\\	
&	&	&	2 &	0.587 &	0.175 &	0.997 &	0.233 &	0.995 &	0.231 &	3.193 &	8.080	\\	
&	&	&	3 &	0.582 &	0.184 &	1.015 &	0.262 &	1.014 &	0.26 &	3.333 &	8.098	\\	\cmidrule(lr){3-4} \cmidrule(lr){5-6} \cmidrule(lr){7-8} \cmidrule(lr){9-10} \cmidrule(lr){11-12}
&	&	$1$ &	1 &	1.86 &	0.37 &	4.446 &	0.584 &	4.438 &	0.584 &	11.622 &	11.228	\\	
&	&	&	2 &	0.734 &	0.27 &	1.39 &	0.453 &	1.389 &	0.453 &	4.192 &	7.281	\\	
&	&	&	3 &	0.727 &	0.265 &	1.436 &	0.432 &	1.433 &	0.432 &	4.457 &	8.761	\\	\cmidrule(lr){1-4} \cmidrule(lr){5-6} \cmidrule(lr){7-8} \cmidrule(lr){9-10} \cmidrule(lr){11-12}
$500$ &	Gaussian &	$0.1$ &	1 &	0.895 &	0.124 &	1.019 &	0.136 &	1.017 &	0.137 &	1.977 &	0.334	\\	
&	&	&	2 &	0.305 &	0.101 &	0.333 &	0.105 &	0.332 &	0.105 &	0.719 &	0.203	\\	
&	&	&	3 &	0.313 &	0.103 &	0.342 &	0.107 &	0.342 &	0.107 &	0.724 &	0.212	\\	\cmidrule(lr){3-4} \cmidrule(lr){5-6} \cmidrule(lr){7-8} \cmidrule(lr){9-10} \cmidrule(lr){11-12}
&	&	$0.5$ &	1 &	0.964 &	0.125 &	1.471 &	0.18 &	1.468 &	0.179 &	2.835 &	0.432	\\	
&	&	&	2 &	0.333 &	0.103 &	0.481 &	0.139 &	0.479 &	0.138 &	1.011 &	0.292	\\	
&	&	&	3 &	0.343 &	0.104 &	0.495 &	0.141 &	0.494 &	0.141 &	1.049 &	0.278	\\	\cmidrule(lr){3-4} \cmidrule(lr){5-6} \cmidrule(lr){7-8} \cmidrule(lr){9-10} \cmidrule(lr){11-12}
&	&	$1$ &	1 &	1.037 &	0.129 &	1.904 &	0.221 &	1.902 &	0.22 &	3.667 &	0.620	\\	
&	&	&	2 &	0.353 &	0.101 &	0.585 &	0.147 &	0.584 &	0.147 &	1.335 &	0.394	\\	
&	&	&	3 &	0.369 &	0.114 &	0.628 &	0.202 &	0.627 &	0.202 &	1.396 &	0.387	\\	\cmidrule(lr){2-4} \cmidrule(lr){5-6} \cmidrule(lr){7-8} \cmidrule(lr){9-10} \cmidrule(lr){11-12}
&	$t_3$ &	$0.1$ &	1 &	0.973 &	0.174 &	1.345 &	0.194 &	1.336 &	0.193 &	2.727 &	0.654	\\	
&	&	&	2 &	0.352 &	0.125 &	0.453 &	0.136 &	0.45 &	0.135 &	0.995 &	0.296	\\	
&	&	&	3 &	0.336 &	0.101 &	0.435 &	0.129 &	0.432 &	0.129 &	0.935 &	0.347	\\	\cmidrule(lr){3-4} \cmidrule(lr){5-6} \cmidrule(lr){7-8} \cmidrule(lr){9-10} \cmidrule(lr){11-12}
&	&	$0.5$ &	1 &	1.202 &	0.32 &	2.475 &	0.399 &	2.466 &	0.396 &	4.987 &	1.673	\\	
&	&	&	2 &	0.493 &	0.238 &	0.809 &	0.263 &	0.806 &	0.264 &	1.823 &	0.606	\\	
&	&	&	3 &	0.457 &	0.195 &	0.752 &	0.197 &	0.75 &	0.198 &	1.792 &	1.293	\\	\cmidrule(lr){3-4} \cmidrule(lr){5-6} \cmidrule(lr){7-8} \cmidrule(lr){9-10} \cmidrule(lr){11-12}
&	&	$1$ &	1 &	1.386 &	0.378 &	3.466 &	0.525 &	3.459 &	0.524 &	7.296 &	4.967	\\	
&	&	&	2 &	0.601 &	0.347 &	1.156 &	0.425 &	1.153 &	0.425 &	2.478 &	0.876	\\	
&	&	&	3 &	0.574 &	0.376 &	1.052 &	0.327 &	1.05 &	0.328 &	2.487 &	1.848	\\	\bottomrule
\end{tabular}
}
\end{table}

\begin{table}[h!t!b!p!]
\caption{\ref{f:two} Common component estimation errors of Trunc, noTrunc, iPE, RTFA and PreAve measured as in~\eqref{eq:err:chi:tensor} with $\mc T = [n]$ (`all') and $\mc T = \{n - 10 + 1, \ldots, n\}$ (`local') scaled by $1000$, over varying $n \in \{100, 200, 500\}$, the distributions for $\mc F_t$ and $\bm\xi_t$ (Gaussian and $t_3$) and the percentages of outliers in the idiosyncratic component under~\ref{o:one} ($\varrho \in \{0.1, 0.5, 1\}$).
We report the mean and the standard deviation over $100$ realisations for each setting.}
\label{tab:tensor:ce:idio:two}
\centering
\resizebox{\textwidth}{!}
{\scriptsize
\begin{tabular}{rrrr cc cc cc cc cc}
\toprule
&	&	&	&	\multicolumn{2}{c}{Trunc} &		\multicolumn{2}{c}{noTrunc} &		\multicolumn{2}{c}{iPE} &		\multicolumn{2}{c}{RTFA} &		\multicolumn{2}{c}{PreAve} 		\\	
$n$ &	Dist &	\% &	Range &	Mean &	SD &	Mean &	SD &	Mean &	SD &	Mean &	SD &	Mean &	SD	\\	\cmidrule(lr){1-4} \cmidrule(lr){5-6} \cmidrule(lr){7-8} \cmidrule(lr){9-10} \cmidrule(lr){11-12} \cmidrule(lr){13-14}
$100$ &	Gaussian &	$0.1$ &	All &	3.386 &	0.987 &	4.331 &	1.217 &	4.447 &	1.25 &	4.445 &	1.25 &	6.358 &	2.078	\\	
&	&	&	Local &	3.394 &	1.085 &	4.359 &	1.371 &	4.474 &	1.401 &	4.473 &	1.4 &	6.377 &	2.181	\\	\cmidrule(lr){3-4} \cmidrule(lr){5-6} \cmidrule(lr){7-8} \cmidrule(lr){9-10} \cmidrule(lr){11-12} \cmidrule(lr){13-14}
&	&	$0.5$ &	All &	3.885 &	1.06 &	8.778 &	2.317 &	9.395 &	2.505 &	9.391 &	2.504 &	13.363 &	4.801	\\	
&	&	&	Local &	3.917 &	1.163 &	8.873 &	2.696 &	9.489 &	2.866 &	9.485 &	2.865 &	13.459 &	5.051	\\	\cmidrule(lr){3-4} \cmidrule(lr){5-6} \cmidrule(lr){7-8} \cmidrule(lr){9-10} \cmidrule(lr){11-12} \cmidrule(lr){13-14}
&	&	$1$ &	All &	4.471 &	1.105 &	14.294 &	3.659 &	15.556 &	4.028 &	15.553 &	4.027 &	21.898 &	6.696	\\	
&	&	&	Local &	4.504 &	1.268 &	14.508 &	4.278 &	15.765 &	4.584 &	15.761 &	4.583 &	22.101 &	7.051	\\	\cmidrule(lr){2-4} \cmidrule(lr){5-6} \cmidrule(lr){7-8} \cmidrule(lr){9-10} \cmidrule(lr){11-12} \cmidrule(lr){13-14}
&	$t_3$ &	$0.1$ &	All &	3.57 &	0.903 &	5.834 &	1.616 &	6.119 &	1.726 &	6.112 &	1.725 &	8.953 &	2.964	\\	
&	&	&	Local &	3.909 &	1.365 &	6.525 &	2.46 &	6.825 &	2.561 &	6.815 &	2.548 &	9.774 &	3.706	\\	\cmidrule(lr){3-4} \cmidrule(lr){5-6} \cmidrule(lr){7-8} \cmidrule(lr){9-10} \cmidrule(lr){11-12} \cmidrule(lr){13-14}
&	&	$0.5$ &	All &	4.781 &	1.204 &	15.554 &	4.009 &	16.924 &	4.448 &	16.913 &	4.446 &	24.693 &	9.733	\\	
&	&	&	Local &	5.222 &	2.285 &	17.558 &	6.896 &	18.974 &	7.245 &	18.961 &	7.237 &	26.983 &	11.547	\\	\cmidrule(lr){3-4} \cmidrule(lr){5-6} \cmidrule(lr){7-8} \cmidrule(lr){9-10} \cmidrule(lr){11-12} \cmidrule(lr){13-14}
&	&	$1$ &	All &	6.218 &	1.975 &	27.79 &	7.025 &	30.669 &	8 &	30.655 &	7.994 &	48.173 &	24.651	\\	
&	&	&	Local &	6.472 &	2.862 &	31.45 &	12.39 &	34.461 &	13.093 &	34.446 &	13.087 &	52.518 &	27.648	\\	\cmidrule(lr){1-4} \cmidrule(lr){5-6} \cmidrule(lr){7-8} \cmidrule(lr){9-10} \cmidrule(lr){11-12} \cmidrule(lr){13-14}
$200$ &	Gaussian &	$0.1$ &	All &	3.456 &	0.878 &	4.533 &	1.078 &	4.6 &	1.091 &	4.599 &	1.091 &	5.618 &	1.447	\\	
&	&	&	Local &	3.481 &	0.984 &	4.566 &	1.151 &	4.633 &	1.161 &	4.631 &	1.161 &	5.654 &	1.475	\\	\cmidrule(lr){3-4} \cmidrule(lr){5-6} \cmidrule(lr){7-8} \cmidrule(lr){9-10} \cmidrule(lr){11-12} \cmidrule(lr){13-14}
&	&	$0.5$ &	All &	3.963 &	0.947 &	9.474 &	2.048 &	9.823 &	2.116 &	9.821 &	2.116 &	11.832 &	2.614	\\	
&	&	&	Local &	3.959 &	1.122 &	9.31 &	2.448 &	9.659 &	2.502 &	9.657 &	2.502 &	11.675 &	2.918	\\	\cmidrule(lr){3-4} \cmidrule(lr){5-6} \cmidrule(lr){7-8} \cmidrule(lr){9-10} \cmidrule(lr){11-12} \cmidrule(lr){13-14}
&	&	$1$ &	All &	4.568 &	1.001 &	15.64 &	3.248 &	16.337 &	3.397 &	16.335 &	3.397 &	19.636 &	4.168	\\	
&	&	&	Local &	4.598 &	1.18 &	15.658 &	3.938 &	16.361 &	4.078 &	16.359 &	4.078 &	19.638 &	4.710	\\	\cmidrule(lr){2-4} \cmidrule(lr){5-6} \cmidrule(lr){7-8} \cmidrule(lr){9-10} \cmidrule(lr){11-12} \cmidrule(lr){13-14}
&	$t_3$ &	$0.1$ &	All &	3.399 &	0.805 &	6.202 &	1.495 &	6.376 &	1.546 &	6.37 &	1.544 &	23.003 &	125.475	\\	
&	&	&	Local &	3.726 &	1.464 &	6.922 &	2.684 &	7.102 &	2.728 &	7.095 &	2.726 &	20.405 &	94.925	\\	\cmidrule(lr){3-4} \cmidrule(lr){5-6} \cmidrule(lr){7-8} \cmidrule(lr){9-10} \cmidrule(lr){11-12} \cmidrule(lr){13-14}
&	&	$0.5$ &	All &	4.761 &	1.722 &	18.484 &	4.462 &	19.375 &	4.702 &	19.368 &	4.7 &	39.278 &	129.314	\\	
&	&	&	Local &	5.254 &	3.575 &	20.459 &	8.403 &	21.363 &	8.526 &	21.356 &	8.525 &	38.494 &	100.490	\\	\cmidrule(lr){3-4} \cmidrule(lr){5-6} \cmidrule(lr){7-8} \cmidrule(lr){9-10} \cmidrule(lr){11-12} \cmidrule(lr){13-14}
&	&	$1$ &	All &	6.795 &	7.052 &	33.871 &	8.471 &	35.764 &	8.993 &	35.755 &	8.992 &	64.334 &	138.470	\\	
&	&	&	Local &	6.376 &	4.543 &	37.113 &	15.608 &	39.062 &	15.908 &	39.051 &	15.908 &	63.372 &	106.823	\\	\cmidrule(lr){1-4} \cmidrule(lr){5-6} \cmidrule(lr){7-8} \cmidrule(lr){9-10} \cmidrule(lr){11-12} \cmidrule(lr){13-14}
$500$ &	Gaussian &	$0.1$ &	All &	3.192 &	0.881 &	4.279 &	1.114 &	4.305 &	1.119 &	4.305 &	1.119 &	4.652 &	1.211	\\	
&	&	&	Local &	3.235 &	1.008 &	4.349 &	1.345 &	4.376 &	1.35 &	4.375 &	1.35 &	4.724 &	1.435	\\	\cmidrule(lr){3-4} \cmidrule(lr){5-6} \cmidrule(lr){7-8} \cmidrule(lr){9-10} \cmidrule(lr){11-12} \cmidrule(lr){13-14}
&	&	$0.5$ &	All &	3.708 &	0.927 &	9.276 &	2.152 &	9.416 &	2.186 &	9.415 &	2.186 &	10.124 &	2.364	\\	
&	&	&	Local &	3.756 &	1.061 &	9.421 &	2.52 &	9.56 &	2.55 &	9.559 &	2.55 &	10.267 &	2.700	\\	\cmidrule(lr){3-4} \cmidrule(lr){5-6} \cmidrule(lr){7-8} \cmidrule(lr){9-10} \cmidrule(lr){11-12} \cmidrule(lr){13-14}
&	&	$1$ &	All &	4.338 &	0.977 &	15.581 &	3.482 &	15.87 &	3.546 &	15.869 &	3.546 &	17.062 &	3.887	\\	
&	&	&	Local &	4.341 &	1.112 &	15.557 &	4.123 &	15.847 &	4.174 &	15.846 &	4.175 &	17.031 &	4.463	\\	\cmidrule(lr){2-4} \cmidrule(lr){5-6} \cmidrule(lr){7-8} \cmidrule(lr){9-10} \cmidrule(lr){11-12} \cmidrule(lr){13-14}
&	$t_3$ &	$0.1$ &	All &	3.73 &	1.269 &	7.698 &	2.127 &	7.794 &	2.153 &	7.791 &	2.152 &	8.468 &	2.494	\\	
&	&	&	Local &	4.467 &	2.449 &	9.169 &	3.951 &	9.269 &	3.965 &	9.265 &	3.963 &	9.96 &	4.217	\\	\cmidrule(lr){3-4} \cmidrule(lr){5-6} \cmidrule(lr){7-8} \cmidrule(lr){9-10} \cmidrule(lr){11-12} \cmidrule(lr){13-14}
&	&	$0.5$ &	All &	5.913 &	3.336 &	26.174 &	7.492 &	26.705 &	7.658 &	26.7 &	7.656 &	29.105 &	10.262	\\	
&	&	&	Local &	7.057 &	5.516 &	31.172 &	13.033 &	31.721 &	13.108 &	31.716 &	13.105 &	34.14 &	14.662	\\	\cmidrule(lr){3-4} \cmidrule(lr){5-6} \cmidrule(lr){7-8} \cmidrule(lr){9-10} \cmidrule(lr){11-12} \cmidrule(lr){13-14}
&	&	$1$ &	All &	10.176 &	21.525 &	49.302 &	14.104 &	50.48 &	14.439 &	50.474 &	14.437 &	57.388 &	38.512	\\	
&	&	&	Local &	8.854 &	6.464 &	59.032 &	26.021 &	60.274 &	26.311 &	60.268 &	26.308 &	67.403 &	44.782	\\	\bottomrule
\end{tabular}
}
\end{table}

\begin{table}[h!t!b!p!]
\caption{\ref{f:three} Loading estimation errors of Trunc, iPE, RTFA and PreAve measured as in~\eqref{eq:err:loading:tensor} for each mode scaled by $100$, over varying $n \in \{100, 200, 500\}$, the distributions for $\mc F_t$ and $\bm\xi_t$ (Gaussian and $t_3$) and the percentages of outliers in the idiosyncratic component under~\ref{o:one} ($\varrho \in \{0.1, 0.5, 1\}$).
We report the mean and the standard deviation over $100$ realisations for each setting.}
\label{tab:tensor:le:idio:three}
\centering
% \resizebox{\textwidth}{!}
{\scriptsize
\begin{tabular}{rrrr cc cc cc cc}
\toprule
&	&	&	&	\multicolumn{2}{c}{Trunc} &		\multicolumn{2}{c}{iPE} &		\multicolumn{2}{c}{RTFA} &		\multicolumn{2}{c}{PreAve} 		\\	
$n$ &	Dist &	\% &	Mode &	Mean &	SD &	Mean &	SD &	Mean &	SD &	Mean &	SD	\\	\cmidrule(lr){1-4} \cmidrule(lr){5-6} \cmidrule(lr){7-8} \cmidrule(lr){9-10} \cmidrule(lr){11-12}
$100$ &	Gaussian &	$0.1$ &	1 &	0.557 &	0.094 &	0.632 &	0.11 &	0.632 &	0.11 &	1.537 &	0.312	\\	
&	&	&	2 &	0.689 &	0.093 &	0.78 &	0.103 &	0.78 &	0.104 &	1.86 &	0.291	\\	
&	&	&	3 &	0.797 &	0.093 &	0.904 &	0.103 &	0.904 &	0.103 &	2.011 &	0.265	\\	\cmidrule(lr){3-4} \cmidrule(lr){5-6} \cmidrule(lr){7-8} \cmidrule(lr){9-10} \cmidrule(lr){11-12}
&	&	$0.5$ &	1 &	0.602 &	0.093 &	0.902 &	0.139 &	0.901 &	0.138 &	2.2 &	0.430	\\	
&	&	&	2 &	0.748 &	0.093 &	1.122 &	0.147 &	1.122 &	0.147 &	2.655 &	0.380	\\	
&	&	&	3 &	0.858 &	0.099 &	1.284 &	0.148 &	1.283 &	0.148 &	2.944 &	0.396	\\	\cmidrule(lr){3-4} \cmidrule(lr){5-6} \cmidrule(lr){7-8} \cmidrule(lr){9-10} \cmidrule(lr){11-12}
&	&	$1$ &	1 &	0.656 &	0.103 &	1.172 &	0.182 &	1.171 &	0.183 &	2.835 &	0.533	\\	
&	&	&	2 &	0.808 &	0.097 &	1.442 &	0.173 &	1.441 &	0.173 &	3.449 &	0.488	\\	
&	&	&	3 &	0.925 &	0.096 &	1.664 &	0.175 &	1.663 &	0.175 &	3.746 &	0.591	\\	\cmidrule(lr){2-4} \cmidrule(lr){5-6} \cmidrule(lr){7-8} \cmidrule(lr){9-10} \cmidrule(lr){11-12}
&	$t_3$ &	$0.1$ &	1 &	0.595 &	0.095 &	0.779 &	0.128 &	0.774 &	0.124 &	1.947 &	0.533	\\	
&	&	&	2 &	0.741 &	0.112 &	0.988 &	0.148 &	0.981 &	0.145 &	2.877 &	5.488	\\	
&	&	&	3 &	0.821 &	0.099 &	1.101 &	0.132 &	1.096 &	0.131 &	2.562 &	0.431	\\	\cmidrule(lr){3-4} \cmidrule(lr){5-6} \cmidrule(lr){7-8} \cmidrule(lr){9-10} \cmidrule(lr){11-12}
&	&	$0.5$ &	1 &	0.723 &	0.158 &	1.373 &	0.201 &	1.37 &	0.2 &	3.423 &	0.688	\\	
&	&	&	2 &	0.873 &	0.181 &	1.659 &	0.248 &	1.656 &	0.247 &	4.603 &	5.479	\\	
&	&	&	3 &	0.969 &	0.159 &	1.904 &	0.221 &	1.9 &	0.22 &	4.507 &	0.730	\\	\cmidrule(lr){3-4} \cmidrule(lr){5-6} \cmidrule(lr){7-8} \cmidrule(lr){9-10} \cmidrule(lr){11-12}
&	&	$1$ &	1 &	0.828 &	0.236 &	1.849 &	0.309 &	1.847 &	0.308 &	5.215 &	5.174	\\	
&	&	&	2 &	0.999 &	0.247 &	2.25 &	0.3 &	2.247 &	0.299 &	6.17 &	5.254	\\	
&	&	&	3 &	1.098 &	0.229 &	2.572 &	0.319 &	2.57 &	0.319 &	6.792 &	5.085	\\	\cmidrule(lr){1-4} \cmidrule(lr){5-6} \cmidrule(lr){7-8} \cmidrule(lr){9-10} \cmidrule(lr){11-12}
$200$ &	Gaussian &	$0.1$ &	1 &	0.398 &	0.067 &	0.446 &	0.07 &	0.446 &	0.07 &	1.068 &	0.175	\\	
&	&	&	2 &	0.493 &	0.06 &	0.558 &	0.061 &	0.558 &	0.062 &	1.284 &	0.195	\\	
&	&	&	3 &	0.578 &	0.071 &	0.657 &	0.078 &	0.656 &	0.078 &	1.481 &	0.237	\\	\cmidrule(lr){3-4} \cmidrule(lr){5-6} \cmidrule(lr){7-8} \cmidrule(lr){9-10} \cmidrule(lr){11-12}
&	&	$0.5$ &	1 &	0.44 &	0.072 &	0.659 &	0.096 &	0.659 &	0.096 &	1.544 &	0.249	\\	
&	&	&	2 &	0.535 &	0.07 &	0.806 &	0.105 &	0.806 &	0.106 &	1.868 &	0.263	\\	
&	&	&	3 &	0.633 &	0.068 &	0.957 &	0.104 &	0.957 &	0.104 &	2.164 &	0.299	\\	\cmidrule(lr){3-4} \cmidrule(lr){5-6} \cmidrule(lr){7-8} \cmidrule(lr){9-10} \cmidrule(lr){11-12}
&	&	$1$ &	1 &	0.48 &	0.074 &	0.853 &	0.126 &	0.852 &	0.126 &	2.023 &	0.335	\\	
&	&	&	2 &	0.587 &	0.068 &	1.062 &	0.131 &	1.062 &	0.13 &	2.375 &	0.347	\\	
&	&	&	3 &	0.683 &	0.069 &	1.223 &	0.124 &	1.222 &	0.124 &	2.77 &	0.404	\\	\cmidrule(lr){2-4} \cmidrule(lr){5-6} \cmidrule(lr){7-8} \cmidrule(lr){9-10} \cmidrule(lr){11-12}
&	$t_3$ &	$0.1$ &	1 &	0.437 &	0.081 &	0.605 &	0.092 &	0.603 &	0.092 &	1.447 &	0.405	\\	
&	&	&	2 &	0.531 &	0.099 &	0.739 &	0.112 &	0.736 &	0.112 &	1.719 &	0.295	\\	
&	&	&	3 &	0.606 &	0.098 &	0.854 &	0.13 &	0.852 &	0.129 &	1.977 &	0.337	\\	\cmidrule(lr){3-4} \cmidrule(lr){5-6} \cmidrule(lr){7-8} \cmidrule(lr){9-10} \cmidrule(lr){11-12}
&	&	$0.5$ &	1 &	0.563 &	0.226 &	1.104 &	0.238 &	1.102 &	0.238 &	2.682 &	0.819	\\	
&	&	&	2 &	0.672 &	0.227 &	1.336 &	0.234 &	1.334 &	0.234 &	3.192 &	0.629	\\	
&	&	&	3 &	0.759 &	0.209 &	1.554 &	0.229 &	1.552 &	0.227 &	3.63 &	0.661	\\	\cmidrule(lr){3-4} \cmidrule(lr){5-6} \cmidrule(lr){7-8} \cmidrule(lr){9-10} \cmidrule(lr){11-12}
&	&	$1$ &	1 &	0.673 &	0.297 &	1.525 &	0.347 &	1.523 &	0.346 &	3.842 &	1.160	\\	
&	&	&	2 &	0.791 &	0.275 &	1.859 &	0.258 &	1.858 &	0.257 &	4.585 &	1.024	\\	
&	&	&	3 &	0.892 &	0.272 &	2.133 &	0.291 &	2.131 &	0.291 &	5.083 &	1.042	\\	\cmidrule(lr){1-4} \cmidrule(lr){5-6} \cmidrule(lr){7-8} \cmidrule(lr){9-10} \cmidrule(lr){11-12}
$500$ &	Gaussian &	$0.1$ &	1 &	0.253 &	0.033 &	0.29 &	0.04 &	0.29 &	0.04 &	0.666 &	0.101	\\	
&	&	&	2 &	0.315 &	0.04 &	0.358 &	0.045 &	0.358 &	0.044 &	0.816 &	0.116	\\	
&	&	&	3 &	0.371 &	0.043 &	0.42 &	0.048 &	0.419 &	0.048 &	0.944 &	0.125	\\	\cmidrule(lr){3-4} \cmidrule(lr){5-6} \cmidrule(lr){7-8} \cmidrule(lr){9-10} \cmidrule(lr){11-12}
&	&	$0.5$ &	1 &	0.275 &	0.033 &	0.416 &	0.053 &	0.416 &	0.052 &	0.976 &	0.143	\\	
&	&	&	2 &	0.345 &	0.041 &	0.524 &	0.06 &	0.524 &	0.06 &	1.191 &	0.143	\\	
&	&	&	3 &	0.407 &	0.045 &	0.611 &	0.064 &	0.611 &	0.064 &	1.359 &	0.174	\\	\cmidrule(lr){3-4} \cmidrule(lr){5-6} \cmidrule(lr){7-8} \cmidrule(lr){9-10} \cmidrule(lr){11-12}
&	&	$1$ &	1 &	0.3 &	0.038 &	0.545 &	0.068 &	0.545 &	0.068 &	1.256 &	0.182	\\	
&	&	&	2 &	0.379 &	0.049 &	0.683 &	0.09 &	0.682 &	0.09 &	1.545 &	0.212	\\	
&	&	&	3 &	0.441 &	0.045 &	0.789 &	0.083 &	0.788 &	0.083 &	1.766 &	0.203	\\	\cmidrule(lr){2-4} \cmidrule(lr){5-6} \cmidrule(lr){7-8} \cmidrule(lr){9-10} \cmidrule(lr){11-12}
&	$t_3$ &	$0.1$ &	1 &	0.285 &	0.056 &	0.427 &	0.059 &	0.425 &	0.059 &	1.01 &	0.182	\\	
&	&	&	2 &	0.337 &	0.056 &	0.513 &	0.059 &	0.511 &	0.059 &	1.163 &	0.167	\\	
&	&	&	3 &	0.396 &	0.065 &	0.604 &	0.064 &	0.601 &	0.064 &	1.327 &	0.195	\\	\cmidrule(lr){3-4} \cmidrule(lr){5-6} \cmidrule(lr){7-8} \cmidrule(lr){9-10} \cmidrule(lr){11-12}
&	&	$0.5$ &	1 &	0.381 &	0.109 &	0.819 &	0.123 &	0.818 &	0.122 &	1.968 &	0.310	\\	
&	&	&	2 &	0.456 &	0.14 &	1.012 &	0.129 &	1.011 &	0.128 &	2.334 &	0.400	\\	
&	&	&	3 &	0.525 &	0.165 &	1.162 &	0.176 &	1.16 &	0.176 &	2.561 &	0.382	\\	\cmidrule(lr){3-4} \cmidrule(lr){5-6} \cmidrule(lr){7-8} \cmidrule(lr){9-10} \cmidrule(lr){11-12}
&	&	$1$ &	1 &	0.468 &	0.169 &	1.158 &	0.193 &	1.157 &	0.193 &	2.776 &	0.535	\\	
&	&	&	2 &	0.552 &	0.213 &	1.419 &	0.181 &	1.418 &	0.181 &	3.288 &	0.523	\\	
&	&	&	3 &	0.63 &	0.234 &	1.623 &	0.246 &	1.622 &	0.246 &	3.687 &	0.618	\\	\bottomrule
\end{tabular}
}
\end{table}

\begin{table}[h!t!b!p!]
\caption{\ref{f:three} Common component estimation errors of Trunc, noTrunc, iPE, RTFA and PreAve measured as in~\eqref{eq:err:chi:tensor} with $\mc T = [n]$ (`all') and $\mc T = \{n - 10 + 1, \ldots, n\}$ (`local') scaled by $1000$, over varying $n \in \{100, 200, 500\}$, the distributions for $\mc F_t$ and $\bm\xi_t$ (Gaussian and $t_3$) and the percentages of outliers in the idiosyncratic component under~\ref{o:one} ($\varrho \in \{0.1, 0.5, 1\}$).
We report the mean and the standard deviation over $100$ realisations for each setting.}
\label{tab:tensor:ce:idio:three}
\centering
\resizebox{\textwidth}{!}
{\scriptsize
\begin{tabular}{rrrr cc cc cc cc cc}
\toprule
&	&	&	&	\multicolumn{2}{c}{Trunc} &		\multicolumn{2}{c}{noTrunc} &		\multicolumn{2}{c}{iPE} &		\multicolumn{2}{c}{RTFA} &		\multicolumn{2}{c}{PreAve} 		\\	
$n$ &	Dist &	\% &	Range &	Mean &	SD &	Mean &	SD &	Mean &	SD &	Mean &	SD &	Mean &	SD	\\	\cmidrule(lr){1-4} \cmidrule(lr){5-6} \cmidrule(lr){7-8} \cmidrule(lr){9-10} \cmidrule(lr){11-12} \cmidrule(lr){13-14}
$100$ &	Gaussian &	$0.1$ &	All &	1.351 &	0.254 &	1.766 &	0.32 &	1.803 &	0.326 &	1.803 &	0.326 &	2.533 &	0.476	\\	
&	&	&	Local &	1.354 &	0.3 &	1.772 &	0.395 &	1.809 &	0.4 &	1.808 &	0.4 &	2.541 &	0.537	\\	\cmidrule(lr){3-4} \cmidrule(lr){5-6} \cmidrule(lr){7-8} \cmidrule(lr){9-10} \cmidrule(lr){11-12} \cmidrule(lr){13-14}
&	&	$0.5$ &	All &	1.607 &	0.277 &	3.741 &	0.627 &	3.929 &	0.66 &	3.929 &	0.66 &	5.465 &	0.967	\\	
&	&	&	Local &	1.621 &	0.32 &	3.749 &	0.763 &	3.937 &	0.788 &	3.937 &	0.788 &	5.468 &	1.078	\\	\cmidrule(lr){3-4} \cmidrule(lr){5-6} \cmidrule(lr){7-8} \cmidrule(lr){9-10} \cmidrule(lr){11-12} \cmidrule(lr){13-14}
&	&	$1$ &	All &	1.938 &	0.297 &	6.204 &	0.968 &	6.594 &	1.036 &	6.593 &	1.036 &	9.135 &	1.563	\\	
&	&	&	Local &	1.948 &	0.352 &	6.188 &	1.246 &	6.578 &	1.298 &	6.578 &	1.298 &	9.126 &	1.810	\\	\cmidrule(lr){2-4} \cmidrule(lr){5-6} \cmidrule(lr){7-8} \cmidrule(lr){9-10} \cmidrule(lr){11-12} \cmidrule(lr){13-14}
&	$t_3$ &	$0.1$ &	All &	1.408 &	0.27 &	2.628 &	0.559 &	2.738 &	0.589 &	2.734 &	0.585 &	6.104 &	22.733	\\	
&	&	&	Local &	1.549 &	0.59 &	2.865 &	0.946 &	2.975 &	0.96 &	2.972 &	0.96 &	6.371 &	22.330	\\	\cmidrule(lr){3-4} \cmidrule(lr){5-6} \cmidrule(lr){7-8} \cmidrule(lr){9-10} \cmidrule(lr){11-12} \cmidrule(lr){13-14}
&	&	$0.5$ &	All &	2.022 &	0.577 &	7.957 &	1.554 &	8.491 &	1.683 &	8.488 &	1.682 &	14.171 &	22.377	\\	
&	&	&	Local &	2.24 &	1.306 &	8.715 &	3.145 &	9.262 &	3.196 &	9.259 &	3.195 &	15.064 &	21.899	\\	\cmidrule(lr){3-4} \cmidrule(lr){5-6} \cmidrule(lr){7-8} \cmidrule(lr){9-10} \cmidrule(lr){11-12} \cmidrule(lr){13-14}
&	&	$1$ &	All &	2.741 &	1.007 &	14.682 &	2.986 &	15.767 &	3.257 &	15.764 &	3.256 &	29.664 &	70.941	\\	
&	&	&	Local &	3.025 &	1.996 &	16.476 &	5.812 &	17.614 &	5.954 &	17.611 &	5.954 &	28.85 &	41.863	\\	\cmidrule(lr){1-4} \cmidrule(lr){5-6} \cmidrule(lr){7-8} \cmidrule(lr){9-10} \cmidrule(lr){11-12} \cmidrule(lr){13-14}
$200$ &	Gaussian &	$0.1$ &	All &	1.342 &	0.273 &	1.788 &	0.342 &	1.807 &	0.345 &	1.807 &	0.344 &	2.184 &	0.434	\\	
&	&	&	Local &	1.357 &	0.343 &	1.818 &	0.446 &	1.837 &	0.449 &	1.837 &	0.449 &	2.217 &	0.528	\\	\cmidrule(lr){3-4} \cmidrule(lr){5-6} \cmidrule(lr){7-8} \cmidrule(lr){9-10} \cmidrule(lr){11-12} \cmidrule(lr){13-14}
&	&	$0.5$ &	All &	1.595 &	0.284 &	3.868 &	0.656 &	3.971 &	0.674 &	3.971 &	0.674 &	4.765 &	0.841	\\	
&	&	&	Local &	1.602 &	0.363 &	3.898 &	0.848 &	4.001 &	0.862 &	4 &	0.862 &	4.8 &	1.006	\\	\cmidrule(lr){3-4} \cmidrule(lr){5-6} \cmidrule(lr){7-8} \cmidrule(lr){9-10} \cmidrule(lr){11-12} \cmidrule(lr){13-14}
&	&	$1$ &	All &	1.937 &	0.307 &	6.538 &	1.04 &	6.749 &	1.075 &	6.749 &	1.075 &	8.05 &	1.343	\\	
&	&	&	Local &	1.947 &	0.394 &	6.617 &	1.354 &	6.828 &	1.378 &	6.827 &	1.378 &	8.136 &	1.602	\\	\cmidrule(lr){2-4} \cmidrule(lr){5-6} \cmidrule(lr){7-8} \cmidrule(lr){9-10} \cmidrule(lr){11-12} \cmidrule(lr){13-14}
&	$t_3$ &	$0.1$ &	All &	1.489 &	0.36 &	3.103 &	0.6 &	3.174 &	0.617 &	3.173 &	0.617 &	3.81 &	0.741	\\	
&	&	&	Local &	1.733 &	0.996 &	3.534 &	1.462 &	3.608 &	1.462 &	3.607 &	1.462 &	4.275 &	1.657	\\	\cmidrule(lr){3-4} \cmidrule(lr){5-6} \cmidrule(lr){7-8} \cmidrule(lr){9-10} \cmidrule(lr){11-12} \cmidrule(lr){13-14}
&	&	$0.5$ &	All &	2.398 &	1.241 &	10.437 &	2.149 &	10.795 &	2.216 &	10.794 &	2.215 &	12.995 &	2.735	\\	
&	&	&	Local &	2.942 &	3.15 &	11.998 &	5.899 &	12.374 &	5.945 &	12.372 &	5.944 &	14.677 &	6.936	\\	\cmidrule(lr){3-4} \cmidrule(lr){5-6} \cmidrule(lr){7-8} \cmidrule(lr){9-10} \cmidrule(lr){11-12} \cmidrule(lr){13-14}
&	&	$1$ &	All &	3.388 &	2.15 &	19.624 &	3.984 &	20.379 &	4.123 &	20.378 &	4.122 &	24.875 &	5.427	\\	
&	&	&	Local &	4.144 &	4.844 &	22.629 &	10.679 &	23.418 &	10.742 &	23.416 &	10.741 &	28.171 &	12.722	\\	\cmidrule(lr){1-4} \cmidrule(lr){5-6} \cmidrule(lr){7-8} \cmidrule(lr){9-10} \cmidrule(lr){11-12} \cmidrule(lr){13-14}
$500$ &	Gaussian &	$0.1$ &	All &	1.299 &	0.219 &	1.766 &	0.278 &	1.774 &	0.279 &	1.774 &	0.279 &	1.923 &	0.301	\\	
&	&	&	Local &	1.322 &	0.294 &	1.793 &	0.401 &	1.801 &	0.402 &	1.801 &	0.402 &	1.95 &	0.424	\\	\cmidrule(lr){3-4} \cmidrule(lr){5-6} \cmidrule(lr){7-8} \cmidrule(lr){9-10} \cmidrule(lr){11-12} \cmidrule(lr){13-14}
&	&	$0.5$ &	All &	1.571 &	0.231 &	3.976 &	0.542 &	4.019 &	0.548 &	4.019 &	0.548 &	4.333 &	0.586	\\	
&	&	&	Local &	1.591 &	0.306 &	4.022 &	0.73 &	4.065 &	0.734 &	4.064 &	0.734 &	4.38 &	0.766	\\	\cmidrule(lr){3-4} \cmidrule(lr){5-6} \cmidrule(lr){7-8} \cmidrule(lr){9-10} \cmidrule(lr){11-12} \cmidrule(lr){13-14}
&	&	$1$ &	All &	1.926 &	0.255 &	6.774 &	0.88 &	6.862 &	0.893 &	6.862 &	0.893 &	7.388 &	0.962	\\	
&	&	&	Local &	1.953 &	0.352 &	6.841 &	1.325 &	6.928 &	1.335 &	6.928 &	1.335 &	7.456 &	1.400	\\	\cmidrule(lr){2-4} \cmidrule(lr){5-6} \cmidrule(lr){7-8} \cmidrule(lr){9-10} \cmidrule(lr){11-12} \cmidrule(lr){13-14}
&	$t_3$ &	$0.1$ &	All &	1.541 &	0.443 &	3.894 &	0.696 &	3.935 &	0.703 &	3.935 &	0.702 &	4.232 &	0.751	\\	
&	&	&	Local &	1.74 &	0.844 &	4.378 &	1.624 &	4.42 &	1.628 &	4.419 &	1.628 &	4.722 &	1.659	\\	\cmidrule(lr){3-4} \cmidrule(lr){5-6} \cmidrule(lr){7-8} \cmidrule(lr){9-10} \cmidrule(lr){11-12} \cmidrule(lr){13-14}
&	&	$0.5$ &	All &	2.776 &	1.715 &	14.781 &	2.767 &	15.002 &	2.81 &	15.001 &	2.81 &	16.144 &	2.983	\\	
&	&	&	Local &	3.131 &	2.438 &	16.71 &	6.14 &	16.938 &	6.181 &	16.937 &	6.181 &	18.099 &	6.319	\\	\cmidrule(lr){3-4} \cmidrule(lr){5-6} \cmidrule(lr){7-8} \cmidrule(lr){9-10} \cmidrule(lr){11-12} \cmidrule(lr){13-14}
&	&	$1$ &	All &	4.107 &	3.145 &	28.423 &	5.344 &	28.894 &	5.426 &	28.893 &	5.426 &	31.221 &	5.884	\\	
&	&	&	Local &	4.518 &	4.241 &	31.646 &	11.523 &	32.147 &	11.617 &	32.145 &	11.617 &	34.526 &	11.974	\\	\bottomrule
\end{tabular}
}
\end{table}

\clearpage 

\paragraph{Outliers in the factors.}

\begin{figure}[h!t!p!]
\centering
\includegraphics[width = 1\textwidth]{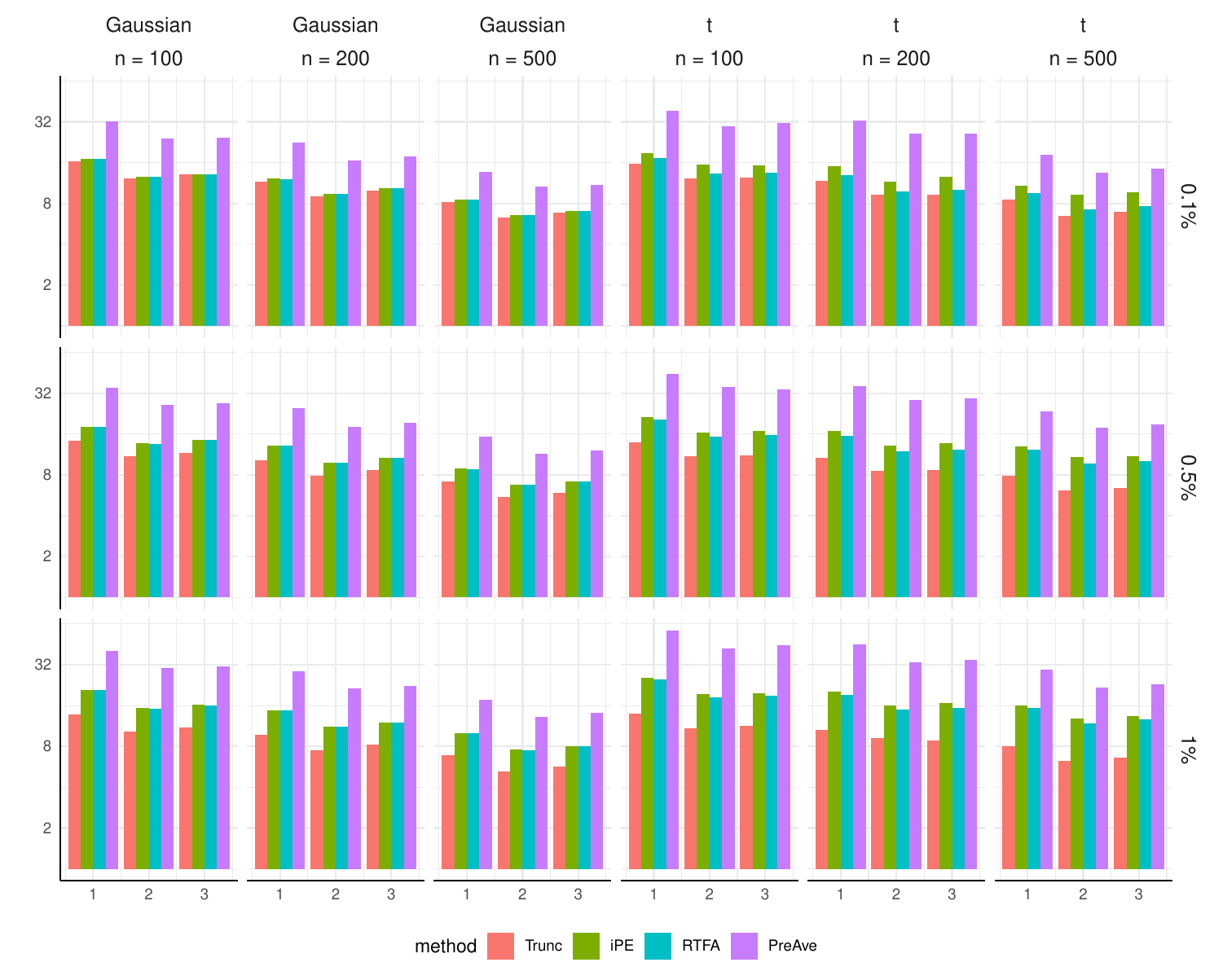}
\caption{\ref{f:one} Loading estimation errors measured as in~\eqref{eq:err:loading:tensor} for each mode ($x$-axis) for Trunc, iPE, RTFA and PreAve over varying $n \in \{100, 200, 500\}$, distributions for $\mc F_t$ and $\bm\xi_t$ (Gaussian and $t_3$) and the percentages of outliers in the factors under~\ref{o:two} ($\varrho \in \{0.1, 0.5, 1\}$, top to bottom), averaged over $100$ realisations per setting. In each plot, the $y$-axis is in the log-scale and all errors have been scaled for the ease of presentation.}
\label{fig:tensor:le:factor:f:one}
\end{figure}

\begin{figure}[h!t!p!]
\centering
\includegraphics[width = 1\textwidth]{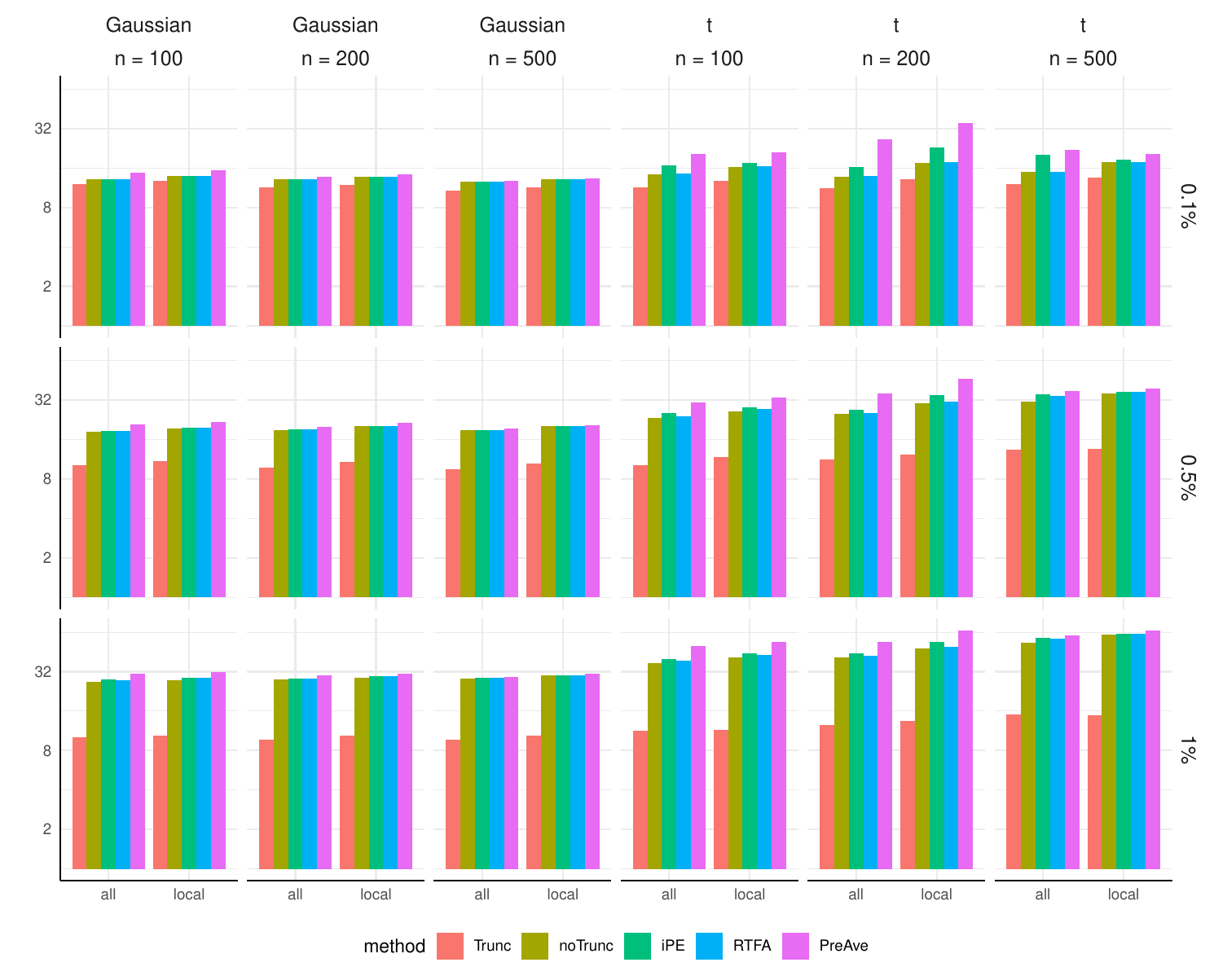}
\caption{\ref{f:one} Common component estimation errors measured as in~\eqref{eq:err:chi:tensor} with $\mc T = [n]$ (`all') and $\mc T = \{n - 10 + 1, \ldots, n \}$ (`local') for Trunc, noTrunc, iPE, RTFA and PreAve over varying $n \in \{100, 200, 500\}$, distributions for $\mc F_t$ and $\bm\xi_t$ (Gaussian and $t_3$) and the percentages of outliers in the factors under~\ref{o:two} ($\varrho \in \{0.1, 0.5, 1\}$, top to bottom), averaged over $100$ realisations per setting. In each plot, the $y$-axis is in the log-scale and all errors have been scaled for the ease of presentation.}
\label{fig:tensor:ce:factor:f:one}
\end{figure}

\begin{figure}[h!t!p!]
\centering
\includegraphics[width = 1\textwidth]{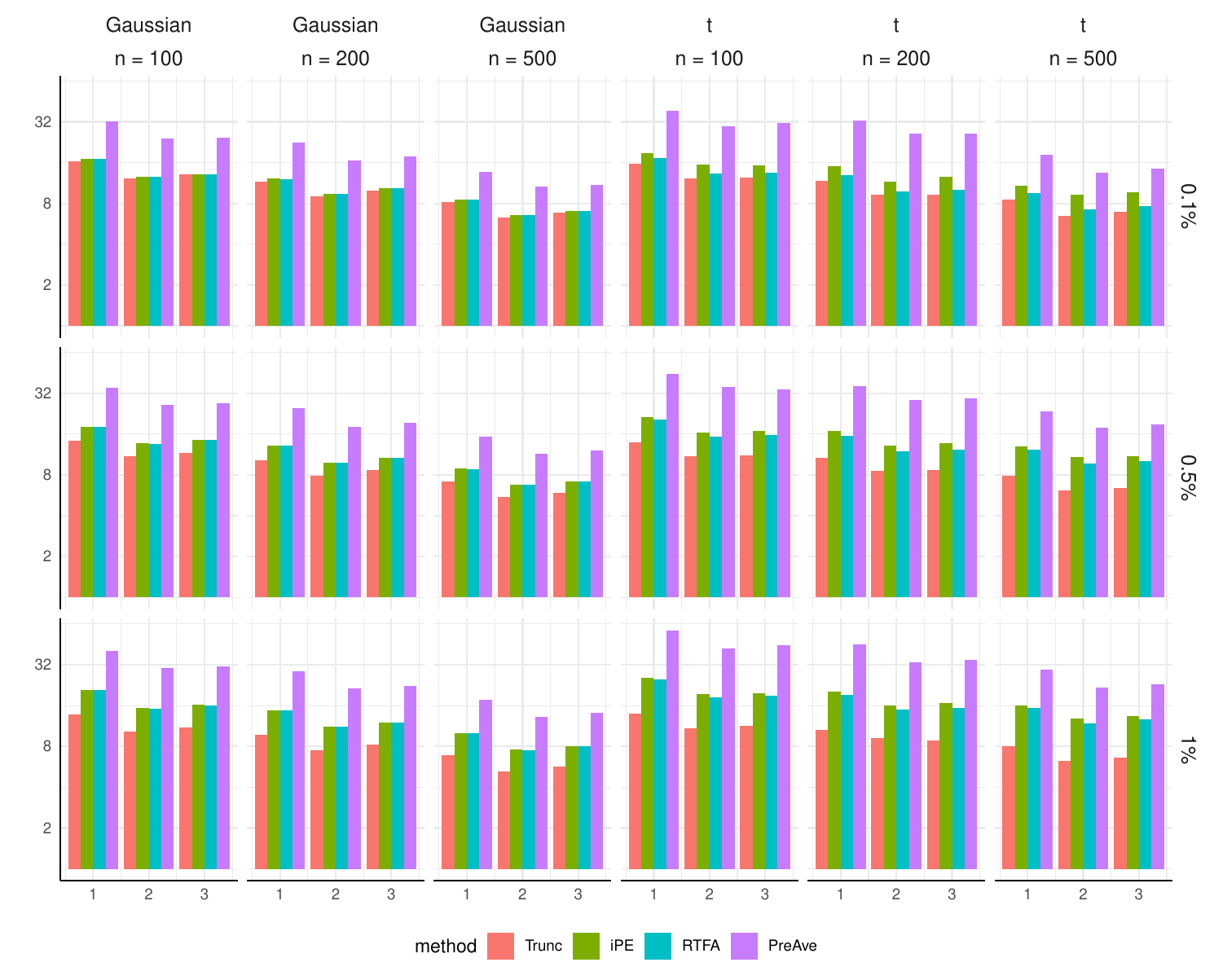}
\caption{\ref{f:two} Loading estimation errors measured as in~\eqref{eq:err:loading:tensor} for each mode ($x$-axis) for Trunc, iPE, RTFA and PreAve over varying $n \in \{100, 200, 500\}$, distributions for $\mc F_t$ and $\bm\xi_t$ (Gaussian and $t_3$) and the percentages of outliers in the factors under~\ref{o:two} ($\varrho \in \{0.1, 0.5, 1\}$, top to bottom), averaged over $100$ realisations per setting. In each plot, the $y$-axis is in the log-scale and all errors have been scaled for the ease of presentation.}
\label{fig:tensor:le:factor:f:two}
\end{figure}

\begin{figure}[h!t!p!]
\centering
\includegraphics[width = 1\textwidth]{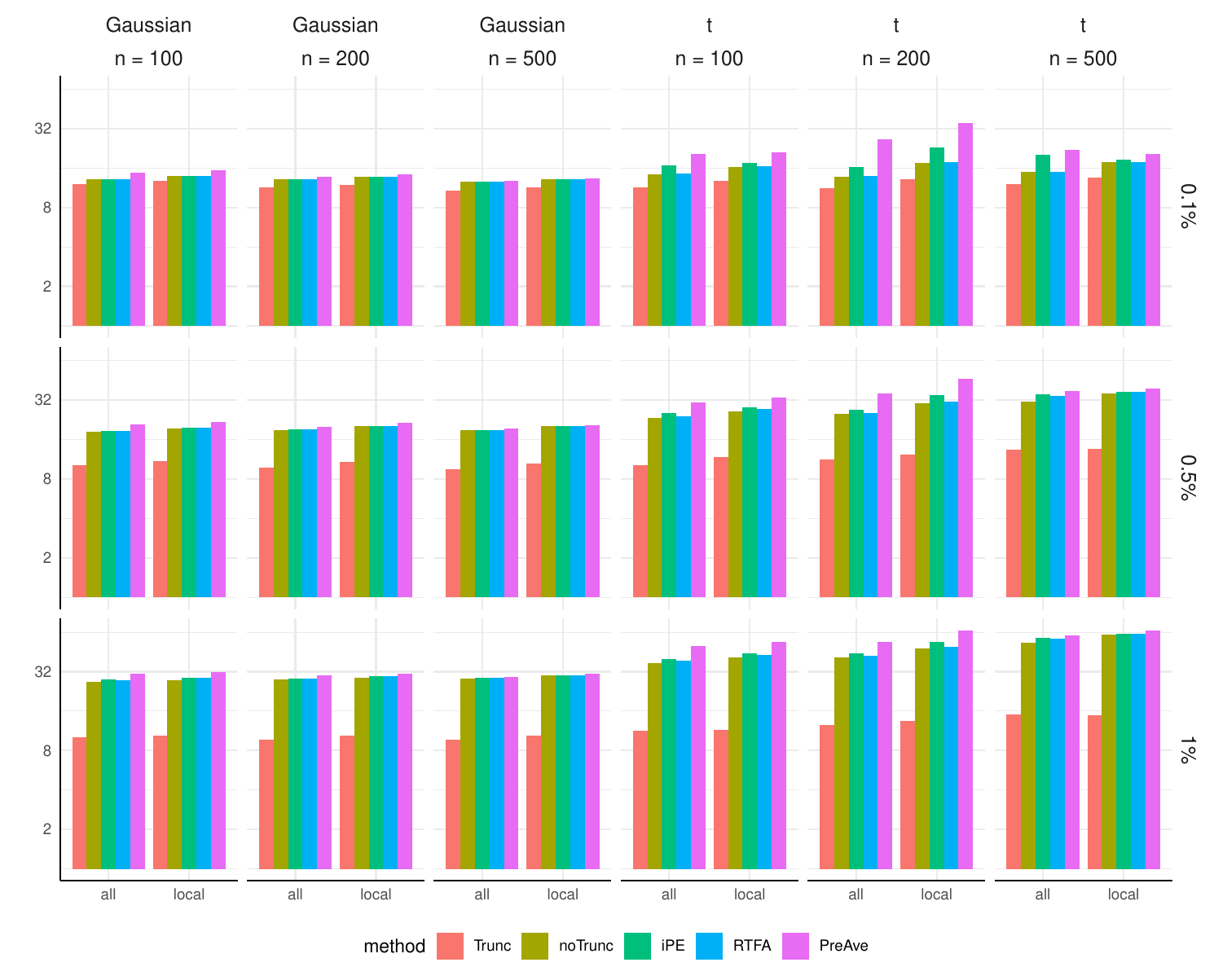}
\caption{\ref{f:two} Common component estimation errors measured as in~\eqref{eq:err:chi:tensor} with $\mc T = [n]$ (`all') and $\mc T = \{n - 10 + 1, \ldots, n \}$ (`local') for Trunc, noTrunc, iPE, RTFA and PreAve over varying $n \in \{100, 200, 500\}$, distributions for $\mc F_t$ and $\bm\xi_t$ (Gaussian and $t_3$) and the percentages of outliers in the factors under~\ref{o:two} ($\varrho \in \{0.1, 0.5, 1\}$, top to bottom), averaged over $100$ realisations per setting. In each plot, the $y$-axis is in the log-scale and all errors have been scaled for the ease of presentation.}
\label{fig:tensor:ce:factor:f:two}
\end{figure}

\begin{figure}[h!t!p!]
\centering
\includegraphics[width = 1\textwidth]{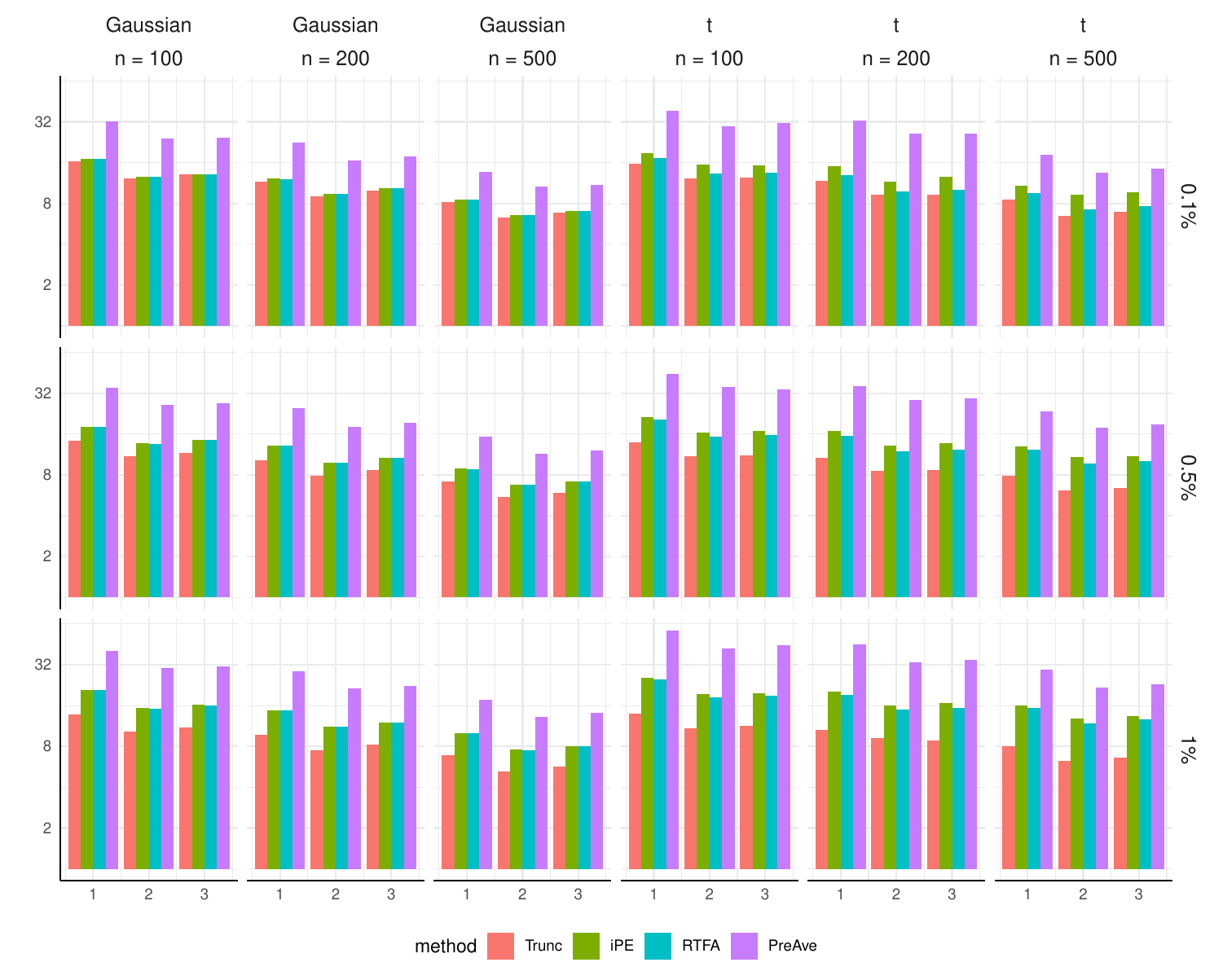}
\caption{\ref{f:three} Loading estimation errors measured as in~\eqref{eq:err:loading:tensor} for each mode ($x$-axis) for Trunc, iPE, RTFA and PreAve over varying $n \in \{100, 200, 500\}$, distributions for $\mc F_t$ and $\bm\xi_t$ (Gaussian and $t_3$) and the percentages of outliers in the factors under~\ref{o:two} ($\varrho \in \{0.1, 0.5, 1\}$, top to bottom), averaged over $100$ realisations per setting. In each plot, the $y$-axis is in the log-scale and all errors have been scaled for the ease of presentation.}
\label{fig:tensor:le:factor:f:three}
\end{figure}

\begin{figure}[h!t!p!]
\centering
\includegraphics[width = 1\textwidth]{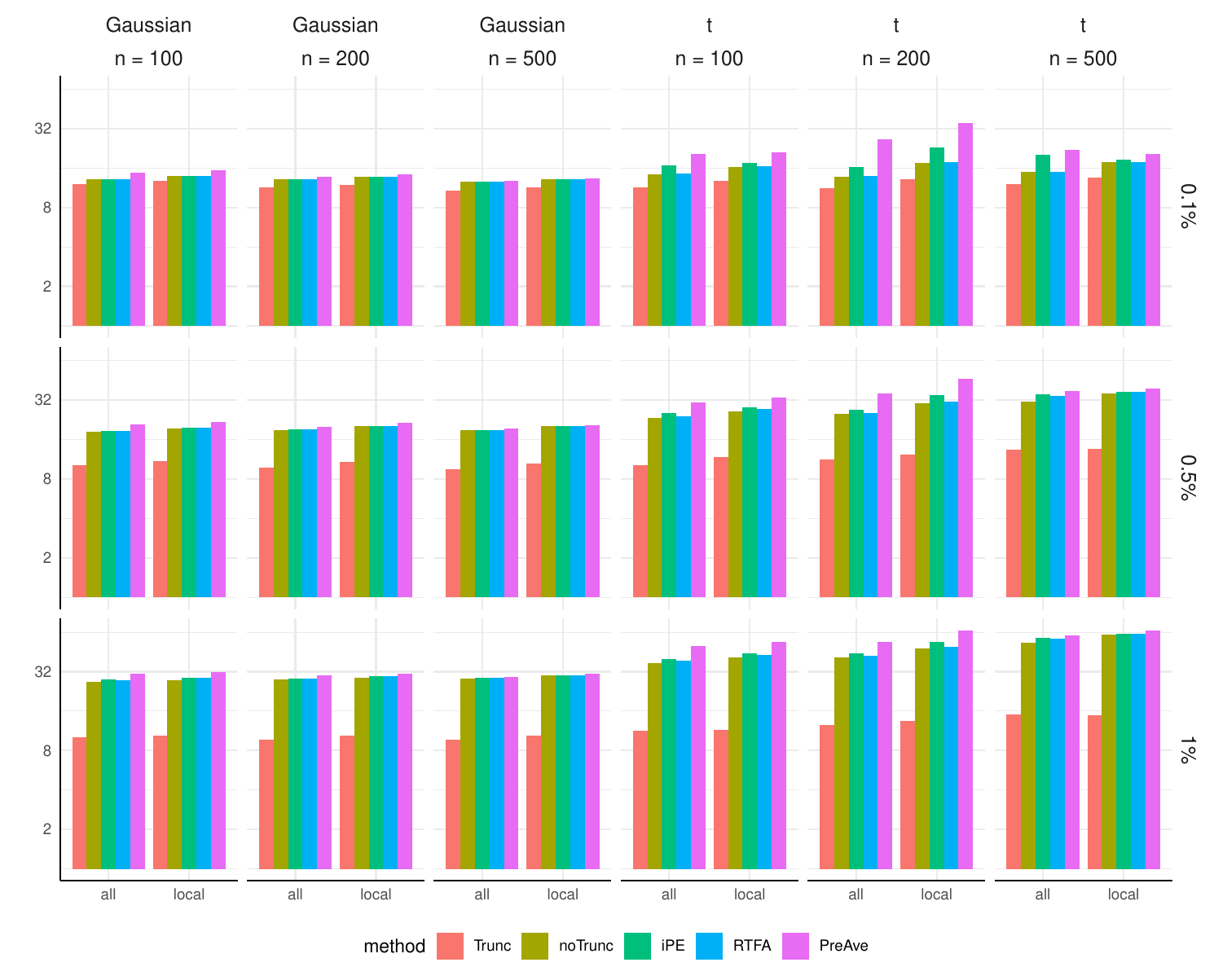}
\caption{\ref{f:three} Common component estimation errors measured as in~\eqref{eq:err:chi:tensor} with $\mc T = [n]$ (`all') and $\mc T = \{n - 10 + 1, \ldots, n \}$ (`local') for Trunc, noTrunc, iPE, RTFA and PreAve over varying $n \in \{100, 200, 500\}$, distributions for $\mc F_t$ and $\bm\xi_t$ (Gaussian and $t_3$) and the percentages of outliers ($\varrho \in \{0.1, 0.5, 1\}$, top to bottom), averaged over $100$ realisations per setting. In each plot, the $y$-axis is in the log-scale and all errors have been scaled for the ease of presentation.}
\label{fig:tensor:ce:factor:f:three}
\end{figure}

See Figures~\ref{fig:tensor:le:factor:f:one}--\ref{fig:tensor:ce:factor:f:three} and Tables~\ref{tab:tensor:le:factor:one}--\ref{tab:tensor:ce:factor:three} for the results from loading and common component estimation obtained under \ref{f:one}--\ref{f:three} with outliers in the factors under~\ref{o:two}.
In this scenario, we cannot recover $\mc F_t^\circ$ prior to the contamination by outliers since all cross-sections of the observed $\mc X_t$ are contaminated by the outliers. 
Nonetheless, we observe that the loadings and $\bm\chi_t$ (post-contamination) are well-estimated by Trunc.
In line with the theory (Theorem~\ref{thm:factor}~\ref{thm:factor:one}), the estimation error tends to increase with the increase in the proportion of outliers.

\begin{table}[h!t!b!p!]
\caption{\ref{f:one} Loading estimation errors of Trunc, iPE, RTFA and PreAve measured as in~\eqref{eq:err:loading:tensor} for each mode scaled by $100$, over varying $n \in \{100, 200, 500\}$, the distributions for $\mc F_t$ and $\bm\xi_t$ (Gaussian and $t_3$) and the percentages of outliers in the factors under~\ref{o:two} ($\varrho \in \{0.1, 0.5, 1\}$).
We report the mean and the standard deviation over $100$ realisations for each setting.}
\label{tab:tensor:le:factor:one}
\centering
% \resizebox{\textwidth}{!}
{\scriptsize
\begin{tabular}{rrrr cc cc cc cc}
\toprule
&	&	&	&	\multicolumn{2}{c}{Trunc} &		\multicolumn{2}{c}{iPE} &		\multicolumn{2}{c}{RTFA} &		\multicolumn{2}{c}{PreAve} 		\\	
$n$ &	Dist &	\% &	Mode &	Mean &	SD &	Mean &	SD &	Mean &	SD &	Mean &	SD	\\	\cmidrule(lr){1-4} \cmidrule(lr){5-6} \cmidrule(lr){7-8} \cmidrule(lr){9-10} \cmidrule(lr){11-12}
$100$ &	Gaussian &	$0.1$ &	1 &	2.415 &	1.876 &	2.278 &	1.034 &	2.279 &	1.037 &	3.875 &	1.545	\\	
&	&	&	2 &	2.326 &	1.32 &	2.238 &	1.08 &	2.238 &	1.079 &	3.792 &	1.253	\\	
&	&	&	3 &	2.552 &	3.733 &	2.198 &	0.966 &	2.197 &	0.964 &	3.764 &	1.434	\\	\cmidrule(lr){3-4} \cmidrule(lr){5-6} \cmidrule(lr){7-8} \cmidrule(lr){9-10} \cmidrule(lr){11-12}
&	&	$0.5$ &	1 &	1.507 &	0.651 &	1.471 &	0.543 &	1.471 &	0.545 &	2.692 &	0.977	\\	
&	&	&	2 &	1.555 &	0.775 &	1.523 &	0.72 &	1.523 &	0.719 &	2.605 &	0.933	\\	
&	&	&	3 &	1.493 &	0.642 &	1.471 &	0.587 &	1.471 &	0.587 &	2.584 &	0.845	\\	\cmidrule(lr){3-4} \cmidrule(lr){5-6} \cmidrule(lr){7-8} \cmidrule(lr){9-10} \cmidrule(lr){11-12}
&	&	$1$ &	1 &	1.145 &	0.431 &	1.14 &	0.43 &	1.14 &	0.433 &	2.073 &	0.741	\\	
&	&	&	2 &	1.121 &	0.414 &	1.116 &	0.412 &	1.116 &	0.412 &	1.889 &	0.572	\\	
&	&	&	3 &	1.121 &	0.502 &	1.119 &	0.501 &	1.119 &	0.501 &	2.05 &	0.750	\\	\cmidrule(lr){2-4} \cmidrule(lr){5-6} \cmidrule(lr){7-8} \cmidrule(lr){9-10} \cmidrule(lr){11-12}
&	$t_3$ &	$0.1$ &	1 &	2.075 &	0.797 &	2.416 &	3.66 &	2.01 &	0.782 &	4.159 &	4.215	\\	
&	&	&	2 &	2.216 &	0.966 &	2.729 &	5.495 &	2.127 &	0.924 &	4.39 &	5.347	\\	
&	&	&	3 &	2.209 &	0.847 &	2.606 &	4.508 &	2.101 &	0.727 &	4.541 &	5.134	\\	\cmidrule(lr){3-4} \cmidrule(lr){5-6} \cmidrule(lr){7-8} \cmidrule(lr){9-10} \cmidrule(lr){11-12}
&	&	$0.5$ &	1 &	1.195 &	0.448 &	1.194 &	0.425 &	1.166 &	0.4 &	2.442 &	1.025	\\	
&	&	&	2 &	1.295 &	0.472 &	1.312 &	0.549 &	1.263 &	0.462 &	2.586 &	1.586	\\	
&	&	&	3 &	1.234 &	0.398 &	1.248 &	0.427 &	1.203 &	0.358 &	2.578 &	1.561	\\	\cmidrule(lr){3-4} \cmidrule(lr){5-6} \cmidrule(lr){7-8} \cmidrule(lr){9-10} \cmidrule(lr){11-12}
&	&	$1$ &	1 &	0.851 &	0.23 &	0.862 &	0.25 &	0.842 &	0.229 &	1.753 &	0.673	\\	
&	&	&	2 &	0.884 &	0.284 &	0.914 &	0.383 &	0.88 &	0.289 &	1.786 &	0.747	\\	
&	&	&	3 &	0.885 &	0.298 &	0.896 &	0.309 &	0.874 &	0.29 &	1.802 &	0.687	\\	\cmidrule(lr){1-4} \cmidrule(lr){5-6} \cmidrule(lr){7-8} \cmidrule(lr){9-10} \cmidrule(lr){11-12}
$200$ &	Gaussian &	$0.1$ &	1 &	1.585 &	0.694 &	1.564 &	0.678 &	1.565 &	0.679 &	2.509 &	0.891	\\	
&	&	&	2 &	1.656 &	0.695 &	1.645 &	0.688 &	1.645 &	0.689 &	2.498 &	0.808	\\	
&	&	&	3 &	1.831 &	1.385 &	1.811 &	1.345 &	1.811 &	1.347 &	2.613 &	1.484	\\	\cmidrule(lr){3-4} \cmidrule(lr){5-6} \cmidrule(lr){7-8} \cmidrule(lr){9-10} \cmidrule(lr){11-12}
&	&	$0.5$ &	1 &	1.009 &	0.397 &	1.005 &	0.396 &	1.005 &	0.397 &	1.784 &	0.525	\\	
&	&	&	2 &	1.072 &	0.387 &	1.068 &	0.384 &	1.069 &	0.385 &	1.71 &	0.538	\\	
&	&	&	3 &	1.156 &	0.76 &	1.15 &	0.756 &	1.151 &	0.757 &	1.809 &	0.658	\\	\cmidrule(lr){3-4} \cmidrule(lr){5-6} \cmidrule(lr){7-8} \cmidrule(lr){9-10} \cmidrule(lr){11-12}
&	&	$1$ &	1 &	0.78 &	0.257 &	0.778 &	0.258 &	0.778 &	0.257 &	1.343 &	0.404	\\	
&	&	&	2 &	0.778 &	0.258 &	0.777 &	0.257 &	0.776 &	0.257 &	1.366 &	0.405	\\	
&	&	&	3 &	0.852 &	0.454 &	0.849 &	0.451 &	0.849 &	0.452 &	1.384 &	0.522	\\	\cmidrule(lr){2-4} \cmidrule(lr){5-6} \cmidrule(lr){7-8} \cmidrule(lr){9-10} \cmidrule(lr){11-12}
&	$t_3$ &	$0.1$ &	1 &	1.576 &	0.758 &	2.038 &	3.59 &	1.534 &	0.72 &	3.826 &	9.825	\\	
&	&	&	2 &	1.593 &	0.718 &	1.918 &	2.637 &	1.533 &	0.644 &	4.183 &	10.317	\\	
&	&	&	3 &	1.545 &	0.581 &	2.19 &	5.65 &	1.505 &	0.565 &	3.518 &	5.420	\\	\cmidrule(lr){3-4} \cmidrule(lr){5-6} \cmidrule(lr){7-8} \cmidrule(lr){9-10} \cmidrule(lr){11-12}
&	&	$0.5$ &	1 &	0.897 &	0.419 &	0.931 &	0.506 &	0.881 &	0.414 &	1.555 &	0.630	\\	
&	&	&	2 &	0.896 &	0.347 &	0.928 &	0.517 &	0.871 &	0.334 &	1.616 &	0.719	\\	
&	&	&	3 &	0.886 &	0.277 &	0.928 &	0.494 &	0.869 &	0.272 &	1.566 &	0.615	\\	\cmidrule(lr){3-4} \cmidrule(lr){5-6} \cmidrule(lr){7-8} \cmidrule(lr){9-10} \cmidrule(lr){11-12}
&	&	$1$ &	1 &	0.637 &	0.192 &	0.649 &	0.237 &	0.625 &	0.186 &	1.136 &	0.390	\\	
&	&	&	2 &	0.648 &	0.229 &	0.667 &	0.3 &	0.638 &	0.225 &	1.216 &	0.560	\\	
&	&	&	3 &	0.65 &	0.195 &	0.662 &	0.235 &	0.64 &	0.194 &	1.135 &	0.358	\\	\cmidrule(lr){1-4} \cmidrule(lr){5-6} \cmidrule(lr){7-8} \cmidrule(lr){9-10} \cmidrule(lr){11-12}
$500$ &	Gaussian &	$0.1$ &	1 &	1.251 &	0.655 &	1.245 &	0.647 &	1.245 &	0.648 &	1.558 &	0.568	\\	
&	&	&	2 &	1.188 &	0.546 &	1.182 &	0.536 &	1.181 &	0.536 &	1.579 &	0.512	\\	
&	&	&	3 &	1.285 &	0.743 &	1.279 &	0.74 &	1.279 &	0.741 &	1.576 &	0.467	\\	\cmidrule(lr){3-4} \cmidrule(lr){5-6} \cmidrule(lr){7-8} \cmidrule(lr){9-10} \cmidrule(lr){11-12}
&	&	$0.5$ &	1 &	0.771 &	0.325 &	0.767 &	0.323 &	0.767 &	0.323 &	1.073 &	0.330	\\	
&	&	&	2 &	0.753 &	0.277 &	0.751 &	0.276 &	0.75 &	0.276 &	1.099 &	0.347	\\	
&	&	&	3 &	0.767 &	0.373 &	0.766 &	0.373 &	0.766 &	0.373 &	1.066 &	0.295	\\	\cmidrule(lr){3-4} \cmidrule(lr){5-6} \cmidrule(lr){7-8} \cmidrule(lr){9-10} \cmidrule(lr){11-12}
&	&	$1$ &	1 &	0.544 &	0.224 &	0.543 &	0.223 &	0.542 &	0.223 &	0.852 &	0.274	\\	
&	&	&	2 &	0.544 &	0.202 &	0.543 &	0.202 &	0.543 &	0.202 &	0.848 &	0.268	\\	
&	&	&	3 &	0.581 &	0.256 &	0.579 &	0.256 &	0.579 &	0.256 &	0.838 &	0.242	\\	\cmidrule(lr){2-4} \cmidrule(lr){5-6} \cmidrule(lr){7-8} \cmidrule(lr){9-10} \cmidrule(lr){11-12}
&	$t_3$ &	$0.1$ &	1 &	1.22 &	0.709 &	1.58 &	3.408 &	1.191 &	0.669 &	2.197 &	6.414	\\	
&	&	&	2 &	1.13 &	0.625 &	1.713 &	5.478 &	1.104 &	0.605 &	2.012 &	5.513	\\	
&	&	&	3 &	1.23 &	0.875 &	1.768 &	5.127 &	1.195 &	0.835 &	2.15 &	5.979	\\	\cmidrule(lr){3-4} \cmidrule(lr){5-6} \cmidrule(lr){7-8} \cmidrule(lr){9-10} \cmidrule(lr){11-12}
&	&	$0.5$ &	1 &	0.667 &	0.299 &	0.944 &	2.692 &	0.655 &	0.29 &	1.415 &	4.459	\\	
&	&	&	2 &	0.607 &	0.278 &	1.168 &	5.422 &	0.599 &	0.277 &	1.456 &	5.517	\\	
&	&	&	3 &	0.651 &	0.376 &	1.049 &	3.83 &	0.64 &	0.355 &	1.48 &	5.144	\\	\cmidrule(lr){3-4} \cmidrule(lr){5-6} \cmidrule(lr){7-8} \cmidrule(lr){9-10} \cmidrule(lr){11-12}
&	&	$1$ &	1 &	0.45 &	0.174 &	0.655 &	2.008 &	0.442 &	0.17 &	0.732 &	0.233	\\	
&	&	&	2 &	0.425 &	0.149 &	0.946 &	5.059 &	0.422 &	0.143 &	1.057 &	3.531	\\	
&	&	&	3 &	0.441 &	0.229 &	0.707 &	2.561 &	0.436 &	0.22 &	0.73 &	0.305	\\	\bottomrule
\end{tabular}
}
\end{table}

\begin{table}[h!t!b!p!]
\caption{\ref{f:one} Common component estimation errors of Trunc, noTrunc, iPE, RTFA and PreAve measured as in~\eqref{eq:err:chi:tensor} with $\mc T = [n]$ (`all') and $\mc T = \{n - 10 + 1, \ldots, n\}$ (`local') scaled by $1000$, over varying $n \in \{100, 200, 500\}$, the distributions for $\mc F_t$ and $\bm\xi_t$ (Gaussian and $t_3$) and the percentages of outliers in the factors under~\ref{o:two} ($\varrho \in \{0.1, 0.5, 1\}$).
We report the mean and the standard deviation over $100$ realisations for each setting.}
\label{tab:tensor:ce:factor:one}
\centering
\resizebox{\textwidth}{!}
{\scriptsize
\begin{tabular}{rrrr cc cc cc cc cc}
\toprule
&	&	&	&	\multicolumn{2}{c}{Trunc} &		\multicolumn{2}{c}{noTrunc} &		\multicolumn{2}{c}{iPE} &		\multicolumn{2}{c}{RTFA} &		\multicolumn{2}{c}{PreAve} 		\\	
$n$ &	Dist &	\% &	Range &	Mean &	SD &	Mean &	SD &	Mean &	SD &	Mean &	SD &	Mean &	SD	\\	\cmidrule(lr){1-4} \cmidrule(lr){5-6} \cmidrule(lr){7-8} \cmidrule(lr){9-10} \cmidrule(lr){11-12} \cmidrule(lr){13-14}
$100$ &	Gaussian &	$0.1$ &	All &	30.938 &	40.311 &	27.567 &	10.331 &	27.09 &	8.464 &	27.089 &	8.464 &	29.024 &	9.080	\\	
&	&	&	Local &	33.127 &	35.111 &	30.493 &	13.661 &	29.976 &	11.932 &	29.976 &	11.932 &	31.914 &	12.370	\\	\cmidrule(lr){3-4} \cmidrule(lr){5-6} \cmidrule(lr){7-8} \cmidrule(lr){9-10} \cmidrule(lr){11-12} \cmidrule(lr){13-14}
&	&	$0.5$ &	All &	14.614 &	5.135 &	14.48 &	4.689 &	14.445 &	4.633 &	14.445 &	4.633 &	15.398 &	4.940	\\	
&	&	&	Local &	17.564 &	8.874 &	17.566 &	8.878 &	17.535 &	8.829 &	17.534 &	8.828 &	18.486 &	8.983	\\	\cmidrule(lr){3-4} \cmidrule(lr){5-6} \cmidrule(lr){7-8} \cmidrule(lr){9-10} \cmidrule(lr){11-12} \cmidrule(lr){13-14}
&	&	$1$ &	All &	9.032 &	2.763 &	9.027 &	2.763 &	9.024 &	2.763 &	9.024 &	2.763 &	9.6 &	2.940	\\	
&	&	&	Local &	10.613 &	5.993 &	10.612 &	5.996 &	10.61 &	5.995 &	10.61 &	5.995 &	11.2 &	6.085	\\	\cmidrule(lr){2-4} \cmidrule(lr){5-6} \cmidrule(lr){7-8} \cmidrule(lr){9-10} \cmidrule(lr){11-12} \cmidrule(lr){13-14}
&	$t_3$ &	$0.1$ &	All &	26.006 &	12.283 &	26.411 &	12.607 &	31.938 &	57.573 &	26.277 &	12.562 &	34.995 &	65.368	\\	
&	&	&	Local &	30.569 &	17.588 &	31.794 &	20.747 &	33.713 &	26.996 &	31.722 &	20.87 &	36.316 &	29.472	\\	\cmidrule(lr){3-4} \cmidrule(lr){5-6} \cmidrule(lr){7-8} \cmidrule(lr){9-10} \cmidrule(lr){11-12} \cmidrule(lr){13-14}
&	&	$0.5$ &	All &	11.163 &	4.834 &	11.368 &	4.935 &	11.401 &	4.953 &	11.342 &	4.904 &	12.705 &	6.752	\\	
&	&	&	Local &	16.349 &	11.237 &	16.642 &	11.726 &	16.658 &	11.737 &	16.616 &	11.731 &	17.893 &	11.960	\\	\cmidrule(lr){3-4} \cmidrule(lr){5-6} \cmidrule(lr){7-8} \cmidrule(lr){9-10} \cmidrule(lr){11-12} \cmidrule(lr){13-14}
&	&	$1$ &	All &	6.49 &	2.787 &	6.589 &	2.847 &	6.603 &	2.856 &	6.583 &	2.845 &	7.15 &	3.130	\\	
&	&	&	Local &	7.743 &	5.667 &	7.827 &	5.783 &	7.835 &	5.789 &	7.821 &	5.781 &	8.407 &	5.899	\\	\cmidrule(lr){1-4} \cmidrule(lr){5-6} \cmidrule(lr){7-8} \cmidrule(lr){9-10} \cmidrule(lr){11-12} \cmidrule(lr){13-14}
$200$ &	Gaussian &	$0.1$ &	All &	26.147 &	10.189 &	26.122 &	10.135 &	26.1 &	10.098 &	26.101 &	10.099 &	26.777 &	10.272	\\	
&	&	&	Local &	28.654 &	14.543 &	28.655 &	14.55 &	28.633 &	14.509 &	28.633 &	14.511 &	29.281 &	14.556	\\	\cmidrule(lr){3-4} \cmidrule(lr){5-6} \cmidrule(lr){7-8} \cmidrule(lr){9-10} \cmidrule(lr){11-12} \cmidrule(lr){13-14}
&	&	$0.5$ &	All &	14.03 &	5.16 &	14.022 &	5.156 &	14.02 &	5.153 &	14.02 &	5.153 &	14.396 &	5.150	\\	
&	&	&	Local &	18.545 &	11.559 &	18.544 &	11.56 &	18.542 &	11.559 &	18.542 &	11.559 &	18.917 &	11.585	\\	\cmidrule(lr){3-4} \cmidrule(lr){5-6} \cmidrule(lr){7-8} \cmidrule(lr){9-10} \cmidrule(lr){11-12} \cmidrule(lr){13-14}
&	&	$1$ &	All &	8.935 &	3.246 &	8.931 &	3.246 &	8.93 &	3.245 &	8.93 &	3.245 &	9.179 &	3.284	\\	
&	&	&	Local &	12.311 &	7.982 &	12.309 &	7.984 &	12.308 &	7.983 &	12.308 &	7.983 &	12.553 &	8.005	\\	\cmidrule(lr){2-4} \cmidrule(lr){5-6} \cmidrule(lr){7-8} \cmidrule(lr){9-10} \cmidrule(lr){11-12} \cmidrule(lr){13-14}
&	$t_3$ &	$0.1$ &	All &	22.249 &	6.565 &	23.004 &	7.664 &	28.682 &	52.623 &	22.982 &	7.797 &	42.115 &	122.632	\\	
&	&	&	Local &	29.79 &	16.019 &	31.991 &	23.092 &	38.13 &	57.329 &	32.165 &	24.722 &	52.687 &	127.384	\\	\cmidrule(lr){3-4} \cmidrule(lr){5-6} \cmidrule(lr){7-8} \cmidrule(lr){9-10} \cmidrule(lr){11-12} \cmidrule(lr){13-14}
&	&	$0.5$ &	All &	10.089 &	3.136 &	10.339 &	3.539 &	10.408 &	3.932 &	10.329 &	3.54 &	10.689 &	3.681	\\	
&	&	&	Local &	15.83 &	12.669 &	16.684 &	15.145 &	17.041 &	17.383 &	16.682 &	15.194 &	17.04 &	15.154	\\	\cmidrule(lr){3-4} \cmidrule(lr){5-6} \cmidrule(lr){7-8} \cmidrule(lr){9-10} \cmidrule(lr){11-12} \cmidrule(lr){13-14}
&	&	$1$ &	All &	6.055 &	1.787 &	6.188 &	2.044 &	6.206 &	2.144 &	6.184 &	2.046 &	6.401 &	2.259	\\	
&	&	&	Local &	9.413 &	8.219 &	10.249 &	12.98 &	10.414 &	14.311 &	10.255 &	13.05 &	10.645 &	14.730	\\	\cmidrule(lr){1-4} \cmidrule(lr){5-6} \cmidrule(lr){7-8} \cmidrule(lr){9-10} \cmidrule(lr){11-12} \cmidrule(lr){13-14}
$500$ &	Gaussian &	$0.1$ &	All &	24.782 &	7.499 &	24.775 &	7.495 &	24.77 &	7.49 &	24.77 &	7.49 &	24.824 &	7.319	\\	
&	&	&	Local &	27.414 &	11.115 &	27.414 &	11.114 &	27.41 &	11.113 &	27.41 &	11.113 &	27.451 &	10.997	\\	\cmidrule(lr){3-4} \cmidrule(lr){5-6} \cmidrule(lr){7-8} \cmidrule(lr){9-10} \cmidrule(lr){11-12} \cmidrule(lr){13-14}
&	&	$0.5$ &	All &	13.626 &	4.076 &	13.621 &	4.075 &	13.621 &	4.074 &	13.62 &	4.074 &	13.704 &	4.042	\\	
&	&	&	Local &	17.421 &	9.794 &	17.42 &	9.795 &	17.419 &	9.795 &	17.419 &	9.795 &	17.499 &	9.777	\\	\cmidrule(lr){3-4} \cmidrule(lr){5-6} \cmidrule(lr){7-8} \cmidrule(lr){9-10} \cmidrule(lr){11-12} \cmidrule(lr){13-14}
&	&	$1$ &	All &	8.63 &	2.561 &	8.626 &	2.561 &	8.626 &	2.561 &	8.626 &	2.561 &	8.7 &	2.564	\\	
&	&	&	Local &	11.029 &	6.947 &	11.028 &	6.948 &	11.027 &	6.949 &	11.027 &	6.949 &	11.098 &	6.936	\\	\cmidrule(lr){2-4} \cmidrule(lr){5-6} \cmidrule(lr){7-8} \cmidrule(lr){9-10} \cmidrule(lr){11-12} \cmidrule(lr){13-14}
&	$t_3$ &	$0.1$ &	All &	21.868 &	7.997 &	22.755 &	9.187 &	36.51 &	141.46 &	22.751 &	9.252 &	40.021 &	176.553	\\	
&	&	&	Local &	29.237 &	16.434 &	29.871 &	16.804 &	32.561 &	29.594 &	29.848 &	16.788 &	36.738 &	68.349	\\	\cmidrule(lr){3-4} \cmidrule(lr){5-6} \cmidrule(lr){7-8} \cmidrule(lr){9-10} \cmidrule(lr){11-12} \cmidrule(lr){13-14}
&	&	$0.5$ &	All &	10.172 &	3.431 &	10.512 &	3.923 &	17.669 &	73.514 &	10.511 &	3.933 &	20.002 &	96.155	\\	
&	&	&	Local &	15.185 &	13.995 &	15.343 &	14.071 &	17.284 &	23.15 &	15.338 &	14.067 &	19.292 &	40.375	\\	\cmidrule(lr){3-4} \cmidrule(lr){5-6} \cmidrule(lr){7-8} \cmidrule(lr){9-10} \cmidrule(lr){11-12} \cmidrule(lr){13-14}
&	&	$1$ &	All &	6.074 &	2.029 &	6.258 &	2.294 &	10.023 &	38.78 &	6.257 &	2.292 &	7.256 &	10.680	\\	
&	&	&	Local &	9.216 &	7.626 &	9.29 &	7.645 &	10.984 &	18.065 &	9.289 &	7.643 &	9.864 &	8.892	\\	\bottomrule
\end{tabular}
}
\end{table}

\begin{table}[h!t!b!p!]
\caption{\ref{f:two} Loading estimation errors of Trunc, iPE, RTFA and PreAve measured as in~\eqref{eq:err:loading:tensor} for each mode scaled by $100$, over varying $n \in \{100, 200, 500\}$, the distributions for $\mc F_t$ and $\bm\xi_t$ (Gaussian and $t_3$) and the percentages of outliers in the factors under~\ref{o:two} ($\varrho \in \{0.1, 0.5, 1\}$).
We report the mean and the standard deviation over $100$ realisations for each setting.}
\label{tab:tensor:le:factor:two}
\centering
% \resizebox{\textwidth}{!}
{\scriptsize
\begin{tabular}{rrrr cc cc cc cc}
\toprule
&	&	&	&	\multicolumn{2}{c}{Trunc} &		\multicolumn{2}{c}{iPE} &		\multicolumn{2}{c}{RTFA} &		\multicolumn{2}{c}{PreAve} 		\\	
$n$ &	Dist &	\% &	Mode &	Mean &	SD &	Mean &	SD &	Mean &	SD &	Mean &	SD	\\	\cmidrule(lr){1-4} \cmidrule(lr){5-6} \cmidrule(lr){7-8} \cmidrule(lr){9-10} \cmidrule(lr){11-12}
$100$ &	Gaussian &	$0.1$ &	1 &	1.766 &	0.275 &	1.765 &	0.274 &	1.765 &	0.274 &	3.565 &	0.777	\\	
&	&	&	2 &	0.562 &	0.164 &	0.558 &	0.163 &	0.558 &	0.163 &	1.331 &	0.437	\\	
&	&	&	3 &	0.56 &	0.155 &	0.556 &	0.154 &	0.556 &	0.154 &	1.307 &	0.454	\\	\cmidrule(lr){3-4} \cmidrule(lr){5-6} \cmidrule(lr){7-8} \cmidrule(lr){9-10} \cmidrule(lr){11-12}
&	&	$0.5$ &	1 &	1.271 &	0.208 &	1.269 &	0.208 &	1.269 &	0.208 &	2.632 &	0.637	\\	
&	&	&	2 &	0.406 &	0.111 &	0.402 &	0.111 &	0.402 &	0.111 &	0.939 &	0.344	\\	
&	&	&	3 &	0.391 &	0.101 &	0.39 &	0.102 &	0.39 &	0.102 &	0.922 &	0.348	\\	\cmidrule(lr){3-4} \cmidrule(lr){5-6} \cmidrule(lr){7-8} \cmidrule(lr){9-10} \cmidrule(lr){11-12}
&	&	$1$ &	1 &	0.994 &	0.161 &	0.993 &	0.161 &	0.993 &	0.161 &	2.032 &	0.515	\\	
&	&	&	2 &	0.32 &	0.086 &	0.319 &	0.085 &	0.319 &	0.085 &	0.759 &	0.351	\\	
&	&	&	3 &	0.302 &	0.078 &	0.3 &	0.077 &	0.3 &	0.077 &	0.682 &	0.207	\\	\cmidrule(lr){2-4} \cmidrule(lr){5-6} \cmidrule(lr){7-8} \cmidrule(lr){9-10} \cmidrule(lr){11-12}
&	$t_3$ &	$0.1$ &	1 &	1.745 &	0.285 &	1.776 &	0.3 &	1.768 &	0.297 &	3.897 &	1.109	\\	
&	&	&	2 &	0.562 &	0.176 &	0.567 &	0.18 &	0.566 &	0.179 &	1.302 &	0.411	\\	
&	&	&	3 &	0.555 &	0.165 &	0.56 &	0.165 &	0.558 &	0.165 &	1.381 &	0.465	\\	\cmidrule(lr){3-4} \cmidrule(lr){5-6} \cmidrule(lr){7-8} \cmidrule(lr){9-10} \cmidrule(lr){11-12}
&	&	$0.5$ &	1 &	1.138 &	0.195 &	1.164 &	0.227 &	1.152 &	0.203 &	2.427 &	0.640	\\	
&	&	&	2 &	0.364 &	0.114 &	0.365 &	0.116 &	0.362 &	0.115 &	0.909 &	0.367	\\	
&	&	&	3 &	0.372 &	0.102 &	0.375 &	0.102 &	0.373 &	0.1 &	0.938 &	0.301	\\	\cmidrule(lr){3-4} \cmidrule(lr){5-6} \cmidrule(lr){7-8} \cmidrule(lr){9-10} \cmidrule(lr){11-12}
&	&	$1$ &	1 &	0.842 &	0.127 &	0.856 &	0.137 &	0.848 &	0.128 &	1.735 &	0.515	\\	
&	&	&	2 &	0.265 &	0.078 &	0.265 &	0.078 &	0.264 &	0.078 &	0.65 &	0.274	\\	
&	&	&	3 &	0.265 &	0.073 &	0.268 &	0.071 &	0.266 &	0.071 &	0.657 &	0.219	\\	\cmidrule(lr){1-4} \cmidrule(lr){5-6} \cmidrule(lr){7-8} \cmidrule(lr){9-10} \cmidrule(lr){11-12}
$200$ &	Gaussian &	$0.1$ &	1 &	1.264 &	0.173 &	1.263 &	0.173 &	1.263 &	0.173 &	2.543 &	0.528	\\	
&	&	&	2 &	0.42 &	0.126 &	0.418 &	0.124 &	0.418 &	0.124 &	0.908 &	0.262	\\	
&	&	&	3 &	0.401 &	0.111 &	0.399 &	0.11 &	0.399 &	0.11 &	0.934 &	0.255	\\	\cmidrule(lr){3-4} \cmidrule(lr){5-6} \cmidrule(lr){7-8} \cmidrule(lr){9-10} \cmidrule(lr){11-12}
&	&	$0.5$ &	1 &	0.91 &	0.122 &	0.909 &	0.122 &	0.909 &	0.122 &	1.796 &	0.346	\\	
&	&	&	2 &	0.292 &	0.081 &	0.29 &	0.08 &	0.29 &	0.08 &	0.624 &	0.197	\\	
&	&	&	3 &	0.294 &	0.081 &	0.293 &	0.08 &	0.293 &	0.08 &	0.653 &	0.201	\\	\cmidrule(lr){3-4} \cmidrule(lr){5-6} \cmidrule(lr){7-8} \cmidrule(lr){9-10} \cmidrule(lr){11-12}
&	&	$1$ &	1 &	0.718 &	0.097 &	0.717 &	0.097 &	0.717 &	0.097 &	1.398 &	0.286	\\	
&	&	&	2 &	0.232 &	0.075 &	0.231 &	0.075 &	0.231 &	0.075 &	0.506 &	0.153	\\	
&	&	&	3 &	0.217 &	0.051 &	0.217 &	0.051 &	0.217 &	0.051 &	0.506 &	0.143	\\	\cmidrule(lr){2-4} \cmidrule(lr){5-6} \cmidrule(lr){7-8} \cmidrule(lr){9-10} \cmidrule(lr){11-12}
&	$t_3$ &	$0.1$ &	1 &	1.167 &	0.155 &	1.186 &	0.16 &	1.18 &	0.159 &	3.033 &	5.680	\\	
&	&	&	2 &	0.355 &	0.095 &	0.36 &	0.098 &	0.358 &	0.096 &	1.36 &	5.206	\\	
&	&	&	3 &	0.365 &	0.105 &	0.37 &	0.107 &	0.368 &	0.105 &	1.558 &	6.641	\\	\cmidrule(lr){3-4} \cmidrule(lr){5-6} \cmidrule(lr){7-8} \cmidrule(lr){9-10} \cmidrule(lr){11-12}
&	&	$0.5$ &	1 &	0.761 &	0.104 &	0.77 &	0.105 &	0.767 &	0.105 &	1.867 &	2.577	\\	
&	&	&	2 &	0.229 &	0.053 &	0.233 &	0.056 &	0.231 &	0.054 &	1.048 &	5.154	\\	
&	&	&	3 &	0.237 &	0.063 &	0.24 &	0.062 &	0.238 &	0.063 &	1.101 &	5.291	\\	\cmidrule(lr){3-4} \cmidrule(lr){5-6} \cmidrule(lr){7-8} \cmidrule(lr){9-10} \cmidrule(lr){11-12}
&	&	$1$ &	1 &	0.585 &	0.077 &	0.588 &	0.078 &	0.586 &	0.078 &	1.183 &	0.244	\\	
&	&	&	2 &	0.173 &	0.047 &	0.173 &	0.044 &	0.172 &	0.043 &	0.411 &	0.139	\\	
&	&	&	3 &	0.184 &	0.044 &	0.184 &	0.043 &	0.183 &	0.043 &	0.411 &	0.125	\\	\cmidrule(lr){1-4} \cmidrule(lr){5-6} \cmidrule(lr){7-8} \cmidrule(lr){9-10} \cmidrule(lr){11-12}
$500$ &	Gaussian &	$0.1$ &	1 &	0.774 &	0.111 &	0.774 &	0.111 &	0.774 &	0.111 &	1.504 &	0.298	\\	
&	&	&	2 &	0.263 &	0.092 &	0.263 &	0.091 &	0.263 &	0.091 &	0.567 &	0.177	\\	
&	&	&	3 &	0.267 &	0.092 &	0.266 &	0.092 &	0.266 &	0.092 &	0.542 &	0.159	\\	\cmidrule(lr){3-4} \cmidrule(lr){5-6} \cmidrule(lr){7-8} \cmidrule(lr){9-10} \cmidrule(lr){11-12}
&	&	$0.5$ &	1 &	0.556 &	0.078 &	0.556 &	0.078 &	0.556 &	0.078 &	1.064 &	0.192	\\	
&	&	&	2 &	0.184 &	0.058 &	0.184 &	0.057 &	0.184 &	0.057 &	0.384 &	0.120	\\	
&	&	&	3 &	0.187 &	0.056 &	0.187 &	0.056 &	0.187 &	0.056 &	0.395 &	0.116	\\	\cmidrule(lr){3-4} \cmidrule(lr){5-6} \cmidrule(lr){7-8} \cmidrule(lr){9-10} \cmidrule(lr){11-12}
&	&	$1$ &	1 &	0.442 &	0.064 &	0.441 &	0.064 &	0.441 &	0.064 &	0.824 &	0.140	\\	
&	&	&	2 &	0.141 &	0.047 &	0.141 &	0.047 &	0.141 &	0.047 &	0.308 &	0.096	\\	
&	&	&	3 &	0.149 &	0.042 &	0.149 &	0.042 &	0.149 &	0.042 &	0.302 &	0.079	\\	\cmidrule(lr){2-4} \cmidrule(lr){5-6} \cmidrule(lr){7-8} \cmidrule(lr){9-10} \cmidrule(lr){11-12}
&	$t_3$ &	$0.1$ &	1 &	0.734 &	0.104 &	0.744 &	0.106 &	0.741 &	0.106 &	1.502 &	0.320	\\	
&	&	&	2 &	0.248 &	0.095 &	0.252 &	0.096 &	0.251 &	0.096 &	0.545 &	0.161	\\	
&	&	&	3 &	0.245 &	0.07 &	0.246 &	0.069 &	0.245 &	0.069 &	0.508 &	0.201	\\	\cmidrule(lr){3-4} \cmidrule(lr){5-6} \cmidrule(lr){7-8} \cmidrule(lr){9-10} \cmidrule(lr){11-12}
&	&	$0.5$ &	1 &	0.485 &	0.068 &	0.489 &	0.068 &	0.488 &	0.068 &	0.959 &	0.200	\\	
&	&	&	2 &	0.157 &	0.055 &	0.158 &	0.055 &	0.157 &	0.056 &	0.343 &	0.093	\\	
&	&	&	3 &	0.152 &	0.043 &	0.154 &	0.043 &	0.153 &	0.042 &	0.33 &	0.105	\\	\cmidrule(lr){3-4} \cmidrule(lr){5-6} \cmidrule(lr){7-8} \cmidrule(lr){9-10} \cmidrule(lr){11-12}
&	&	$1$ &	1 &	0.368 &	0.054 &	0.371 &	0.055 &	0.369 &	0.055 &	0.727 &	0.151	\\	
&	&	&	2 &	0.118 &	0.039 &	0.119 &	0.04 &	0.119 &	0.04 &	0.262 &	0.078	\\	
&	&	&	3 &	0.119 &	0.031 &	0.12 &	0.031 &	0.12 &	0.031 &	0.252 &	0.071	\\	\bottomrule
\end{tabular}
}
\end{table}

\begin{table}[h!t!b!p!]
\caption{\ref{f:two} Common component estimation errors of Trunc, noTrunc, iPE, RTFA and PreAve measured as in~\eqref{eq:err:chi:tensor} with $\mc T = [n]$ (`all') and $\mc T = \{n - 10 + 1, \ldots, n\}$ (`local') scaled by $1000$, over varying $n \in \{100, 200, 500\}$, the distributions for $\mc F_t$ and $\bm\xi_t$ (Gaussian and $t_3$) and the percentages of outliers in the factors under~\ref{o:two} ($\varrho \in \{0.1, 0.5, 1\}$).
We report the mean and the standard deviation over $100$ realisations for each setting.}
\label{tab:tensor:ce:factor:two}
\centering
\resizebox{\textwidth}{!}
{\scriptsize
\begin{tabular}{rrrr cc cc cc cc cc}
\toprule
&	&	&	&	\multicolumn{2}{c}{Trunc} &		\multicolumn{2}{c}{noTrunc} &		\multicolumn{2}{c}{iPE} &		\multicolumn{2}{c}{RTFA} &		\multicolumn{2}{c}{PreAve} 		\\	
$n$ &	Dist &	\% &	Range &	Mean &	SD &	Mean &	SD &	Mean &	SD &	Mean &	SD &	Mean &	SD	\\	\cmidrule(lr){1-4} \cmidrule(lr){5-6} \cmidrule(lr){7-8} \cmidrule(lr){9-10} \cmidrule(lr){11-12} \cmidrule(lr){13-14}
$100$ &	Gaussian &	$0.1$ &	All &	2.734 &	0.833 &	2.732 &	0.833 &	2.731 &	0.832 &	2.731 &	0.832 &	3.872 &	1.314	\\	
&	&	&	Local &	3.017 &	1.114 &	3.016 &	1.116 &	3.014 &	1.116 &	3.014 &	1.116 &	4.19 &	1.584	\\	\cmidrule(lr){3-4} \cmidrule(lr){5-6} \cmidrule(lr){7-8} \cmidrule(lr){9-10} \cmidrule(lr){11-12} \cmidrule(lr){13-14}
&	&	$0.5$ &	All &	1.452 &	0.449 &	1.451 &	0.449 &	1.45 &	0.448 &	1.45 &	0.448 &	2.046 &	0.700	\\	
&	&	&	Local &	1.758 &	0.925 &	1.758 &	0.925 &	1.758 &	0.925 &	1.758 &	0.925 &	2.38 &	1.098	\\	\cmidrule(lr){3-4} \cmidrule(lr){5-6} \cmidrule(lr){7-8} \cmidrule(lr){9-10} \cmidrule(lr){11-12} \cmidrule(lr){13-14}
&	&	$1$ &	All &	0.911 &	0.279 &	0.91 &	0.279 &	0.91 &	0.279 &	0.91 &	0.279 &	1.258 &	0.418	\\	
&	&	&	Local &	1.078 &	0.813 &	1.078 &	0.813 &	1.078 &	0.813 &	1.078 &	0.813 &	1.431 &	0.887	\\	\cmidrule(lr){2-4} \cmidrule(lr){5-6} \cmidrule(lr){7-8} \cmidrule(lr){9-10} \cmidrule(lr){11-12} \cmidrule(lr){13-14}
&	$t_3$ &	$0.1$ &	All &	2.552 &	0.799 &	2.636 &	0.843 &	2.648 &	0.852 &	2.644 &	0.85 &	3.98 &	1.580	\\	
&	&	&	Local &	3.16 &	1.347 &	3.297 &	1.508 &	3.314 &	1.534 &	3.308 &	1.52 &	4.779 &	2.235	\\	\cmidrule(lr){3-4} \cmidrule(lr){5-6} \cmidrule(lr){7-8} \cmidrule(lr){9-10} \cmidrule(lr){11-12} \cmidrule(lr){13-14}
&	&	$0.5$ &	All &	1.149 &	0.363 &	1.175 &	0.374 &	1.181 &	0.377 &	1.178 &	0.375 &	1.689 &	0.557	\\	
&	&	&	Local &	1.788 &	1.246 &	1.835 &	1.272 &	1.843 &	1.277 &	1.838 &	1.274 &	2.4 &	1.368	\\	\cmidrule(lr){3-4} \cmidrule(lr){5-6} \cmidrule(lr){7-8} \cmidrule(lr){9-10} \cmidrule(lr){11-12} \cmidrule(lr){13-14}
&	&	$1$ &	All &	0.669 &	0.202 &	0.681 &	0.208 &	0.684 &	0.208 &	0.682 &	0.208 &	0.954 &	0.315	\\	
&	&	&	Local &	1.1 &	1.043 &	1.12 &	1.045 &	1.123 &	1.045 &	1.121 &	1.045 &	1.4 &	1.092	\\	\cmidrule(lr){1-4} \cmidrule(lr){5-6} \cmidrule(lr){7-8} \cmidrule(lr){9-10} \cmidrule(lr){11-12} \cmidrule(lr){13-14}
$200$ &	Gaussian &	$0.1$ &	All &	2.684 &	0.714 &	2.683 &	0.713 &	2.683 &	0.713 &	2.683 &	0.713 &	3.263 &	0.885	\\	
&	&	&	Local &	2.886 &	1.158 &	2.884 &	1.158 &	2.883 &	1.158 &	2.883 &	1.158 &	3.472 &	1.291	\\	\cmidrule(lr){3-4} \cmidrule(lr){5-6} \cmidrule(lr){7-8} \cmidrule(lr){9-10} \cmidrule(lr){11-12} \cmidrule(lr){13-14}
&	&	$0.5$ &	All &	1.465 &	0.391 &	1.465 &	0.391 &	1.465 &	0.391 &	1.465 &	0.391 &	1.743 &	0.475	\\	
&	&	&	Local &	1.844 &	1.174 &	1.844 &	1.174 &	1.844 &	1.173 &	1.844 &	1.173 &	2.128 &	1.228	\\	\cmidrule(lr){3-4} \cmidrule(lr){5-6} \cmidrule(lr){7-8} \cmidrule(lr){9-10} \cmidrule(lr){11-12} \cmidrule(lr){13-14}
&	&	$1$ &	All &	0.938 &	0.247 &	0.938 &	0.247 &	0.938 &	0.247 &	0.938 &	0.247 &	1.11 &	0.300	\\	
&	&	&	Local &	1.22 &	0.929 &	1.22 &	0.929 &	1.22 &	0.929 &	1.22 &	0.929 &	1.397 &	0.961	\\	\cmidrule(lr){2-4} \cmidrule(lr){5-6} \cmidrule(lr){7-8} \cmidrule(lr){9-10} \cmidrule(lr){11-12} \cmidrule(lr){13-14}
&	$t_3$ &	$0.1$ &	All &	2.209 &	0.555 &	2.279 &	0.584 &	2.285 &	0.586 &	2.283 &	0.585 &	12.353 &	95.291	\\	
&	&	&	Local &	2.849 &	1.459 &	3.014 &	1.679 &	3.021 &	1.69 &	3.018 &	1.685 &	10.568 &	69.896	\\	\cmidrule(lr){3-4} \cmidrule(lr){5-6} \cmidrule(lr){7-8} \cmidrule(lr){9-10} \cmidrule(lr){11-12} \cmidrule(lr){13-14}
&	&	$0.5$ &	All &	1.025 &	0.263 &	1.05 &	0.274 &	1.051 &	0.274 &	1.051 &	0.274 &	6.444 &	51.611	\\	
&	&	&	Local &	1.436 &	1.032 &	1.49 &	1.076 &	1.491 &	1.076 &	1.491 &	1.076 &	8.338 &	66.068	\\	\cmidrule(lr){3-4} \cmidrule(lr){5-6} \cmidrule(lr){7-8} \cmidrule(lr){9-10} \cmidrule(lr){11-12} \cmidrule(lr){13-14}
&	&	$1$ &	All &	0.62 &	0.16 &	0.63 &	0.165 &	0.631 &	0.165 &	0.63 &	0.165 &	0.755 &	0.210	\\	
&	&	&	Local &	0.77 &	0.378 &	0.794 &	0.39 &	0.794 &	0.39 &	0.794 &	0.39 &	0.919 &	0.412	\\	\cmidrule(lr){1-4} \cmidrule(lr){5-6} \cmidrule(lr){7-8} \cmidrule(lr){9-10} \cmidrule(lr){11-12} \cmidrule(lr){13-14}
$500$ &	Gaussian &	$0.1$ &	All &	2.453 &	0.716 &	2.453 &	0.716 &	2.453 &	0.716 &	2.453 &	0.716 &	2.654 &	0.781	\\	
&	&	&	Local &	2.65 &	1.021 &	2.65 &	1.021 &	2.65 &	1.021 &	2.65 &	1.021 &	2.853 &	1.057	\\	\cmidrule(lr){3-4} \cmidrule(lr){5-6} \cmidrule(lr){7-8} \cmidrule(lr){9-10} \cmidrule(lr){11-12} \cmidrule(lr){13-14}
&	&	$0.5$ &	All &	1.352 &	0.382 &	1.352 &	0.382 &	1.352 &	0.382 &	1.352 &	0.382 &	1.449 &	0.407	\\	
&	&	&	Local &	1.693 &	0.947 &	1.693 &	0.947 &	1.693 &	0.947 &	1.693 &	0.947 &	1.793 &	0.956	\\	\cmidrule(lr){3-4} \cmidrule(lr){5-6} \cmidrule(lr){7-8} \cmidrule(lr){9-10} \cmidrule(lr){11-12} \cmidrule(lr){13-14}
&	&	$1$ &	All &	0.864 &	0.248 &	0.864 &	0.248 &	0.864 &	0.248 &	0.864 &	0.248 &	0.923 &	0.264	\\	
&	&	&	Local &	1.158 &	0.862 &	1.158 &	0.862 &	1.158 &	0.862 &	1.158 &	0.862 &	1.217 &	0.868	\\	\cmidrule(lr){2-4} \cmidrule(lr){5-6} \cmidrule(lr){7-8} \cmidrule(lr){9-10} \cmidrule(lr){11-12} \cmidrule(lr){13-14}
&	$t_3$ &	$0.1$ &	All &	2.158 &	0.646 &	2.224 &	0.671 &	2.226 &	0.672 &	2.225 &	0.672 &	2.425 &	0.730	\\	
&	&	&	Local &	3.015 &	1.523 &	3.132 &	1.717 &	3.134 &	1.72 &	3.134 &	1.718 &	3.344 &	1.762	\\	\cmidrule(lr){3-4} \cmidrule(lr){5-6} \cmidrule(lr){7-8} \cmidrule(lr){9-10} \cmidrule(lr){11-12} \cmidrule(lr){13-14}
&	&	$0.5$ &	All &	1.02 &	0.316 &	1.043 &	0.324 &	1.043 &	0.324 &	1.043 &	0.324 &	1.124 &	0.348	\\	
&	&	&	Local &	1.842 &	1.528 &	1.889 &	1.646 &	1.889 &	1.647 &	1.889 &	1.646 &	1.973 &	1.658	\\	\cmidrule(lr){3-4} \cmidrule(lr){5-6} \cmidrule(lr){7-8} \cmidrule(lr){9-10} \cmidrule(lr){11-12} \cmidrule(lr){13-14}
&	&	$1$ &	All &	0.614 &	0.191 &	0.625 &	0.194 &	0.625 &	0.194 &	0.625 &	0.194 &	0.672 &	0.208	\\	
&	&	&	Local &	1.021 &	0.988 &	1.079 &	1.289 &	1.08 &	1.29 &	1.079 &	1.289 &	1.128 &	1.298	\\	\bottomrule
\end{tabular}
}
\end{table}

\begin{table}[h!t!b!p!]
\caption{\ref{f:three} Loading estimation errors of Trunc, iPE, RTFA and PreAve measured as in~\eqref{eq:err:loading:tensor} for each mode scaled by $100$, over varying $n \in \{100, 200, 500\}$, the distributions for $\mc F_t$ and $\bm\xi_t$ (Gaussian and $t_3$) and the percentages of outliers in the factors under~\ref{o:two} ($\varrho \in \{0.1, 0.5, 1\}$).
We report the mean and the standard deviation over $100$ realisations for each setting.}
\label{tab:tensor:le:factor:three}
\centering
% \resizebox{\textwidth}{!}
{\scriptsize
\begin{tabular}{rrrr cc cc cc cc}
\toprule
&	&	&	&	\multicolumn{2}{c}{Trunc} &		\multicolumn{2}{c}{iPE} &		\multicolumn{2}{c}{RTFA} &		\multicolumn{2}{c}{PreAve} 		\\	
$n$ &	Dist &	\% &	Mode &	Mean &	SD &	Mean &	SD &	Mean &	SD &	Mean &	SD	\\	\cmidrule(lr){1-4} \cmidrule(lr){5-6} \cmidrule(lr){7-8} \cmidrule(lr){9-10} \cmidrule(lr){11-12}
$100$ &	Gaussian &	$0.1$ &	1 &	0.5 &	0.088 &	0.5 &	0.089 &	0.5 &	0.089 &	1.196 &	0.253	\\	
&	&	&	2 &	0.618 &	0.081 &	0.618 &	0.081 &	0.618 &	0.081 &	1.467 &	0.238	\\	
&	&	&	3 &	0.715 &	0.09 &	0.715 &	0.091 &	0.715 &	0.091 &	1.609 &	0.275	\\	\cmidrule(lr){3-4} \cmidrule(lr){5-6} \cmidrule(lr){7-8} \cmidrule(lr){9-10} \cmidrule(lr){11-12}
&	&	$0.5$ &	1 &	0.359 &	0.06 &	0.359 &	0.06 &	0.359 &	0.06 &	0.853 &	0.190	\\	
&	&	&	2 &	0.45 &	0.068 &	0.449 &	0.068 &	0.449 &	0.068 &	1.07 &	0.235	\\	
&	&	&	3 &	0.513 &	0.068 &	0.512 &	0.068 &	0.512 &	0.068 &	1.186 &	0.273	\\	\cmidrule(lr){3-4} \cmidrule(lr){5-6} \cmidrule(lr){7-8} \cmidrule(lr){9-10} \cmidrule(lr){11-12}
&	&	$1$ &	1 &	0.282 &	0.047 &	0.281 &	0.047 &	0.281 &	0.047 &	0.681 &	0.172	\\	
&	&	&	2 &	0.341 &	0.046 &	0.341 &	0.046 &	0.341 &	0.046 &	0.829 &	0.198	\\	
&	&	&	3 &	0.396 &	0.049 &	0.396 &	0.049 &	0.396 &	0.049 &	0.902 &	0.184	\\	\cmidrule(lr){2-4} \cmidrule(lr){5-6} \cmidrule(lr){7-8} \cmidrule(lr){9-10} \cmidrule(lr){11-12}
&	$t_3$ &	$0.1$ &	1 &	0.498 &	0.087 &	0.509 &	0.094 &	0.504 &	0.087 &	1.249 &	0.311	\\	
&	&	&	2 &	0.614 &	0.096 &	0.625 &	0.099 &	0.622 &	0.097 &	1.503 &	0.364	\\	
&	&	&	3 &	0.695 &	0.094 &	0.707 &	0.097 &	0.704 &	0.095 &	1.675 &	0.290	\\	\cmidrule(lr){3-4} \cmidrule(lr){5-6} \cmidrule(lr){7-8} \cmidrule(lr){9-10} \cmidrule(lr){11-12}
&	&	$0.5$ &	1 &	0.321 &	0.055 &	0.329 &	0.057 &	0.326 &	0.057 &	0.895 &	0.271	\\	
&	&	&	2 &	0.397 &	0.063 &	0.404 &	0.063 &	0.402 &	0.063 &	1.025 &	0.292	\\	
&	&	&	3 &	0.449 &	0.064 &	0.455 &	0.064 &	0.453 &	0.064 &	1.117 &	0.287	\\	\cmidrule(lr){3-4} \cmidrule(lr){5-6} \cmidrule(lr){7-8} \cmidrule(lr){9-10} \cmidrule(lr){11-12}
&	&	$1$ &	1 &	0.238 &	0.037 &	0.241 &	0.038 &	0.24 &	0.037 &	0.599 &	0.176	\\	
&	&	&	2 &	0.292 &	0.046 &	0.296 &	0.047 &	0.294 &	0.046 &	0.75 &	0.219	\\	
&	&	&	3 &	0.333 &	0.04 &	0.337 &	0.04 &	0.335 &	0.04 &	0.828 &	0.197	\\	\cmidrule(lr){1-4} \cmidrule(lr){5-6} \cmidrule(lr){7-8} \cmidrule(lr){9-10} \cmidrule(lr){11-12}
$200$ &	Gaussian &	$0.1$ &	1 &	0.352 &	0.06 &	0.351 &	0.06 &	0.351 &	0.06 &	0.823 &	0.144	\\	
&	&	&	2 &	0.429 &	0.058 &	0.429 &	0.057 &	0.429 &	0.057 &	0.995 &	0.177	\\	
&	&	&	3 &	0.505 &	0.061 &	0.505 &	0.061 &	0.505 &	0.061 &	1.138 &	0.211	\\	\cmidrule(lr){3-4} \cmidrule(lr){5-6} \cmidrule(lr){7-8} \cmidrule(lr){9-10} \cmidrule(lr){11-12}
&	&	$0.5$ &	1 &	0.25 &	0.038 &	0.25 &	0.038 &	0.25 &	0.038 &	0.585 &	0.117	\\	
&	&	&	2 &	0.309 &	0.041 &	0.309 &	0.041 &	0.309 &	0.041 &	0.708 &	0.129	\\	
&	&	&	3 &	0.366 &	0.048 &	0.365 &	0.048 &	0.365 &	0.048 &	0.838 &	0.149	\\	\cmidrule(lr){3-4} \cmidrule(lr){5-6} \cmidrule(lr){7-8} \cmidrule(lr){9-10} \cmidrule(lr){11-12}
&	&	$1$ &	1 &	0.202 &	0.029 &	0.202 &	0.029 &	0.202 &	0.029 &	0.471 &	0.099	\\	
&	&	&	2 &	0.242 &	0.035 &	0.242 &	0.035 &	0.242 &	0.035 &	0.555 &	0.113	\\	
&	&	&	3 &	0.285 &	0.036 &	0.285 &	0.036 &	0.285 &	0.036 &	0.64 &	0.116	\\	\cmidrule(lr){2-4} \cmidrule(lr){5-6} \cmidrule(lr){7-8} \cmidrule(lr){9-10} \cmidrule(lr){11-12}
&	$t_3$ &	$0.1$ &	1 &	0.332 &	0.054 &	0.336 &	0.055 &	0.335 &	0.055 &	0.826 &	0.219	\\	
&	&	&	2 &	0.401 &	0.059 &	0.406 &	0.059 &	0.405 &	0.06 &	0.975 &	0.190	\\	
&	&	&	3 &	0.471 &	0.064 &	0.479 &	0.065 &	0.477 &	0.065 &	1.135 &	0.230	\\	\cmidrule(lr){3-4} \cmidrule(lr){5-6} \cmidrule(lr){7-8} \cmidrule(lr){9-10} \cmidrule(lr){11-12}
&	&	$0.5$ &	1 &	0.217 &	0.037 &	0.219 &	0.038 &	0.219 &	0.038 &	0.563 &	0.183	\\	
&	&	&	2 &	0.268 &	0.039 &	0.27 &	0.039 &	0.27 &	0.039 &	0.651 &	0.148	\\	
&	&	&	3 &	0.308 &	0.044 &	0.312 &	0.044 &	0.31 &	0.045 &	0.73 &	0.166	\\	\cmidrule(lr){3-4} \cmidrule(lr){5-6} \cmidrule(lr){7-8} \cmidrule(lr){9-10} \cmidrule(lr){11-12}
&	&	$1$ &	1 &	0.169 &	0.028 &	0.171 &	0.029 &	0.17 &	0.029 &	0.397 &	0.098	\\	
&	&	&	2 &	0.202 &	0.027 &	0.203 &	0.027 &	0.203 &	0.027 &	0.479 &	0.096	\\	
&	&	&	3 &	0.237 &	0.033 &	0.238 &	0.033 &	0.238 &	0.033 &	0.548 &	0.100	\\	\cmidrule(lr){1-4} \cmidrule(lr){5-6} \cmidrule(lr){7-8} \cmidrule(lr){9-10} \cmidrule(lr){11-12}
$500$ &	Gaussian &	$0.1$ &	1 &	0.216 &	0.029 &	0.216 &	0.029 &	0.216 &	0.029 &	0.502 &	0.082	\\	
&	&	&	2 &	0.27 &	0.036 &	0.27 &	0.036 &	0.27 &	0.036 &	0.61 &	0.091	\\	
&	&	&	3 &	0.32 &	0.036 &	0.319 &	0.036 &	0.319 &	0.036 &	0.703 &	0.086	\\	\cmidrule(lr){3-4} \cmidrule(lr){5-6} \cmidrule(lr){7-8} \cmidrule(lr){9-10} \cmidrule(lr){11-12}
&	&	$0.5$ &	1 &	0.155 &	0.022 &	0.155 &	0.022 &	0.155 &	0.022 &	0.363 &	0.062	\\	
&	&	&	2 &	0.196 &	0.026 &	0.195 &	0.026 &	0.195 &	0.026 &	0.445 &	0.075	\\	
&	&	&	3 &	0.229 &	0.027 &	0.229 &	0.027 &	0.229 &	0.027 &	0.508 &	0.066	\\	\cmidrule(lr){3-4} \cmidrule(lr){5-6} \cmidrule(lr){7-8} \cmidrule(lr){9-10} \cmidrule(lr){11-12}
&	&	$1$ &	1 &	0.126 &	0.019 &	0.126 &	0.019 &	0.126 &	0.019 &	0.292 &	0.054	\\	
&	&	&	2 &	0.156 &	0.021 &	0.156 &	0.021 &	0.156 &	0.021 &	0.355 &	0.058	\\	
&	&	&	3 &	0.18 &	0.022 &	0.18 &	0.022 &	0.18 &	0.022 &	0.403 &	0.057	\\	\cmidrule(lr){2-4} \cmidrule(lr){5-6} \cmidrule(lr){7-8} \cmidrule(lr){9-10} \cmidrule(lr){11-12}
&	$t_3$ &	$0.1$ &	1 &	0.205 &	0.031 &	0.208 &	0.031 &	0.207 &	0.031 &	0.486 &	0.083	\\	
&	&	&	2 &	0.242 &	0.027 &	0.245 &	0.029 &	0.244 &	0.028 &	0.573 &	0.089	\\	
&	&	&	3 &	0.284 &	0.035 &	0.288 &	0.036 &	0.288 &	0.036 &	0.642 &	0.116	\\	\cmidrule(lr){3-4} \cmidrule(lr){5-6} \cmidrule(lr){7-8} \cmidrule(lr){9-10} \cmidrule(lr){11-12}
&	&	$0.5$ &	1 &	0.135 &	0.019 &	0.137 &	0.019 &	0.136 &	0.019 &	0.316 &	0.053	\\	
&	&	&	2 &	0.164 &	0.02 &	0.166 &	0.02 &	0.165 &	0.02 &	0.377 &	0.068	\\	
&	&	&	3 &	0.188 &	0.021 &	0.19 &	0.022 &	0.19 &	0.022 &	0.426 &	0.069	\\	\cmidrule(lr){3-4} \cmidrule(lr){5-6} \cmidrule(lr){7-8} \cmidrule(lr){9-10} \cmidrule(lr){11-12}
&	&	$1$ &	1 &	0.103 &	0.015 &	0.104 &	0.015 &	0.103 &	0.015 &	0.242 &	0.043	\\	
&	&	&	2 &	0.125 &	0.015 &	0.126 &	0.015 &	0.125 &	0.015 &	0.289 &	0.045	\\	
&	&	&	3 &	0.146 &	0.016 &	0.147 &	0.017 &	0.146 &	0.017 &	0.324 &	0.051	\\	\bottomrule
\end{tabular}
}
\end{table}

\begin{table}[h!t!b!p!]
\caption{\ref{f:three} Common component estimation errors of Trunc, noTrunc, iPE, RTFA and PreAve measured as in~\eqref{eq:err:chi:tensor} with $\mc T = [n]$ (`all') and $\mc T = \{n - 10 + 1, \ldots, n\}$ (`local') scaled by $1000$, over varying $n \in \{100, 200, 500\}$, the distributions for $\mc F_t$ and $\bm\xi_t$ (Gaussian and $t_3$) and the percentages of outliers in the factors under~\ref{o:two} ($\varrho \in \{0.1, 0.5, 1\}$).
We report the mean and the standard deviation over $100$ realisations for each setting.}
\label{tab:tensor:ce:factor:three}
\centering
\resizebox{\textwidth}{!}
{\scriptsize
\begin{tabular}{rrrr cc cc cc cc cc}
\toprule
&	&	&	&	\multicolumn{2}{c}{Trunc} &		\multicolumn{2}{c}{noTrunc} &		\multicolumn{2}{c}{iPE} &		\multicolumn{2}{c}{RTFA} &		\multicolumn{2}{c}{PreAve} 		\\	
$n$ &	Dist &	\% &	Range &	Mean &	SD &	Mean &	SD &	Mean &	SD &	Mean &	SD &	Mean &	SD	\\	\cmidrule(lr){1-4} \cmidrule(lr){5-6} \cmidrule(lr){7-8} \cmidrule(lr){9-10} \cmidrule(lr){11-12} \cmidrule(lr){13-14}
$100$ &	Gaussian &	$0.1$ &	All &	1.082 &	0.221 &	1.082 &	0.221 &	1.082 &	0.221 &	1.082 &	0.221 &	1.527 &	0.340	\\	
&	&	&	Local &	1.169 &	0.389 &	1.169 &	0.389 &	1.169 &	0.389 &	1.169 &	0.389 &	1.622 &	0.500	\\	\cmidrule(lr){3-4} \cmidrule(lr){5-6} \cmidrule(lr){7-8} \cmidrule(lr){9-10} \cmidrule(lr){11-12} \cmidrule(lr){13-14}
&	&	$0.5$ &	All &	0.579 &	0.115 &	0.578 &	0.114 &	0.578 &	0.114 &	0.578 &	0.114 &	0.816 &	0.189	\\	
&	&	&	Local &	0.773 &	0.414 &	0.772 &	0.414 &	0.772 &	0.414 &	0.772 &	0.414 &	1.012 &	0.458	\\	\cmidrule(lr){3-4} \cmidrule(lr){5-6} \cmidrule(lr){7-8} \cmidrule(lr){9-10} \cmidrule(lr){11-12} \cmidrule(lr){13-14}
&	&	$1$ &	All &	0.359 &	0.068 &	0.359 &	0.068 &	0.359 &	0.068 &	0.359 &	0.068 &	0.498 &	0.103	\\	
&	&	&	Local &	0.472 &	0.297 &	0.472 &	0.297 &	0.471 &	0.297 &	0.471 &	0.297 &	0.613 &	0.326	\\	\cmidrule(lr){2-4} \cmidrule(lr){5-6} \cmidrule(lr){7-8} \cmidrule(lr){9-10} \cmidrule(lr){11-12} \cmidrule(lr){13-14}
&	$t_3$ &	$0.1$ &	All &	0.983 &	0.226 &	1.018 &	0.239 &	1.021 &	0.242 &	1.02 &	0.239 &	1.475 &	0.364	\\	
&	&	&	Local &	1.172 &	0.523 &	1.204 &	0.539 &	1.208 &	0.538 &	1.207 &	0.538 &	1.696 &	0.671	\\	\cmidrule(lr){3-4} \cmidrule(lr){5-6} \cmidrule(lr){7-8} \cmidrule(lr){9-10} \cmidrule(lr){11-12} \cmidrule(lr){13-14}
&	&	$0.5$ &	All &	0.441 &	0.096 &	0.453 &	0.1 &	0.454 &	0.101 &	0.453 &	0.1 &	0.681 &	0.180	\\	
&	&	&	Local &	0.619 &	0.492 &	0.63 &	0.497 &	0.631 &	0.497 &	0.631 &	0.497 &	0.863 &	0.538	\\	\cmidrule(lr){3-4} \cmidrule(lr){5-6} \cmidrule(lr){7-8} \cmidrule(lr){9-10} \cmidrule(lr){11-12} \cmidrule(lr){13-14}
&	&	$1$ &	All &	0.258 &	0.055 &	0.264 &	0.058 &	0.265 &	0.059 &	0.264 &	0.058 &	0.383 &	0.098	\\	
&	&	&	Local &	0.325 &	0.251 &	0.329 &	0.252 &	0.33 &	0.252 &	0.33 &	0.252 &	0.458 &	0.290	\\	\cmidrule(lr){1-4} \cmidrule(lr){5-6} \cmidrule(lr){7-8} \cmidrule(lr){9-10} \cmidrule(lr){11-12} \cmidrule(lr){13-14}
$200$ &	Gaussian &	$0.1$ &	All &	1.032 &	0.223 &	1.032 &	0.223 &	1.032 &	0.223 &	1.032 &	0.223 &	1.253 &	0.285	\\	
&	&	&	Local &	1.116 &	0.413 &	1.116 &	0.413 &	1.116 &	0.413 &	1.116 &	0.413 &	1.339 &	0.463	\\	\cmidrule(lr){3-4} \cmidrule(lr){5-6} \cmidrule(lr){7-8} \cmidrule(lr){9-10} \cmidrule(lr){11-12} \cmidrule(lr){13-14}
&	&	$0.5$ &	All &	0.565 &	0.123 &	0.564 &	0.123 &	0.564 &	0.123 &	0.564 &	0.123 &	0.679 &	0.153	\\	
&	&	&	Local &	0.699 &	0.439 &	0.699 &	0.439 &	0.699 &	0.439 &	0.699 &	0.439 &	0.812 &	0.460	\\	\cmidrule(lr){3-4} \cmidrule(lr){5-6} \cmidrule(lr){7-8} \cmidrule(lr){9-10} \cmidrule(lr){11-12} \cmidrule(lr){13-14}
&	&	$1$ &	All &	0.361 &	0.079 &	0.361 &	0.079 &	0.361 &	0.079 &	0.361 &	0.079 &	0.431 &	0.102	\\	
&	&	&	Local &	0.451 &	0.321 &	0.451 &	0.321 &	0.451 &	0.322 &	0.451 &	0.322 &	0.521 &	0.340	\\	\cmidrule(lr){2-4} \cmidrule(lr){5-6} \cmidrule(lr){7-8} \cmidrule(lr){9-10} \cmidrule(lr){11-12} \cmidrule(lr){13-14}
&	$t_3$ &	$0.1$ &	All &	0.9 &	0.189 &	0.928 &	0.199 &	0.929 &	0.2 &	0.929 &	0.2 &	1.142 &	0.253	\\	
&	&	&	Local &	1.104 &	0.473 &	1.138 &	0.487 &	1.139 &	0.487 &	1.139 &	0.487 &	1.358 &	0.516	\\	\cmidrule(lr){3-4} \cmidrule(lr){5-6} \cmidrule(lr){7-8} \cmidrule(lr){9-10} \cmidrule(lr){11-12} \cmidrule(lr){13-14}
&	&	$0.5$ &	All &	0.416 &	0.088 &	0.425 &	0.091 &	0.426 &	0.091 &	0.426 &	0.091 &	0.52 &	0.116	\\	
&	&	&	Local &	0.576 &	0.401 &	0.588 &	0.406 &	0.588 &	0.406 &	0.588 &	0.406 &	0.684 &	0.419	\\	\cmidrule(lr){3-4} \cmidrule(lr){5-6} \cmidrule(lr){7-8} \cmidrule(lr){9-10} \cmidrule(lr){11-12} \cmidrule(lr){13-14}
&	&	$1$ &	All &	0.252 &	0.054 &	0.257 &	0.056 &	0.257 &	0.056 &	0.257 &	0.056 &	0.308 &	0.068	\\	
&	&	&	Local &	0.314 &	0.239 &	0.319 &	0.241 &	0.32 &	0.241 &	0.32 &	0.241 &	0.372 &	0.245	\\	\cmidrule(lr){1-4} \cmidrule(lr){5-6} \cmidrule(lr){7-8} \cmidrule(lr){9-10} \cmidrule(lr){11-12} \cmidrule(lr){13-14}
$500$ &	Gaussian &	$0.1$ &	All &	0.984 &	0.176 &	0.984 &	0.176 &	0.984 &	0.176 &	0.984 &	0.176 &	1.066 &	0.189	\\	
&	&	&	Local &	1.146 &	0.334 &	1.146 &	0.334 &	1.146 &	0.334 &	1.146 &	0.334 &	1.229 &	0.344	\\	\cmidrule(lr){3-4} \cmidrule(lr){5-6} \cmidrule(lr){7-8} \cmidrule(lr){9-10} \cmidrule(lr){11-12} \cmidrule(lr){13-14}
&	&	$0.5$ &	All &	0.545 &	0.097 &	0.545 &	0.097 &	0.545 &	0.097 &	0.545 &	0.097 &	0.588 &	0.105	\\	
&	&	&	Local &	0.73 &	0.376 &	0.73 &	0.376 &	0.73 &	0.376 &	0.73 &	0.376 &	0.774 &	0.379	\\	\cmidrule(lr){3-4} \cmidrule(lr){5-6} \cmidrule(lr){7-8} \cmidrule(lr){9-10} \cmidrule(lr){11-12} \cmidrule(lr){13-14}
&	&	$1$ &	All &	0.349 &	0.062 &	0.349 &	0.062 &	0.349 &	0.062 &	0.349 &	0.062 &	0.377 &	0.067	\\	
&	&	&	Local &	0.472 &	0.301 &	0.472 &	0.301 &	0.472 &	0.301 &	0.472 &	0.301 &	0.5 &	0.303	\\	\cmidrule(lr){2-4} \cmidrule(lr){5-6} \cmidrule(lr){7-8} \cmidrule(lr){9-10} \cmidrule(lr){11-12} \cmidrule(lr){13-14}
&	$t_3$ &	$0.1$ &	All &	0.829 &	0.151 &	0.854 &	0.156 &	0.854 &	0.156 &	0.854 &	0.156 &	0.925 &	0.171	\\	
&	&	&	Local &	1.104 &	0.521 &	1.136 &	0.54 &	1.136 &	0.54 &	1.136 &	0.54 &	1.208 &	0.546	\\	\cmidrule(lr){3-4} \cmidrule(lr){5-6} \cmidrule(lr){7-8} \cmidrule(lr){9-10} \cmidrule(lr){11-12} \cmidrule(lr){13-14}
&	&	$0.5$ &	All &	0.391 &	0.069 &	0.4 &	0.07 &	0.4 &	0.07 &	0.4 &	0.07 &	0.431 &	0.076	\\	
&	&	&	Local &	0.631 &	0.465 &	0.648 &	0.489 &	0.648 &	0.489 &	0.648 &	0.489 &	0.679 &	0.492	\\	\cmidrule(lr){3-4} \cmidrule(lr){5-6} \cmidrule(lr){7-8} \cmidrule(lr){9-10} \cmidrule(lr){11-12} \cmidrule(lr){13-14}
&	&	$1$ &	All &	0.235 &	0.041 &	0.239 &	0.042 &	0.239 &	0.042 &	0.239 &	0.042 &	0.257 &	0.046	\\	
&	&	&	Local &	0.354 &	0.312 &	0.36 &	0.316 &	0.36 &	0.316 &	0.36 &	0.316 &	0.378 &	0.316	\\	\bottomrule
\end{tabular}
}
\end{table}

\restoregeometry

\clearpage 

\subsubsection{Factor number estimation}
\label{app:sim:tensor:r}

See Tables~\ref{tab:tensor:r:one:100}--\ref{tab:tensor:r:three:500} for the results from factor number estimation obtained under \ref{f:one}--\ref{f:three}, with outliers introduced to the idiosyncratic component under~\ref{o:one}.

\begin{table}[h!t!p!]
\caption{\ref{f:one} with $n = 100$. Factor number estimation results from Trunc, iPE, RTFA and PreAve over varying $n \in \{100, 200, 500\}$,  the distribution for $\mc F_t$ and $\bm\xi_t$ (Gaussian and $t_3$) and the percentage of outliers in the idiosyncratic component under~\ref{o:one} ($\varrho \in \{0, 0.1, 0.5, 1\}$).
We report the mean and the standard deviation over $100$ realisations per each setting.}
\label{tab:tensor:r:one:100}
\centering
%\resizebox{\textwidth}{!}
{\footnotesize
\begin{tabular}{rrr cc cc cc cc cc}
\toprule
&	&	&	\multicolumn{2}{c}{Trunc} &		\multicolumn{2}{c}{iPE} &		\multicolumn{2}{c}{RTFA} & 		\multicolumn{2}{c}{PreAve} 		\\	
Dist &	$\%$ &	Mode &	Mean &	SD &	Mean &	SD &	Mean &	SD &	Mean &	SD 	\\	\cmidrule(lr){1-3} \cmidrule(lr){4-5} \cmidrule(lr){6-7} \cmidrule(lr){8-9} \cmidrule(lr){10-11} \cmidrule(lr){12-13}
Gaussian &	0 &	1 &	2.55 &	0.657 &	2.92 &	0.307 &	2.92 &	0.307 &	2.81 &	0.394	\\	
&	&	2 &	2.54 &	0.688 &	2.97 &	0.223 &	2.97 &	0.223 &	2.81 &	0.394	\\	
&	&	3 &	2.59 &	0.683 &	2.96 &	0.197 &	2.96 &	0.197 &	2.86 &	0.349	\\	
&	0.1 &	1 &	2.51 &	0.689 &	2.89 &	0.345 &	2.89 &	0.345 &	2.82 &	0.386	\\	
&	&	2 &	2.54 &	0.688 &	2.95 &	0.261 &	2.96 &	0.243 &	2.81 &	0.394	\\	
&	&	3 &	2.59 &	0.683 &	2.95 &	0.219 &	2.95 &	0.219 &	2.88 &	0.327	\\	
&	0.5 &	1 &	2.53 &	0.674 &	2.78 &	0.484 &	2.77 &	0.489 &	2.78 &	0.416	\\	
&	&	2 &	2.53 &	0.688 &	2.88 &	0.409 &	2.87 &	0.418 &	2.8 &	0.402	\\	
&	&	3 &	2.59 &	0.683 &	2.89 &	0.373 &	2.89 &	0.373 &	2.84 &	0.368	\\	
&	1 &	1 &	2.49 &	0.703 &	2.69 &	0.563 &	2.7 &	0.56 &	2.67 &	0.493	\\	
&	&	2 &	2.51 &	0.689 &	2.75 &	0.539 &	2.76 &	0.534 &	2.8 &	0.402	\\	
&	&	3 &	2.57 &	0.7 &	2.8 &	0.492 &	2.8 &	0.492 &	2.76 &	0.452	\\	\cmidrule(lr){1-3} \cmidrule(lr){4-5} \cmidrule(lr){6-7} \cmidrule(lr){8-9} \cmidrule(lr){10-11} \cmidrule(lr){12-13}
t&	0 &	1 &	2.56 &	0.729 &	3 &	0.246 &	2.99 &	0.225 &	2.8 &	0.426	\\	
&	&	2 &	2.53 &	0.703 &	2.96 &	0.425 &	2.91 &	0.404 &	2.68 &	0.490	\\	
&	&	3 &	2.49 &	0.759 &	3 &	0.284 &	2.97 &	0.223 &	2.74 &	0.441	\\	
&	0.1 &	1 &	2.53 &	0.745 &	2.96 &	0.315 &	2.96 &	0.315 &	2.83 &	0.403	\\	
&	&	2 &	2.51 &	0.718 &	2.91 &	0.452 &	2.91 &	0.404 &	2.69 &	0.506	\\	
&	&	3 &	2.47 &	0.758 &	2.98 &	0.284 &	2.96 &	0.243 &	2.71 &	0.456	\\	
&	0.5 &	1 &	2.49 &	0.772 &	2.75 &	0.657 &	2.79 &	0.591 &	2.73 &	0.489	\\	
&	&	2 &	2.48 &	0.731 &	2.77 &	0.601 &	2.81 &	0.526 &	2.61 &	0.549	\\	
&	&	3 &	2.45 &	0.77 &	2.82 &	0.557 &	2.81 &	0.545 &	2.69 &	0.486	\\	
&	1 &	1 &	2.43 &	0.807 &	2.59 &	0.726 &	2.6 &	0.725 &	2.68 &	0.530	\\	
&	&	2 &	2.48 &	0.731 &	2.64 &	0.659 &	2.64 &	0.674 &	2.53 &	0.594	\\	
&	&	3 &	2.43 &	0.782 &	2.71 &	0.64 &	2.71 &	0.64 &	2.62 &	0.528	\\	\bottomrule
\end{tabular}
}
\end{table}

\begin{table}[h!t!p!]
\caption{\ref{f:one} with $n = 200$.}
\label{tab:tensor:r:one:200}
\centering
%\resizebox{\textwidth}{!}
{\footnotesize
\begin{tabular}{rrr cc cc cc cc cc}
\toprule
&	&	&	\multicolumn{2}{c}{Trunc} &		\multicolumn{2}{c}{iPE} &		\multicolumn{2}{c}{RTFA} & 		\multicolumn{2}{c}{PreAve} 		\\	
Dist &	$\%$ &	Mode &	Mean &	SD &	Mean &	SD &	Mean &	SD &	Mean &	SD 	\\	\cmidrule(lr){1-3} \cmidrule(lr){4-5} \cmidrule(lr){6-7} \cmidrule(lr){8-9} \cmidrule(lr){10-11} \cmidrule(lr){12-13}
Gaussian &	0 &	1 &	2.61 &	0.665 &	2.92 &	0.273 &	2.92 &	0.273 &	2.89 &	0.314	\\	
&	&	2 &	2.55 &	0.73 &	2.95 &	0.261 &	2.95 &	0.261 &	2.9 &	0.302	\\	
&	&	3 &	2.57 &	0.685 &	2.95 &	0.261 &	2.95 &	0.261 &	2.84 &	0.368	\\	
&	0.1 &	1 &	2.59 &	0.683 &	2.92 &	0.273 &	2.92 &	0.273 &	2.88 &	0.327	\\	
&	&	2 &	2.53 &	0.745 &	2.95 &	0.261 &	2.95 &	0.261 &	2.91 &	0.288	\\	
&	&	3 &	2.55 &	0.702 &	2.91 &	0.351 &	2.91 &	0.351 &	2.84 &	0.368	\\	
&	0.5 &	1 &	2.59 &	0.683 &	2.85 &	0.386 &	2.87 &	0.338 &	2.9 &	0.302	\\	
&	&	2 &	2.53 &	0.731 &	2.87 &	0.442 &	2.88 &	0.433 &	2.87 &	0.338	\\	
&	&	3 &	2.55 &	0.702 &	2.85 &	0.435 &	2.85 &	0.435 &	2.83 &	0.378	\\	
&	1 &	1 &	2.54 &	0.731 &	2.76 &	0.534 &	2.79 &	0.498 &	2.86 &	0.349	\\	
&	&	2 &	2.52 &	0.731 &	2.79 &	0.537 &	2.8 &	0.532 &	2.84 &	0.368	\\	
&	&	3 &	2.55 &	0.702 &	2.79 &	0.498 &	2.79 &	0.498 &	2.81 &	0.394	\\	\cmidrule(lr){1-3} \cmidrule(lr){4-5} \cmidrule(lr){6-7} \cmidrule(lr){8-9} \cmidrule(lr){10-11} \cmidrule(lr){12-13}
t&	0 &	1 &	2.66 &	0.623 &	2.91 &	0.404 &	2.91 &	0.379 &	2.75 &	0.435	\\	
&	&	2 &	2.5 &	0.732 &	2.97 &	0.332 &	2.95 &	0.297 &	2.72 &	0.473	\\	
&	&	3 &	2.4 &	0.778 &	2.95 &	0.359 &	2.95 &	0.33 &	2.63 &	0.506	\\	
&	0.1 &	1 &	2.63 &	0.63 &	2.89 &	0.399 &	2.88 &	0.433 &	2.71 &	0.478	\\	
&	&	2 &	2.43 &	0.756 &	2.95 &	0.33 &	2.94 &	0.312 &	2.67 &	0.514	\\	
&	&	3 &	2.41 &	0.793 &	2.91 &	0.429 &	2.92 &	0.367 &	2.63 &	0.506	\\	
&	0.5 &	1 &	2.61 &	0.65 &	2.79 &	0.537 &	2.79 &	0.537 &	2.71 &	0.478	\\	
&	&	2 &	2.44 &	0.756 &	2.8 &	0.512 &	2.85 &	0.479 &	2.66 &	0.517	\\	
&	&	3 &	2.35 &	0.796 &	2.78 &	0.561 &	2.83 &	0.493 &	2.61 &	0.530	\\	
&	1 &	1 &	2.57 &	0.671 &	2.62 &	0.708 &	2.61 &	0.695 &	2.69 &	0.486	\\	
&	&	2 &	2.44 &	0.743 &	2.74 &	0.597 &	2.7 &	0.595 &	2.66 &	0.497	\\	
&	&	3 &	2.34 &	0.794 &	2.56 &	0.715 &	2.59 &	0.698 &	2.56 &	0.574	\\	\bottomrule
\end{tabular}
}
\end{table}

\begin{table}[h!t!p!]
\caption{\ref{f:one} with $n = 500$.}
\label{tab:tensor:r:one:500}
\centering
%\resizebox{\textwidth}{!}
{\footnotesize
\begin{tabular}{rrr cc cc cc cc cc}
\toprule
&	&	&	\multicolumn{2}{c}{Trunc} &		\multicolumn{2}{c}{iPE} &		\multicolumn{2}{c}{RTFA} & 		\multicolumn{2}{c}{PreAve} 		\\	
Dist &	$\%$ &	Mode &	Mean &	SD &	Mean &	SD &	Mean &	SD &	Mean &	SD 	\\	\cmidrule(lr){1-3} \cmidrule(lr){4-5} \cmidrule(lr){6-7} \cmidrule(lr){8-9} \cmidrule(lr){10-11} \cmidrule(lr){12-13}
Gaussian &	0 &	1 &	2.56 &	0.671 &	2.91 &	0.288 &	2.91 &	0.288 &	2.84 &	0.368	\\	
&	&	2 &	2.56 &	0.641 &	2.91 &	0.321 &	2.91 &	0.321 &	2.82 &	0.386	\\	
&	&	3 &	2.57 &	0.714 &	2.97 &	0.223 &	2.97 &	0.223 &	2.88 &	0.327	\\	
&	0.1 &	1 &	2.53 &	0.688 &	2.85 &	0.359 &	2.85 &	0.359 &	2.85 &	0.359	\\	
&	&	2 &	2.55 &	0.642 &	2.9 &	0.333 &	2.9 &	0.333 &	2.82 &	0.386	\\	
&	&	3 &	2.57 &	0.714 &	2.96 &	0.243 &	2.96 &	0.243 &	2.88 &	0.327	\\	
&	0.5 &	1 &	2.49 &	0.718 &	2.75 &	0.539 &	2.75 &	0.539 &	2.84 &	0.368	\\	
&	&	2 &	2.55 &	0.642 &	2.81 &	0.443 &	2.83 &	0.428 &	2.81 &	0.394	\\	
&	&	3 &	2.56 &	0.715 &	2.91 &	0.351 &	2.91 &	0.351 &	2.88 &	0.327	\\	
&	1 &	1 &	2.48 &	0.717 &	2.67 &	0.604 &	2.67 &	0.604 &	2.81 &	0.394	\\	
&	&	2 &	2.55 &	0.642 &	2.73 &	0.51 &	2.73 &	0.51 &	2.8 &	0.402	\\	
&	&	3 &	2.54 &	0.731 &	2.84 &	0.465 &	2.86 &	0.427 &	2.87 &	0.338	\\	\cmidrule(lr){1-3} \cmidrule(lr){4-5} \cmidrule(lr){6-7} \cmidrule(lr){8-9} \cmidrule(lr){10-11} \cmidrule(lr){12-13}
t&	0 &	1 &	2.61 &	0.68 &	2.86 &	0.427 &	2.84 &	0.443 &	2.8 &	0.426	\\	
&	&	2 &	2.55 &	0.716 &	2.91 &	0.379 &	2.91 &	0.379 &	2.86 &	0.377	\\	
&	&	3 &	2.63 &	0.646 &	2.89 &	0.399 &	2.88 &	0.409 &	2.83 &	0.403	\\	
&	0.1 &	1 &	2.61 &	0.68 &	2.83 &	0.451 &	2.83 &	0.451 &	2.79 &	0.433	\\	
&	&	2 &	2.55 &	0.716 &	2.86 &	0.472 &	2.86 &	0.472 &	2.87 &	0.367	\\	
&	&	3 &	2.58 &	0.684 &	2.85 &	0.458 &	2.86 &	0.45 &	2.78 &	0.440	\\	
&	0.5 &	1 &	2.59 &	0.653 &	2.76 &	0.534 &	2.75 &	0.539 &	2.78 &	0.462	\\	
&	&	2 &	2.55 &	0.716 &	2.75 &	0.575 &	2.73 &	0.601 &	2.83 &	0.403	\\	
&	&	3 &	2.58 &	0.699 &	2.78 &	0.543 &	2.79 &	0.518 &	2.77 &	0.468	\\	
&	1 &	1 &	2.54 &	0.702 &	2.56 &	0.729 &	2.57 &	0.714 &	2.73 &	0.489	\\	
&	&	2 &	2.57 &	0.7 &	2.55 &	0.744 &	2.59 &	0.712 &	2.8 &	0.426	\\	
&	&	3 &	2.53 &	0.731 &	2.7 &	0.644 &	2.7 &	0.644 &	2.78 &	0.440	\\	\bottomrule
\end{tabular}
}
\end{table}

\begin{table}[h!t!p!]
\caption{\ref{f:two} with $n = 100$. Factor number estimation results from Trunc, iPE, RTFA and PreAve over varying $n \in \{100, 200, 500\}$,  the distribution for $\mc F_t$ and $\bm\xi_t$ (Gaussian and $t_3$) and the percentage of outliers in the idiosyncratic component under~\ref{o:one} ($\varrho \in \{0, 0.1, 0.5, 1\}$).
We report the mean and the standard deviation over $100$ realisations per each setting.}
\label{tab:tensor:r:two:100}
\centering
%\resizebox{\textwidth}{!}
{\footnotesize
\begin{tabular}{rrr cc cc cc cc cc}
\toprule
&	&	&	\multicolumn{2}{c}{Trunc} &		\multicolumn{2}{c}{iPE} &		\multicolumn{2}{c}{RTFA} & 		\multicolumn{2}{c}{PreAve} 		\\	
Dist &	$\%$ &	Mode &	Mean &	SD &	Mean &	SD &	Mean &	SD &	Mean &	SD 	\\	\cmidrule(lr){1-3} \cmidrule(lr){4-5} \cmidrule(lr){6-7} \cmidrule(lr){8-9} \cmidrule(lr){10-11} \cmidrule(lr){12-13}
Gaussian &	0 &	1 &	3 &	0 &	3 &	0 &	3 &	0 &	3.02 &	0.141	\\	
&	&	2 &	2.73 &	0.548 &	2.99 &	0.1 &	2.99 &	0.1 &	2.88 &	0.327	\\	
&	&	3 &	2.81 &	0.465 &	3 &	0 &	3 &	0 &	2.89 &	0.314	\\	
&	0.1 &	1 &	3 &	0 &	3 &	0 &	3 &	0 &	3.02 &	0.141	\\	
&	&	2 &	2.73 &	0.548 &	2.99 &	0.1 &	2.99 &	0.1 &	2.87 &	0.338	\\	
&	&	3 &	2.81 &	0.465 &	3 &	0 &	3 &	0 &	2.87 &	0.338	\\	
&	0.5 &	1 &	3 &	0 &	3 &	0 &	3 &	0 &	3 &	0.000	\\	
&	&	2 &	2.71 &	0.574 &	2.98 &	0.141 &	2.98 &	0.141 &	2.87 &	0.338	\\	
&	&	3 &	2.8 &	0.471 &	3 &	0 &	3 &	0 &	2.87 &	0.338	\\	
&	1 &	1 &	3 &	0 &	3 &	0 &	3 &	0 &	3 &	0.000	\\	
&	&	2 &	2.7 &	0.577 &	2.98 &	0.141 &	2.98 &	0.141 &	2.87 &	0.338	\\	
&	&	3 &	2.8 &	0.471 &	3 &	0 &	3 &	0 &	2.83 &	0.378	\\	\cmidrule(lr){1-3} \cmidrule(lr){4-5} \cmidrule(lr){6-7} \cmidrule(lr){8-9} \cmidrule(lr){10-11} \cmidrule(lr){12-13}
t&	0 &	1 &	3 &	0 &	3 &	0 &	3 &	0 &	3 &	0.000	\\	
&	&	2 &	2.733 &	0.546 &	3 &	0 &	3 &	0 &	2.713 &	0.476	\\	
&	&	3 &	2.683 &	0.615 &	3 &	0 &	3 &	0 &	2.713 &	0.476	\\	
&	0.1 &	1 &	3 &	0 &	3 &	0 &	3 &	0 &	3 &	0.000	\\	
&	&	2 &	2.713 &	0.554 &	3 &	0 &	3 &	0 &	2.693 &	0.485	\\	
&	&	3 &	2.683 &	0.615 &	3 &	0 &	3 &	0 &	2.693 &	0.485	\\	
&	0.5 &	1 &	3 &	0 &	3 &	0 &	3 &	0 &	3 &	0.000	\\	
&	&	2 &	2.693 &	0.579 &	2.98 &	0.14 &	2.98 &	0.14 &	2.693 &	0.485	\\	
&	&	3 &	2.663 &	0.621 &	2.99 &	0.1 &	2.99 &	0.1 &	2.663 &	0.496	\\	
&	1 &	1 &	3 &	0 &	3 &	0 &	3 &	0 &	3.455 &	0.933	\\	
&	&	2 &	2.693 &	0.579 &	2.96 &	0.196 &	2.96 &	0.196 &	2.683 &	0.488	\\	
&	&	3 &	2.634 &	0.644 &	2.96 &	0.196 &	2.96 &	0.196 &	2.673 &	0.492	\\	\bottomrule
\end{tabular}
}
\end{table}

\begin{table}[h!t!p!]
\caption{\ref{f:two} with $n = 200$.}
\label{tab:tensor:r:two:200}
\centering
%\resizebox{\textwidth}{!}
{\footnotesize
\begin{tabular}{rrr cc cc cc cc cc}
\toprule
&	&	&	\multicolumn{2}{c}{Trunc} &		\multicolumn{2}{c}{iPE} &		\multicolumn{2}{c}{RTFA} & 		\multicolumn{2}{c}{PreAve} 		\\	
Dist &	$\%$ &	Mode &	Mean &	SD &	Mean &	SD &	Mean &	SD &	Mean &	SD 	\\	\cmidrule(lr){1-3} \cmidrule(lr){4-5} \cmidrule(lr){6-7} \cmidrule(lr){8-9} \cmidrule(lr){10-11} \cmidrule(lr){12-13}
Gaussian &	0 &	1 &	3 &	0 &	3 &	0 &	3 &	0 &	3 &	0.000	\\	
&	&	2 &	2.79 &	0.478 &	2.99 &	0.1 &	2.99 &	0.1 &	2.89 &	0.314	\\	
&	&	3 &	2.72 &	0.57 &	3 &	0 &	3 &	0 &	2.89 &	0.314	\\	
&	0.1 &	1 &	3 &	0 &	3 &	0 &	3 &	0 &	3 &	0.000	\\	
&	&	2 &	2.79 &	0.478 &	2.99 &	0.1 &	2.99 &	0.1 &	2.89 &	0.314	\\	
&	&	3 &	2.71 &	0.591 &	3 &	0 &	3 &	0 &	2.89 &	0.314	\\	
&	0.5 &	1 &	3 &	0 &	3 &	0 &	3 &	0 &	3 &	0.000	\\	
&	&	2 &	2.78 &	0.484 &	2.99 &	0.1 &	2.99 &	0.1 &	2.88 &	0.327	\\	
&	&	3 &	2.7 &	0.595 &	3 &	0 &	3 &	0 &	2.86 &	0.349	\\	
&	1 &	1 &	3 &	0 &	3 &	0 &	3 &	0 &	3 &	0.000	\\	
&	&	2 &	2.77 &	0.489 &	2.97 &	0.171 &	2.97 &	0.171 &	2.87 &	0.338	\\	
&	&	3 &	2.69 &	0.598 &	3 &	0 &	3 &	0 &	2.85 &	0.359	\\	\cmidrule(lr){1-3} \cmidrule(lr){4-5} \cmidrule(lr){6-7} \cmidrule(lr){8-9} \cmidrule(lr){10-11} \cmidrule(lr){12-13}
t&	0 &	1 &	3 &	0 &	3.02 &	0.141 &	3 &	0 &	3 &	0.000	\\	
&	&	2 &	2.798 &	0.494 &	3.01 &	0.175 &	2.99 &	0.101 &	2.828 &	0.379	\\	
&	&	3 &	2.758 &	0.454 &	3.02 &	0.141 &	3 &	0 &	2.737 &	0.442	\\	
&	0.1 &	1 &	3 &	0 &	3.02 &	0.141 &	3 &	0 &	3.01 &	0.101	\\	
&	&	2 &	2.798 &	0.494 &	3.01 &	0.175 &	2.99 &	0.101 &	2.798 &	0.404	\\	
&	&	3 &	2.758 &	0.454 &	3.02 &	0.141 &	3 &	0 &	2.727 &	0.448	\\	
&	0.5 &	1 &	3 &	0 &	3.01 &	0.101 &	3 &	0 &	3.091 &	0.380	\\	
&	&	2 &	2.778 &	0.526 &	2.97 &	0.224 &	2.96 &	0.198 &	2.798 &	0.428	\\	
&	&	3 &	2.747 &	0.459 &	3.01 &	0.101 &	3 &	0 &	2.717 &	0.453	\\	
&	1 &	1 &	3.01 &	0.101 &	3.02 &	0.141 &	3.01 &	0.101 &	4.596 &	1.911	\\	
&	&	2 &	2.768 &	0.531 &	2.97 &	0.266 &	2.96 &	0.244 &	2.788 &	0.435	\\	
&	&	3 &	2.747 &	0.459 &	3 &	0 &	3 &	0 &	2.667 &	0.535	\\	\bottomrule
\end{tabular}
}
\end{table}

\begin{table}[h!t!p!]
\caption{\ref{f:two} with $n = 500$.}
\label{tab:tensor:r:two:500}
\centering
%\resizebox{\textwidth}{!}
{\footnotesize
\begin{tabular}{rrr cc cc cc cc cc}
\toprule
&	&	&	\multicolumn{2}{c}{Trunc} &		\multicolumn{2}{c}{iPE} &		\multicolumn{2}{c}{RTFA} & 		\multicolumn{2}{c}{PreAve} 		\\	
Dist &	$\%$ &	Mode &	Mean &	SD &	Mean &	SD &	Mean &	SD &	Mean &	SD 	\\	\cmidrule(lr){1-3} \cmidrule(lr){4-5} \cmidrule(lr){6-7} \cmidrule(lr){8-9} \cmidrule(lr){10-11} \cmidrule(lr){12-13}
Gaussian &	0 &	1 &	3 &	0 &	3 &	0 &	3 &	0 &	3.01 &	0.100	\\	
&	&	2 &	2.8 &	0.426 &	3 &	0 &	3 &	0 &	2.87 &	0.338	\\	
&	&	3 &	2.66 &	0.623 &	3 &	0 &	3 &	0 &	2.88 &	0.327	\\	
&	0.1 &	1 &	3 &	0 &	3 &	0 &	3 &	0 &	3.01 &	0.100	\\	
&	&	2 &	2.8 &	0.426 &	3 &	0 &	3 &	0 &	2.88 &	0.327	\\	
&	&	3 &	2.66 &	0.623 &	3 &	0 &	3 &	0 &	2.89 &	0.314	\\	
&	0.5 &	1 &	3 &	0 &	3 &	0 &	3 &	0 &	3 &	0.000	\\	
&	&	2 &	2.78 &	0.462 &	2.99 &	0.1 &	2.99 &	0.1 &	2.87 &	0.338	\\	
&	&	3 &	2.64 &	0.644 &	3 &	0 &	3 &	0 &	2.89 &	0.314	\\	
&	1 &	1 &	3 &	0 &	3 &	0 &	3 &	0 &	3.02 &	0.141	\\	
&	&	2 &	2.77 &	0.468 &	2.97 &	0.171 &	2.97 &	0.171 &	2.88 &	0.327	\\	
&	&	3 &	2.64 &	0.644 &	3 &	0 &	3 &	0 &	2.89 &	0.314	\\	\cmidrule(lr){1-3} \cmidrule(lr){4-5} \cmidrule(lr){6-7} \cmidrule(lr){8-9} \cmidrule(lr){10-11} \cmidrule(lr){12-13}
t&	0 &	1 &	3 &	0 &	3 &	0 &	3 &	0 &	3 &	0.000	\\	
&	&	2 &	2.69 &	0.563 &	3 &	0 &	3 &	0 &	2.85 &	0.359	\\	
&	&	3 &	2.86 &	0.427 &	3 &	0 &	3 &	0 &	2.88 &	0.327	\\	
&	0.1 &	1 &	3 &	0 &	3 &	0 &	3 &	0 &	3.01 &	0.100	\\	
&	&	2 &	2.69 &	0.563 &	3 &	0 &	3 &	0 &	2.84 &	0.368	\\	
&	&	3 &	2.85 &	0.435 &	2.99 &	0.1 &	2.99 &	0.1 &	2.88 &	0.327	\\	
&	0.5 &	1 &	3 &	0 &	3 &	0 &	3 &	0 &	3.4 &	0.791	\\	
&	&	2 &	2.64 &	0.595 &	2.98 &	0.141 &	2.97 &	0.171 &	2.81 &	0.394	\\	
&	&	3 &	2.79 &	0.537 &	2.99 &	0.1 &	2.99 &	0.1 &	2.89 &	0.314	\\	
&	1 &	1 &	3 &	0 &	3 &	0 &	3 &	0 &	6.62 &	3.296	\\	
&	&	2 &	2.63 &	0.597 &	2.93 &	0.256 &	2.93 &	0.256 &	2.82 &	0.386	\\	
&	&	3 &	2.79 &	0.537 &	2.96 &	0.243 &	2.96 &	0.243 &	2.86 &	0.349	\\	\bottomrule
\end{tabular}
}
\end{table}

\begin{table}[h!t!p!]
\caption{\ref{f:three} with $n = 100$. Factor number estimation results from Trunc, iPE, RTFA and PreAve over varying $n \in \{100, 200, 500\}$,  the distribution for $\mc F_t$ and $\bm\xi_t$ (Gaussian and $t_3$) and the percentage of outliers in the idiosyncratic component under~\ref{o:one} ($\varrho \in \{0, 0.1, 0.5, 1\}$).
We report the mean and the standard deviation over $100$ realisations per each setting.}
\label{tab:tensor:r:three:100}
\centering
%\resizebox{\textwidth}{!}
{\footnotesize
\begin{tabular}{rrr cc cc cc cc cc}
\toprule
&	&	&	\multicolumn{2}{c}{Trunc} &		\multicolumn{2}{c}{iPE} &		\multicolumn{2}{c}{RTFA} & 		\multicolumn{2}{c}{PreAve} 		\\	
Dist &	$\%$ &	Mode &	Mean &	SD &	Mean &	SD &	Mean &	SD &	Mean &	SD 	\\	\cmidrule(lr){1-3} \cmidrule(lr){4-5} \cmidrule(lr){6-7} \cmidrule(lr){8-9} \cmidrule(lr){10-11} \cmidrule(lr){12-13}
Gaussian &	0 &	1 &	2.98 &	0.2 &	3 &	0 &	3 &	0 &	3 &	0.000	\\	
&	&	2 &	3 &	0 &	3 &	0 &	3 &	0 &	3 &	0.000	\\	
&	&	3 &	3 &	0 &	3 &	0 &	3 &	0 &	3 &	0.000	\\	
&	0.1 &	1 &	2.97 &	0.223 &	3 &	0 &	3 &	0 &	3 &	0.000	\\	
&	&	2 &	3 &	0 &	3 &	0 &	3 &	0 &	3 &	0.000	\\	
&	&	3 &	3 &	0 &	3 &	0 &	3 &	0 &	3 &	0.000	\\	
&	0.5 &	1 &	2.97 &	0.223 &	3 &	0 &	3 &	0 &	3 &	0.000	\\	
&	&	2 &	3 &	0 &	3 &	0 &	3 &	0 &	3 &	0.000	\\	
&	&	3 &	3 &	0 &	3 &	0 &	3 &	0 &	3 &	0.000	\\	
&	1 &	1 &	2.97 &	0.223 &	2.98 &	0.2 &	2.98 &	0.2 &	3 &	0.000	\\	
&	&	2 &	3 &	0 &	3 &	0 &	3 &	0 &	3 &	0.000	\\	
&	&	3 &	3 &	0 &	3 &	0 &	3 &	0 &	3 &	0.000	\\	\cmidrule(lr){1-3} \cmidrule(lr){4-5} \cmidrule(lr){6-7} \cmidrule(lr){8-9} \cmidrule(lr){10-11} \cmidrule(lr){12-13}
t&	0 &	1 &	2.99 &	0.1 &	3.03 &	0.223 &	3.03 &	0.223 &	3 &	0.000	\\	
&	&	2 &	3 &	0 &	3.03 &	0.223 &	3.03 &	0.223 &	3 &	0.000	\\	
&	&	3 &	3 &	0 &	3.03 &	0.223 &	3.03 &	0.223 &	3 &	0.000	\\	
&	0.1 &	1 &	2.97 &	0.223 &	3.02 &	0.2 &	3.02 &	0.2 &	2.97 &	0.171	\\	
&	&	2 &	2.98 &	0.2 &	3.03 &	0.223 &	3.02 &	0.2 &	3 &	0.000	\\	
&	&	3 &	3 &	0 &	3.03 &	0.223 &	3.02 &	0.2 &	3 &	0.000	\\	
&	0.5 &	1 &	2.99 &	0.301 &	2.99 &	0.301 &	2.99 &	0.301 &	2.98 &	0.141	\\	
&	&	2 &	2.99 &	0.225 &	3 &	0.284 &	3 &	0.284 &	2.99 &	0.100	\\	
&	&	3 &	3.02 &	0.2 &	3.02 &	0.2 &	3.02 &	0.2 &	3 &	0.000	\\	
&	1 &	1 &	2.96 &	0.243 &	2.92 &	0.394 &	2.94 &	0.343 &	2.99 &	0.100	\\	
&	&	2 &	2.98 &	0.2 &	2.99 &	0.225 &	3.01 &	0.1 &	3.02 &	0.200	\\	
&	&	3 &	3 &	0 &	3.01 &	0.1 &	3.01 &	0.1 &	3.01 &	0.100	\\	\bottomrule
\end{tabular}
}
\end{table}

\begin{table}[h!t!p!]
\caption{\ref{f:three} with $n = 200$.}
\label{tab:tensor:r:three:200}
\centering
%\resizebox{\textwidth}{!}
{\footnotesize
\begin{tabular}{rrr cc cc cc cc cc}
\toprule
&	&	&	\multicolumn{2}{c}{Trunc} &		\multicolumn{2}{c}{iPE} &		\multicolumn{2}{c}{RTFA} & 		\multicolumn{2}{c}{PreAve} 		\\	
Dist &	$\%$ &	Mode &	Mean &	SD &	Mean &	SD &	Mean &	SD &	Mean &	SD 	\\	\cmidrule(lr){1-3} \cmidrule(lr){4-5} \cmidrule(lr){6-7} \cmidrule(lr){8-9} \cmidrule(lr){10-11} \cmidrule(lr){12-13}
Gaussian &	0 &	1 &	3 &	0 &	3 &	0 &	3 &	0 &	3 &	0.000	\\	
&	&	2 &	3 &	0 &	3 &	0 &	3 &	0 &	3 &	0.000	\\	
&	&	3 &	3 &	0 &	3 &	0 &	3 &	0 &	3 &	0.000	\\	
&	0.1 &	1 &	3 &	0 &	3 &	0 &	3 &	0 &	3 &	0.000	\\	
&	&	2 &	3 &	0 &	3 &	0 &	3 &	0 &	3 &	0.000	\\	
&	&	3 &	3 &	0 &	3 &	0 &	3 &	0 &	3 &	0.000	\\	
&	0.5 &	1 &	3 &	0 &	3 &	0 &	3 &	0 &	3 &	0.000	\\	
&	&	2 &	3 &	0 &	3 &	0 &	3 &	0 &	3 &	0.000	\\	
&	&	3 &	3 &	0 &	3 &	0 &	3 &	0 &	3 &	0.000	\\	
&	1 &	1 &	3 &	0 &	3 &	0 &	3 &	0 &	3 &	0.000	\\	
&	&	2 &	3 &	0 &	3 &	0 &	3 &	0 &	3 &	0.000	\\	
&	&	3 &	3 &	0 &	3 &	0 &	3 &	0 &	3 &	0.000	\\	\cmidrule(lr){1-3} \cmidrule(lr){4-5} \cmidrule(lr){6-7} \cmidrule(lr){8-9} \cmidrule(lr){10-11} \cmidrule(lr){12-13}
t&	0 &	1 &	2.98 &	0.141 &	3 &	0 &	3 &	0 &	2.99 &	0.100	\\	
&	&	2 &	3 &	0 &	3 &	0 &	3 &	0 &	2.99 &	0.100	\\	
&	&	3 &	3 &	0 &	3 &	0 &	3 &	0 &	3 &	0.000	\\	
&	0.1 &	1 &	2.96 &	0.243 &	2.97 &	0.223 &	2.97 &	0.223 &	2.99 &	0.100	\\	
&	&	2 &	3 &	0 &	3 &	0 &	3 &	0 &	2.99 &	0.100	\\	
&	&	3 &	3 &	0 &	3 &	0 &	3 &	0 &	3 &	0.000	\\	
&	0.5 &	1 &	2.96 &	0.243 &	2.93 &	0.326 &	2.93 &	0.326 &	2.99 &	0.100	\\	
&	&	2 &	3 &	0 &	2.98 &	0.2 &	2.98 &	0.2 &	2.99 &	0.100	\\	
&	&	3 &	3 &	0 &	2.98 &	0.2 &	2.98 &	0.2 &	3 &	0.000	\\	
&	1 &	1 &	2.95 &	0.297 &	2.91 &	0.379 &	2.92 &	0.367 &	2.98 &	0.141	\\	
&	&	2 &	2.98 &	0.2 &	2.98 &	0.2 &	2.98 &	0.2 &	2.99 &	0.100	\\	
&	&	3 &	2.98 &	0.2 &	2.98 &	0.2 &	2.98 &	0.2 &	3.02 &	0.141	\\	\bottomrule
\end{tabular}
}
\end{table}

\begin{table}[h!t!p!]
\caption{\ref{f:three} with $n = 500$.}
\label{tab:tensor:r:three:500}
\centering
%\resizebox{\textwidth}{!}
{\footnotesize
\begin{tabular}{rrr cc cc cc cc cc}
\toprule
&	&	&	\multicolumn{2}{c}{Trunc} &		\multicolumn{2}{c}{iPE} &		\multicolumn{2}{c}{RTFA} & 		\multicolumn{2}{c}{PreAve} 		\\	
Dist &	$\%$ &	Mode &	Mean &	SD &	Mean &	SD &	Mean &	SD &	Mean &	SD 	\\	\cmidrule(lr){1-3} \cmidrule(lr){4-5} \cmidrule(lr){6-7} \cmidrule(lr){8-9} \cmidrule(lr){10-11} \cmidrule(lr){12-13}
Gaussian &	0 &	1 &	3 &	0 &	3 &	0 &	3 &	0 &	3 &	0.000	\\	
&	&	2 &	3 &	0 &	3 &	0 &	3 &	0 &	3 &	0.000	\\	
&	&	3 &	3 &	0 &	3 &	0 &	3 &	0 &	3 &	0.000	\\	
&	0.1 &	1 &	3 &	0 &	3 &	0 &	3 &	0 &	3 &	0.000	\\	
&	&	2 &	3 &	0 &	3 &	0 &	3 &	0 &	3 &	0.000	\\	
&	&	3 &	3 &	0 &	3 &	0 &	3 &	0 &	3 &	0.000	\\	
&	0.5 &	1 &	3 &	0 &	3 &	0 &	3 &	0 &	3 &	0.000	\\	
&	&	2 &	3 &	0 &	3 &	0 &	3 &	0 &	3 &	0.000	\\	
&	&	3 &	3 &	0 &	3 &	0 &	3 &	0 &	3 &	0.000	\\	
&	1 &	1 &	3 &	0 &	3 &	0 &	3 &	0 &	3 &	0.000	\\	
&	&	2 &	3 &	0 &	3 &	0 &	3 &	0 &	3 &	0.000	\\	
&	&	3 &	3 &	0 &	3 &	0 &	3 &	0 &	3 &	0.000	\\	\cmidrule(lr){1-3} \cmidrule(lr){4-5} \cmidrule(lr){6-7} \cmidrule(lr){8-9} \cmidrule(lr){10-11} \cmidrule(lr){12-13}
t&	0 &	1 &	2.99 &	0.1 &	3 &	0 &	3 &	0 &	2.98 &	0.141	\\	
&	&	2 &	3 &	0 &	3 &	0 &	3 &	0 &	3 &	0.000	\\	
&	&	3 &	3 &	0 &	3 &	0 &	3 &	0 &	3 &	0.000	\\	
&	0.1 &	1 &	2.99 &	0.1 &	3 &	0 &	3 &	0 &	2.98 &	0.141	\\	
&	&	2 &	3 &	0 &	3 &	0 &	3 &	0 &	3 &	0.000	\\	
&	&	3 &	3 &	0 &	3 &	0 &	3 &	0 &	3 &	0.000	\\	
&	0.5 &	1 &	2.98 &	0.141 &	2.96 &	0.197 &	2.96 &	0.197 &	2.98 &	0.141	\\	
&	&	2 &	3 &	0 &	3 &	0 &	3 &	0 &	3 &	0.000	\\	
&	&	3 &	3 &	0 &	3 &	0 &	3 &	0 &	3 &	0.000	\\	
&	1 &	1 &	2.97 &	0.171 &	2.89 &	0.373 &	2.89 &	0.373 &	2.98 &	0.141	\\	
&	&	2 &	3 &	0 &	2.99 &	0.1 &	2.99 &	0.1 &	3 &	0.000	\\	
&	&	3 &	3 &	0 &	3 &	0 &	3 &	0 &	3.01 &	0.100	\\	\bottomrule
\end{tabular}
}
\end{table}

\clearpage

\subsubsection{Asymptotic normality}
\label{app:sim:infer}

We investigate the asymptotic normality of the loading estimator $\wc{\bm\Lambda}^{[2]}_k$.
For this, we generate the tensor time series as described in Section~\ref{app:sim:tensor:setting} but to avoid the estimation of the rotation matrix $\wc{\mbf H}^{[2]}_k$, set $(r_1, r_2, r_3) = (1, 1, 1)$ and scale (now vectors) $\bm\Lambda_k = (\lambda_{k, 11}, \ldots, \lambda_{k, p_k1})^\top \in \R^{p_k}$, to have $\vert \bm\Lambda_k \vert_2 = 1$ for all $k \in [K]$.
Also, to study the validity of the asymptotic normality without being hampered by the difficulties from estimating the long-run variance in the limit in Theorem~\ref{thm:third}~\ref{thm:third:two}, particularly in the presence of heavy tails, we set $\phi = \psi = 0$.

\begin{figure}[h!t!p!]
\centering
\includegraphics[width = 1\textwidth]{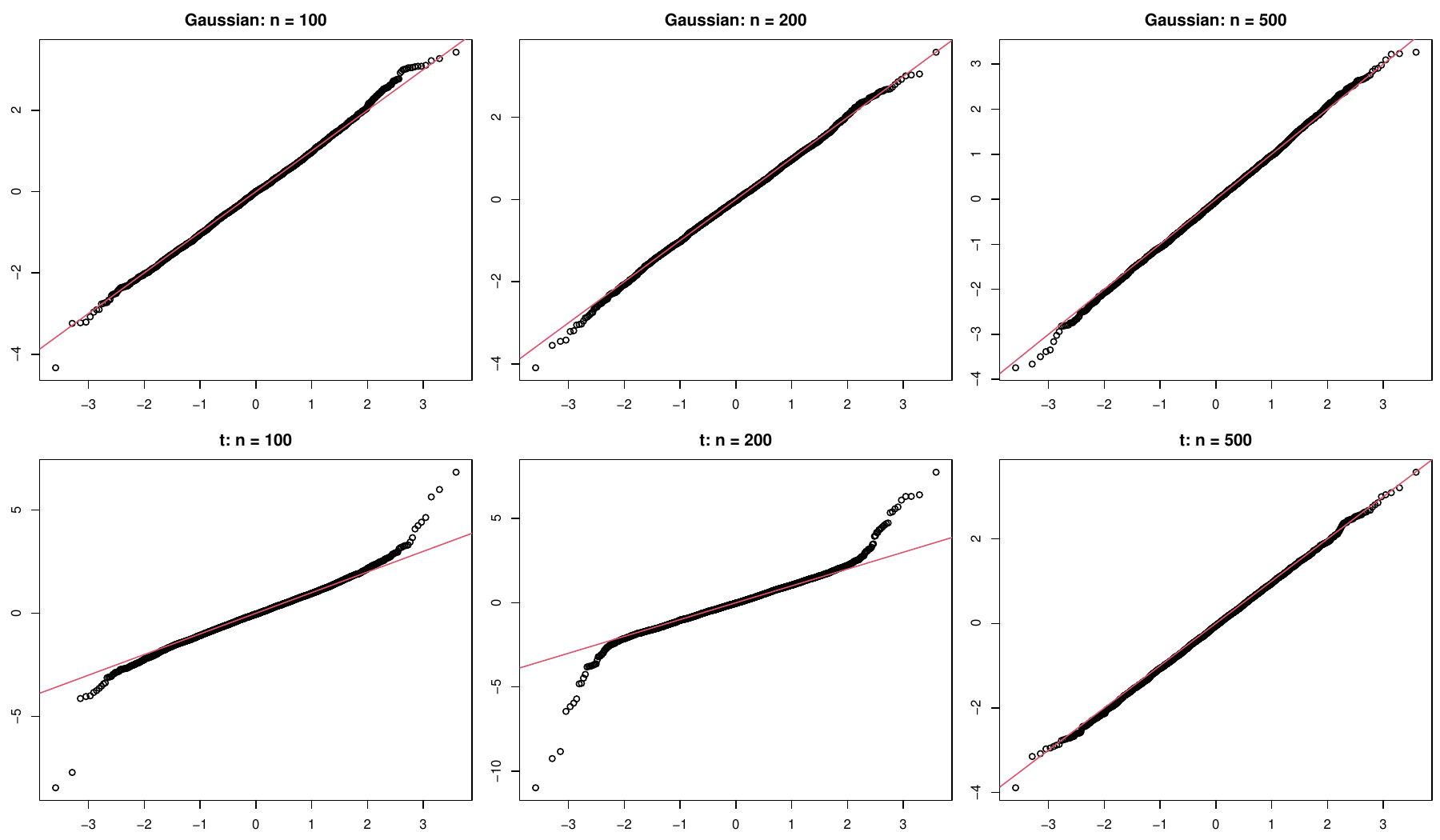}
\caption{\ref{f:one} Q-Q plot of the sample quantiles of $Z_{b, k, i}, \, i \in [p_k], \, k \in [K], \, b \in [B]$ ($y$-axis), against the quantiles from the standard normal distribution ($x$-axis) over varying $n \in \{100, 200, 500\}$ (left to right) and the distributions for $\mc F_t$ and $\bm\xi_t$ (Gaussian and $t_3$, top to bottom) over $B = 100$ realisations per setting. In each plot, the $y = x$ line is given in red. }
\label{fig:tensor:asymp:one}
\end{figure}

\begin{figure}[h!t!p!]
\centering
\includegraphics[width = 1\textwidth]{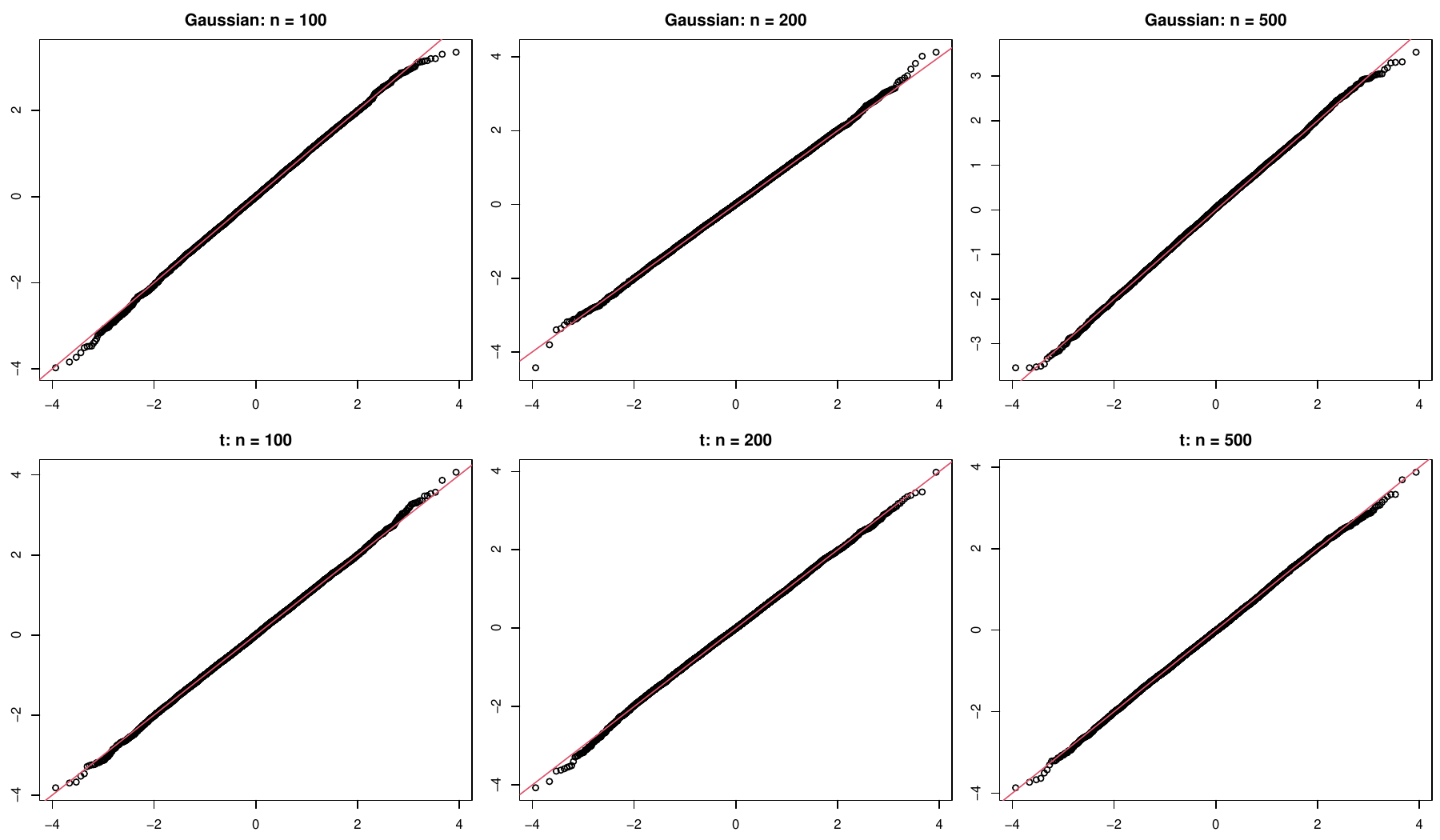}
\caption{\ref{f:two} Q-Q plot of the sample quantiles of $Z_{b, k, i}, \, i \in [p_k], \, k \in [K], \, b \in [B]$ ($y$-axis), against the quantiles from the standard normal distribution ($x$-axis) over varying $n \in \{100, 200, 500\}$ (left to right) and the distributions for $\mc F_t$ and $\bm\xi_t$ (Gaussian and $t_3$, top to bottom) over $B = 100$ realisations per setting. In each plot, the $y = x$ line is given in red. }
\label{fig:tensor:asymp:two}
\end{figure}

\begin{figure}[h!t!p!]
\centering
\includegraphics[width = 1\textwidth]{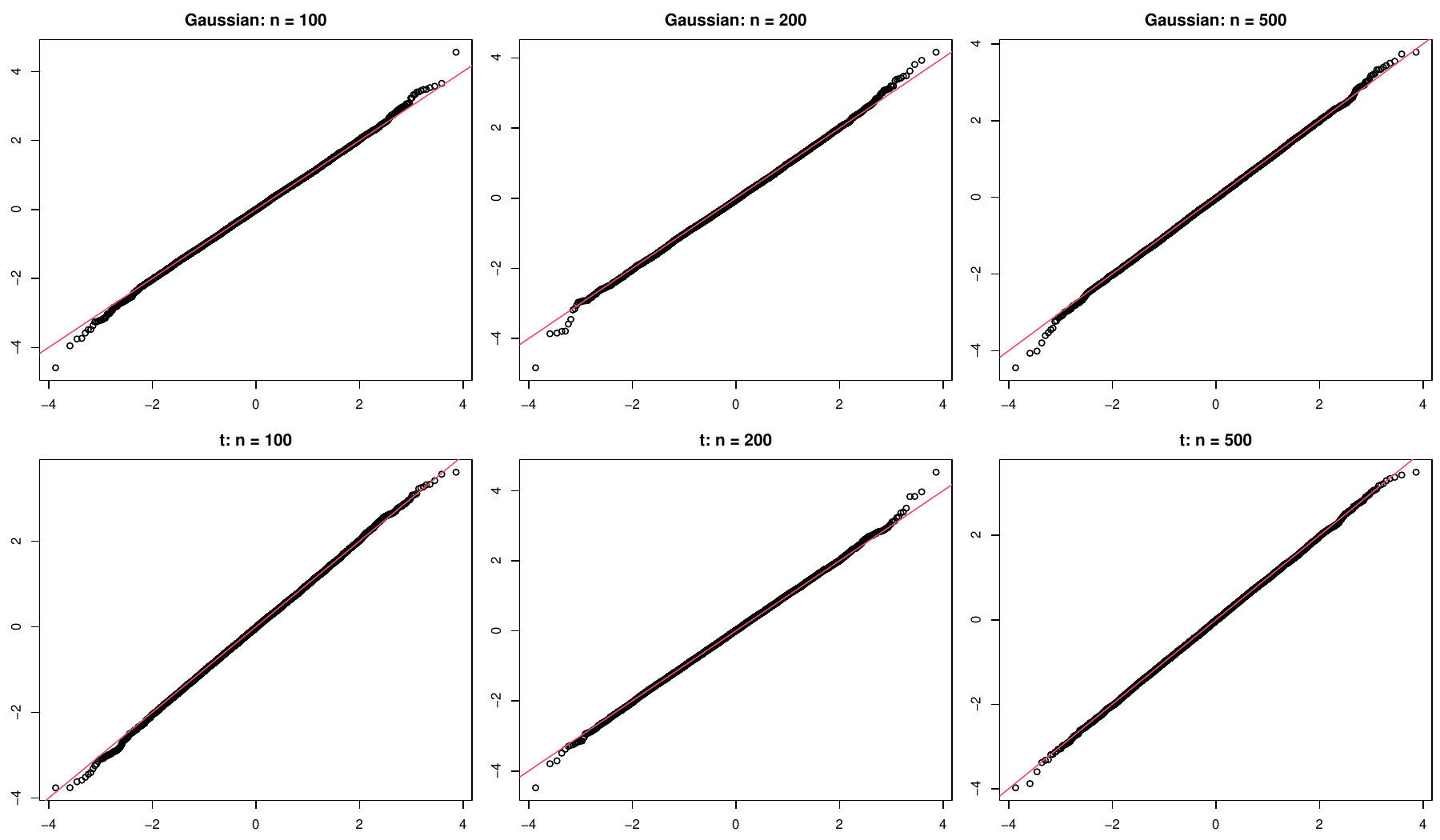}
\caption{\ref{f:three} Q-Q plot of the sample quantiles of $Z_{b, k, i}, \, i \in [p_k], \, k \in [K], \, b \in [B]$ ($y$-axis), against the quantiles from the standard normal distribution ($x$-axis) over varying $n \in \{100, 200, 500\}$ (left to right) and the distributions for $\mc F_t$ and $\bm\xi_t$ (Gaussian and $t_3$, top to bottom) over $B = 100$ realisations per setting. In each plot, the $y = x$ line is given in red. }
\label{fig:tensor:asymp:three}
\end{figure}

For each realisation, we compute
\begin{align*}
Z_{k, i} = \frac{\sqrt{np_{-k}}(\wc{\lambda}^{[2]}_{k, i}(\tau) - s_k \lambda_{k, i1})}{(\wh{\Phi}^\kk_i(\tau))^{1/2}} \text{ \ for all \ } i \in [p_k] \text{ \ and \ } k \in [K], 
\end{align*}
where $s_k \in \{-1, 1\}$ denotes the sign set to be $\text{median}_{i \in [p_k]} \, \mathsf{sign}(\wc{\lambda}^{[2]}_{k, i}(\tau) \cdot \lambda_{k, i1})$.
The variance estimator $\wh\Phi^\kk_i(\tau)$ is obtained as
\begin{align*}
\wh\Phi^\kk_i(\tau) = (\wh{\bm\Gamma}^\kk_f)^{-1} \wh{\Var}\l[ \l( \mat_k(\mc X^\trunc_t(\tau))_{i \cdot} - \mat_k(\wh{\bm\chi}^\trunc_t(\tau))_{i \cdot} \r) \wc{\mbf D}^{[2]}_k \r] (\wh{\bm\Gamma}^\kk_f)^{-1},
\end{align*}
where $\wh{\bm\Gamma}^\kk_f = n^{-1} \sum_{t \in [n]} \wh{\mc F}_t^2$, $\wh{\bm\chi}_t$ denotes the estimator of the common component and $\wh{\bm\chi}^\trunc_t(\tau)$ its truncated version, $\wc{\mbf D}^{[2]}_k = p_{-k}^{-1/2} \wc{\bm\Lambda}^{[2]}_K \otimes \ldots \otimes \wc{\bm\Lambda}^{[2]}_{k + 1} \otimes \wc{\bm\Lambda}^{[2]}_{k - 1} \otimes \ldots \otimes \wc{\bm\Lambda}^{[2]}_1$, and $\wh{\Var}(\cdot)$ denotes the sample variance operator.
Repeating the experiments over $B = 100$ realisations, we generate the Q-Q plot of $Z_{b, k, i}, \, i \in [p_k], \, k \in [K], \, b \in [B]$, against the standard normal distribution for each setting while varying $(p_1, p_2, p_3)$ under \ref{f:one}--\ref{f:three}, distributions of $\bm\xi_t$ and $\mc F_t$ (Gaussian and $t_3$) and $n \in \{100, 200, 500\}$, see Figures~\ref{fig:tensor:asymp:one}--\ref{fig:tensor:asymp:three}.

We observe that as $n$ (within each figure) and $p_k$'s (from~\ref{f:one} to~\ref{f:three}) increase, the approximation by the standard normal distribution improves, even when the data are generated from the $t_3$ distribution.
While the asymptotic normality of the loading estimator in Theorem~\ref{thm:third}~\ref{thm:third:two} is derived under spatial independence, the mild cross-sectional dependence introduced as described in Appendix~\ref{app:sim:tensor:setting} does not influence the approximation greatly.

\clearpage 

\subsubsection{Additional simulation results}
\label{app:sim:add}

\begin{figure}[h!t!b!]
\begin{center}
\begin{tabular}{c}
\includegraphics[width = .9\textwidth]{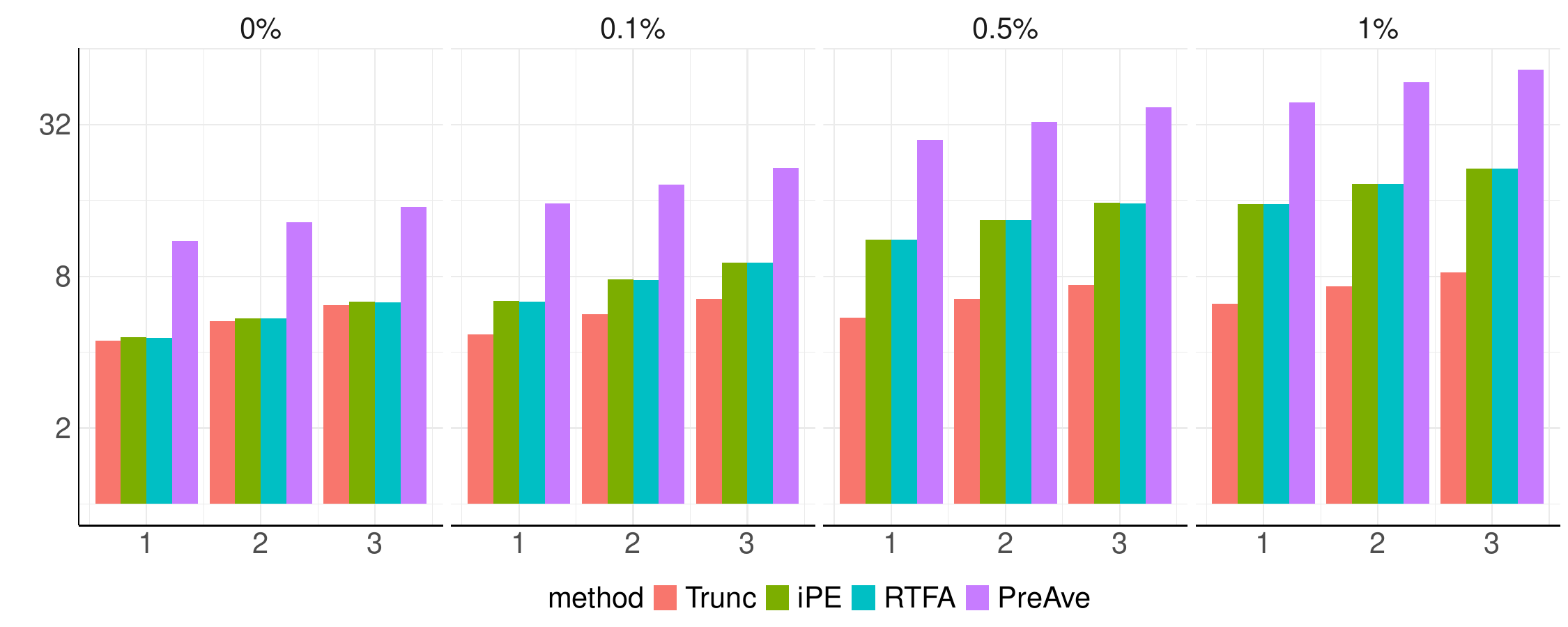}
\\
\includegraphics[width = .9\textwidth]{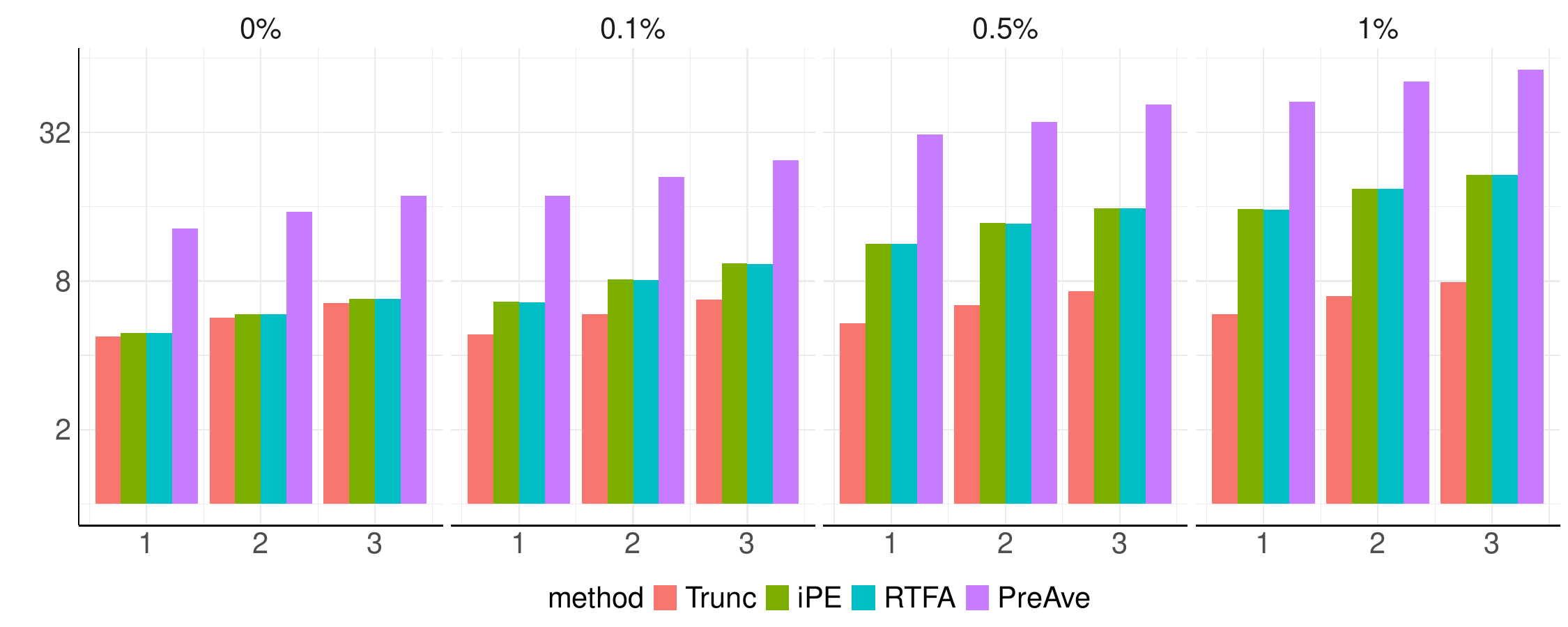}
\end{tabular}
\end{center}
\caption{\ref{f:three} with $(p_1, p_2, p_3) = (20, 30, 40)$: Loading estimation errors measured as in~\eqref{eq:err:loading:tensor} for each mode ($x$-axis) for Trunc, iPE, RTFA and PreAve averaged over $100$ realisations per setting, over varying $\varrho \in \{0, 0.1, 0.5, 1\}$ when $n = 200$ and $\mc F_t$ and $\bm\xi_t$ are generated from scaled $t_3$ distributions, with $\phi$ in~\eqref{eq:ar:f} set to be $\phi = 0.7$ (top) and $\phi = 0.9$ (bottom).
The $y$-axis is on a log scale and all errors have been scaled for better presentation.}
\label{fig:phi:loading:err}
\end{figure}

\paragraph{Stronger serial dependence.}
We investigate the performance of Trunc against its competitors when the degree of serial dependence in the tensor time series is stronger.
For this, we focus on the model~\ref{f:three} with fixed $n = 200$ and $\mc F_t$ and $\bm\xi_t$ generated from scaled $t_3$ distributions, where stronger serial dependence is introduced to $\mc F_t$ by considering $\phi \in \{0.7, 0.9\}$ in~\eqref{eq:ar:f}.
See Figures~\ref{fig:phi:loading:err} and~\ref{fig:phi:loading:err} where we observe that the proposed Trunc exhibits good performance regardless of $\phi$ and thus is insensitive to the degree of serial dependence.  

\begin{figure}[h!t!b!]
\begin{center}
\begin{tabular}{c}
\includegraphics[width = .9\textwidth]{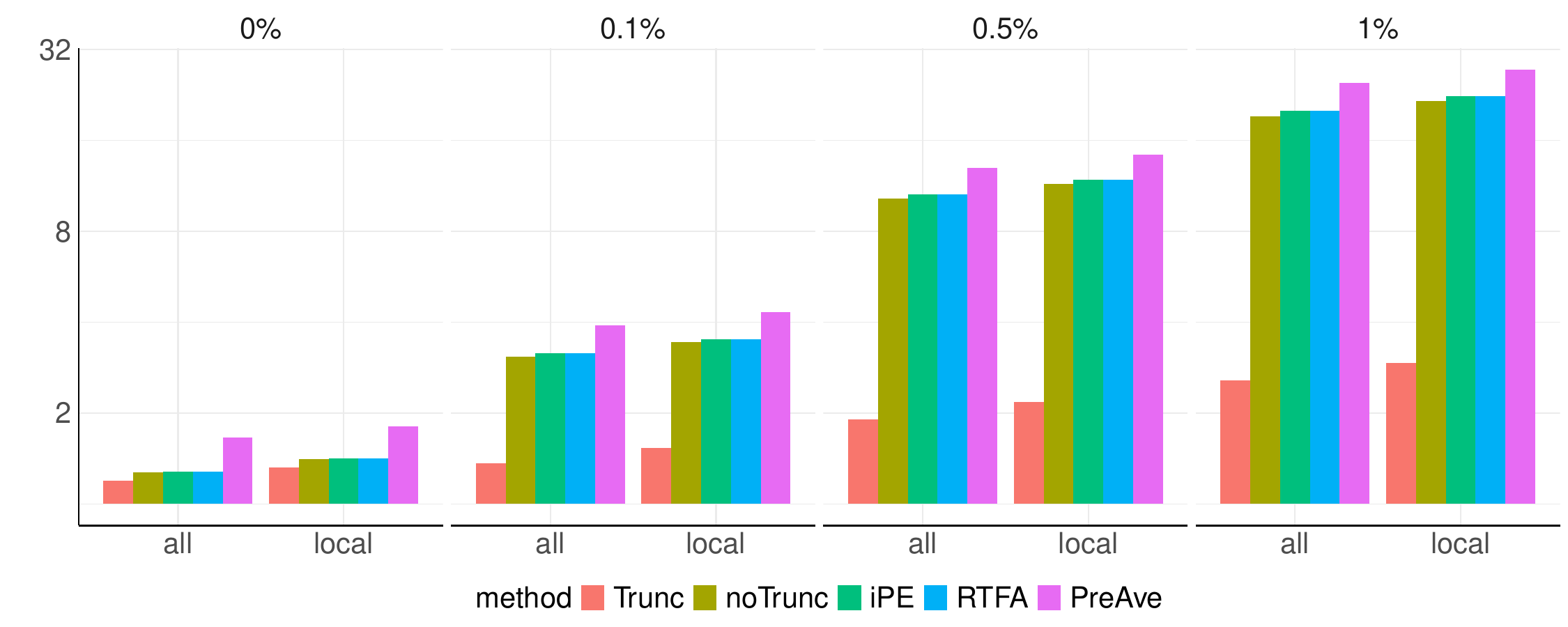}
\\
\includegraphics[width = .9\textwidth]{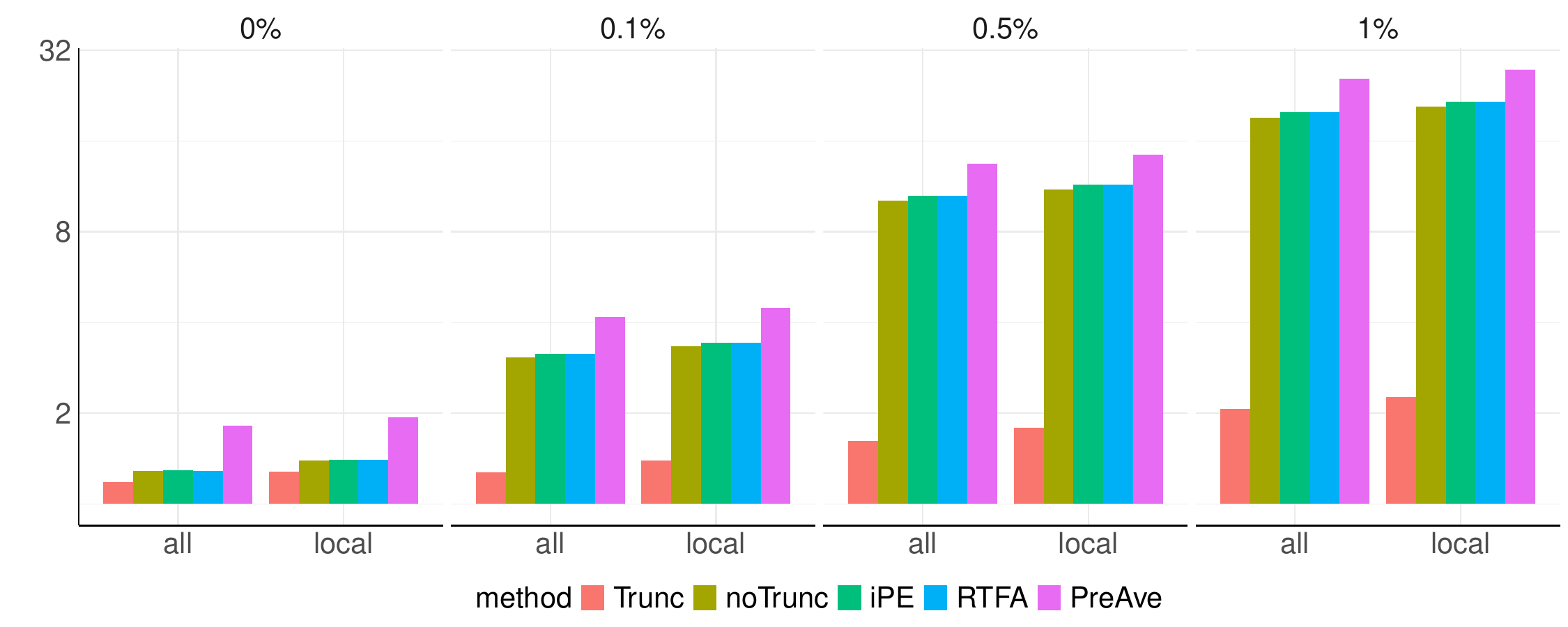}
\end{tabular}
\end{center}
\caption{\ref{f:three} with $(p_1, p_2, p_3) = (20, 30, 40)$: Common component estimation errors measured as in~\eqref{eq:err:chi:tensor} for each mode ($x$-axis) for Trunc, noTrunc, iPE, RTFA and PreAve averaged over $100$ realisations per setting, over varying $\varrho \in \{0, 0.1, 0.5, 1\}$ when $n = 200$ and $\mc F_t$ and $\bm\xi_t$ are generated from scaled $t_3$ distributions, with $\phi$ in~\eqref{eq:ar:f} set to be $\phi = 0.7$ (top) and $\phi = 0.9$ (bottom).
The $y$-axis is on a log scale and all errors have been scaled for better presentation.}
\label{fig:phi:common:err}
\end{figure}

\paragraph{Convergence of the iteratively projected loading estimators.}

\begin{figure}[h!t!b!]
\begin{center}
\includegraphics[width = 1\textwidth]{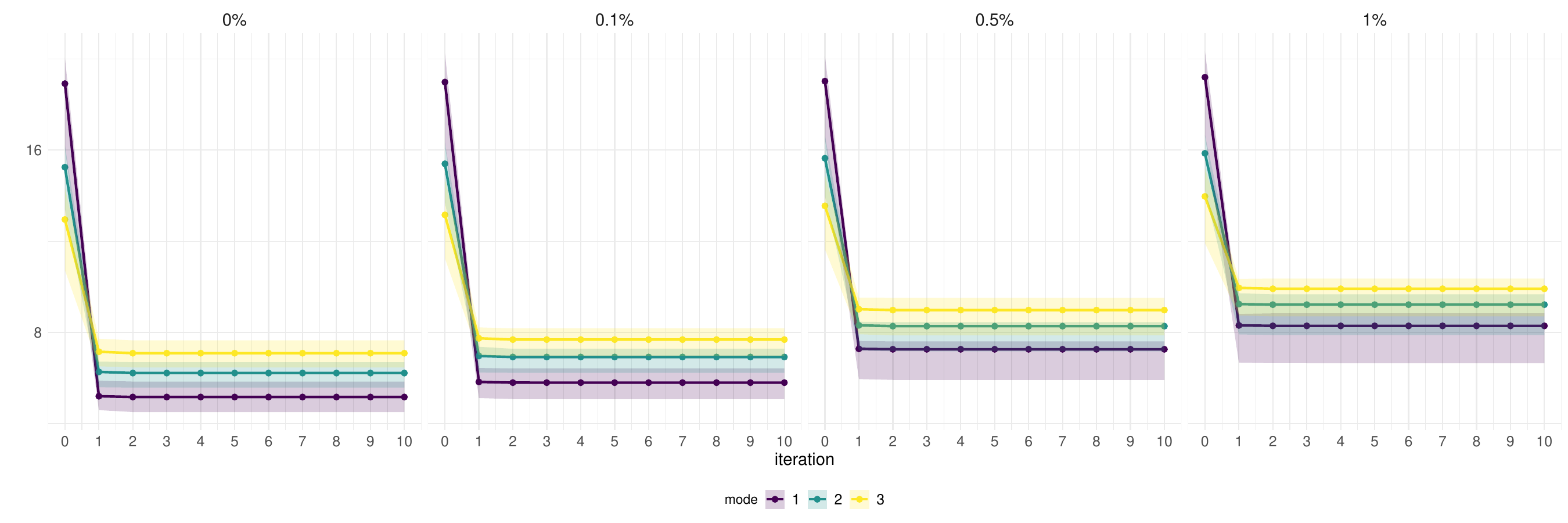}
\end{center}
\caption{\ref{f:three} with $(p_1, p_2, p_3) = (20, 30, 40)$, $n = 200$ and $\mc F_t$ and $\bm\xi_t$ generated from scaled $t_3$ distributions: Loading estimation errors measured as in~\eqref{eq:err:loading:tensor} for $\wc{\bm\Lambda}^{[\iota]}_k$ for $\iota \in \{0, 1, \ldots, 10\}$ ($x$-axis) over $100$ realisations per setting, with varying $\varrho \in \{0, 0.1, 0.5, 1\}$ (left to right).
We display mean error curves with the shaded regions representing the interquartile range. 
The $y$-axis is on a log scale and all errors have been scaled for better presentation.}
\label{fig:iter}
\end{figure}

\begin{figure}[h!t!b!]
\begin{center}
\includegraphics[width = 1\textwidth]{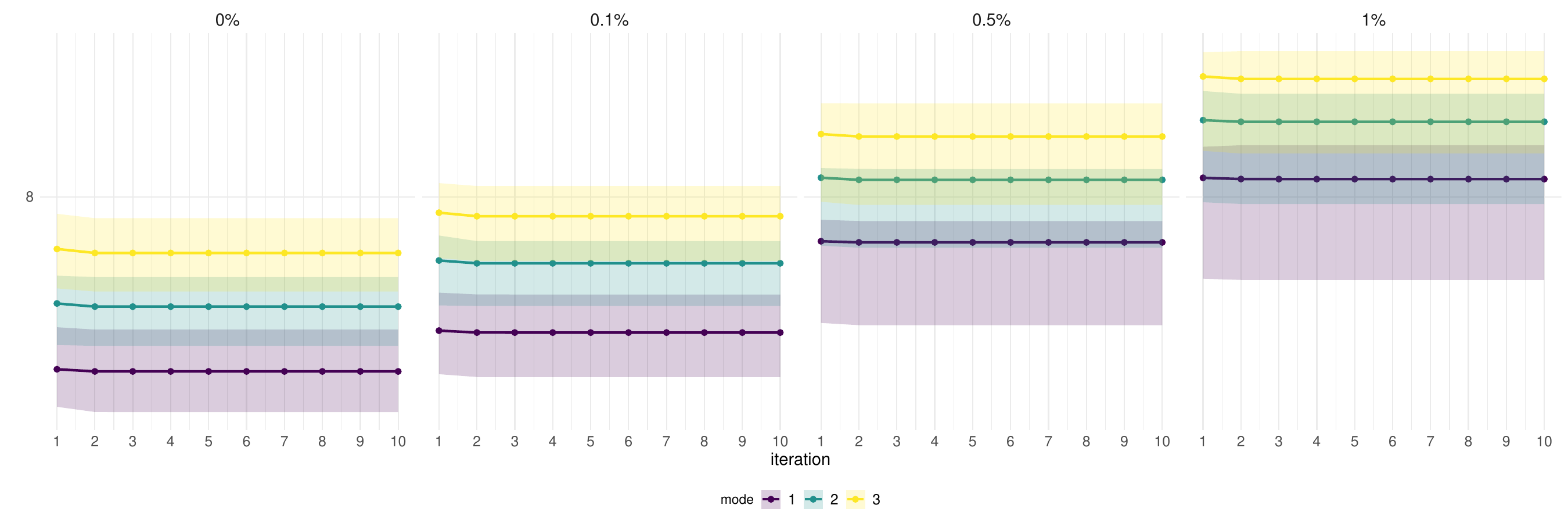}
\end{center}
\caption{Zoomed in version of Figure~\ref{fig:iter} where we display the estimation errors from $\wc{\bm\Lambda}^{[\iota]}_k$ for $\iota \in \{1, \ldots, 10\}$ ($x$-axis).}
\label{fig:iter:zoom}
\end{figure}

We investigate the convergence behaviour of the iterative estimator proposed in Section~\ref{sec:pc}.
Focusing on the model~\ref{f:three}, we fix $n = 200$ and the distributions of $\mc F_t$ and $\bm\xi_t$ to be the (scaled) $t_3$, and report the loading estimation errors as measured in~\eqref{eq:err:loading:tensor} for $\wc{\bm\Lambda}^{[0]}_k = \wh{\bm\Lambda}_k$ (initial estimator) and $\wc{\bm\Lambda}^\ii_k$ for $\iota \in \{1, 2, \ldots, 10\}$ (we omit the dependence on $\tau$ for simplicity), see Figures~\ref{fig:iter}--\ref{fig:iter:zoom}.
The results indicate that after the initial large drop between the errors of $\wc{\bm\Lambda}^{[0]}_k$ and $\wc{\bm\Lambda}^{[1]}_k$, and the slight one between those of $\wc{\bm\Lambda}^{[1]}_k$ and $\wc{\bm\Lambda}^{[2]}_k$, there is little difference in the performance of $\wc{\bm\Lambda}^\ii_k, \, \iota \in \{2, \ldots, 10\}$, which supports our proposal of using the twice iterated estimator.

\begin{figure}[h!t!b!]
\begin{center}
\begin{tabular}{c}
\includegraphics[width = 1\textwidth]{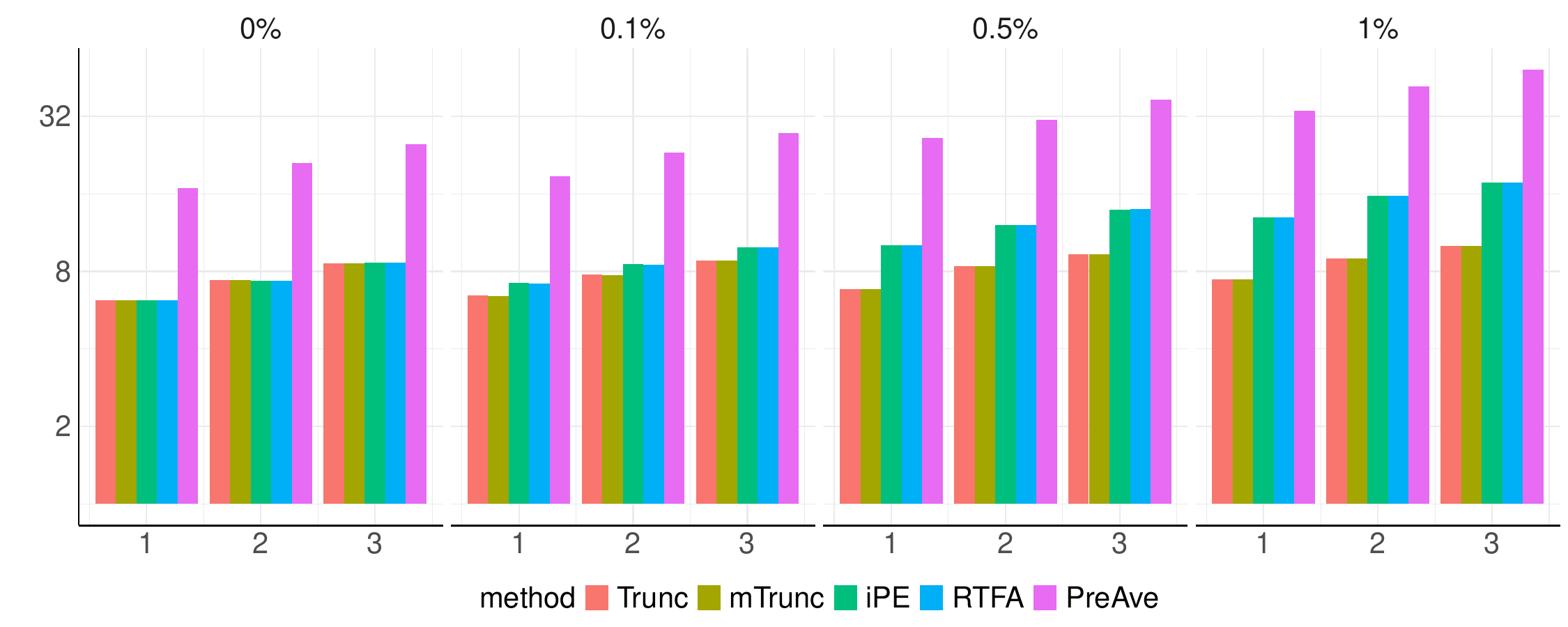}
\\
\includegraphics[width = 1\textwidth]{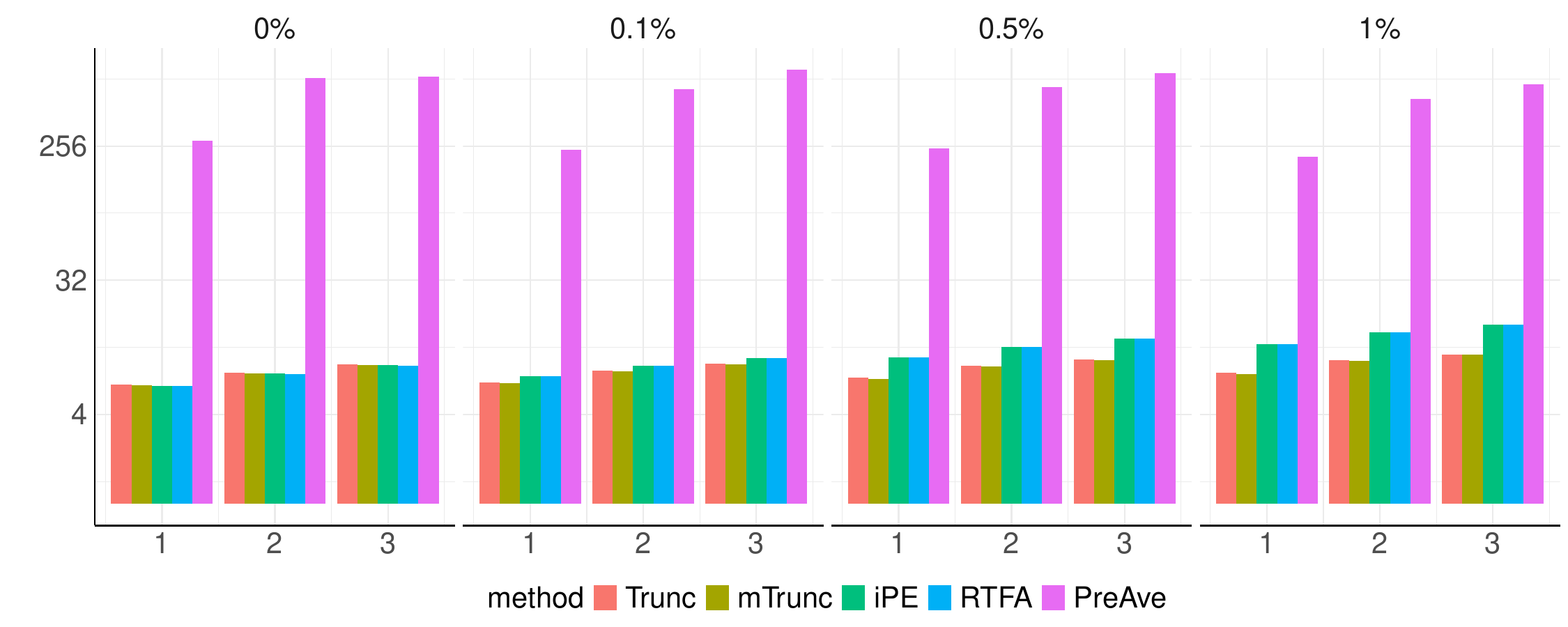}
\\
\includegraphics[width = 1\textwidth]{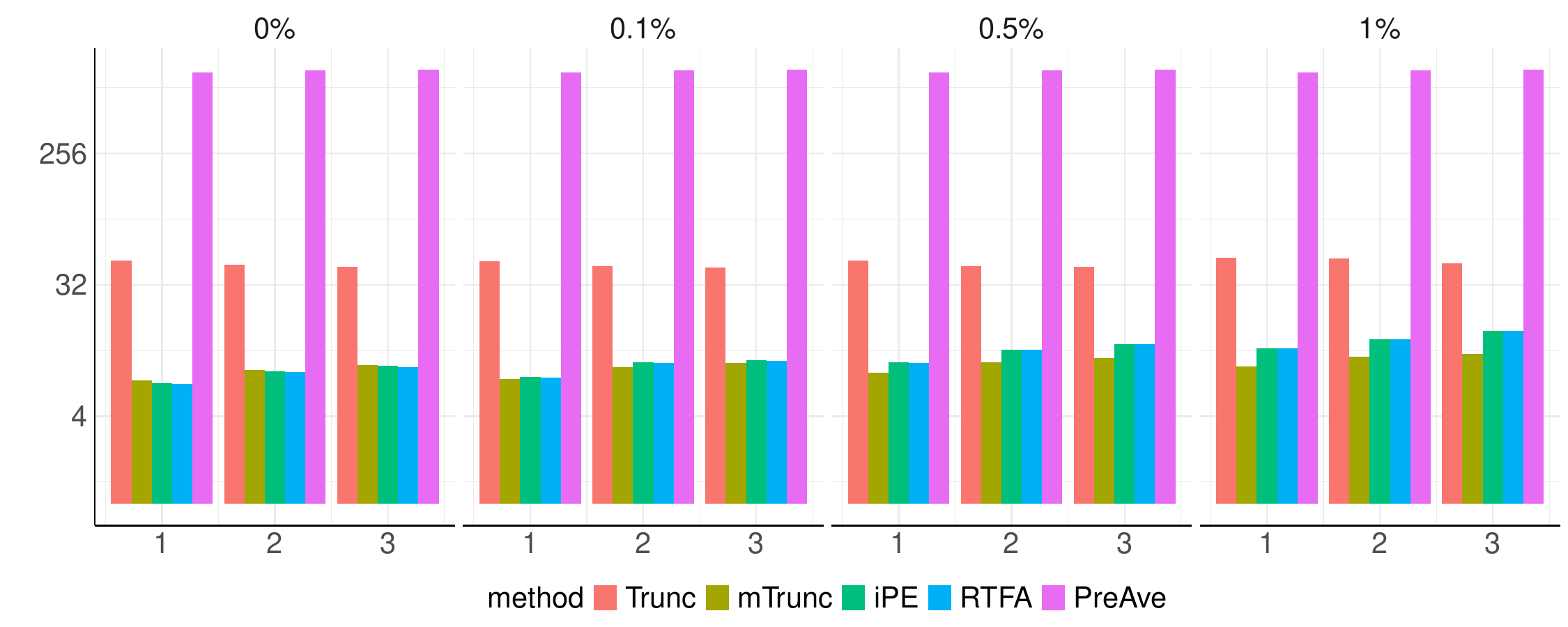}
\end{tabular}
\end{center}
\caption{Toeplitz scenarios with $n = 200$, $(p_1, p_2, p_3) = (20, 30, 40)$ and $\mc F_t$ and $\bm\xi_t$ are generated from scaled $t_3$ distributions: Loading estimation errors measured as in~\eqref{eq:err:loading:tensor} for each mode ($x$-axis) for Trunc, mTrunc, iPE, RTFA and PreAve averaged over $100$ realisations per setting, over varying $\varrho \in \{0, 0.1, 0.5, 1\}$ (left to right) and $\varphi \in \{0.5, 0.7, 0.9\}$ (top to bottom).
The $y$-axis is on a log scale and all errors have been scaled for better presentation.}
\label{fig:toep}
\end{figure}

\paragraph{$\bm\Sigma_k$ as Toeplitz matrix.} 
Recall the data generating process for $\bm\xi_t$ from Appendix~\ref{app:sim:tensor:setting} (when no outlier is present):
\begin{align}
\vecop(\bm\xi_t) &= \psi \cdot \vecop(\bm\xi_{t - 1}) + \sqrt{1 - \psi^2}  \cdot \vecop(\mc V_t), 
\nn
\end{align}
where $\mc V_t \in \R^{p_1 \times p_2 \times p_3}$ such that $\vecop(\mc V_t) = \otimes_{k = K}^1 \bm\Sigma_k^{1/2} \mbf v_t$ and the entries of $\mbf v_t$ are i.i.d.
To further understand the impact of cross-sectional correlations on our estimators, we consider the situations where $\bm\Sigma_k$ is a Toeplitz matrix, i.e.\ $\bm\Sigma_k = [\varphi^{\vert i - i' \vert}, \, i, i' \in [p_k]]$, with $\varphi \in \{0.5, 0.7, 0.9\}$; such a scenario better conforms to the strongly mixing random field condition in Assumption~\ref{assum:rf}, compared to the one considered in the main text.
We fix $n = 200$, $(p_1, p_2, p_3) = (20, 30, 40)$, $(r_1, r_2, r_3) = (3, 3, 3)$ and generate $\mc F_t$ and $\mbf v_t$ from scaled $t_3$ distributions.
In addition to `Trunc', which refers to the twice-iterated estimator $\wc{\bm\Lambda}^{[2]}_k(\tau)$, we also present the results from the ten-times-iterated estimator $\wc{\bm\Lambda}^{[10]}_k(\tau)$ which is referred to as `mTrunc', see Figure~\ref{fig:toep} for the results.

\begin{figure}[h!t!b!]
\begin{center}
\includegraphics[width = 1\textwidth]{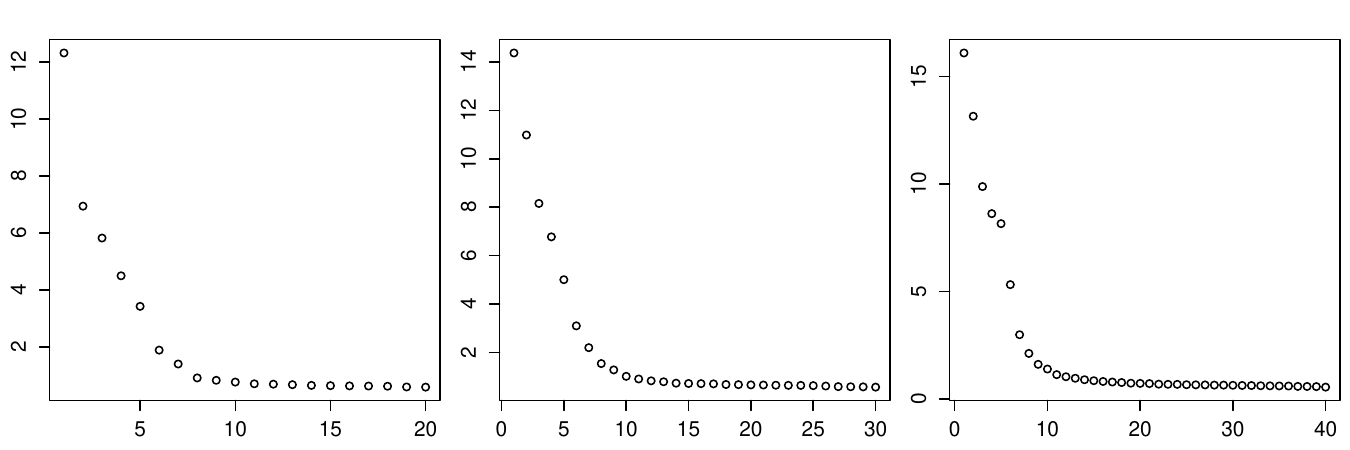}
\end{center}
\caption{Toeplitz scenarios with $n = 200$, $(p_1, p_2, p_3) = (20, 30, 40)$ and $\mc F_t$ and $\bm\xi_t$ are generated from scaled $t_3$ distributions: The ordered eigenvalues from $\wh{\bm\Gamma}^\kk(\tau)$ (initial estimator) % and $\wc{\bm\Gamma}^{\kk, [10]}(\tau)$ obtained with the true knowledge of $(r_1, r_2, r_3) = (3, 3, 3)$ (bottom) 
for $k \in \{1, 2, 3\}$ (left to right) on a single realisation when $\varphi = 0.9$ and $\varrho = 0.1$.}
\label{fig:toep:eigval}
\end{figure}

We observe that in the presence of weaker cross-sectional correlations in $\bm\xi_t$ with $\varphi \in \{0.5, 0.7\}$, Trunc is performing as competitively as, or better than the competitors including mTrunc.
This supports that in the presence of weak cross-sectional dependence, the additional iterations performed by mTrunc do not appear to make much difference, thus confirming the earlier observation on the fast convergence of the iteratively projected estimator.
On the other hand, with $\varphi = 0.9$, we see a large difference between the outputs from Trunc and mTrunc.
This is attributed to that such strong cross-sectional dependence in $\bm\xi_t$ may be regarded as being driven by extra `factors'; indeed all factor number estimators including ours tend to over-estimate the number of factors with $\wh r_k \ge 5$, when the true factor numbers are $(r_1, r_2, r_3) = (3, 3, 3)$, see also Figure~\ref{fig:toep:eigval}.
In other words, the data generating process with $\varphi = 0.9$ may be considered as an extreme situation and Trunc shows invariably good performance in the presence of sufficiently weak cross-sectional correlations.

\clearpage

\subsection{Vector time series}
\label{sec:sim:vec}

\subsubsection{Set-up}

\paragraph{Data generation.} We adopt the following vector time series factor model ($K = 1$) from \cite{ahn2013}:
\begin{align*}
X^\circ_{it} &= \sum_{j = 1}^r \lambda_{ij} f_{jt} + \xi_{it}, \quad \xi_{it} = \sqrt{\frac{1 - \rho^2}{1 + 2 J \beta^2}} e_{it}, 
\\
e_{it} &= \rho e_{i, t - 1} + (1 - \beta) v_{it} + \beta \cdot \sum_{\ell = \max(i - J, 1)}^{\min(i + J, p)} v_{\ell t}, \quad i \in [p], \, t \in [n],
\end{align*}
where $r = 3$ and $\lambda_{ij} \sim_{\iid} \mc N(0, 1)$. 
We set $\rho = \beta = J = 0$ in the `independent' scenario while $(\rho, \beta, J) = (0.5, 0.2, \max(10, p/20))$ in the `dependent' scenario. 
Following \cite{he2023huber}, we consider the five models for the generation of $f_{jt}$ and $v_{it}$:
\begin{enumerate}[label = (V\arabic*)]
\item \label{d:one} $f_{jt}$ and $v_{it}$ are i.i.d.\ random variables from the standard normal distribution.
% $\mbf f_t$ (resp.\ $\mbf v_t$) are i.i.d.\ random vectors from the multivariate Gaussian distribution $\mc N_r(\mbf 0, \mbf I_r)$ (resp.\ $\mc N_p(\mbf 0, \mbf I_p)$).

\item \label{d:two} $f_{jt}$ and $v_{it}$ are i.i.d.\ random variables from the scaled $t_3$ distribution such that $\Var(f_{jt}) = \Var(v_{it}) = 1$.

\item \label{d:three} $f_{jt}$ are generated as in~\ref{d:one} while $v_{it}$ are generated as in~\ref{d:two}.

\item \label{d:four} $f_{jt}$ and $v_{it}$ are i.i.d.\ random variables from symmetric $\alpha$-stable distribution with the skewness parameter set at $0$, the scale parameter at $1$, the location parameter at $0$ and the index parameter $\alpha = 1.9$ (we use the R package \verb+stabledist+ \citep{stabledist} for data generation).

\item \label{d:five} $f_{jt}$ are i.i.d.\ random variables from skewed $t_3$ distribution with the slant parameter $20$ (we use the R package \verb+sn+ \citep{sn} for data generation), while $v_{it}$ are generated as in~\ref{d:four}.
\end{enumerate}
For the stable distribution considered in~\ref{d:four} and~\ref{d:five}, the second moment does not exist with the choice of $\alpha = 1.9$.

With $X^\circ_{it}$ generated as above, we consider the situation where the observed data $X_{it}$ are contaminated by outliers. 
For this, we randomly select $\mc O \subset [p] \times [n]$ with its cardinality $\vert \mc O \vert = [\varrho n p]$.
Then for $(i, t) \in \mc O$, we set $X_{it} = s_{it} \cdot U_{it}$ with $s_{it} \sim_{\iid} \{-1, 1\}$ and $U_{it} \sim_{\iid} \text{Unif}[Q + 12, Q + 15]$ with $Q$ set to be the $\max(1 - 100/(np), 0.999)$-quantile of $\vert X^\circ_{it} \vert$, while $X_{it} = X^\circ_{it}$ otherwise; if $\varrho = 0$, we have $X_{it} = X^\circ_{it}$ for all $i$ and $t$.
%\cite{raymaekers2024challenges} remark on the difference between the contamination model and the concept of heavy-tailedness, but also note that `in general, we expect that estimators for heavy-tailed data can still perform reasonably well under cellwise contamination', which motivates our investigation into the effect of outliers on the performance of the proposed method.
We vary $n, p \in \{100, 200, 500\}$ and $\varrho \in \{0, 0.1, 0.5, 1\} \times 10^{-2}$.
 
\paragraph{Performance assessment.}
To assess the factor loadings space estimation performance,  for any estimator $\wh{\bm\Lambda}$ we compute
\begin{align}
\label{eq:err:loading:vec}
\text{Err}_\Lambda = \sqrt{1 - \text{tr}\l( \Pi_{\wh{\bm\Lambda}} \Pi_{\bm\Lambda} \r) / r}, \text{ \ where \ } \Pi_{\mbf A} = \mbf A (\mbf A^\top \mbf A)^{-1} \mbf A^\top.
\end{align}
To evaluate the quality in common component estimation,  for any estimator $\wh{\bm\chi}_t$ we evaluate
\begin{align}
\label{eq:err:chi:vec}
\text{Err}_\chi(\mc T) = \frac{\sum_{t \in \mc T} \vert \wh{\bm\chi}_t - \bm\chi_t \vert_2^2}{\sum_{t \in \mc T} \vert \bm\chi_t \vert_2^2} 
\end{align}
with $\mc T = [n]$ (`all') and $\mc T = \{n - 10 + 1, \ldots, n\}$ (`local'). 
% When the aim is to control the global error $\text{Err}_\chi([n])$, Theorem~\ref{thm:factor}~\ref{thm:factor:two} indicates that no truncation is required, and thus we set $\kappa = \max_{i, t} \vert X_{it} \vert$.
%For the controlling of the local estimation error, we perform the proposed cross validation.

\paragraph{Competitors.} In applying the proposed truncation-based estimator (hereafter referred to as `Trunc'), we select the truncation parameters $\tau$ and $\kappa$ as described in Section~\ref{sec:tuning}, and obtain the estimators $\wh{\bm\Lambda}(\tau) = \sqrt{p}\, \wh{\mbf E}(\tau)$ and $\wh{\bm\chi}_t(\tau, \kappa) = \wh{\mbf E}(\tau) \wh{\mbf E}(\tau)^\top \mbf X^\trunc_t(\kappa)$.
For comparison, we consider an `oracle' version of the proposed truncation-based methodology where we utilise the true (unobservable) loading matrix and common component for the selection of truncation parameters.
Additionally, we include the classical PC-based estimator (`PCA') as well as the method proposed by \cite{he2022large} for high-dimensional elliptical factor model (`RTS'), and the two methods based on minimising vector-variate (`HPCA') and element-wise (`IHR') Huber losses proposed by \cite{he2023huber}, all implemented in the R package \verb+HDRFA+ \citep{hdrfa}.

\subsubsection{Results}

\begin{figure}[h!t!b!p!]
\centering
\includegraphics[width = 1\textwidth]{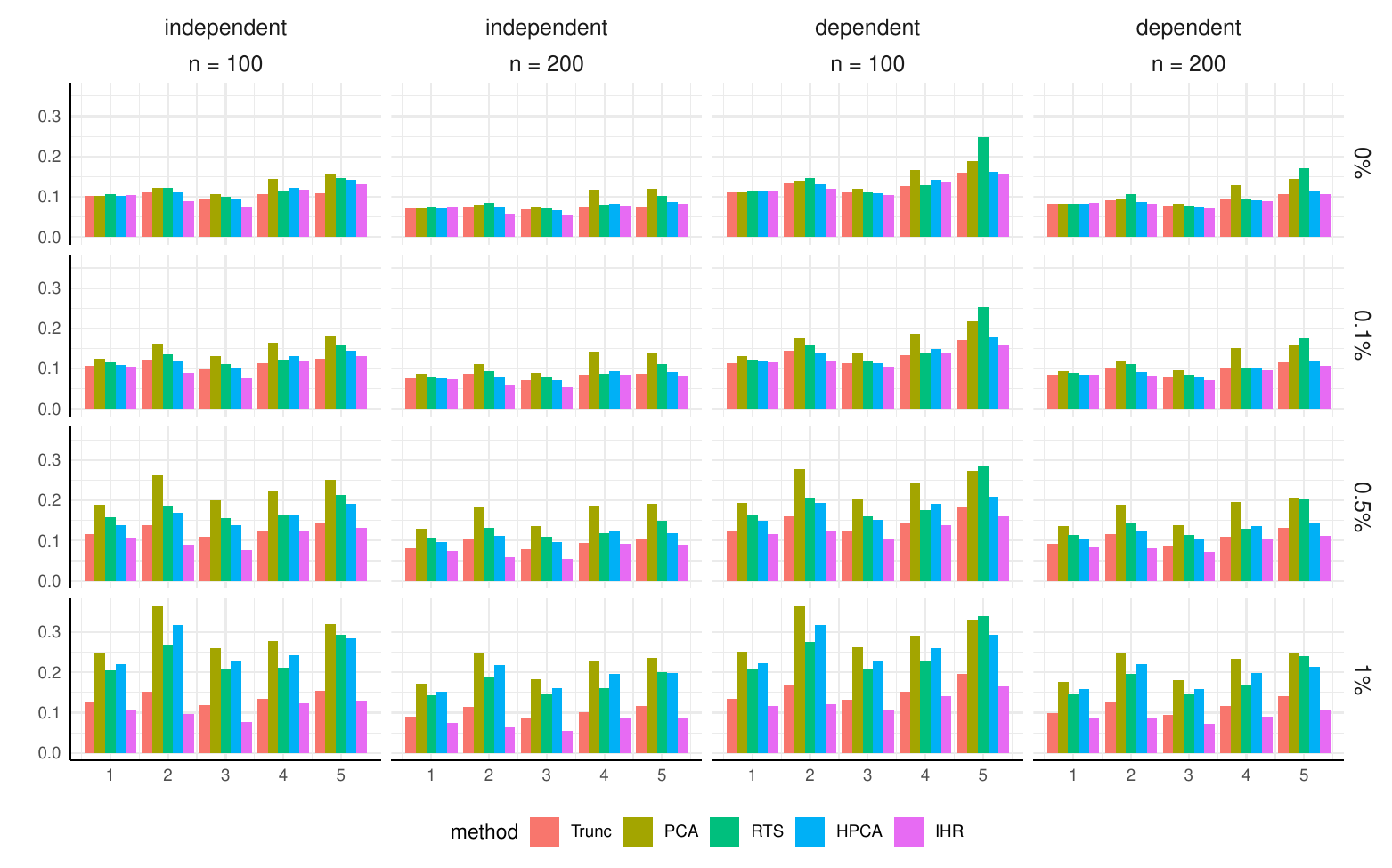}
\caption{Loading estimation errors measured as in~\eqref{eq:err:loading:vec} in different scenarios \ref{d:one}--\ref{d:five} ($x$-axis) for Trunc, PCA, RTS, HPCA and IHR over varying $n$ ($\{100, 200\}$), with and without temporal dependence in the idiosyncratic component and the percentage of outliers ($\{0, 0.1, 0.5, 1\}$), averaged over $100$ realisations per setting. Here, $p = 100$.}
\label{fig:vector:le:one}
\end{figure}

\begin{figure}[h!t!b!p!]
\centering
\includegraphics[width = 1\textwidth]{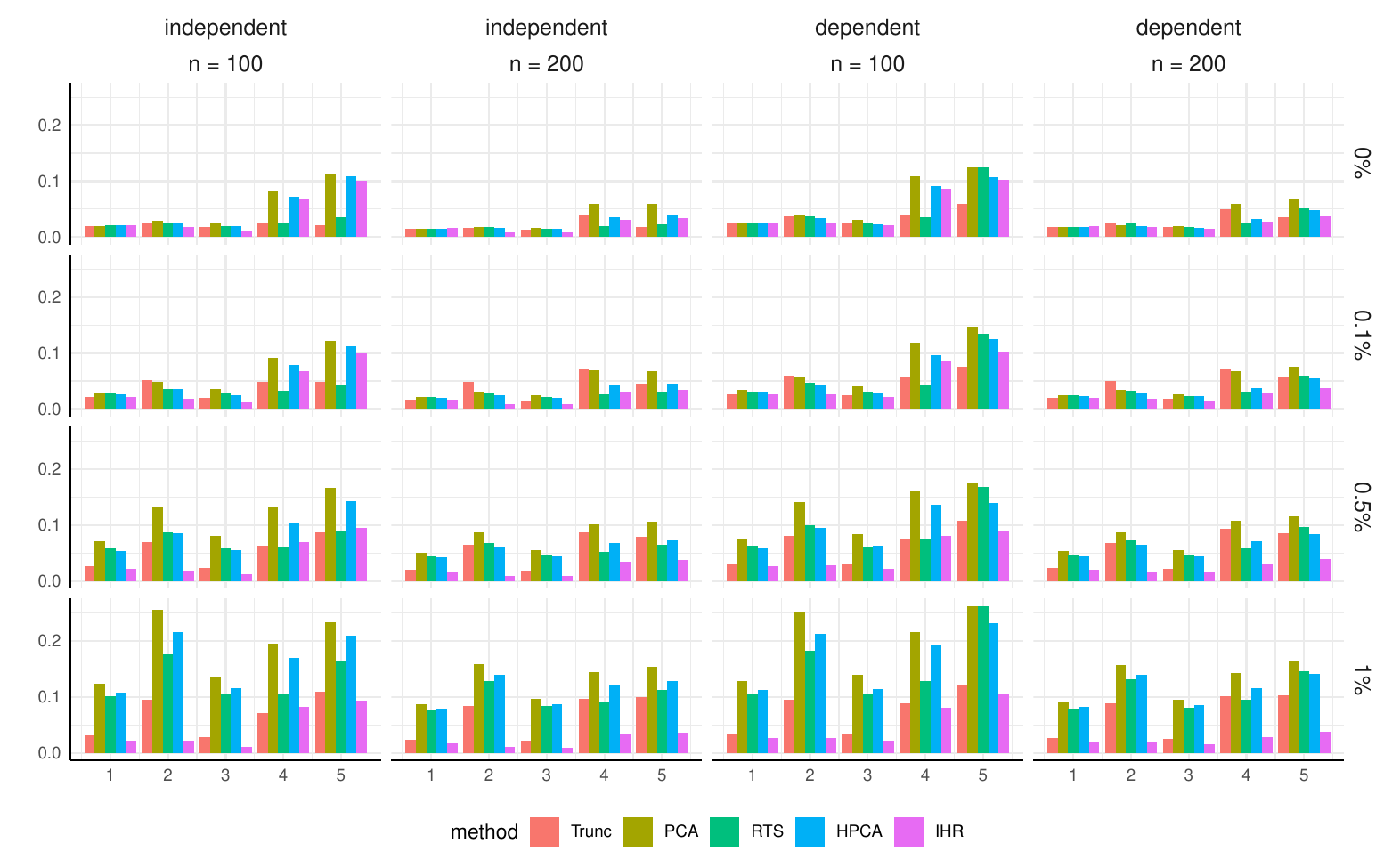}
\caption{Common component estimation errors measured as in~\eqref{eq:err:chi:vec} with $\mc T = [n]$ (`all') in different scenarios \ref{d:one}--\ref{d:five} ($x$-axis) for Trunc, PCA, RTS, HPCA and IHR over varying $n$ ($\{100, 200\}$), with and without temporal dependence in the idiosyncratic component and the percentage of outliers ($\{0, 0.1, 0.5, 1\}$), averaged over $100$ realisations per setting. Here, $p = 100$.}
\label{fig:vector:ce:all:one}
\end{figure}

\begin{figure}[h!t!b!p!]
\centering
\includegraphics[width = 1\textwidth]{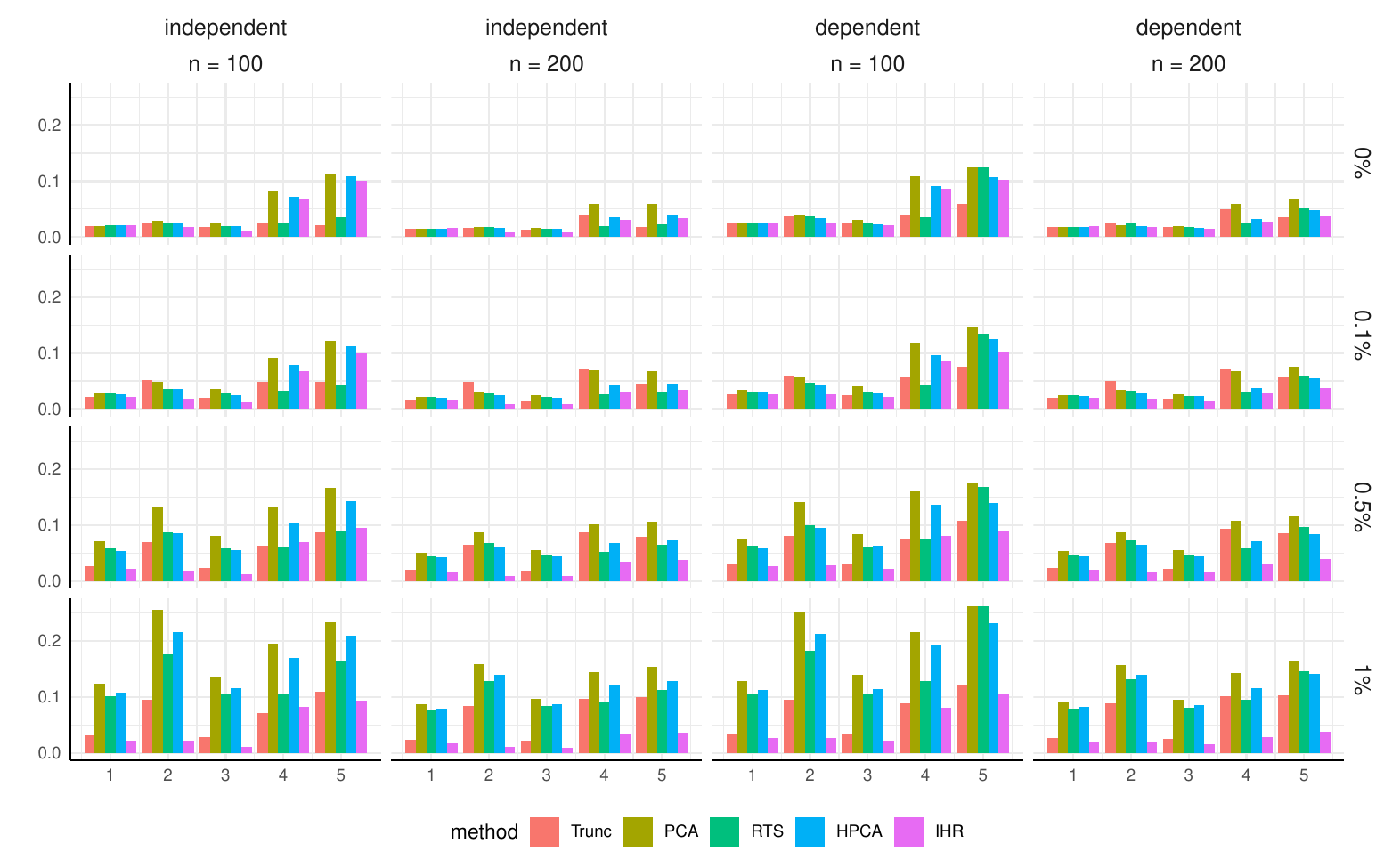}
\caption{Common component estimation errors measured as in~\eqref{eq:err:chi:vec} with $\mc T = \{n - 10 + 1, \ldots, n\}$ (`local') in different scenarios \ref{d:one}--\ref{d:five} ($x$-axis) for Trunc, PCA, RTS, HPCA and IHR over varying $n$ ($\{100, 200\}$), with and without temporal dependence in the idiosyncratic component and the percentage of outliers ($\{0, 0.1, 0.5, 1\}$), averaged over $100$ realisations per setting. Here, $p = 100$.}
\label{fig:vector:ce:local:one}
\end{figure}

\begin{figure}[h!t!b!p!]
\centering
\includegraphics[width = 1\textwidth]{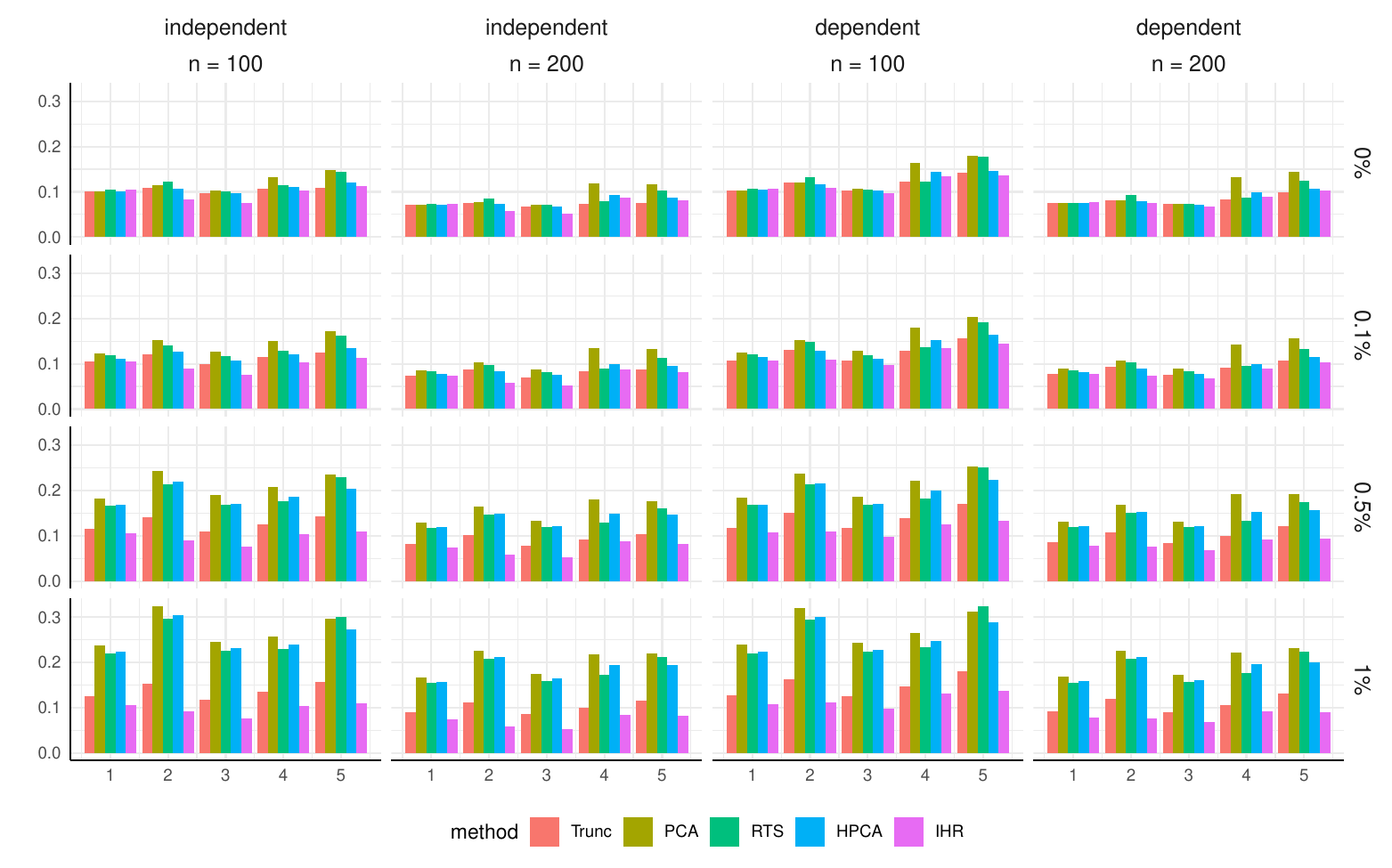}
\caption{Loading estimation errors measured as in~\eqref{eq:err:loading:vec} in different scenarios \ref{d:one}--\ref{d:five} ($x$-axis) for Trunc, PCA, RTS, HPCA and IHR over varying $n$ ($\{100, 200\}$), with and without temporal dependence in the idiosyncratic component and the percentage of outliers ($\{0, 0.1, 0.5, 1\}$), averaged over $100$ realisations per setting. Here, $p = 200$.}
\label{fig:vector:le:two}
\end{figure}

\begin{figure}[h!t!b!p!]
\centering
\includegraphics[width = 1\textwidth]{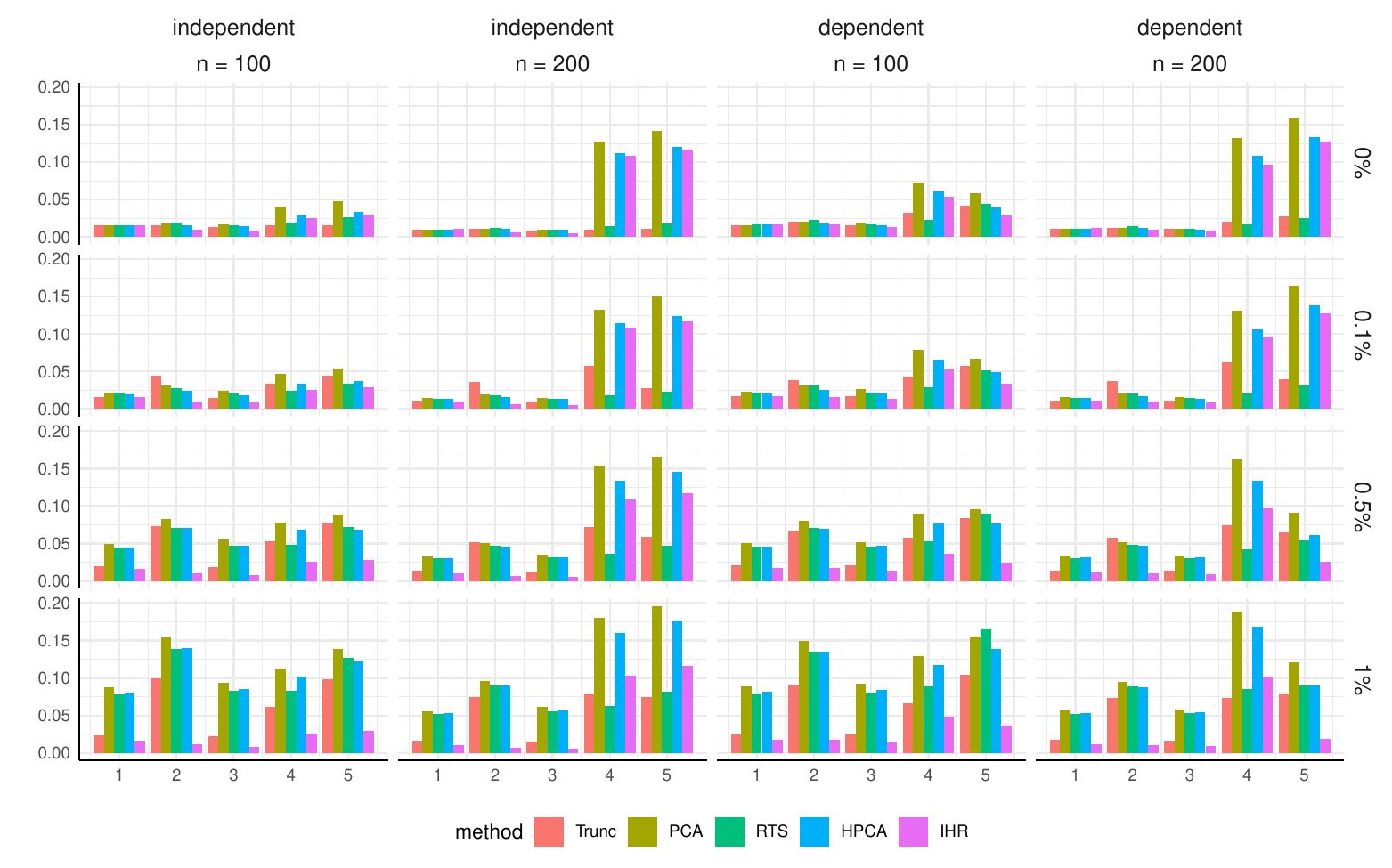}
\caption{Common component estimation errors measured as in~\eqref{eq:err:chi:vec} with $\mc T = [n]$ (`all') in different scenarios \ref{d:one}--\ref{d:five} ($x$-axis) for Trunc, PCA, RTS, HPCA and IHR over varying $n$ ($\{100, 200\}$), with and without temporal dependence in the idiosyncratic component and the percentage of outliers ($\{0, 0.1, 0.5, 1\}$), averaged over $100$ realisations per setting. Here, $p = 200$.}
\label{fig:vector:ce:all:two}
\end{figure}

\begin{figure}[h!t!b!p!]
\centering
\includegraphics[width = 1\textwidth]{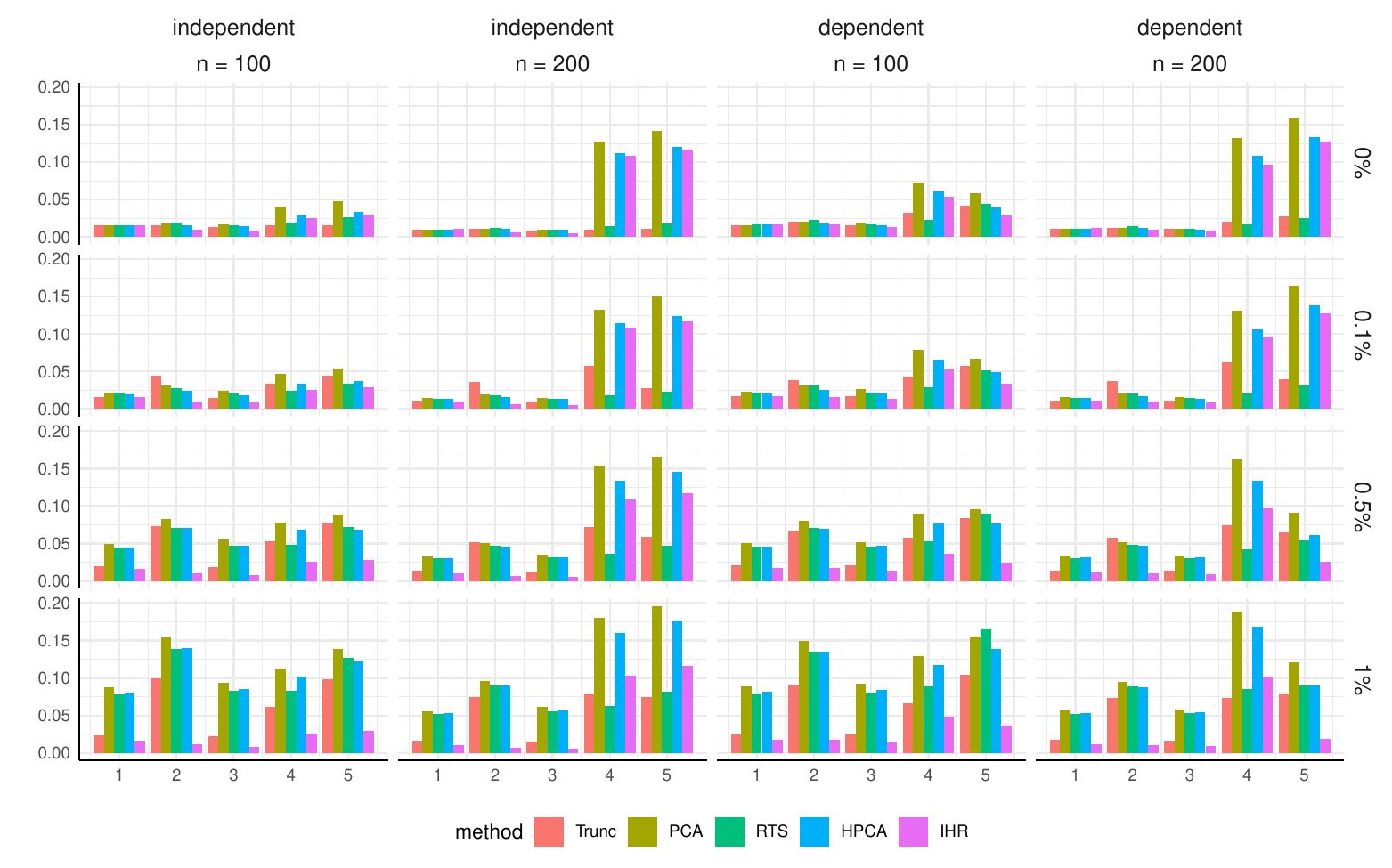}
\caption{Common component estimation errors measured as in~\eqref{eq:err:chi:vec} with $\mc T = \{n - 10 + 1, \ldots, n\}$ (`local') in different scenarios \ref{d:one}--\ref{d:five} ($x$-axis) for Trunc, PCA, RTS, HPCA and IHR over varying $n$ ($\{100, 200\}$), with and without temporal dependence in the idiosyncratic component and the percentage of outliers ($\{0, 0.1, 0.5, 1\}$), averaged over $100$ realisations per setting. Here, $p = 200$.}
\label{fig:vector:ce:local:two}
\end{figure}

For each setting, we generate $100$ realisations and report the average and the standard deviation (in brackets) of the evaluation metrics in~\eqref{eq:err:loading:vec}--\eqref{eq:err:chi:vec}, see Figures~\ref{fig:vector:le:one}--\ref{fig:vector:ce:local:two}.
We observe that different robust methods perform the best in different scenarios; IHR shows good performance across many scenarios, closely followed by Trunc.
On the other hand, for common component estimation, we see overwhelming evidence favouring Trunc over other competitors in some scenarios such as \ref{d:four} and \ref{d:five} which supports the use of data truncation for factor estimation.
% We see little difference between Trunc and Oracle, which indicates that the CV procedure proposed in Section~\ref{sec:tuning} works well in truncation parameter selection.

\clearpage

\section{Additional empirical results}

\subsection{Euro Area macroeconomic data}
\label{app:euro}

\begin{figure}[h!t!b!]
\centering
\includegraphics[width = .7\textwidth]{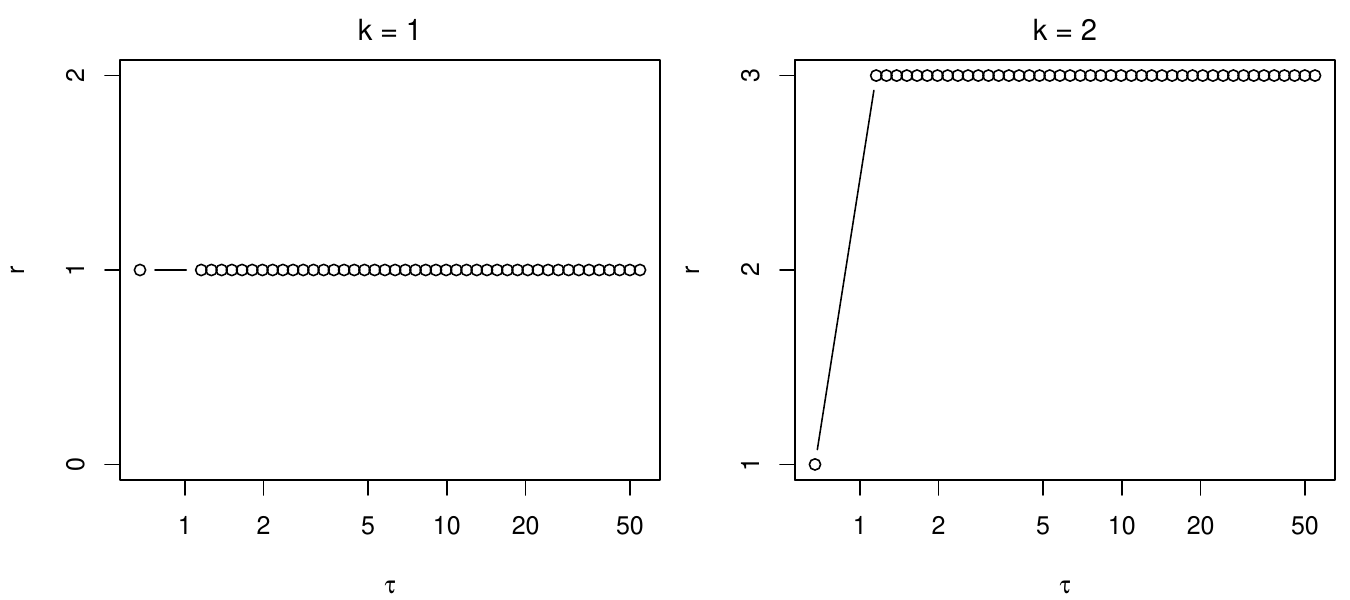}
\caption{EA-MD: Factor number estimators ($y$-axis) against the varying values of the truncation parameter $\tau$ ($x$-axis, in log scale) % obtained as in~\eqref{eq:r:est}, 
for $k = 1$ (left) and $k = 2$ (right).}
\label{fig:er:r}
\end{figure}

Continuing with EA-MD analysed in Section~\ref{sec:euro}, Figure~\ref{fig:er:r} plots the output from the ratio-based factor number estimation as $\tau$ varies. 

\begin{figure}[h!t!b!]
\centering
\includegraphics[width = .5\textwidth]{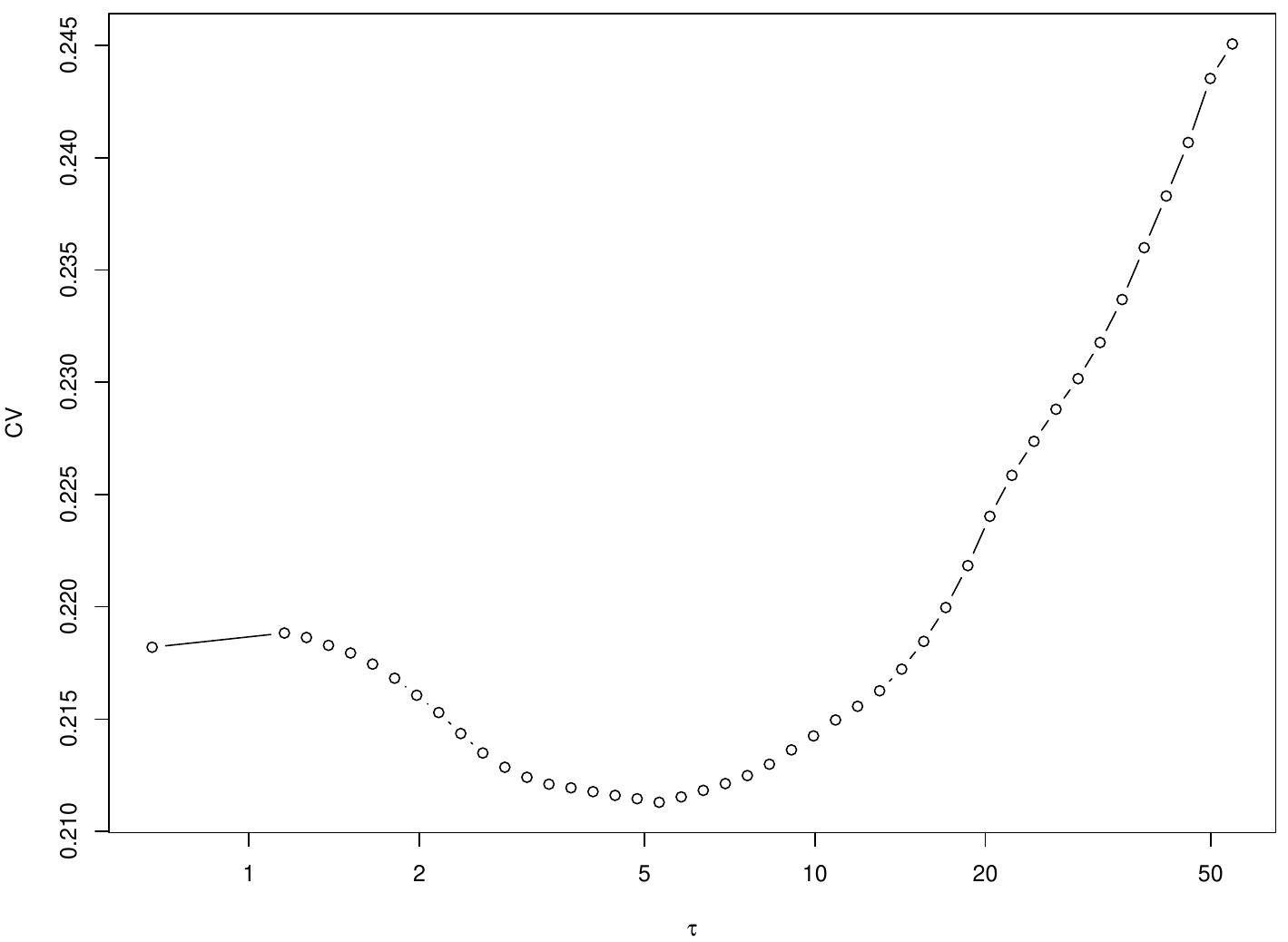}
\caption{EA-MD: CV measure in~\eqref{eq:tau:cv} with varying $\tau$ ($x$-axis, in log scale); $\tau_{\text{\tiny CV}} \approx 5.306$.}
\label{fig:er:cv}
\end{figure}

Figure~\ref{fig:er:cv} plots the output from the CV procedure described in Section~\ref{sec:tuning} for the truncation parameter selection.

% Figure~\ref{fig:er:factor} plots the three factor time series $\{\wh f_{1j, t}\}_{t \in [n]}$ from the EA-MD. 
% Figure~\ref{fig:er:loading:kron:cr} visualises the matrix $\wc{\bm\Lambda}^{[2]}_1 \otimes \wc{\bm\Lambda}^{[2]}_2 \in \R^{296 \times 3}$ which gives the loadings for the $p_1 p_2$-dimensional vector $\vecop(\mc X_t^\top)$.

% \begin{figure}[h!t!b!]
% \centering
% \includegraphics[width = .9\textwidth]{figs/euro_factor3_trunc.pdf}
% \caption{EA-MD: Factor time series $\wh f_{1j, t}$ for $j = 1, 2, 3$ (top to bottom).}
% \label{fig:er:factor}
% \end{figure}

% \begin{figure}[h!t!b!]
% \centering
% \includegraphics[width = .9\textwidth]{figs/euro_alt_loading_kron_rc.pdf}
% \caption{EA-MD: The Kronecker product of the estimated loading matrices $\wc{\bm\Lambda}^{[2]}_1 \otimes \wc{\bm\Lambda}^{[2]}_2$ of dimensions $(p_1p_2, \wh r_1\wh r_2) = (296, 3)$.
% For ease of visualisation, we only give the names of the countries in the $y$-axis; for each country, the $37$ indicators are arranged in the order given in the right panel of Figure~\ref{fig:er:loading}.}
% %For ease of visualisation, we only give the names of the indicators in the $y$-axis; for each indicator, the $8$ countries indicators are arranged in the order given in the right panel of Figure~\ref{fig:er:loading}.}
% % are arranged in the order AT, BE, DE, EL, ES, FR, IT, NL.
% \label{fig:er:loading:kron:cr}
% \end{figure}

Complementing Figure~\ref{fig:ea:loading:trunc} that visualises the loading matrices obtained with and without truncation, Table~\ref{tab:ea:loading:trunc} summarises their ranges.
Focusing on the estimators from Trunc, we observe that the elements of $\wc{\bm\Lambda}^{[2]}_1(\tau)$ are of the same sign across the $8$ countries, indicating that the three factor time series aggregating the $37$ indicators, are loaded onto the $8$ countries in the same direction. 
This shows the evidence of a common EA factor, and that the correlations among the economic variables are similar within the $8$ countries considered. 
Overall, Italy, Spain, Netherlands, and France have the larger loadings out of the 8 countries.
The elements of $\wc{\bm\Lambda}^{[2]}_2(\tau)$ exhibit clusterings based on the grouping of the macroeconomic indicators. 
Specifically, the first factor 
% in the sense that f_{11, t} is loaded by the first column of $\wc{\bm\Lambda}_2$
is strongly related to various confidence indicators such as the Economic Sentiment Indicator (ESENTIX) and, to a lesser extent, to the unemployment rate (UNETOT). 
The second factor is strongly related to Industrial Production (IDs starting with IP), and thus to real economic activity.
The third factor is a nominal one, strongly related to price indexes such as the overall Harmonized Index of Consumer Prices (HICPOV).

% -1.2656789 -0.6218955
%  -2.176932  2.645170
% -2.303639  2.659295

% 0.4319545 1.6375641
% -2.945406  2.916807
% -2.928145  2.917794

\begin{table}[h!t!]
\caption{EA-MD: The range of the elements of $\wc{\bm\Lambda}^{[2]}_k(\tau), \, k \in \{1, 2\}$, and that from the Varimax rotated $\wc{\bm\Lambda}^{[2]}_2(\tau)$, for Trunc and noTrunc.}
\label{tab:ea:loading:trunc}
\centering
\begin{tabular}{r ccc}
\toprule
& $\wc{\bm\Lambda}^{[2]}_1(\tau)$ & $\wc{\bm\Lambda}^{[2]}_2(\tau)$ & $\wc{\bm\Lambda}^{[2]}_2(\tau)$ after rotation \\ 
\cmidrule(lr){1-1} \cmidrule(lr){2-2} \cmidrule(lr){3-3} \cmidrule(lr){4-4}
Trunc & $(0.622, 1.266)$ & $(-1.103, 2.645)$ & $(-1.158, 2.659)$  \\
noTrunc & $(0.432, 1.638)$ & $(-1.145, 2.945)$ & $(-1.172, 2.928)$ \\
\bottomrule
\end{tabular}
\end{table}

\paragraph{Forecasting exercise.}
We perform a forecast exercise to further illustrate the effect of truncation, whose result is summarised in Table~\ref{tab:ea:forecast}. 
Denoting by $N = 236$ the number of observations prior to 2022-01 and $n = 257$ the total number of observations, we sequentially produce a one-step ahead forecast for the two indicators, the growth rate of the industrial production manufacturing index (IPMN) and the difference of the core consumer prices inflation (HICPNEF), on $t \in \{N + 1, \ldots, n\}$, each time using all the preceding observations as the training data.
For this, fixing $(\wh r_1, \wh r_2) = (1, 3)$, we estimate the loadings and the factors with and without the proposed truncation, fit a vector autoregressive (VAR) model to the vectorised factors with its order selected by Akaike information criterion, generate their one-step ahead forecast and then combine the latter with the loading estimates to produce the final forecast. 
With some abuse of notation, we denote the forecast at given $\mbf i \in [p_1] \times \{\text{IPMN, HICPNEF}\}$ and time $t$ by $\wh{X}^\star_{\mbf i, t \vert t - 1}$, where $\star \in \{\text{Trunc, noTrunc}\}$ denotes whether the truncation is performed or not throughout the estimation procedure.
Then, we inspect the forecast error
\begin{align}
\label{eq:ea:fe}
\text{Err}_{\mbf i, t}(\star) = \l\vert \wh{X}^\star_{\mbf i, t \vert t - 1} - X_{\mbf i, t} \r\vert, \quad \star \in \{\text{Trunc, noTrunc}\}.
\end{align}
As each new observation arrives, we update the loading and the factor estimates and the VAR fitted to the estimated factors, which is repeated until the end of the dataset is reached.

\subsection{US macroeconomic data}
\label{sec:fredmd}

FRED-MD is a large, monthly frequency, macroeconomic database maintained by the Federal Reserve Bank of St.~Louis \citep{mccracken2016fred} and has been analysed frequently in the time series factor modelling literature as a benchmark.
We use the dataset spanning the period from 1960 to 2023 ($n = 767$).
Removing the time series with missing observations,
we have $p = 111$ variables in total.
All time series are individually transformed to stationarity as suggested in \cite{mccracken2016fred}, see their Appendix~I for full details of variable-specific transformations. % using the R package \verb+fbi+ \citep{fbi}.
Further, to handle the heterogeneity in the scale of the variables, we center and standardise each time series.
We opt to use the mean and the standard deviation in place of the more robust statistics such as the median and the mean absolute deviation (MAD), as some indicators have MAD very close to zero over some rolling windows considered in the forecasting exercise described below. % due to the presence of many zeros {\matt maybe small values rather than exact zeros} in some time periods. 

\begin{figure}[h!t!b!]
\begin{center}
\begin{tabular}{cc}
\includegraphics[width = .4\textwidth]{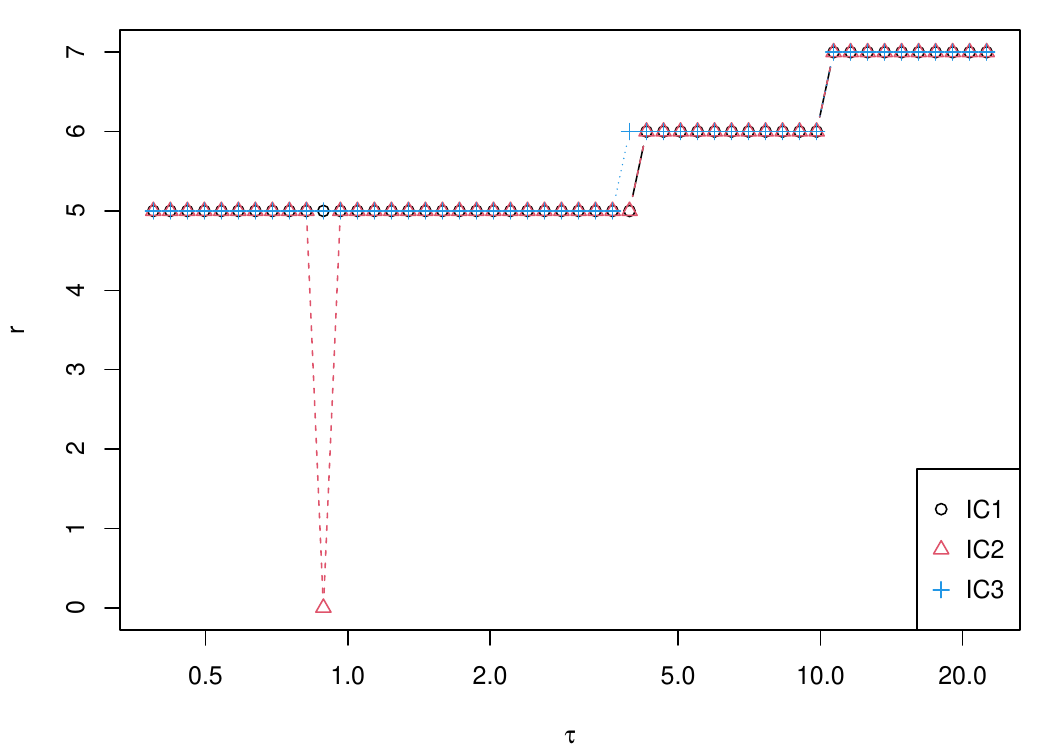}
&
\includegraphics[width = .4\textwidth]{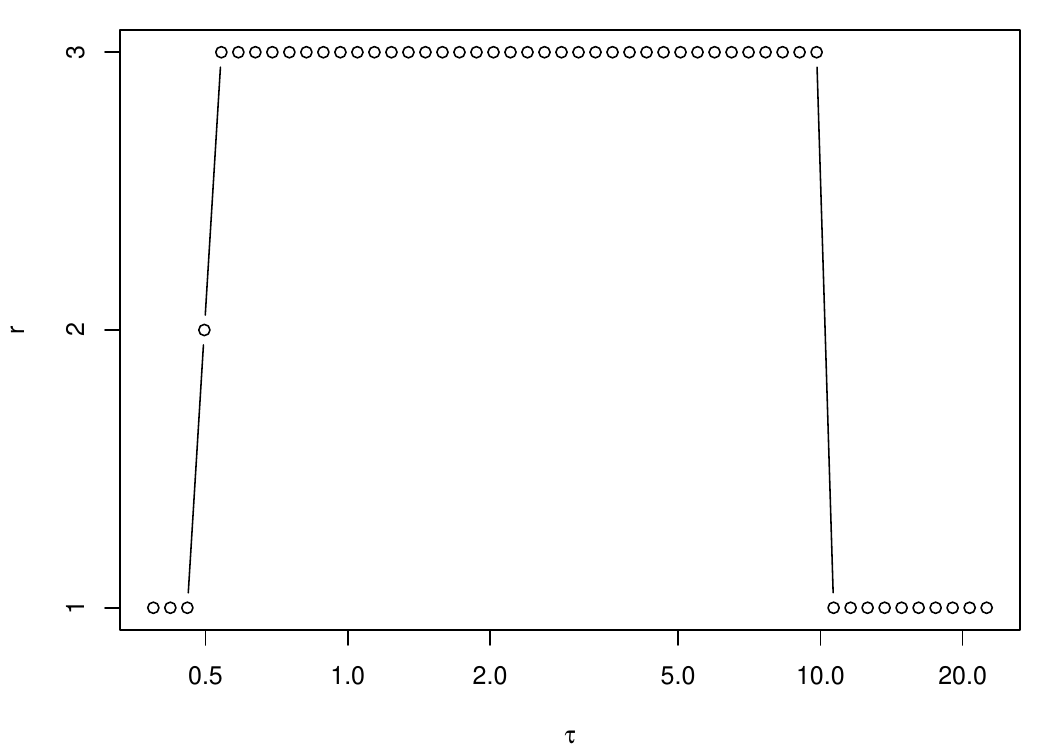}
\end{tabular}
\end{center}
\caption{FRED-MD: Factor number estimators ($y$-axis) against the varying values of the truncation parameter $\tau$ ($x$-axis, in log scale) generated by the information criterion-based method of \cite{alessi2010improved} (left) and by the ratio-based estimator in Section~\ref{sec:r:est} (right).}
\label{fig:fredmd:r}
\end{figure}

We consider two approaches for factor number estimation: The first one, proposed in \cite{alessi2010improved}, applies the information criteria of \cite{bai2002} to the subsets of the data of varying dimensions and sample sizes in order to mitigate the arbitrariness in the choice of constants applied to the penalty. The second is the ratio-based estimator in Section~\ref{sec:r:est} with $\bar{r} = \min(\lfloor p/2 \rfloor, 20) = 20$.
Since their performance depends on the choice of the truncation parameter $\tau$, we examine the outputs from the two methods with varying values of~$\tau$, see Figure~\ref{fig:fredmd:r}.
For the first approach, the three information criteria of \cite{bai2002} with different penalties % (denoted by IC$_1$--IC$_3$) 
attain consensus over the most $\tau$ values at $\wh r = 5$, while the second one favours $\wh r = 3$.
For comparison, the Huber PCA-based estimator \citep{he2023huber} gives $\wh r = 4$.
Below we present the forecasting results with $\wh r \in \{3, 5\}$ and demonstrate that regardless of its choice, the proposed estimator performs competitively.

\paragraph{Forecasting exercise.}
We compare the forecasting performance of the proposed truncation-based estimator against the non-robust counterpart.
We also considered the method proposed by \cite{he2023huber} but do not report the results due to its numerical convergence issues.

Let $X_{i, t}$ denote the variable of interest to forecast, which may be any one of the $p$ variables.
For given window size $T$ and forecasting horizon $h \ge 1$, let $\wh{X}_{i, t + h \vert T}$ denote the $h$-step ahead forecast of the $i$-th variable based on the past observations at $\{t - T + 1, \ldots, t\}$, obtained as an estimator of the best linear predictor $\text{Proj}(X_{i, t + h} \vert \mc X_u, \, u \le t)$, where $\text{Proj}(\cdot \vert \mbf z)$ denotes the linear projection operator onto the space spanned by $\mbf z$ \citep{stock2002forecasting}.
In line with the tail-robust method described in Section~\ref{sec:method}, we propose the estimator % the best linear predictor of $X_{i, t + h}$ given $\mc X_u, \, t - n + 1 \le u \le t$, 
$$
\wh{X}_{i, t + h \vert T}(\tau, \kappa) = \bm\varphi_i^\top (\wh{\bm\Gamma}_t(\tau, h))^\top \wh{\mbf E}_t(\tau) (\wh{\mbf M}_t(\tau))^{-1} \wh{\mc F}_t(\kappa),
$$
where $\wh{\bm\Gamma}_t(\tau, h) = T^{-1} \sum_{u = t - T + 1}^{t - h} \mc X^{\trunc}_u(\tau) (\mc X^{\trunc}_{u + h}(\tau))^\top$, $\wh{\mbf E}_t(\tau)$ and $\wh{\mbf M}_t(\tau)$ contain the leading $\wh r$ eigenvectors and eigenvalues of $\wh{\bm\Gamma}_t(\tau, 0)$, respectively, and $\bm\varphi_i$ is the unit vector with its $i$-th element set to one; see also \cite{barigozzi2022fnets}.
We refer to such an estimator combined with the truncation parameters chosen via CV (see Section~\ref{sec:tuning}) as `Trunc', while the one combined with $\tau = \kappa = \infty$ (i.e.\ no truncation) by `noTrunc'; in fact, noTrunc amounts to the standard PC-based approach taken e.g.\ in \cite{bai2003}. 
Following \cite{trucios2021robustness}, we compare their forecasting performance in a rolling window-based exercise with $T = 12 \times 10$ ($10$ years).

\begin{figure}[h!t!b!]
\centering
\begin{tabular}{c}
\includegraphics[width = 1\textwidth]{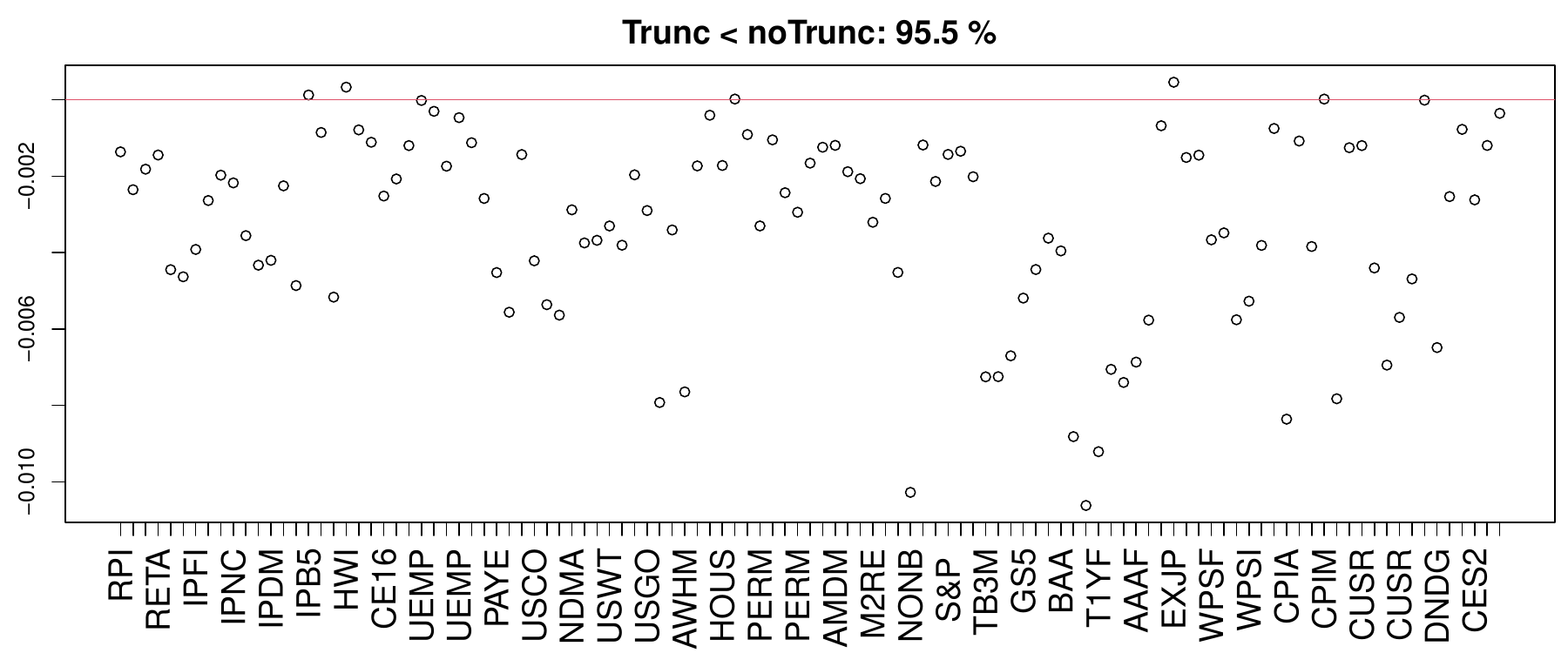} \\
\includegraphics[width = 1\textwidth]{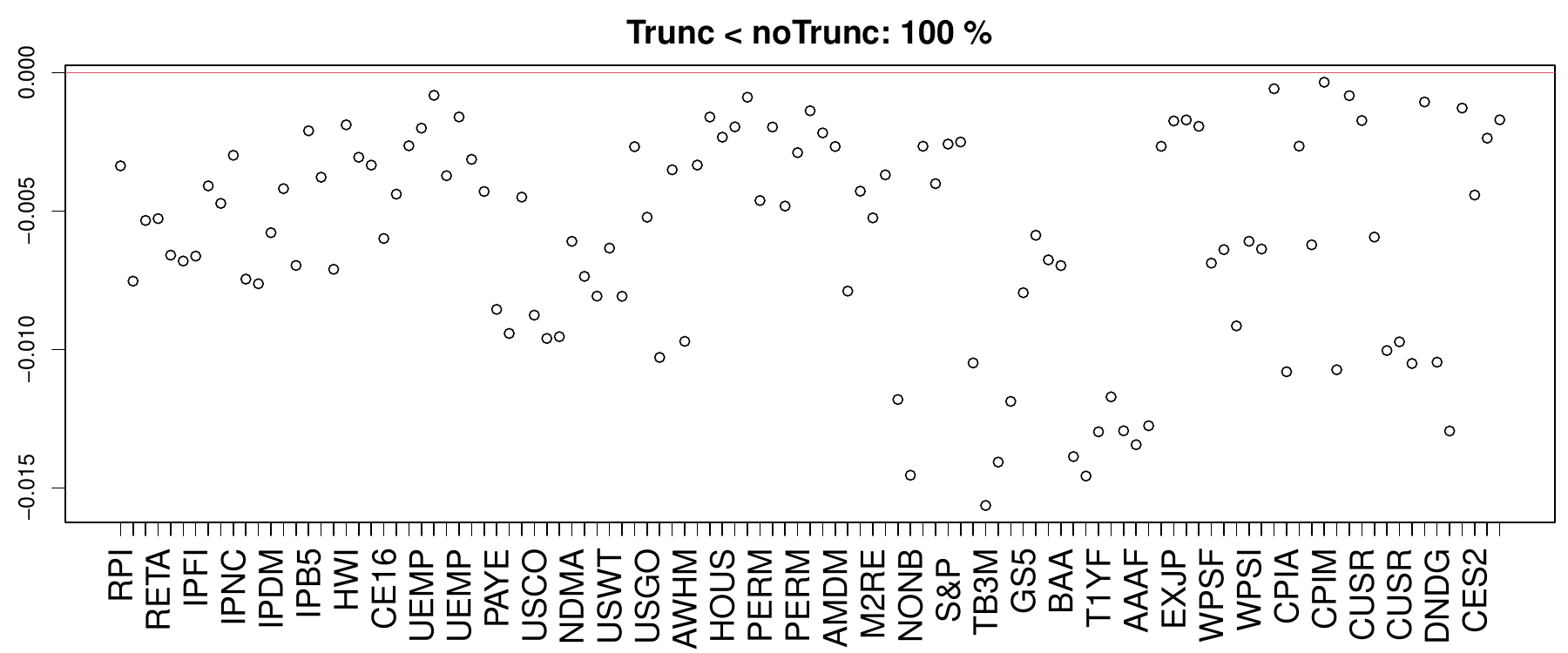}
\end{tabular}
\caption{$\overline{\text{Err}}_i(\text{Trunc}) - \overline{\text{Err}}_i(\text{noTrunc})$ for all $i \in [p]$ ($x$-axis) after standardisation by the overall standard deviation, with $\wh r = 3$ (top) and $\wh r = 5$ (bottom). The first four letters for each variable's ticker are given as the $x$-axis labels. 
Horizontal lines are drawn at $y = 0$.}
\label{fig:fredmd:all}
\end{figure}

At given $t$, we aggregate the forecasting error over the forecasting horizons $h \in [24]$, as
\begin{align*}
\text{Err}_{it}(\star) = \frac{1}{24} \sum_{h = 1}^{24} \l\vert \wh{X}^\star_{i, t + h \vert T} - X_{i, t + h} \r\vert, \quad \star \in \{ \text{Trunc}, \text{noTrunc} \},
\end{align*}
and also write $\overline{\text{Err}}_i(\star) = (n - T - 24)^{-1} \sum_{t = T + 1}^{n - 24} \text{Err}_{it}(\star)$.
Out of the $p = 111$ variables, we have $\overline{\text{Err}}_i(\text{Trunc}) < \overline{\text{Err}}_i(\text{noTrunc})$ for $106$ variables with $\wh r = 3$, and the inequality holds for all $p$ variables with $\wh r = 5$, see Figure~\ref{fig:fredmd:all}.

\begin{figure}[t!b!p!]
\centering
\begin{tabular}{cc}
\includegraphics[width = .45\textwidth]{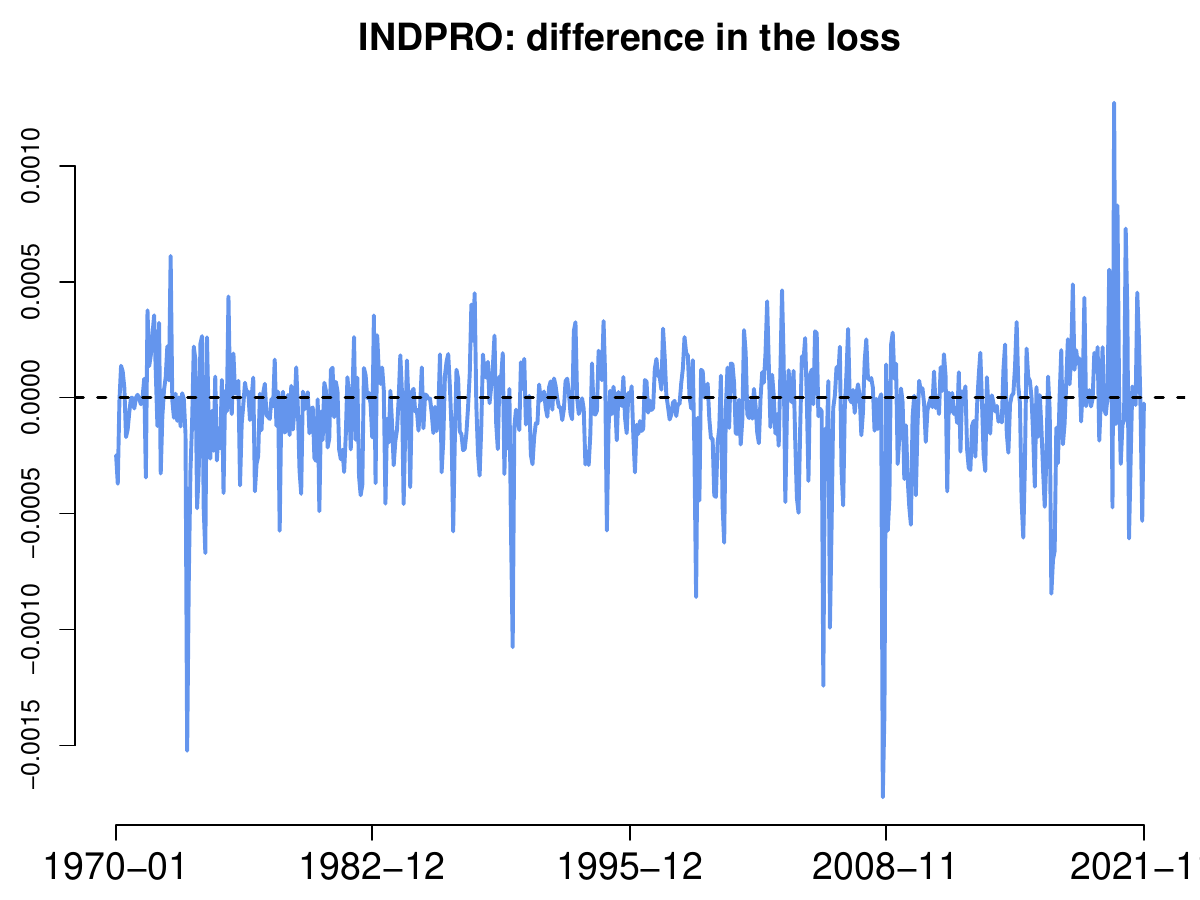} &
\includegraphics[width = .45\textwidth]{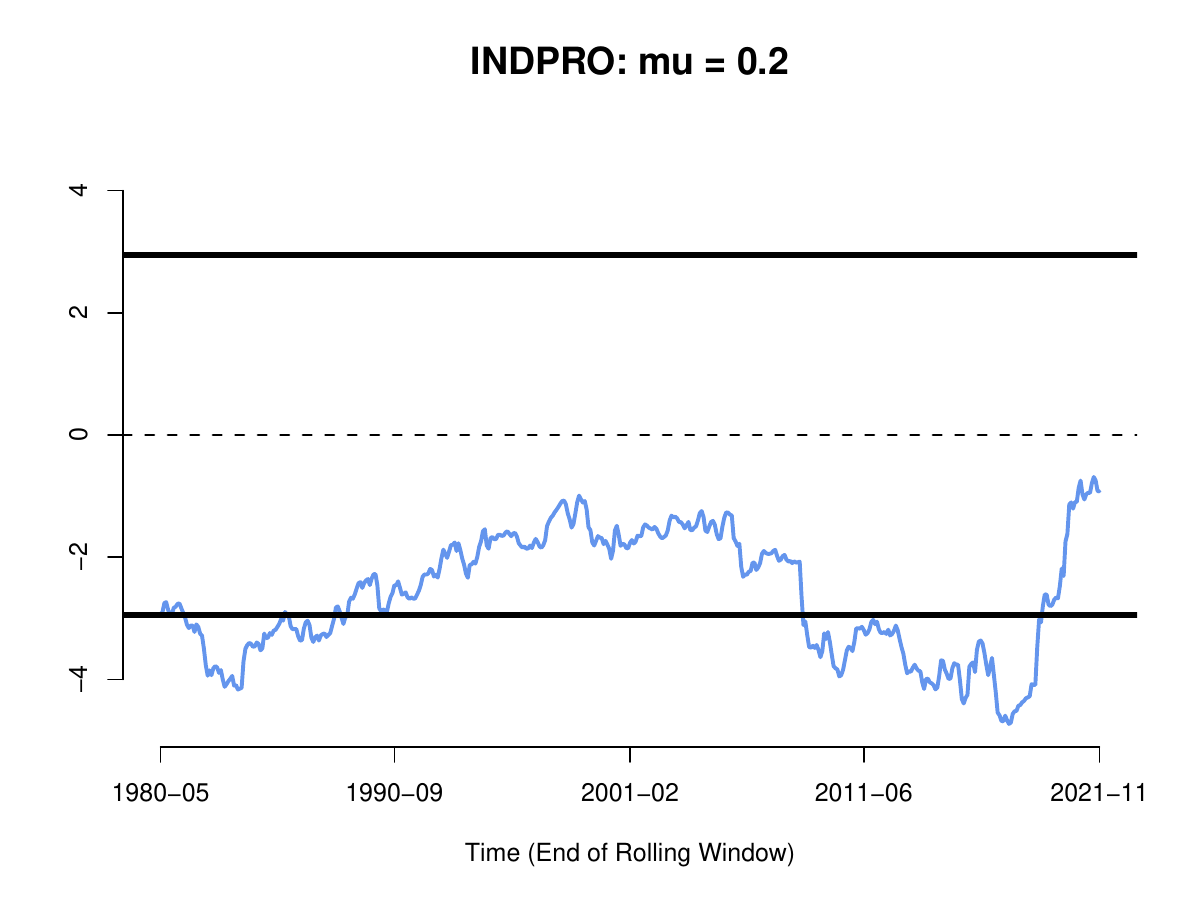} \\
\includegraphics[width = .45\textwidth]{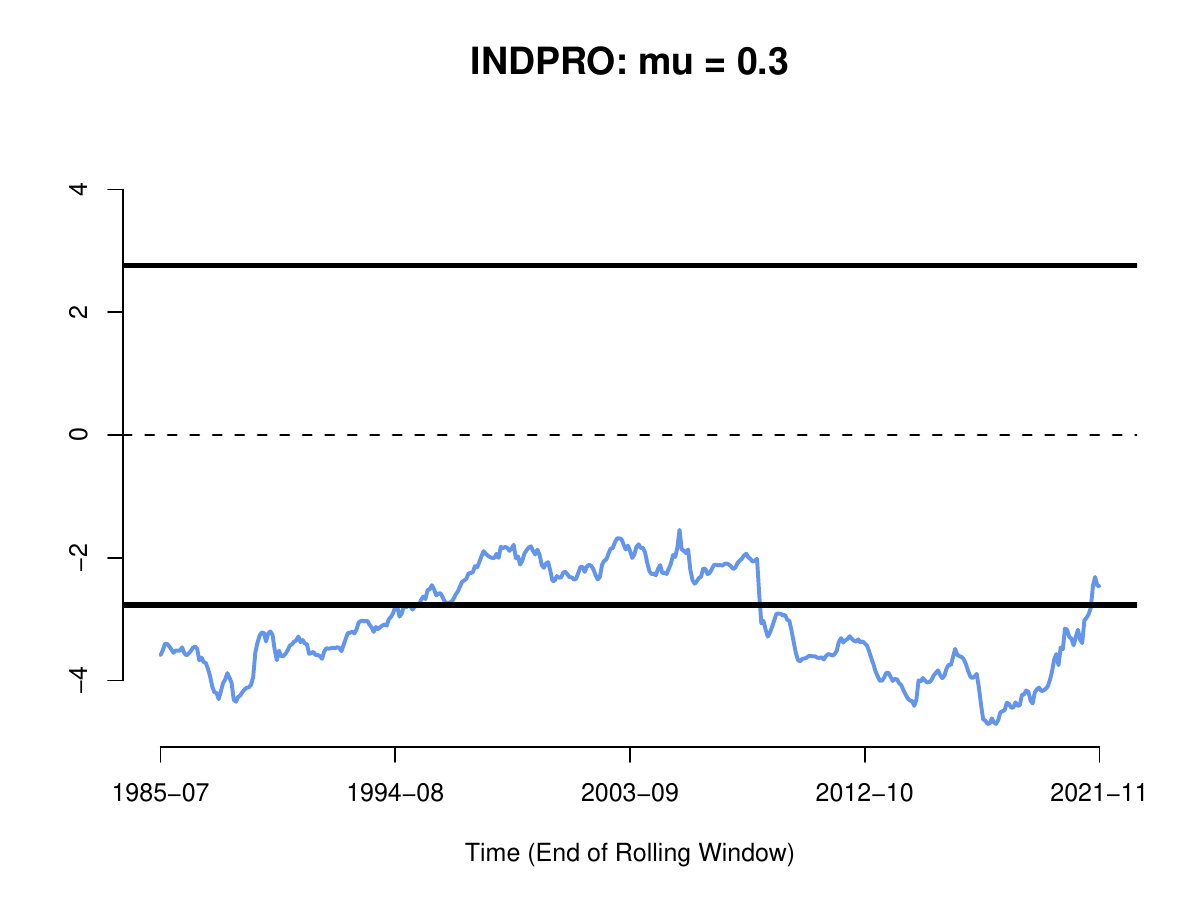} &
\includegraphics[width = .45\textwidth]{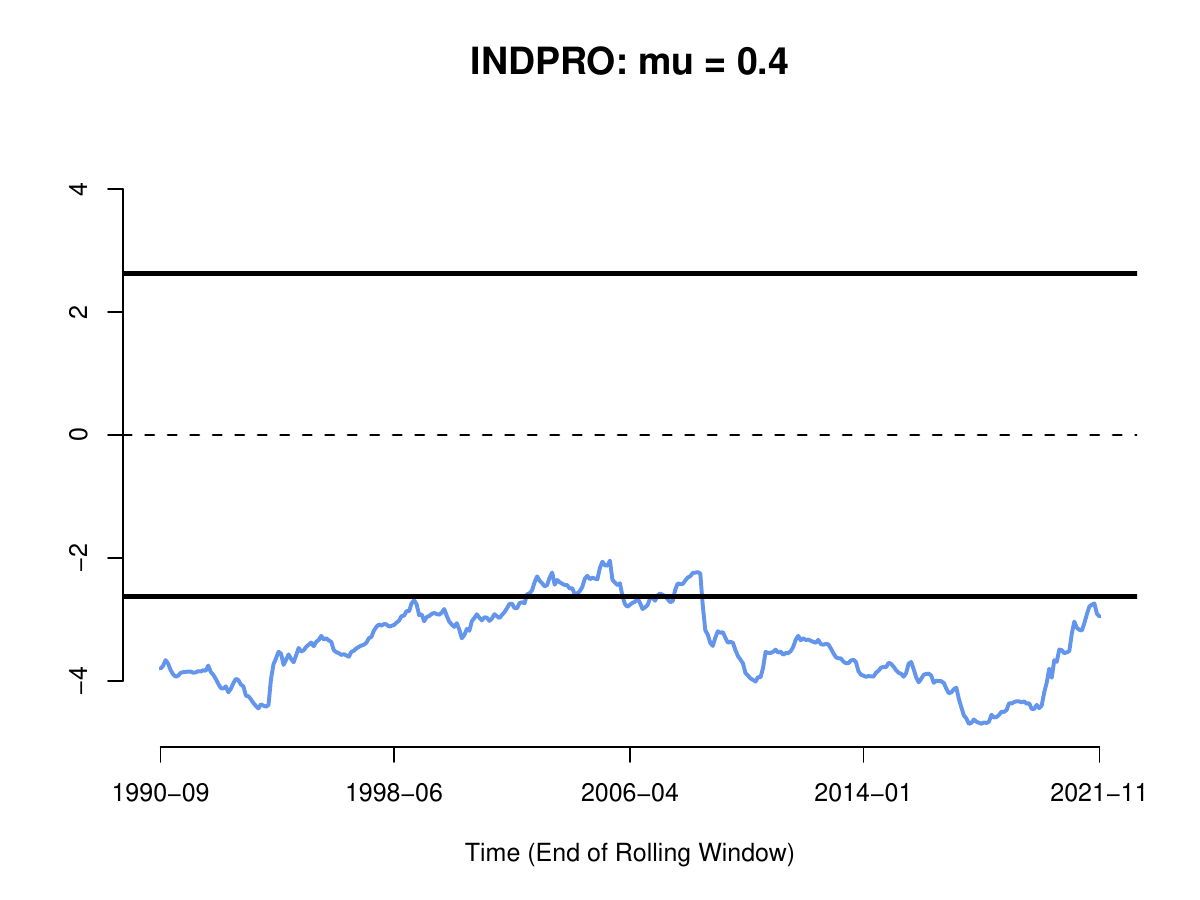}
\end{tabular}
\caption{Industrial production total index: $\text{Err}_{it}(\text{Trunc}) - \text{Err}_{it}(\text{noTrunc}), \, T + 1 \le t \le n - 24$ (top left) and the corresponding fluctuation test statistics computed with $\mu \in \{0.2, 0.3, 0.4\}$ along with the two-sided critical values at the significance level $\alpha = 0.1$. When the fluctuation test statistic falls below the lower solid line, Trunc outperforms the noTrunc and vice versa. Here, we set $\wh r = 3$.}
\label{fig:fredmd:indpro:6}
\end{figure}

\begin{figure}[t!b!p!]
\centering
\begin{tabular}{cc}
\includegraphics[width = .45\textwidth]{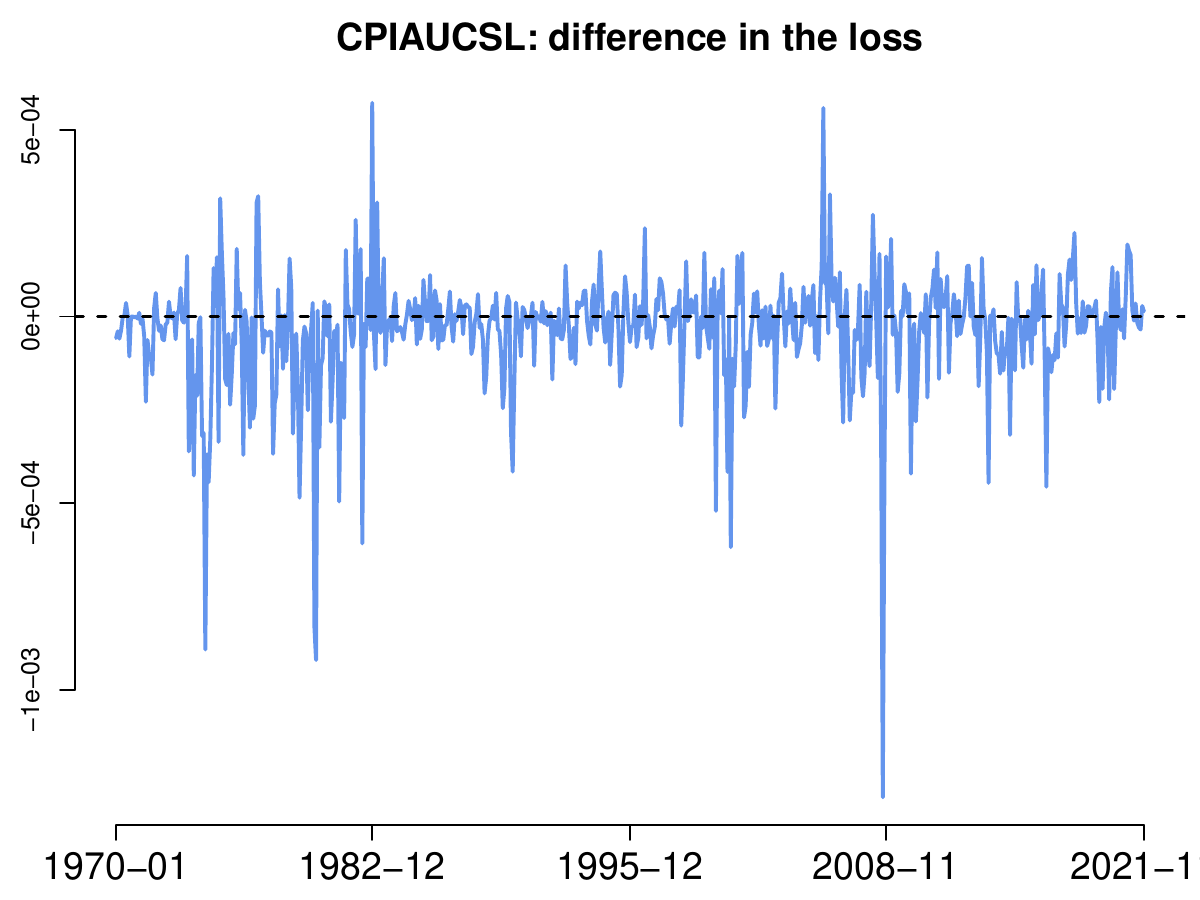} &
\includegraphics[width = .45\textwidth]{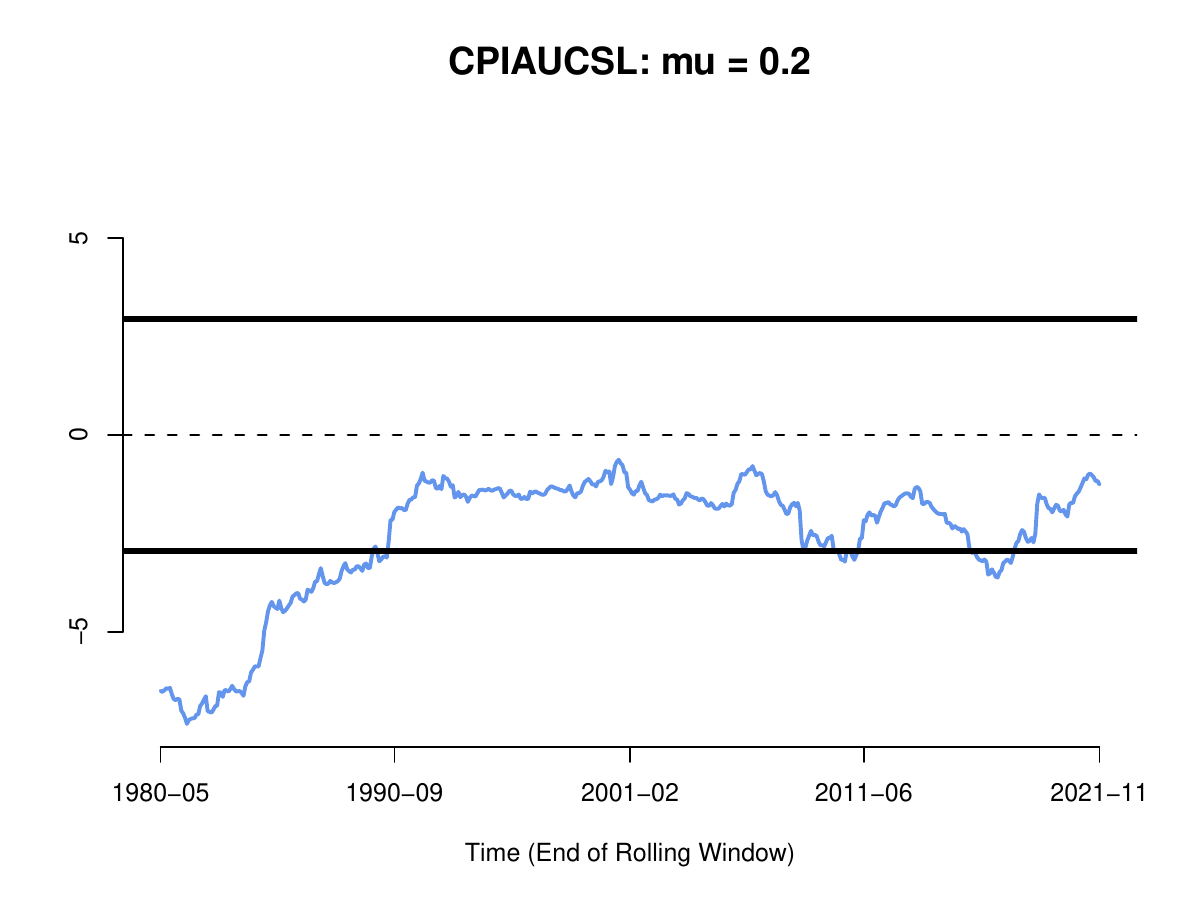} \\
\includegraphics[width = .45\textwidth]{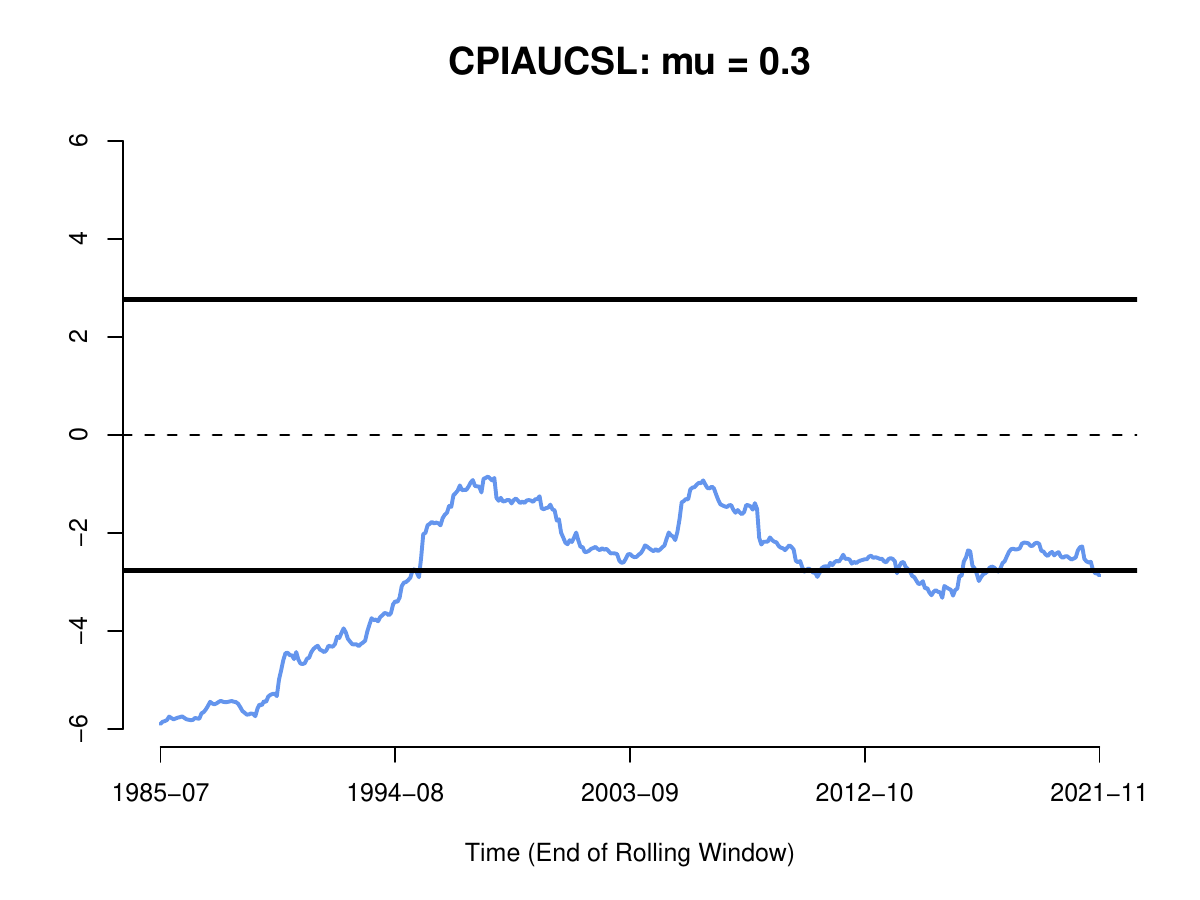} &
\includegraphics[width = .45\textwidth]{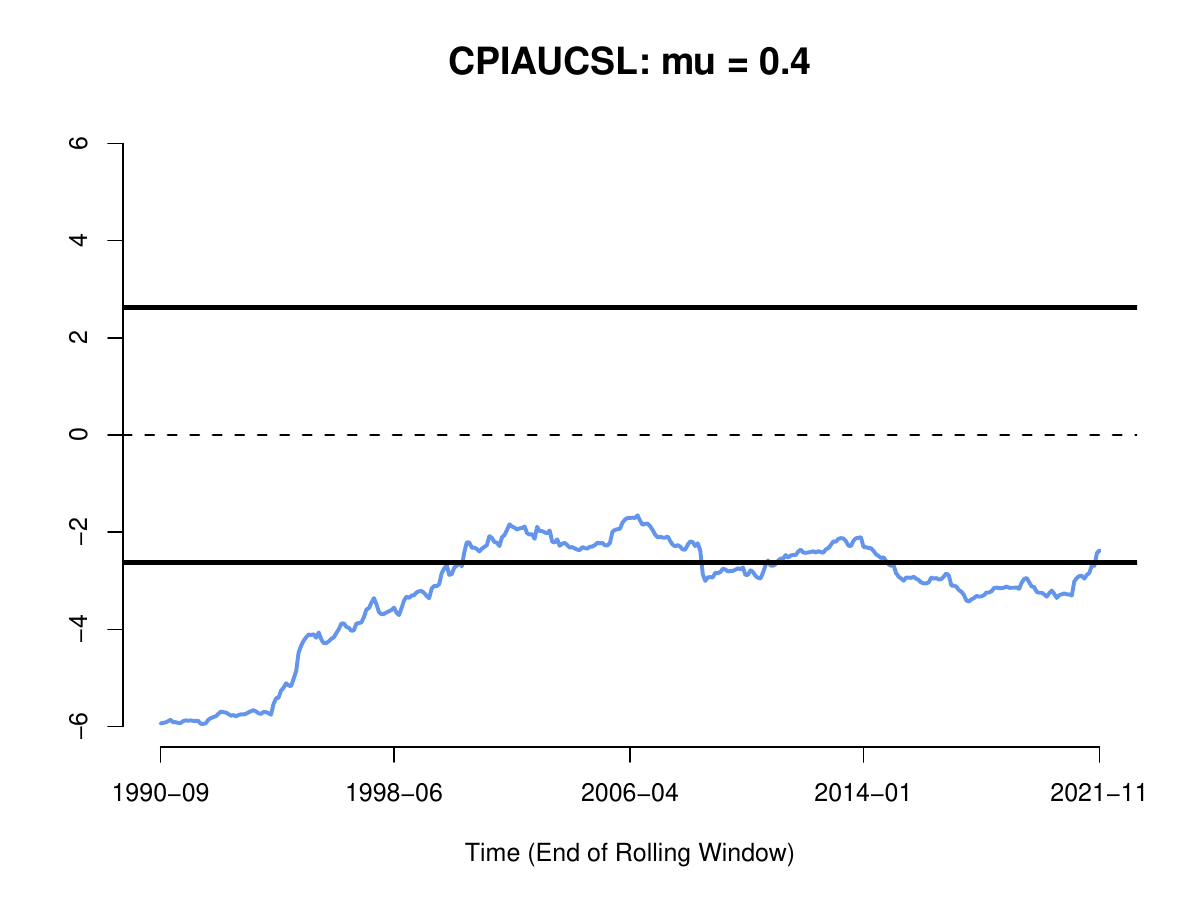}
\end{tabular}
\caption{Consumer prices index (all items): $\text{Err}_{it}(\text{Trunc}) - \text{Err}_{it}(\text{noTrunc}), \, T + 1 \le t \le n - 24$ (top left) and the corresponding fluctuation test statistics computed with $\mu \in \{0.2, 0.3, 0.4\}$ along with the two-sided critical values at the significance level $\alpha = 0.1$. When the fluctuation test statistic falls below the lower solid line, Trunc outperforms the noTrunc and vice versa. Here, we set $\wh r = 3$.}
\label{fig:fredmd:cpi:6}
\end{figure}

\begin{figure}[t!b!p!]
\centering
\begin{tabular}{cc}
\includegraphics[width = .45\textwidth]{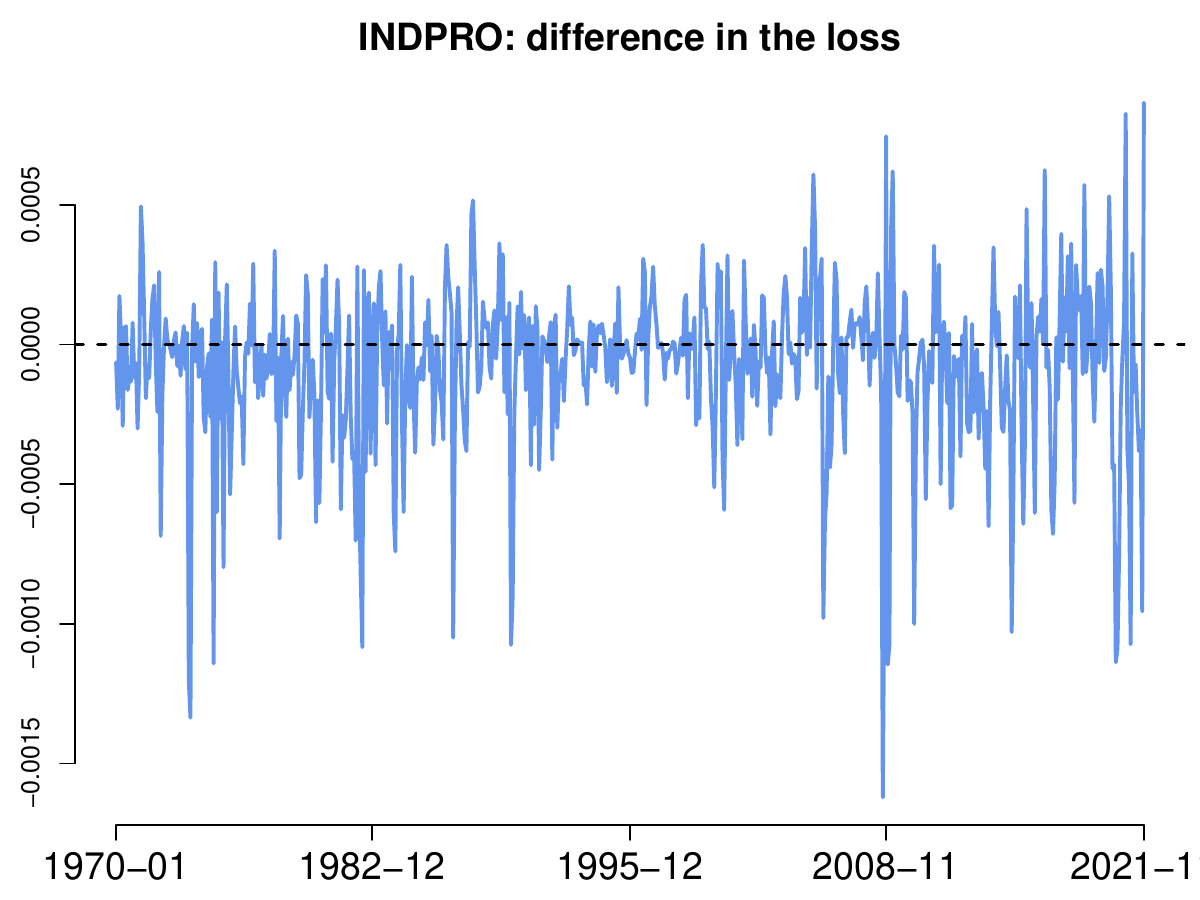} &
\includegraphics[width = .45\textwidth]{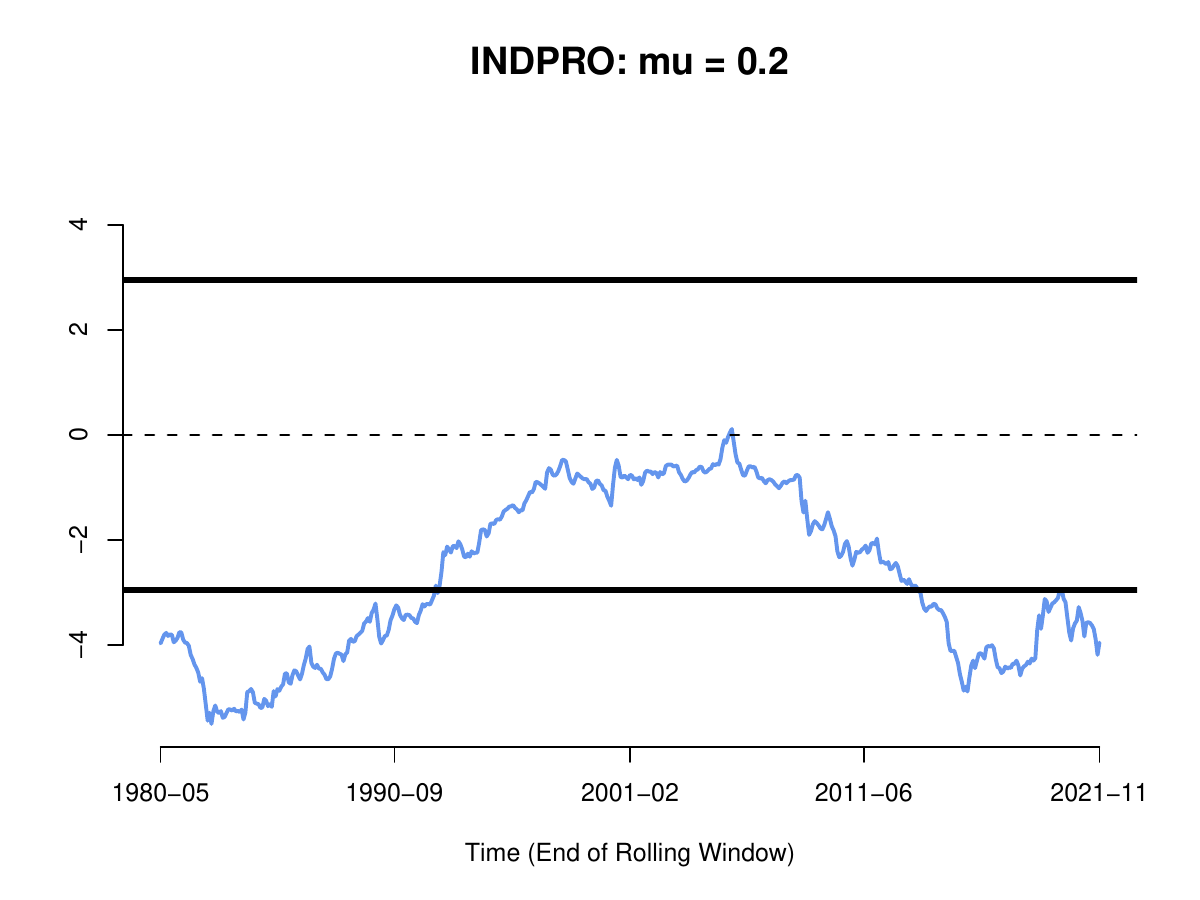} \\
\includegraphics[width = .45\textwidth]{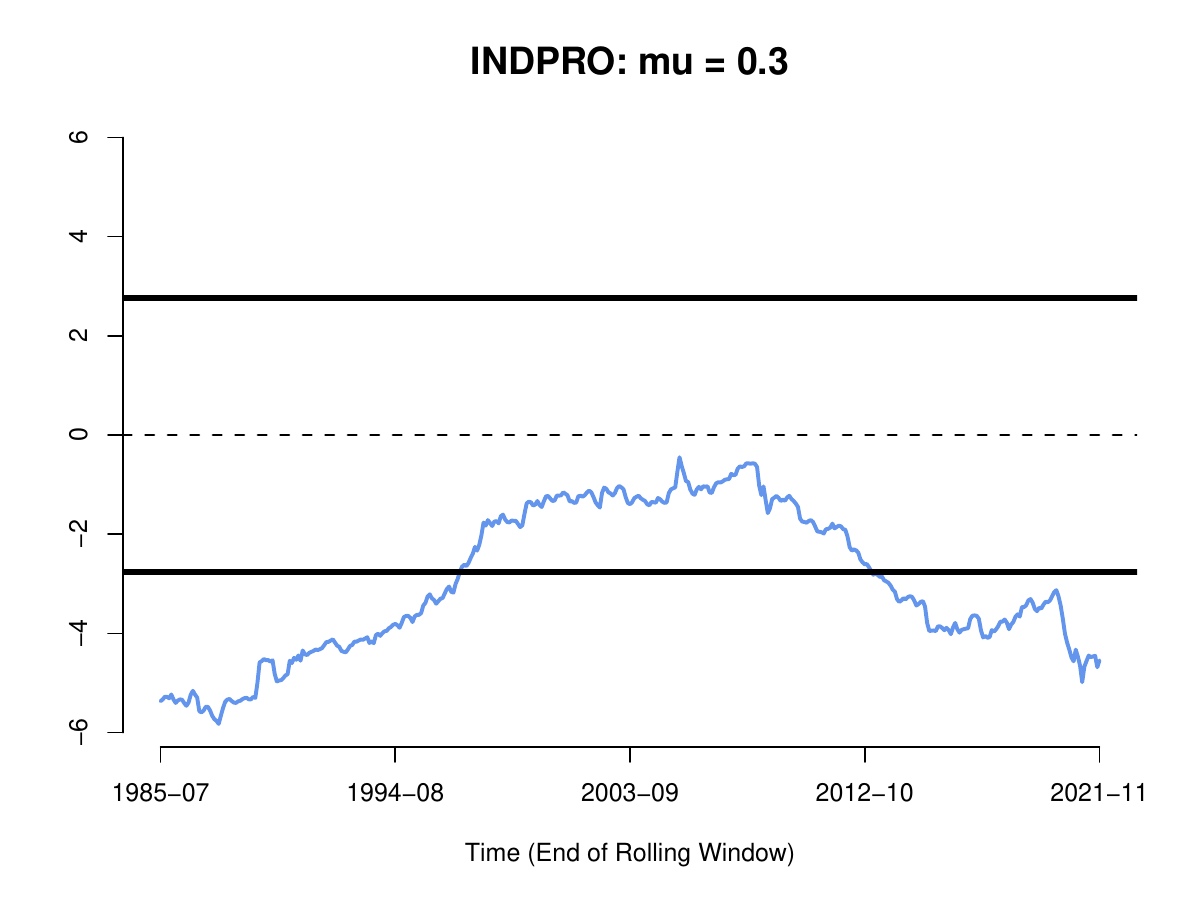} &
\includegraphics[width = .45\textwidth]{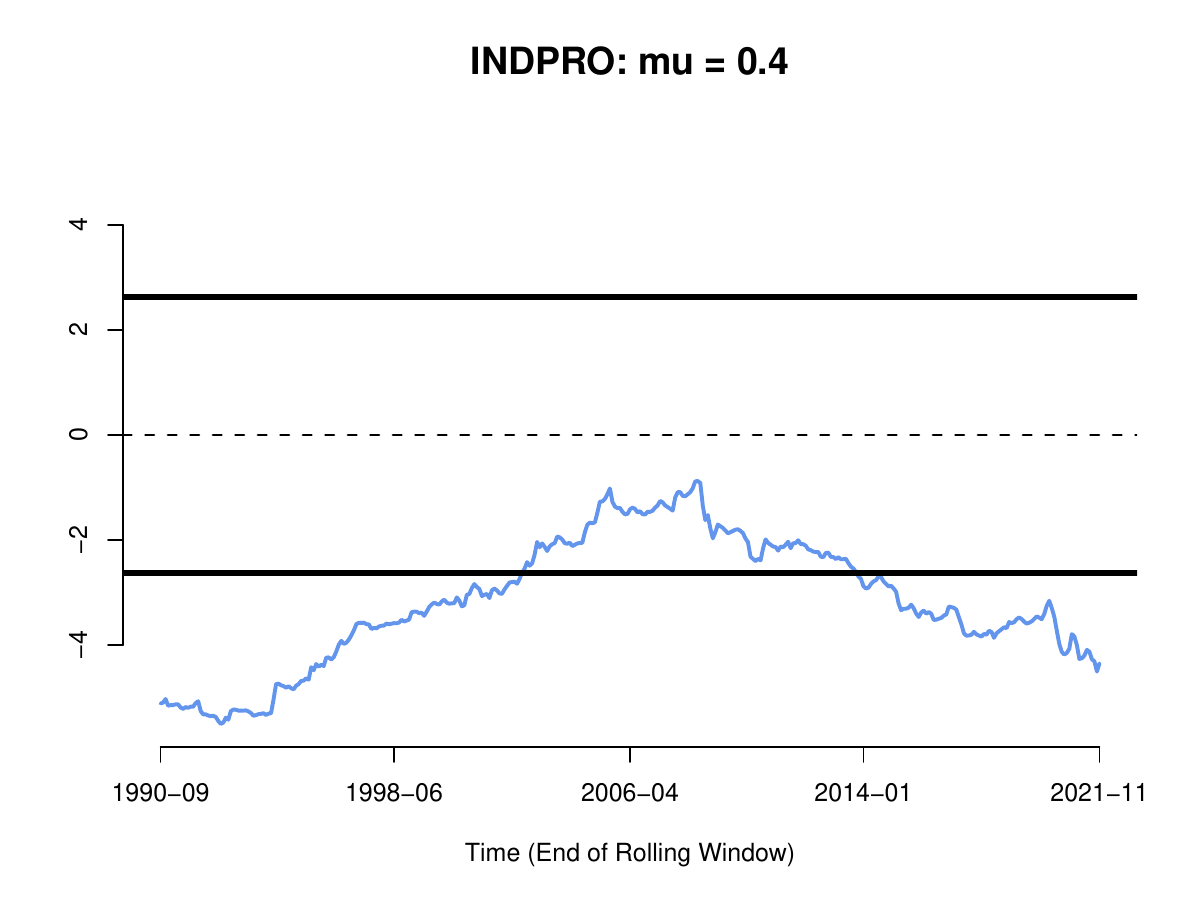}
\end{tabular}
\caption{Industrial production total index: $\text{Err}_{it}(\text{Trunc}) - \text{Err}_{it}(\text{noTrunc}), \, T + 1 \le t \le n - 24$ (top left) and the corresponding fluctuation test statistics computed with $\mu \in \{0.2, 0.3, 0.4\}$ along with the two-sided critical values at the significance level $\alpha = 0.1$. When the fluctuation test statistic falls below the lower solid line, Trunc outperforms the noTrunc and vice versa. Here, we set $\wh r = 5$.}
\label{fig:fredmd:indpro:5}
\end{figure}

\begin{figure}[t!b!p!]
\centering
\begin{tabular}{cc}
\includegraphics[width = .45\textwidth]{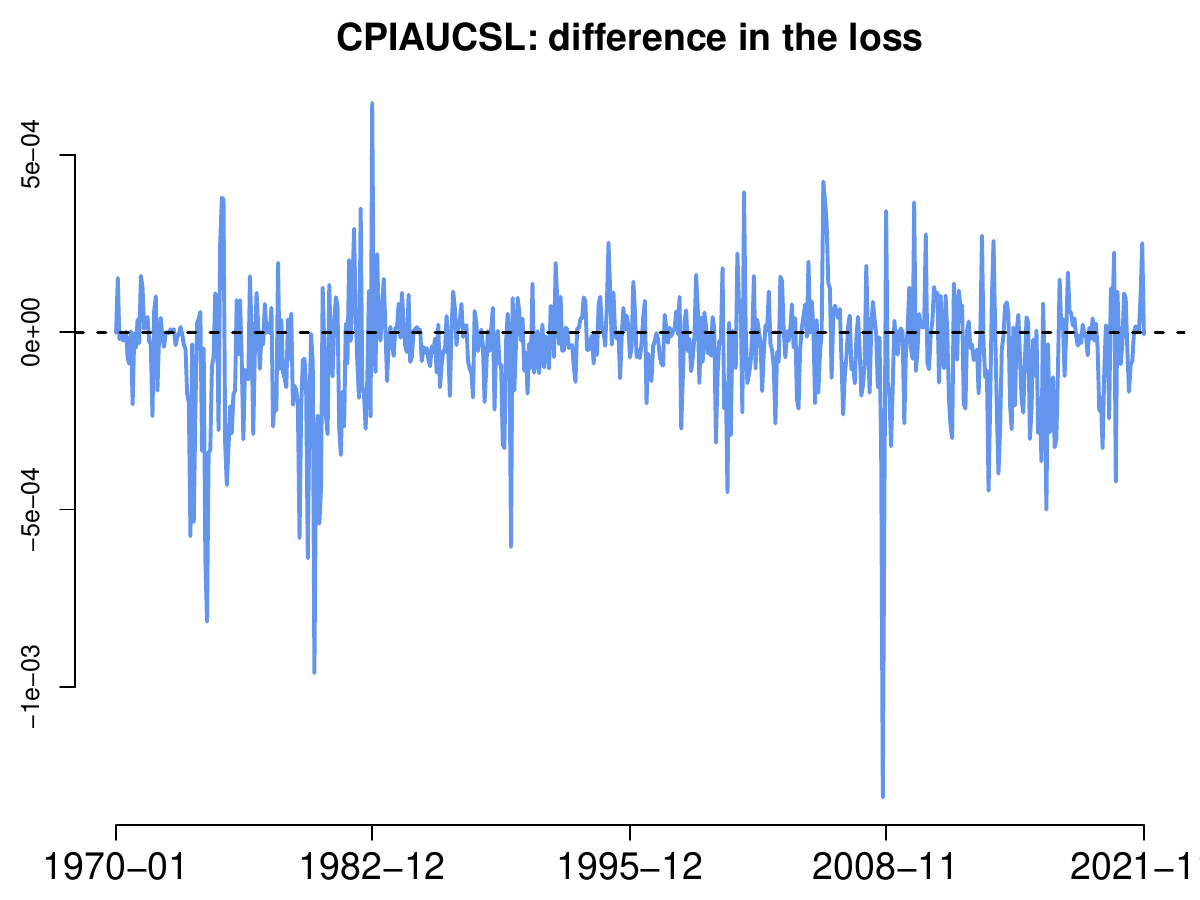} &
\includegraphics[width = .45\textwidth]{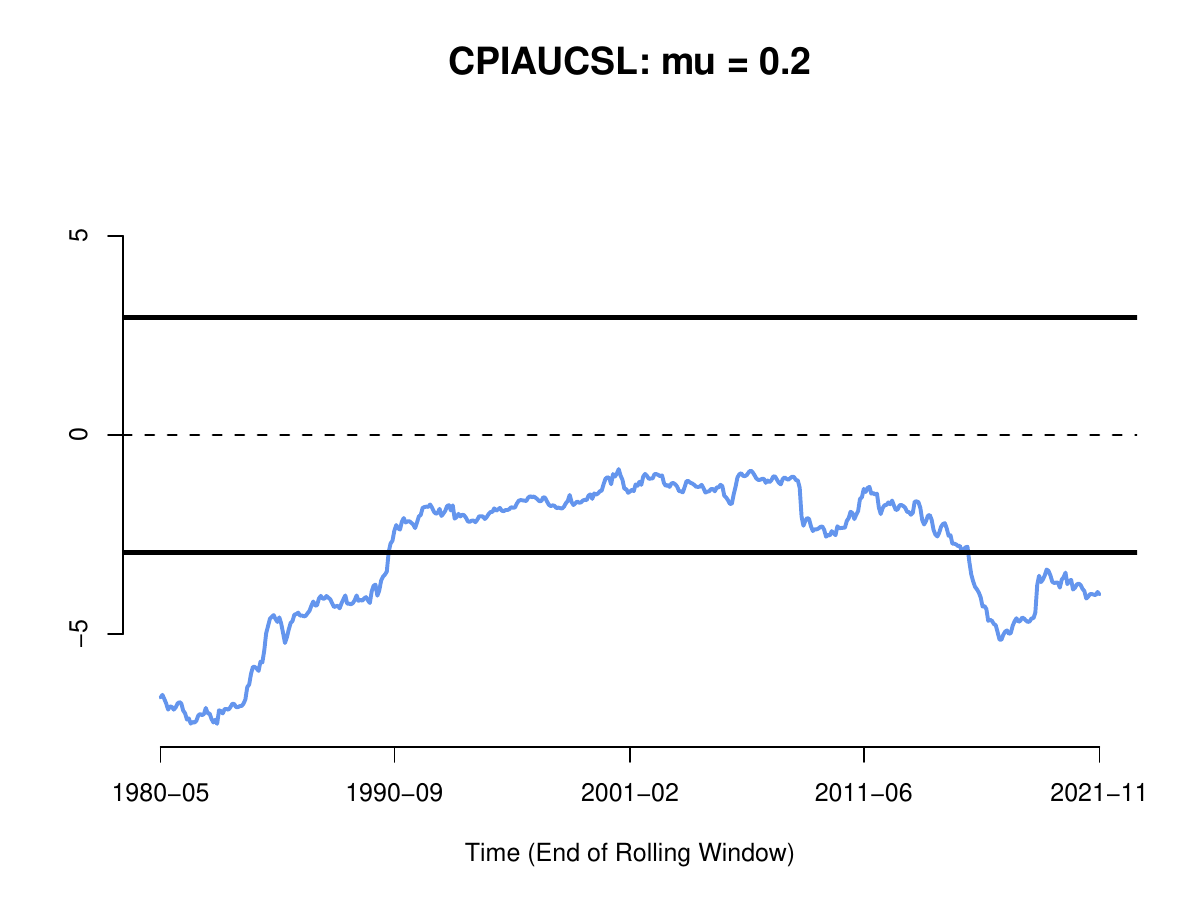} \\
\includegraphics[width = .45\textwidth]{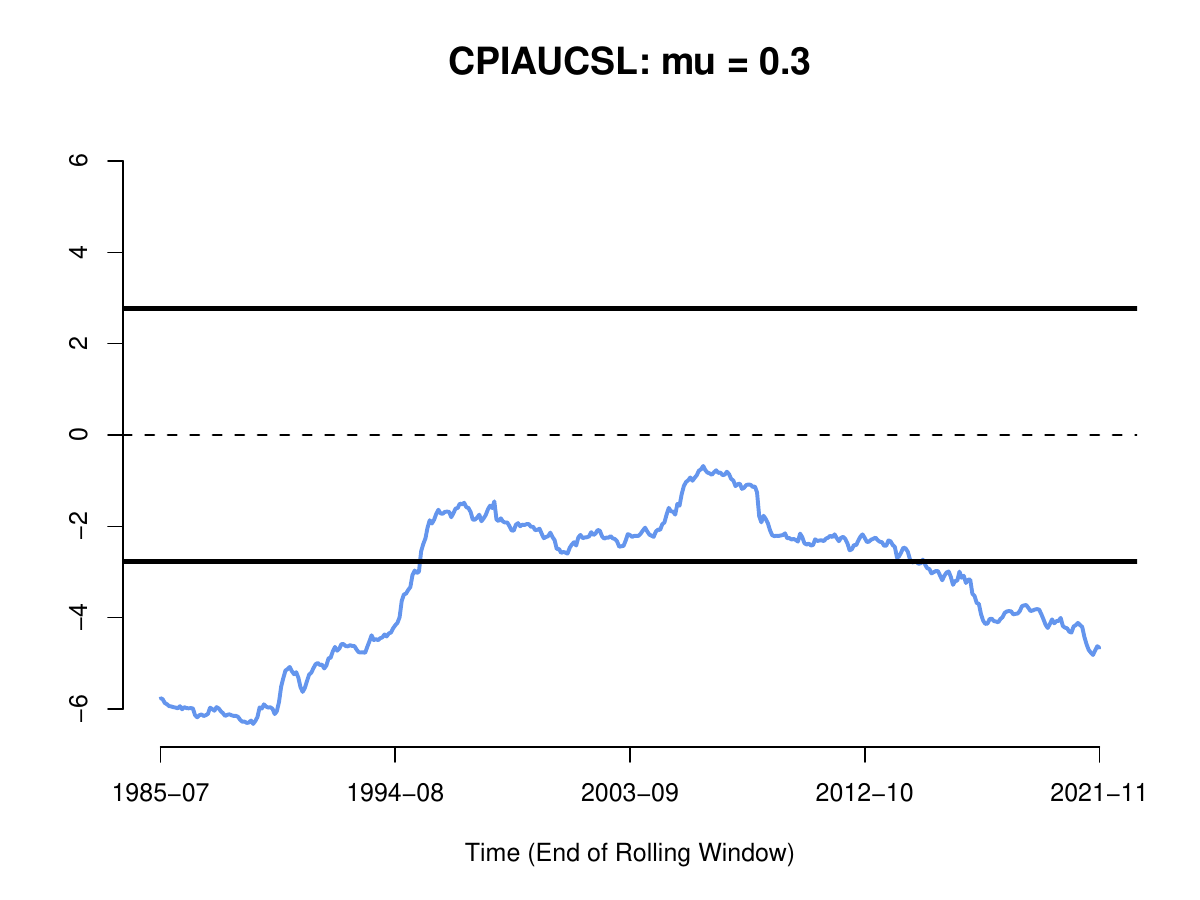} &
\includegraphics[width = .45\textwidth]{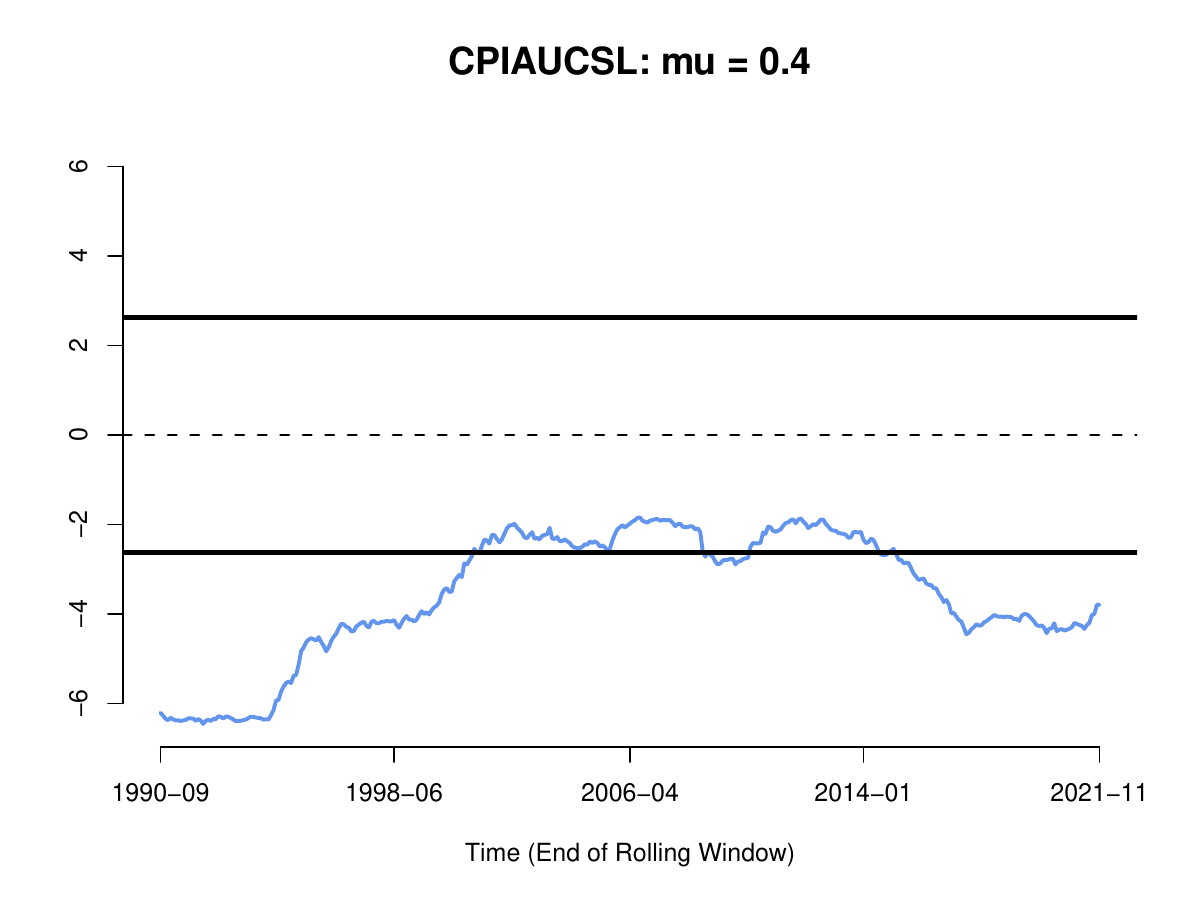}
\end{tabular}
\caption{Consumer prices index (all items): $\text{Err}_{it}(\text{Trunc}) - \text{Err}_{it}(\text{noTrunc}), \, T + 1 \le t \le n - 24$ (top left) and the corresponding fluctuation test statistics computed with $\mu \in \{0.2, 0.3, 0.4\}$ along with the two-sided critical values at the significance level $\alpha = 0.1$. When the fluctuation test statistic falls below the lower solid line, Trunc outperforms the noTrunc and vice versa. Here, we set $\wh r = 5$.}
\label{fig:fredmd:cpi:5}
\end{figure}

To further investigate whether the two methods perform significantly differently at any point during the forecasting exercise, we employ the fluctuation test proposed by \cite{giacomini2010forecast} (implemented in the R package \verb+murphydiagram+ \citep{murphydiagram}), 
and focus on two variables, the growth rate of the industrial production total index (INDPRO) and the difference of the consumer prices inflation index (all items, CPIAUCSL).
The test requires a tuning parameter $\mu \in (0, 1)$ which determines the bandwidth $\lceil \mu (n - 24 - T) \rceil$ over which the moving average of the forecasting error differences is produced, and it is recommended that $\mu \ge 0.2$; we consider $\mu \in \{0.2, 0.3, 0.4\}$.
Figures~\ref{fig:fredmd:indpro:6}--\ref{fig:fredmd:cpi:5} display the difference in the forecasting errors and the results from the fluctuation test with $\wh r \in \{3, 5\}$ and $\mu \in \{0.2, 0.3, 0.4\}$.
Regardless of the choice of $\wh r$ or~$\mu$, we observe that the two approaches perform significantly differently in forecasting over certain periods where Trunc is favoured over noTrunc. 
Such intervals correspond to where the estimation involves the data from the Great Financial Crisis (2007--09) and Covid-19 pandemic (2020--21). 
The raw difference in forecasting error shows that entering the crisis period, $\text{Err}_{it}(\text{Trunc}) - \text{Err}_{it}(\text{noTrunc})$ may initially increase as the forecasting target itself is highly anomalous, but it is followed by a sharp decrease indicating that the robust approach benefits from truncating such anomalous observations in estimation.

\end{document}